\documentclass[10pt,oneside,onecolumn]{article}  

\usepackage{cuted}

\usepackage[body={19.cm,25.cm}]{geometry}                		
\usepackage{graphicx}			
\usepackage{color}

\usepackage{amsmath,amsfonts,amssymb}
\usepackage{mathtools}
\usepackage{epstopdf}
\usepackage{fancyhdr}
\usepackage{empheq}
\usepackage{epstopdf}
\usepackage{pdfpages}

\usepackage{titlesec}
\usepackage{titletoc}
\renewcommand{\thesection}{\S\arabic{section}}
\renewcommand{\thesubsection}{\arabic{section}.\arabic{subsection}}
\titleformat{\section}{\fontfamily{put}\selectfont\large\bfseries}{\thesection.}{.5em}{}

\usepackage{enumitem}

\definecolor{xll}{rgb}{0.85,0.44,0.84}
\usepackage{hyperref}
\hypersetup{
  colorlinks=true,
  citecolor=brown,
  linkcolor=magenta,
  urlcolor=blue,
  urlcolor=cyan,
  pdfborder={0 0 0.5},
}

\DeclareMathAlphabet{\mathcal}{OMS}{cmsy}{m}{n}
\renewcommand{\d}{\mathrm{d}}

\renewcommand{\v}[1]{\mathbf{#1}}

\renewcommand{\rm}[1]{\mathrm{#1}}

\renewcommand{\today}{\number\month/\number\day/\number\year}

\numberwithin{equation}{section}

\usepackage{orcidlink} 

\usepackage{setspace}

\makeatletter
\renewcommand\@makefntext[1]{%
  \noindent\textcolor{gray}{\@thefnmark.\ #1}%
}
\makeatother

\usepackage{utopia}
\usepackage[T1]{fontenc}
\usepackage[utopia]{mathdesign}

\usepackage{authblk}
\usepackage{cite}

\usepackage{titletoc}
\titlecontents{section}
              [1.cm]
              {\normalsize\bfseries{\textcolor{magenta}{\thecontentslabel}}}
              {\normalsize\makebox[.25cm][l]{}}
              {}
              {\normalsize\titlerule*[.5pc]{$\cdot$}\contentspage
              \hspace*{1.cm}}

\titlecontents{subsection}
              [1.75cm]
              {\textcolor{magenta}{\small\thecontentslabel}}
              {\small\makebox[0.5cm][l]{}}
              {}
              {\small\titlerule*[.5pc]{$\cdot$}{\textcolor{gray}{\contentspage
              \hspace*{1.cm}}       }}  

\titleformat{\section}{\large\bfseries}{\thesection}{.5em}{}
\titleformat{\subsection}{\color{teal}\normalsize\bfseries}{\thesubsection}{.5em}{}

\usepackage{tikz,xcolor,hyperref}

\definecolor{lime}{HTML}{A6CE39}
\DeclareRobustCommand{\orcidicon}{
	\begin{tikzpicture}
	\draw[lime, fill=lime] (0,0) 
	circle [radius=0.16] 
	node[white] {{\fontfamily{qag}\selectfont \tiny ID}};
	\draw[white, fill=white] (-0.0625,0.095) 
	circle [radius=0.007];
	\end{tikzpicture}
	\hspace{-2mm}
}
\foreach \x in {A, ..., Z}{%
	\expandafter\xdef\csname orcid\x\endcsname{\noexpand\href{https://orcid.org/\csname orcidauthor\x\endcsname}{\noexpand\orcidicon}}
}

\usepackage{multicol}

\allowdisplaybreaks
\usepackage{stfloats}
\usepackage{microtype}
\usepackage{caption}
\DeclareCaptionFont{captionstretch}{\linespread{0.9}\selectfont}
\usepackage{balance}

\begin{document}

\title{\Large\bfseries Nucleon Short-Range Correlations and High-Momentum Dynamics:\\ Implications on the Equation of State of Dense Matter}

\author[1,2]{\normalsize Bao-Jun Cai\orcidlink{0000-0002-8150-1020}\thanks{\textcolor{gray}{bjcai@fudan.edu.cn}}}
\affil[1]{\small \it
Key Laboratory of Nuclear Physics and Ion-beam Application (MOE), Institute of Modern Physics, Fudan University, Shanghai 200433, China
}
\affil[2]{\small\it Shanghai Research Center for Theoretical Nuclear Physics, NSFC and Fudan University, Shanghai 200438, China}

\author[3]{\normalsize Bao-An Li\orcidlink{0000-0001-7997-4817}\thanks{\textcolor{gray}{Bao-An.Li$@$etamu.edu}}}
\affil[3]{\small\it Department of Physics and Astronomy, East Texas A\&M University, Commerce, TX 75429-3011, USA}

\affil[4]{\small\it College of Physics, East China Normal University, Shanghai 200241, China}

\author[1,2,4]{\normalsize Yu-Gang Ma\orcidlink{0000-0002-0233-9900}\thanks{\textcolor{gray}{mayugang$@$fudan.edu.cn}}}

\date{\small\today}
\maketitle

\vspace{-0.25cm}

\begin{abstract}

Nucleon short-range correlations (SRCs) and their associated high-momentum tails (HMTs) in the
single-nucleon momentum distribution $n_{\mathbf{k}} = n(k)$ have emerged as key manifestations of
strong, short-range dynamics in nuclear many-body systems. Despite substantial recent progress, our
understanding of these correlations, and their implications for finite nuclei, nuclear reactions, and
dense matter, remains incomplete and continues to evolve. In this review, we offer a necessarily
selective overview of several aspects of SRC physics that directly influence the Equation of State
(EOS) of dense matter, particularly in regimes of large isospin asymmetry and high baryon density.
We first summarize the empirical and theoretical features of the momentum distribution $n_{\mathbf{k}}
= n(k)$, including its isospin dependence, microscopic origins, and representative parameterizations.
Special emphasis is placed on the strong neutron--proton (np) tensor force at intermediate momenta,
which drives the dominance of correlated np pairs and enhances the minority-species HMT in
asymmetric nuclei and nuclear matter. We further discuss connections to nucleon effective masses,
quasi-deuteron components, and orbital entanglement entropy, providing a broader microscopic
foundation that links SRCs to single-particle and two-body structure.
We then examine how SRC-induced HMTs modify the EOS of asymmetric nuclear matter within both
non-relativistic and relativistic frameworks. The depletion of low-momentum states and the repopulation
of high-momentum components alter kinetic and potential contributions to the EOS. We additionally
consider generalizations to arbitrary spatial dimensions and estimates involving very high-momentum
components, which help clarify the sensitivity of the EOS to the detailed structure of $n(k)$. Particular
attention is devoted to the softening of the kinetic symmetry energy and to deviations from the standard
parabolic approximation of isospin-asymmetric nuclear matter EOS, effects that grow increasingly important with isospin asymmetry.
In the context of heavy-ion reactions, we summarize the influence of SRCs on isospin-sensitive
observables including particle yields, nuclear collective flows, and neutron-proton bremsstrahlung
gamma rays. These effects arise from both modified initial momentum distributions and the increased
availability of high relative-momentum np pairs, which can strongly affect threshold behavior and
transport dynamics. We also briefly comment on experimental probes of high-momentum nucleon components,
including electron- and proton-induced knockout reactions and meson production channels in heavy-ion reactions.
Finally, we discuss implications for neutron-star matter, wherein extreme densities and large isospin
asymmetries amplify many SRC-induced effects known from finite nuclei. Topics include consequences
for mass--radius relations, tidal deformabilities, proton fractions, Migdal--Luttinger $Z$-factors, cooling
processes, and the core-crust transition. We also highlight potential connections between
SRC-modified nucleon momentum distributions and dark-matter interactions in dense astrophysical
environments.

\vspace{0.25cm}



\end{abstract}

{
\begin{spacing}{1.}
\tableofcontents
\end{spacing}
}

\vspace{1cm}

{\large\it Notations of main quantities used in this review:}

\begin{enumerate}[label=\null,itemsep=1.pt]
\item FFG: free Fermi gas
\item HMT: high-momentum tail
\item SRC: short-range correlation
\item NS: neutron star
\item TOV: Tolman--Oppenheimer--Volkoff
\item $\rho_J$: number density of nucleon $J=\rm{n,p}$
\item $\rho=\rho_{\rm n}+\rho_{\rm p}$: total number density of ANM
\item $\delta=(\rho_{\rm n}-\rho_{\rm p})/(\rho_{\rm n}+\rho_{\rm p})$: isospin asymmetry between neutrons and protons
\item $k_{\rm F}=(3\pi^2\rho/2)^{1/3}$: nucleon Fermi momentum in SNM
\item $k_{\rm F}^J=k_{\rm F}(1+\tau_3^J\delta)^{1/3}$: Fermi momentum of nucleon $J$ with $\tau_3^{\rm{n}}=+1$ and $\tau_3^{\rm p}=-1$
\item $E(\rho,\delta)$: Equation of State (EOS) of asymmetric nuclear matter (ANM)
\item $E_0(\rho)\equiv E(\rho,0)$: EOS of symmetric nuclear matter (SNM)
\item $E_{\rm n}(\rho)\equiv E(\rho,1)$: EOS of pure neutron matter (PNM)
\item $E_{\rm{sym}}(\rho)$: nuclear symmetry energy
\item $E^{\rm{kin}}(\rho,\delta)$: kinetic EOS encapsulating the momentum distribution effects
\item $E_{\rm{sym}}^{\rm{kin}}(\rho)$: the kinetic part of the symmetry energy
\item $\Delta E^{\rm{kin}}_{\rm{np}}=\langle E_{\rm{p}}^{\rm{kin}}\rangle-\langle E_{\rm{n}}^{\rm{kin}}\rangle$: proton skin in $k$-space
\item $\Delta r_{\rm{np}}=\langle r_{\rm{n}}^2\rangle^{1/2}-\langle r_{\rm p}^2\rangle^{1/2}$: neutron skin in $r$-space
\item $\varepsilon(\rho,\delta)$: energy density as a function of density and isospin asymmetry
\item $P(\rho,\delta)$: pressure as a function of density and isospin asymmetry
    \item $a_2(A)$: probability of finding a high-momentum nucleon pair in nucleus $A$ relative to that in deuteron
    \item $n_{\v k}=n(k)$: single nucleon momentum distribution function (suppressing the $\rho$- and $\delta$-dependence)
    \item $C_J$: strength of the HMT of nucleon $J=\rm{n,p}$
    \item $\Delta_J$: depletion of the nucleon momentum distribution for $k<k_{\rm F}^J$
    \item $\kappa_J$: average depletion of the nucleon momentum distribution
    \item $\mu_J$: nucleon chemical potential
    \item $\phi_J$: high-momentum cutoff of the SRC in $n_{\v k}=n(k)$
    \item $x^{\rm{HMT}}_J=3C_J(1-\phi_J^{-1})$: high-momentum fraction of $J=\rm{n,p}$ in nucleus $A$
    \item $Z^J_{\rm F}$: Migdal--Luttinger jump of $n_{\v k}=n(k)$ at the Fermi momentum
    \item $x_{\rm p}=\rho_{\rm p}/\rho$: proton fraction in NSs
\end{enumerate}

\twocolumn

{
\section{Introduction}

\indent 

\indent 

The Equation of State (EOS) of dense nucleonic matter lies at the heart of contemporary nuclear physics and astrophysics\cite{Walecka1974,Collins1975,Chin1977,Freedman1977-1,Freedman1977-2,Freedman1977-3,Baluni1978,Wiringa1988,Akmal1998,Migdal1978,Morley1979,Shuryak1980,Bailin1984,Lattimer2001,Dan02,Steiner2005,LCK08,Alford2008,Watts2016,Ozel2016,Oertel2017,Vidana2018,Bur2021,Dri2021,Lov2022,Sor2024,Kumar2024,Baym2018,Bai2019,Ors2019,Li2019,Dex2021,Lattimer:2021emm,ChenJH24}. It governs the thermodynamic, transport, and microscopic properties of matter across a vast range of densities and isospin asymmetries, from terrestrial heavy-ion reactions\cite{LCK08,Sor2024} to the interiors of neutron stars (NSs)\cite{Shapiro1983,Haen2007,Baym2018,Bai2019,Alford2008,Watts2016,Ozel2016,Oertel2017,Vidana2018}. Despite decades of progress, major uncertainties persist, especially in the high-density behavior of EOS, owing to the complex interplay between short-range nuclear interactions, many-body correlations, and the nonperturbative nature of quantum chromodynamics (QCD) in the strongly coupled regime\cite{Bram14QCD}. Among these ingredients, short-range correlations (SRCs) and the associated high-momentum tail (HMT) in the single nucleon momentum distribution $n_{\v{k}}=n(k)$ have emerged as indispensable components in developing an accurate, microscopically grounded EOS for dense matter.

\renewcommand*\figurename{\small FIG.}
\begin{figure}[h!]
\centering
  \includegraphics[width=8.cm]{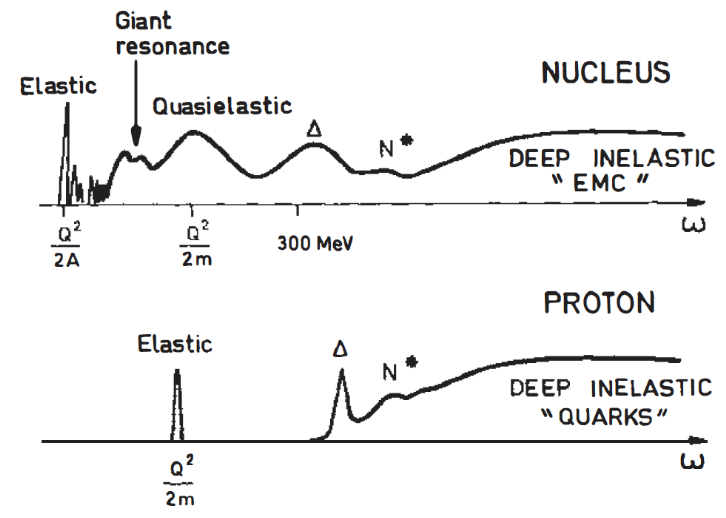}\\[0.25cm]
  \includegraphics[width=8.cm]{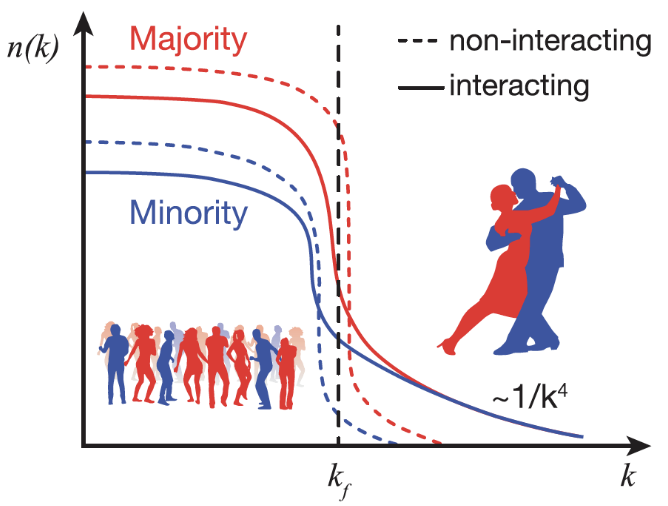}
\caption{(Color Online). Upper: schematic electron energy-loss spectrum, showing elastic scattering, excitation of discrete nuclear levels, excitation of giant resonances, quasi-elastic scattering, pion production and $\Delta(1232)$ formation, $\rm{N}^\ast$ resonance formation and the deep inelastic region. Figure taken from Ref.\cite{Frois1987}. Lower: sketch of the nucleon momentum distribution $n(k)$ in finite nuclei or in ANM. Majority nucleons dominate the $n(k)$ below the Fermi momentum $k_{\rm F}$, while a relatively larger fraction of minority nucleons reside in the HMT above $k_{\rm F}$ where neutron-proton pairs spatially close, having zero total momentum but large relative momentum are formed by SRCs. Figure taken from Ref.\cite{Hen14}.}
  \label{fig_QEfig}
\end{figure}

SRCs originate from the strong short-range repulsion and tensor components of the nucleon-nucleon interaction\cite{bethe}. They deplete the Fermi sea, generate a substantial population of high-momentum nucleons, and give rise to a characteristic power-law tail $n(k) \sim k^{-4}$ at large $k$\cite{Tan08-a,Tan08-b,Tan08-c}, as shown in the lower part of FIG.\,\ref{fig_QEfig}, and observed in nuclei\cite{Hen14,Sub08}, nuclear matter\cite{Ben93}, and even universal cold-atom systems\cite{Gio08RMP,Blo08RMP}.
In finite nuclei or asymmetric nuclear matter (ANM), majority nucleons dominate the occupation of $n(k)$ below the Fermi momentum $k_{\rm F}$, while above $k_{\rm F}$ in the HMT induced by SRCs, minority nucleons are preferentially populated relative to their respective total populations as it takes equal numbers of neutrons and protons to make SRC pairs\cite{Hen14}. This inversion of the usual occupation pattern is a distinctive feature of SRC-driven dynamics. Moreover, the correlated nucleon pairs responsible for the HMT are characterized by a large relative momentum while maintaining a comparatively small center-of-mass momentum, a hallmark of the tensor-dominated short-range interaction\cite{Hen14};
in other words, they are predominantly back-to-back, see FIG.\,\ref{fig_bb} for the sketch of such a configuration in finite nuclei\cite{Pias23}.
Because the EOS depends sensitively on the momentum distribution, even modest SRC-induced changes can significantly modify the kinetic contribution to the EOS and the symmetry energy\cite{Li15PRC}, the validity of the parabolic approximation of ANM EOS\cite{Bom91,Cai12PRC-S4,Seif14PRC-S4,Gonz17PRC-S4,Pu17PRC-S4}, the pressure of neutron-rich matter, and the composition of dense matter.
The isospin dependence of SRCs in ANM further leads to a proton-neutron fractional population imbalance in the HMT, influencing effective masses\cite{LCCX18}, single-particle properties\cite{Ding16PRC} and the overall isovector response. 
It is well known that nuclear force has a tensor component that is attractive at intermediate range due do pion exchange and a repulsive one at short distance due to $\rho$-meson exchange.  
The SRC-driven features, rooted largely in strong spin-isospin dependence of neutron-proton tensor correlations tied to in-medium pion and rho-meson dynamics\cite{Ericson1988}, also play a central role in phenomena such as the European Muon Collaboration (EMC) effect\cite{EMC83}. The latter refers to the finding that the structure function (quark momentum distribution function) in nucleons depends on whether the nucleon is in free-space or in a nuclear medium. In fact, it has been found that only the momentum distributions of quarks in SRC neutron-proton pairs are modified when a local temporal high-density is formed in nuclei\cite{Weinstein:2010rt}. Moreover, because a larger fraction of protons populate the HMT in neutron-rich nuclei, 
the structure function of protons are modified more compared to neutrons in neutron-richer nuclei\cite{CLAS:2019vsb}. 

The richness of the nuclear medium effects can be revealed by not only the traditional electron-nucleus and proton-nucleus scatterings\cite{Boffi96} (upper panel of FIG.\,\ref{fig_QEfig}) but also nucleus-nucleus scatterings (some involving inverse kinematics in reactions with radioactive beams) over a wide range of energies\cite{Frois1987}. These experiments have largely verified the approximately $k^{-4}$ behavior of HMT with momentum up to about $2k_{\rm F}$\cite{Hen14}. Further investigating the microscopic origin and properties of SRCs, their connections with EMC effects as well as quantifying their broad impact in nuclear physics and astrophysics, e.g., nuclear EOS, therefore remain key goals of several future experimental programs\cite{Hauenstein02012021,Hen:2025rlk}, see, e.g., SRC proposals\cite{Tu:2020ymk,Hauenstein:2021zql} at the electron-ion collider (EIC) at BNL in the US\cite{Boer:2011fh}, those by the R$^3$B Collaboration\cite{SRC-r3b,r3b} at GSI-FAIR (Facility for Anti-proton and Ion Research in Europe) in Germany\cite{myref}, as well as plans at HIAF (High-Intensity Heavy-Ion Accelerator Facility) and EicC (Electron-Ion Collider in China)\cite{Ye24,EicC-ref}. Moreover, ongoing SRC experiments at facilities like the Jefferson Lab (JLab), GSI, JINR (Joint Institute for Nuclear Research) and IMP/Lanzhou (Institute of Modern Physics) are continuously producing excitingly new results, see, e.g., Refs.\cite{Kahlbow:2023mtc,2023EPJA...59..188A,2023EPJA...59..205F,JHXu25PRR}.

\begin{figure}[h!]
\centering
  \includegraphics[width=8.cm]{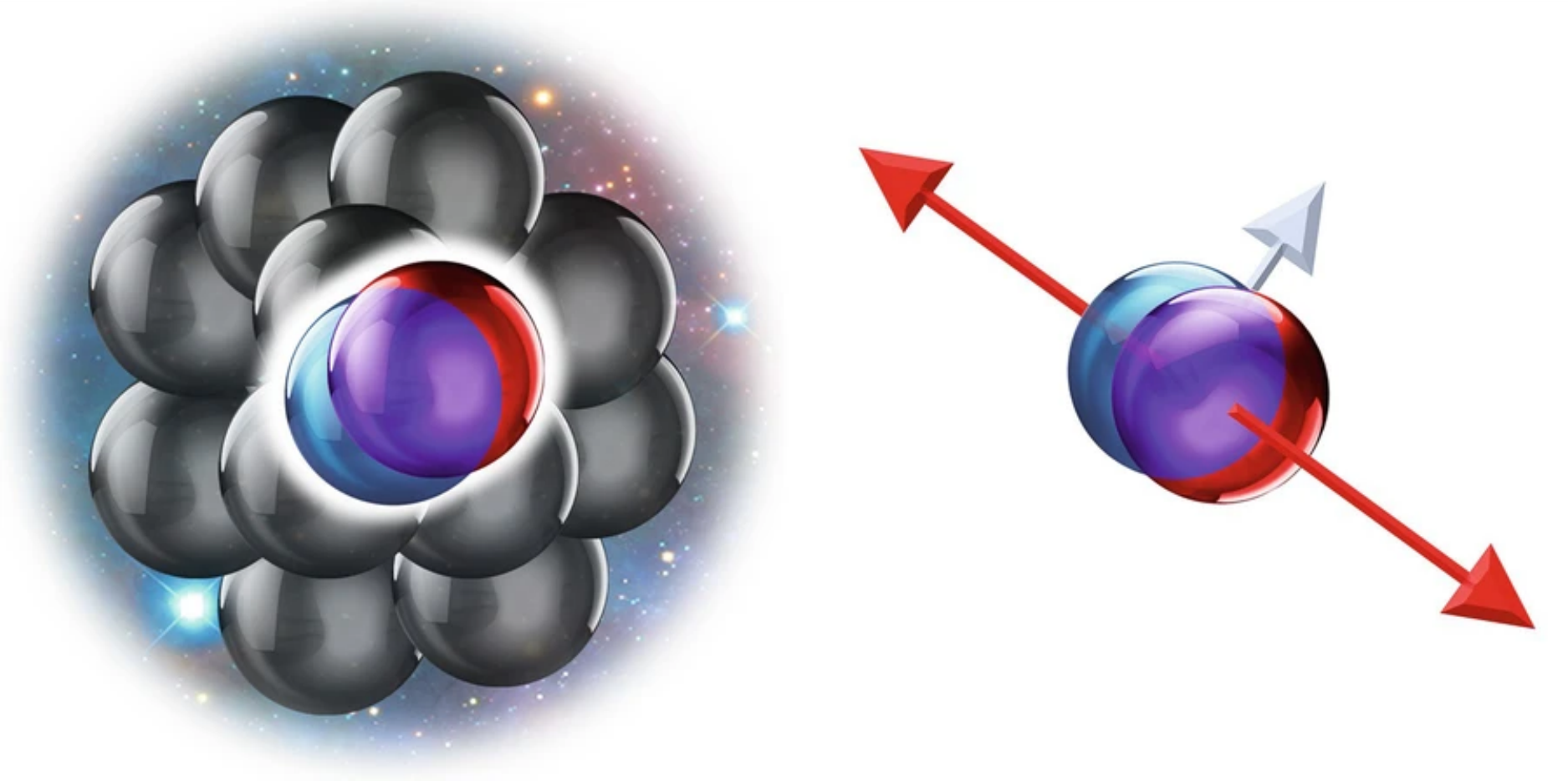}
\caption{(Color Online). Back-to-back configuration of a SRC pair in finite nuclei. Figure taken from Ref.\cite{Pias23}.}
  \label{fig_bb}
\end{figure}

The relevance of the EOS extends far beyond nuclear structure. In heavy-ion reactions\cite{LCCX18}, where high densities, large isospin asymmetries, and transient nonequilibrium conditions are created, the EOS dictates collective flows, particle production thresholds, meson yield ratios, bremsstrahlung signals, and sub-threshold production phenomena. SRC-induced HMTs modify the initial nucleon momentum distributions and reaction dynamics, leading to observable consequences such as enhanced high-energy particle emission, altered isospin-sensitive observables, and novel coexistence patterns between coordinate-space neutron skins and momentum-space proton skins\cite{Cai16b}. These observables are actively used to constrain the symmetry energy near and above saturation density\cite{LCK08,Xiao09,LiX25,CaiLi21Aux,WangSP24,WangSP25,YeJT25}, making SRC physics a vital link between microscopic nuclear dynamics and experimental probes of the EOS\cite{Li2019}.
In astrophysics, the EOS of superdense matter under strong-field gravity governs nearly every macroscopic property of NSs\cite{Vidana2018,Baym2018,CaiLi25-IPADTOV}. SRC-induced changes in the nucleon momentum distribution influence the pressure-energy density relation, the proton fraction, Urca thresholds, superfluid gaps, core-crust transition densities, and the emergence of complex nuclear structures in low-density regions. Consequently, they affect the mass-radius (M-R) relation, tidal deformabilities, cooling curves, neutrino opacities, and the thermal and compositional evolution of NSs\cite{Yak01}. The discovery of massive NSs\cite{Riley19,Miller19,Riley21,Miller21,Fon21,Choud24,Reard24,Ditt24,Salmi22,Salmi24,Vin24}, together with multimessenger observations from gravitational-wave detections\cite{Abbott2017,Abbott2018,Abbott2020-a,Abbott2020}, has elevated the importance of understanding dense-matter EOS effects with unprecedented precision. SRC-HMT physics enriches this picture by providing a microscopic mechanism that tightens the connection between terrestrial experiments, modern nuclear many-body theories, and astrophysical observables of dense matter.

This review aims to provide a coherent examination of how SRC-induced nucleon high-momentum structures in $n(k)$ modify the EOS of dense matter and how these modifications propagate into nuclear reactions and astrophysical systems. Section \ref{SEC_nk} lays the microscopic foundation by discussing the structure, strength, and isospin dependence of $n(k)$, including its effects on nucleon effective masses, nuclear energy density functionals, and entanglement properties. Section \ref{SEC_EOS} focuses on how the SRC-HMT affects the EOS in relativistic and non-relativistic frameworks, examines kinetic contributions to the symmetry energy, and generalizes these concepts to arbitrary dimensions. Section \ref{SEC_HIC} explores the broad impact of SRC-HMT physics on isospin-sensitive observables in heavy-ion collisions and scattering experiments. Section \ref{SEC_NS} examines NS properties under the influence of SRC-modified EOSs, covering macroscopic structure, cooling, composition, and the potential presence of exotic components. Finally, Section \ref{SEC_OUTLOOK} provides a broader perspective and outlines potential problems for future exploration.

Through this unified treatment, the review highlights how SRCs and the HMT in $n(k)$ serve as a microscopic bridge connecting the EOS of dense matter to its manifestations in both laboratory experiments and NSs. Understanding these connections is essential for constraining the behavior of strongly interacting matter under extreme conditions and for advancing the global effort to determine the dense-matter EOS with high fidelity.
As this review addresses only selected aspects related to the EOS of dense matter,  it is necessarily incomplete and limited; for other studies on SRCs and HMTs, we refer the reader to comprehensive reviews, e.g., Refs.\cite{Hen17RMP,Dick04,Fa17,Arr12,Arr21,Arr22,Dal22,Bur23,Aum21,Fra81,Fra88,Att15,Fra08,fomin2026}. 
In addition, this review is intended to be introductory in nature. We include essential formulas along with detailed and pedagogical explanations, with the hope that it may serve as a helpful starting point for newcomers to this exciting field.

\section{Nucleon Momentum Distribution $n_{\v k}=n(k)$ with a HMT in Nuclear Matter}\label{SEC_nk}

\indent

In this section, we review several foundational aspects of the single-nucleon momentum distribution function $n_{\mathbf{k}} = n(k)$. Subsection \ref{sub_EOSANM} outlines the basic definitions of the characteristic quantities of the EOS of  ANM. Subsection \ref{sub_nk} discusses $n_{\mathbf{k}}$ in detail, including its functional form, the strength of the nuclear contact, and its isospin structure from both theoretical and experimental perspectives. Subsection \ref{sub_Emass} examines the nucleon E-effective mass, with emphasis on the $Z$-factor and its physical implications. Subsection \ref{sub_quasideuteron} briefly introduces the quasi-deuteron picture used to study SRCs within relativistic energy density functional approaches. Finally, Subsection \ref{sub_entropy} presents the orbital entanglement entropy associated with SRCs and highlights its connection to the nuclear contact.

\subsection{Characteristics of EOS of ANM at Zero Temperature}\label{sub_EOSANM}

\indent 

The EOS of ANM at zero temperature is defined as the energy per nucleon and is written as $E(\rho,\delta)$, where $\rho=\rho_{\rm{n}}+\rho_{\rm{p}}$ is the total nucleon density and $\delta\equiv(\rho_{\rm{n}}-\rho_{\rm{p}})/\rho$ is the isospin asymmetry associated with the neutron and proton densities $\rho_{\rm{n}}$ and $\rho_{\rm{p}}$. Since $-1\leq\delta\leq1$, the EOS of ANM can be naturally expanded around $\delta=0$, and isospin (neutron-proton exchange) symmetry ensures that no odd powers of $\delta$ appear\cite{LCK08,LCCX18},
\begin{equation}
E(\rho,\delta)\approx E_0(\rho)+E_{\rm{sym}}(\rho)\delta^2+E_{\rm{sym,4}}(\rho)\delta^4+\cdots,
\end{equation}
where $E_0(\rho)\equiv E(\rho,0)$ is the EOS of symmetric nucleonic matter (SNM). The functions $E_{\rm{sym}}(\rho)$ and $E_{\rm{sym},4}(\rho)$, the second- and fourth-order symmetry energies, are defined by
\begin{equation}
E_{\rm{sym}}(\rho)\equiv\left.\frac{1}{2}\frac{\partial^2E(\rho,\delta)}{\partial\delta^2}\right|_{\delta=0},~
E_{\rm{sym},4}(\rho)\equiv\left.\frac{1}{24}\frac{\partial^4E(\rho,\delta)}{\partial\delta^4}\right|_{\delta=0},
\end{equation}
and represent the corresponding coefficients in the Taylor expansion of $E(\rho,\delta)$ over $\delta$.

The EOS of SNM, the symmetry energy, and the fourth-order symmetry energy may in turn be expanded around a reference density $\rho_{\rm{rf}}$, commonly chosen as the saturation density $\rho_0$, giving\cite{LCK08,LCCX18}
\begin{align}
E_0(\rho)\approx& E_0(\rho_0)+\frac{K_0}{2!}\chi^2+\frac{J_0}{3!}\chi^3+\frac{I_0}{4!}\chi^4+\mathcal{O}(\chi^5),\label{def_E0_exp}\\
E_{\rm{sym}}(\rho)\approx& E_{\rm{sym}}(\rho_0)+L\chi+\frac{K_{\rm{sym}}}{2!}\chi^2+\frac{J_{\rm{sym}}}{3!}\chi^3\notag\\
&+\frac{I_{\rm{sym}}}{4!}\chi^4+\mathcal{O}(\chi^5),\label{def_Esym_exp}\\
E_{\rm{sym},4}(\rho)\approx& E_{\rm{sym,4}}(\rho_0)+L_{\rm{sym,4}}\chi+
\frac{K_{\rm{sym,4}}}{2!}\chi^2+\frac{J_{\rm{sym,4}}}{3!}\chi^3\notag\\
&+\frac{I_{\rm{sym,4}}}{4!}\chi^4+\mathcal{O}(\chi^5),\label{def_Esym4_exp}
\end{align}
where the dimensionless expansion variables are $
\chi\equiv\chi_0=({\rho-\rho_{0}})/{3\rho_{0}}$ and $
\chi_{\rm{rf}}=({\rho-\rho_{\rm{rf}}})/{3\rho_{\rm{rf}}}$.
The coefficients appearing in Eq.\,(\ref{def_Esym_exp}) are
\begin{align}
K_{\rm{sym}}
&=\left.9\rho_0^2\frac{\partial^2E_{\rm{sym}}(\rho)}{\partial\rho^2}\right|_{\rho=\rho_0},~
J_{\rm{sym}}
=\left.27\rho_0^3\frac{\partial^3E_{\rm{sym}}(\rho)}{\partial\rho^3}\right|_{\rho=\rho_0},
\end{align}
and analogous definitions apply to $L$, $I_{\rm{sym}}$, $K_0$, $J_0$, $I_0$, $K_{\rm{sym},4}$, $J_{\rm{sym},4}$ and $I_{\rm{sym},4}$.
For nucleon in ANM, the free Fermi gas (FFG) model single-nucleon momentum distribution function takes the step function $n_{\v{k}}^J(\rho,\delta)=\Theta(k_{\rm{F}}^J-|\v{k}|)$, where $J=\rm{n,p}$. Consequently the relation between the nucleon density $\rho_J$ and the corresponding Fermi momentum $k_{\rm{F}}^J$ could be obtained,
\begin{equation}
k_{\rm{F}}^J=k_{\rm{F}}\left(1+\tau_3^J\delta\right)^{1/3},
\end{equation}
where $\tau_3^{\rm{n}}=+1$ and $\tau_3^{\rm{p}}=-1$.
For a pure neutron matter (PNM), $\delta=1$.

\subsection{Form, Strength and the Isospin Structure of $n_{\v k}=n(k)$}\label{sub_nk}

\indent 

In ANM, the single-nucleon momentum distribution $n_{\mathbf{k}}^{J}$ exhibits a HMT above the Fermi momentum $k_{\rm F}^J$ and a corresponding depletion below it, originating from the repulsive short-range core (or the SRCs) and the tensor force in nuclear interactions\cite{Migdal57,bethe,Pan92,Pan99}, as sketched in FIG.\,\ref{fig_Hen17}\cite{Hen17RMP}.
Over the past two decades, extensive theoretical and experimental studies have quantified the nucleon HMT in ANM\cite{Hen15a,Hen14,Egi06,Shn07,Wei11,Kor14,Fa17,Pia06,Wei15}. As an illustration, the upper panel of FIG.\,\ref{fig_Benh08} presents calculated momentum distributions for $^{2}\rm{H}$, $^{4}\rm{He}$, $^{16}\rm{O}$, and (symmetric) nuclear matter\cite{Ben93}, where the distributions $n_{\mathbf{k}}^{0}$ (with the superscript ``0'' denoting SNM) for the light nuclei were obtained using the Argonne $v(ij)$ and model VII $v(ijk)$, while those for SNM\cite{Fan84} were calculated using the Urbana $v(ij)$ supplemented with density-dependent terms that effectively incorporate $v(ijk)$. A key feature of all these results is the universal scaling of the HMT, highlighting its common short-range origin in all nuclear systems.
As another example, the $n(k)$ for $^4\rm{He}$ nucleus using the tensor-optimized high-momentum antisymmetrized molecular dynamics (TO-HMAMD) is shown in the lower panel of FIG.\,\ref{fig_Benh08}.
In their calculation\cite{Lyu20PLB}, they found that the correlated terms contribute to about 23\% of the total wave function. Furthermore, the contribution of tensor correlation term is about 12\%, which is roughly double as compared to the central contribution; the many-body contribution of about 4\% is comparably small.
In particular, Ref.\cite{Lyu20PLB} found that the high-momentum components ($k >2\,\rm{fm}^{-1}$) of $^4\rm{He}$ nucleus are mostly contributed by the tensor and the short-range correlated terms, where the tensor correlation dominates around $k \approx2\,\rm{fm}^{-1}$ and the short-range repulsion dominates around $k \approx4\,\rm{fm}^{-1}$, as denoted by the blue and yellow regions in the figure, respectively.

\begin{figure}[h!]
\centering
  \includegraphics[width=8.cm]{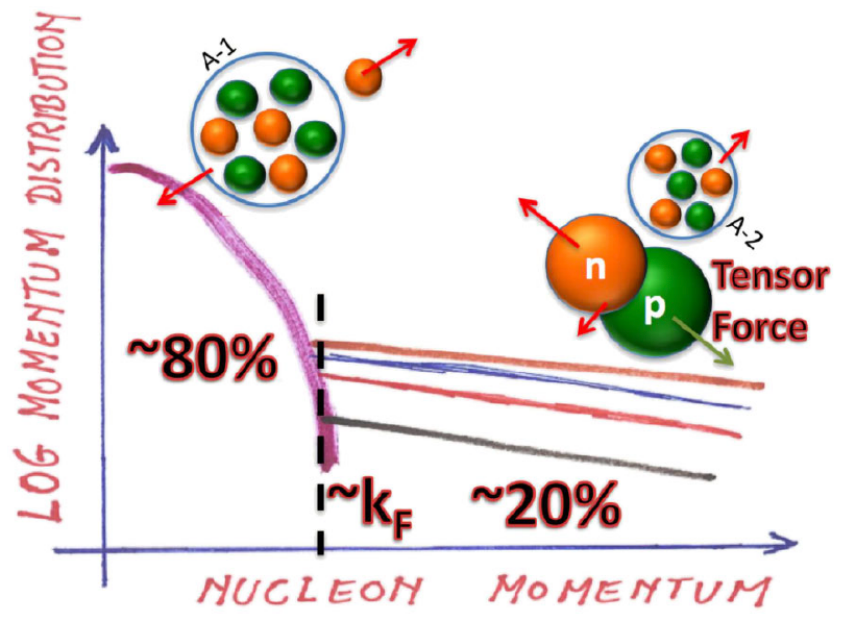}
  \caption{(Color Online). A cartoon on the $n_{\v{k}}$ in finite nuclei, figure taken from Ref.\cite{Hen17RMP}. }
  \label{fig_Hen17}
\end{figure}

\begin{figure}[h!]
\centering
  \includegraphics[width=7.5cm]{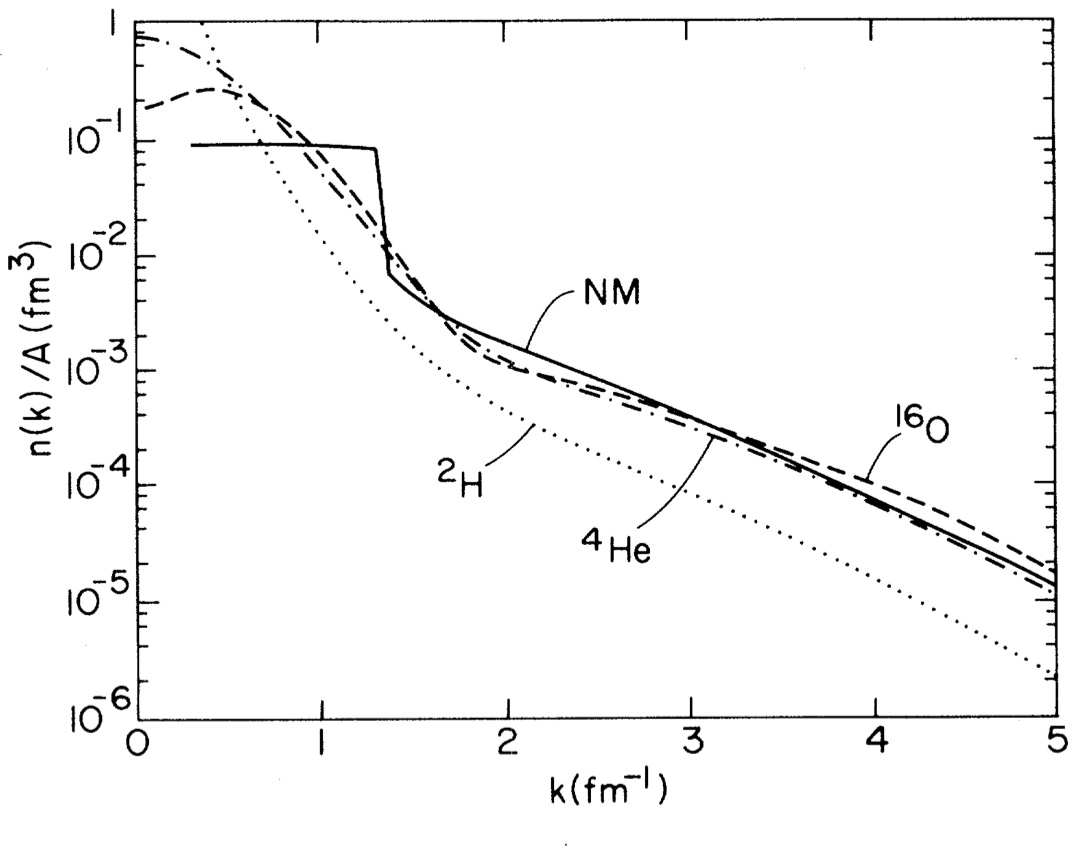}\\
\hspace{-0.3cm}
  \includegraphics[width=7.6cm]{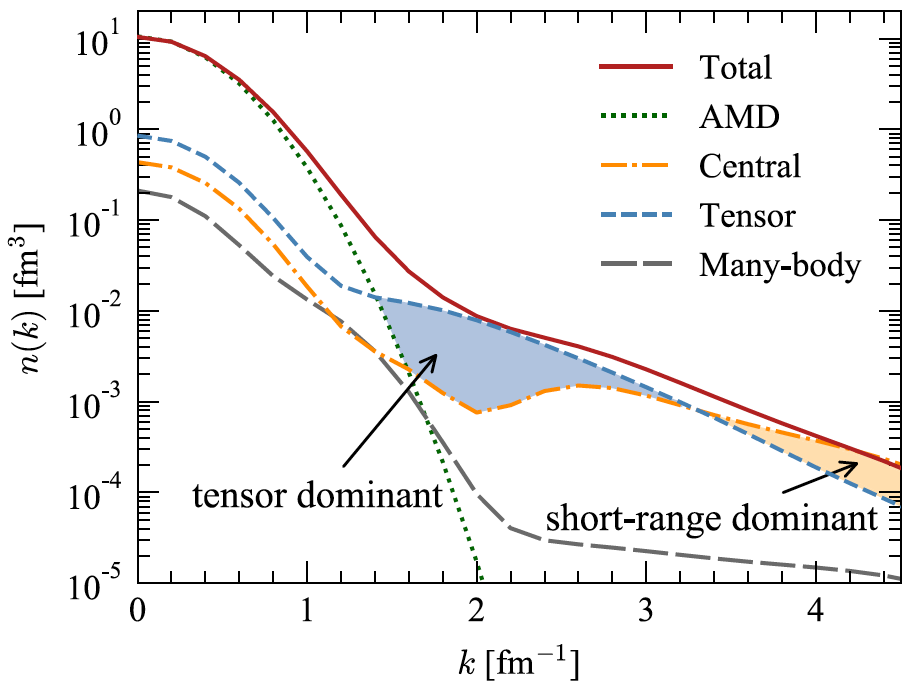}
  \caption{(Color Online). Upper: the nucleon momentum distribution in $^{2}\rm{H}$, $^4{\rm{He}}$, $^{16}\rm{O}$ and that in infinite SNM; figure taken from Ref.\cite{Ben93}.
  Lower: the component of $n(k)$ for $^4\rm{He}$ obtained from the tensor-optimized high-momentum antisymmetrized molecular dynamics (TO-HMAMD). Figures taken from Ref.\cite{Lyu20PLB}.}
  \label{fig_Benh08}
\end{figure}

The characteristic SRC length scale in finite nuclei is roughly about $1\,\mathrm{fm}$ (the size of nucleon). 
In FIG.\,\ref{fig_Torr}, we show an example of nucleon--nucleon correlation functions 
$F\sim\rho(\v{x},\v{x}')/\rho(\v{x})\rho(\v{x}')$ from a microscopic calculation\cite{Torr18PLB}. 
The Fourier transform of the two-body density gives the single-nucleon momentum distribution, 
$n_{\v k}\sim\int\d\v{x}\d\v{x}'\,e^{i\v{k}\cdot(\v{x}-\v{x}')}\rho(\v{x},\v{x}')$, so that 
short-distance features of $F(r)$ translate directly into high-momentum components of $n_{\v k}$. 
The pronounced bump of $F(r)$ around $r\sim1\,\mathrm{fm}$ thus reflects the spatial scale at which SRCs 
are most active.
As can be seen from the figure, the correlation functions for ${}^{16}\mathrm{O}$ and 
${}^{40}\mathrm{Ca}$ share a very similar global pattern, indicating that the dominant features of 
SRCs are largely nucleus-independent across these medium-mass systems. 
At the same time, a clear isospin dependence emerges at short distances ($r\lesssim1.5\,\mathrm{fm}$), where the difference between proton-proton and proton-neutron 
pairs becomes noticeable. The proton-neutron pairs exhibit stronger correlations, 
a consequence of the dominant spin-triplet components in that channel.
The isospin-dependence of the two-body pair density is produced by short-ranged interactions, driven by the two-nucleon tensor force\cite{Torr18PLB}.
In these calculations, protons and neutrons are treated symmetrically; the correlations for 
proton-proton and neutron-neutron pairs are identical in both ${}^{16}\mathrm{O}$ and ${}^{40}\mathrm{Ca}$.

\begin{figure}[h!]
\centering
  \includegraphics[width=9.cm]{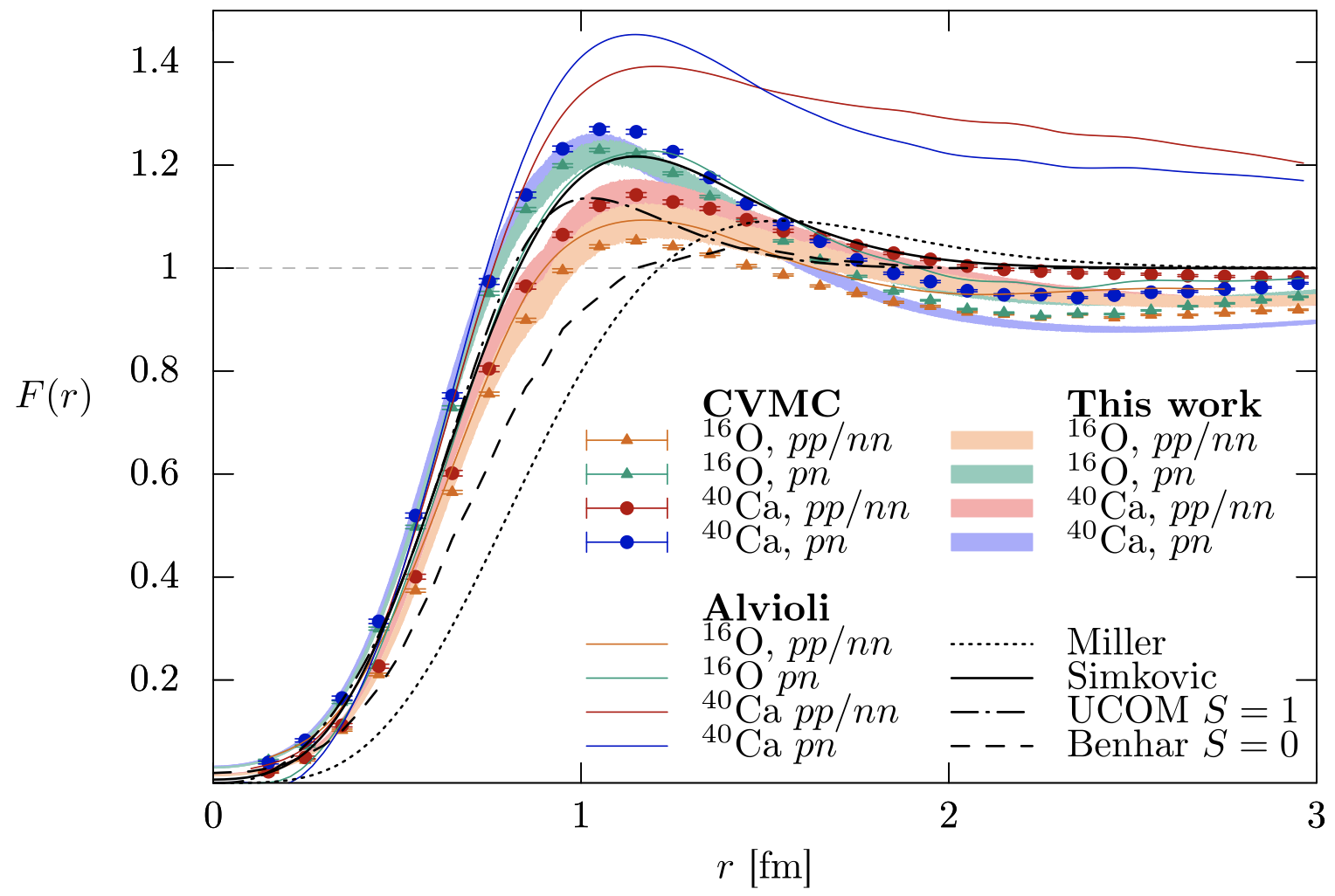}
  \caption{(Color Online). An example on nucleon-nucleon correlation functions for $^{16}\rm{O}$ and $^{40}\rm{Ca}$ for different coupled channels. Figure taken from Ref.\cite{Torr18PLB}.}
  \label{fig_Torr}
\end{figure}

Motivated by both microscopic calculations and several well-known earlier predictions from nuclear many-body theories\cite{Ant88}, together with recent experimental findings\cite{Wei15,Hen15a,Hen14}, one may describe the single-nucleon momentum distribution in ANM as\cite{Cai15a}
\begin{equation}\label{MDGen}
n^J_{\v{k}}(\rho,\delta)=\left\{\begin{array}{ll} \Delta_J+\beta_J{I}\left(\displaystyle\frac{|\v{k}|}{k_{\rm{F}}^J}\right),~~&0<|\v{k}|<k_{\rm{F}}^J,\\
\displaystyle{C}_J\left(\frac{k_{\rm{F}}^{J}}{|\v{k}|}\right)^4,~~&k_{\rm{F}}^J<|\v{k}|<\phi_Jk_{\rm{F}}^J. \end{array}\right. \end{equation}
Here, the depletion at zero momentum $\Delta_J$ is defined relative to the free Fermi gas (FFG), whereas $\beta_J$ encodes the strength of the momentum dependence $I(|\v{k}|/k_{\rm{F}}^J)$\cite{Bel61,Czy60,Sar80} near (and at) the Fermi surface.
For momenta $|\v{k}|\lesssim2\,\rm{fm}^{-1}$, the function $I$ behaves approximately as $\exp[{-\alpha|\v{k}|^2}]$ where $\alpha\approx0.12\,\rm{fm}^2$. Since $\alpha k_{\rm{F}}^2\approx0.21\ll1$ at $\rho_0$, this allows one to employ $\exp[{-\alpha|\v{k}|^2}]\approx1-\alpha|\v{k}|^2+\mathcal{O}(|\v{k}|^4)$ for $0<|\v{k}|<k_{\rm{F}}^J$, making the quadratic form\cite{Cai15a}
\begin{equation}\label{def-I}
{I}\left({|\v{k}|}/{k_{\rm{F}}^J}\right)=\left({|\v{k}|}/{k_{\rm{F}}^J}\right)^2
\end{equation}
a reasonable and effective representation of the depletion.
The jump $Z^J_{\rm{F}}$ at $k_{\rm{F}}^J$, namely the renormalization (strength) function, contains rich information about the nucleon effective E-mass via the Migdal--Luttinger theorem\cite{Migdal57,Lut60} and also reflects its nontrivial isospin dependence\cite{Jeu76,Mah85}; the related issue on the E-mass, e.g., the effects of the function $I(x)$, will be discussed in the Subsection \ref{sub_Emass}, see FIG.\,\ref{fig_nkANM} for a sketch of the $n_{\v k}$ in ANM\cite{CaiLi16a}.  The parameters ${C}_J$ and $\phi_J$ determine the HMT fraction through
\begin{equation}\label{def_xJHMT}
\boxed{
x_J^{\rm{HMT}}=\left.\int_{k_{\rm{F}}^J}^{\phi_Jk_{\rm{F}}^J} n_{\v{k}}^J\d\v{k}\right/{\displaystyle\int_0^{\phi_Jk_{\rm{F}}^J} n_{\v{k}}^J\d\v{k}}
=3C_{{J}}\left(1-\frac{1}{\phi_{{J}}}\right).}
\end{equation}
The normalization condition
\begin{equation}\label{def_NC}
\frac{2}{(2\pi)^3}\int_0^{\infty}n^J_{\v{k}}(\rho,\delta)\d\v{k}
=\rho_J=\frac{(k_{\rm{F}}^{J})^3}{3\pi^2}
\end{equation}
implies that among the four parameters $\beta_J$, ${C}_J$, $\phi_J$, and $\Delta_J$, only three are independent; one may, for instance, choose $\beta_J$, ${C}_J$, and $\phi_J$ to be independent and then determine $\Delta_J$ from\footnote{In Ref.\cite{Hen15b}, the normalization condition is set as
$[2/(2\pi)^3]\int_0^{\infty}n_{\v{k}}^0\d\v{k}=\rho$, differs by a factor 2 from the definition of Eq.\,(\ref{def_NC}),
from which the corresponding relation reads $[4/(2\pi)^3]\int_0^{\infty}n_{\v{k}}^0\d\v{k}=\rho$ for SNM.}
\begin{equation}\label{DeltaJ}
\Delta_J=1-\frac{3\beta_J}{(k_{\rm{F}}^{J})^3}\int_0^{k_{\rm{F}}^J}{I}\left(\frac{k}{k_{\rm{F}}^J}\right)k^2\d k
-3{C}_J\left(1-\frac{1}{\phi_J}\right).
\end{equation}

\begin{figure}[h!]
\centering
\includegraphics[width=7.5cm]{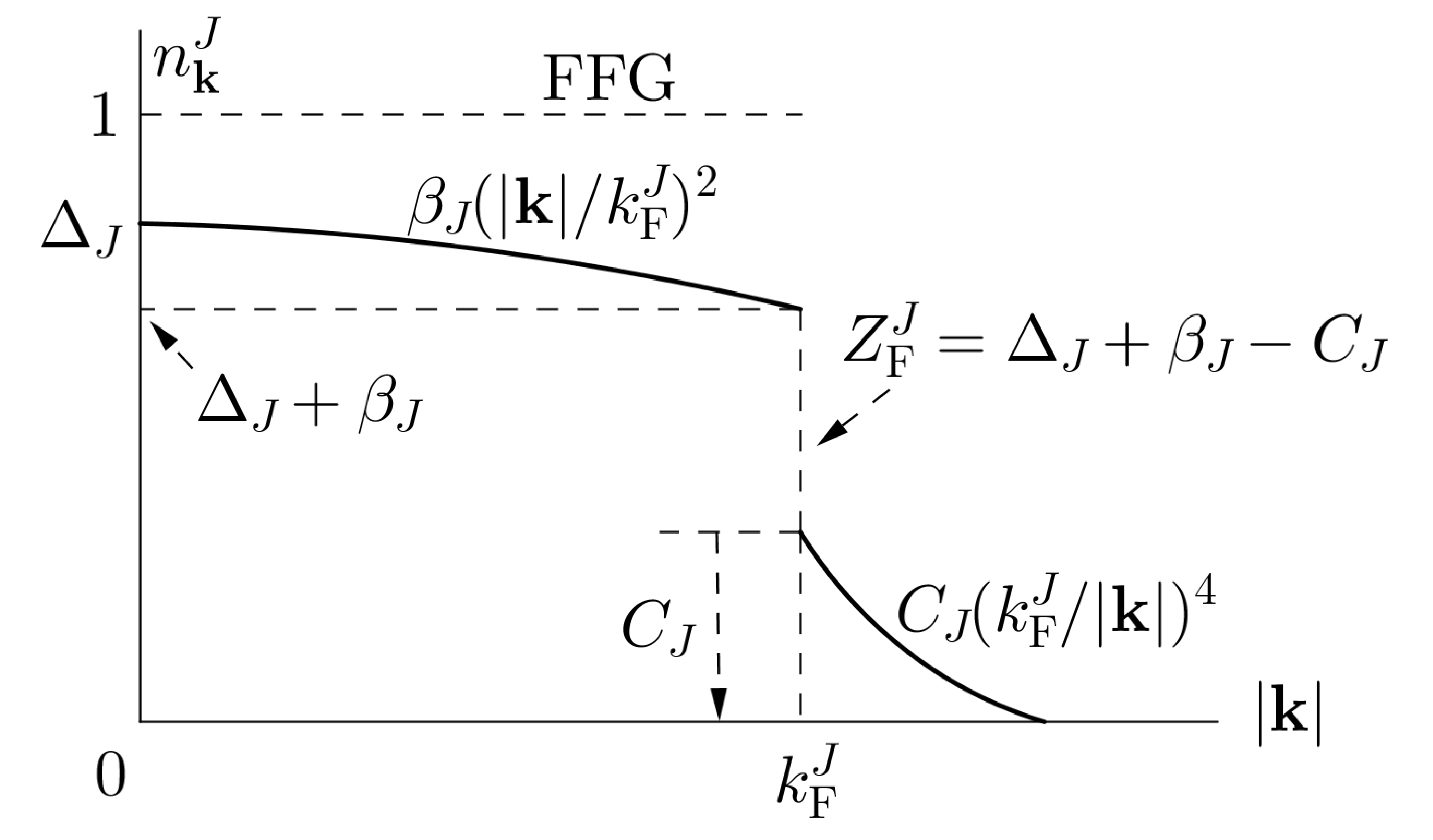}
\caption{A sketch of the single-nucleon momentum distribution function $n_{\v k}$ in ANM. Figure taken from Ref.\cite{CaiLi16a}.}
\label{fig_nkANM}
\end{figure}

Since the self-consistent Green's function (SCGF)\cite{Rio09,Rio20,Dris22} and the Brueckner--Hartree--Fock (BHF)\cite{Yin13,Yang19PRC} calculations both indicate that $\Delta_J$ varies approximately linearly with $\delta$ but in opposite directions for neutrons and protons, it was further assumed that all four parameters in Eq.\,(\ref{MDGen}) can be expanded as $Y_J=Y_0(1+Y_1^J\delta)$\cite{Cai15a}. This assumption can be justified by examining the resulting average kinetic energy per nucleon in ANM,
\begin{equation}\label{kinE}
\boxed{
E^{\rm{kin}}(\rho,\delta)=\frac{1}{\rho}\frac{2}{(2\pi)^3}
\sum_{J=\rm{n,p}}\int_0^{\phi_Jk_{\rm{F}}^J}\frac{\v{k}^2}{2M_{\rm N}}n_{\v{k}}^J(\rho,\delta)\d\v{k},}
\end{equation}
which, under Eq.\,(\ref{MDGen}), acquires a linear term in $\delta$:
\begin{align}\label{Ekin1}
E_1^{\rm{kin}}(\rho)\equiv&
\left.\frac{\partial E^{\rm{kin}}(\rho,\delta)}{\partial\delta} \right|_{\delta=0}
=\frac{3}{5}\frac{k_{\rm{F}}^2}{2M_{\rm N}}\Bigg[
\frac{5}{2}C_0\phi_0(\phi_1^{\rm{n}}+\phi_1^{\rm{p}})\notag\\
&
+\frac{5}{2}C_0(\phi_0-1)(C_1^{\rm{n}}+C_1^{\rm{p}}) 
+\frac{1}{2}\Delta_0(\Delta_1^{\rm{n}}+\Delta_1^{\rm{p}})\notag\\
&
+\frac{5\beta_0(\beta_1^{\rm{n}}+\beta_1^{\rm{p}})}{2k_{\rm{F}}^5}
\int_0^{k_{\rm{F}}}I\left(\frac{k}{k_{\rm{F}}}\right)k^4\d k\Bigg].
\end{align}
Imposing neutron-proton exchange symmetry of the EOS by requiring $E_1^{\rm{kin}}(\rho)=0$ leads directly to
$\Delta_1^{\rm{n}}=-\Delta_1^{\rm{p}}$,
$\beta_1^{\rm{n}}=-\beta_1^{\rm{p}}$,
${C}_1^{\rm{n}}=-{C}_1^{\rm{p}}$,
and $\phi_{1}^{\rm{n}}=-\phi_1^{\rm{p}}$,
i.e., each parameter satisfies 
\begin{equation}
Y_J/Y_0=1+Y_1\tau_3^J\delta,~~\tau_3^{\rm n}=+1,~\tau_3^{\rm p}=-1.
\end{equation}

\begin{figure}[h!]
\centering
  \includegraphics[width=7.5cm]{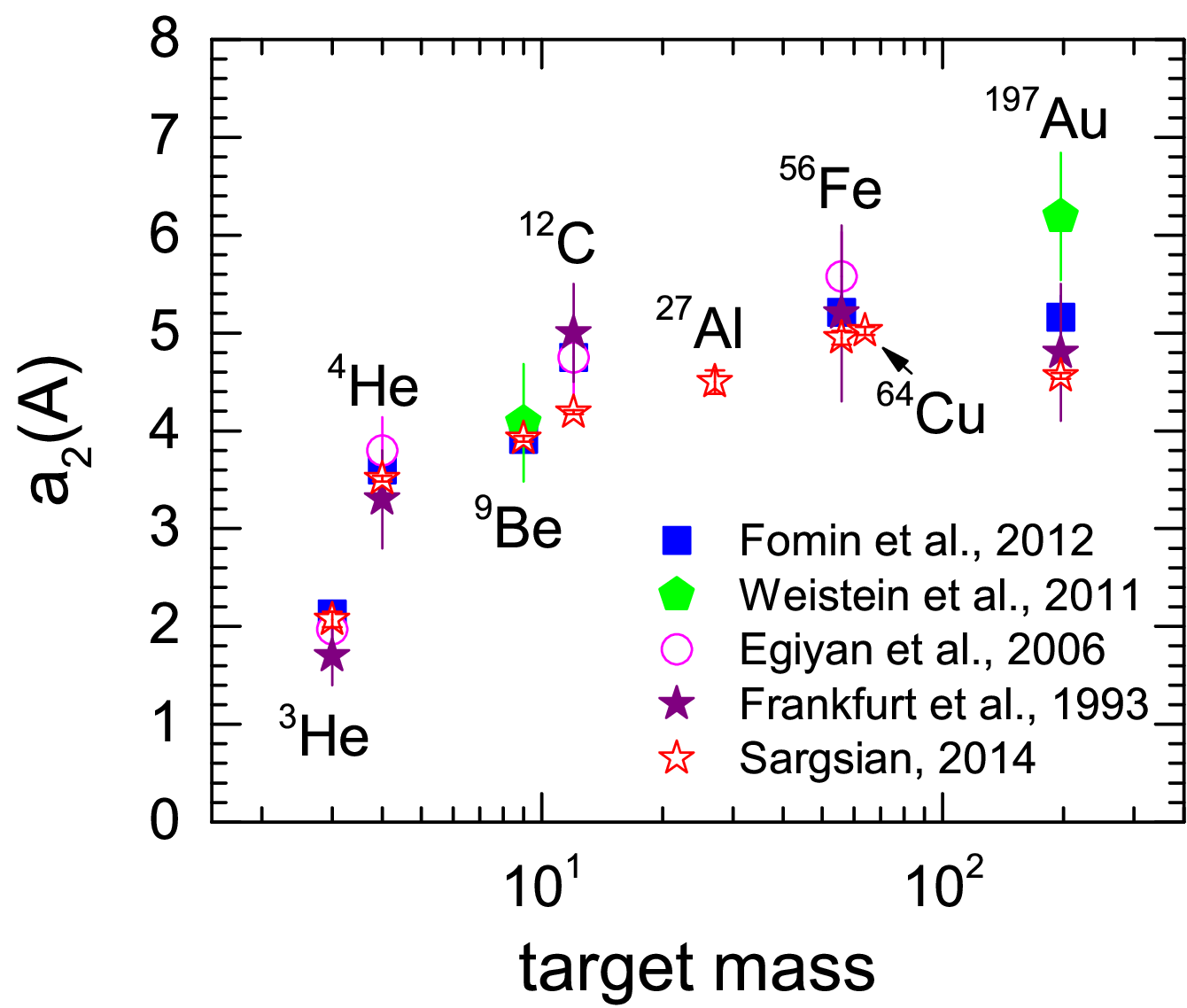}
  \hspace{0.5cm}
  \caption{(Color Online). Experimental $a_2(A)$ for several typical nuclei, i.e., $^3\rm{He}$,
 $^4\rm{He}$, $^9\rm{Be}$, $^{12}\rm{C}$, $^{56}\rm{Fe}$ ($^{63}\rm{Cu}$) and $^{197}\rm{Au}$\cite{Fom12,Wei11,Egi06,Fra93,Sar14}.
 Figure adopted from Ref.\cite{LCCX18}.
 }
 \label{fig_a2exp}
\end{figure}

\begin{figure}[h!]
\centering
  \includegraphics[width=9.cm]{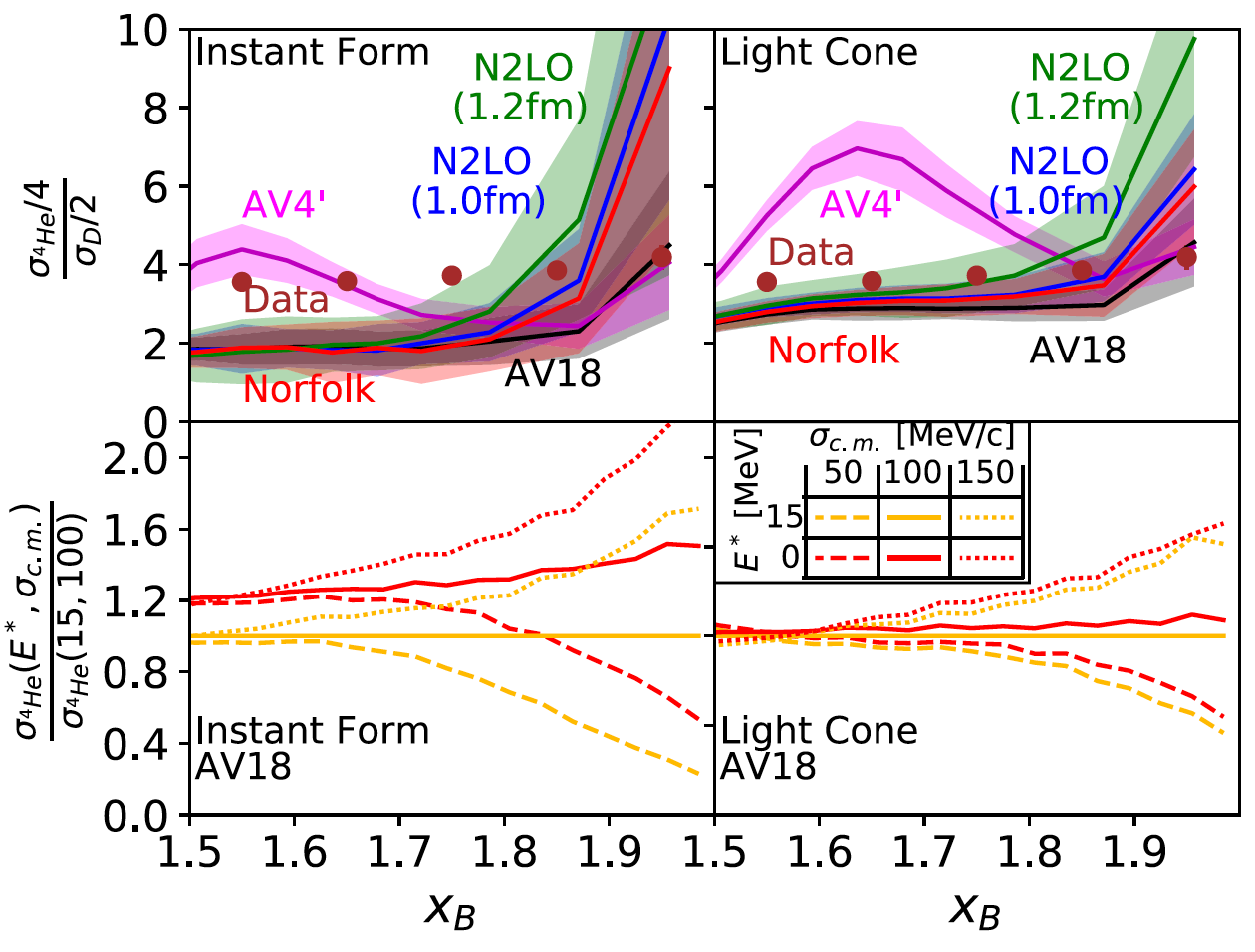}
  \caption{(Color Online). Top: Measured per-nucleon $(\rm{e},\rm{e}^{\prime})$ cross-section ratios $(\sigma_{^4\rm{He}}/4) / (\sigma_{\rm d}/2)$ as a function of $x_{\rm B}=x_{\rm{Bj}}$. The data are compared with generalized contact formalism (GCF) calculations using both instant form (left) and light-cone (right) formulations with different NN interaction models. The widths of the bands indicate the 68\% confidence interval due to uncertainties in the model parameters. Figure taken from Ref.\cite{Weiss21PRC}.
 }
 \label{fig_a2exp-21}
\end{figure}

As shown in the upper panel of FIG.\,\ref{fig_Benh08}, the nucleon HMT exhibits an approximate scaling from the deuteron to nuclear matter $n_{\v{k}}^A = a_2(A) n_{\v{k}}^{\rm{d}}$, consistent with the observations\cite{Fan84,Pie92,Cio96}, here $n_{\v{k}}^A$ and $n_{\v{k}}^{\rm{d}}$ denote the high-momentum components of the nucleon momentum distribution for a nucleus with mass number $A$ and for deuterium\cite{Fra81,WangR21CPC}, respectively. The factor $a_2(A)$ is independent of momentum and represents the probability of finding a high-momentum nucleon pair in nucleus $A$ relative to that in the deuteron\cite{Fra88}. Experimentally, $a_2(A)$ is determined from the plateau in the per-nucleon inclusive $(\rm{e},\rm{e}^{\prime})$ cross-section ratios for heavy nuclei relative to deuteron in the Bjorken scaling region $1.5 \lesssim x_{\rm{Bj}} \lesssim 1.9$\cite{Arr12}, i.e.,
\begin{equation}
\boxed{
a_2(A) = \frac{\sigma_A/A}{\sigma_{\rm{d}}/2}.}
\end{equation}
Representative experimental $a_2(A)$ values for nuclei such as $^3\rm{He}$, $^4\rm{He}$, $^9\rm{Be}$, $^{12}\rm{C}$, $^{56}\rm{Fe}$ ($^{63}\rm{Cu}$), and $^{197}\rm{Au}$ are displayed in FIG.\,\ref{fig_a2exp}\cite{Fom12,Wei11,Egi06,Fra93}. Extrapolating these results to infinite nuclear matter leads to $a_2(\infty) \approx 7 \pm 1$\cite{McG11,Att91,Hen15b,Dai17}.
In a recent study, the relative abundances of SRC nucleon pairs have been investigated using inclusive electron scattering and the generalized contact formalism (GCF) with various nuclear interaction models. GCF calculations successfully reproduce the observed scaling of cross-section ratios $a_2$ at high $x_{\rm B}=x_{\rm{Bj}}$ and large momentum transfer $Q^2$. They found that ratios of similar-mass isotopes, such as $^{40}\rm{Ca}$ and $^{48}\rm{Ca}$, are influenced not only by the isospin dependence of SRCs but also by low-energy nuclear structure. Consequently, the empirical extraction of SRC pair abundances from measured $a_2$ values is accurate only to about 20\%, and further improvement requires cross-section calculations that consistently incorporate both nuclear structure and relativistic effects.
GCF is an important framework for investigating SRC-related issues, see Refs.\cite{Weiss15PRC,Cruz21NP,Cosyn21PLB,Liang24PLB,Liang24xxx,Weiss17PRC,Weiss18xxx,Weiss19PLB,Yank25PRC} for more discussions.
It should be noted, as emphasized in Ref.\cite{Hen17RMP}, that the substantial uncertainties in the SRC ratios reported in Ref.\cite{Fra93} are mainly due to combining data obtained at different experimental kinematics. Additionally, the SRC ratios in Ref.\cite{Egi06} were used in the original EMC-SRC correlation analysis of Ref.\cite{Wei11}. Despite these efforts, a satisfactory theoretical understanding of the nearly linear correlation between the slope of the EMC effect and the $a_2$ factor\cite{Wei11} is lacking\cite{Wang20PRL,LRP2015}.

\begin{figure}[h!]
\centering
  \hspace*{0.cm}
  \includegraphics[width=7.5cm]{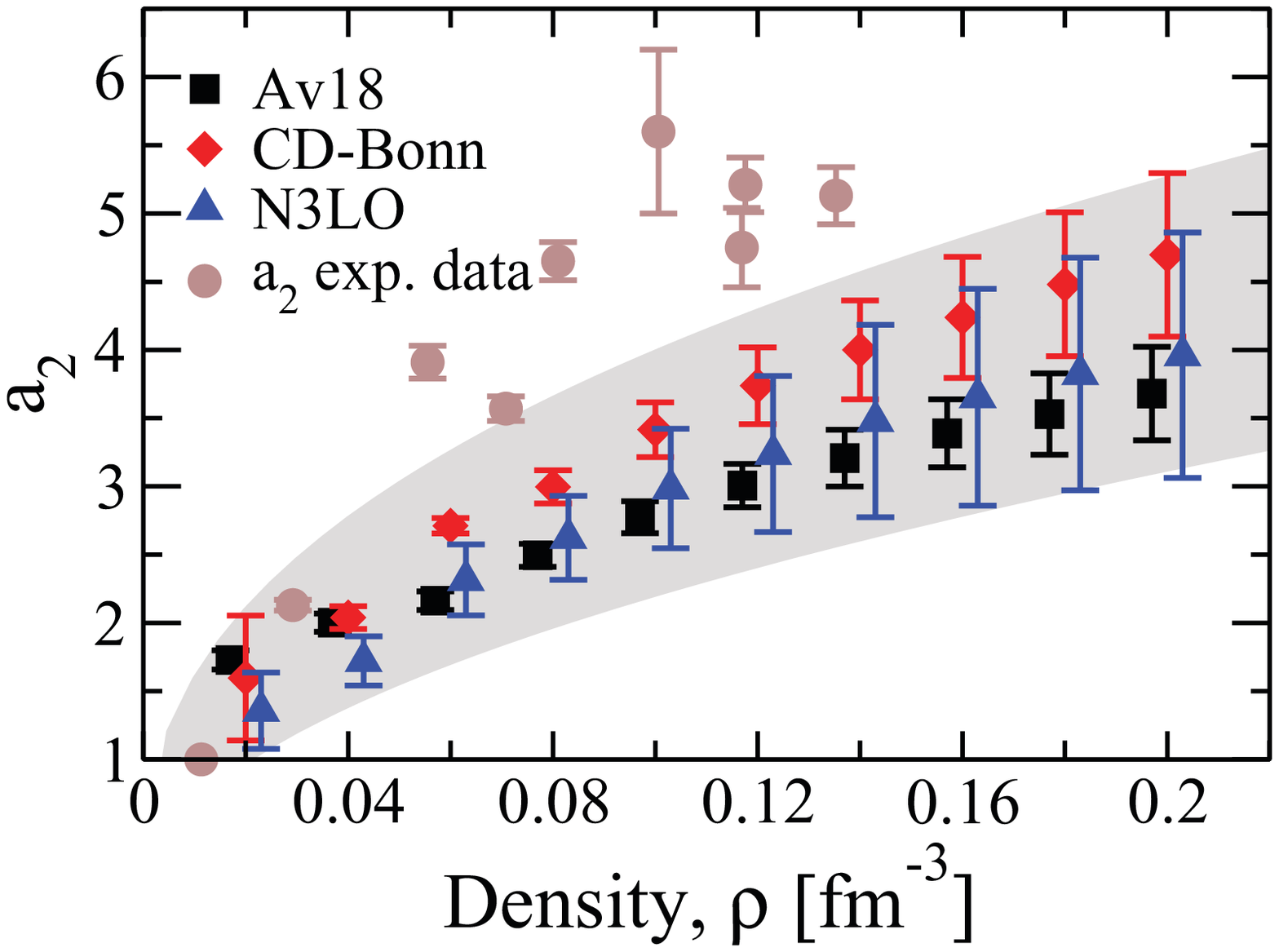}\\
  \hspace{-0.5cm}
  \includegraphics[width=8.cm]{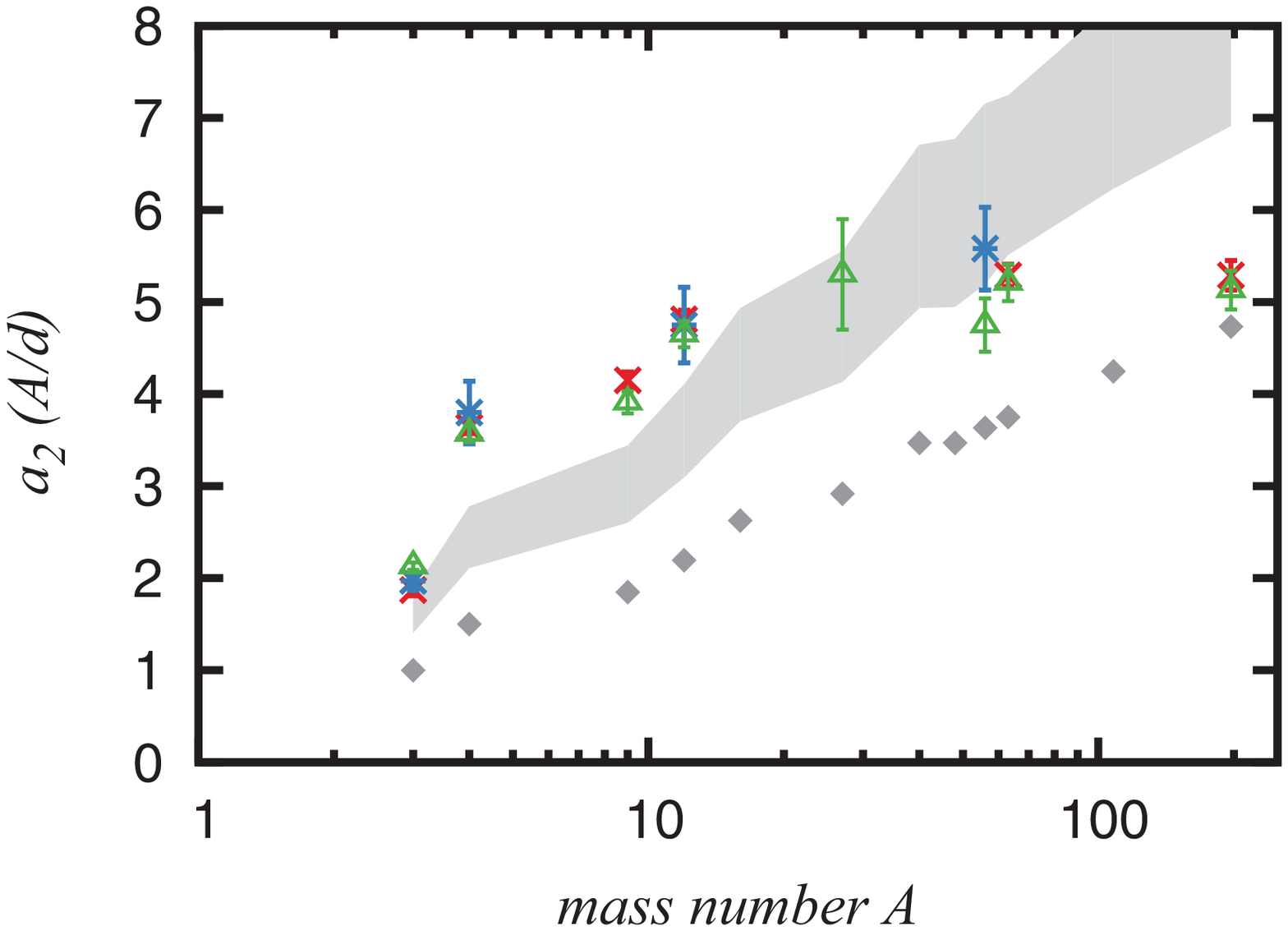}
  \caption{(Color Online). Comparisons of the $a_2(A)$ data with predictions using the SCGF theory by Ref.\cite{Rio14} (upper) and
  a independent particle model (IPM) by Refs.\cite{Van11,Van12} (lower).}
  \label{fig_a2the}
\end{figure}

\begin{figure}[h!]
\centering
  \hspace*{0.cm}
  \includegraphics[width=8cm]{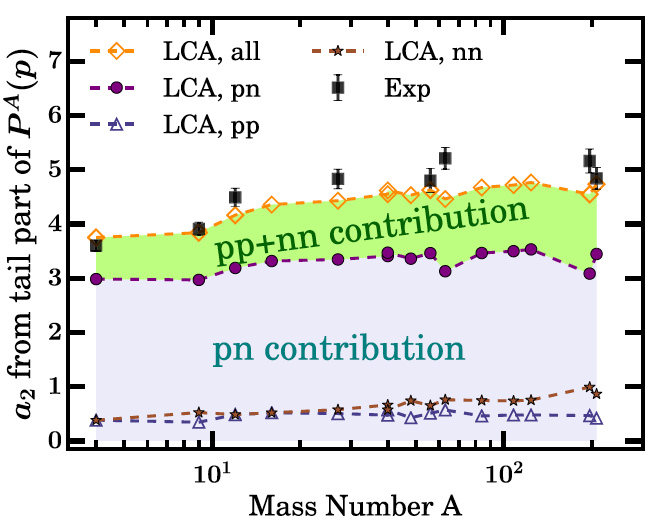}\\[0.25cm]
  \includegraphics[width=8.cm]{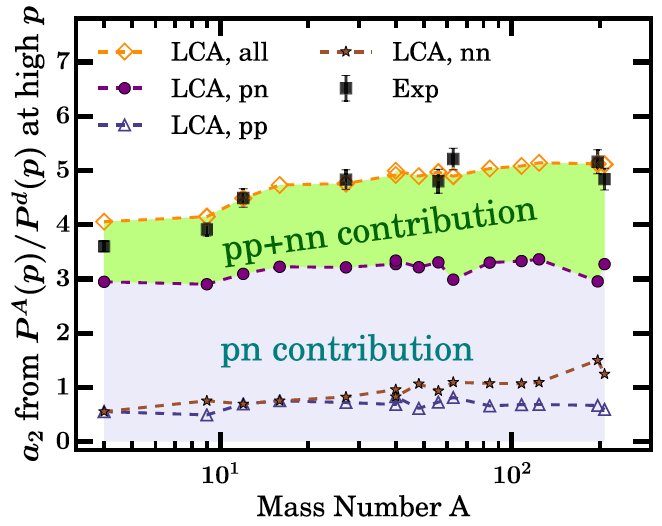}
  \caption{(Color Online). The $a_2$ obtained from the low-order correlation operator approximation (LCA) adopting two estimate schemes. Figure taken from Ref.\cite{Ryck19PRC}.}
  \label{fig_a2the-ab}
\end{figure}

\begin{figure}[h!]
\centering
  \includegraphics[width=8.5cm]{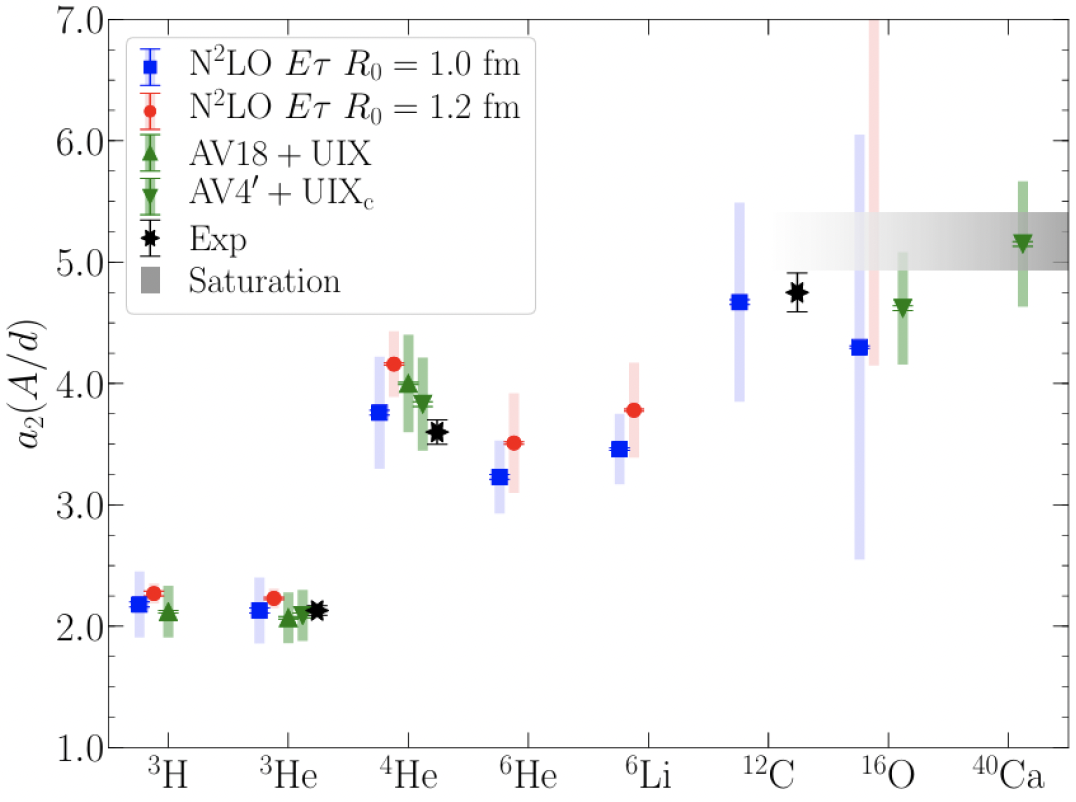}\\[0.5cm]
  \includegraphics[width=8.5cm]{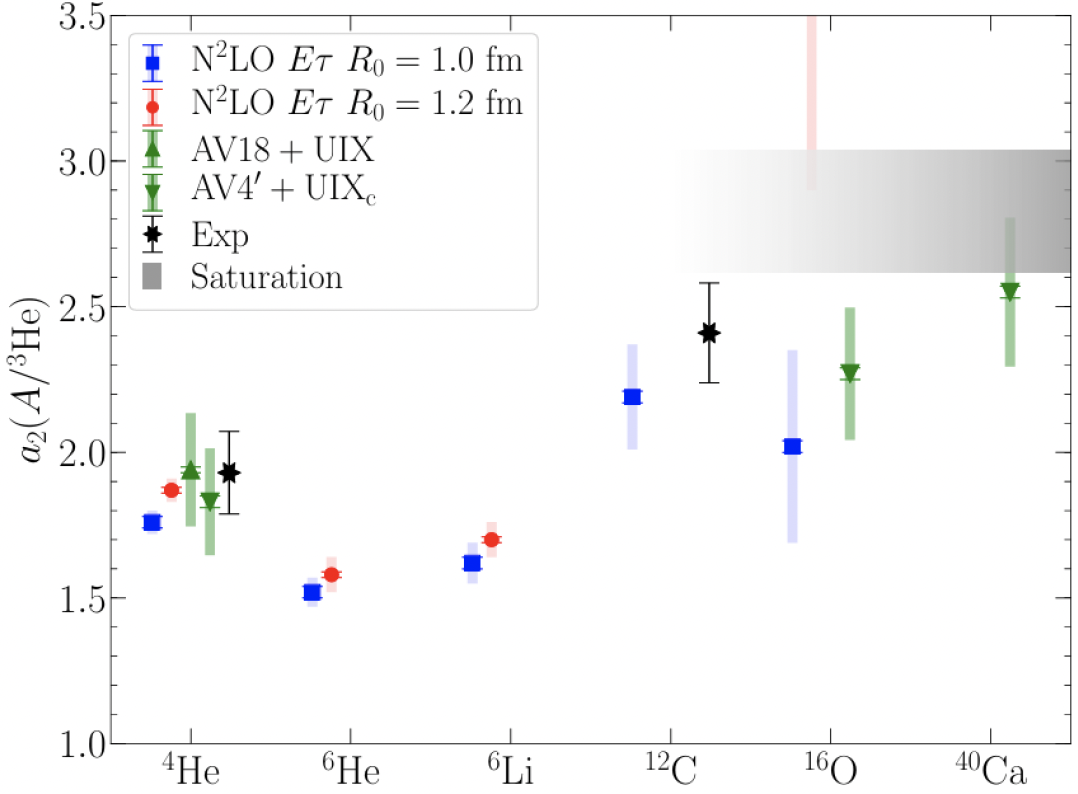}
  \caption{(Color Online). SRC scaling factors $a_2$ for nuclei from $A = 3$ to $A = 40$, shown relative to the deuteron (upper) and $^3$He (lower). Results are obtained using Quantum Monte Carlo (QMC) calculations based on chiral interactions within chiral effective field theory (ChPT), with cutoffs $R_0 = 10\,\mathrm{fm}$ (blue squares) and $R_0 = 12\,\mathrm{fm}$ (red circles). Predictions using the AV18+UIX (green upward triangles) and simplified AV4+UIX$_{\rm c}$ (green downward triangles) phenomenological potentials are also shown. Experimental data are indicated by black stars. Gray bands denote the expected saturation region inferred from measurements on heavier nuclei. Dark error bars represent Monte Carlo statistical uncertainties, while lighter bands show combined systematic uncertainties from chiral truncation and phenomenological fits.
  Figure taken from Ref.\cite{Lynn20}.
  }
  \label{fig_Lynn-nk}
\end{figure}

\begin{figure}[h!]
\centering
  \hspace*{-0.5cm}
  \includegraphics[width=8cm]{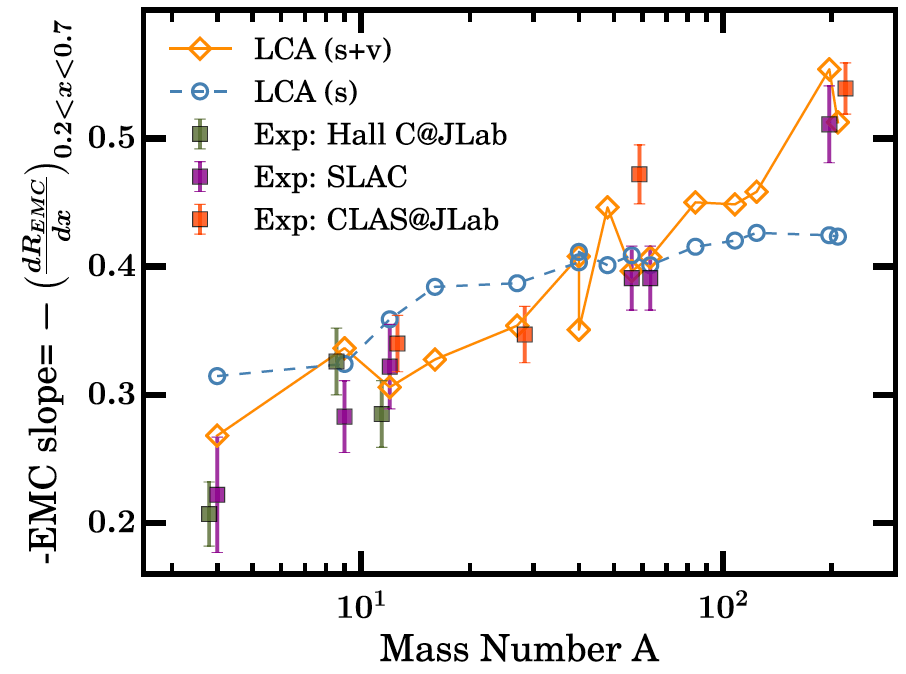}\\[0.5cm]
  \includegraphics[width=8.5cm]{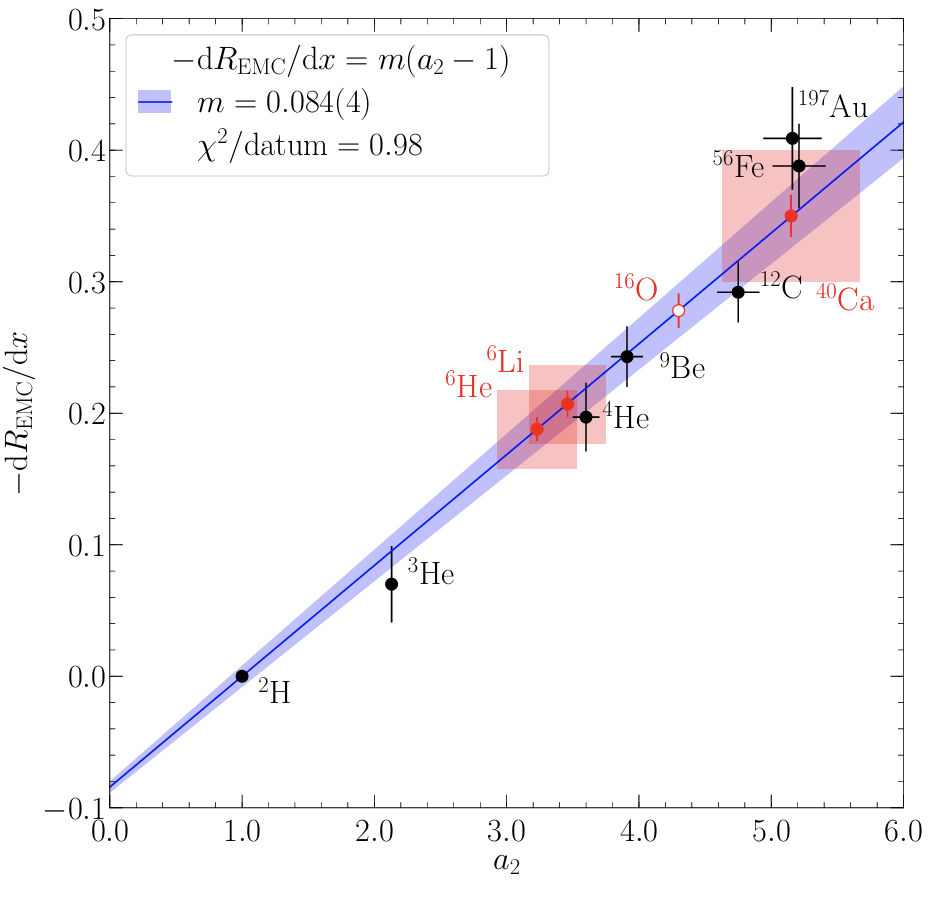}
  \caption{(Color Online). Upper: the EMC-SRC relation calculated using the LCA method; figure taken from Ref.\cite{Ryck19PRC}. Lower: the slope of the EMC effect and the $a_2$ factor obtained by QMC with ChPT theories; figure taken from Ref.\cite{Lynn20}.}
  \label{fig_EMC}
\end{figure}

Four theoretical calculations are compared with the experimental $a_2(A)$ in FIGs.\,\ref{fig_a2the}-\ref{fig_Lynn-nk}\cite{Rio14,Van12,Van11,Ryck19PRC,Lynn20}. In Ref.\cite{Rio14}, $a_2$ was estimated as $
a_2 = \langle n_{\v{k}}^0 / n_{\v{k}}^{\rm{d}} \rangle_{|\v{k}|=400\sim550\,\rm{MeV}}$,
and empirically parameterized as $
a_2(\rho) \approx b_1 \rho^{b_2}$ with $b_1 \approx 7\sim10$ and $b_2 \approx 0.4\sim0.5$.
Considering that SCGF calculations generally predict a saturation density $\rho_0^{\rm{SCGF}}$ higher than the empirical value, the extrapolated $a_2(\rho_0^{\rm{SCGF}}) \approx 5\pm 1$ underestimates the data, but remains roughly consistent with $a_2\approx7\pm1$ given in the above, see the upper panel of FIG.\,\ref{fig_a2the}. The relatively low high-momentum cutoff used in SCGF calculations (400-550\,MeV) also contributes to this underestimation, since the experimental momentum window for two-body SRC pairs spans approximately 300-600\,MeV\cite{Hen14}.
In contrast, Refs.\cite{Van12,Van11} evaluated $a_2$ within the independent-particle model (IPM, lower panel of FIG.\,\ref{fig_a2the}):
\begin{equation}
a_2(A/\rm{d}) = \frac{2}{A} N_{\rm{pn}(S=1)}(A,Z) \int_{\rm{PS}} \d\v{P}_{12} F^{\rm{pn}}(P_{12}),
\end{equation}
where $N_{\rm{pn}(S=1)}(A,Z)$ counts the number of correlated  pn pairs in the spin-triplet state, the integral runs over the relevant center-of-mass phase space, and $F^{\rm{pn}}$ is the center-of-mass distribution of correlated pairs. Ref.\cite{Van11} found that $a_2(A/\rm{d})$ increases nearly linearly with $A$ for $A \lesssim 40$, and then saturates for heavier nuclei. Their predictions for $^3\rm{He}$, $^4\rm{He}$, $^9\rm{Be}$, $^{12}\rm{C}$, $^{27}\rm{Al}$, $^{40}\rm{Ca}$, $^{48}\rm{Ca}$, $^{56}\rm{Fe}$, $^{108}\rm{Ag}$, and $^{197}\rm{Au}$ are compared with the experimental data in FIG.\,\ref{fig_a2exp}.
Similarly, Ref.\cite{Ryck19PRC} estimates the $a_2(A)$ factor by using the low-order correlation operator approximation (LCA).
For the studied sample of fifteen nuclei, the total SRC scaling factor is roughly in the range 4.05-5.14, of which roughly 3 can be attributed to pn correlations, and the SRC scaling factors receive sizable contributions from pp and nn correlations; their results are shown in FIG.\,\ref{fig_a2the-ab} where the upper and lower panels adopt two different schemes for estimating $a_2(A)$:
\begin{equation}
    a_2(A)\approx\frac{\int_{p>2\,\rm{fm}^{-1}}\d pP^A(p)}{\int_{p>2\,\rm{fm}^{-1}}\d pP^{\rm{d}}(p)};\quad\text{and }~ a_2(A)\approx\lim_{\rm{high}\,p}\frac{P^A(p)}{P^{\rm{d}}(p)},
\end{equation}
here $P^A(p)=p^2n^A(p)/A$ is the probability distribution normalized by the condition $\int\d pP^A(p)=1$\cite{Ryck19PRC}.

In Ref.\cite{Lynn20}, the scaling factor $a_2$ is extracted using Quantum Monte Carlo (QMC) methods combined with chiral effective field theory (ChPT). The authors exploit these advances to confirm the linear EMC-SRC correlation in light nuclei up to $^{12}$C by comparing ab initio calculations with existing experimental data. They then extended the analysis to additional light systems ($^6$He, $^6$Li, and $^{16}$O) as well as the medium-mass nucleus $^{40}$Ca. The prediction is shown in FIG.\,\ref{fig_Lynn-nk}.
For $a_2(A/d)$, the gray band at large $A$ represents the expected saturation range, determined from the difference (including uncertainties) between the measurements in $^{197}$Au and $^{63}$Cu. This gives an estimate $\lim_{A \to \infty} a_2(A/d) \approx 4.94$-$5.41$. The lower panel shows the corresponding results for $a_2(A/^3\mathrm{He})$ from $^4$He to $^{40}$Ca, where the saturation region is guided by the experimental value from $^{56}$Fe. 
Overall, the calculated SRC scaling factors show excellent agreement with experiment wherever data exist. Using chiral interactions at N$^2$LO with the cutoff $R_0 = 1.0$\,fm, the relative deviations from experiment for $a_2(A/d)$ are $0.0\%$, $4.4\%$, and $1.7\%$ for $^3$He, $^4$He, and $^{12}$C, respectively. The softer cutoff $R_0 = 1.2$\,fm yields values typically larger by $5$-$10\%$, but still consistent with the estimated systematic uncertainties. Interestingly, the predicted $a_2$ values for the $A=6$ nuclei lie below that of $^4$He, placing them between $^3$He and $^4$He along the fitted trend. This supports the picture that the local nuclear density, rather than the overall mass number, controls both the size of the EMC effect and the height of the SRC scaling plateau.
As further discussed in Ref.\cite{Lynn20}, these predictions are experimentally testable. For $^6$Li, inclusive $(\rm{e,e'})$ measurements at Jefferson Lab in quasi-elastic kinematics already offer a direct path for verification. For $^6$He, inverse-kinematics experiments at future rare-isotope facilities such as FAIR, using $(\rm{p,2p})$ reactions with a $^6$He beam, provide a promising route toward direct measurement.
The slope of the EMC effect and the $a_2$ factor calculated using the LCA and QMC methods are shown in FIG.\,\ref{fig_EMC}.
Specifically, it was pointed out that the obtained theory-experiment
comparisons for the EMC slopes cannot shed light
on underlying mechanisms that are due to non-SRC related
medium modifications\cite{Ryck19PRC}.

\begin{figure}[h!]
\centering
\includegraphics[width=4.5cm]{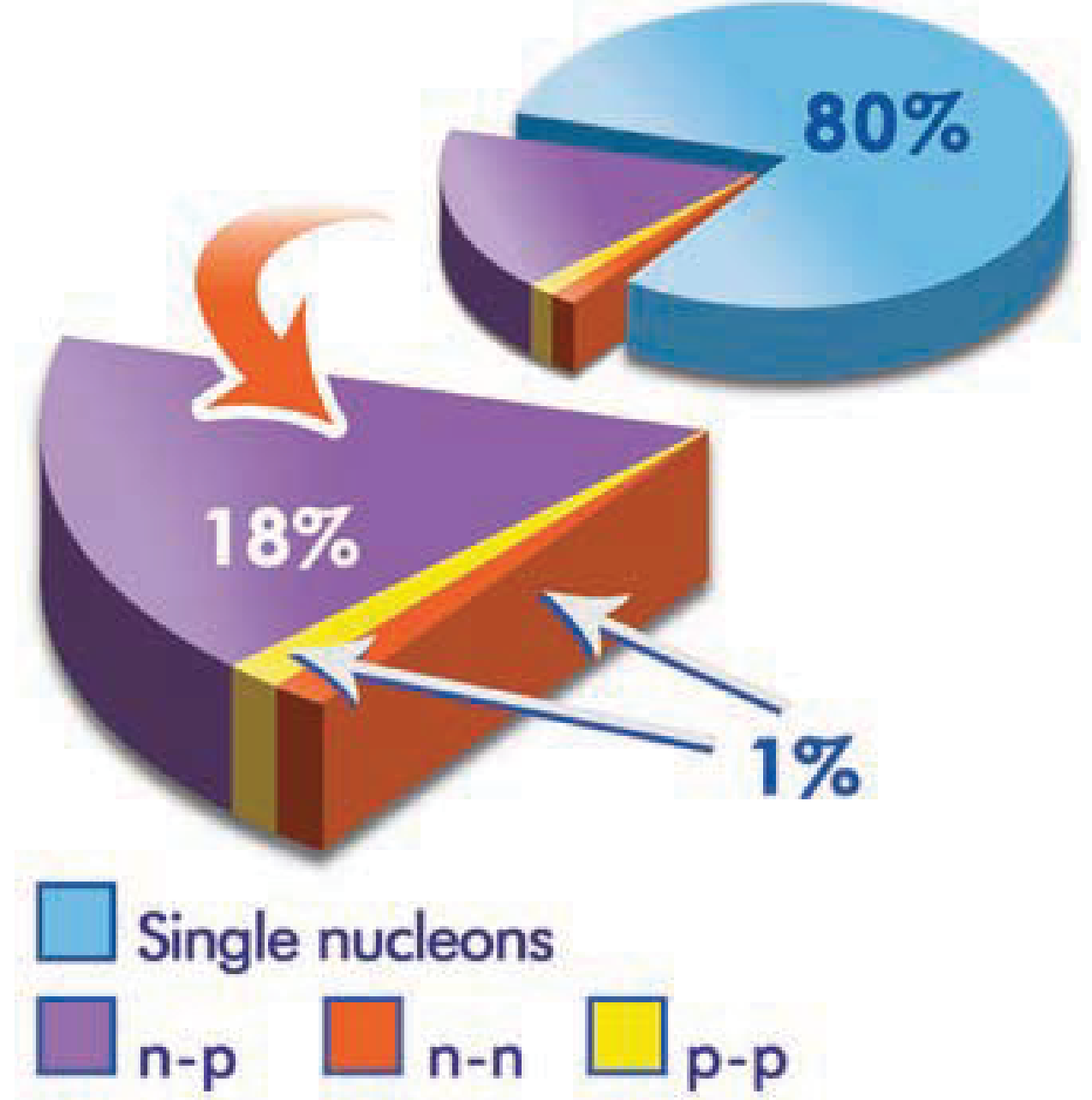}\\[0.5cm]
\includegraphics[width=9.cm]{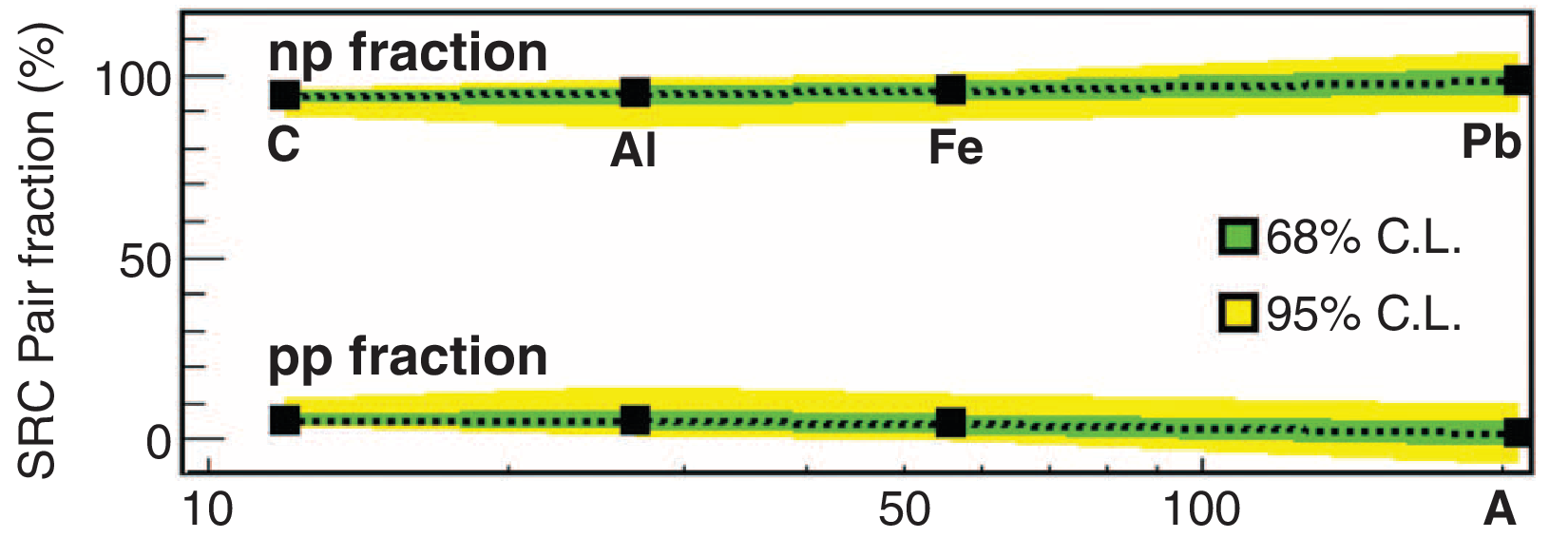}
\caption{(Color Online). Upper: the average fraction of nucleons in various initial-state configurations of $^{12}\rm{C}$\cite{Sub08}. Lower: extracted fractions of np (top) and pp (bottom) SRC pairs\cite{Hen14}.}
\label{fig_Sub}
\end{figure}

\begin{figure*}[h!]
\centering
\includegraphics[width=15.cm]{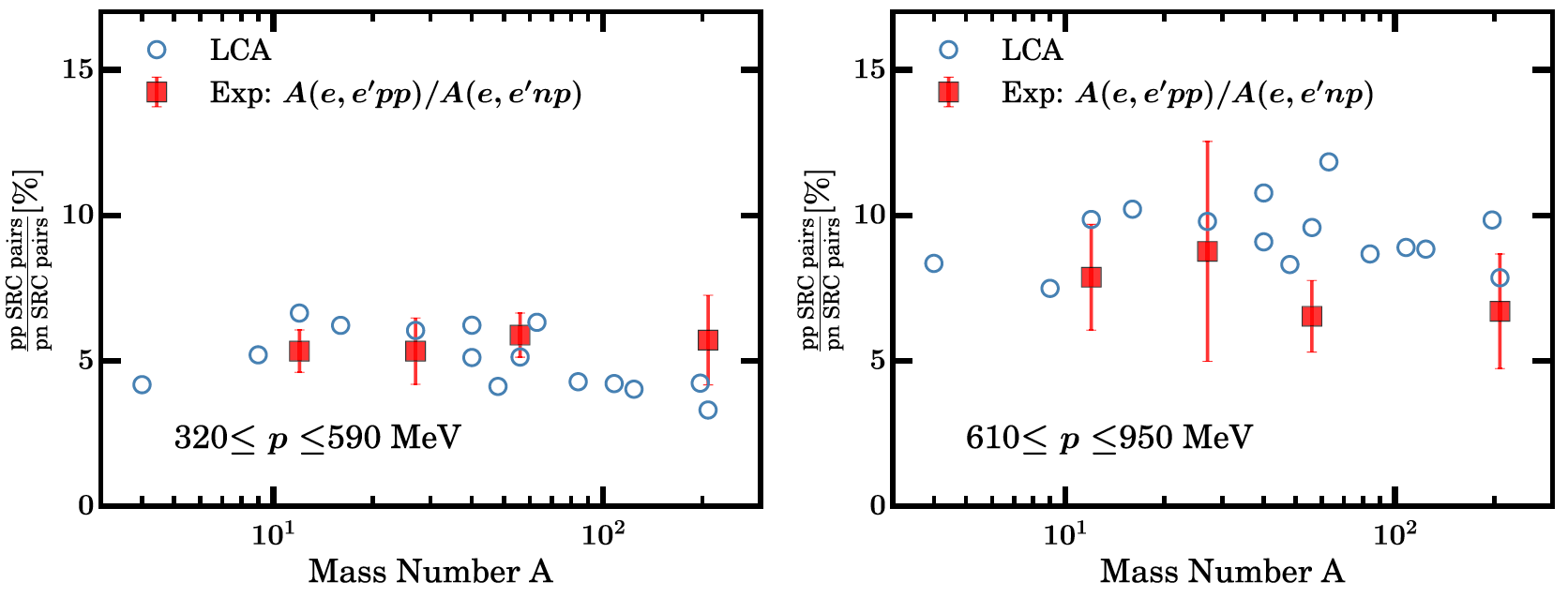}
\caption{(Color Online).The nuclear mass dependence of the ratios  of pp-to-pn SRC correlated pairs in two momentum ranges above the Fermi momentum; the blue circles are LCA predictions.
Figure taken from Ref.\cite{Ryck19PRC}.}
\label{fig_pp-np-ratio}
\end{figure*}

Experimental data show that reproducing the measured deuteron quadrupole moment and the spin correlation parameter requires approximately $4$-$5\%$ tensor-force induced D-wave contribution\cite{Fra88}. Combining this with the extrapolated $a_2(\infty)$, the HMT fraction in infinite SNM is estimated as $x_{\rm{SNM}}^{\rm{HMT}} \approx 28\% \pm 4\%$\cite{Hen15a,Hen14,Egi06,Pia06,Shn07,Wei11,Kor14,Hen15b}. According to Ref.\cite{Sub08}, roughly $80\%$ of nucleons in $^{12}\rm{C}$ behave independently (or as described by the shell model), while the remaining $20\%$ form correlated pairs, of which $90\pm10\%$ are pn SRC pairs and $5\pm1.5\%$ are pp pairs. This indicates an approximate 20-fold excess of pn over pp pairs\cite{Egi06,Sub08}, a trend that persists in heavier nuclei from C to $^{208}\rm{Pb}$\cite{Hen14} (see FIG.\,\ref{fig_Sub}, lower panel). By isospin symmetry, nn SRC pairs are also estimated to account for $5\pm1.5\%$, yielding an HMT fraction in PNM of $x_{\rm{PNM}}^{\rm{HMT}} \approx 1.5\% \pm 0.5\%$\cite{Hen15a,Hen14,Egi06,Pia06,Shn07,Wei11,Kor14,Hen15b}. The observed dominance of pn over pp pairs reflects the spin-isospin dependence of underlying tensor correlations, resulting in approximately equal numbers of protons and neutrons in the HMT. The prediction on the pp-to-pn SRC pairs using the LCA approximation\cite{Ryck19PRC} is shown in FIG.\,\ref{fig_pp-np-ratio}. In line with the data,
the LCA predictions for the number of pp-to-pn SRC pairs
are fairly constant among the fifteen nuclei in their sample\cite{Ryck19PRC}
and increase with increasing nucleon momentum. For a fixed
momentum range, the variation in the predicted pp-to-pn SRC
pair ratios across nuclei is of the order of few percent, in line
with the experimental observations\cite{Duer19PRL}.

\begin{figure}[h!]
\centering
\includegraphics[width=7.cm]{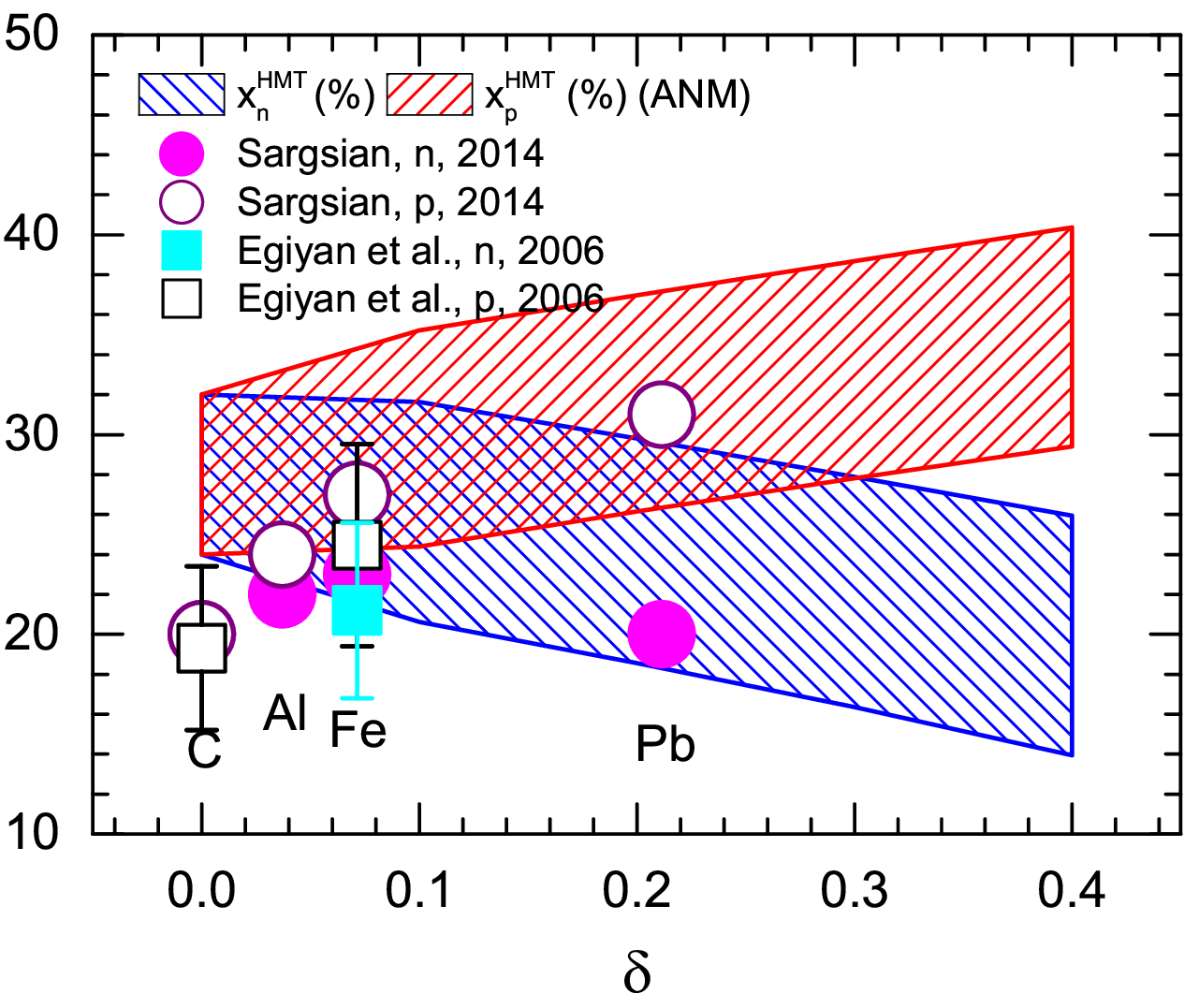}
\caption{(Color Online). High-momentum nucleon fractions in heavy nuclei and nuclear matter. Figure taken from Ref.\cite{LCCX18}.}
\label{fig_xnxpdelta}
\end{figure}

\begin{figure}[h!]
\centering
\includegraphics[width=8.5cm]{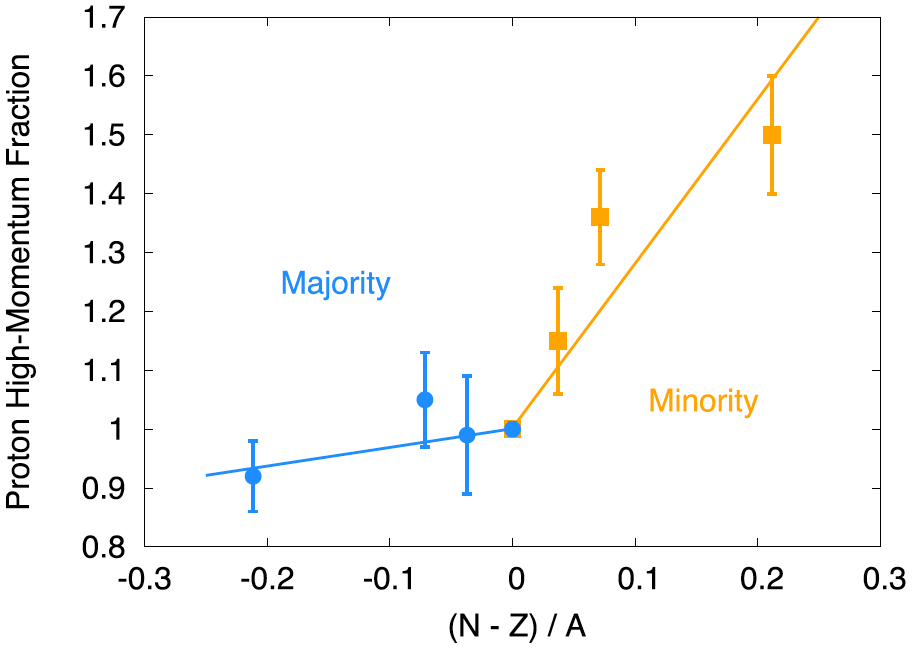}\\[0.25cm]
\includegraphics[width=8.5cm]{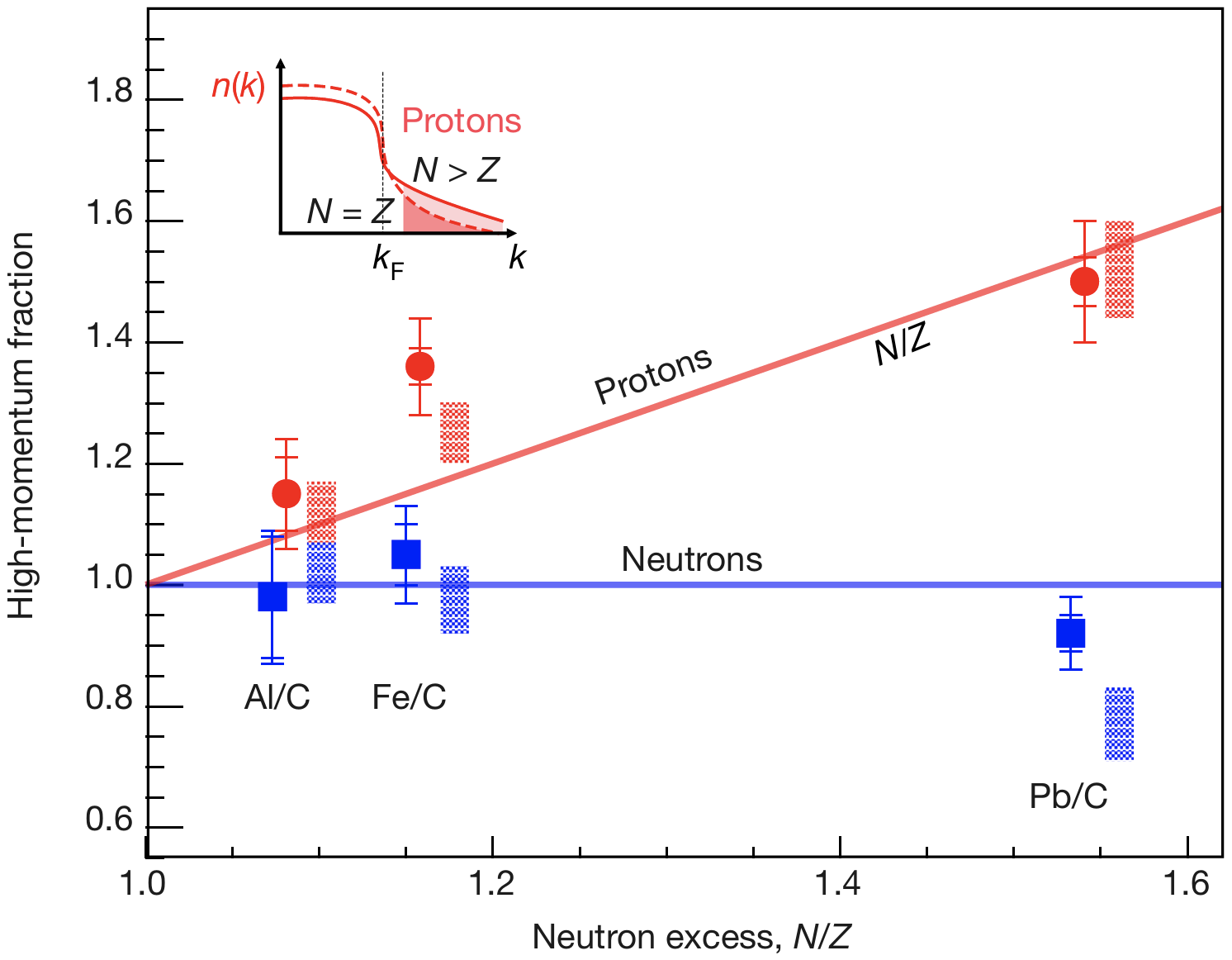}
\caption{(Color Online). Upper: the proton high-momentum fraction relative to that of $^{12}\rm{C}$ as a function of $\delta$ using a phenomenological model connecting the quenching of spectroscopic factors with the JLab experimental data; figure taken from Ref.\cite{Pas20}. Lower: relative high-momentum fractions for neutrons and protons using the quasi-elastic knock-out data\cite{Duer18Nature}.}
\label{fig_HMT-fr}
\end{figure}

Consequently, in neutron-rich matter, the fraction of protons in the HMT exceeds that of neutrons\cite{Hen14,Sar14}. The isospin dependence of $x_J^{\rm{HMT}}$ is illustrated in FIG.\,\ref{fig_xnxpdelta} (blue band for neutrons, red band for protons) based on parameters from Ref.\cite{Cai15a}.
For finite nuclei, the high-momentum fraction can be approximated by\cite{Sar14,Egi06}
\begin{equation}
\boxed{
x_J^{\rm{HMT}}(A,y) \approx \frac{1}{2 x_J} a_2(A,y) \int_{k_{\rm{F}}}^{\infty} n_{\v{k}}^{\rm{d}} \d\v{k},}
\end{equation}
where $a_2(A,y)$ is taken from Refs.\cite{Egi06,McG11,Fra93}, $x_{\rm{p}}=Z/A$, $x_{\rm{n}}=N/A$, and $y=|x_{\rm{p}}-x_{\rm{n}}|$. As FIG.\,\ref{fig_xnxpdelta} shows, the fractional proton-neutron imbalance in the HMT grows with increasing isospin asymmetry. For instance, in $^{208}\rm{Pb}$, the fraction of high-momentum protons exceeds that of neutrons by roughly $55\%$; see also Ref.\cite{Pas20} for a phenomenological study on the high-momentum fraction using the quasi-elastic knock-out data\cite{Duer18Nature} as shown in FIG.\,\ref{fig_HMT-fr}; the overall linear dependence of the fraction on either $\delta$ or $N/Z$ is obvious.

\begin{figure}[h!]
\centering
\hspace{-0.5cm}
\includegraphics[width=7.5cm]{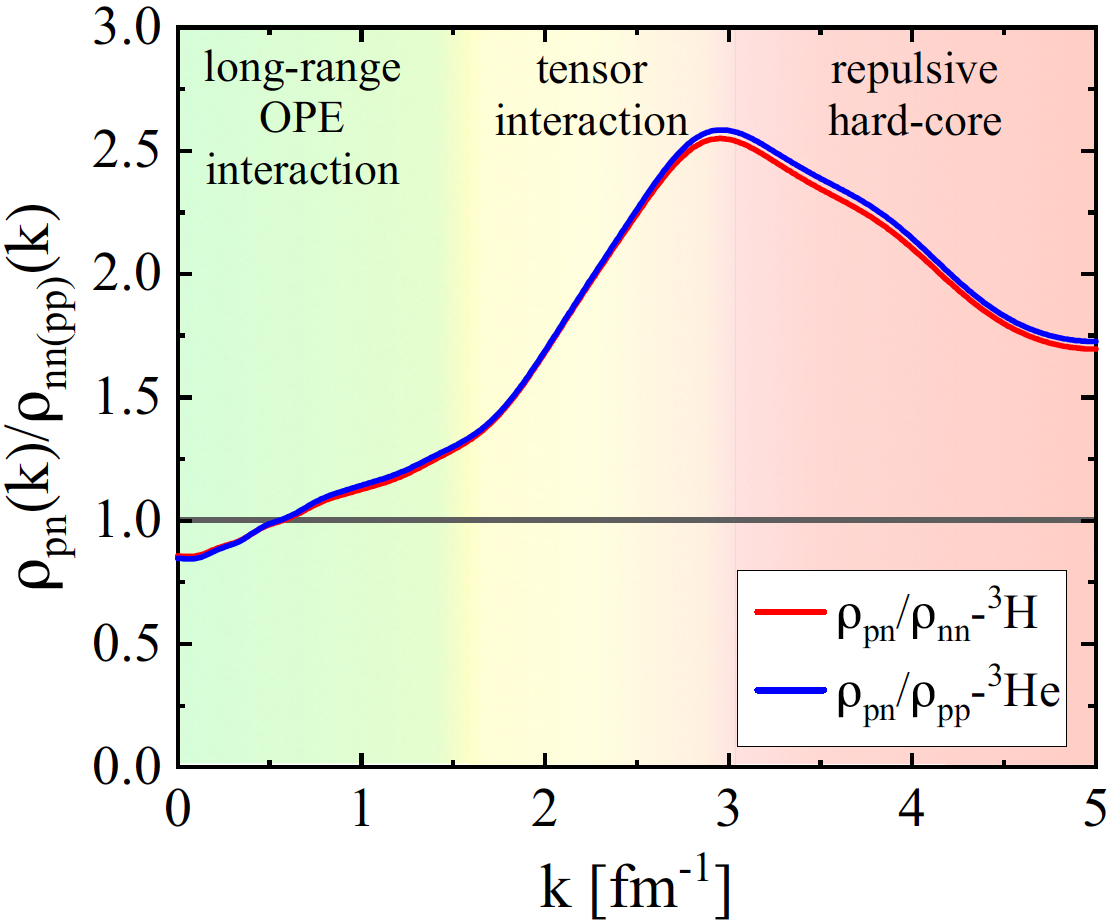}\\[0.25cm]
\includegraphics[width=8.cm]{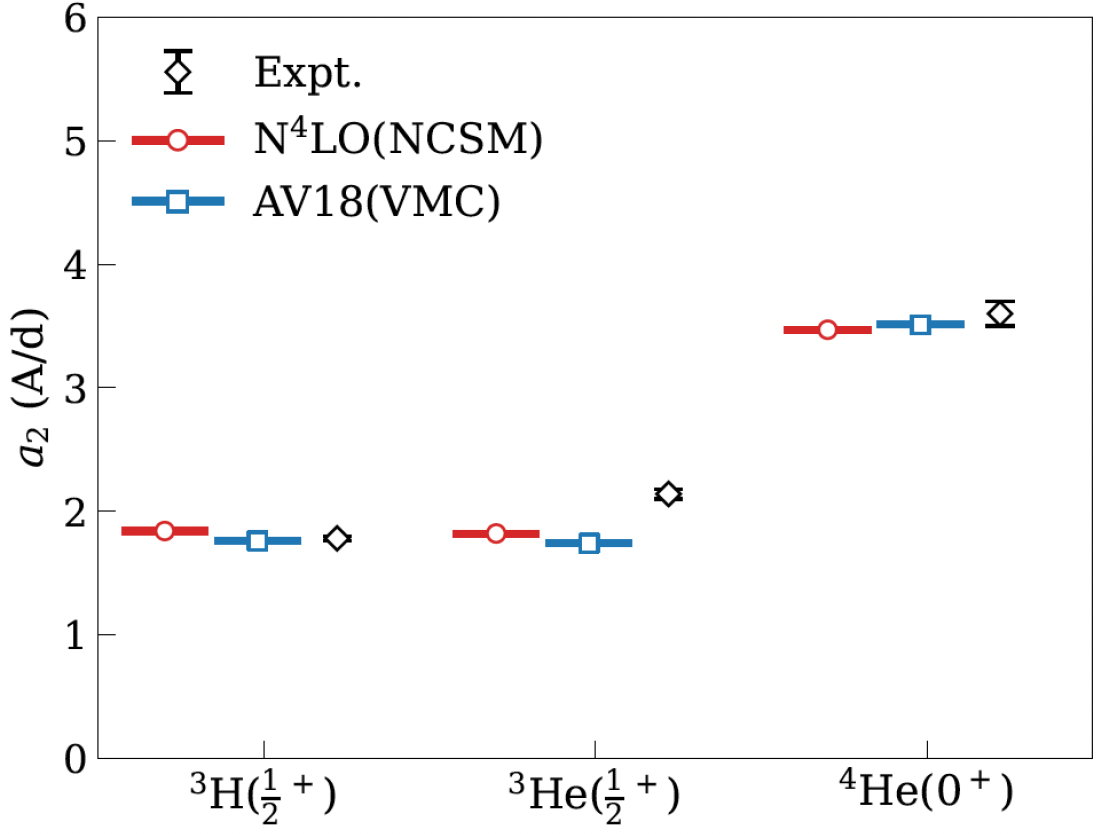}
\caption{(Color Online). Upper: ratios of pn to nn pairs for $^3\rm{H}$ and pn to pp
pairs for $^3\rm{He}$ as function of the relative momentum $k$, from which the enhancement factor could be calculated; figure taken from Ref.\cite{Meng23PRC}. Lower: the estimated scaling factor $a_2$ in ground states of
$^3\rm{H}$, $^3\rm{He}$ and $^4\rm{He}$ by no-core shell-model (NCSM) calculations using the N$^4$LO chiral
nuclear force; figure taken from Ref.\cite{Shang25}.}
\label{fig_Meng}
\end{figure}

While heavy nuclei are dominated by np SRCs, precise measurements in light mirror nuclei ($^3$H and $^3$He) reveal a significant deviation from np dominance, indicating unexpected structure in their high-momentum wavefunctions\cite{Li22Nature,Sch19Nature}.
The experimental data show that the ratio of pn-SRCs to pp-SRCs over the pair-counting prediction $P_{\rm{np}/\rm{pp}} = NZ/[Z(Z -1)/2]$ for $A = 3$ nuclei is about $2.17^{+0.25}_{
-0.20}$ (the enhancement factor), which is much smaller
than that in heavy nuclei; these results provide new insight into the short-range part of the nucleon-nucleon interaction.
For example, in Ref.\cite{Meng23PRC}, the wave functions of $^3$H and $^3$He were calculated using the Gaussian expansion method to solve the three-body Schrödinger equation. Their analysis of one- and two-nucleon momentum distributions shows that np-SRC pairs are significantly more prevalent than nn- or pp- pairs, with an enhancement factor of about 1.6, in agreement with experimental observations, see the upper panel of FIG.\,\ref{fig_Meng} for the ratios of pn- to nn-pairs for $^3\rm{H}$ and pn to pp
pairs for $^3\rm{He}$ as function of the relative momentum $k$, from which the enhancement factor could be calculated\cite{Meng23PRC}.
Similarly, Ref.\cite{Shang25} employed a fifth-order N$^4$LO chiral nuclear force without softening, enabling the isolation of nuclear-state effects with different quantum numbers on SRCs. Notably, the $a_2$ values are reduced and nearly equal in triplet isobaric analog states of neighboring nuclei, demonstrating that SRC abundances cannot be reliably estimated from mean-field shell structures alone. This is attributed to specific nuclear states suppressing the formation of deuteron-like components, offering insights into the connection between high-energy partonic properties and low-energy nuclear structure, see the lower panel of FIG.\,\ref{fig_Meng}.
The scaling factors of different nuclear states in
$^6\rm{Li}$ was also calculated in Ref.\cite{Shang25}, see also Ref.\cite{shang2026}.

\begin{figure}[h!]
\centering
\hspace{-0.5cm}
\includegraphics[width=8.cm]{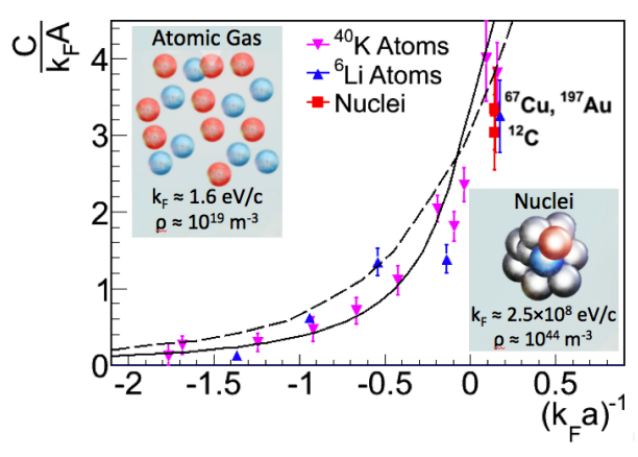}\\[0.5cm]
\hspace{.75cm}
\includegraphics[width=7.cm]{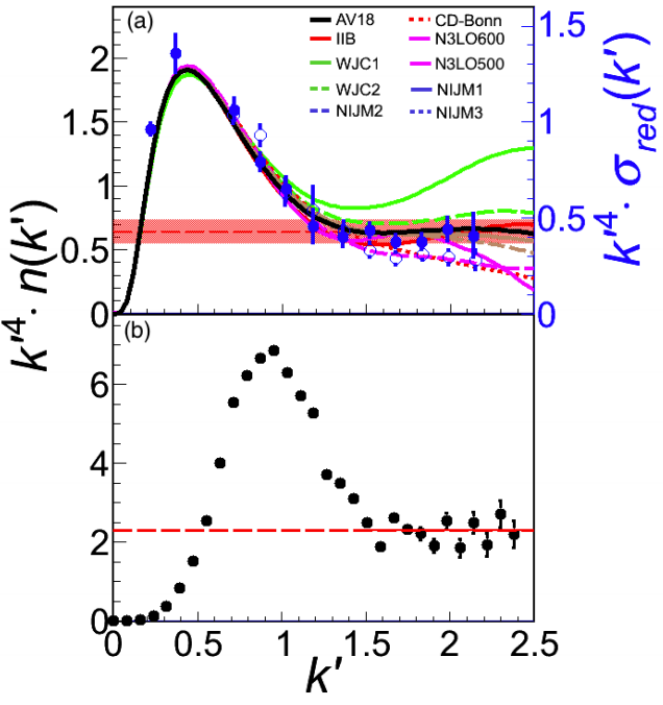}
\caption{(Color Online). Upper: the ultra-cold atomic Fermi gas and the atomic nuclei share very similar behavior although the densities of these two systems are different by 25 orders of magnitude. Lower: the scaled momentum distribution, $k'^4n_{k'}$ where $k'
= |\v{k}|/k_{\rm{F}}$ for (a) deuteron and (b) atomic systems. Figures taken from Ref.\cite{Hen15a}.}\label{fig_nk4univ}
\end{figure}

We now discuss briefly the $1/|\v{k}|^4$ form of the HMT as well as the strength of contact of the tail, and suggest interested reader to Ref.\cite{LCCX18} for more detailed discussions on related issues. The $1/|\v{k}|^4$ scaling of the HMT is a universal feature in quantum many-body systems\cite{Gio08RMP,Blo08RMP}. 
For example, the upper panel of FIG.\,\ref{fig_nk4univ} shows that the ultra-cold atomic Fermi gas and the atomic nuclei share the very similarity although the densities of these two systems are different by 25 orders of magnitude\cite{Hen15a}.
For deuterons, on the other hand, variational many-body calculations reproduce this behavior within $\sim10\%$, consistent with $\rm{d(e,e'p)}$ data\cite{Hen15a}. In SNM at $\rho_0$, the HMT magnitude is
$C_{\rm{SNM}} = C_0 = R_{\rm{d}} a_2(\infty) k_{\rm{F}} \rho / k_{\rm{F}}^4$, where $R_{\rm{d}}$ characterizes the plateau shown in the lower panel of FIG.\,\ref{fig_nk4univ}. Moreover, $\rm{d(e,e'p)}$ cross sections in the region $1.3 k_{\rm{F}} \lesssim |\v{k}| \lesssim 2.5 k_{\rm{F}}$ also follow the $|\v{k}|^{-4}$ scaling. From these theoretical and experimental analyses, $C_0 \approx 0.15 \pm 0.03$ is obtained\cite{Hen15a}, after correcting for normalization differences.
Independent photonuclear absorption studies further support the $C/|\v{k}|^4$ behavior and connect $C_0$ to the Levinger constant $L$: $
C_0 = \rho \pi L / a_{\rm{t}} k_{\rm{F}}^4 = 2 L / 3 \pi k_{\rm{F}} a_{\rm{t}}$,
with $a_{\rm{t}}$ being the scattering length for the $^3\rm{S}_1$ channel, $\rho=\rho_{\rm n}+\rho_{\rm p}=2k_{\rm F}^3/3\pi^2$ is nucleon density including both protons and neutrons (therefore a factor of 2).
Using $1/k_{\rm{F}} a_{\rm{t}} \approx 0.15$ and $L \approx 5.50 \pm 0.21$\cite{Wei15} gives $C_0 \approx 0.172 \pm 0.007$\cite{Wei15}, consistent with previous determinations. Averaging these results yields $C_0 \approx 0.161 \pm 0.015$\cite{Cai15a,CaiLi16a}. Combining this $C_0$ with $x_{\rm{SNM}}^{\rm{HMT}}$, the HMT cutoff in SNM is then obtained as $
\phi_0 = \left(1 - x_{\rm{SNM}}^{\rm{HMT}} / 3 C_0\right)^{-1} \approx 2.38 \pm 0.56$, which is
well within the SRC momentum range of $\sim 300$-$600\,\rm{MeV}$.
To fix all isospin-dependent parameters in Eq.\,(\ref{MDGen}), as boundary conditions one also needs information about the EOS and single-neutron momentum distribution in PNM.
Some of the information can be obtained from the state-of-the-art microscopic many-body calculations at sub-saturation densities, see Ref.\cite{LCCX18} for some detailed discussion on this issue, here we only list the main results.
In particular, we have $
C_{\rm{n}}^{\rm{PNM}}\approx0.12$, consequently a value of
$C_1\approx-0.25\pm0.07$ is obtained. Similarly, the high momentum
cutoff parameter for PNM is found to be about
$\phi_{\rm{n}}^{\rm{PNM}}\equiv
\phi_0(1+\phi_1)=(1-x_{\rm{PNM}}^{\rm{HMT}}/3C_{\rm{n}}^{\rm{PNM}})^{-1}\approx1.04\pm0.02$.
It is easy to understand that the $\phi_{\rm{n}}^{\rm{PNM}}$ is very
close to unity, since the high momentum neutron fraction in PNM is only about 1.5\%.
Finally, inserting the $\phi_0$ obtained earlier for SNM,
$\phi_1\approx-0.56\pm0.10$ is found. 

\begin{figure}[h!]
\centering
\includegraphics[height=5.5cm]{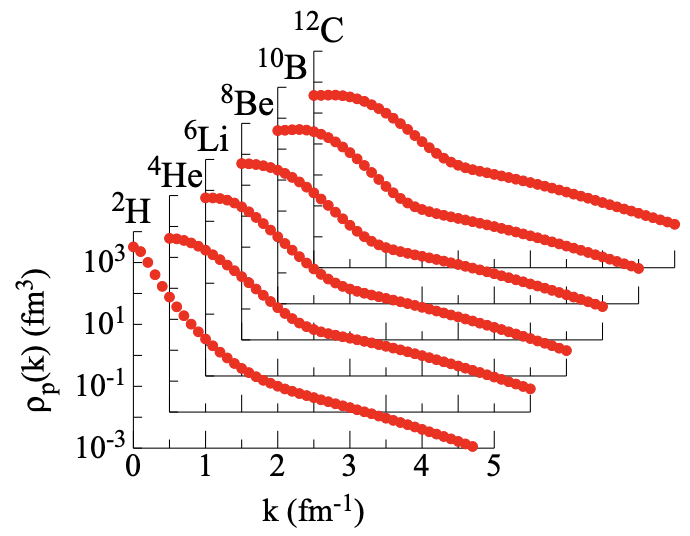}\\[0.5cm]
\includegraphics[height=6.5cm]{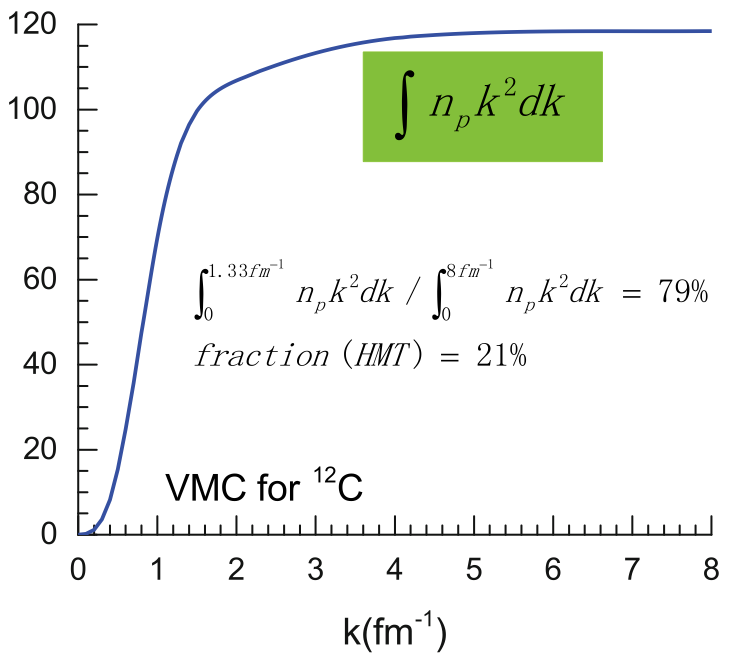}
\caption{(Color Online). Upper: the proton momentum distributions
in all nuclei from $A = 2-12$ calculated by QMC simulations at $T=0$; figure taken from Ref.\cite{Wir14}. Lower: the un-normalized population of nucleons in $^{12}\rm{C}$ up to momentum $k$ (blue) from the prediction shown in the upper panel.}\label{fig_Wir-nk}
\end{figure}

It is worth commenting on the uncertainties associated with the HMT fractions, particularly in PNM. While various many-body approaches consistently predict the qualitative effects of SRCs on the HMT in agreement with experimental observations, the predicted magnitude of the HMT depends on both the model and the nuclear interaction employed. For instance, SCGF calculations using the AV18 interaction yield an HMT fraction of (11-13)\% for SNM at saturation density $\rho_0$\cite{Rio09,Rio14}. In contrast, recent BHF results span a wider range, from roughly 10\% with the N$^3$LO450 interaction to over 20\% with AV18, Paris, or Nij93 interactions\cite{ZHLi16PRC}, whereas variational Monte Carlo (VMC) calculations for $^{12}$C indicate an HMT of about 21\% in SNM\cite{Wir14}; see FIG.\,\ref{fig_Wir-nk} for the predictions.
Direct experimental information on the isospin dependence of the HMT remains scarce. As discussed earlier, np SRC pairs occur roughly 18-20 times more frequently than pp pairs\cite{Hen14,Sub08}, leading to an estimated HMT in PNM of approximately 1-2\%\cite{Hen15a}. However, some theoretical studies, such as SCGF calculations, suggest a significantly larger HMT in PNM, around 4-5\%\cite{Rio09,Rio14}. This uncertainty in the HMT magnitude for PNM affects certain quantitative predictions in model calculations. Future calculations of the HMT in PNM using the same models and interactions applied to the PNM EOS are highly desirable to improve accuracy.  
A better understanding of the HMT in PNM is crucial. For example, in addition to the HMT-exp parameter set described in the above, the HMT-SCGF parameter set is estimated as $x_{\rm{SNM}}^{\rm{HMT}}\approx12\%$, $x_{\rm{PNM}}^{\rm{HMT}}\approx4\%$, $\phi_0\approx1.49$, $\phi_1\approx-0.25$, $C_0\approx0.121$ and $C_1\approx-0.01$\cite{Cai16b}.
Overall, the fraction of HMT nucleons in SNM is generally expected to exceed about 12\%, whereas in PNM it is much smaller, typically $\lesssim 5\%$. In astrophysical processes such as core collapse supernovae, where matter spans a wide range of proton fractions $x_{\rm p}$ (or equivalently isospin asymmetries $\delta$), the impact of SRC-HMT components may therefore exhibit non-trivial and composition-dependent features, warranting more detailed and systematic investigations. We will use these HMT sets (namely the HMT-exp and HMT-SCGF) in the following discussions.

As illustrated in FIG.\,\ref{fig_nk4univ}, the HMT of nucleons exhibits a ${1}/{|\v{k}|^4}$ behavior that closely resembles that observed in two-component (spin-up and spin-down) cold atomic Fermi gases. For the latter systems, this behavior was first predicted by Tan\cite{Tan08-a,Tan08-b,Tan08-c} and subsequently confirmed experimentally\cite{Gio08RMP}. It is important to note that Tan's general result applies to all two-component Fermi systems with a contact interaction, provided the s-wave scattering length $a$ is much larger than the inter-particle distance $d \sim \rho^{-1/3}$, which itself must greatly exceed the effective interaction range $r_{\rm{eff}}$, i.e., $r_{\rm{eff}} \ll d \ll a$. In the unitary limit, $k_{\rm{F}}a \rightarrow \pm \infty$, Tan's prediction becomes universal for all Fermi systems.  
In nuclei and SNM, however, the HMT is predominantly generated by tensor-force-induced np pairs with $a \approx 5.4\,\rm{fm}$ and $d \approx 1.8\,\rm{fm}$ at $\rho_0$, as noted in Refs.\cite{Hen15a,Wei15}. Consequently, the stringent conditions required for unitary Fermi gases are not fully satisfied in ordinary nuclei and SNM. Nevertheless, the nearly identical ${1}/{|\v{k}|^4}$ behavior observed in both nuclear systems and cold atomic Fermi gases suggests the presence of underlying universal physics, which merits further theoretical and experimental exploration\cite{Hen15a}.

\begin{figure}[h!]
\centering
\includegraphics[width=8.5cm]{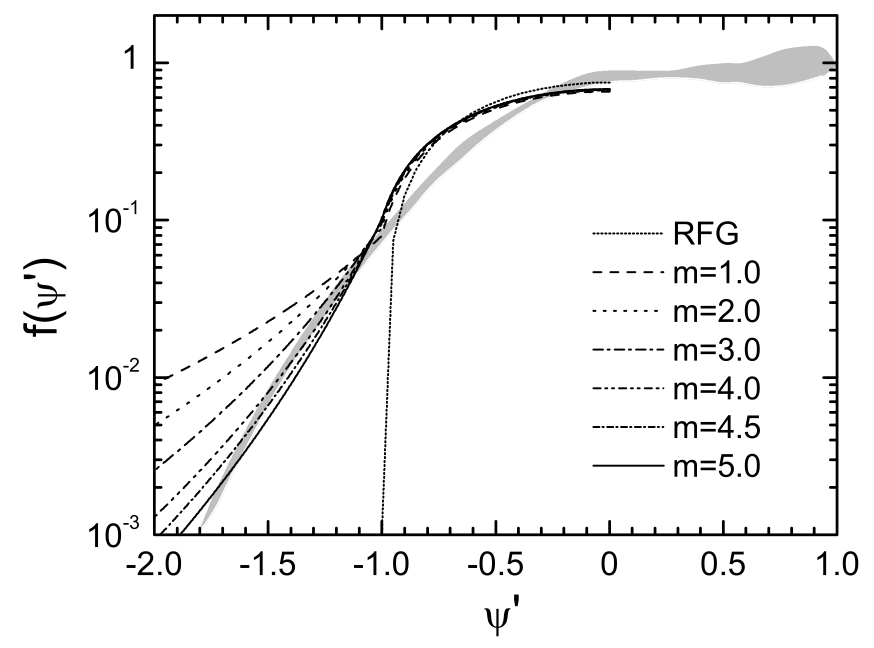}
\caption{Scaling function in a dilute Fermi gas for different values of $m$ in the asymptotics of the momentum
distribution $n(k)\sim 1/k^{4+m}$ given in comparison with the relativistic Fermi gas (RFG) result; grey area shows experimental data from Ref.\cite{Donn99PRC}. Figure taken from Ref.\cite{Ant07PRC}.}
\label{fig_psipr}
\end{figure}

\begin{figure*}[h!]
\centering
\includegraphics[width=14.cm]{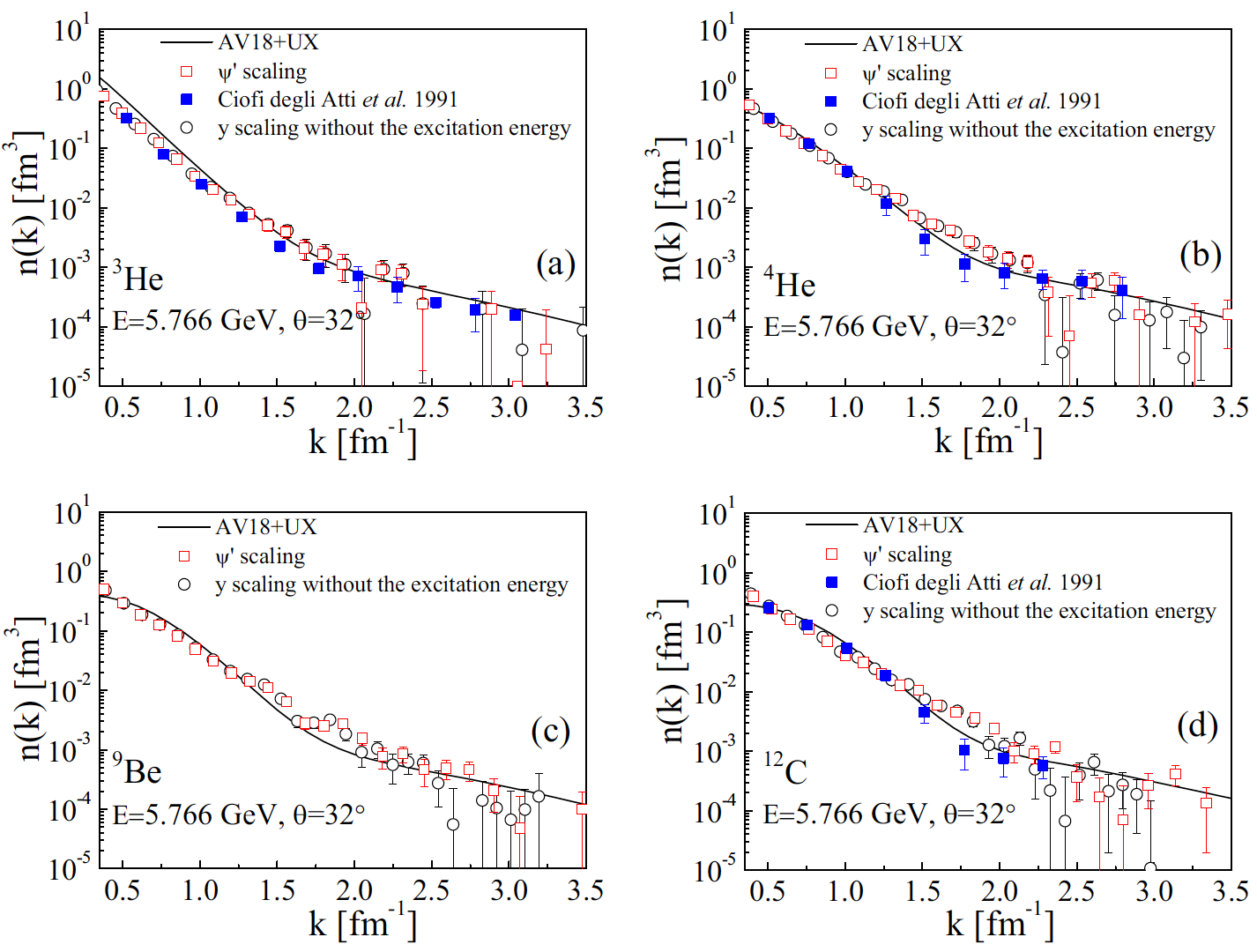}
\caption{(Color Online). Nucleon momentum distributions $n(k)$ of $^3\rm{He}$, $^4\rm{He}$, $^9\rm{Be}$ and $^{12}\rm{C}$ extracted from the latest inclusive electron scattering data using $\psi'$-scaling (red squares), $y$-scaling (circles), and results from Ref.\cite{Att91} (blue solid squares). Figure taken from Ref.\cite{Liang22PRC}.
}
\label{fig_Liang-nk}
\end{figure*}

As discussed in Ref.\cite{Hen15a}, although the density of ANM differs from that of ultra-cold Fermi gases by nearly 25 orders of magnitude, the contact that quantifies the probability of finding particle pairs turns out to be surprisingly similar. In a two-component Fermi gas, the two-body correlation function has the general structure $
\langle\psi_1^{\dag}(\v{r}')\psi_2^{\dag}(\v{0})
\psi_2(\v{0})\psi_1(\v{r})\rangle
=\sum_i\gamma_i\phi_i(\v{r})\phi_i^{\ast}(\v{r}')$,
treating it as a Hermitian operator in $\v{r}$ and $\v{r}'$. For separations $r\ll d$, where $d$ denotes the inter-particle distance, the functions $\phi_i$, which act as s-wave Jastrow factors describing the short-distance structure of the many-body wave function, are determined by two-body physics. They take the form $\sin[kr+\delta_0(k)]/r$, where $\delta_0(k)$ is the s-wave phase shift and $k$ is the wave vector of the scattering state. The s-wave scattering amplitude
\begin{equation}
f_0(k)=[-k\cot\delta_0(k)+ik]^{-1}
\end{equation}
is independent of the scattering angle and, in the low-energy limit $k\to 0$, approaches $f_0(k\to0)=-a$, with $a$ being the s-wave scattering length that governs low-energy scattering\cite{Gio08RMP}. Including contributions up to order $k^2$ in the small-$k$ expansion of $\delta_0(k)$ leads to
\begin{equation}
f_0(k)=-\frac{1}{a^{-1}-k^2r_{\rm{eff}}/2+ik},
\end{equation}
where $r_{\rm{eff}}$ is the effective range of the interaction. In the unitary limit $a\to\infty$, and for wave vectors $k\ll r_{\rm{eff}}^{-1}$, the above expression reduces to the universal form $f_0(k)=i/k$, independent of the interaction details. Under these conditions, the wave function $\sin(kr+\delta_0)/r$ simplifies to $\sin\delta_0\chi(r)/r$ with $\chi(r)=1-r/a$, and thus one obtains
$
\langle\psi_1^{\dag}(\v{r}')\psi_2^{\dag}(\v{0})
\psi_2(\v{0})\psi_1(\v{r})\rangle
=C({\chi(r)}/{r})^2
=C(r^{-1}-a^{-1})^2$,
where $C$ is Tan's contact parameter. At unitarity, the Fourier transform of this short-distance structure generates a high-momentum tail proportional to $|\v{k}|^{-4}$ in the momentum distribution.
Many interesting physical issues on Tan's contact theory and implications are given in Refs.\cite{Gio08RMP,Blo08RMP,BECBOOK}.

While it is intriguing that both nuclei and cold atomic systems near the unitary limit exhibit similar ${1}/{|\v{k}|^{4}}$ HMT shapes, it should be emphasized that, for the nuclear case, most supporting evidence remains theoretical. Only limited indirect experimental indications exist so far, in clear contrast to the situation in ultra-cold atomic gases. As noted earlier, a more thorough investigation of the HMT shape in nuclei and nuclear matter, both theoretically and experimentally, remains necessary. 
In fact, the study of the form and range of the HMT for Fermions with $k>k_{\rm F}$ has a long history, see, e.g., Refs.\cite{Migdal57,Bel61,Sar80,Czy60,Lut60,Gal58}. These early works, however, focused almost exclusively on SNM. In such analyses, the HMT of $n(k)=n_{\v k}$ is often obtained by expanding relevant many-body quantities for a dilute gas of hard spheres in powers of $ak_{\rm F}$\cite{HY,LY}. In general, the asymptotic momentum distribution is related to the Fourier transform $\widetilde{V}_{\rm{NN}}(k)$ of the nucleon-nucleon interaction via
\begin{equation}
n(k) \xrightarrow{k\to\infty} \left[\frac{\widetilde{V}_{\rm{NN}}(k)}{k^{2}}\right]^{2}.
\end{equation}
However, whether the required condition for this asymptotic behavior should be $k\to\infty$ or $k/A\to\infty$ with $A$ being the mass number remains unclear\cite{Ant07PRC,Ant04PRC,Ant05PRC,Ant06PRC-a,Ant06PRC-b,nkB-88,nkB-12}. Moreover, the spin-isospin structure of the short-range NN interaction and the resulting isospin dependence of the HMT are still not well understood.
When a contact interaction is assumed (so $\widetilde{V}_{\rm{NN}}(k)\to\rm{const.}$), the asymptotic tail naturally reduces to $n(k)\sim 1/k^{4}$\cite{Amado76PRC}, consistent with the behavior in ultra-cold atoms near the unitary limit or with the HMT in SNM within the np dominance picture using a zero-range nuclear force. Systematic analyses of scaling functions, such as $y$-scaling and superscaling, in the quasi-elastic (QE) region of inclusive electron-nucleus scattering\cite{Ant07PRC} indicate that the momentum distribution follows $n(k)\sim k^{-4-m}$ with $m\approx 4\sim5$ for $k$ up to $(1.59\sim 1.97)k_{\rm F}$, where $k_{\rm F}\approx250\,\mathrm{MeV}/c$ is the Fermi momentum. The corresponding NN potentials extracted via inverse Fourier transformation behave as $V_{\rm{NN}}(r)\sim 1/r$ for $m=4$ and $V_{\rm{NN}}(r)\sim (1/r)^{1/2}$ for $m=5$.

As an illustration, FIG.\,\ref{fig_psipr} compares scaling functions calculated for different values of $m$ with both the relativistic Fermi Gas (RFG) model and the experimental QE scaling function from Ref.\cite{Ant07PRC}. A good description of the experimental data is obtained with $m\approx 4.5$, implying a momentum distribution of the form $n(k)\sim k^{-8.5}$. This example demonstrates how studying the HMT can provide insights into the short-range behavior of nuclear forces. 
In a recent study, the authors of Ref.\cite{Liang22PRC} extracted the nucleon momentum distributions for the deuteron, $^3\mathrm{He}$, $^4\mathrm{He}$, $^9\mathrm{Be}$ and $^{12}\mathrm{C}$ using the $\psi'$-scaling approach and the latest high-precision data. Their analysis indicates that, for all these nuclei, the $\psi'$-scaling method yields results consistent with modern \textit{ab initio} calculations up to $3.5\,\mathrm{fm}^{-1}$. See FIG.\,\ref{fig_Liang-nk} for the results for $n(k)$ obtained in this way.

\begin{figure}[h!]
\centering
\includegraphics[width=9.cm]{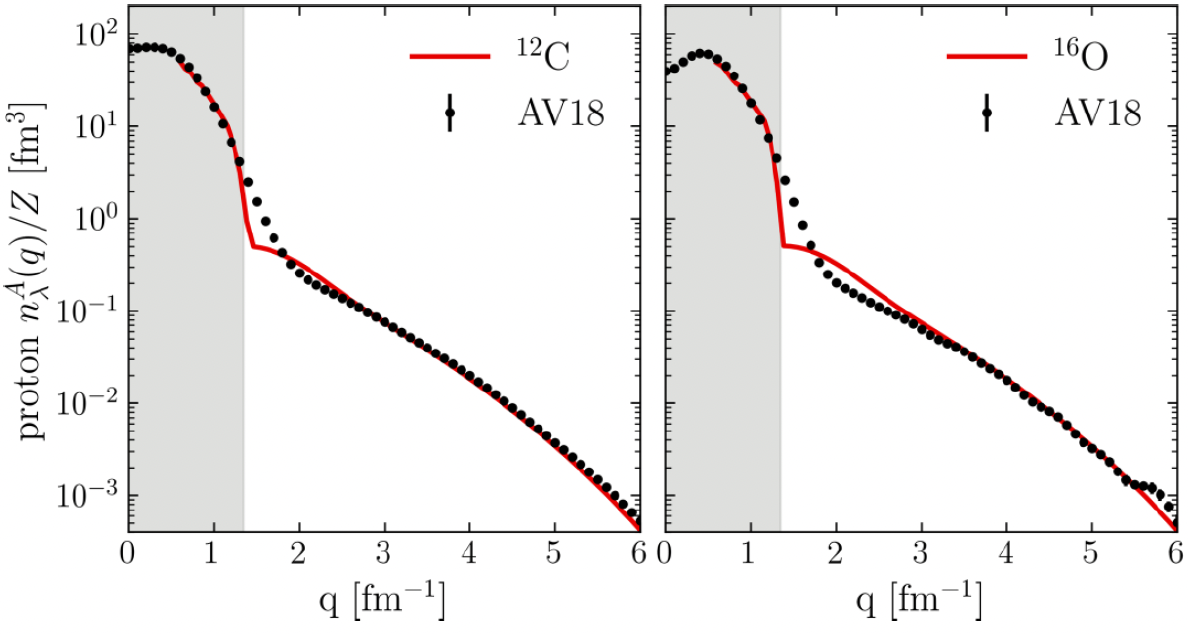}
\caption{(Color Online). Proton momentum distributions for $^{12}\rm{C}$ and $^{16}\rm{O}$ calculated in the LDA for $q > 0.6\,\rm{fm}^{-1}$. It used AV18 to evolve
the operator setting $\lambda = 1.35\,\rm{fm}^{-1}$ using the similarity renormalization group (SRG), and divided each distribution by the proton number $Z$. The gray-shaded sections are where $q < \lambda$. Black
dots correspond to AV18 QMC calculations\cite{QMC-W}. Figure taken from Ref.\cite{Trop21PRC}. }
\label{fig_SRG-nk}
\end{figure}

\begin{figure}[h!]
\centering
\includegraphics[width=7.cm]{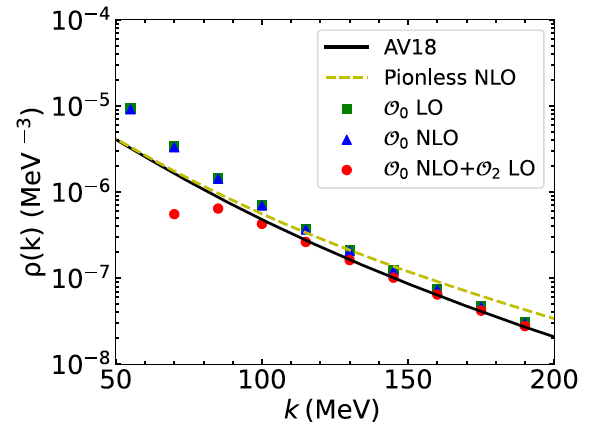}\\[0.25cm]
\includegraphics[width=7.cm]{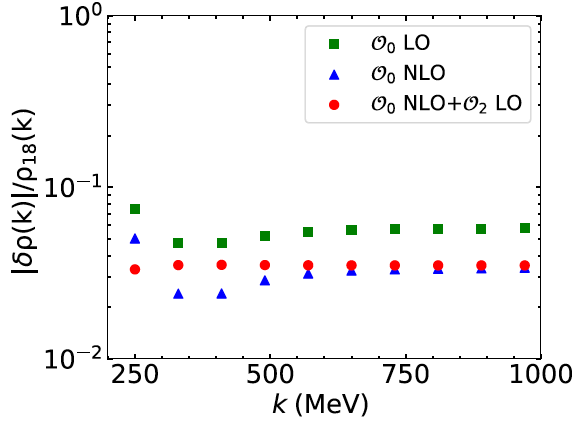}
\caption{(Color Online). Upper: the deuteron single nucleon momentum distributions
calculated using the OPE-Pionless EFT approach. Lower: the relative difference between $\rho_{\rm{d}}(k)$ from the OPE-EFT
and the AV18 potential at large momenta. Figures taken from Ref.\cite{Yu25PRC}.}
\label{fig_OPE-nk}
\end{figure}

An alternative interpretation, however, arises from renormalization-group (RG) evolution to low RG resolution: while SRCs at high resolution appear as large relative-momentum components in the nuclear wave function, RG evolution shifts this physics into the reaction operators without altering observables. This scale separation leads to wave-function factorization, enabling descriptions based on the generalized contact formalism or the low-order correlation operator approximation. At low resolution, the same SRC-driven experimental signatures can be reproduced using simple two-body operators and local-density approximations (LDA) built from uncorrelated wave functions, with the underlying dynamics traced to well-established features of the nucleon-nucleon interaction, such as the tensor force. As an illustration, FIG.\,\ref{fig_SRG-nk} shows the LDA proton momentum distributions obtained with the AV18 potential using proton and neutron densities from the Skyrme functional, and compared with QMC results\cite{QMC-W}. For momenta above the gray-shaded region, the LDA describes the high-momentum tails reasonably well, particularly at the largest $q$, while deviations appear below $q=0.6\,\mathrm{fm}^{-1}$ where the approximations break down. The nearly identical high-momentum behavior across nuclei reflects the universal nature of the HMT, whereas the sharp cutoff near $k_{\rm F}$ is expected to be smoothed once higher-order density-matrix-expansion terms and long-range correlations are included\cite{Trop21PRC}.
In a very recent work, the authors of Ref.\cite{Yu25PRC} applied the operator product expansion (OPE) within the framework of pionless effective field theory (EFT) to investigate the short-range structure of nuclei. By matching the OPE to selected nuclear potentials for nucleon-nucleon scattering states, the corresponding Wilson coefficients are extracted. The nucleon momentum distribution in the deuteron is subsequently employed to benchmark the OPE predictions against those of these nuclear potentials. To enable a more systematic separation of short- and long-range dynamics, this study further discusses how the OPE approximation can be refined through the inclusion of higher-order EFT interactions and higher-dimension local operators. The upper panel of FIG.\,\ref{fig_OPE-nk} shows the deuteron single nucleon distribution function obtained using the OPE-EFT approach, while the lower panel is the relative deviation\cite{Yu25PRC}:
\begin{equation}
    \left|{\delta\rho(k)}/{\rho_{18}(k)}\right|
    \equiv\left|{\rho_{\rm{OPE}}(k)}/{\rho_{18}(k)}-1\right|,
\end{equation}
where $\rho_{18}(k)$ is the $\rho_{\rm d} (k)$ evaluated directly with the AV18 potential.
This comparison provides a concise benchmark of the OPE-EFT method, illustrating how accurately the extracted Wilson coefficients reproduce the deuteron momentum distribution. Importantly, because the Wilson coefficients are independent of the external states, the success of this approach demonstrates that short-range information obtained from two-body scattering states can be systematically applied to many-body bound systems. In this way, the OPE-EFT framework provides a controlled and model-independent method to separate short-range nuclear dynamics from long-range structure, enabling predictive calculations of nucleon momentum distributions in larger nuclei while maintaining consistency with underlying NN interactions.

\begin{figure}[h!]
\centering
\includegraphics[width=7.5cm]{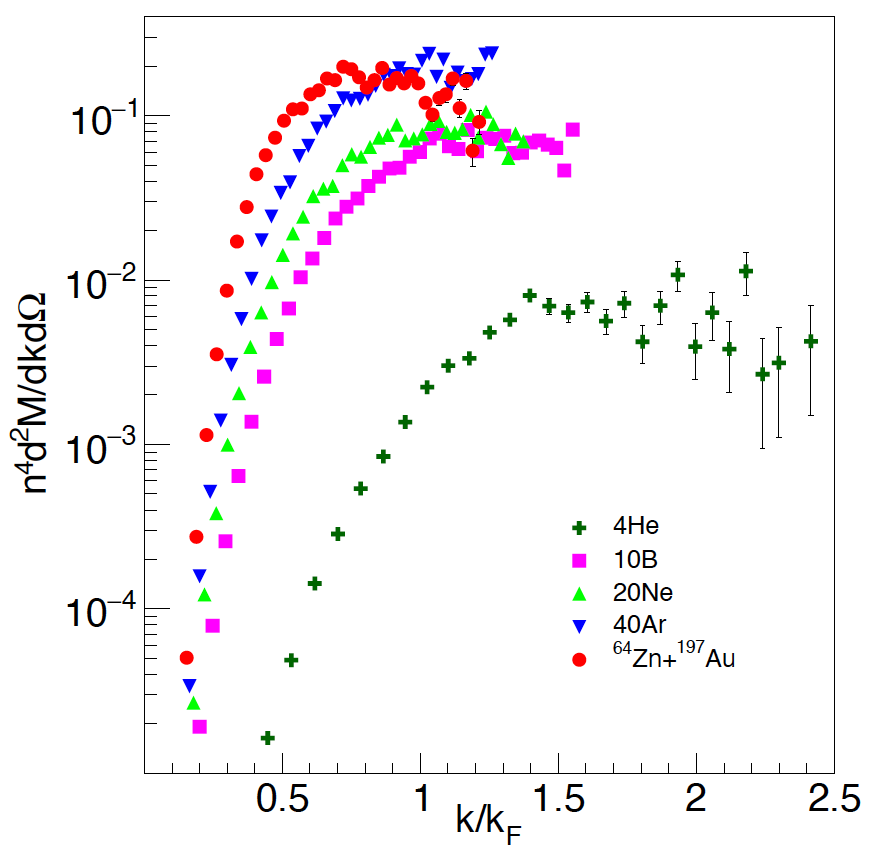}
\caption{(Color Online). Differential cross sections $\d^2M/\d k\d
\Omega$ for proton emission, multiplied by $k^4$ and plotted against
$k/k_{\rm F}$. Figure taken from Ref.\cite{Hagel2021}.}
\label{fig_Hagel-nk}
\end{figure}

\begin{figure}[h!]
\centering
\includegraphics[width=9.cm]{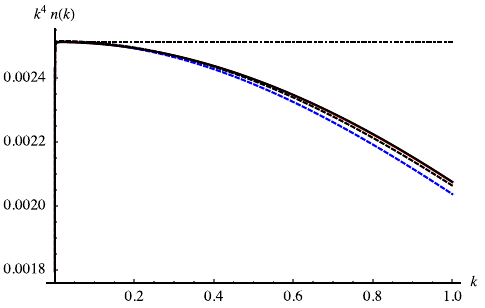}
\caption{(Color Online). Comparison of $k^4n(k)$, for five different wave functions with $n(k) ≡ \widetilde{\phi}^2(k)$. Square well (red) solid, Hulthein (blue) dashed, zero-range (black) dot-dashed, surface delta (black) solid, and exponential (black). Figure taken from Ref.\cite{Miller18PLB}.}
\label{fig_k4n}
\end{figure}

We have previously discussed that the high-momentum fraction in SNM is experimentally found to be about $
x_{\rm{SNM}}^{\rm{HMT}} \approx 28\pm4\%$\cite{Hen14}.
Correspondingly, the high-momentum cutoff parameter $\phi_0$ is constrained as $\phi_0\approx2.4$ through the relation $\phi_0 = (1 - x_{\rm{SNM}}^{\rm{HMT}}/3C_0)^{-1}$. This implies that high-momentum nucleons are confined to the momentum region $k_{\rm{F}} \lesssim k \lesssim 2.4 k_{\rm{F}}$.
As shown in (\ref{kinE}), when calculating the kinetic contribution to the EOS, the integral formally diverges due to the $k^{-4}$ form of the HMT, i.e.,  
$
\int^{\phi_0 k_{\rm{F}}} k^{-4} k^2 \, d\v{k} \sim \int^{\phi_0k_{\rm{F}}} dk \sim \phi_0$,
which diverges linearly as $\phi_0 \to \infty$. This divergence motivates the introduction of the cutoff $\phi_0$. Another motivation is the tension between the experimentally extracted fraction $x_{\rm{HMT}}$ and the contact $C$, since taking $\phi_0 \to \infty$ would yield $x_{\rm{SNM}}^{\rm{HMT}} = 3C_0$, inconsistent with constraints on $x_{\rm{SNM}}^{\rm{HMT}}$. Furthermore, at momenta beyond about $2.4 k_{\rm{F}}$, three-nucleon correlations may become relevant, though these are significantly more complex than two-nucleon interactions.
Beyond the cutoff $\phi_0$, several issues arise concerning the $k^{-4}$ form of the HMT:
\begin{enumerate}[label=(\alph*),leftmargin=*]
\item Although $\phi_0 \approx 2.4$ derived from $\phi_0=(1-x_{\rm{SNM}}^{\rm{HMT}}/3C_0)^{-1}$ is consistent with some experimental analyses, studies of scaling functions such as $y$-scaling in the QE region of inclusive electron-nucleus scattering suggest that $
n_{\v{k}}=n(k) \sim k^{-4-m} $ as we discussed earlier\cite{Ant07PRC}.
This indicates that approximating the HMT as $k^{-4}$ over $k_{\rm{F}} \lesssim k \lesssim 2.4 k_{\rm{F}}$ is not fully accurate. Naturally, using a $k^{-4-m}$ form would resolve the divergence issue in the EOS.

\item Recent analyses of proton energy spectra from reactions with 47\,MeV/u projectiles on Sn and Au targets provide experimental support for the $1/k^4$ behavior of the momentum distributions, and the observed plateau in the differential cross section only extends from $k \gtrsim k_{\rm{F}}$ to about $1.3 k_{\rm{F}}$\cite{Hagel2021}. Data for higher momenta are currently unavailable, see FIG.\,\ref{fig_Hagel-nk}.

\item The $k^{-4}$ form originates from the zero-range assumption of the two-body interaction. However, the momentum distribution cannot be approximated by a zero-range model for 
$
k > k_{\rm{eff}} = (a r_{\rm{eff}}^2)^{-1/3}$,
where $a \approx 5.4\,\rm{fm}$ is the $s$-wave scattering length and $r_{\rm{eff}} \approx m_\phi^{-1}$ is the effective range of the interaction\cite{Miller18PLB}. For example, $r_{\rm{eff}} \approx 1.42\,\rm{fm}$ for $ m_\pi \approx 139\,\rm{MeV}$ or $r_{\rm{eff}} \approx 0.39\,\rm{fm}$ for $m_\phi \approx 500\,\rm{MeV}$. FIG.\,\ref{fig_k4n} illustrates that $k^4 n(k)$ exhibits a plateau in which the contact $
\lim_{k\to\infty} k^4 n(k)$ is approximately constant. Beyond this plateau, $k^4 n(k)$ decreases, implying that $n(k)$ decays faster than $k^{-4}$.

\item Comparisons with ultra-cold atomic Fermi gases show that $k^4 n(k)$ maintains a plateau for $k_{\rm{F}} \lesssim k \lesssim 2.5 k_{\rm{F}}$. In contrast, nuclear matter with realistic interactions (e.g., $\rm{N}^3\rm{LO}$ chiral forces) exhibits a plateau only for smaller $k$, i.e., $k^4 n(k)$ decreases for $k \gtrsim \phi_0 k_{\rm{F}}$ with $\phi_0\approx 1.5$–2\cite{Hen15a} (see FIG.\,\ref{fig_nk4univ}). This highlights that, unlike ultra-cold atoms tuned to the unitary limit with the Feshbach technique\cite{Chin10RMP}, nuclear matter generally cannot be treated as unitary, and the relevant HMT cutoff $\phi_0$ must be smaller.
\end{enumerate}

All these considerations indicate that the $k^{-4}$ HMT is only an approximate description valid for momenta moderately above $k_{\rm{F}}$, but not for arbitrarily large $k$. A natural question then arises: what is the effective form of the HMT at very high momenta? In principle, this requires detailed microscopic calculations incorporating all types of nucleon-nucleon interactions and many-body correlations.
As a first-order estimate, the upper limit of contributions to $E_0^{\rm{kin}}(\rho_0)$ from higher-momentum nucleons can be obtained from the Lippmann--Schwinger equation\cite{Miller18PLB}:  
\begin{equation}\label{LS-nk}
n(k) = \frac{8\pi a^3}{(k^2 a^2 + 1)^2}\left(1 - \Lambda k^2 r_{\rm{eff}}^2\right), 
\end{equation}  
where $\Lambda \sim \mathcal{O}(1)$. Writing $a^{-1} = \xi k$ and $k = \xi' r_{\rm{eff}}^{-1}$, for $a^{-1} \ll k \ll r_{\rm{eff}}^{-1}$ we have $0 < \xi, \xi' \ll 1$, yielding  
\begin{equation}
n(k) \approx \frac{8 \pi a^3}{k^4 a^4} \cdot \left(1 - \Lambda \xi'^2 - 2 \xi^2 + 2 \Lambda \xi^2 \xi'^2 \right) \approx \frac{8 \pi a^3}{k^4 a^4} \sim \frac{1}{k^4},
\end{equation}  
which recovers the conventional $k^{-4}$ form. At even higher $k$, $\xi' \sim \mathcal{O}(1)$, giving  
\begin{align}\label{for-kn6}
n(k)\approx &\frac{8 \pi a^3}{k^4 a^4} \cdot \left[1 - \Lambda \xi'^2 + 2(\Lambda \xi'^2 - 1)\xi^2 + 3(1 - \Lambda \xi'^2) \xi^4 \right]\notag\\
\sim& \frac{\xi^2}{k^4 a^4} \sim \frac{1}{k^6}, \quad \xi = \frac{1}{k a},
\end{align}  
indicating that for $k \sim r_{\rm{eff}}^{-1}$, the HMT can be reasonably modeled as $k^{-6}$. Higher-order corrections scale as $k^{-8}, k^{-10}, \cdots$.

Finally, the contact coefficient $C_0$ exhibits universality in both nuclear systems and ultra-cold gases, and it is reasonable to assume that $C_0$ is independent of the cutoff $\phi_0$. QMC simulations in 3D and 2D using various interactions also support this universality\cite{BECBOOK }. The zero-momentum depletion $\Delta_0$ can in principle be computed via Monte Carlo or other theoretical methods; in this review, it is determined via $
\Delta_0 = 1 - x_{\rm{SNM}}^{\rm{HMT}}$,
directly linking it to the experimental $x_{\rm{SNM}}^{\rm{HMT}} \approx 28\%$. Recent studies further investigated the relation between the contact and mean-field distributions, showing that SRCs are present even if $n(k)$ does not scale as $ 1/k^4$, regardless of whether the system is superfluid or a Fermi-like liquid\cite{Bulgac2023-a,Bulgac2023-b}.
In Subsection \ref{sub_k6_EOSkin} we estimate the contribution of a $k^{-6}$ HMT to the kinetic EOS of nucleonic matter.

\subsection{Nucleon E-effective Mass and its Isospin Splitting}\label{sub_Emass}

\indent 

In this subsection, we discuss the nucleon E-effective mass (or E-mass for brevity). The concept of nucleon effective mass $M_{\rm N}^*$ was originally developed by Brueckner\cite{Bru58} to describe equivalently the motion of nucleons in a momentum-dependent potential with the motion of a quasi-nucleon of mass
$M_{\rm N}^{\ast}$ in a momentum-independent potential. It has been generalized later and widely used to characterize the momentum and/or energy dependence of the single-nucleon potential or the real part of the nucleon self-energy in nuclear medium\cite{Jeu76,Mah85}. Basically, the nucleon effective mass reflects leading effects of the space-time non-locality of underlying nuclear interactions and the Pauli exchange principle. It is a fundamental quantity characterizing the nucleon's propagation in nuclear medium\cite{Jam89,Sjo76,Neg81,Neg82RMP}. In particular, the so-called k-mass and E-mass describe, respectively, the momentum and energy dependence of the  single-nucleon potential while the total effective mass is the product of the two\cite{LCCX18}.

\begin{figure}[h!]
\centering
\includegraphics[width=6.5cm]{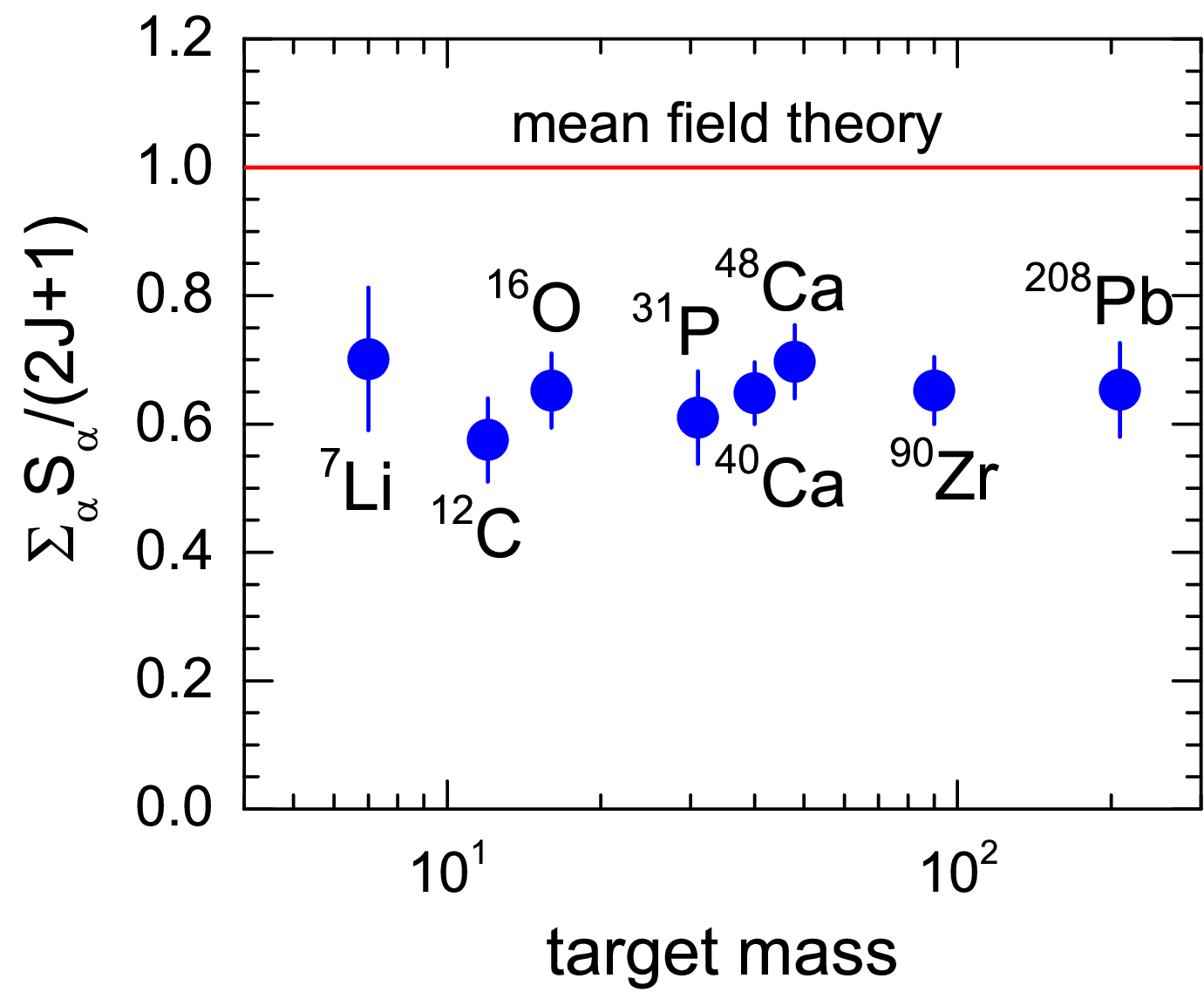}
\caption{(Color Online). Spectroscopic factor as a function of target mass. Taken from Ref.\cite{Lap93}.}\label{fig_SF}
\end{figure}

\begin{figure}[h!]
\centering
\includegraphics[width=9.cm]{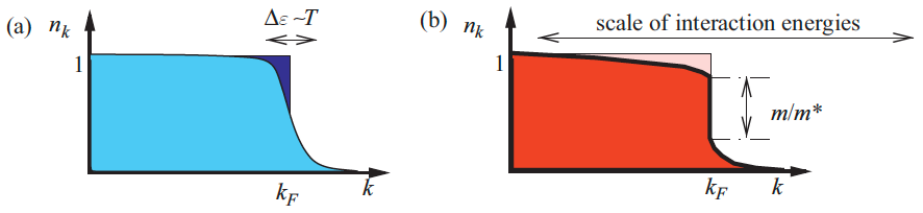}\\[.0cm]
\hspace{-0.5cm}
\includegraphics[width=8.5cm]{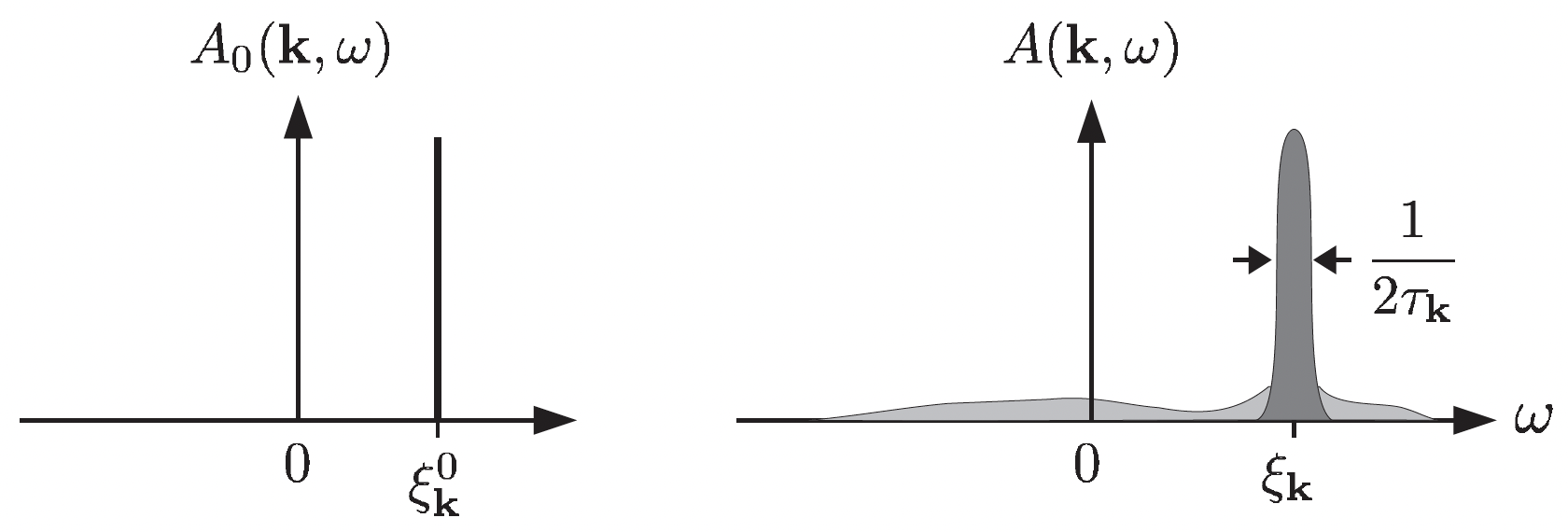}
\caption{(Color Online). Upper: momentum distribution function in a FFG model (left panel) and the one including interactions (right panel); here $m^{\ast}/m$ in panel (b) is the (reduced) effective E-mass, which equals to $Z_{\rm F}^{-1}$.
Figure taken from Ref.\cite{Coleman15B}.
Lower: spectral function in an ideal Fermi gas and in a Fermi liquid;
notation conversion: $A\leftrightarrow S$, $\xi_{\v k}\leftrightarrow\epsilon_{\v k}$, $\gamma_{\v k}\leftrightarrow\Gamma_{\v k}$.
Figure taken from Ref.\cite{Dupuis23}.
}
\label{fig_nkZ}
\end{figure}

\begin{figure*}[h!]
\centering
\includegraphics[width=8.cm]{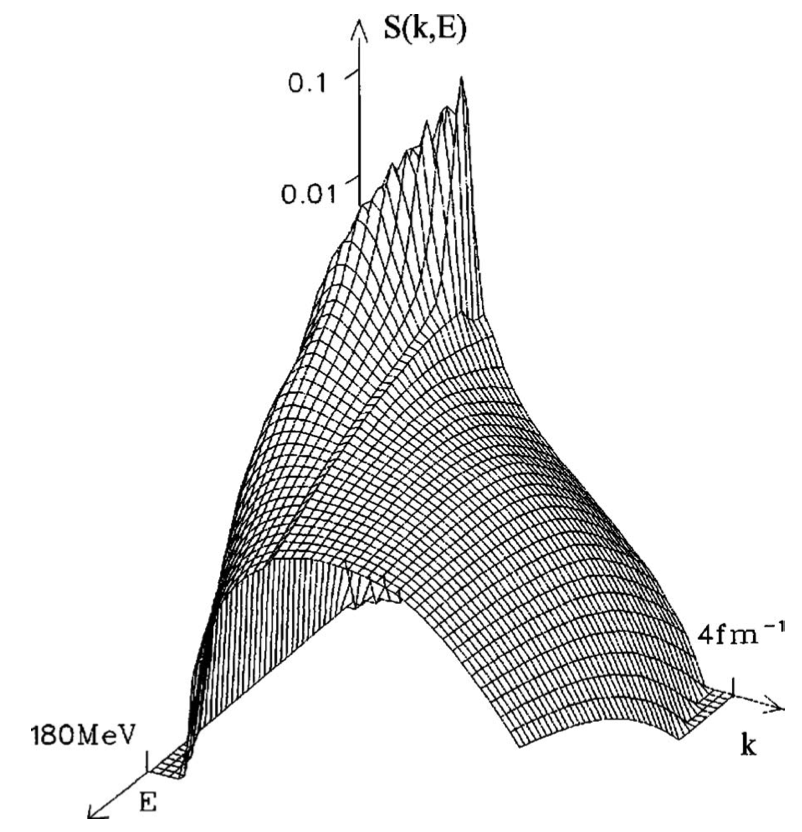}\qquad
\includegraphics[width=6.cm]{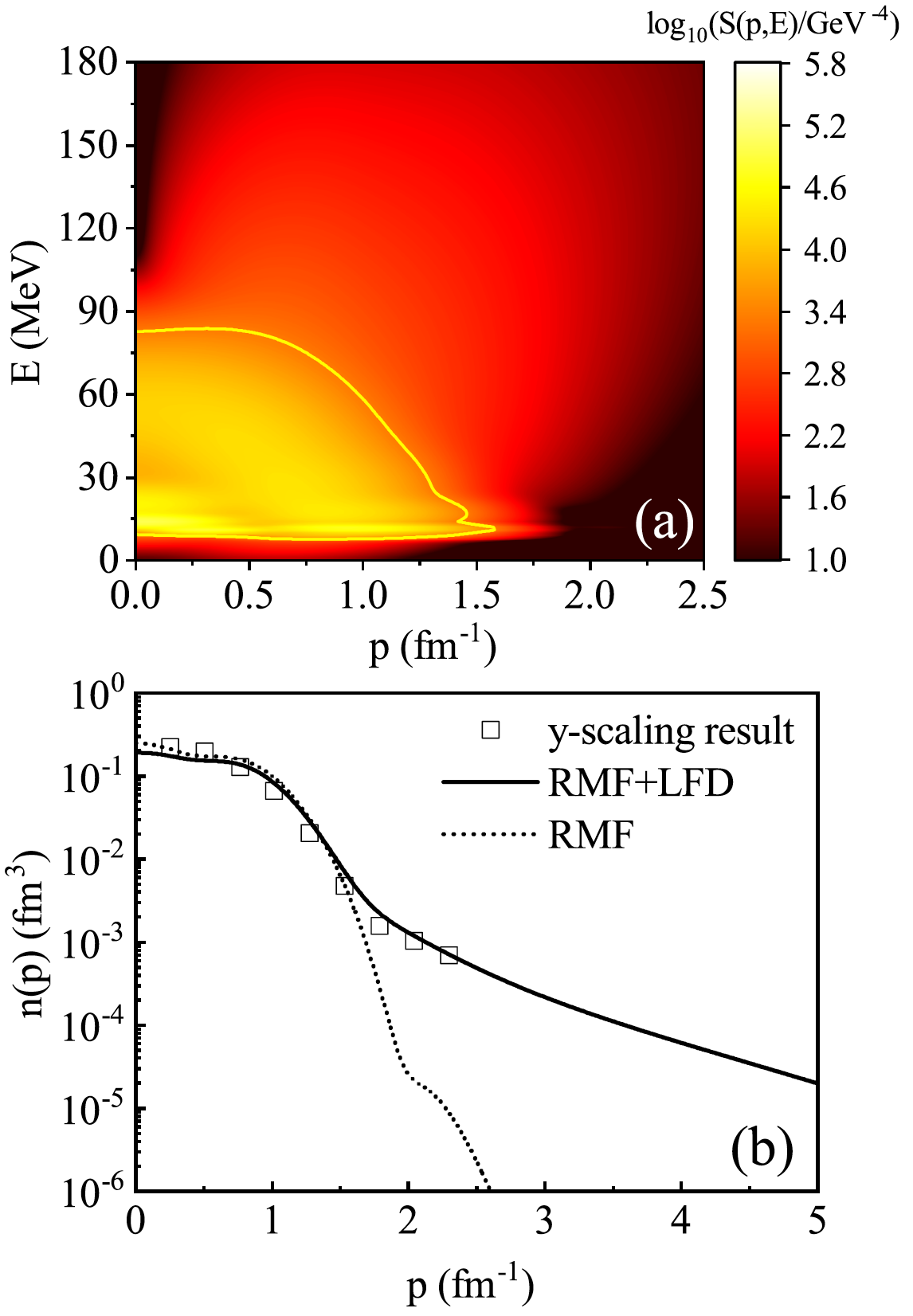}
\caption{(Color Online). 
Left: a typical nucleon spectral function $S(\v{k},E)$; figure taken from Ref.\cite{Benhar08RMP}.
Right: spectral function $S(\v{p},E)$ (upper) and the momentum distribution $n(p)$ (lower) of $^{56}\rm{Fe}$. In panel (a), the logarithm of $S(\v{p}, E)$ is presented to highlight the NN-SRC part, and the region enclosed by a curve describes the mean-field part. The open squares represent the $n(p)$ obtained from $y$-scaling analyses on $\rm{(e, e')}$ cross sections; figure taken from Ref.\cite{Niu22PRC}.}
\label{fig_Niu-nk}
\end{figure*}

\begin{figure}[h!]
\centering
\includegraphics[width=7.5cm]{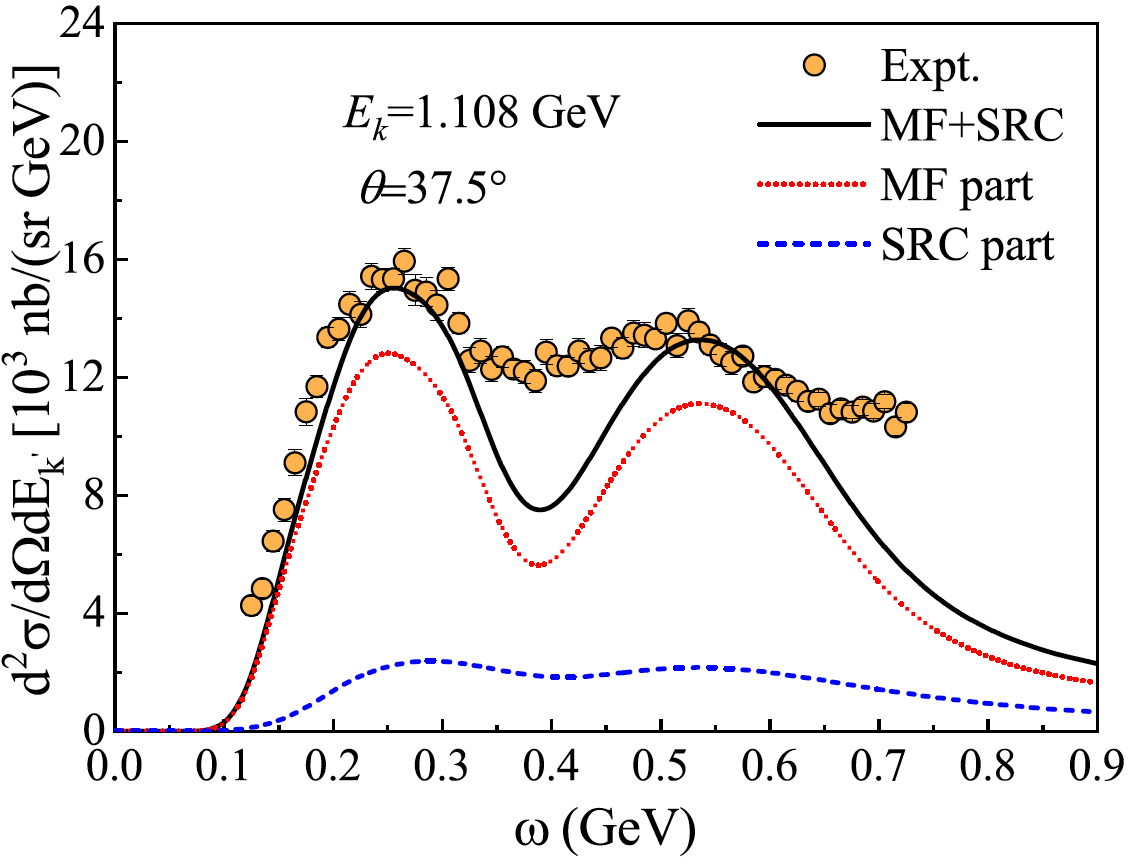}
\caption{(Color Online). 
The $\rm{(e, e')}$ cross sections on $^{56}\rm{Fe}$ using the $S(\v{p}, E)$ of right panel of FIG.\,\ref{fig_Niu-nk}.
Solid line: total inclusive cross sections; dotted line: contributions of mean field part; dashed line: contributions of SRCs.
Figure taken from Ref.\cite{Niu22PRC}.}
\label{fig_Niu-a}
\end{figure}

The E-mass is directly related to the discontinuity of the single-nucleon momentum distribution at the Fermi surface of nuclear matter\cite{Jeu76,Lut60,Czy60,Sar80}, that is\cite{CaiLi16a}
\begin{equation}
    \frac{M^{\ast,\rm{E}}_{J}}{M_{\rm N}}=1-\frac{\partial U_{J}}{\partial E}\equiv \frac{1}{Z_{\rm F}^J},
\text{ with }
Z_{\rm{F}}^J\equiv n_{\v{k}}^J(k^J_{\rm{F}-0})-n_{\v{k}}^J(k^J_{\rm{F}+0}),
\end{equation}
is the Migdal--Luttinger jump\cite{Migdal57,Lut60} of the
single-nucleon momentum distribution $n_{\v{k}}^J$ at the Fermi
momentum $k^J_{\rm{F}}$, see FIG.\,\ref{fig_nkANM}.
Moreover, the $U_J$ here corresponds to the real part of the nucleon self-energy. The role of the $Z$-factor in NS cooling will be discussed in Subsection \ref{sub_NScooling}. In fact, the $Z$-factor renormalizes the density of states at the Fermi surface, and therefore enters any process involving phase-space integrals over fermions in neutron stars.
Over the past decades, the nuclear physics community has devoted significant effort to probing the depletion of the nucleon Fermi sphere, mainly through transfer, pick-up, and $\rm{(e,e'p)}$ reactions. These measurements, usually expressed in terms of nucleon spectroscopic factors, provide direct constraints on $n_{\v{k}}^J(k_{\rm{F}-0}^J)$\cite{Jeu76,Mah85,Jam89,Lap93}. As shown in FIG.\,\ref{fig_SF}, where the spectroscopic factors are displayed as a function of target mass, it is well established that mean-field theory typically overpredicts the data by about 20$\sim$40\%, reflecting its neglect of dynamical correlations. 
In parallel, more recent analyses of inclusive and exclusive electron-nucleus scattering\cite{Hen15a,Hen14,Col15,Egi06,Shn07,Wei11,Kor14,Fa17}, nucleon-nucleus reactions\cite{Tang03,Pia06}, and medium-energy photon-nuclear absorption\cite{Wei15} have provided quantitative information on the size, shape, and isospin dependence of the HMT above the Fermi surface, as discussed in detail in Subsection \ref{sub_nk},  thereby constraining $n_{\v{k}}^J(k_{\rm{F}+0}^J)$. Taken together, these constraints, on both the depletion below $k_{\rm F}^J$ and the HMT above it, offer an important empirical determination of the Migdal--Luttinger jump and hence of the nucleon E-mass in neutron-rich nucleonic matter.

The $Z$-factor is a fundamental quantity in many-body physics and is closely related to the spectral function, which provides a direct link to experimental data\cite{Dick25B}. For instance, the reaction $(\rm{e},\rm{e'}\rm{p})$ can probe the hole spectral function in typical finite nuclei. Because the electron interacts weakly with nucleons, it serves as an ideal probe for studying nuclear structure. In electron scattering, the dominant excitation operators are primarily one-body in nature. When the incident electron has sufficient energy, the knocked-out proton can be energetic enough for the impulse approximation to be applicable. If the electron transfers a substantial amount of energy, the resulting excited state is expected to be dominated by a simple particle-hole configuration, corresponding to an outgoing particle influenced by the surrounding medium and a valence hole selected by the reaction kinematics. Additional interactions between this particle and the hole can then be neglected. In the impulse approximation, the energy $\omega$ and momentum $\v{q}$ lost by the electron are transferred to a proton with binding (missing) energy $E_{\rm{m}}$ and (missing) momentum $-\v{p}_{\rm{m}}$, which exits the nucleus with momentum $\v{p}'=\v{q}-\v{p}_{\rm{m}}$ and kinetic energy $T_{\v{p}'}$. The proton knockout cross section is proportional to the spectral function,
$
\d\sigma\sim S(\v{p}_{\rm{m}},E_{\rm{m}})$,
where $S(\v{p}_{\rm{m}},E_{\rm{m}})$ is the hole spectral function, i.e., the probability of finding a proton with binding energy $E_{\rm{m}}$ and momentum $\v{p}_{\rm{m}}$ in the parent nucleus. The spectroscopic factor of the last valence orbit across different nuclei is shown in FIG.\,\ref{fig_SF}, indicating a roughly 30\% global reduction of single-particle strength for these valence holes in most nuclei. This significant deviation from mean-field predictions requires a detailed explanation based on nuclear correlations. In quantum field theory, the cross section can similarly be computed from time-ordered correlation functions ($S$-matrix) under appropriate conditions.
The nucleon (or other fermions like electrons) momentum distribution can be derived from the Green's function or spectral function as\cite{Ding16PRC}\footnote{Spin and other degrees of freedom are suppressed, so $n_{\v{k}}\to n_{\v{k}\sigma}$, etc.}
\begin{equation}
n_{\v{k}}=\int_{-\infty}^0\frac{\d\omega}{2\pi}{S(\v{k},\omega)},
\end{equation}
where $\omega$ is measured below the Fermi surface. 
The spectral function in nuclear system is discussed in details in Ref.\cite{Rio20}.

When interactions are included, a sharp quasi-particle peak appears in the spectral function\cite{Coleman15B,Ding16PRC},
\begin{align}
{S(\v{k},\omega)}/{2\pi}=&{\pi}^{-1}\rm{Im}\,G(\v{k},\omega-i0^+)
=Z_{\v{k}}\delta(\omega-\epsilon_{\v{k}})+\cdots,
\end{align}
where $\epsilon_{\v{k}}=E_{\v{k}}-\mu$ and $Z_{\v{k}}$ is the wave-function renormalization, interpreted as the residue of the single-nucleon pole in the exact propagator. In a FFG, the spectral function is a sharp delta function at $\epsilon_{\v{k}}$ (lower-left panel of FIG.\,\ref{fig_nkZ}) while in an interacting Fermi liquid the quasi-particle forms a broadened peak of width $\Gamma_{\v{k}}$ at $\epsilon_{\v{k}}$ (lower-right panel of FIG.\,\ref{fig_nkZ}). At the Fermi momentum, this peak becomes infinitely sharp, corresponding to a long-lived quasi-particle on the Fermi surface\cite{KB_BK62}. The quasi-particle peak weight is $Z_{\v{k}}^{-1} \sim M_{\rm N}^{\ast,\rm{E}}$, with $M_{\rm N}^{\ast,\rm{E}}$ the nucleon E-effective mass. The nucleon momentum distribution then reads
\begin{equation}
n_{\v{k}}=Z_{\v{k}}\Theta(-\epsilon_{\v{k}})+\text{smooth background}.
\end{equation}
Finite temperature induces a continuous HMT in $n_{\v{k}}$ (upper-left panel of FIG.\,\ref{fig_nkZ}), while interactions at zero temperature produce a discrete HMT (upper-right panel of FIG.\,\ref{fig_nkZ})\cite{Coleman15B}. Although interactions smear the momentum distribution, the jump at $k_{\rm{F}}$ survives in a reduced form, which is just the Migdal--Luttinger jump\cite{Migdal57,Lut60}. In a Landau Fermi liquid, the Fermi surface volume measures the particle density, and since quasi-particle and unrenormalized particle Fermi surfaces coincide, the Fermi surface volume is an adiabatic invariant under interactions, known as the Luttinger theorem (Luttinger sum rule)\cite{Lut60,Lut60-a}.
A typical nucleon spectral function is shown in the left panel of FIG.\,\ref{fig_Niu-nk}, from which one can see its peaked structure around certain energies and momenta.
Shown in the right panel of FIG.\,\ref{fig_Niu-nk} is the spectral function $S(\v{p},E)$ (upper panel) and the momentum distribution $n(p)$ (lower panel) of $^{56}\rm{Fe}$ using the light-front dynamics (LFD). The correlated part of the spectral function $S_{\rm{corr}}(p)\sim n_{\rm{corr}}(p)$ with the latter being deduced from the LFD wave functions\cite{Carb95}.
In panel (a), the logarithm of $S(\v{p}, E)$ is presented to highlight the NN-SRC part, and the region enclosed by a curve describes the mean-field part. The open squares represent the $n(p)$ obtained from $y$-scaling analyses of $\rm{(e, e')}$ cross sections.
One can see that the ``RMF+LFD'' prediction for the $n(p)$ is quite consistent with the $y$-scaling result.
NN-SRC nucleons account for about  $25\%$ of nucleons, but contribute $\sim 16\%$ to the $\rm{(e,e')}$ cross section at the kinematics of FIG.\,\ref{fig_Niu-a}; their contribution grows at higher $Q^2$, making such kinematics ideal for extracting SRC strength.

\begin{figure}[h!]
\centering
\includegraphics[width=8.cm]{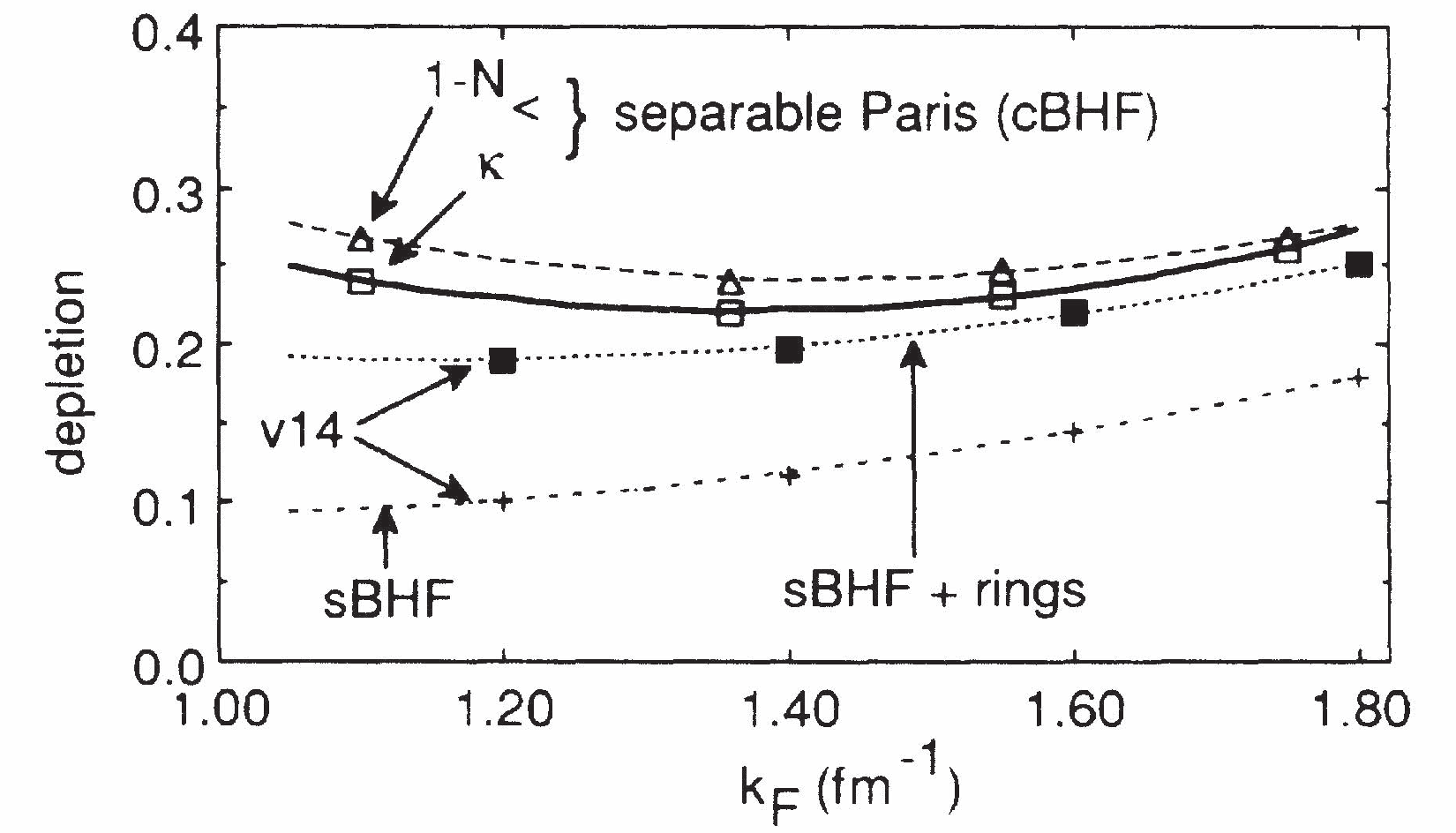}\\[0.25cm]
\hspace{-0.2cm}
\includegraphics[width=8.3cm]{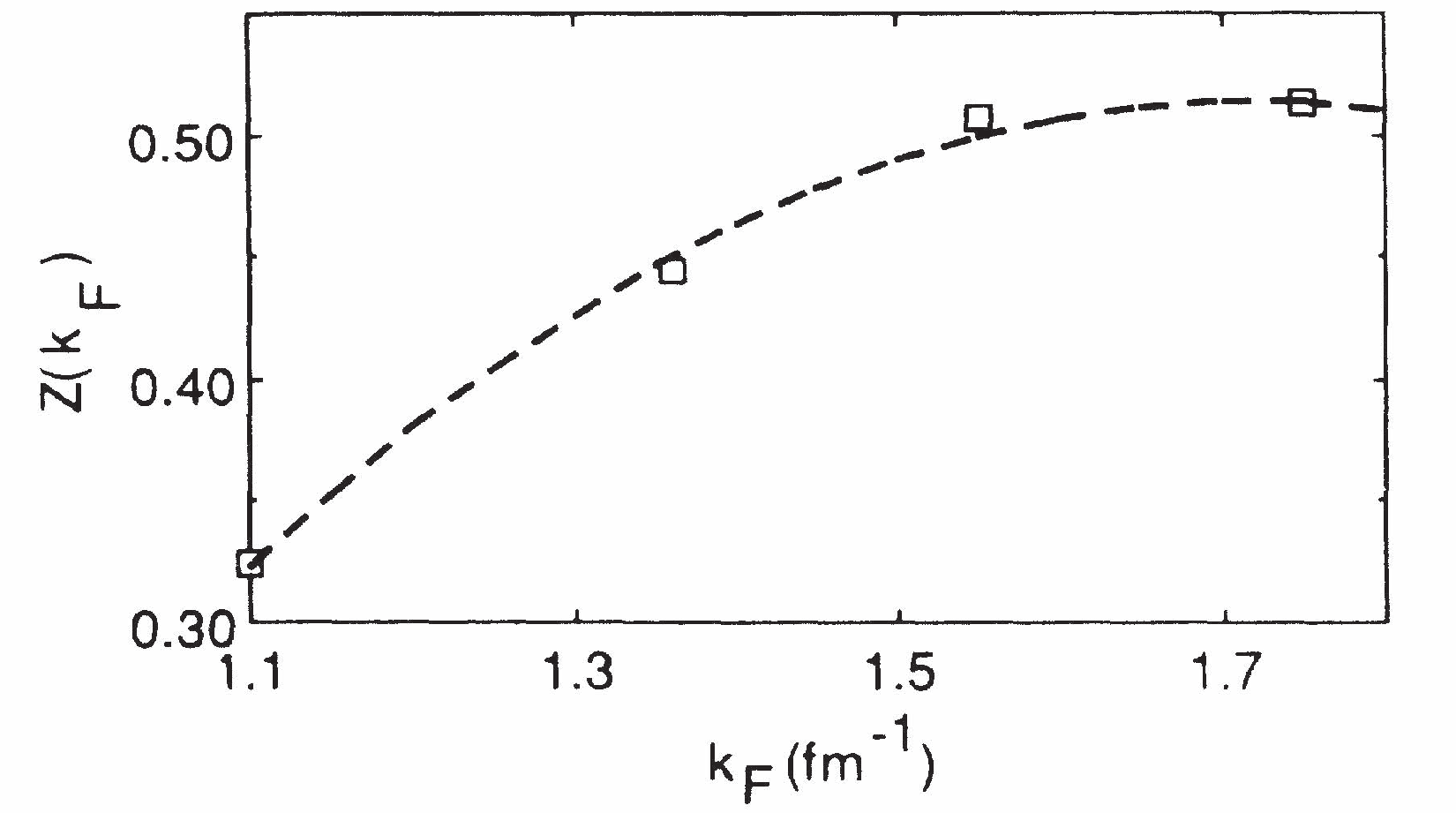}
\caption{Dependence of depletion $\kappa$ (upper) and $Z$-factor (lower) upon $k_{\rm{F}}$ in SNM obtained using the BHF approach with a separable representation of the Paris interaction. Figures taken from Ref.\cite{Bal90}.}
\label{Fig-kappa-Z}
\end{figure}

\begin{figure}[h!]
\centering
\includegraphics[width=9.cm]{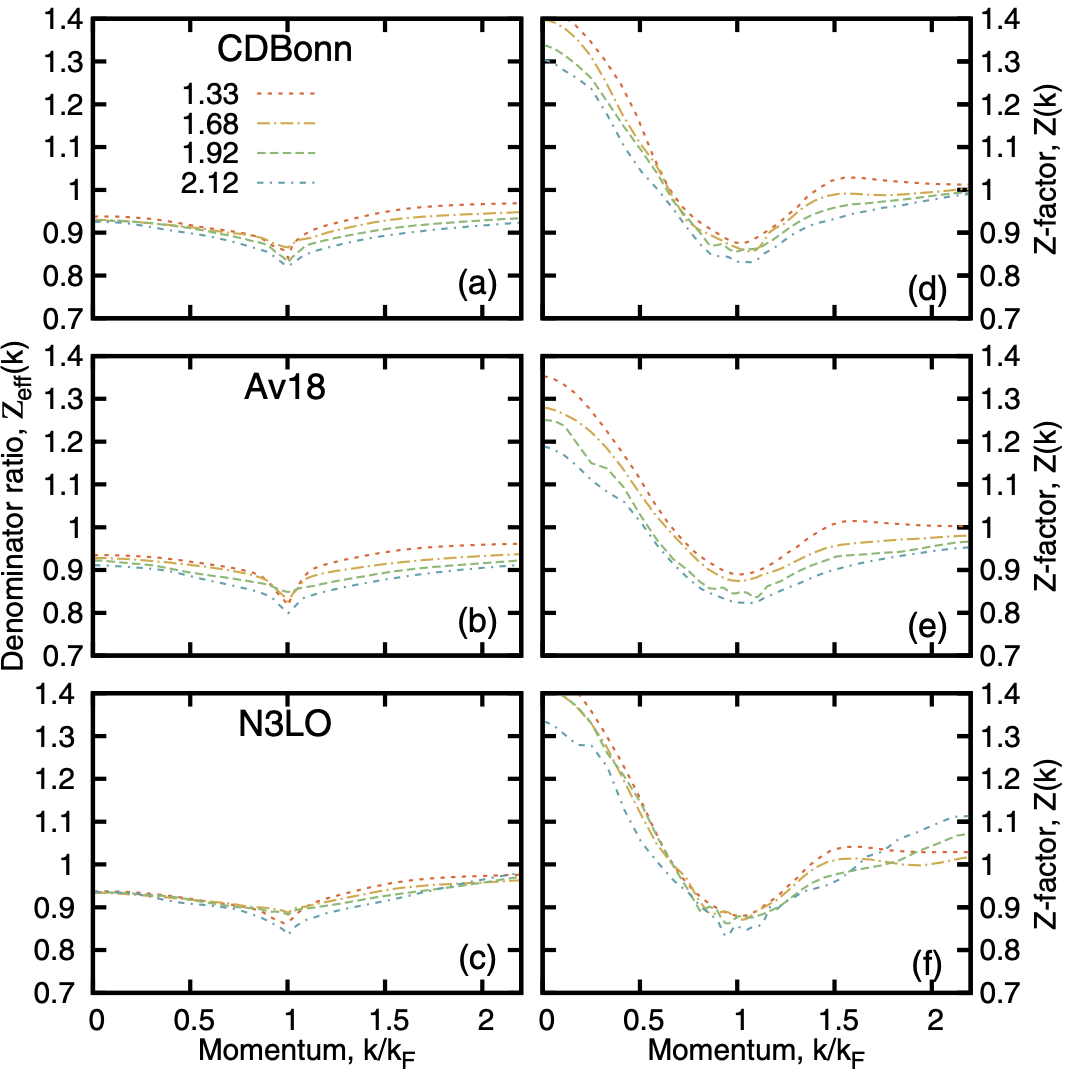}
\caption{(Color Online). The effective renormalization factor $Z_{\rm{eff}}$ at zero temperature as a 
function of momentum for different Fermi momenta, shown in panels (a)–(c), corresponding to the (a) CDBonn, (b) AV18, and (c) N$^3$LO interactions. Panels (d)–(f) display the standard renormalization $Z$-factor under the same conditions. Figure taken from Ref.\cite{Ding16PRC}.
}
\label{fig_Ding16-Z}
\end{figure}

We now discuss the numerical implication on the $Z$-factor and related E-mass.
Most phenomenological models approximate the depletion of the Fermi sea by introducing a constant reduction factor $\Delta_J$, as illustrated in FIG.\,\ref{fig_nkANM}. To explore the role of the two parameters entering $\beta_J=\beta_0(1+\tau_3^J\beta_1\delta)$ in the single-nucleon momentum distribution $n^J_{\v{k}}(\rho,\delta)$ of Eq.\,(\ref{MDGen}), we first recall that these parameters encode how the depletion behaves at finite momenta below and near the Fermi surface, thereby influencing the E-mass obtained from the Migdal--Luttinger theorem. A convenient way to estimate their impact is to adopt a widely used parametrization of the momentum distribution e.g., from Ref.\cite{Cio96}, which is based on microscopic many-body calculations; consequently, the function $I$ adopted in Ref.\cite{Cai15a} takes the form of Eq.\,(\ref{def-I}). The constants in the parametrization of Ref.\cite{Cio96} are then absorbed into $\Delta_J$ and $\beta_J$, and Eq.\,(\ref{DeltaJ}) gives $\Delta_J=1-3\beta_J/5-3{C}_J(1-1/\phi_J)$. This leads in SNM to $\beta_0=(5/3)[1-\Delta_0-x_{\rm{SNM}}^{\rm{HMT}}]$; using the predicted $\Delta_0\approx0.88\pm0.03$\cite{Yin13,Pan99,Fan84} together with $x_{\rm{SNM}}^{\rm{HMT}}\approx0.28\pm0.04$ yields $\beta_0\approx-0.27\pm0.08$. Furthermore, demanding $\beta_J=\beta_0(1+\beta_1\tau_3^J\delta)<0$ so that $n_{\v{k}}^J$ decreases monotonically towards $k_{\rm{F}}^J$ imposes $|\beta_1|\leq1$.
A nonzero $\beta_J$ directly influences the Migdal--Luttinger ``renormalization function'' $Z^J_{\rm{F}}$. For SNM, one has $Z_{\rm{F}}^0=1+2\beta_0/5-C_0-x_{\rm{SNM}}^{\rm{HMT}}\approx0.45\pm0.07$ (or $0.56\pm0.04$ when $\beta_0$ is omitted).

\begin{figure}[h!]
\centering
\includegraphics[width=7.5cm]{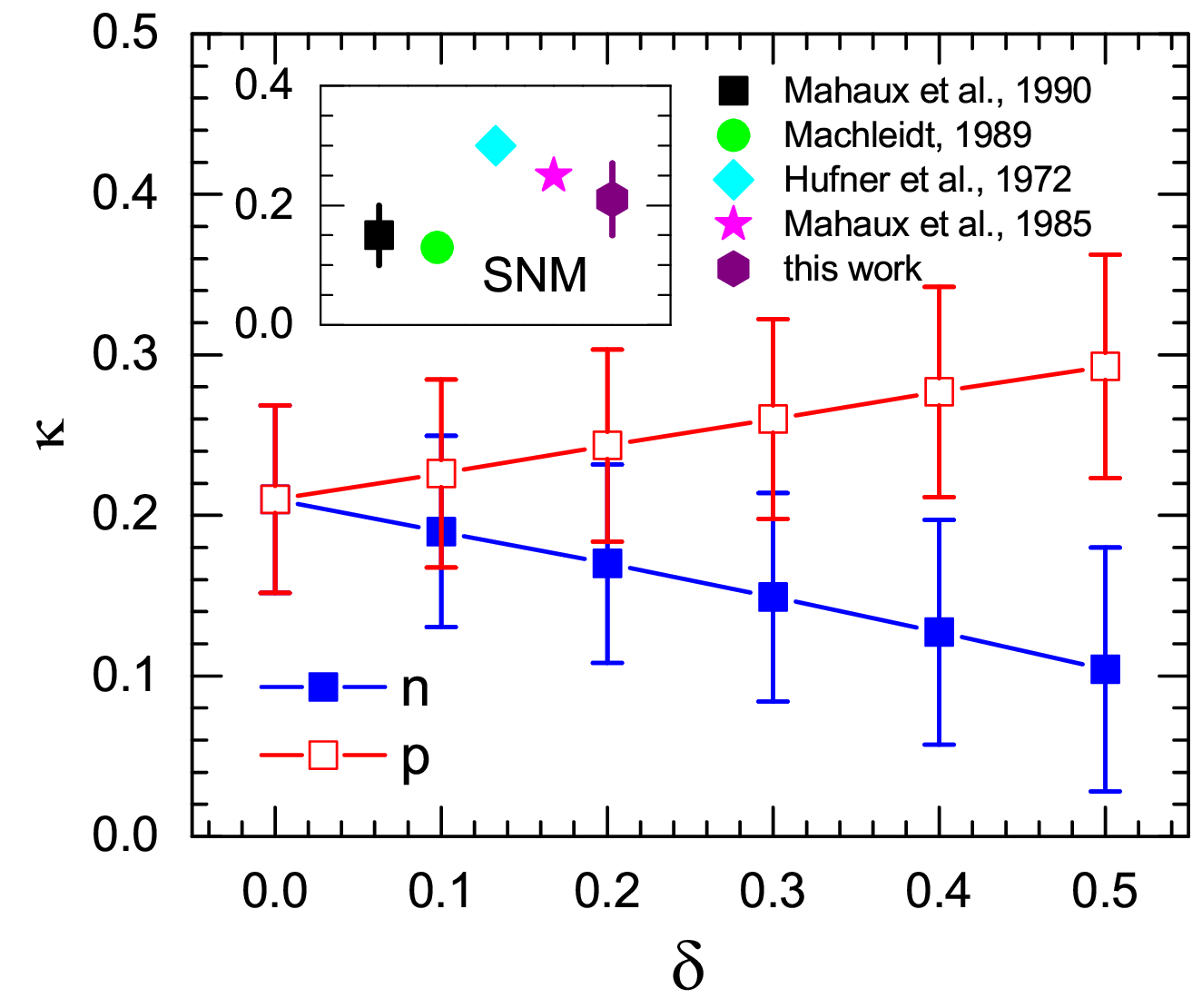}\\[0.25cm]
\includegraphics[width=9.cm]{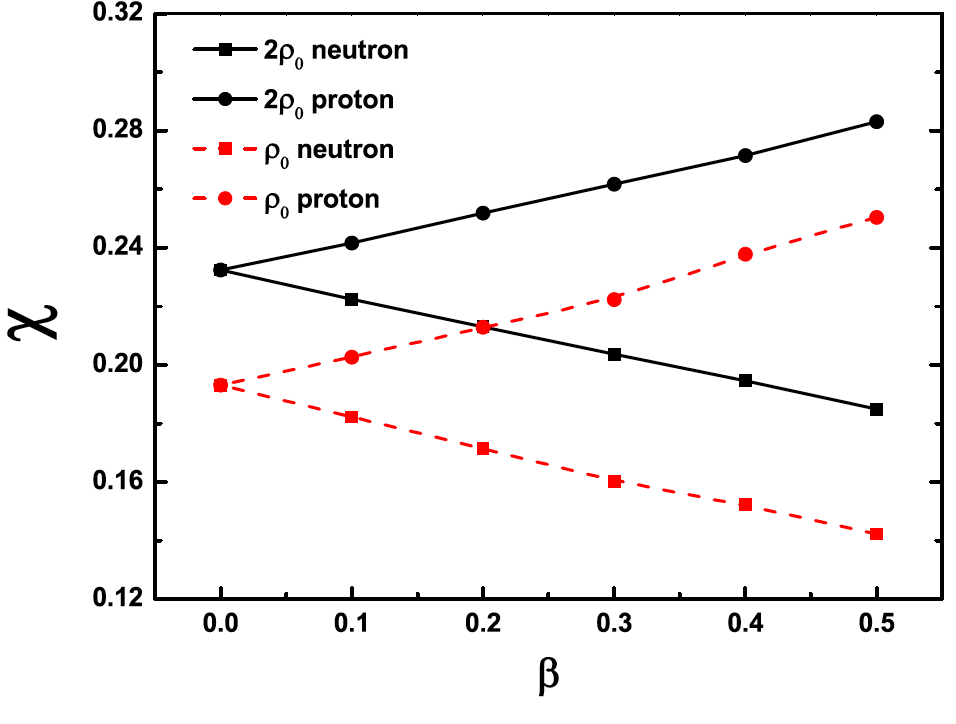}
\caption{(Color Online). Upper: average depletion of neutron and proton Fermi surfaces versus isospin asymmetry. Inset: comparison of average nucleon depletion in SNM from several studies\cite{Mac89,Huf72,Mah85,Mah90,Jon91}. Taken from Ref.\cite{CaiLi16a}.
Lower: the depletion obtained from the BHF theories for protons and neutrons, here $\chi\leftrightarrow\kappa$ and $\beta\leftrightarrow\delta$. Figure taken from Ref.\cite{Yang19PRC}.}
\label{Cai-kappa}
\end{figure}

To apply the Migdal--Luttinger theorem\cite{Migdal57,Lut60} and extract the E-mass, the nucleon momentum distribution must be known both below and above the Fermi surface. In the previous subsection, we focused on the contact and HMT contributions around and above $k_{\rm{F}}^J$; here, we first summarize the depletion below the Fermi surface and then assemble all information for the E-mass extraction.
The depletion parameter $\kappa_J$, which quantifies the occupation deficit below $k_{\rm{F}}^J$, provides a useful indicator of the extent to which the Hugenholtz-van Hove (HVH) theorem\cite{Hug58} and independent-particle pictures may fail. Larger depletion corresponds to stronger deviations\cite{Mah90,Mac89,Huf72,Jon91}. The sensitivity of $\kappa_J$ to the tensor force was established long ago\cite{Fan84,Ram90}. For the parametrization in Eq.\,(\ref{MDGen}), the average depletion at $\rho_0$ in ANM is\cite{CaiLi16a}
\begin{align}
\kappa_J
=&\frac{4}{15}\beta_J+3C_J\left(1-\frac{1}{\phi_J}\right).
\end{align}
In SNM, one finds $\kappa \equiv \kappa_0 = 4\beta_0/15 + x_{\rm{SNM}}^{\rm{HMT}} \approx 0.21 \pm 0.06$. FIG.\,\ref{Fig-kappa-Z} shows the $k_{\rm F}$-dependence of the depletion $\kappa$ (upper) and the corresponding $Z$-factor (lower) calculated within the BHF approach using a separable representation of the Paris interaction\cite{Bal90}. Our extracted Fermi-surface discontinuity $Z_{\rm F}^0$ and the depletion $\kappa$ are found to be very close to the result of Ref.\cite{Bal90}.
Similarly, FIG.\,\ref{fig_Ding16-Z} shows the $Z$-factor obtained in the SCGF approach for the three interactions considered (CDBonn, AV18, and N$^3$LO), see panels (d)-(f). Overall, $Z(k)$ reaches its maximum near $k\approx0$ with values of approximately $1.3$-$1.5$, decreases toward a minimum around $k_{\rm F}$, and then rises again to approach $\approx 1$ at high momenta; see Ref.\cite{Ding16PRC} for more related discussions.
In contrast, the situation in ANM is more uncertain, primarily because the isovector parameter $\beta_1$ remains poorly constrained, despite its direct impact on both $Z_{\rm F}^J$ and the neutron-proton E-mass splitting.
The result is shown in the upper panel of FIG.\,\ref{Cai-kappa}, and the decrease (increase) of neutron (proton) depletion with isospin asymmetry $\delta$ follows an approximate linear trend, in agreement with microscopic studies\cite{Ram90,Yin13}, phenomenological models\cite{Sar14}, and experimental indications from spectroscopic factors\cite{Gade}, dispersive optical analyses\cite{Bob06}, and SRC studies\cite{Hen14}. 
For example, in the lower panel of FIG.\,\ref{Cai-kappa} the depletions for protons and neutrons in the BHF framework are shown at $\rho_0$ and $2\rho_0$, respectively; the overall linearity of $\kappa$ on $\delta$ is clearly indicated.
The splitting
$\kappa_\rm{n}-\kappa_\rm{p}\approx[8\beta_0\beta_1/15+6C_0\phi_1/\phi_0+6C_0C_1(1-\phi_0^{-1})]\delta\approx(-0.37\pm0.16)\delta$.

\begin{figure}[h!]
\centering
\includegraphics[height=6.5cm]{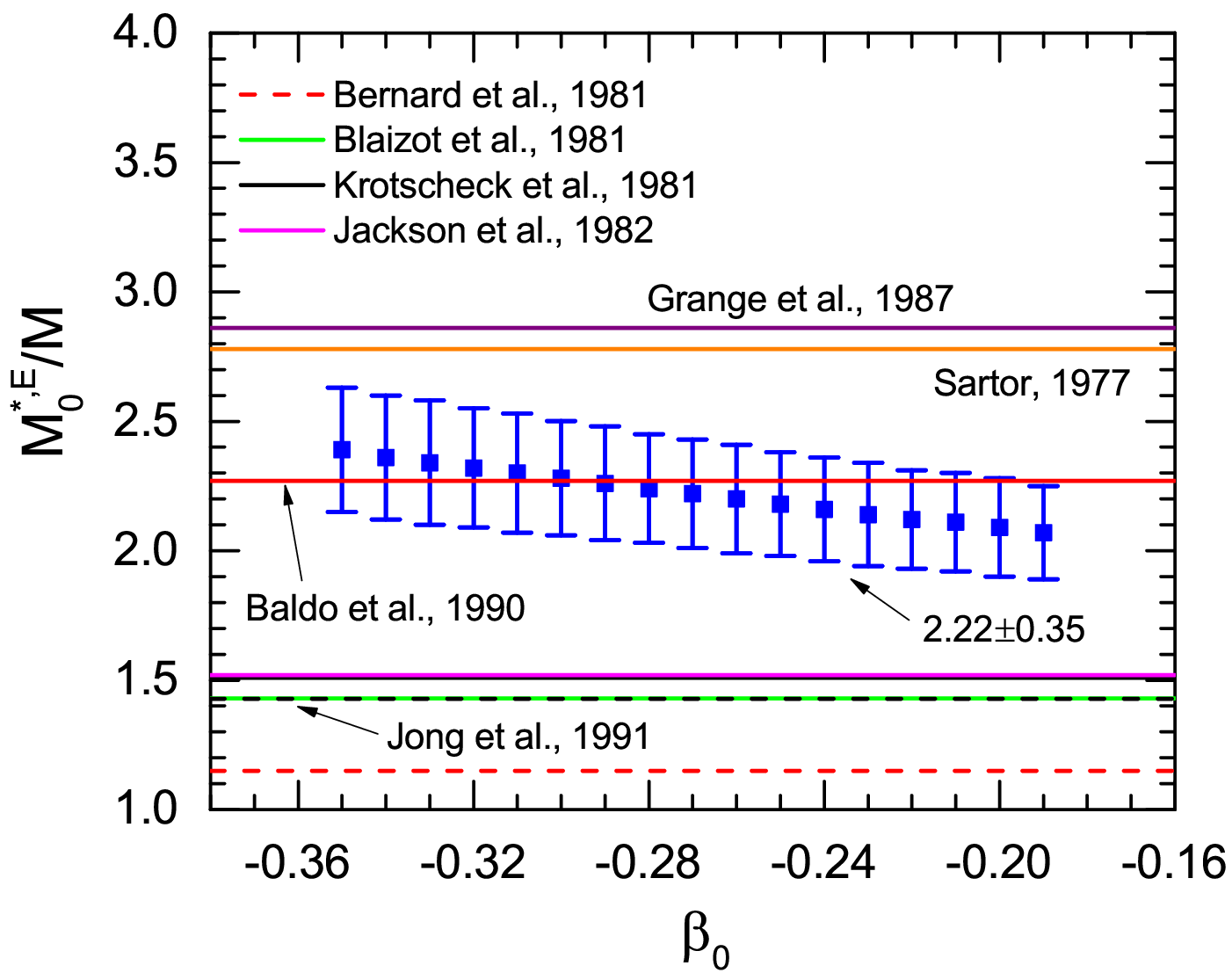}\\[0.25cm]
\hspace{0.5cm}
\includegraphics[height=6.cm]{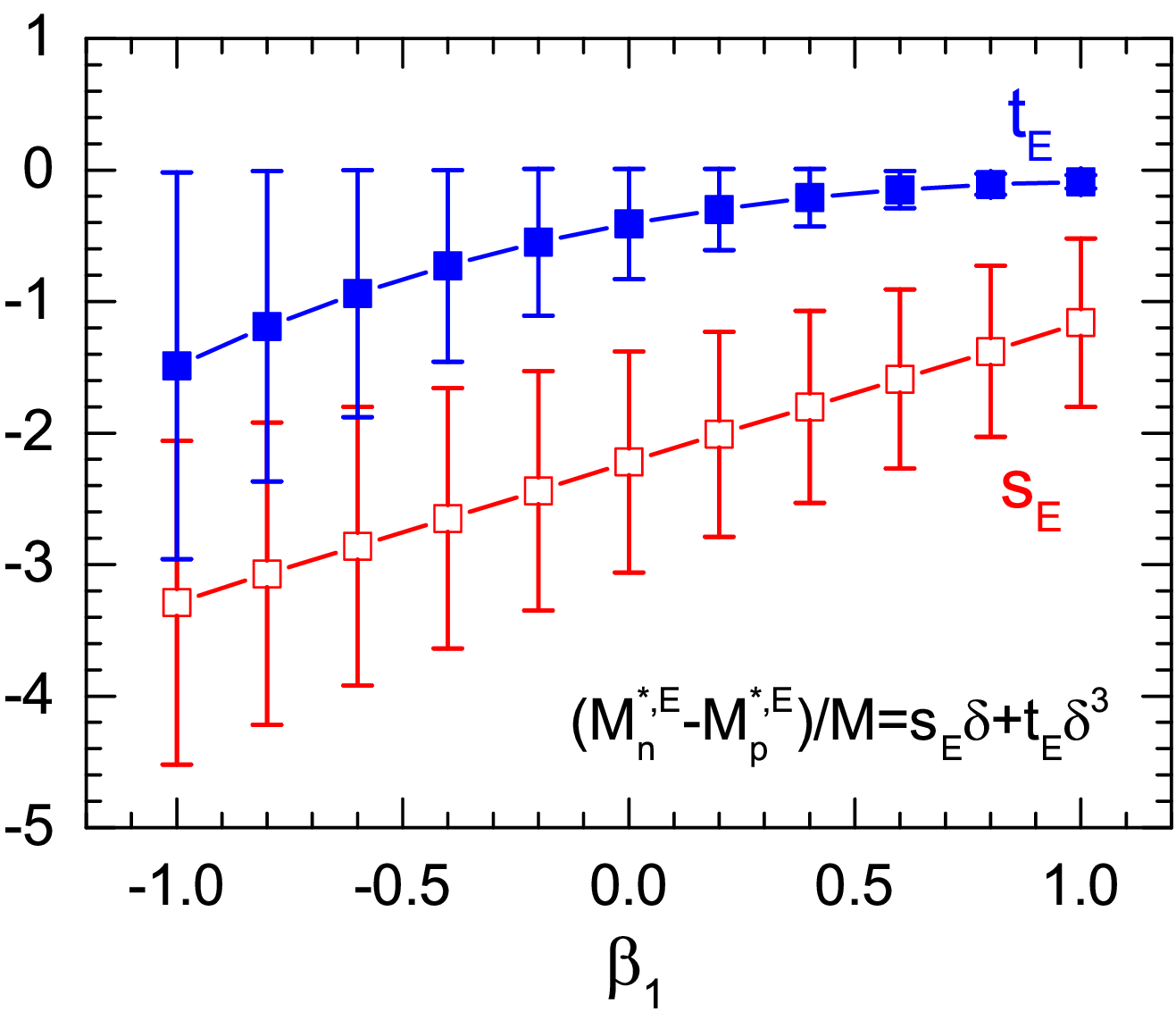}
\caption{(Color Online). Upper: nucleon E-mass in SNM at $\rho_0$ from the Migdal--Luttinger theorem using the uncertain $\beta_0$ range, compared with earlier predictions\cite{Ber81,Bla81,Kro81,Jac82,Gra87,Bal90,Sar77,Jon91}.
Lower: linear and cubic splitting functions $s_{\rm{E}}$ and $t_{\rm{E}}$ at $\rho_0$ within the uncertainty of $\beta_1$. Figures taken from Ref.\cite{CaiLi16a}.}
\label{Fig-Emass0}
\end{figure}

\begin{figure*}[h!]
\centering
\includegraphics[width=17.5cm]{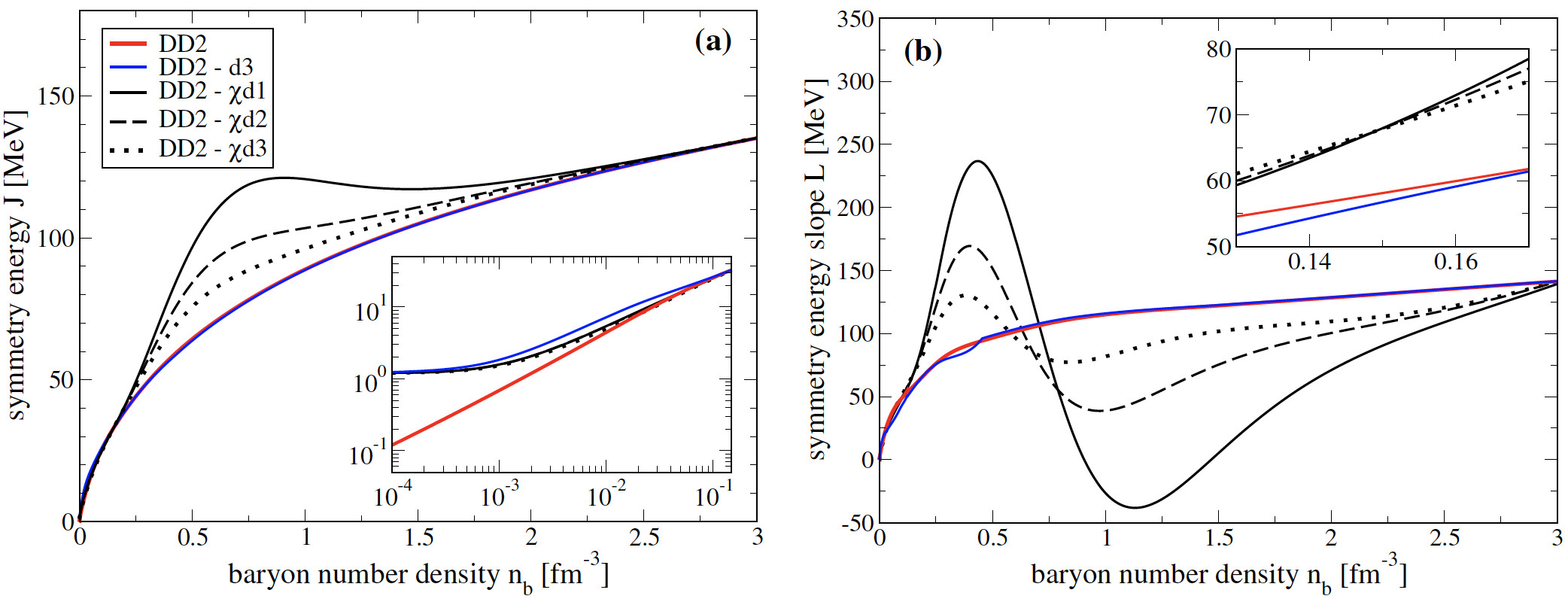}
\caption{(Color Online). Left: symmetry energy as function of the baryon density for four selected parameterizations
employed. The inset shows an enlarged view at sub-saturation densities.
Right: slope of the symmetry energy as function of the baryon
density for the same parameterizations
as in panel (a). The inset shows an enlarged view around the saturation density. Figure taken from Ref.\cite{Burr22}.}
\label{fig_quasi-d}
\end{figure*}

Using the parametrization of Eq.\,(\ref{MDGen}),
$Z_{\rm{F}}^J=\Delta_J+\beta_J-C_J=1+2\beta_J/5-4C_J+3C_J\phi_J^{-1}$.
In SNM, this becomes
$Z_{\rm{F}}^0=1+2\beta_0/5-C_0-x_{\rm{SNM}}^{\rm{HMT}}$,
which leads to $M_0^{\ast,\rm{E}}/M\approx2.22\pm0.35$ when combined with the empirical values of $\beta_0$, $\phi_0$, $C_0$, and $x_{\rm{SNM}}^{\rm{HMT}}$. The dependence on $\beta_0$ is modest, as shown in the upper panel of FIG.\,\ref{Fig-Emass0}. Comparisons with a broad spectrum of earlier microscopic and phenomenological results are also displayed there\cite{Ber81,Bla81,Kro81,Jac82,Gra87,Bal90,Sar77,Jon91}.
The term $x_{\rm{SNM}}^{\rm{HMT}}$ directly reflects the SRC strength; thus stronger SRC increase the E-mass. This is consistent with the contrast between BHF (with SRC via ladder diagrams and TBF) and RHF (without explicit SRC) predictions\cite{ALi16}. In PNM, the corresponding jump
$Z_{\rm{F}}^{\rm{PNM}}=1+2\beta_{\rm{n}}^{\rm{PNM}}/5-x_{\rm{HMT}}^{\rm{PNM}}-C_{\rm{n}}^{\rm{PNM}}$
is reduced mainly by $C_{\rm{n}}^{\rm{PNM}}\approx0.12$, because $x_{\rm{HMT}}^{\rm{PNM}}\approx1.5\%$ is extremely small.

In neutron-rich matter, the neutron-proton E-mass splitting can be written as $
({M_{\rm{n}}^{\ast,\rm{E}}-M_{\rm{p}}^{\ast,\rm{E}}})/{M_{\rm N}}
=s_{\rm{E}}\delta+t_{\rm{E}}\delta^3+\mathcal{O}(\delta^5)$.
The dependence of $s_{\rm{E}}$ and $t_{\rm{E}}$ on $\beta_1$ is displayed in the lower panel of FIG.\,\ref{Fig-Emass0}. For $\beta_1=-1,0,1$, one finds, respectively,
$s_{\rm{E}}(\beta_1=-1)\approx-3.29\pm1.23$, $t_{\rm{E}}(\beta_1=-1)\approx-1.49\pm1.47$;
$s_{\rm{E}}(\beta_1=0)\approx-2.22\pm0.84$, $t_{\rm{E}}(\beta_1=0)\approx-0.41\pm0.42$;
$s_{\rm{E}}(\beta_1=1)\approx-1.16\pm0.64$, $t_{\rm{E}}(\beta_1=1)\approx-0.09\pm0.05$.
A robust qualitative conclusion is that in neutron-rich matter one always finds $M_{\rm{n}}^{\ast,\rm{E}}<M_{\rm{p}}^{\ast,\rm{E}}$, consistent with predictions of most microscopic many-body models. However, the magnitude of the splitting is quite sensitive to the isovector parameter $\beta_1$, which remains experimentally unconstrained. Future measurements of the isospin dependence of SRC, via proton or electron scattering on isotopic chains, and of nucleon spectroscopic factors using radioactive beams have strong potential to constrain $\beta_1$ and thus to pin down the neutron-proton E-mass splitting more precisely.

\subsection{Quasi-deuteron: Embedding SRCs into Relativistic Density Functionals}\label{sub_quasideuteron}

\indent 

Recently, Burrello and Typel develop an extended relativistic energy density functional (EDF) to embed SRCs directly into the nuclear EOS\cite{Burr22}. The motivation stems from a fundamental inconsistency: modern electron-scattering experiments show that a significant fraction of nucleons (approximately $\gtrsim 20\%$) participate in strongly correlated np pairs even at and above the saturation density $\rho_0$, while standard EDFs based on mean-field theory predict that all light clusters, including the deuteron, should dissolve as $\rho_0$ is approached. In other words, traditional EDFs assume that nuclear matter becomes an uncorrelated Fermi gas at $\rho_0$, whereas experiments demonstrate that SRCs remain an important and persistent mode of many-body correlation at these densities.

To resolve this conflict, the authors reinterpret SRC pairs as \emph{in-medium quasi-deuterons}, effective composite degrees of freedom introduced into the functional alongside nucleons.
Quasi-deuterons
immersed in dense matter are used as surrogate for correlations.
The key innovation is the parametrization of the \emph{in-medium deuteron binding energy (mass shift)}, which is allowed to evolve with density such that correlated np pairs do not disappear at saturation density, instead they survive in dense matter as SRC-like quasiparticles. The mass shift is constrained by (i) microscopic few-body calculations at low densities, (ii) empirical SRC fractions near $\rho_0$, and (iii) physical requirements such as maintaining a positive Dirac effective mass for nucleons at high densities.
A central quantity in the formulation is the meson--cluster coupling scaling factor $\chi$, which controls the interaction strength between the quasi-deuterons and the background mean fields. The authors demonstrate that conventional choices ($\chi = 1$) lead to unphysical behavior, such as the impossibility of defining a unique deuteron fraction across the full density range, thereby forcing the model toward the traditional result where clusters disappear below or near saturation. However, for more realistic values ($\chi \le 1/\sqrt{2}$), the model supports smooth and physically consistent deuteron fractions for all densities, allowing SRC-like correlations to persist naturally in the EOS.

The presence of quasi-deuterons modifies the macroscopic properties of nuclear matter. Near $\rho_0$, the additional attraction from correlated pairs increases the binding energy compared with standard mean-field functionals. At higher densities, the stiffness of the EOS and the behavior of the symmetry energy become sensitive to the value of $\chi$, leading to a range of physically plausible scenarios, as shown in FIG.\,\ref{fig_quasi-d} for the symmetry energy (the solid red line is the conventional
model without the quasi-deuteron mechanism).
Some differences emerge among the parameterizations
which account for the presence of the deuterons
below $\rho_0$.
Moreover, the model recovers the correct low-density limit: as $\rho\to0$, the energy per nucleon approaches one half of the vacuum deuteron binding energy\cite{Burr22}, unlike traditional mean-field EDFs that neglect clustering.
Overall, this work represents the first systematic attempt to incorporate SRC physics not by heuristic adjustments to single-nucleon momentum distributions, but by embedding the correlations directly into the EDF through modified composite degrees of freedom. This framework may provide a starting point for future extensions, including momentum-dependent mass shifts, heavier cluster species, and applications to NS matter.

\begin{figure*}[h!]
\centering
\includegraphics[width=16cm]{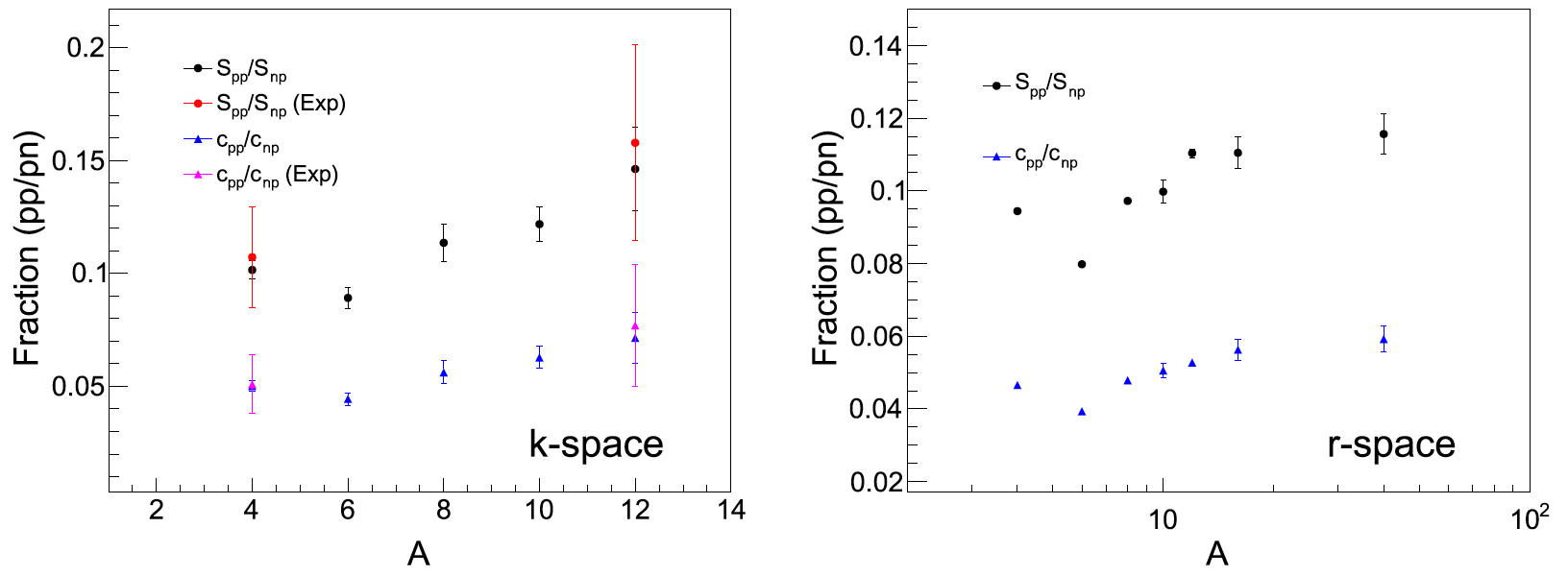}\\[0.25cm]
\hspace{-0.4cm}
\includegraphics[width=16.2cm]{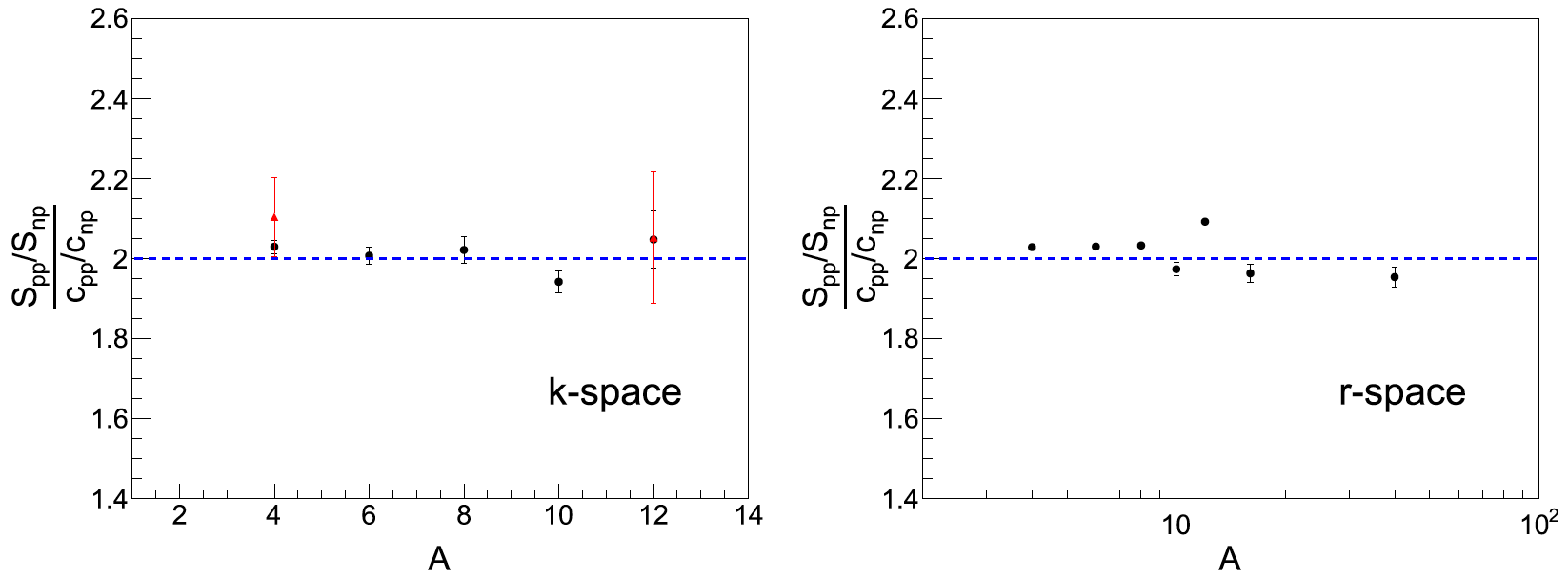}
\caption{(Color Online). Upper: the ratio between the nuclear contacts of pp and np channels and the ratio of the SRC orbital entanglement entropies.
Nuclear contacts are extracted from the two-body wave function in $k$-space (left) and $r$-space (right)\cite{Weiss18PLB}. 
Lower: calculated ratios of reduced nuclear contacts extracted from the $k$-space (left) and from the $r$-space (right) nuclear two-body density distribution as a function of the nuclear mass number. The blue dashed line indicates a value of 2 on the vertical axis.
Figures taken from Ref.\cite{Kou24}.}
\label{fig_Kou-1}
\end{figure*}

\subsection{Orbital Entanglement Entropy of SRC}\label{sub_entropy}

\indent 

In this subsection, we briefly discuss the orbital entanglement entropy associated with SRCs in quantum many-body systems. A key new insight is the connection between the strength of SRCs, the emergence of high-momentum components in single-particle occupations, and the resulting entanglement structure\cite{Bulgac2023-a,Bulgac2023-b}. The universal SRC-induced HMT $n(k)\sim C/k^{4}$ tails, observed in nuclei and cold atomic gases, significantly alter the standard relation between kinetic energy and occupation probabilities far from the Fermi surface, particularly in dynamical situations. Consequently, SRCs increase the orbital entanglement entropy of fermionic systems\cite{Bulgac2023-a}, with important implications for nonequilibrium phenomena such as heavy-ion collisions and nuclear fission.  
The orbital entanglement entropy provides a well-defined and intrinsic measure of complexity of many-fermion wave functions, vanishing for a single Slater determinant and increasing with correlations and superpositions of multiple determinants. Its evolution in nonequilibrium processes, such as nuclear fission, illustrates the localization of the many-body wave function in Fock and Hilbert space, and captures both short-range and long-range correlations. Thus, orbital entanglement entropy offers a natural and quantitative link between SRC-induced high-momentum components and the complexity of strongly interacting quantum systems\cite{Bulgac2023-a}.

Building on the preceding discussion of SRCs, we now focus on their impact on single-particle momentum distributions and entanglement in quantum many-body systems. Even for momenta near or slightly above the local Fermi momentum, SRCs significantly modify the standard mean-field occupation probabilities, 
\begin{equation}
n_{\rm{mf}}(k) = \frac{1}{1 + \exp[\beta\epsilon(k)]},
\end{equation}
where $\beta = 1/k_{\rm B}T$, $\epsilon(k)=E(k)-\mu=k^2/2M_{\rm N}-\mu$ is the single nucleon kinetic energy, where $\mu$ is the chemical potential. While this formula applies in a narrow energy window around the Fermi surface for quasi-equilibrium and weakly interacting systems, decades of studies in nuclei and cold atomic gases demonstrate that SRCs require more sophisticated treatments.  

Recent work has emphasized the intimate link between SRCs and quantum entanglement\cite{Bulgac2023-b}: particles interacting at distances shorter than the average interparticle separation naturally become entangled, as in paradigmatic examples such as the EPR scenario or Young's double-slit experiment. Consequently, many-body systems with SRCs are entangled across all energy scales, independent of equilibrium. The corresponding entanglement entropy directly influences the momentum distribution and nonequilibrium dynamics of the system. For isolated quantum systems in a pure state, entanglement entropy evolves in time and, in the long run, can describe equilibration processes in a manner analogous to Boltzmann entropy in dilute, weakly interacting systems. Thus, entanglement entropy provides a unified measure capturing both mean-field and SRCs in strongly interacting many-body systems.

In Ref.\cite{Bulgac2023-b}, the nucleon momentum distribution accounting for SRCs was parametrized as
\begin{equation}
n(k) = 
\eta(k_0)\begin{cases} 
n_{\rm{mf}}(k), & k \le k_0, \\
n_{\rm{mf}}(k_0)\left(\dfrac{k_0}{k}\right)^4, & k_0 < k < \Lambda,
\end{cases}
\end{equation}
where \(n_{\rm{mf}}(k)\) is the mean-field distribution, \(k_0\) marks the transition to the HMT, and \(\eta(k_0)\) is a normalization factor reflecting the depletion of the Fermi sea due to correlations. The HMT is characterized by the Tan's contact\cite{Tan08-a,Tan08-b,Tan08-c}
\begin{equation}
C(k_0) = \eta(k_0)\, n_{\rm{mf}}(k_0)k_0^4,
\end{equation}
which measures the strength of SRCs, while the total distribution is normalized to the saturation density:
\begin{equation}
\rho_0=A/V = g \int \frac{\d \v{k}}{(2\pi)^3} n(k) = \frac{g k_{\rm F}^3}{6 \pi^2},
\end{equation}
here $g$ is the degeneracy of the system.
The factor \(\eta(k_0)\) is obtained from particle number conservation\cite{Bulgac2023-b}
\begin{equation}
\eta(k_0) = \frac{\int_0^{\Lambda} dk \, k^2 n_{\rm{mf}}(k)}{k_0^3 n_{\rm{mf}}(k_0) + \int_0^{\Lambda} dk \, k^2 n_{\rm{mf}}(k)}.
\end{equation}

Assuming that $\Lambda = \infty$ and $
\rho_0 = g \int [{\d\v{k}}/{(2\pi)^3}] n_{\rm{mf}}(k)$,
a lower limit for $\eta(k_0)$ can be obtained in the case of a FFG at zero temperature by choosing $k_0 = k_{\rm F}$. By putting the HMT part of the momentum distribution into the (orbital) entanglement entropy\cite{Bulgac2023-b}, one then obtains\cite{Pazy23}
\begin{align}
    S_{\rm{SRC}}=&-\frac{g}{(2\pi)^3}\left[
    \d\v{k}n(k)\ln n(k)+\int\d\v{k}\left[1-n(k)\right]\ln\left[1-n(k)\right]
    \right]\notag\\
    \approx&-\frac{g}{2\pi^2}\left[\frac{C}{k_{\rm{F}}}
    \ln\left(\frac{C/k_{\rm F}^4}{1-C/k_{\rm F}^4}\right)
    +k_{\rm F}^3\ln\left(1-\frac{C}{k_{\rm F}^4}\right)
    \right]\notag\\
    =&-3\rho_0\left[\frac{C}{k_{\rm{F}}^4}
    \ln\left(\frac{C/k_{\rm F}^4}{1-C/k_{\rm F}^4}\right)
    +\ln\left(1-\frac{C}{k_{\rm F}^4}\right)
    \right],
\end{align}
where $C/k_{\rm F}^4=c_{\rm{np}}$ is the reduced np contact. Consequently\cite{Pazy23}
\begin{equation}\label{ENTS}
\boxed{
    S_{\rm{SRC}}\approx-3\rho_0\left[c_{\rm{np}}
    \ln\left(\frac{c_{\rm{np}}}{1-c_{\rm{np}}}\right)
    +\ln\left(1-c_{\rm{np}}\right)
    \right],}
\end{equation}
which clearly shows that the orbital entanglement entropy is proportional to the mass number $A$, since $\rho_0 = A/V$. The nuclear structure orbital entanglement entropy of SRCs is calculated based on the nuclear scale separation, specifically considering the entanglement between the SRC orbitals and the rest of the system. It should be emphasized that this is a single-nucleon entanglement entropy, not a pairwise entanglement between a proton and a neutron. The entanglement arises from the probability for a nucleon to occupy a momentum state above the Fermi momentum $k_{\rm F}$. The nuclear momentum space separates into two parts: nucleons occupy either the mean-field part of the wave function, i.e., the Fermi sea, or the high-momentum SRC part. The orbital entropy is computed between these two subspaces, where one contains all low-momentum Fermi sea states and the other contains the high-momentum SRC states, treated as a distinct SRC ``orbital'' that can be multiply occupied. Since the occupation probability of a single SRC is given by the nuclear contact, a general expression for the SRC orbital entanglement entropy can be derived using the GCF\cite{Weiss15PRC,Cruz21NP,Cosyn21PLB,Liang24PLB,Liang24xxx,Weiss17PRC,Weiss18xxx,Weiss19PLB,Yank25PRC}. The GCF approach allows one to obtain the scaling of the entropy with the mass number, and it shows that, unlike typical entanglement entropies in quantum systems which scale with the surface area $\sim A^{2/3}$, the SRC orbital entanglement entropy in large nuclei is extensive and scales linearly with $A$\cite{Pazy23}.

SRC increase the orbital entanglement entropy because nucleons acquire a finite probability of occupying the high-momentum states above the Fermi sea. Without SRCs, all nucleons reside in the low-momentum Fermi sea, resulting in minimal entropy. The presence of SRCs introduces uncertainty about whether a nucleon occupies a Fermi sea or an SRC state, directly enhancing the entanglement entropy. This contribution scales with the nuclear density and thus grows linearly with the mass number, making the entropy extensive in large nuclei. SRC-induced HMT of single-particle occupation probabilities not only raise the entanglement entropy of fermionic systems but may also influence the dynamics of nuclear reactions such as heavy-ion collisions and fission\cite{Bulgac2023-b}.

Using Eq.\,(\ref{ENTS}), one can define a ratio that describes the relative SRC entanglement entropies and reduced nuclear contacts in different channels\cite{Kou24}:
\begin{equation}\label{Rcpp_cnp}
R(c_{\rm{pp}}, c_{\rm{np}}) = \frac{S_{\rm{SRC}}^{\rm{pp}}/S_{\rm{SRC}}^{\rm{np}}}{c_{\rm{pp}}/c_{\rm{np}}},
\end{equation}
where for the $\rm{pp}$ channel one considers the contribution from spin-0 pairs, while for the $\rm{np}$ channel the total contribution from both spin-0 and spin-1 pairs are included. The nuclear contacts of symmetric nuclei were extracted in Ref.\cite{Weiss18PLB}.
The study of Ref.\cite{Kou24} reveals that the entanglement entropies and nuclear contacts for pp and np SRC pairs follow a scaling relation. In particular, the proportionality of entanglement entropy between pp and np pairs is directly related to the ratio of the corresponding nuclear contacts, which is found to be approximately 2.0, as shown in FIG.\,\ref{fig_Kou-1}. Although the exact reason why this ratio converges to 2 in all examined nuclei is not yet understood, the introduction of entanglement entropy provides an additional constraint on the values of nuclear contacts. The authors suggest that this scaling behavior is universal for all symmetric nuclei and provide a possible explanation for this phenomenon\cite{Kou24}.

\section{Effects of SRC-HMT on Cold Dense Matter EOS}\label{SEC_EOS}

\indent

This section discusses the SRC-HMT effects on the EOS of cold ANM. In Subsection~\ref{sub_RFFG}, we introduce the relativistic free Fermi gas (RFFG) model and emphasize its behavior in a general spatial dimension~$d$. In Subsection \ref{sub_kinEsym}, we examine the reduction of the kinetic symmetry energy caused by the SRC-induced HMT and also provide the relativistic corrections to the kinetic symmetry energies; in addition, the validity of the parabolic approximation for the EOS of ANM is analyzed at both the kinetic level and the full level. In Subsection \ref{sub_WaleckaSRC}, the SRC-induced HMT is incorporated into the Walecka model (using the nonlinear version as an example) and into the modified Gogny-type energy density functional, and the resulting overall reduction (softening) of the nuclear symmetry energy at supra-saturation densities is highlighted. In Subsection \ref{sub_dEOS}, we generalize the EOS of ANM to a general dimension $d$, following the strategy used in Subsection \ref{sub_RFFG} and employing the HVH theorem as the main tool; in particular, we examine the EOS in 2D and outline several general features. We also explore the $\epsilon$-expansion of the EOS around a reference dimension and highlight several interesting implications. Finally, we estimate the impact of an additional HMT of the form $k^{-\sigma}$ with $\sigma>4$ on the kinetic part of the nuclear EOS in Subsection \ref{sub_k6_EOSkin}; since such a tail has been suggested by microscopic studies, our estimate provides a useful reference for assessing its potential significance.

\subsection{EOS of a Relativistic Free Fermi Gas}\label{sub_RFFG}

\indent 

In this subsection, we briefly review the non-relativistic and relativistic FFG models, providing insights relevant for later discussions. As a reminder of the basic formalisms and notations used in the present work, we first recall the predictions of the non-relativistic FFG model in a 3D coordinate space.  
The average kinetic energy per nucleon in ANM is given by\cite{CaiLi22PRCFFG}
\begin{align}\label{def_EOSANMFFG}
E^{\rm{kin}}(\rho,\delta)=&\left.\left[\int_0^{k_{\rm{F}}^{\rm{n}}}\frac{\v{k}^2}{2M_{\rm N}}\d\v{k}+\int_0^{k_{\rm{F}}^{\rm{p}}}\frac{\v{k}^2}{2M_{\rm N}}\d\v{k}\right]\right/2\int_0^{k_{\rm{F}}}\d\v{k}\notag\\
=&\frac{3}{5}\frac{k_{\rm{F}}^2}{2M_{\rm N}}\frac{1}{2}\left[(1+\delta)^{5/3}+(1-\delta)^{5/3}\right]\notag\\
\approx&\frac{3k_{\rm{F}}^2}{10M_{\rm N}}+\frac{k_{\rm{F}}^2}{6M_{\rm N}}\delta^2+\frac{k_{\rm{F}}^2}{162M_{\rm N}}\delta^4
+\frac{7k_{\rm{F}}^2}{4374M_{\rm N}}\delta^6.
\end{align}  
From this, one obtains $E_{\rm{sym,4}}^{\rm{kin}}(\rho)={k_{\rm{F}}^2}/{162M_{\rm N}}$ and $E_{\rm{sym,6}}^{\rm{kin}}(\rho)=7k_{\rm{F}}^2/4374M_{\rm N}$ besides the well-known expressions $E_0^{\rm{kin}}(\rho)=3k_{\rm{F}}^2/10M_{\rm N}$, $E_{\rm{sym}}^{\rm{kin}}(\rho)={k_{\rm{F}}^2}/{6M_{\rm N}}$.
Using $k_{\rm{F}}=(3\pi^2\rho/2)^{1/3}\approx263\,\rm{MeV}$ corresponding to the saturation density $\rho_0\approx0.16\,\rm{fm}^{-3}$, one finds $E_0^{\rm{kin}}(\rho_0)\approx 22.2\,\rm{MeV}$, $E_{\rm{sym}}^{\rm{kin}}(\rho_0)\approx 12.3\,\rm{MeV}$, $ E_{\rm{sym,4}}^{\rm{kin}}(\rho_0)\approx 0.45\,\rm{MeV}$, and $E_{\rm{sym,6}}^{\rm{kin}}(\rho_0)\approx 0.12\,\rm{MeV}$.  
Beside, we obtain:
\begin{equation}\label{def-Psi-FFG}
\Psi\equiv {E^{\rm{kin}}_{\rm{sym,4}}(\rho)}/{E^{\rm{kin}}_{\rm{sym}}(\rho)}=27^{-1}.
\end{equation}

These predictions serve as useful references for evaluating effects of relativistic kinematics, SRCs, and nuclear interactions missing in the model. For instance, to reproduce the empirical binding energy of about $-16\,\rm{MeV}$ at $\rho_0$, an interaction contribution of about $-38.2\,\rm{MeV}$ is required. Similarly, with a kinetic symmetry energy of about 12.3\,MeV, an interaction contribution of $\sim 20$\,MeV is needed to reproduce the total empirical nuclear symmetry energy of 32\,MeV\cite{LCCX18}.  
While the FFG model predicts a fourth-order kinetic symmetry energy of $\sim 0.45$\,MeV, several microscopic theories and phenomenological models estimate empirically a total $E_{\rm{sym,4}}(\rho_0)$ below $\sim 2$\,MeV\cite{Lee98, Bom91,Steiner06,Cai12PRC-S4,Seif14PRC-S4,Gonz17PRC-S4,Pu17PRC-S4}. There is, however, no consensus on the empirical value of $E_{\rm{sym,4}}(\rho_0)$, and the situation is even less certain for the sixth-order symmetry energy $E_{\rm{sym,6}}(\rho_0)$. Thus, we cannot determine whether the relative kinetic contribution to the total symmetry energy decreases from $\delta^2$ to higher-order terms.

The relativistic kinetic energy $T(\v{k})=\sqrt{\v{k}^2+M_{\rm N}^2}-M_{\rm N}$ can be expanded as  
\begin{align}\label{def_nonexp}
T(\v{k})\approx\frac{\v{k}^2}{2M_{\rm N}}\left(1-\frac{\v{k}^2}{4M_{\rm N}^2}+\frac{\v{k}^4}{8M_{\rm N}^4}-\frac{5\v{k}^6}{64M_{\rm N}^6}+\frac{7\v{k}^8}{128M_{\rm N}^8}\right),
\end{align}  
to order $\v{k}^{10}$. Considering ANM in a coordinate space of arbitrary dimension $d$, define  
\begin{align}\label{k-sigma}
\langle k^{\sigma}(\rho,\delta)\rangle\equiv&\left.\left[\int_0^{k_{\rm{F}}^{\rm{n}}}k^{\sigma}\d^d\v{k}+\int_0^{k_{\rm{F}}^{\rm{p}}}k^{\sigma}\d^d\v{k}\right]\right/2\int_0^{k_{\rm{F}}}\d^d\v{k}\notag\\
=&\Upsilon\left[(1+\delta)^{1+\sigma/d}+(1-\delta)^{1+\sigma/d}\right],
\end{align}  
where $\Upsilon$ is independent of $\delta$, and $\sigma/d$ controls the expansion. Here $\sigma=2$ is the non-relativistic kinetic energy, $\sigma=4$ the first relativistic correction, and $\sigma=6$ the second, etc. The Fermi momentum in ANM is $k_{\rm{F}}^J=k_{\rm{F}}(1+\tau_3^J\delta)^{1/d}$, with $k_{\rm{F}}(\rho)=[\rho2^{d-2}\pi^{d/2}\Gamma(d/2+1)]^{1/d}\sim\rho^{1/d}$.  
Expanding around $\delta=0$ gives  
\begin{align}
\langle k^{\sigma}(\rho,\delta)\rangle=&\sum_{j=0}k_{\rm{sym},2j}^{\sigma}(\rho)\delta^{2j}\notag\\
\approx &k_0^{\sigma}(\rho)+k_{\rm{sym}}^{\sigma}(\rho)\delta^2+k^{\sigma}_{\rm{sym,4}}(\rho)\delta^4
+\mathcal{O}(\delta^{6}),
\end{align}  
with $k_0^{\sigma}(\rho)\equiv k_{\rm{sym},0}^{\sigma}(\rho)$ and $k_{\rm{sym}}^{\sigma}(\rho)\equiv k_{\rm{sym,2}}^{\sigma}(\rho)$. The ratio of quartic to quadratic coefficients is  
\begin{align}\label{def_pp1}
\Psi^{\sigma}_{d}(j=2)\equiv\frac{k^{\sigma}_{\rm{sym,4}}(\rho)}{k^{\sigma}_{\rm{sym}}(\rho)}
=\frac{1}{12}\left(\frac{\sigma}{d}-2\right)\left(\frac{\sigma}{d}-1\right).
\end{align}  
Depending on the value of $\sigma/d$, this ratio is not necessarily as small as the $E^{\rm{kin}}_{\rm{sym,4}}(\rho)/E^{\rm{kin}}_{\rm{sym}}(\rho)=1/27$ in the FFG model in 3D. For example, if one takes $\sigma=4$ and $d=1$,  i.e., the average of the momentum to the fourth power (corresponding to the first relativistic correction) in dimension 1,  one obtains $\Psi^4_1(j=2)=1/2$.
As indicated by Eq.\,(\ref{def_pp1}), the 
ratio $E^{\rm{kin}}_{\rm{sym,4}}(\rho)/E^{\rm{kin}}_{\rm{sym}}(\rho)$ is zero in both 1D and 2D (in fact the quartic terms in 1D and 2D are identically zero), while it is $1/27$ in 3D as normally used in preparing the EOS for modeling NSs and various studies of heavy-ion reactions. Studies of nuclear EOSs in 1D and 2D and their observational effects in collectively moving sub-systems in heavy-ion reactions and/or NSs might be interesting, we will come back to discuss this issue in Subsection \ref{sub_dEOS}.

\begin{figure}[h!]
\centering
\includegraphics[width=2.1cm]{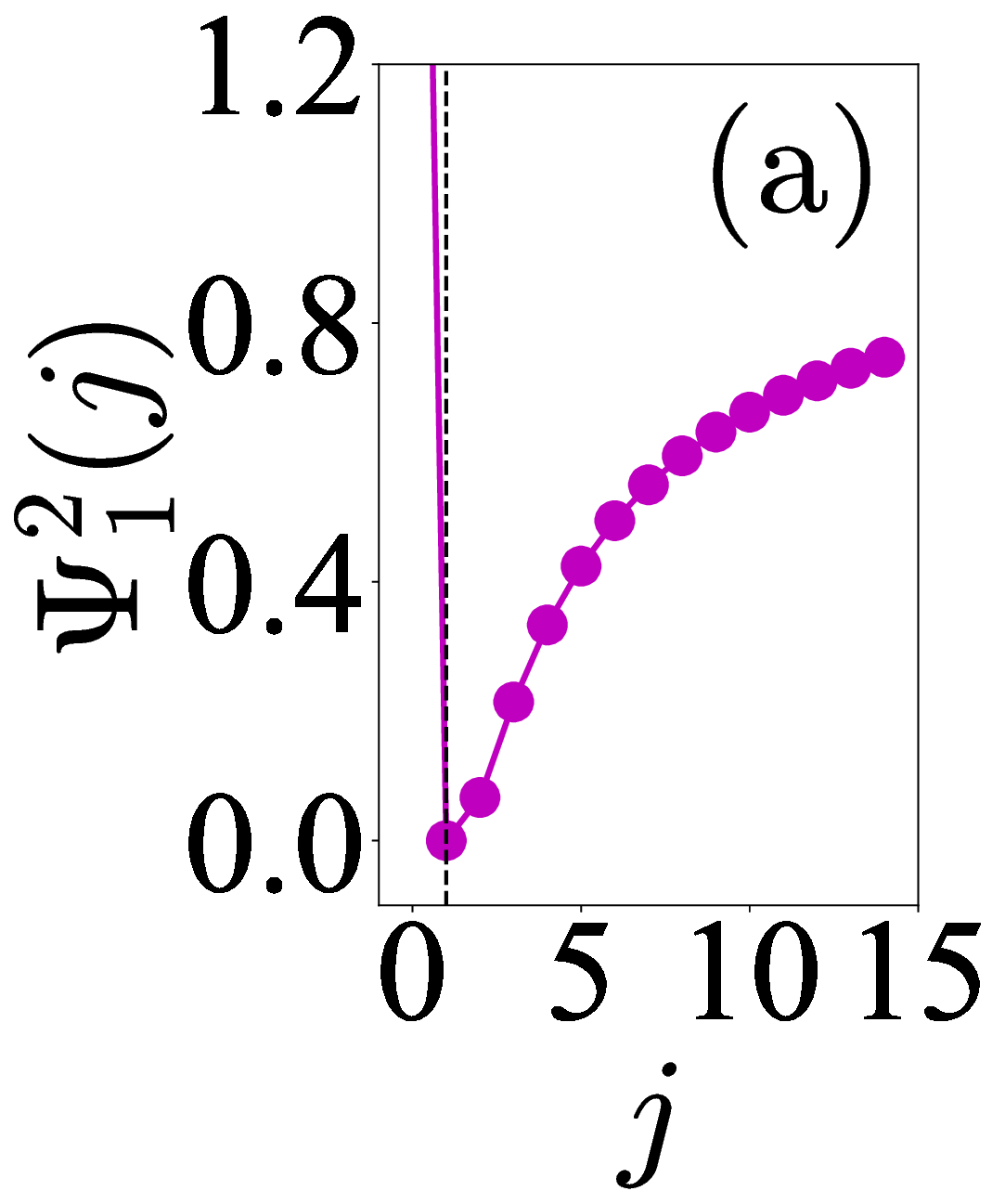}
\includegraphics[width=2.1cm]{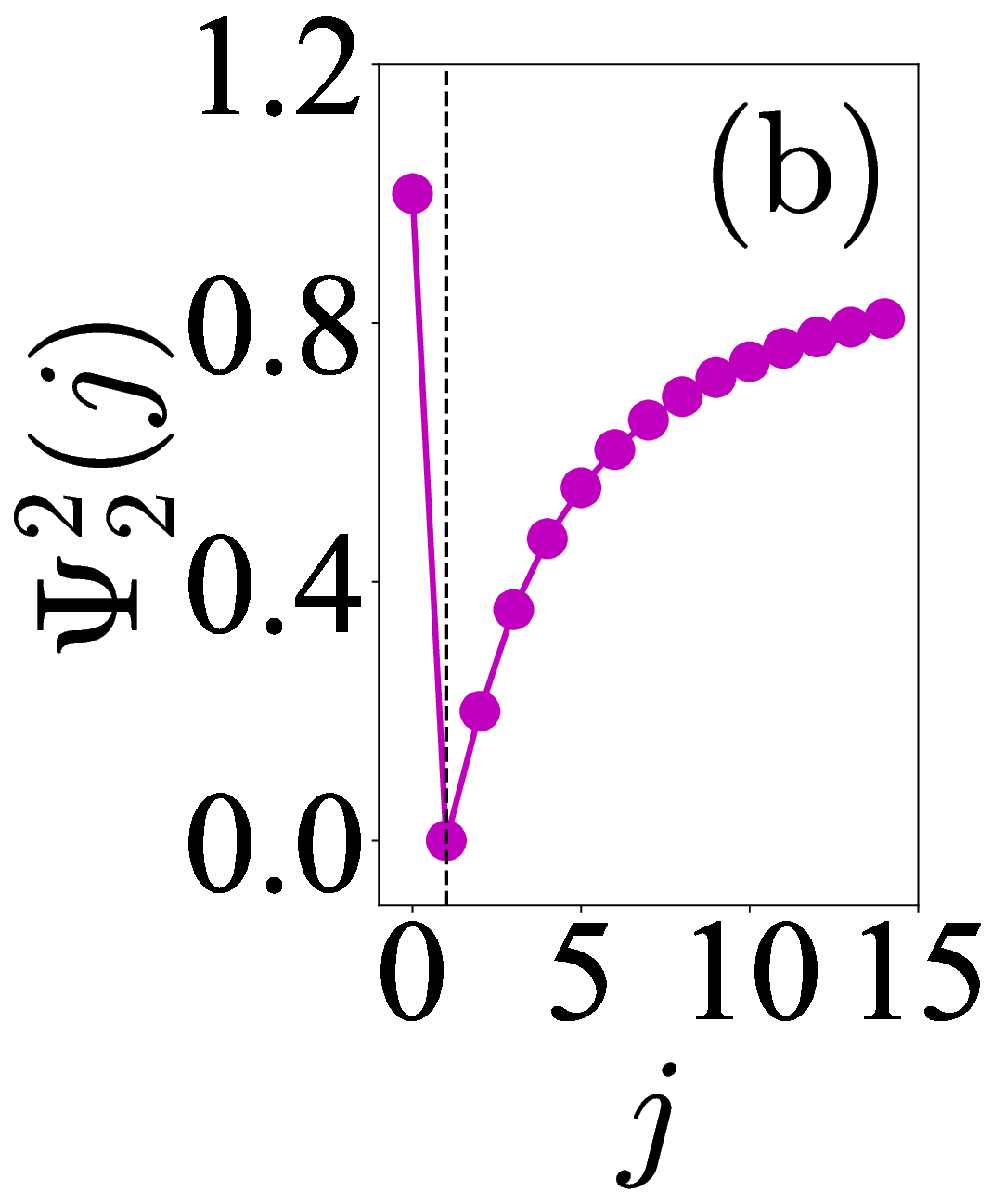}
\includegraphics[width=2.1cm]{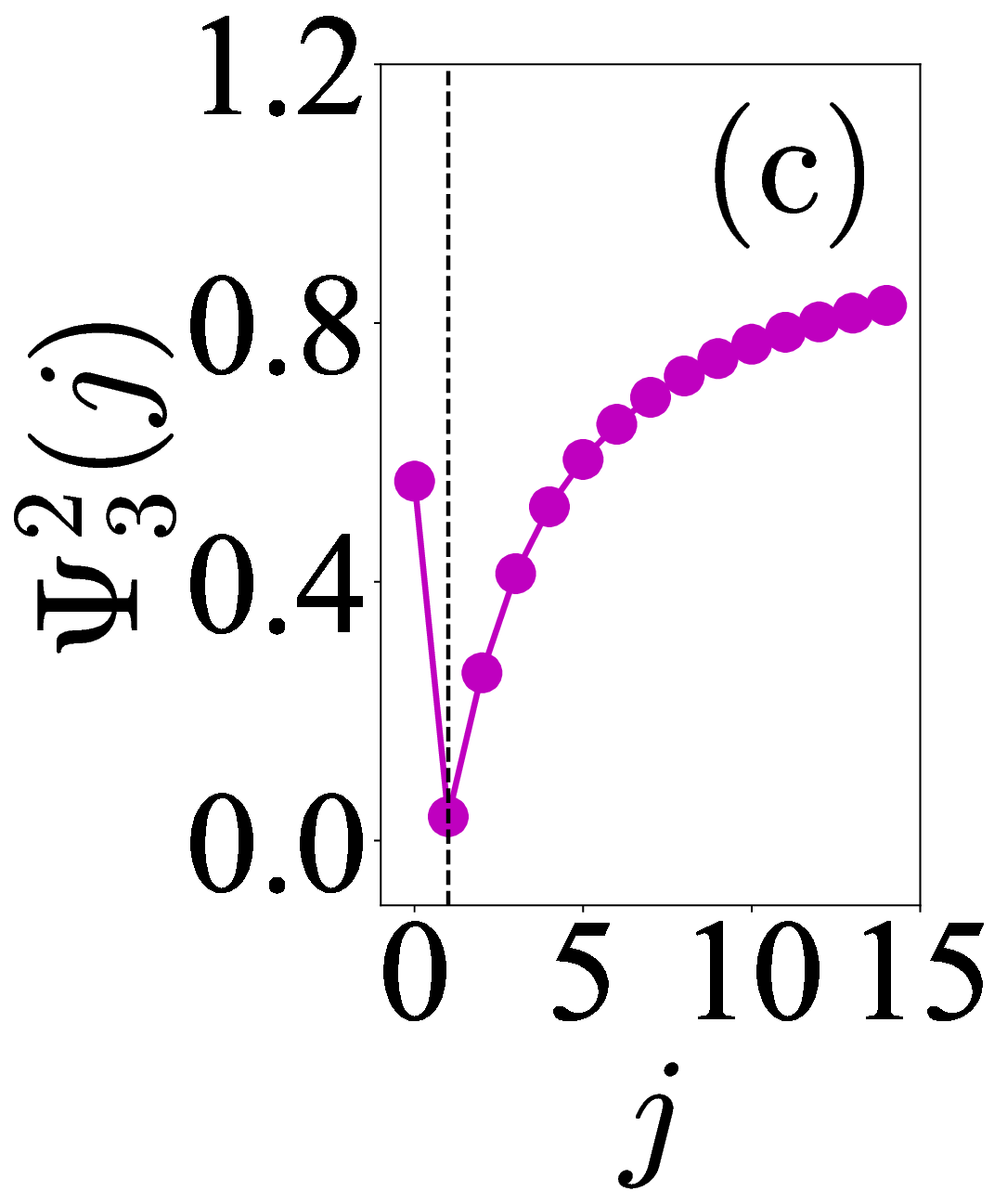}
\includegraphics[width=2.1cm]{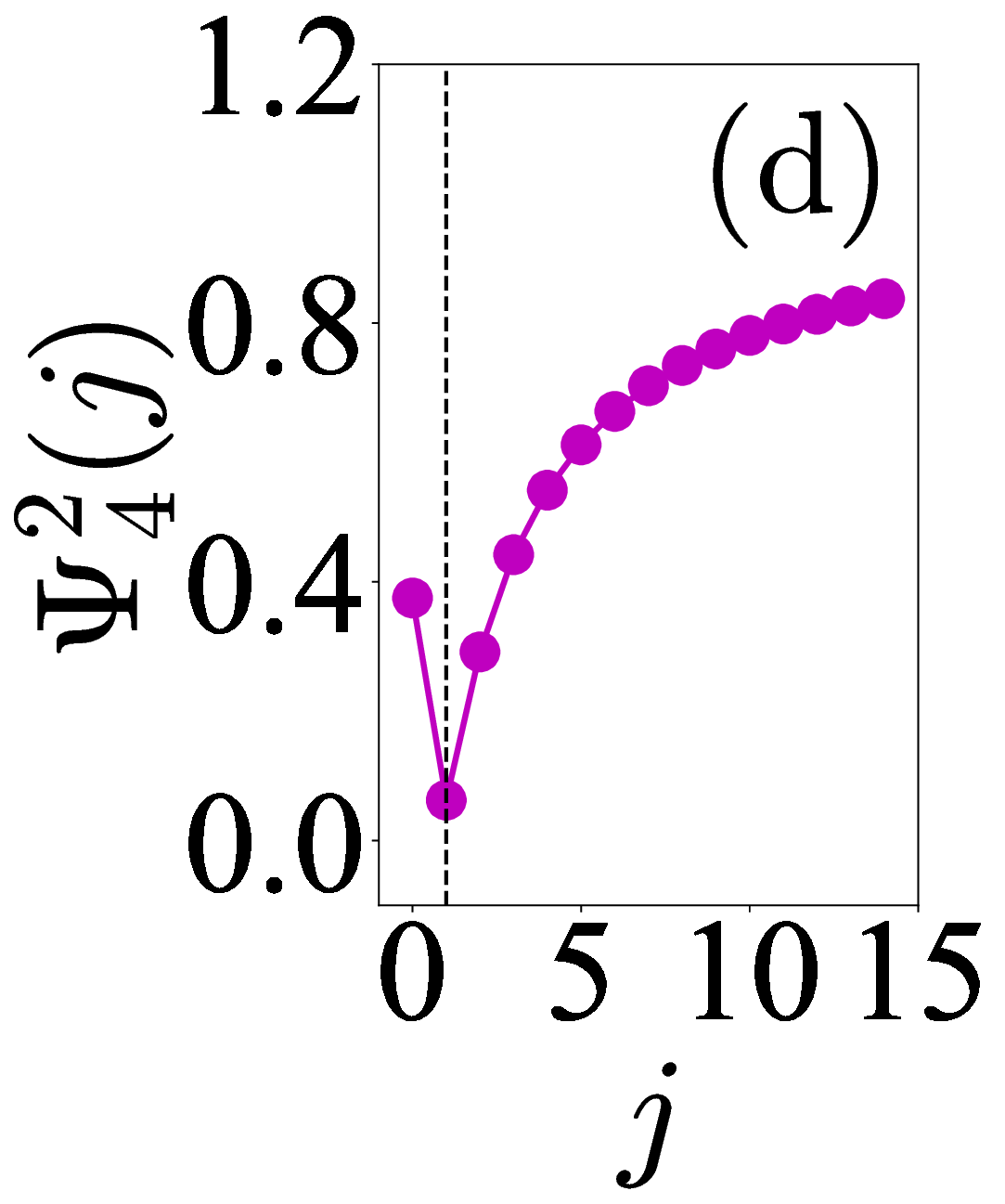}\\
\includegraphics[width=2.1cm]{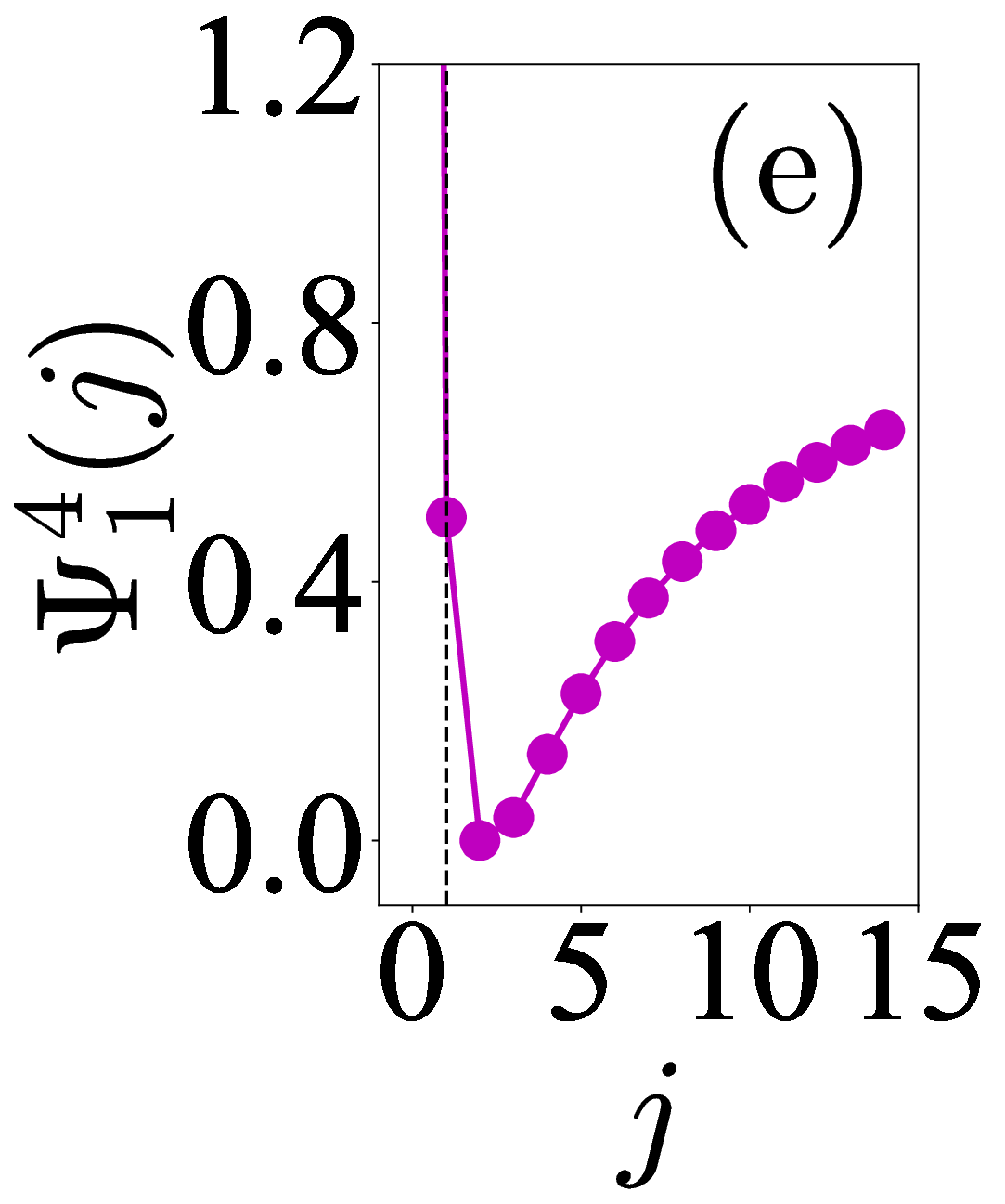}
\includegraphics[width=2.1cm]{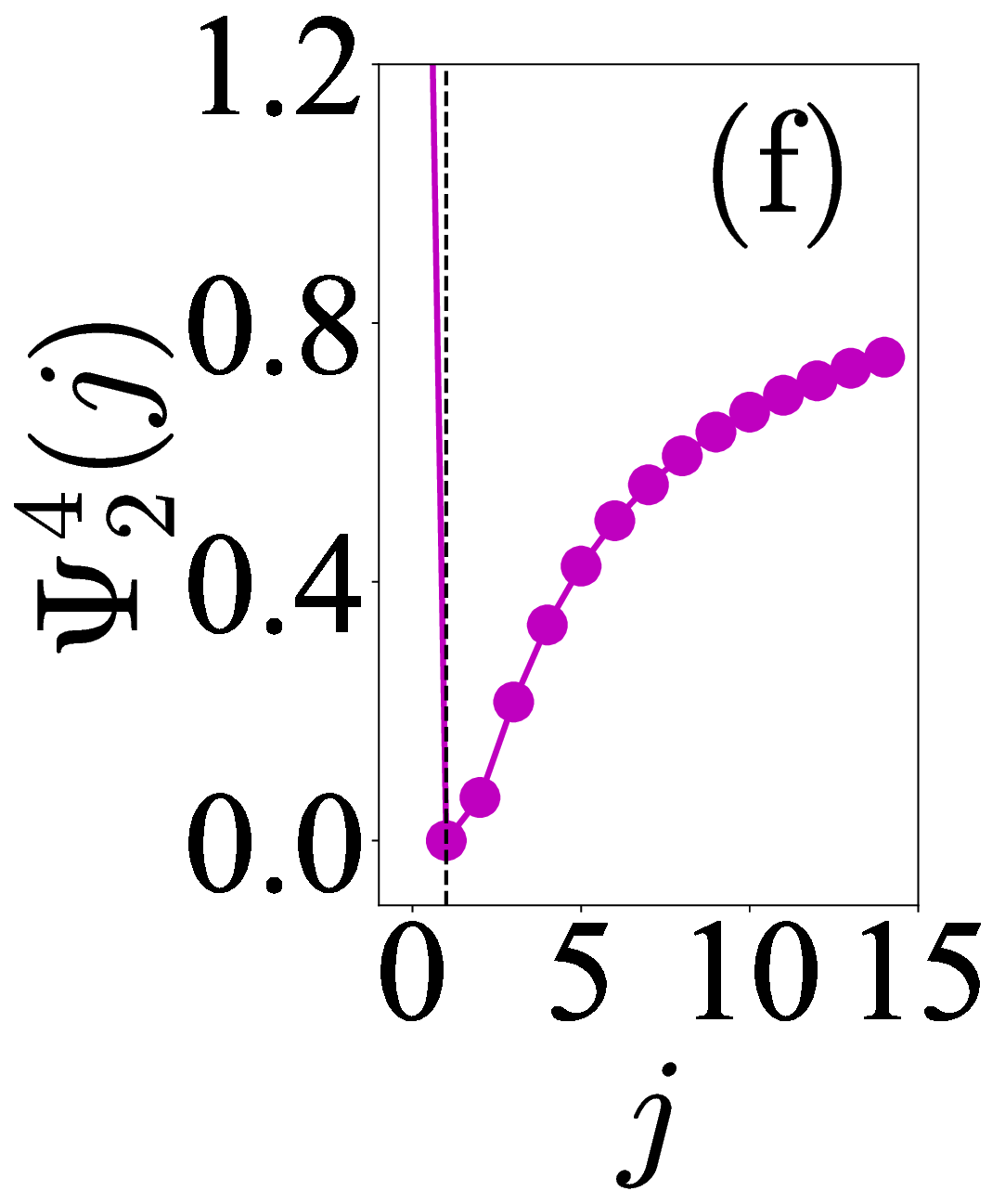}
\includegraphics[width=2.1cm]{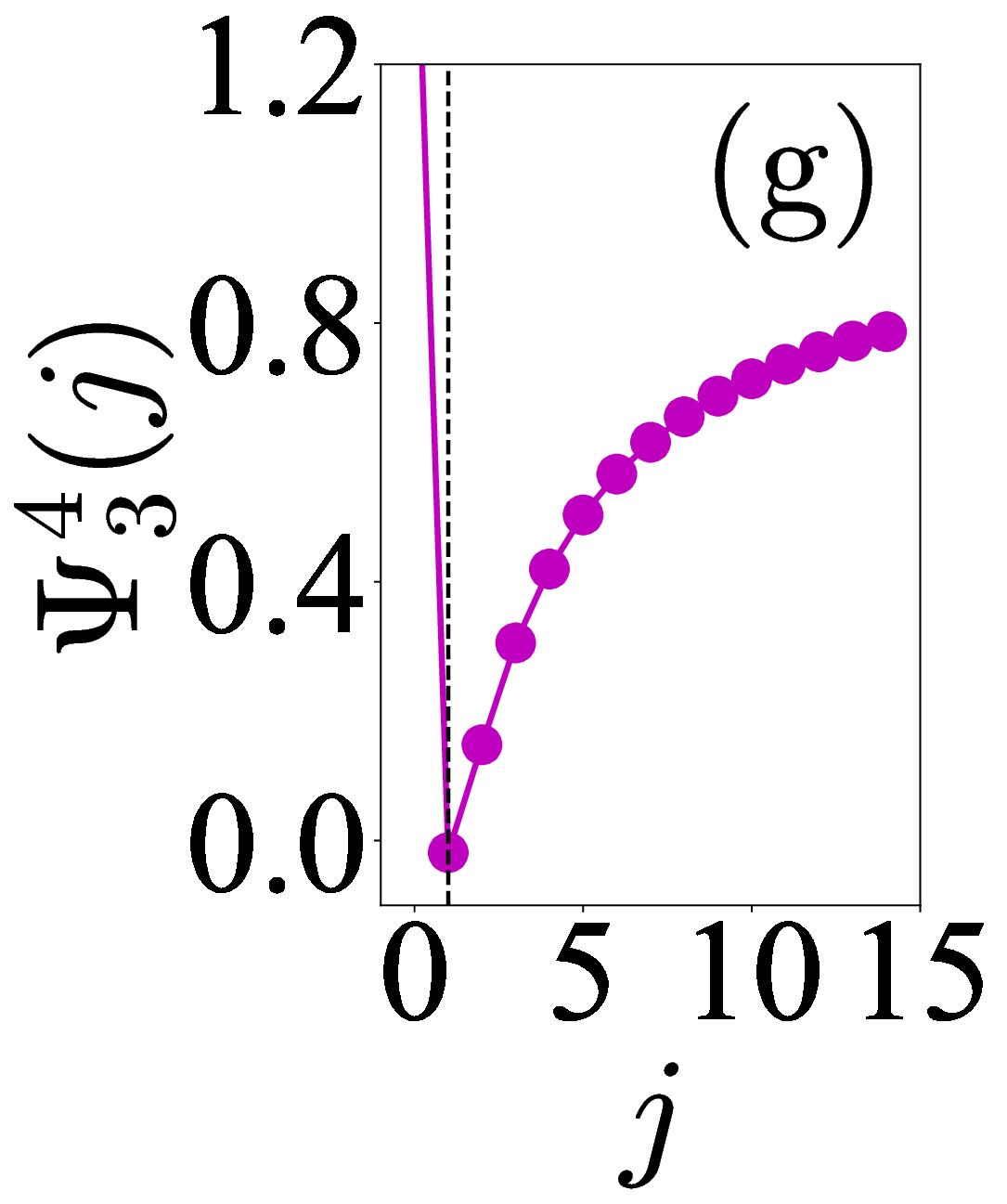}
\includegraphics[width=2.1cm]{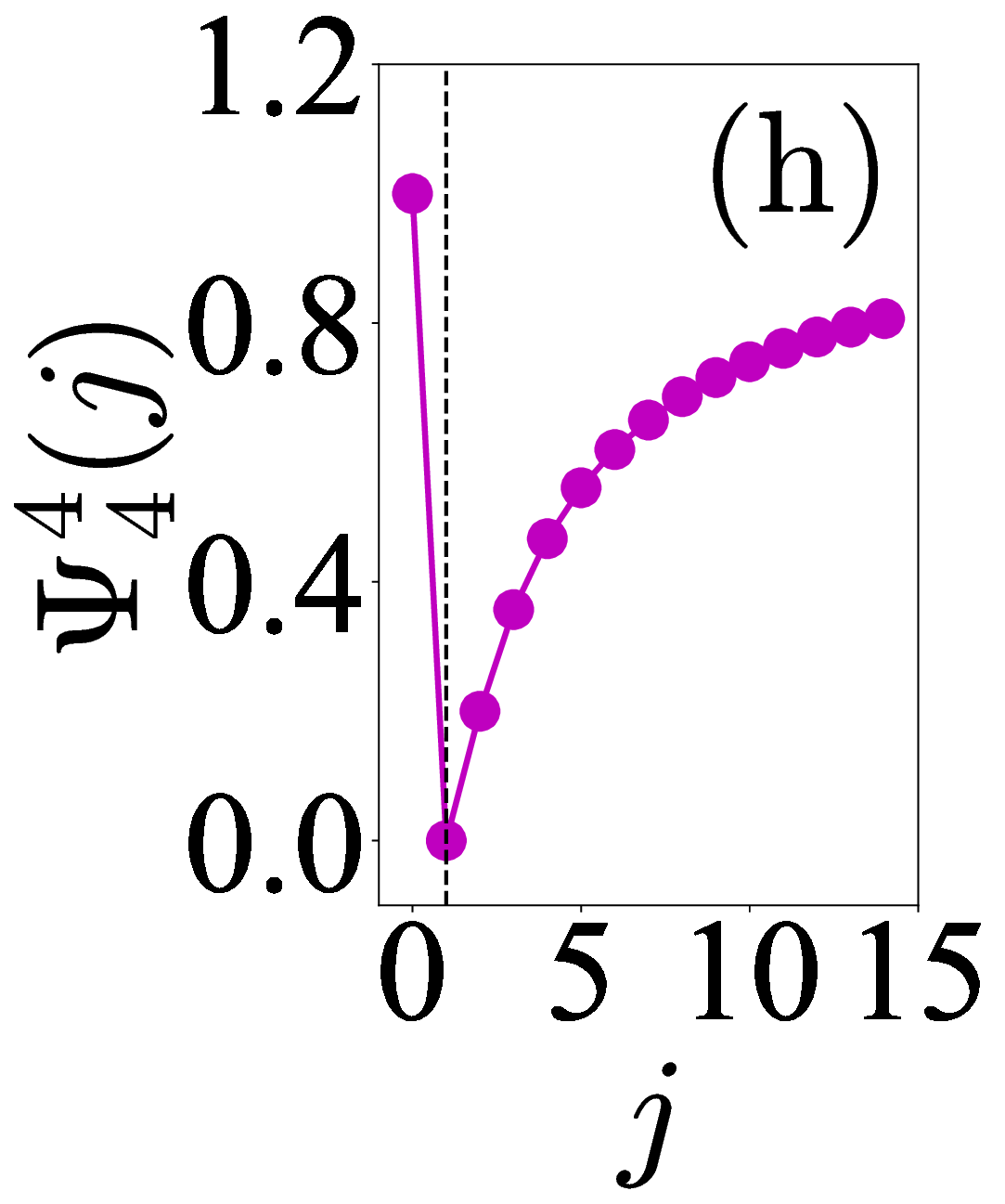}\\
\includegraphics[width=2.1cm]{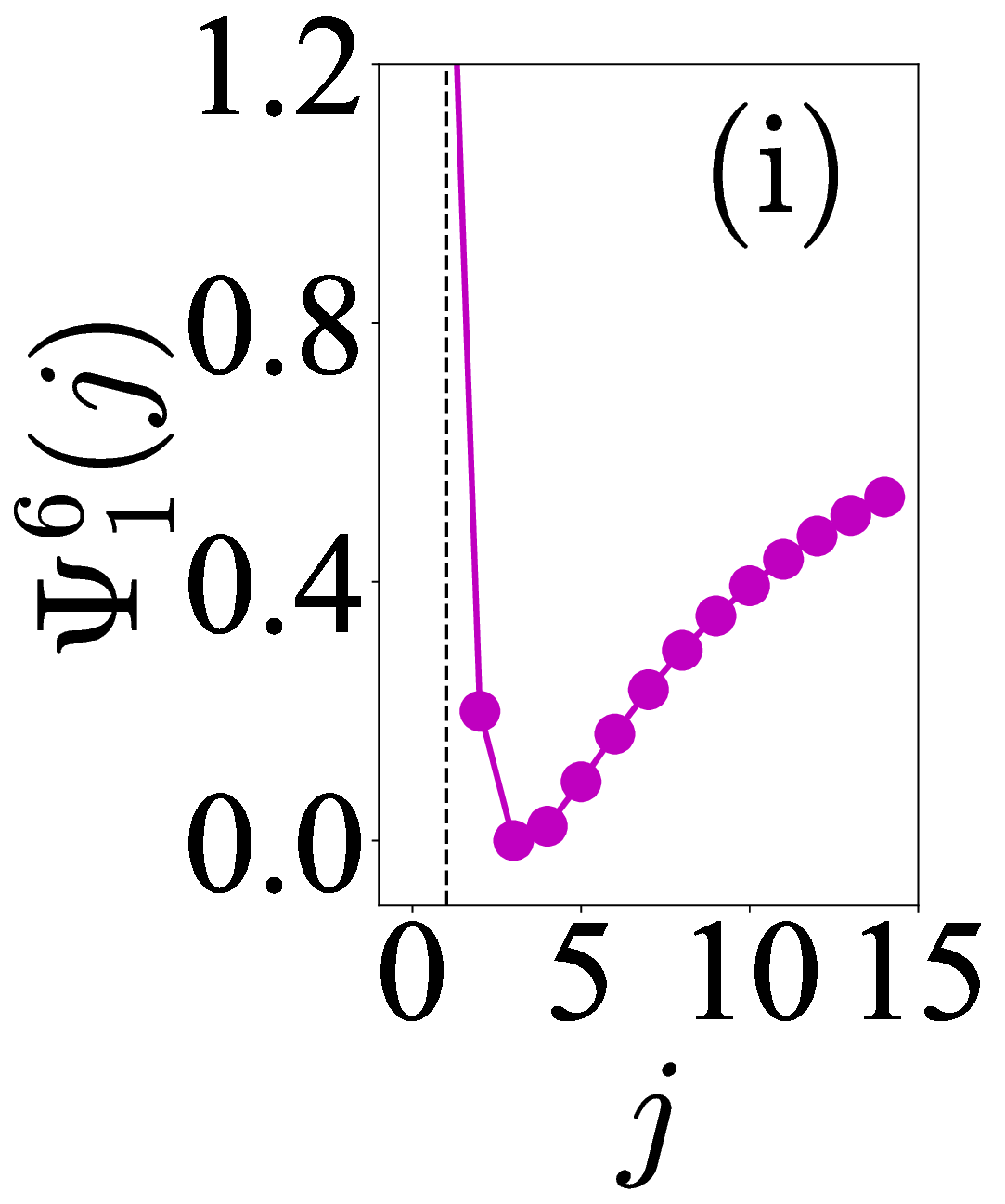}
\includegraphics[width=2.1cm]{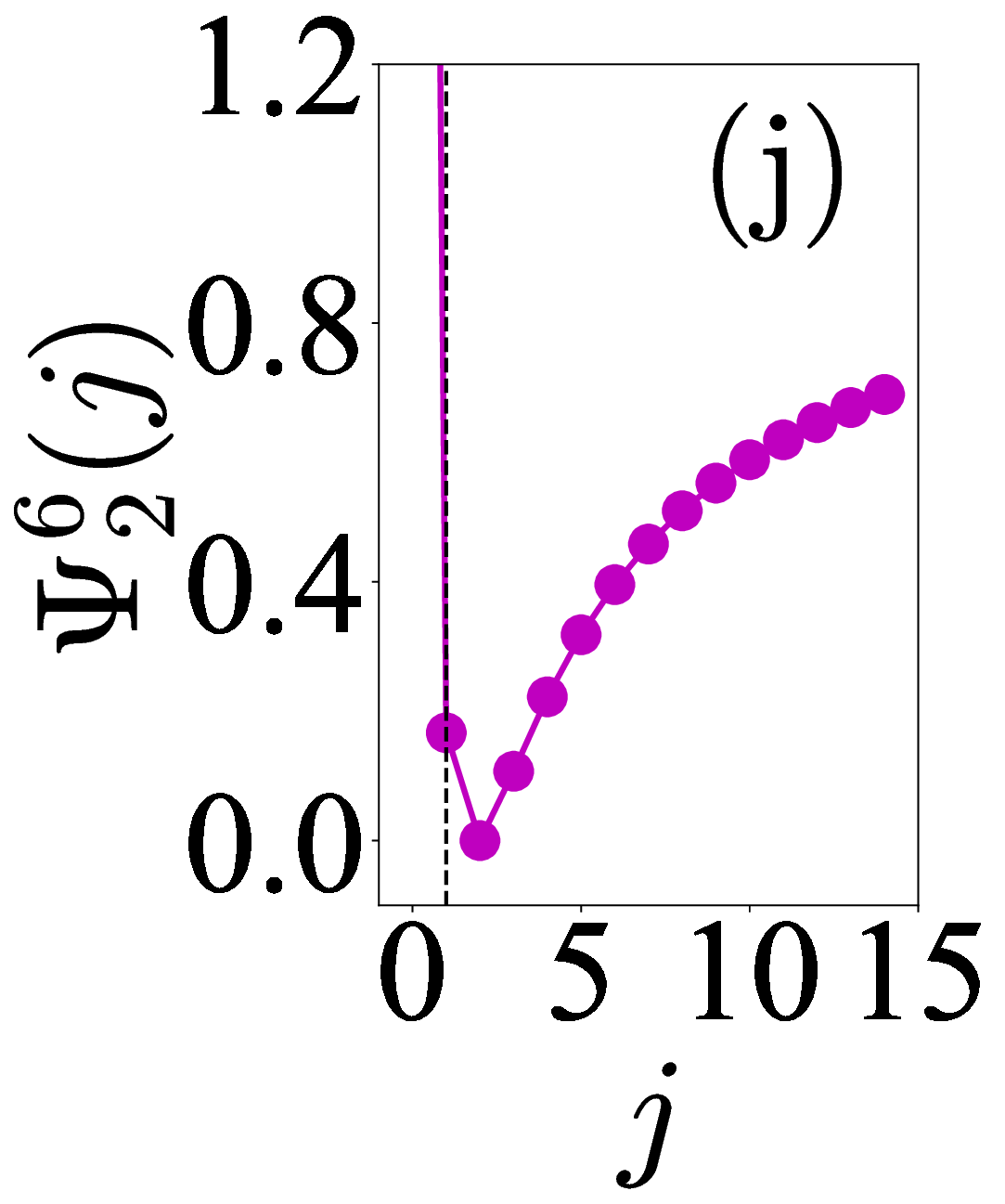}
\includegraphics[width=2.1cm]{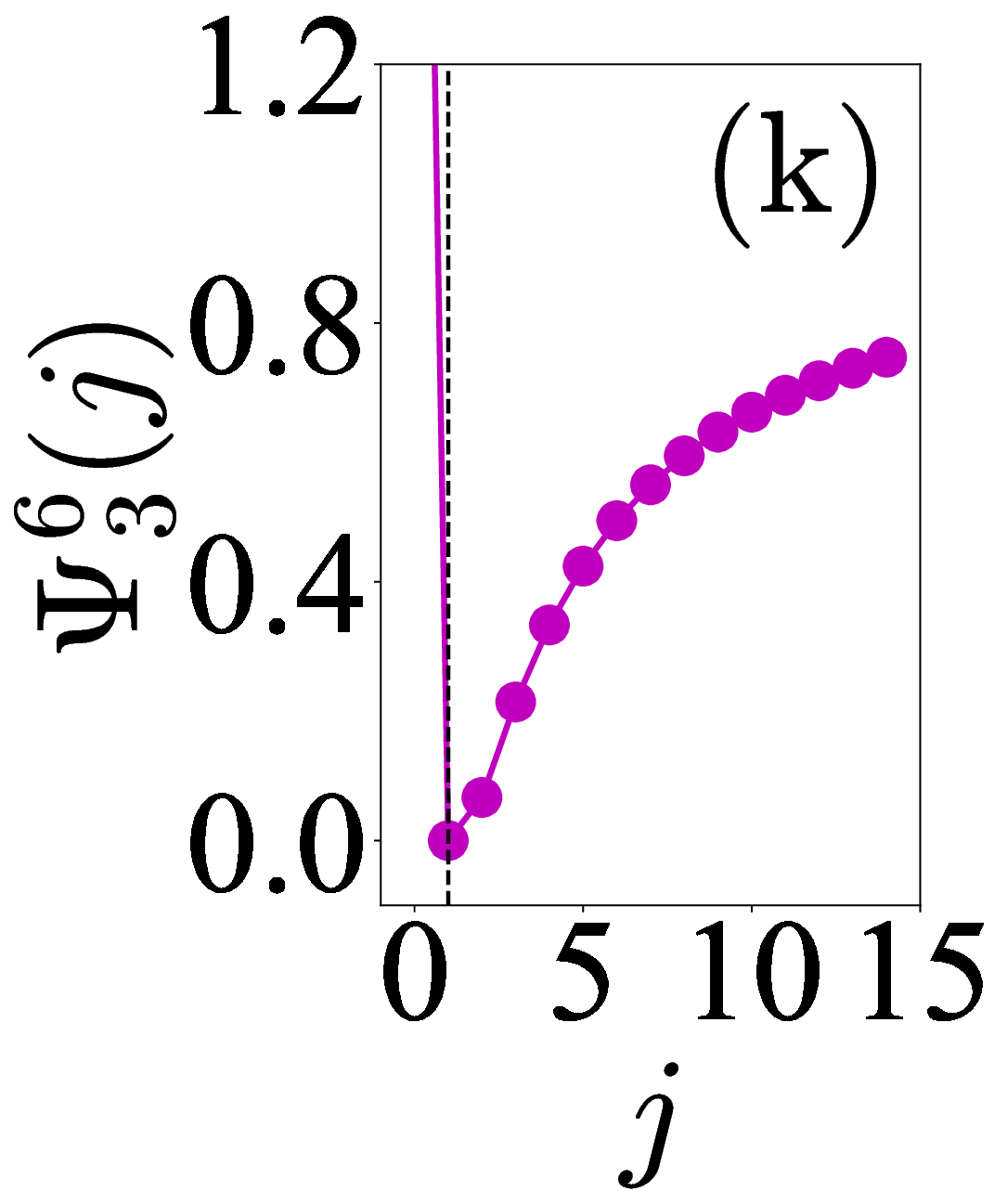}
\includegraphics[width=2.1cm]{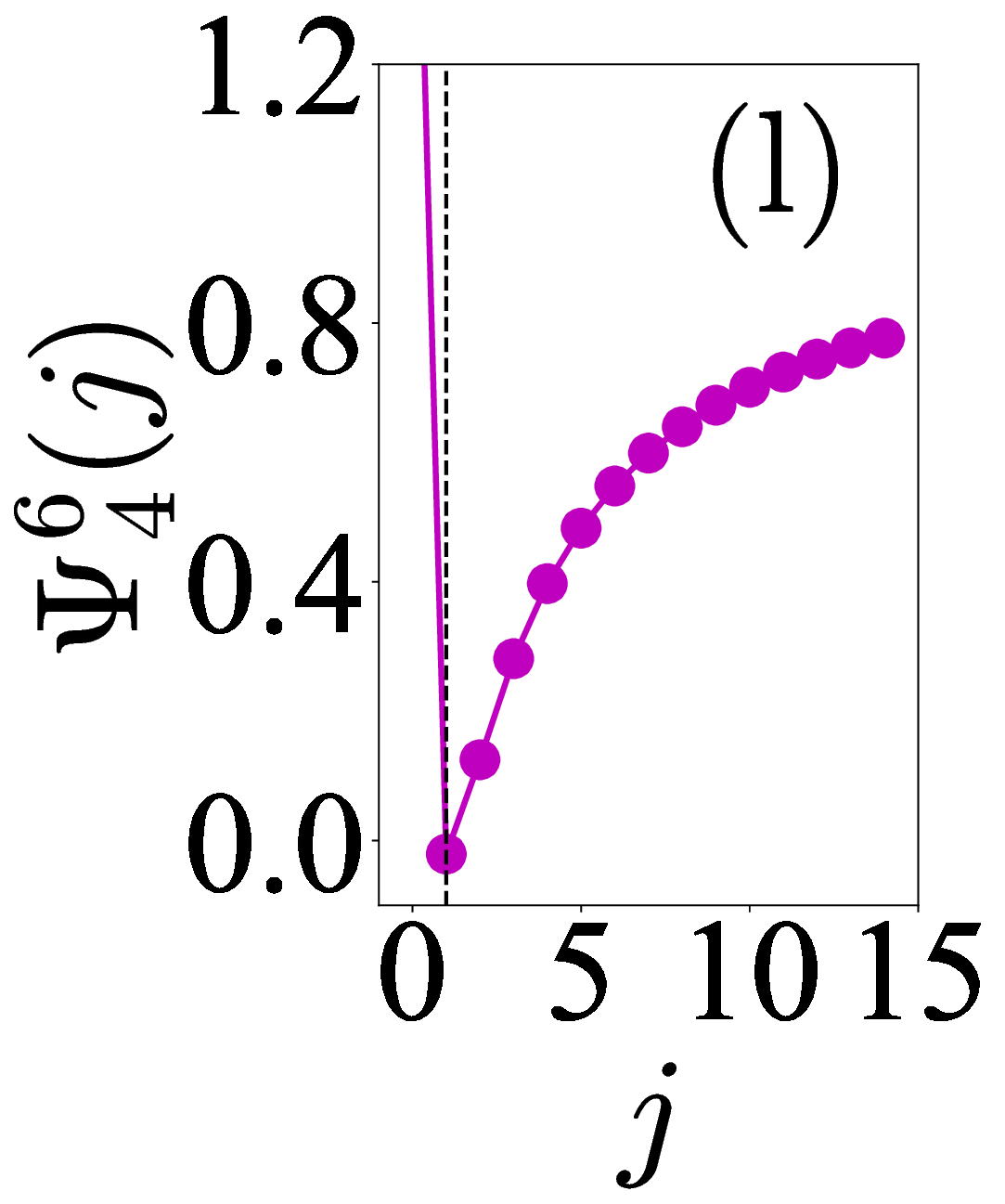}
\caption{(Color Online). The ratio $\Psi_d^{\sigma}(j)=k^{\sigma}_{\rm{sym},2j+2}/k^{\sigma}_{\rm{sym},2j}$ as a function of $j$ for different $d$. Here, panels (a-f) have $\sigma=2$, panels (e-h) have $\sigma=4$, panels (i-l) have $\sigma=6$. The vertical dashed line corresponds to $j=1$.
Figure taken from Ref.\cite{CaiLi22PRCFFG}.
}\label{fig_RatioPsi}
\end{figure}

As $d\to\infty$ while keeping $\sigma$ fixed, the isospin dependence of $\langle k^{\sigma}(\rho,\delta)\rangle$ becomes weak (approaching zero) as indicated by Eq.\,(\ref{k-sigma}). However, the isospin-expansion terms are not strictly zero. Specifically, one finds  
\begin{equation}
k_{\rm{sym,4}}^{\sigma}(\rho)=\Upsilon\left(\frac{g}{6}-\frac{g^2}{12}-\frac{g^3}{6}+\frac{g^4}{12}\right),\quad
k_{\rm{sym}}^{\sigma}(\rho)=\Upsilon(g+g^2),
\end{equation}
with $g=\sigma/d$. Both quantities approach zero as $g\to 0$ in the infinite-$d$ limit, yet their ratio remains finite:  
\begin{equation}
{k_{\rm{sym,4}}^{\sigma}(\rho)}/{k_{\rm{sym}}^{\sigma}(\rho)} \to 6^{-1},
\end{equation}
i.e., in an imagined space of infinite dimensions the ratio $\Psi$ takes the value of $1/6\approx16.7$\%, which is independent of the power $\sigma$ and is about five times the FFG model prediction in 3D. 
More generally, the ratio of adjacent coefficients is  
\begin{align}\label{def_Psi}
\Psi_d^{\sigma}(j)=\frac{1}{(2j+1)(2j+2)}\left(\frac{\sigma}{d}-2j+1\right)\left(\frac{\sigma}{d}-2j\right),
\end{align}  
shown in FIG.\,\ref{fig_RatioPsi} for various $\sigma$ and $d$. The minimum typically occurs at $j=1$, a special point for $\sigma$ and $d$. As $j\to\infty$,  
\begin{equation}
\lim_{j\to\infty}{k^{\sigma}_{\rm{sym},2j+2}}/{k^{\sigma}_{\rm{sym},2j}}=1,
\end{equation}
indicating a convergence radius $R_\delta=1$\cite{Wel16}, irrespective of the integer $\sigma$.
For $\sigma=0$, the expression (\ref{def_Psi}) gives a nonzero value, i.e.,  $\lim_{\sigma\to0}\Psi_d^{\sigma}(j)=(2j-1)(2j)/(2j+1)(2j+2)$, although in this case both the $k_{\rm{sym},2j+2}^{\sigma}$ and $k_{\rm{sym},2j}^{\sigma}$ are approaching zero (but their ratio is not necessarily zero).

The relativistic kinetic energy per nucleon in ANM in 3D can be straightforwardly and similarly written as:
\begin{align}
\label{def_EOSANMReFFG}
E^{\rm{kin}}(\rho,\delta)=\left[2\int_0^{k_{\rm{F}}}\d\v{k}\right]^{-1}\cdot
\sum_{J=\rm{n,p}}\int_0^{k_{\rm{F}}^{J}}\d\v{k}\left[\sqrt{\v{k}^2+M_{\rm N}^2}-M_{\rm N}\right],
\end{align}
where the nucleon dispersion relation changes from the non-relativistic $\v{k}^2/2M_{\rm N}$ to the relativistic form $\sqrt{\v{k}^2+M_{\rm N}^2}-M_{\rm N}$.
For small deviations $\delta x$, useful mathematical approximations are
\begin{align}
&\sum_{\Delta=\pm1}\rm{arcsinh}\,(x+\Delta\delta x)\approx2\rm{arcsinh}\,x
-\frac{x}{(x^2+1)^{3/2}}\delta x^2\notag\\
&\hspace{1cm}-\frac{1}{4}\frac{x(2x^2-3)}{(x^2+1)^{7/2}}\delta x^4
-\frac{1}{24}\frac{x(8x^4-40x^2+15)}{(x^2+1)^{11/2}}\delta x^6,\\
&\sqrt{1+\delta x}\approx1+\frac{1}{2}\delta x-\frac{1}{8}\delta x^2
+\frac{1}{16}\delta x^3
-\frac{5}{128}\delta x^4\notag\\
&\hspace{1cm}+\frac{7}{256}\delta x^5-\frac{21}{1024}\delta x^6.
\end{align}
Using these expansions, the kinetic energy and its symmetry components can be expressed analytically as
\begin{align}
E_0^{\rm{kin}}(\rho)=&\frac{3M_{\rm N}}{8\nu^2}\left[\left(1+2\nu^2\right)\sqrt{1+\nu^2}-\frac{\rm{arcsinh}\,\nu}{\nu}\right]-M_{\rm N},\\
E_{\rm{sym}}^{\rm{kin}}(\rho)=&\frac{M_{\rm N}}{6}\frac{\nu^2}{\sqrt{1+\nu^2}},\\
E_{\rm{sym,4}}^{\rm{kin}}(\rho)=&\frac{M_{\rm N}}{648}\frac{\nu^2(10\nu^4+11\nu^2+4)}{(1+\nu^2)^{5/2}},\\
E_{\rm{sym,6}}^{\rm{kin}}(\rho)=&\frac{M_{\rm N}}{34992}\frac{\nu^2(176\nu^8+428\nu^6+477\nu^4+260\nu^2+56)}{(1+\nu^2)^{9/2}},
\end{align}
with the dimensionless parameter $
\nu=k_{\rm{F}}/M_{\rm N}$.

{\it Non-relativistic limit with $\nu=k_{\rm F}/M_{\rm N}\ll1$.} Expanding for small $\nu$, one recovers
\begin{align}
E_0^{\rm{kin}}(\rho)\approx&\frac{3k_{\rm{F}}^2}{10M_{\rm N}}\left(1-\frac{5}{28}\nu^2
+\frac{5}{72}\nu^4-\frac{25}{704}\nu^6\right),\label{non0}\\
E^{\rm{kin}}_{\rm{sym}}(\rho)\approx&\frac{k_{\rm{F}}^2}{6M_{\rm N}}\left(1-\frac{1}{2}\nu^2+\frac{3}{8}\nu^4-\frac{5}{16}\nu^6\right),\label{non2}\\
E^{\rm{kin}}_{\rm{sym,4}}(\rho)\approx &\frac{k_{\rm{F}}^2}{162M_{\rm N}}\left(1+\frac{1}{4}\nu^2-\frac{25}{32}\nu^6\right),\label{non4}\\
E^{\rm{kin}}_{\rm{sym,6}}(\rho)\approx &\frac{7k_{\rm{F}}^2}{4374M_{\rm N}}\left(1+\frac{1}{7}\nu^2-\frac{5}{112}\nu^6\right),\label{non6}
\end{align}
where the prefactors correspond to the familiar non-relativistic FFG expressions.  
Notably, the first-order relativistic correction to the kinetic symmetry energy, $-k_{\rm{F}}^4/(12M^3)$, was previously given in Ref.\cite{Fri05} and is about $-0.48\,\rm{MeV}$ at $\rho_0$.

{\it Ultra-relativistic limit with $\nu=k_{\rm F}/M_{\rm N}\gg1$.}
In the opposite extreme, defining $\mu=\nu^{-1}$, one obtains
\begin{align}
E^{\rm{kin}}_0(\rho)\approx&\frac{3k_{\rm{F}}}{4}\left(1-\frac{4\mu}{3}+\mu^2
+\frac{1-4\ln(2/\mu)}{8}\mu^4\right),\label{rr0}\\
E^{\rm{kin}}_{\rm{sym}}(\rho)\approx&\frac{k_{\rm{F}}}{6}\left(1-\frac{1}{2}\mu^2+\frac{3}{8}\mu^4\right),\label{rr2}\\
E^{\rm{kin}}_{\rm{sym,4}}(\rho)\approx&\frac{5k_{\rm{F}}}{324}\left(1-\frac{7}{5}\mu^2+\frac{81}{40}\mu^4\right),\label{rr4}\\
E^{\rm{kin}}_{\rm{sym,6}}(\rho)\approx&\frac{11k_{\rm{F}}}{2187}\left(1-\frac{91}{44}\mu^2+\frac{729}{176}\mu^4\right).\label{rr6}
\end{align}
The density dependence is distinct in the two limits. For example, in the non-relativistic regime $E_{\rm{sym}}^{\rm{kin}}\sim \rho^{2/3}\ll k_{\rm{F}}$, while in the ultra-relativistic limit $E_{\rm{sym}}^{\rm{kin}}(\rho)\sim \rho^{1/3}\gg M_{\rm N}$.

The ratio of fourth- to second-order kinetic symmetry energies is then given by:
\begin{equation}\label{Psi_nu}
\Psi=\frac{E_{\rm{sym,4}}^{\rm{kin}}(\rho)}{E_{\rm{sym}}^{\rm{kin}}(\rho)}
=\frac{1}{108}\frac{4+11\nu^2+10\nu^4}{1+2\nu^2+\nu^4}.
\end{equation}
It is noteworthy that in the ultra-relativistic regime, the Fermi momentum $k_{\rm{F}}$ becomes the dominant scale since the nucleon rest mass $M_{\rm N}$ is much smaller than $k_{\rm{F}}$ and can be safely neglected. In this limit, the kinetic symmetry energies simplify to $E_{\rm{sym}}^{\rm{kin}}(\rho)\approx k_{\rm{F}}/6$ and $E_{\rm{sym,4}}^{\rm{kin}}(\rho)\approx 5k_{\rm{F}}/324$, indicating that relativistic effects enhance the quartic term relative to the conventional quadratic contribution. Specifically, this reduces to the familiar FFG result $\Psi_{\rm{NR}}=1/27$ in the non-relativistic limit, and to $\Psi_{\rm{UR}}=5/54$ in the ultra-relativistic limit; hence $1/27\leq\Psi\leq5/54$.
The enhancement of $\Psi$ reflects relativistic effects on the quartic contribution relative to the quadratic kinetic symmetry energy. For conventional nuclear densities ($\rho\lesssim 10\rho_0$), the system remains effectively non-relativistic, providing a partial explanation for the smallness of $E_{\rm{sym,4}}^{\rm{kin}}$ in the conventional FFG model.  
Furthermore, if one roughly considers $\nu=k_{\rm{F}}/M_{\rm N}\approx1$ as the transition point between non-relativistic and relativistic behavior, this corresponds to a density of $\rho\approx 45\rho_0$.

\begin{figure*}[h!]
\centering
\includegraphics[width=13.cm]{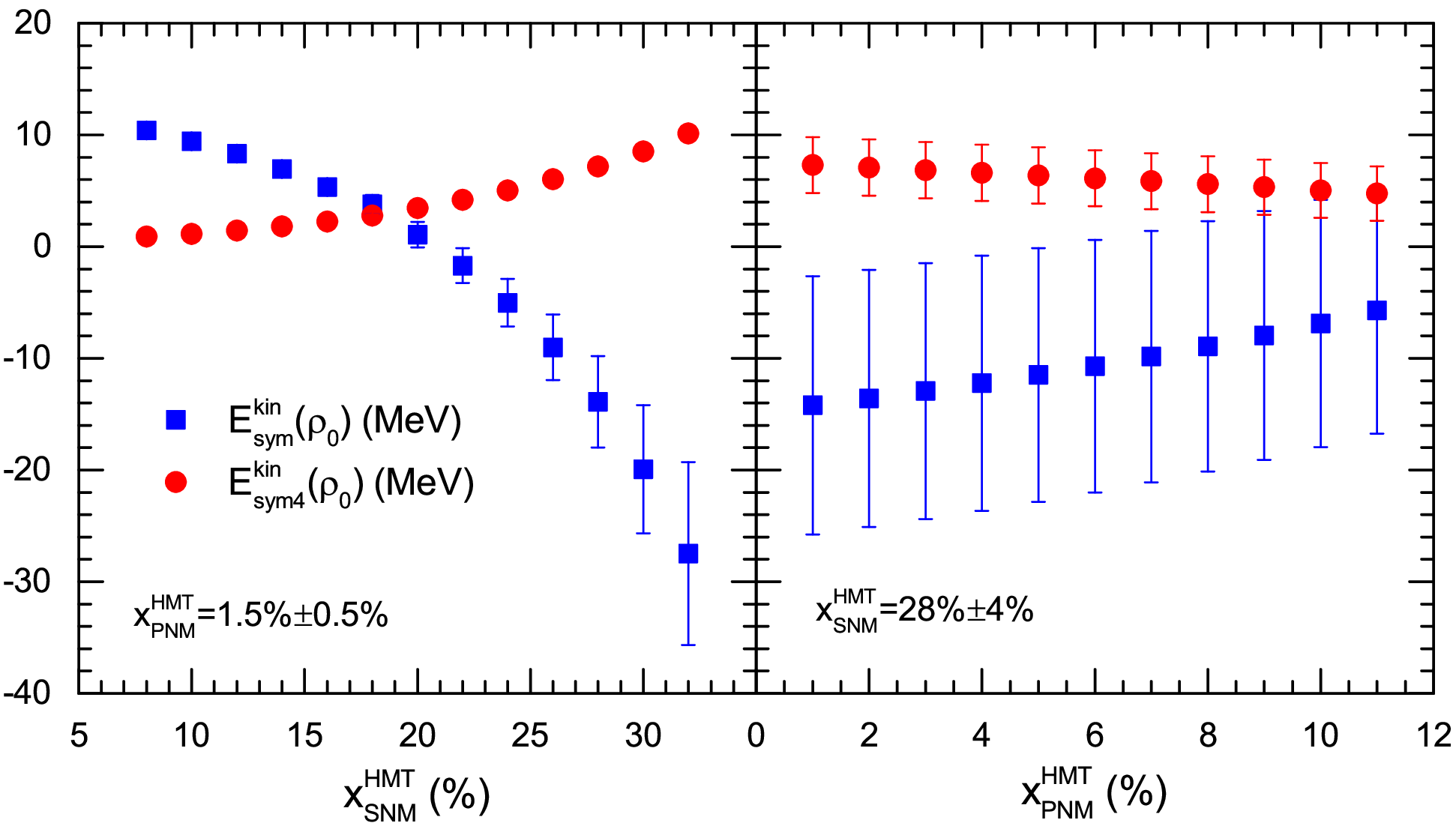}
\caption{(Color Online). Dependence of the kinetic symmetry energy and the isospin-quartic term on the fractions of high-momentum nucleons in SNM ($x_{\rm{SNM}}^{\rm{HMT}}$) and PNM ($x_{\rm{PNM}}^{\rm{HMT}}$).
Figure taken from Ref.\cite{LCCX18}.
}
\label{fig_E2E4withFraction}
\end{figure*}

\begin{figure}[h!]
\centering
  \includegraphics[height=9.cm]{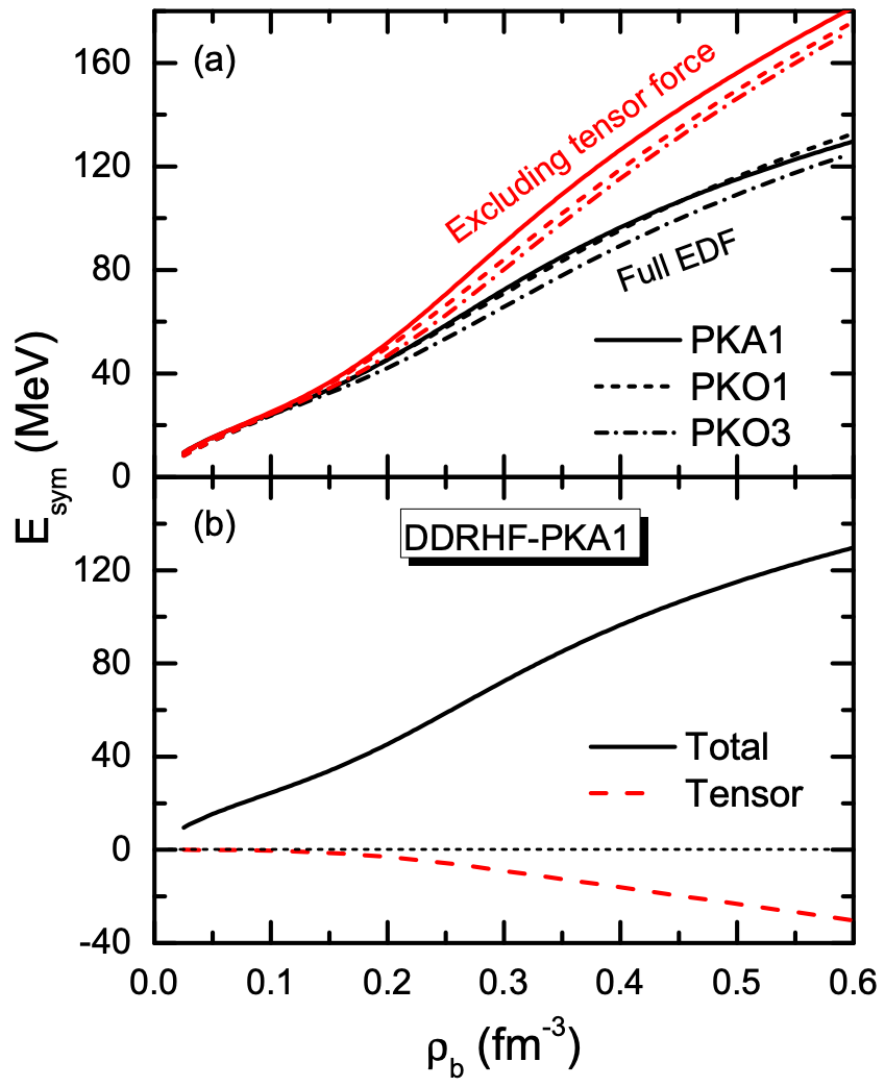}\\[0.25cm]
  \includegraphics[height=8.cm]{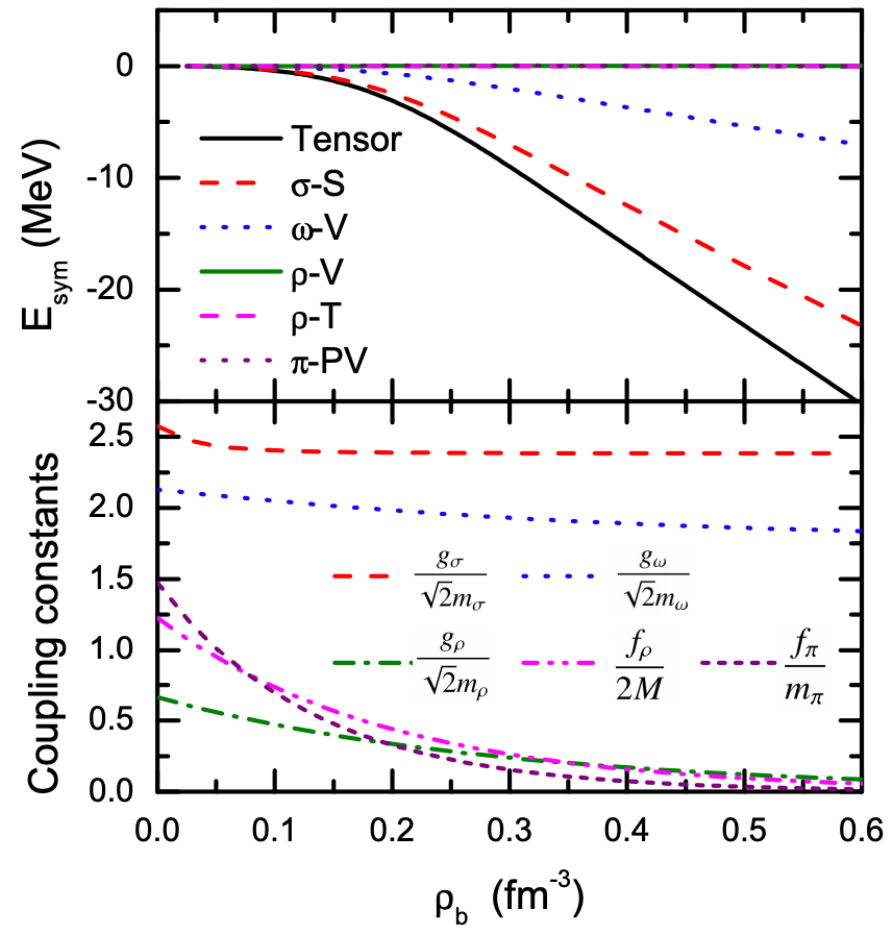}
 \caption{(Color Online). Upper: symmetry energy with/without tensor contribution obtained in the RHF model. Lower: different contributions of the the tensor part. Figures taken from Ref.\cite{LJia15}.}
  \label{fig_tensor-12}
\end{figure}

\begin{figure}[h!]
\centering
  \includegraphics[width=6.5cm]{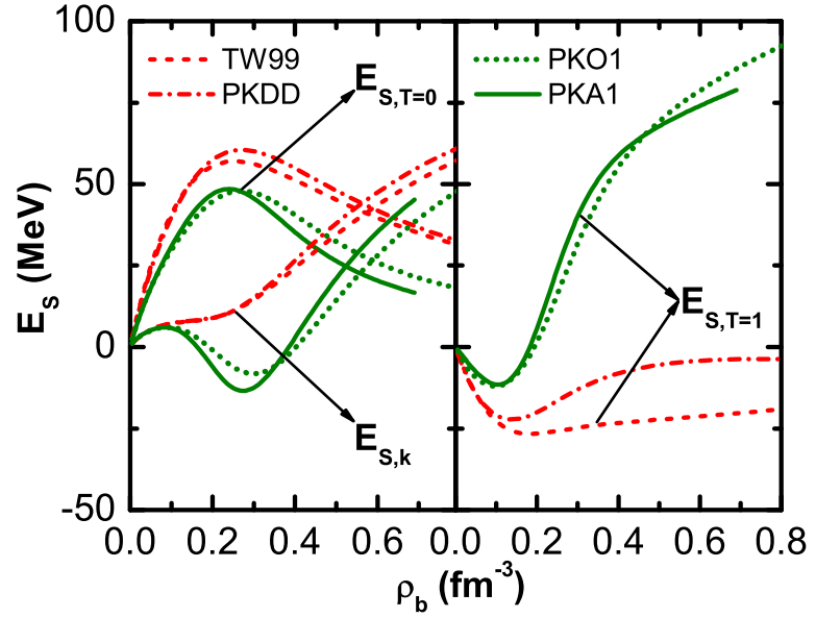}
 \caption{(Color Online). Kinetic part, the isospin-single and the isospin-triplet parts of the symmetry energy obtained in the RHF theory. Figure taken from Ref.\cite{QZha15}.}
 \label{fig_tensor-3}
\end{figure}

\begin{figure}[h!]
\centering
  \includegraphics[width=9.cm]{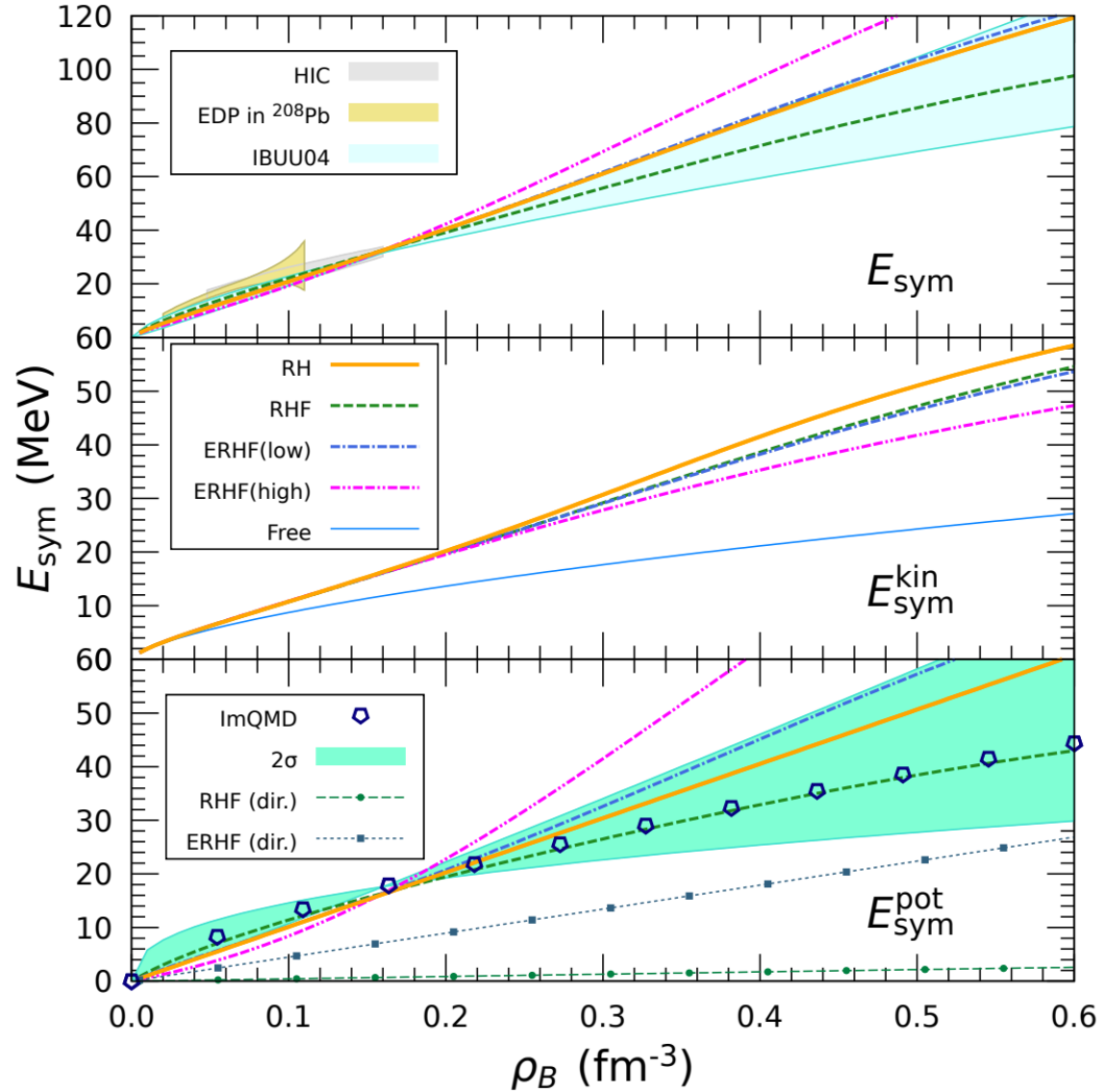}
 \caption{(Color Online). Symmetry energy obtained in the RHF theory via the HVH theorem. Figure taken from Ref.\cite{Miy19}.}
 \label{fig_tensor-4}
\end{figure}

\subsection{Reduction of Kinetic Symmetry Energy and the Parabolic Approximation Issue}\label{sub_kinEsym}

\indent

The SRC-induced HMT modifies the single-nucleon momentum distribution $n_{\v{k}}^J$, which has direct consequences on the kinetic EOS of ANM\cite{Cai15a}. In this subsection, we discuss these effects. The kinetic EOS can be expanded in powers of $\delta$:
\begin{equation}
E^{\rm{kin}}(\rho,\delta)\approx E_0^{\rm{kin}}(\rho)+E_{\rm{sym}}^{\rm{kin}}(\rho)\delta^2+E_{\rm{sym,4}}^{\rm{kin}}(\rho)\delta^4+\mathcal{O}(\delta^6).
\end{equation}
Using the $n^J_{\v{k}}(\rho,\delta)$ in Eq.\,(\ref{MDGen}) with $\beta_J=0$, the coefficients from Eq.\,(\ref{kinE}) are
\begin{align}
E^{\rm{kin}}_0(\rho)=&\frac{3}{5}E_{\rm{F}}(\rho)\left[
1+{C}_0\left(5\phi_0+\frac{3}{\phi_0}-8\right)\right],\label{E0kin}\\
E_{\rm{sym}}^{\rm{kin}}(\rho)=&\frac{1}{3}E_{\rm{F}}(\rho)\Bigg[1+{C}_0\left(1+3{C}_1\right)\left(5\phi_0+\frac{3}{\phi_0}-8\right)
\notag\\
&+3{C}_0\phi_1\left(1+\frac{3}{5}{C}_1\right)\left(5\phi_0-\frac{3}{\phi_0}\right)+\frac{27{C}_0\phi_1^2}{5\phi_0}\Bigg],\label{Esymkin}\\
E_{\rm{sym,4}}^{\rm{kin}}(\rho)=&\frac{1}{81}E_{\rm{F}}(\rho)\Bigg[1+{C}_0(1-3{C}_1)\left(5\phi_0+\frac{3}{\phi_0}
-8\right)\notag\\
&+3{C}_0\phi_1(9{C}_1-1)\left(5\phi_0-\frac{3}{\phi_0}\right)\notag\\
&+\frac{81{C}_0\phi_1^2(9\phi_1^2-9{C}_1\phi_1-15\phi_1+15{C}_1+5)}{5\phi_0}
\Bigg],\label{Esymkin4}
\end{align}
where $E_{\rm F}(\rho)=k_{\rm F}^2/2M_{\rm N}$ is the nucleon Fermi energy in SNM.
Because $\phi_0>1$ in the presence of HMT, one has 
\begin{equation}\label{def-Phi0}
\Phi_0\equiv 5\phi_0+\frac{3}{\phi_0}-8>0.
\end{equation}

Consequently, the kinetic EOS of SNM is enhanced relative to the FFG prediction\cite{CXu11,Hen14}. This enhancement arises naturally since neutron-proton SRC dominates in SNM, and the HMT significantly increases the kinetic energy through the $k^4$ factor in Eq.\,(\ref{kinE}). In the absence of HMT, i.e., the FFG case with $\phi_0=1$ and $\phi_1=0$, $\Phi_0=0$, and the expressions reduce to the well-known results
$
E^{\rm{kin}}_0(\rho)=3E_{\rm{F}}(\rho)/5$, $E_{\rm{sym}}^{\rm{kin}}(\rho)=E_{\rm{F}}(\rho)/3$, as well as
$E_{\rm{sym,4}}^{\rm{kin}}(\rho)=E_{\rm{F}}(\rho)/81$.
For interacting nucleons in ANM, using the  momentum distribution parameters, it was found\cite{CaiLi16a} that
$E_0^{\rm{kin}}(\rho_0)\approx40.45\pm8.15\,\rm{MeV}$,
$E_{\rm{sym}}^{\rm{kin}}(\rho_0)\approx-13.90\pm11.54\,\rm{MeV}$, and
$E_{\rm{sym,4}}^{\rm{kin}}(\rho_0)\approx7.19\pm2.52\,\rm{MeV}$.
The relativistic corrections to these quantities would be briefly discussed in the end of this subsection.
Compared to FFG values, the isospin-dependent HMT significantly increases $E_0^{\rm{kin}}(\rho_0)$ while reducing $E_{\rm{sym}}^{\rm{kin}}(\rho_0)$ to a negative value, consistent qualitatively with several recent studies using phenomenological and microscopic many-body approaches\cite{Rio14,CXu11,Vid11,Lov11,Car12,Car14}.
The validity of the empirical isospin parabolic law for the kinetic EOS in the presence of isospin-dependent HMT was previously unknown, but it was found to be seriously broken\cite{CaiLi16a}. Since the corrections in the square brackets for $E_0^{\rm{kin}}(\rho)$, $E_{\rm{sym}}^{\rm{kin}}(\rho)$ and $E_{\rm{sym,4}}^{\rm{kin}}(\rho)$ are not small, the parabolic approximation of the EOS of ANM at the kinetic level is expected to fail. Quantitatively, the ratio
$|E_{\rm{sym,4}}^{\rm{kin}}(\rho_0)/E_{\rm{sym}}^{\rm{kin}}(\rho_0)| \approx 52\%\pm26\%$
is much larger than the FFG value of $3.7\%$\cite{CaiLi16a}. The large quartic term is mainly due to the strong isospin dependence of the HMT cutoff parameter $\phi_1$. Setting $\phi_1=0$ yields
$E^{\rm{kin}}_{\rm{sym}}(\rho_0)\approx14.68\pm2.80$\,MeV and
$E^{\rm{kin}}_{\rm{sym},4}(\rho_0)\approx1.12\pm0.27$\,MeV, close to the FFG values.

FIG.\,\ref{fig_E2E4withFraction} illustrates that a large quartic term arises when the difference between $x_{\rm{SNM}}^{\rm{HMT}}$ and $x_{\rm{PNM}}^{\rm{HMT}}$ is large. Previous studies that assumed the parabolic approximation evaluated $E_{\rm{sym}}^{\rm{kin}}(\rho)$ as the difference between the kinetic energies of PNM and SNM, effectively summing the kinetic symmetry and quartic energies:
$E_{\rm{sym}}^{\rm{kin,app}}(\rho_0)\approx E_{\rm{sym}}^{\rm{kin}}(\rho_0)+E_{\rm{sym,4}}^{\rm{kin}}(\rho_0)\approx-6.71\pm9.11$\,MeV.
This explains the quantitative consistency with $E_{\rm{sym}}^{\rm{kin}}(\rho_0)$ reported in Ref.\cite{Hen15b} under the parabolic approximation.
Using the HVH theorem\cite{Hug58} at the mean-field level, an appreciable isospin-quartic term in the EOS may arise from high-order derivatives of the momentum-dependent symmetry potential or differences between nucleon isoscalar and isovector effective masses. SRC modifies the nucleon momentum distribution, producing quasi-particles whose distribution differs from the FFG step function. Since SRC is mainly tensor-force induced, the reduction in kinetic symmetry energy and enhancement of the quartic term can also be attributed to the tensor force, which does not contribute at the lowest order of the mean-field. These SRC-induced modifications can be considered additions to the mean-field EOS, though further quantitative studies are needed.
The enhancement of the kinetic quartic term is also observed in relativistic Hartree-Fock (RHF) theories. In Ref.\cite{LiuZW18PRC}, the authors further showed that the Fock terms, particularly those from the isoscalar meson coupling channels, significantly affect the kinetic contribution to $S_4(\rho)\equiv E_{\rm{sym,4}}(\rho)$, systematically leading to an increased kinetic term in RHF especially for $\rho\gtrsim0.2\,\rm{fm}^{-3}$, see the upper panel of FIG.\,\ref{fig_RHF-S4}.
The authors of Ref.\cite{LJia15} found that including the tensor correlations reduces the symmetry energy as a function of density, see the upper panel of FIG.\,\ref{fig_tensor-12} where the three interaction sets namely PKA1, PKO1, and PKO3 are considered.
Although the tensor correlations contributes very little at/around the normal density, its effect could be sizable at large densities to be an amount of 10-20$\%$ of the total symmetry energy compared with other channels. Moreover, based on the interaction set PKA1, it was found that the tensor component in the $\sigma$-S coupling channel dominates the tensor contributions to
the symmetry energy, followed by the $\omega$-V couplings, while
those in the $\pi$- and $\rho$- and exchanges are relatively small. This result can be well understood from the tensor coupling constants shown in the lower panel of the lower part of FIG.\,\ref{fig_tensor-12} namely that the tensor coupling constants in $\sigma$-S and $\omega$-V channels tend to certain values at high density, whereas due to the
exponential density-dependent behavior of $g_{\rho},f_{\rho}$ and $f_{\pi}$, those
from the isovector $\rho$-V, $\rho$-T and $\pi$-PV channels vanish at the supra-normal density region where the tensor effects become notable.
Based on the RHF framework, the kinetic-potential decomposition of the symmetry energy was studied in Ref.\cite{QZha15}, and the kinetic part of the symmetry energy was found to be reduced, see FIG.\,\ref{fig_tensor-3} for the decomposition of the symmetry energy.
For example the kinetic symmetry energy in the PKO1 (PKA1) with the tensor correlations included is found to be about 3.7\,MeV (0.5\,MeV), which is much smaller than the one no tensor correlations is included, i.e., the 8.0\,MeV in TW99 (8.1\,MeV in PKDD).
Parallel to the work given in Ref.\cite{LJia15}, the Fock contribution to the symmetry energy was recently investigated in Ref.\cite{Miy19} in the RHF framework via the HVH theorem.
See FIG.\,\ref{fig_tensor-4} for the density dependence of the symmetry energy where the reduction on the symmetry energy from the RHF set is obvious, here the $E_{\rm{sym}}(\rho_0)$ in different models (i.e., RH, RHF, ERHF (low) and ERHF (high)) is fixed at 32.5\,MeV. The two sets namely ERHF(low) and ERHF(high) are designed to well describe the single-nucleon potential obtained from optical models at low energies and high energies, respectively.
In the RHF theory the potential part is decomposed into two parts $E_{\rm{sym}}^{\rm{pot}}(\rho)=E_{\rm{sym}}^{\rm{pot,dir}}(\rho)+E_{\rm{sym}}^{\rm{pot,ex}}(\rho)$ where the direct term is simply given by $E_{\rm{sym}}^{\rm{pot,dir}}(\rho)=g_{\rho}^2\rho/2Q_{\rho}$ with $Q_\rho=m_\rho^2+g_\rho^2g_{\omega}^2\overline{\omega}_0^2$, and once the negative Fock contribution to the symmetry energy is included the direct term decreases to a very small value, and the final density dependence of the symmetry energy is competition result from these difference channels.

\begin{figure}[h!]
\centering
\hspace{0.2cm}
\includegraphics[width=8.5cm]{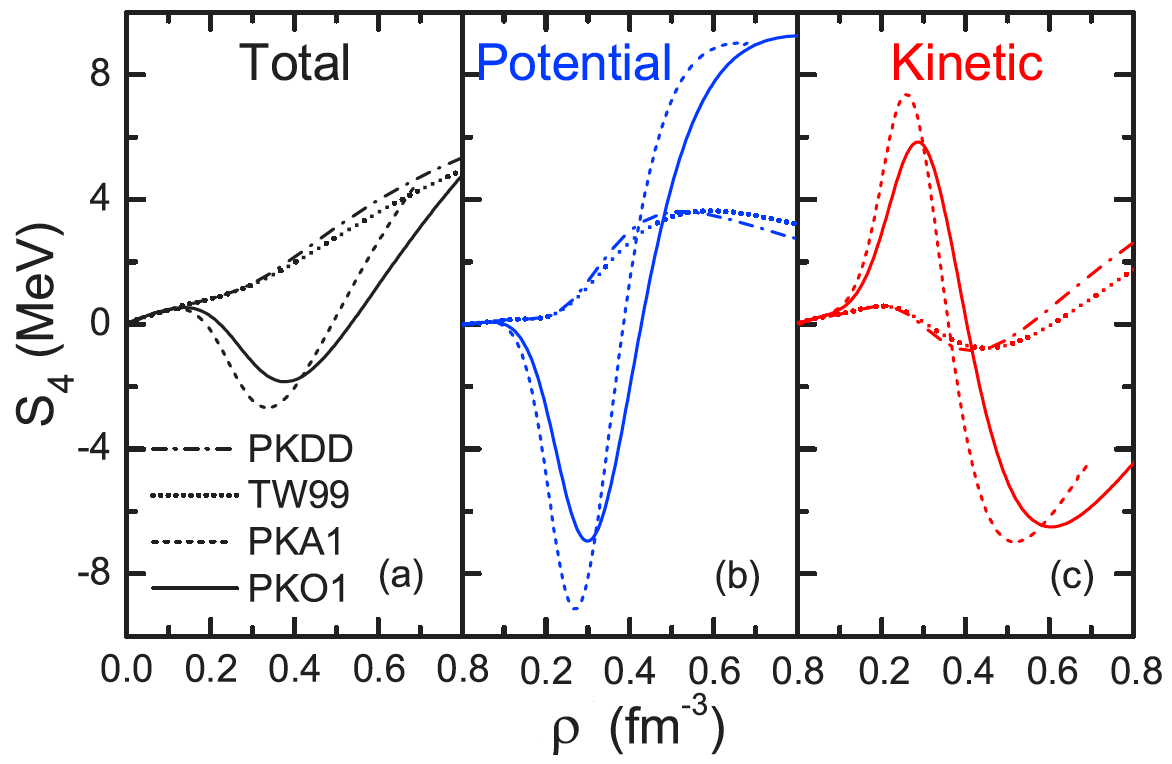}\\[0.25cm]
\includegraphics[width=9.cm]{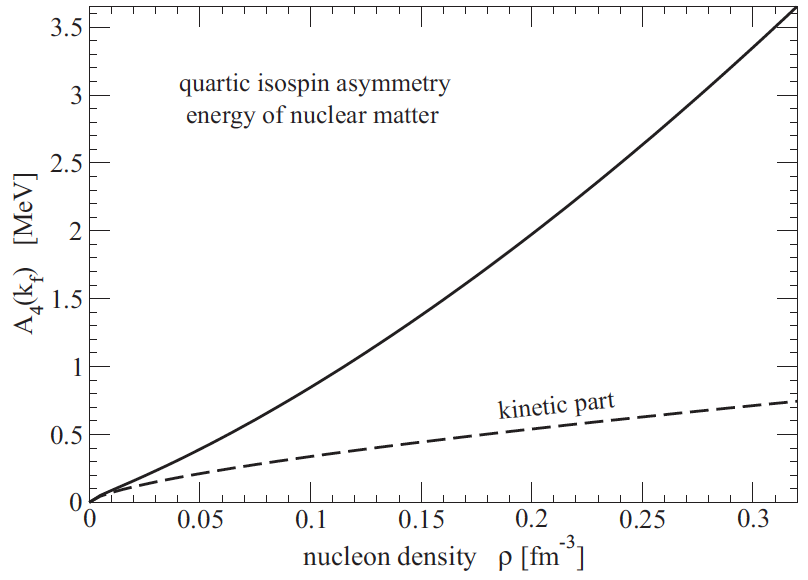}
\caption{(Color Online). Upper: the nuclear fourth-order symmetry energy, its potential part and kinetic part as functions of the baryon density. The results are calculated with RHF interaction sets PKA1 and PKO1, in comparison with the RMF ones PKDD and TW99; figure taken from Ref.\cite{LiuZW18PRC}.
Lower: the isospin quartic term obtained in the chiral effective theory; figure taken from Ref.\cite{Kai16PRC}.
}
\label{fig_RHF-S4}
\end{figure}

\begin{figure}[h!]
\centering
\includegraphics[width=8.cm]{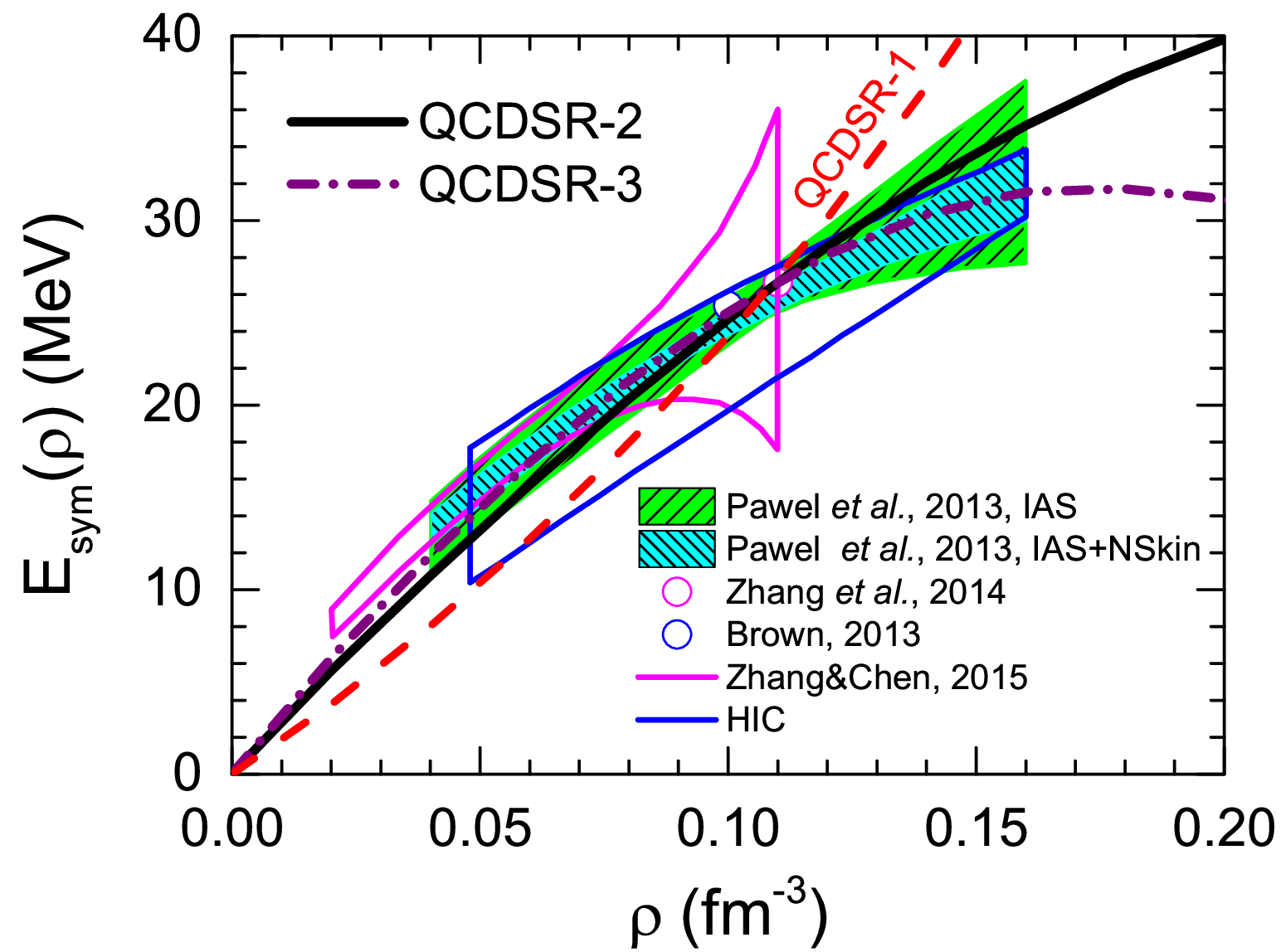}\quad
\includegraphics[width=8.cm]{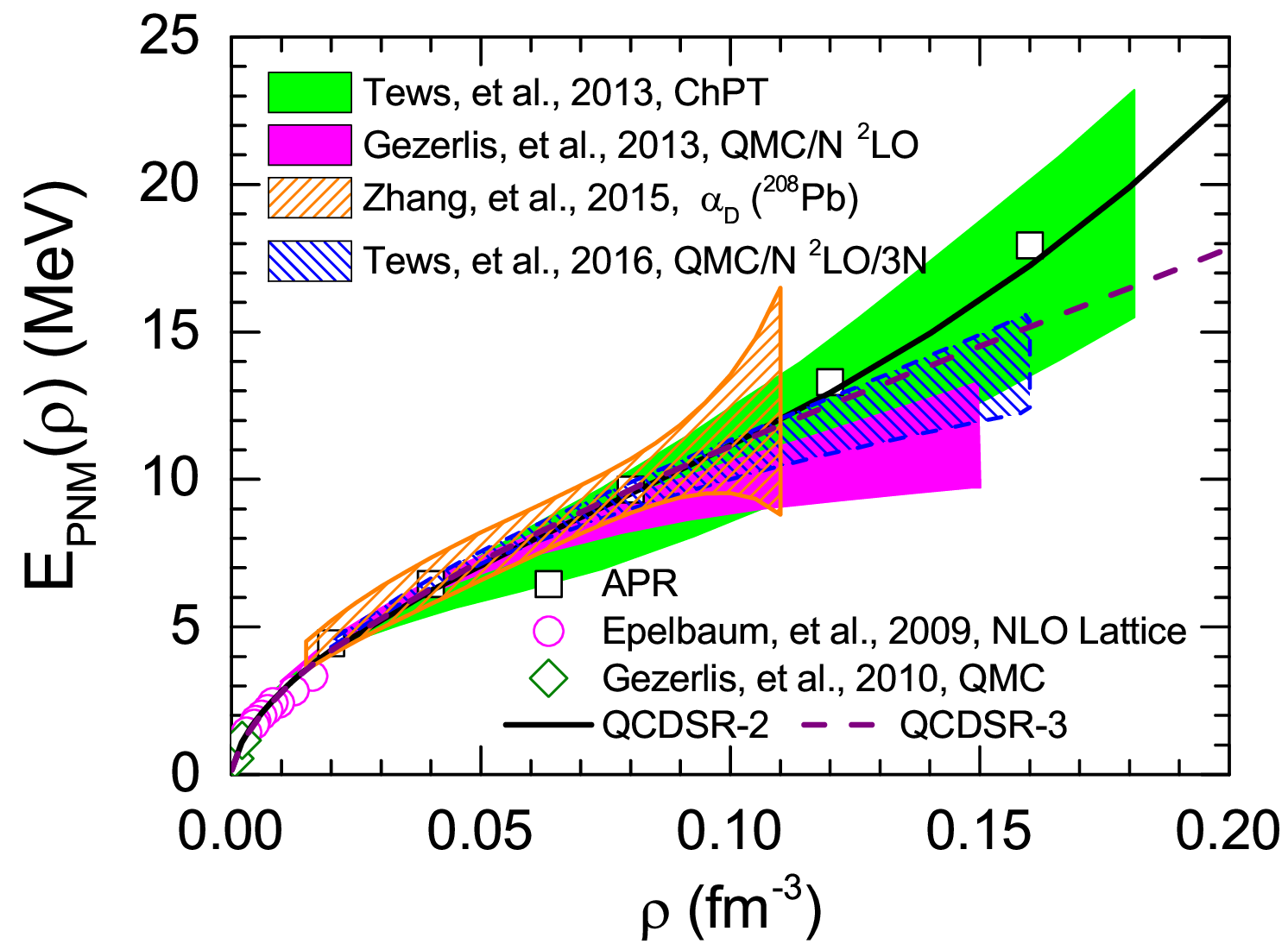}
\caption{(Color Online). Symmetry energy (upper) and the EOS of PNM (lower) obtained in QCDSR. Figures taken from Ref.\cite{Cai19-QSR2}.}
\label{fig_QCDSR}
\end{figure}

Although the parabolic approximation for the EOS may break down at the kinetic level, the full EOS of ANM, including the potential part, is still expected to satisfy this approximation (especially near the saturation density), as illustrated, for example, by the RHF calculations shown in the upper panel of FIG.\,\ref{fig_RHF-S4}. The parabolic behavior of the full EOS of ANM has also been confirmed by chiral effective field theory studies; for instance, Ref.\cite{Kai16PRC} found that $E_{\rm{sym,4}}(\rho_0)\lesssim1.5\,\rm{MeV}$ (see the lower panel of FIG.\,\ref{fig_RHF-S4}), consistent with other investigations\cite{Cai12PRC-S4,Seif14PRC-S4,Gonz17PRC-S4,Pu17PRC-S4}.
One exception comes from the QCD sum rules (QCDSR) calculations for the EOS of ANM\cite{Cai18-QSR1,Cai19-QSR2}.
In QCDSR, the EOS of PNM and the symmetry energy are well described, see, e.g., FIG.\,\ref{fig_QCDSR}\cite{Cai19-QSR2}.
In order to estimate the contributions of higher-order terms in the EOS, we write out the EOS of SNM and that of PNM as\cite{Cai19-QSR2},
\begin{align}\label{E0s}
E_0(\rho)
\approx&E_0^{\rm{FFG}}(\rho)+
\frac{1}{2}\frac{M_{\rm N}\rho}{\langle\overline{q}q\rangle_{\rm{vac}}}\notag\\
&\times\left[\frac{\sigma_{\rm{N}}}{2m_{\rm{q}}}\left(1+\frac{16\pi^2f}{3}\frac{M_{\rm N}\langle\overline{q}q\rangle_{\rm{vac}}}{\mathscr{M}^4}\right)-4\right],\\
E_{\rm{n}}(\rho)
\approx&E_{\rm{n}}^{\rm{FFG}}(\rho)+\frac{1}{2}\frac{M_{\rm N}\rho}{\langle\overline{q}q\rangle_{\rm{vac}}}\notag\\
&\times\left[\left(1-\xi\right)\frac{\sigma_{\rm{N}}}{2m_{\rm{q}}}\left(1+\frac{16\pi^2f}{3}\frac{M_{\rm N}\langle\overline{q}q\rangle_{\rm{vac}}}{\mathscr{M}^4}\right)-5\right]
,\label{Ens}
\end{align}
here
$\xi\approx0.1$\cite{Cai19-QSR2} is the 
coefficient characterizing the isospin splitting of the chiral condensate.
Since the EOS of PNM at very low densities (say densities smaller than $0.01\,\rm{fm}^{-3}$) could be
determined very accurate by simulations or microscopic calculations, there exists a relation between several fundamental quantities and the model parameters in QCDSR,
such as the four-quark effective parameter $f$\cite{Cohen95}, the nucleon-sigma term $\sigma_{\rm{N}}$\cite{Gas91}, the chiral condensate $\langle\overline{q}q\rangle_{\rm{vac}}$ in vacuum\cite{Cohen95}, the light quark mass $m_{\rm q}$, and the Borel mass $\mathscr{M}$\cite{Cohen95,Cai19-QSR2}.

The symmetry energy obtained in the parabolic approximation is roughly given by
\begin{align}\label{Esyms}
E_{\rm{sym}}^{\rm{para}}(\rho)
\approx&E_{\rm{sym,para}}^{\rm{FFG}}(\rho)
-\frac{1}{2}\frac{M_{\rm N}\rho}{\langle\overline{q}q\rangle_{\rm{vac}}}\notag\\
&\times\left[\frac{\xi\sigma_{\rm{N}}}{2m_{\rm{q}}}\left(1+\frac{16\pi^2f}{3}\frac{M_{\rm N}\langle\overline{q}q\rangle_{\rm{vac}}}{\mathscr{M}^4}\right)+1\right].
\end{align}
The kinetic symmetry energy
in the parabolic approximation is 
$E_{\rm{sym,para}}^{\rm{FFG}}(\rho)=(2^{2/3}-1)E_0^{\rm{FFG}}(\rho)\approx0.59E_0^{\rm{FFG}}(\rho)$.
If one takes $f=0$ in Eq.\,(\ref{Ens}), then the $E_{\rm{n}}(\rho)$ could be written as\cite{Cai18-QSR1}
\begin{align}
\boxed{
E_{\rm{n}}(\rho)
\approx E_{\rm{n}}^{\rm{FFG}}(\rho)+\frac{1}{2}\frac{M_{\rm N}\rho}{\langle\overline{q}q\rangle_{\rm{vac}}}\left[\left(1-\xi\right)\frac{\sigma_{\rm{N}}}{2m_{\rm{q}}}-5\right]
,}\label{Ens-1}
\end{align}
which depends only on the fundamental quantities such as $m_{\rm{q}}$, $\langle\overline{q}q\rangle_{\rm{vac}}$,
and $\sigma_{\rm{N}}$, and not on the effective parameters $f$ and $\mathscr{M}^2$.
Despite its simplicity, Eq.\,(\ref{Ens-1}) has already the power of quantitative predictions at very low densities.
The high-order (HO) term of the EOS is simply obtained as $E_{\rm{HO}}(\rho)=E_{\rm{sym}}^{\rm{para}}(\rho)-E_{\rm{sym}}(\rho)$.
The numerical results on the $E_{\rm{sym}}(\rho)$ and $E_{\rm{HO}}(\rho)$ using the QCDSR equations are shown in the upper part of FIG.\,\ref{fig_fEHOEsym}.
For example, the high order effects in the EOS at  $\rho_0=0.16\,\rm{fm}^{-3}$ are found to be about 8.8\,MeV, 7.4\,MeV, and 6.4\,MeV for $f=0$, 0.25 and 0.5, respectively; clearly showing that the high order term $E_{\rm{HO}}(\rho)$ at the saturation density $\rho_0$ are generally not small (e.g., $\lesssim2\,\rm{MeV}$)\cite{Cai12PRC-S4,Gonz17PRC-S4,Pu17PRC-S4,Seif14PRC-S4}.

\begin{figure}[h!]
\centering
  \includegraphics[width=7.5cm]{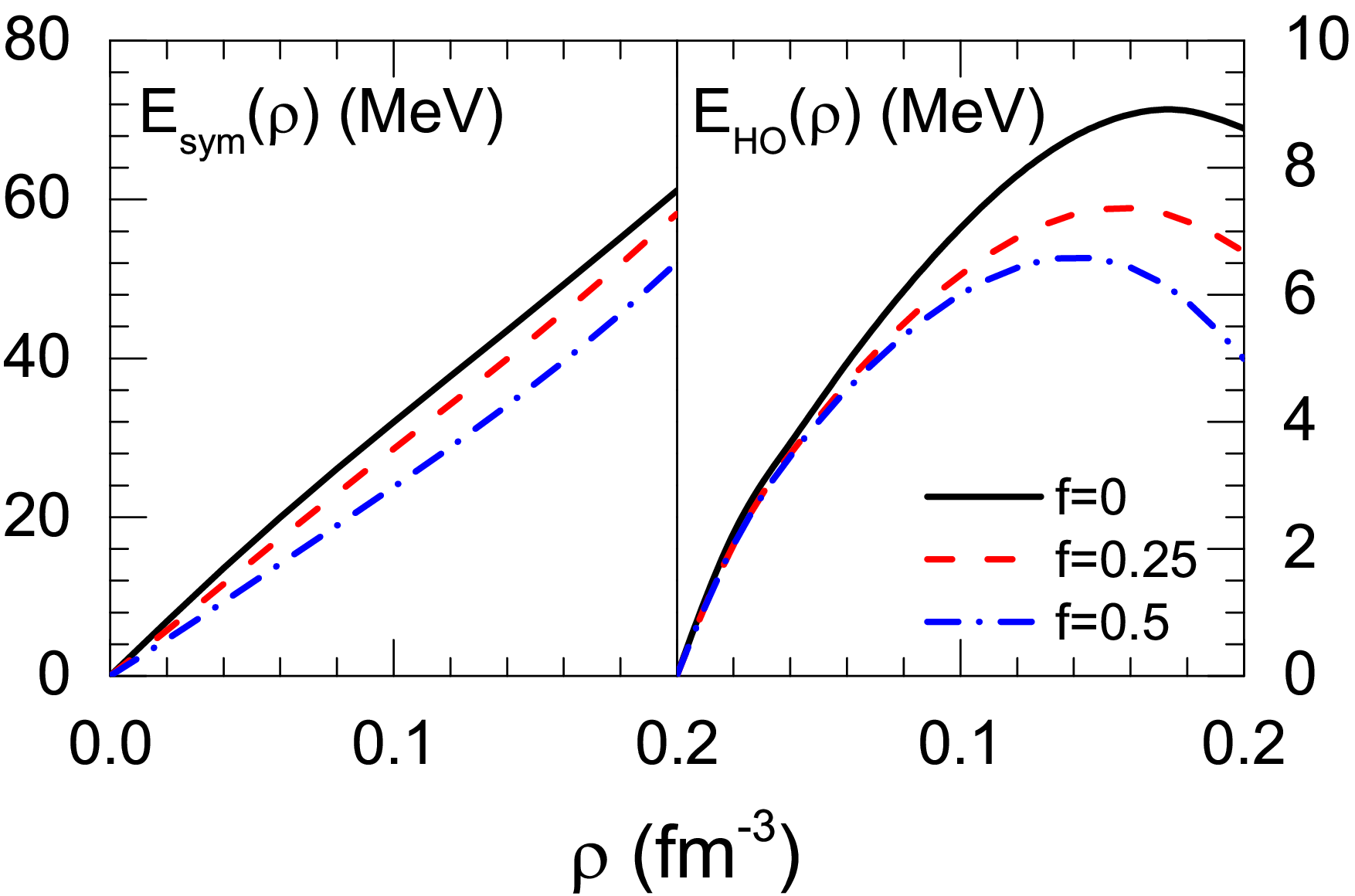}\\[0.25cm]
  \hspace{-.85cm}
  \includegraphics[width=7.5cm]{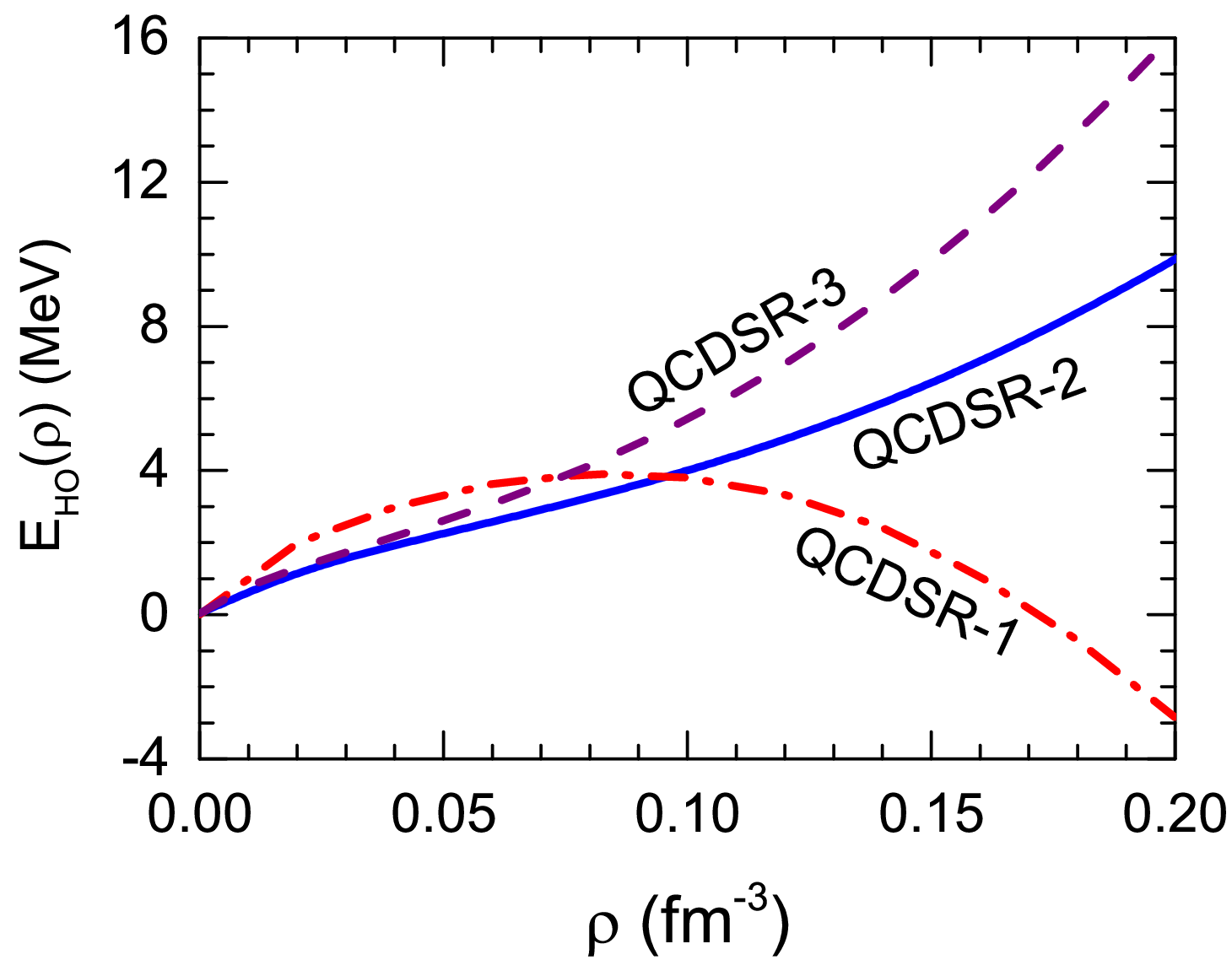}
  \caption{(Color Online). Upper:  the $E_{\rm{sym}}(\rho)$ and $E_{\rm{HO}}(\rho)$ as functions of density with different $f$ using the QCDSR. Lower: high order EOS obtained in QCDSR for three different calculation schemes. Figures taken from Ref.\cite{Cai19-QSR2}.}
  \label{fig_fEHOEsym}
\end{figure}

It is useful to generally analyze the origin of the possible breakdown of the parabolic approximation of the EOS of ANM in QCDSR framework.
In the simplest QCDSR situation, the density dependence of the chiral condensates has the structure of
$
\langle\overline{q}q\rangle_{\rho,\delta}^{\rm{u,d}}
=\langle\overline{q}q\rangle_{\rm{vac}}+a(1\mp\xi\delta)\rho$,
where $a=\sigma_{\rm{N}}/2m_{\rm{q}}$.
The nucleon Dirac effective mass and the vector self-energy are given as $M_J^{\ast}\sim\langle\overline{q}q\rangle_{\rho,\delta}^{\rm{u,d}}$
and $\Sigma_{\rm{V}}^J\sim\langle q^{\dag}q\rangle_{\rho,\delta}^{\rm{u,d}}$, respectively.
Then the $E_{\rm{HO}}(\rho)$ could be estimated as\cite{Cai19-QSR2}:
\begin{equation}\label{zz_EHOes}
E_{\rm{HO}}(\rho)\sim-2^{-1}a\xi\rho\left(1-{M_0^{\ast}}/{{E}_{\rm{F}}^{\ast}}\right),
\end{equation}
where the overall factor is unimportant for the qualitative demonstration.
At low densities, $E_{\rm{F}}^{\ast}=(M_0^{\ast,2}+k_{\rm{F}}^2)^{1/2}\approx k_{\rm{F}}^2/2M_0^{\ast}+M_0^{\ast}
\approx M_0^{\ast}$, indicating that the expression in the parentheses in Eq.\,(\ref{zz_EHOes}) is small.
As density increases, the $E_{\rm{HO}}(\rho)$ by Eq.\,(\ref{zz_EHOes}) increases eventually, leading to the increasing of the
high order effects eventually. For example, keeping only the leading order term in $(k_{\rm{F}}/M_0^{\ast})^2$, one obtains the high order effect
as $
E_{\rm{HO}}(\rho)\sim-a\xi k_{\rm{F}}^2\rho/4M_0^{\ast,2}$,
which is negative.
When considering high order terms in density in the
chiral condensates, e.g., $
\langle\overline{q}q\rangle_{\rho,\delta}^{\rm{u,d}}
=\langle\overline{q}q\rangle_{\rm{vac}}+a(1\mp\xi\delta)\rho+b(1\mp\zeta\delta)\rho^{\vartheta}$ with $\vartheta>1$,
and assuming other quark/gluon condensates remain
unchanged, one may obtain\cite{Cai19-QSR2}:
\begin{align}
E_{\rm{HO}}^{\vartheta\rm{\,term}}(\rho)\approx
\frac{1}{2}b\zeta\rho^{\vartheta}\frac{\vartheta-1}{\vartheta+1}
-\frac{1}{4}b\zeta\left(\frac{k_{\rm{F}}}{M_0^{\ast}}\right)^2\rho^{\vartheta}
\label{EHOXX}.\end{align}
It is obvious from this expression the $E_{\rm{HO}}(\rho)$ will not be small even at low densities, and
the high order effects become more and more important as density increases.
By adopting the three fitting schemes, the EOS of ANM is systematically calculated\cite{Cai19-QSR2} and the corresponding results are shown in the lower part of FIG.\,\ref{fig_fEHOEsym}.
We find, for example, the $E_{\rm{HO}}(\rho)$ in QCDSR-2
is about 7.1\,MeV at $\rho_0$, and an even larger value about 10.9\,MeV in the QCDSR-3 scheme could be inferred.
Similarly in the QCDSR-1 set the fourth-order symmetry at $\rho_0$ is about 1.1\,MeV.
As a concise summary, in QCDSR calculations, the high-order term $E_{\rm{HO}}(\rho)$ is found to be sizable at densities $\lesssim \rho_0$, implying that the traditional parabolic approximation for the EOS of ANM may be violated. At $\rho_0$, the full QCDSR calculation yields $E_{\rm{HO}}(\rho_0)\approx 1.1\sim11.9\,\rm{MeV}$\cite{Cai19-QSR2}, demonstrating a much larger uncertainty than in phenomenological models. Nevertheless, caution is required when treating these higher-order contributions, as emphasized in Ref.\cite{Wel16}, where it was noted that the expansion in $\delta$ may converge slowly, become nonanalytic, or even display reversed term magnitudes in complex scenarios.

When the QCDSR case is set aside, the quartic term's small size may seem somewhat surprising in light of finite nuclei calculations.
In particular, the fourth-order term $a_{\rm{sym},4}(A)$ extracted from nuclear mass formulas has been found to be sizable. Specifically, values of $a_{\rm{sym},4}(A)\approx3.28 \pm 0.50$\,MeV and $a_{\rm{sym},4}(A)\approx8.33 \pm 1.21$\,MeV were reported in Refs.\cite{Jiang14PRC} and \cite{Tian15CPC}, respectively. Furthermore, in Ref.\cite{Jiang15PRC}, the fourth-order symmetry energy of finite nuclei was analyzed by fitting nuclear mass data using the nuclear mass formula with two different forms of the Wigner energy, resulting in a constraint of $a_{\rm{sym},4}(A)\approx3.91 \pm 0.10$\,MeV. These relatively large values of $a_{\rm{sym},4}(A)$ have led to suggestions that the full $E_{\rm{sym},4}(\rho_0)$ in infinite nuclear matter might also be large.

A careful analysis revisiting the relation between the fourth-order symmetry energy of nuclear matter at saturation density and its counterpart $a_{\rm{sym},4}(A)$ in finite nuclei shows the importance of the high-order isospin-dependent surface tension contribution. In Ref.\cite{Cai22PRC-asym4}, the full expression for $a_{\rm{sym},4}(A)$, explicitly including high-order surface tension effects, was derived. It was found that $E_{\rm{sym,4}}(\rho_0)$ cannot be reliably extracted from measured $a_{\rm{sym},4}(A)$ without proper knowledge of the high-order surface tension. These results indicate that large $a_{\rm{sym},4}(A)$ values of several MeV obtained from nuclear mass analyses can be fully compatible with the empirical constraint $E_{\rm{sym,4}}(\rho_0)\lesssim2\,\rm{MeV}$ from chiral effective field theory\cite{Kai16PRC} and mean-field models\cite{Cai12PRC-S4,Seif14PRC-S4,Gonz17PRC-S4,Pu17PRC-S4}, and do not necessarily imply a large $E_{\rm{sym,4}}(\rho_0)$ as previously inferred when high-order surface tension effects were neglected.
Specifically, one has
\begin{align}
a_{\rm{sym}}(A)=&\frac{\alpha}{1+(\alpha/\beta)A^{-1/3}},\\
a_{\rm{sym,4}}(A)=&\left(1+\frac{E_{\rm{sym}}(\rho_0)}{\beta A^{1/3}}\right)^{-4}\notag\\
&\times\left(E_{\rm{sym,4}}(\rho_0)
-\frac{L^2}{2K_0}+\frac{4\theta\kappa E_{\rm{sym}}^4(\rho_0)}{\beta A^{1/3}}\right),\label{def_asym4}
\end{align}
as well as the surface contribution:
\begin{align}\label{def_asym4_surf}
\boxed{
a^{\rm{s}}_{\rm{sym,4}}(A)=\frac{4\theta\kappa E_{\rm{sym}}^4(\rho_0)}{\beta A^{1/3}}\cdot\left(1+\frac{E_{\rm{sym}}(\rho_0)}{\beta A^{1/3}}\right)^{-4},}
\end{align}
here $\alpha\equiv E_{\rm{sym}}(\rho_0)$ is the nuclear matter symmetry energy at $\rho_0$, $\beta$ denotes the surface symmetry energy, and $\theta$ and $\kappa$ appear in the surface tension expressed as $\sigma\approx\sigma_0-\gamma\mu_{\rm a}^2-\gamma'\mu_{\rm a}^4$ or equivalently $\sigma/\sigma_0\approx1+y+\kappa y^2$ with $y=-\theta\mu_{\rm a}^2$, where $\theta=\gamma/\sigma_0$ and $\kappa=-\gamma'\sigma_0/\gamma^2$; evidently, $\kappa$ represents a higher-order effect.

\begin{figure}[h!]
\centering
\includegraphics[width=7.5cm]{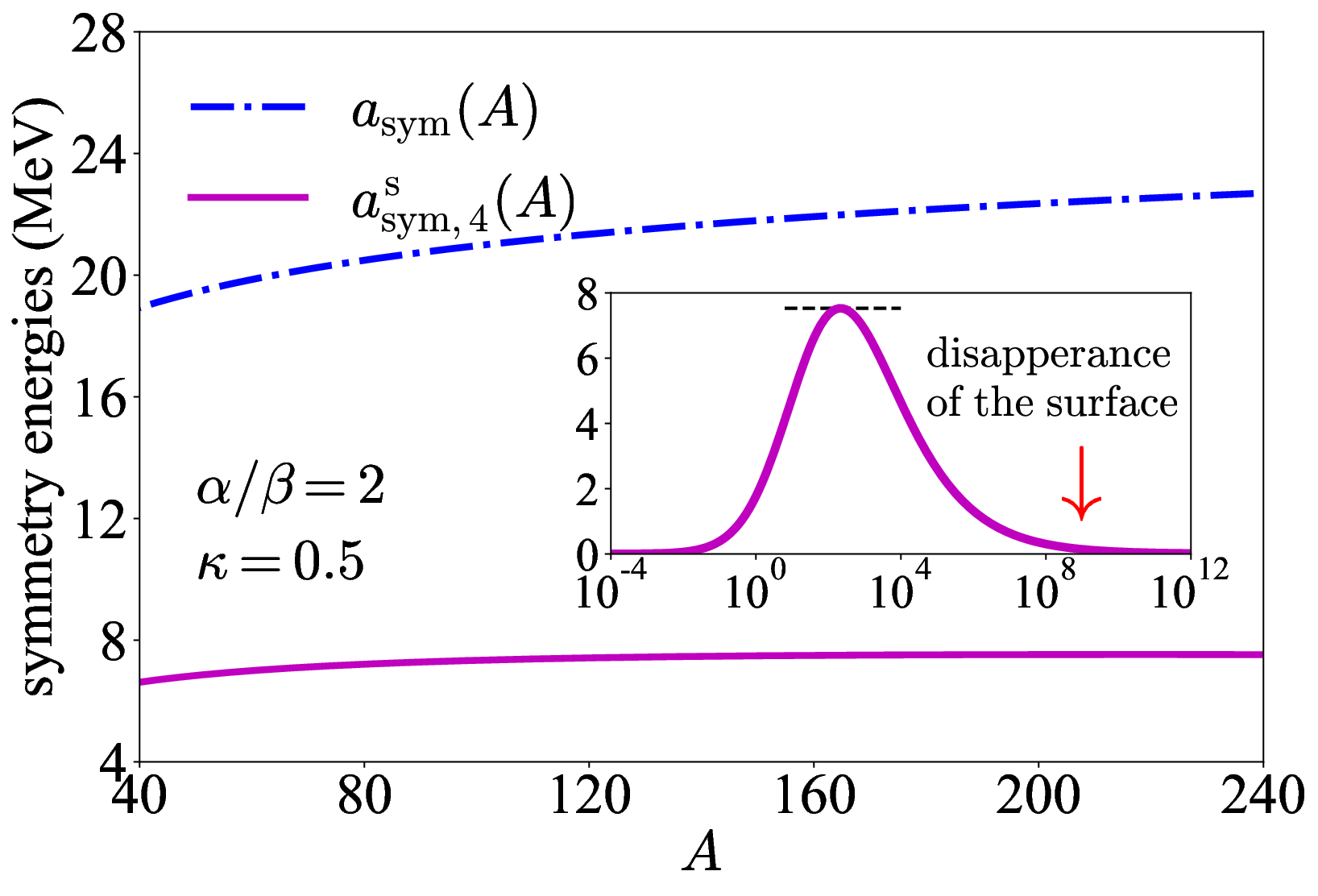}
\caption{(Color Online). The symmetry energy $a_{\rm{sym}}(A)$ and the surface fourth-order symmetry energy $a^{\rm s}_{\rm{sym,4}}(A)$ of finite nuclei as functions of nuclear mass number $A$ . Here $\alpha/\beta=2$ and $\kappa=0.5$ are adopted for illustration, $\alpha\equiv E_{\rm{sym}}(\rho_0)$. Figure taken from Ref.\cite{Cai22PRC-asym4}.}
\label{fig_a4exact}
\end{figure}

In FIG.\,\ref{fig_a4exact}, we show the $A$-dependence of the symmetry energy $a_{\rm{sym}}(A)$ (blue dash-dot line) and the surface fourth-order symmetry energy $a^{\rm s}_{\rm{sym,4}}(A)$ (magenta line) of finite nuclei, adopting $\kappa=0.5$. For illustration, we further use $\alpha\approx 30\,\rm{MeV}$ and $\beta\approx 15\,\rm{MeV}$\cite{Cai22PRC-asym4}. Within the considered range of $A$, the surface contribution to the fourth-order symmetry energy exhibits only a weak dependence on the mass number.
In the small-$A$ limit, one finds $ a_{\rm{sym}}(A)\to\beta A^{1/3}, a^{\rm s}_{\rm{sym},4}(A)\to4\kappa\theta\beta^3A$, and $a^{\rm s}_{\rm{sym},4}(A)/a_{\rm{sym}}(A)\to 4\kappa\theta\beta^2A^{2/3}$, all independent of the bulk coefficient $\alpha$ and vanishing as $A\to0$. For heavier nuclei, the parameter $\varphi=\alpha/\beta A^{1/3}$ is generally smaller than unity, leading to the approximation $ a^{\rm s}_{\rm{sym,4}}(A)\approx4\alpha^3\theta\kappa\varphi= {\alpha^3A^{2/3}\kappa\varphi}/{S\beta\sigma_0}$. Notably, this approximation shows that $a^{\rm s}_{\rm{sym,4}}(A)$ is proportional to $\varphi$, which approaches zero in the infinite matter limit, i.e., $ \lim_{A\to\infty}a^{\rm s}_{\rm{sym,4}}(A)=0\,\rm{MeV}$.
The vanishing of $a^{\rm s}_{\rm{sym},4}(A)$ for very large $A$ is expected, as it represents only the surface part of the fourth-order symmetry energy. Physically, the surface disappears in the limit $A\to\infty$, so the associated contribution naturally goes to zero, as illustrated in the inset of FIG.\,\ref{fig_a4exact}. However, this does not imply that the fourth-order symmetry energy of infinite matter vanishes\cite{Cai22PRC-asym4}. Hence, the finite-nucleus fourth-order symmetry energy discussed here is still characterized by $\alpha$ and $\beta$, rather than $E_{\rm{sym,4}}(\rho_0)$ (see Ref.\cite{RWang17}). 
The mass number at which $a^{\rm s}_{\rm{sym},4}(A)$ reaches its maximum can be determined from $\partial a^{\rm s}_{\rm{sym},4}(A)/\partial A=0$, giving $A_{\max}=27\alpha^3/\beta^3\approx216$. At this point, $a^{\rm s}_{\rm{sym},4}(A_{\max})=27\alpha^3\theta\kappa/64\approx7.5\,\rm{MeV}$, as indicated by the black dashed line in the inset of FIG.\,\ref{fig_a4exact}. Interestingly, the empirical ratio $\alpha/\beta$ (around 2-3) coincidentally predicts that the fourth-order surface symmetry energy peaks near $^{208}\rm{Pb}$. This highlights the fundamental difference between finite nuclei ($A\sim208$) and infinite matter with $A\to\infty$\cite{RWang17}.

Similarly, if one includes higher-order isospin-dependent terms in the surface tension coefficient as $\sigma/\sigma_0\approx1+y+\kappa y^2+s y^3$, with $s$ representing an effective parameter beyond $\kappa$, the sixth-order symmetry energy $a_{\rm{sym,6}}(A)$ of finite nuclei can be directly obtained as a function of $A$, incorporating both volume and surface contributions\cite{Cai22PRC-asym4}:
\begin{align}\label{ee6}
a_{\rm{sym,6}}(A)
=&\left(1+\frac{E_{\rm{sym}}(\rho_0)}{\beta A^{1/3}}\right)^{-6}\cdot\Bigg[E_{\rm{sym,6}}(\rho_0)
-L_{\rm{sym,4}}\left(\frac{L}{K_0}\right)
\notag\\&+\frac{K_{\rm{sym}}}{2}\left(\frac{L}{K_0}\right)^2
-\frac{J_0}{6}\left(\frac{L}{K_0}\right)^3
\notag\\&+\frac{16\theta^2E_{\rm{sym}}^6(\rho_0)}{\beta A^{1/3}}\left(\frac{4\kappa^2}{1+\beta A^{1/3}/E_{\rm{sym}}(\rho_0)}-s\right)
\Bigg].
\end{align}
The surface contribution in Eq.\,(\ref{ee6}) (the last line) originates from two sources: a higher-order term derived from the lower-order coefficient $\kappa$ (proportional to $\kappa^2$) and a higher-order term from the coefficient $s$. Naturally, this surface term vanishes in the limit $A\to\infty$.

Parallel to the discussion in the previous paragraphs, it remains very challenging to extract the sixth-order symmetry energy of infinite matter, $E_{\rm{sym,6}}(\rho_0)$, from this expression, even if the coefficient $a_{\rm{sym,6}}(A)$ appearing in the mass formula is constrained, because the coefficient $s$ can be adjusted to make $E_{\rm{sym,6}}(\rho_0)$ small or large without significantly affecting finite nuclei properties such as the nuclear surface tension. For instance, neglecting the surface term, a ``small'' $a_{\rm{sym,6}}(A)\approx1\,\rm{MeV}$ could correspond to an $E_{\rm{sym,6}}(\rho_0)$ around 7-8\,MeV, potentially conflicting with empirical constraints on the EOS of ANM.
In this regard, the surface contribution to $a_{\rm{sym,6}}(A)$ is essential. Furthermore, by imposing conditions such as $E_{\rm{sym,6}}(\rho_0)\lesssim2\,\rm{MeV}$ and $a_{\rm{sym,6}}(A)\lesssim2\,\rm{MeV}$, Eq.\,(\ref{ee6}) establishes a connection among several characteristics with sizable magnitudes, which could be used to explore correlations among them\cite{Cai22PRC-asym4}.

\begin{figure}[h!]
\centering
  \includegraphics[width=6.5cm]{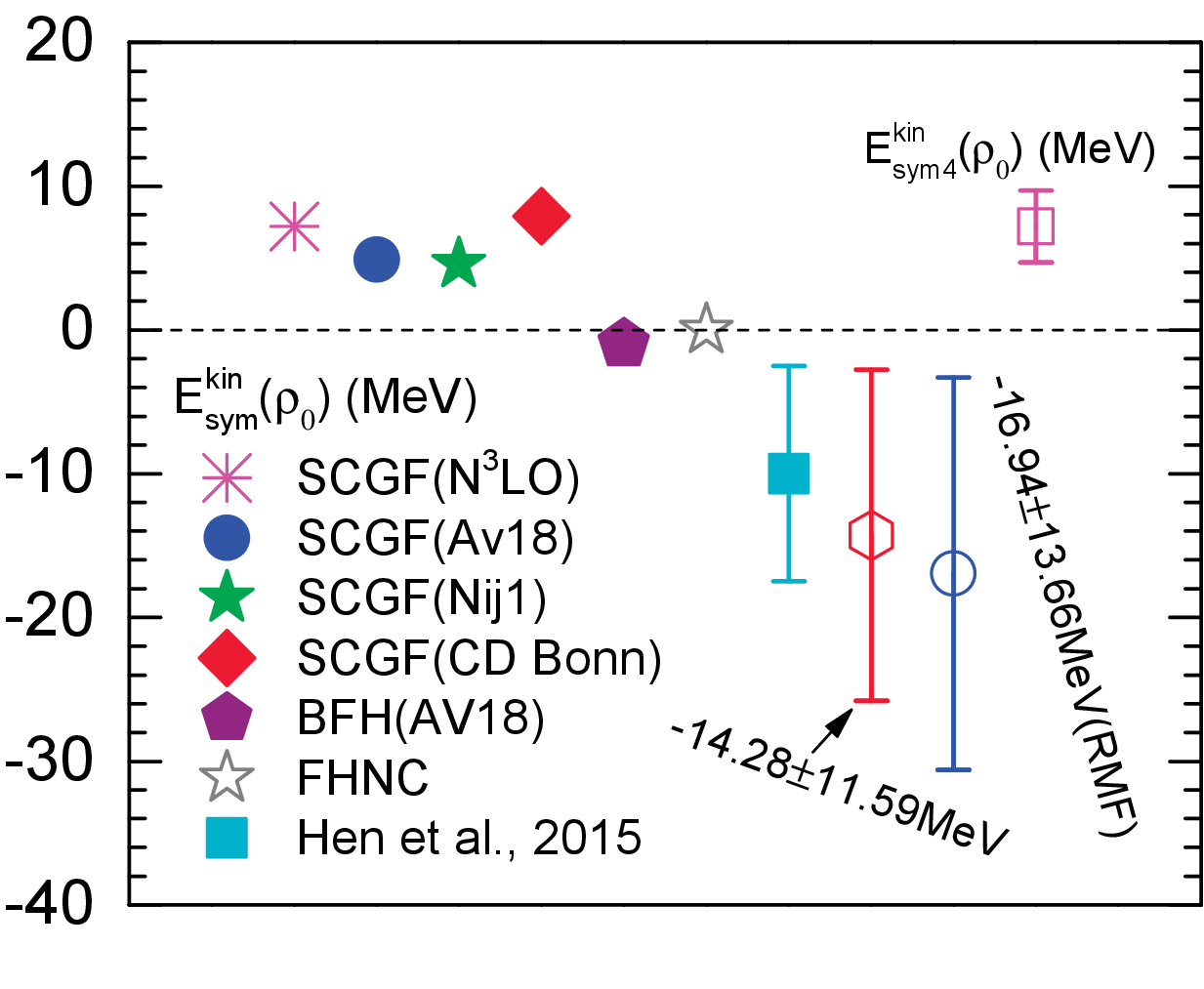}
  \caption{(Color Online). Kinetic symmetry energy and the isospin-quartic term including the SRC-induced HMT in comparison with predictions by other theories. Figure taken from Ref.\cite{LCCX18}. }
  \label{fig_esym0}
\end{figure}

The SRC-induced reduction of the kinetic symmetry energy relative to the FFG value is a common feature in many-body theories that include SRC effects. As shown in FIG.\,\ref{fig_esym0}, comparisons with microscopic many-body and phenomenological models, including SCGF (N$^3$LO, AV18, Nij1, CD Bonn)\cite{Car12}, BHF with AV18 plus Urbana IX three-body-force (TBF)\cite{Vid11}, the Fermi hypernetted chain (FHNC) model\cite{Lov11}, and a phenomenological parametrization\cite{Hen15b}, all confirm qualitatively that SRC reduces the kinetic symmetry energy, though quantitative differences exist; the relativistic prediction on the reduction of the kinetic symmetry energy will be discussed in Subsection \ref{sub_WaleckaSRC}.
The SRC-reduced kinetic symmetry energy affects not only the understanding of the origin of symmetry energy but also isovector observables, such as free neutron/proton and $\pi^-/\pi^+$ ratios in heavy-ion collisions\cite{Hen15b,Li15PRC}, and properties of NSs, where isospin asymmetry is large, see Section \ref{SEC_NS}. To date, possible effects of a large quartic term on heavy-ion collisions remain unexplored due to the small asymmetry in such reactions.
However, in neutron-rich systems such as rare isotope reactions or peripheral collisions of heavy nuclei with thick neutron skins, the quartic term may play a significant role.
Typically, the EOS of neutron-rich nucleonic matter is parameterized as the sum of a FFG kinetic term and a potential energy up to the isospin-quadratic term. Findings in Ref.\cite{Cai15a} indicate the importance of including the isospin-quartic term in both kinetic and potential parts. Accurate extraction of the potential isospin-quartic term $E^{\rm{pot}}_{\rm{sym,4}}(\rho)\delta^4$ requires using the kinetic EOS of quasi-particles with reduced symmetry energy and enhanced quartic term due to the isospin-dependent HMT.
For isovector observables in heavy-ion collisions, the nucleon isovector potential is critical. Beyond the Lane potential $\pm 2\rho E^{\rm{pot}}_{\rm{sym}}(\rho)\delta$, the $E^{\rm{pot}}_{\rm{sym,4}}(\rho)\delta^4$ term adds $\pm 4 \rho E^{\rm{pot}}_{\rm{sym,4}}(\rho)\delta^3$. 
In neutron-rich systems, such as nuclear reactions induced by rare
isotopes and peripheral collisions between two heavy nuclei having thick neutron-skins, the latter may play a significant role in understanding the isovector observables or extracting the sizes of neutron-skins of the neutron-rich nuclei involved.
On the NS side, the kinetic part as well as the potential part of the symmetry energy were found to separately affect the formation density of the $\Delta(1232)$ state, e.g.,
\begin{align}\label{1232-fd}
\rho_{\Delta^-}^{\rm{crit}}:\;&k_{\rm{F}}^{\rm{n},2}/2M_{\rm{D}}^{\ast}
\approx \Phi_{\Delta}+g_{\sigma\rm{N}}(1-x_{\sigma})\overline{\sigma}-
g_{\omega\rm{N}}(1-x_{\omega})\overline{\omega}_0\notag\\
&-6(1-x_{\rho})E_{\rm{sym}}^{\rm{pot}}(\rho)\delta
-4E_{\rm{sym}}^{\rm{kin}}(\rho)\delta
\end{align}
for the formation density of $\Delta^-$\cite{Cai15Delta}, 
where $\Phi_{\Delta}\approx293\,\rm{MeV}$ is the mass difference between the $\Delta(1232)$ and the nucleon, $M_{\rm{D}}^{\ast}$ is the Dirac nucleon effective mass, $g_{\sigma\rm{N}},g_{\omega\rm{N}},g_{\rho\rm{N}}$ are the meson-nucleon coupling constants and $x_{\sigma},x_{\omega}$ and $x_{\rho}$ characterize the couplings between the $\Delta$'s and the mesons.
The relation (\ref{1232-fd}) is obtained using the nonlinear RMF model incorporating the couplings between the $\Delta$'s and the mesons.
In this sense one could study the separable effects of the kinetic part and the potential part of the symmetry energy on the $\rho_{\Delta^-}^{\rm{crit}}$.
It could be possible to investigate the effects of the large kinetic quartic term if one could find similar relations in NSs.

We offer an explanation for the reduction of the kinetic symmetry energy, from a mathematical viewpoint. To illustrate this, it is convenient to consider the difference of the symmetry energy at a reference density $\rho_{\rm{rf}}\neq \rho_0$, defined as 
$
\mathcal{O}\equiv E_{\rm{sym}}(\rho_{\rm{rf}})[\rm{FFG}]-E_{\rm{sym}}(\rho_{\rm{rf}})[\rm{HMT}]$, 
under the constraints that $E_{\rm{sym}}(\rho_0)$ and $L$ are kept fixed. If $\mathcal{O}>0$, the SRC-induced HMT reduces the symmetry energy, whereas if $\mathcal{O}<0$, the symmetry energy is enhanced. 
For simplicity and to highlight the essential features, one may parametrize $E_{\rm{sym}}(\rho)$ as
\begin{equation}\label{CS}
E_{\rm{sym}}(\rho,\Phi)=\Phi
\left(\frac{\rho}{\rho_0}\right)^{\nu_0}+A\left(\frac{\rho}{\rho_0}\right)^{\nu_1}+B\left(\frac{\rho}{\rho_0}\right)^{\nu_2},
\end{equation}
where $\nu_0=2/3$ corresponds to the FFG prediction\cite{BALi25rho} for the density dependence of the kinetic symmetry energy, representing the smallest power in $E_{\rm{sym}}(\rho)$. The exponents $\nu_1$ and $\nu_2$ are larger than $2/3$, and without loss of generality, we take $\nu_2>\nu_1$. Moreover, $\Phi=(1/6M_{\rm N})(3\pi^2/2)^{2/3}\Upsilon_{\rm{sym}}^{\rm{NR}}$, with $\Upsilon_{\rm{sym}}^{\rm{NR}}$ defined in Eq.\,(\ref{Esymkin}). The coefficients $A$ and $B$ are determined by the constraints that $E_{\rm{sym}}(\rho_0)$ and $L$ remain fixed.
After straightforward calculations, one finds
\begin{align}\label{CCC}
\mathcal{O}\approx&\delta\Phi
\left(\frac{\rho_{\rm{rf}}}{\rho_0}\right)^{\nu_2}\cdot\frac{\nu_1-\nu_0}{\nu_2-\nu_1}\cdot\Pi\left(\rho_{\rm{rf}},\nu_2\right)\notag\\
&+\mbox{higher-order terms in }\left(\frac{\nu_0}{\nu_2}\right)\mbox{ or
}\left(\frac{\nu_1}{\nu_2}\right),
\end{align}
where $\delta\Phi=\Phi_{\rm{FFG}}-\Phi_{\rm{HMT}}$ and
\begin{equation}
\Pi\left(\rho,\nu_2\right)=1-\left(\frac{\rho}{\rho_0}\right)^{-\nu_2}
\left[1+\nu_2\ln\left(\frac{\rho}{\rho_0}\right)\right].
\end{equation}
The function $\Pi$ increases monotonically with $\nu_2$ at fixed density, and for fixed $\nu_2$, $\Pi$ increases (decreases) for $\rho/\rho_0>1$ ($0<\rho/\rho_0<1$). Therefore, $\Pi(\rho,\nu_2)>\Pi(\rho,0)=0$ and $\Pi(\rho,\nu_2)>\Pi(\rho_0,\nu_2)=0$, i.e., $\Pi$ is positive-definite. The pre-factor of $\Pi(\rho_{\rm{rf}},\nu_2)$ in Eq.\,(\ref{CCC}) is also positive because $\delta\Phi>0$, implying that $\mathcal{O}>0$. Consequently, the symmetry energy is softened (hardened) at supra- (sub-) saturation densities.
Furthermore, as the difference between the high-momentum nucleon fractions in SNM and PNM increases (for example, comparing the HMT-exp set with the HMT-SCGF set), the factor $\Phi_{\rm{HMT}}$ can become much smaller or even negative, thereby enhancing $\delta\Phi$. This leads to a more pronounced softening (hardening) of the symmetry energy at supra- (sub-) saturation densities.

As shown in Eq.\,(\ref{def_nonexp}), the first four relativistic corrections to the nucleon kinetic energy are ``$-\v{k}^2/4M_{\rm N}^2$'', ``$\v{k}^4/8M_{\rm N}^4$'', ``$-5\v{k}^6/64M_{\rm N}^6$'' and ``$7\v{k}^8/128M_{\rm N}^8$'', respectively. In calculating the average energy per nucleon in ANM, after integrating over momentum $k$, the $k$-dependence transforms into a dependence on the Fermi momentum $k_{\rm{F}}\sim\rho^{1/3}$. Specifically, for general $\sigma$, the term $\v{k}^{\sigma}/M_{\rm N}^{\sigma-1}$ leads to $3k_{\rm{F}}^{\sigma}/(\sigma+3)M_{\rm N}^{\sigma-1}$ in the kinetic EOS if the step function for the momentum distribution $n_{\v{k}}^J$ is adopted. Similarly, the corresponding term for the kinetic symmetry energy is $\sigma k_{\rm{F}}^{\sigma}/6M_{\rm N}^{\sigma-1}$. For example, the first-order relativistic correction contributes $-3k_{\rm{F}}^4/56M_{\rm N}^3$ to the SNM kinetic EOS and $-k_{\rm{F}}^4/12M_{\rm N}^3$ to the kinetic symmetry energy\cite{Fri05}, see also Eqs.\,(\ref{non0}) and (\ref{non2}).
When the SRC-induced HMT is considered, analytical expressions for $\langle \v{k}^{\sigma}\rangle$ become non-trivial. The $\langle \v{k}^\sigma\rangle$ at orders $\delta^0$, $\delta^2$, and $\delta^4$ are given (assuming $\beta_J=0$ in Eq.\,(\ref{MDGen}))

\begin{strip}
\begin{align}
\langle\v{k}^{\sigma}\rangle\left(\mbox{at order }\delta^0\right)\equiv&\left(\frac{1}{\rho}\frac{2}{(2\pi)^3}\sum_{J=\rm{n,p}}\int_0^{\phi_Jk_{\rm{F}}^J}\v{k}^\sigma n_{\v{k}}^J(\rho,\delta)\d\v{k}
\right)\left(\mbox{at order }\delta^0\right)\notag\\
=&\frac{2[(1-3C_0)\phi_0+3C_0]k_{\rm{F}}^{\sigma+1}}{(\sigma+1)\phi_0}
+\frac{2C_0k_{\rm{F}}^{\sigma+1}(\phi_0^{\sigma-3}-1)}{\sigma-3},\label{ct-0}
\end{align}
and similarly the quadratic and the quartic contributions for $\v{k}^{\sigma}$,
\begin{align}
&\langle
\v{k}^{\sigma}\rangle\left(\mbox{at order }\delta^2\right)\equiv\left(\frac{1}{\rho}\frac{2}{(2\pi)^3}\sum_{J=\rm{n,p}}\int_0^{\phi_Jk_{\rm{F}}^J}\v{k}^\sigma n_{\v{k}}^J(\rho,\delta)\d\v{k}
\right)\left(\mbox{at order }\delta^2\right)\notag\\
=&\frac{3k_{\rm{F}}^\sigma}{\sigma+3}\Bigg[\frac{3C_0\phi_1}{\phi_0}(\phi_1-C_1)+(\sigma+3)C_0C_1\phi_1e^{(\sigma-1)\ln\phi_0}
+\frac{1}{2}\frac{\sigma+3}{\sigma-1}C_0\phi_1^2(\sigma-1)(\sigma-2)e^{(\sigma-1)\ln\phi_0}
+(\sigma+3)C_0C_1\phi_1e^{(\sigma-1)\ln\phi_0}\notag\\
&\hspace*{1.cm}-\left(1+\frac{\sigma}{3}\right)
\Bigg[\frac{3C_0\phi_1}{\phi_0}+3C_0C_1\left(1-\frac{1}{\phi_0}\right)
-(\sigma+3)C_0\phi_1e^{(\sigma-1)\ln\phi_0}
-\frac{\sigma+3}{\sigma-1}C_0C_1\left[e^{(\sigma-1)\ln\phi_0}-1\right]\Bigg]\notag\\
&\hspace*{1cm}+\frac{\sigma(\sigma+3)}{18}\Bigg[1-3C_0\left(1-\frac{1}{\phi_0}\right)+
\frac{\sigma+3}{\sigma-1}C_0\left[e^{(\sigma-1)\ln\phi_0}-1\right]\Bigg]\Bigg],\label{ct-2}\\
&\langle \v{k}^\sigma\rangle\left(\mbox{at order }\delta^4\right)\equiv\left(\frac{1}{\rho}\frac{2}{(2\pi)^3}\sum_{J=\rm{n,p}}\int_0^{\phi_Jk_{\rm{F}}^J}\v{k}^\sigma n_{\v{k}}^J(\rho,\delta)\d\v{k}
\right)\left(\mbox{at order }\delta^4\right)\notag\\
=&\frac{3k_{\rm{F}}^\sigma}{\sigma+3}
\Bigg[\frac{3C_0\phi_1^3}{\phi_0}(\phi_1-C_1)+\frac{\sigma+3}{24}C_0\phi_1^4(\sigma-2)(\sigma-3)(\sigma-4)e^{(\sigma-1)\ln\phi_0}+\frac{\sigma+3}{6}C_0C_1\phi_1^3(\sigma-2)(\sigma-3)e^{(\sigma-1)\ln\phi_0}\notag\\
&\hspace*{1.cm}+\left(1+\frac{\sigma}{3}\right)\Bigg[\frac{3C_0\phi_1^2}{\phi_0}(C_1-\phi_1)
+\frac{\sigma+3}{6}C_0\phi_1^3(\sigma-2)(\sigma-3)e^{(\sigma-1)\ln\phi_0}
+\frac{\sigma+3}{2}C_0C_1\phi_1^2(\sigma-2)e^{(\sigma-1)\ln\phi_0}\Bigg]\notag\\
&\hspace*{1.cm}+\frac{\sigma(\sigma+3)}{18}\Bigg[\frac{3C_0\phi_1}{\phi_0}(\phi_1-C_1)+\frac{\sigma+3}{2}C_0\phi_1^2(\sigma-2)e^{(\sigma-1)\ln\phi_0}
+(\sigma+3)C_0C_1\phi_1e^{(\sigma-1)\ln\phi_0}\Bigg]\notag\\
&\hspace*{1.cm}+\left(\frac{\sigma}{18}-\frac{\sigma^3}{162}\right)
\Bigg[3C_0\left[\frac{\phi_1}{\phi_0}+C_1\left(1-\frac{1}{\phi_0}\right)\right]-(\sigma+3)C_0\phi_1e^{(\sigma-1)\ln\phi_0}-\frac{\sigma+3}{\sigma-1}C_0C_1\left[e^{(\sigma-1)\ln\phi_0}-1\right]\Bigg]\notag\\
&\hspace*{1.cm}+\frac{1}{1944}\sigma(\sigma-6)(\sigma-3)(\sigma+3)\left[1-3C_0\left(1-\frac{1}{\phi_0}\right)
+\frac{\sigma+3}{\sigma-1}C_0\left[e^{(\sigma-1)\ln\phi_0}-1\right]\right]\Bigg].\label{ct-4}
\end{align}
For our purposes, numerical evaluations are performed for $\sigma=2,4,6,8,10$. Sixth-order terms ($\delta^6$) can also be obtained numerically, though the explicit expressions are very complicated.
\end{strip}

\begin{figure}
\centering
\includegraphics[width=4cm]{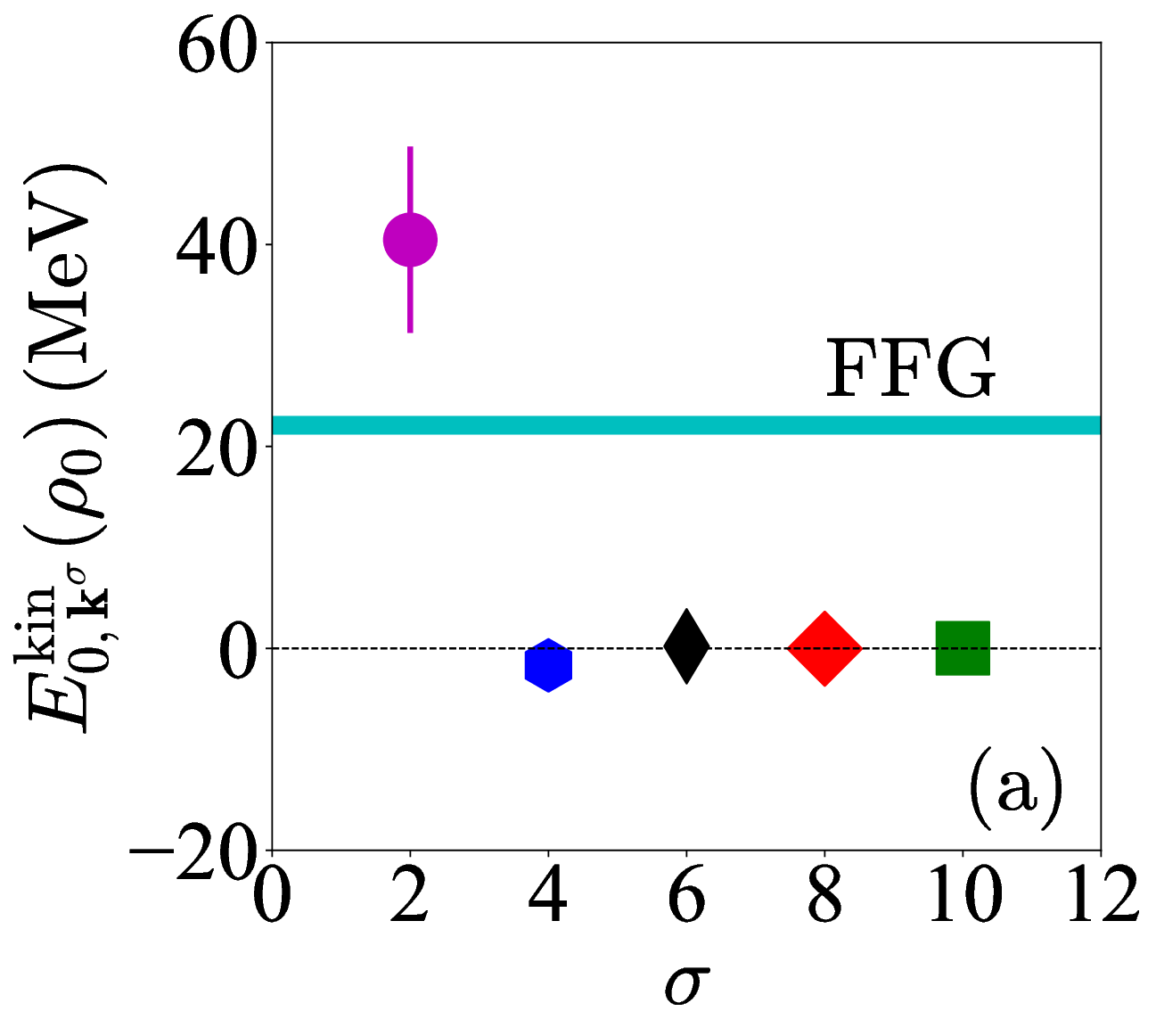}\quad
\includegraphics[width=4cm]{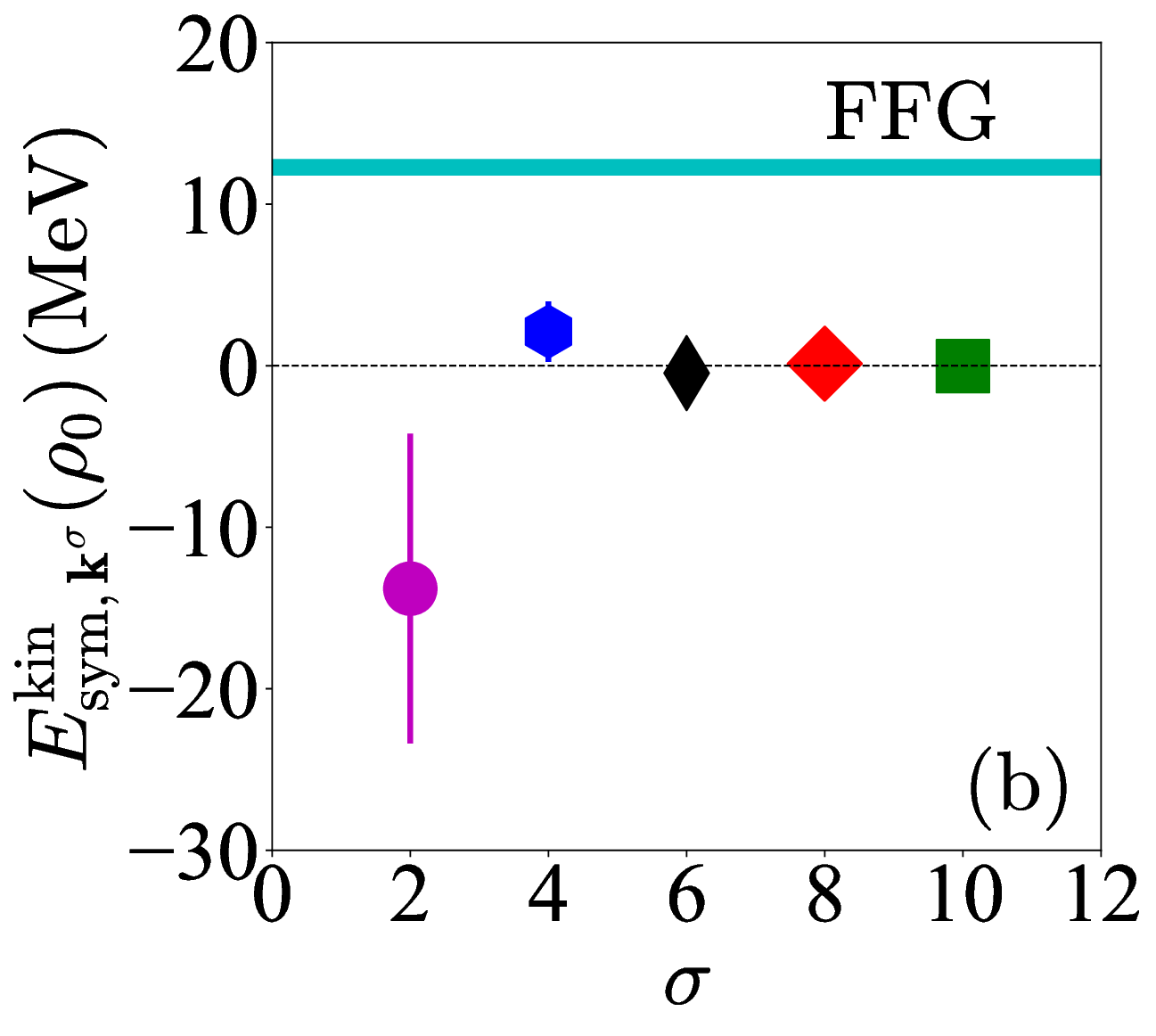}\\
\includegraphics[width=4cm]{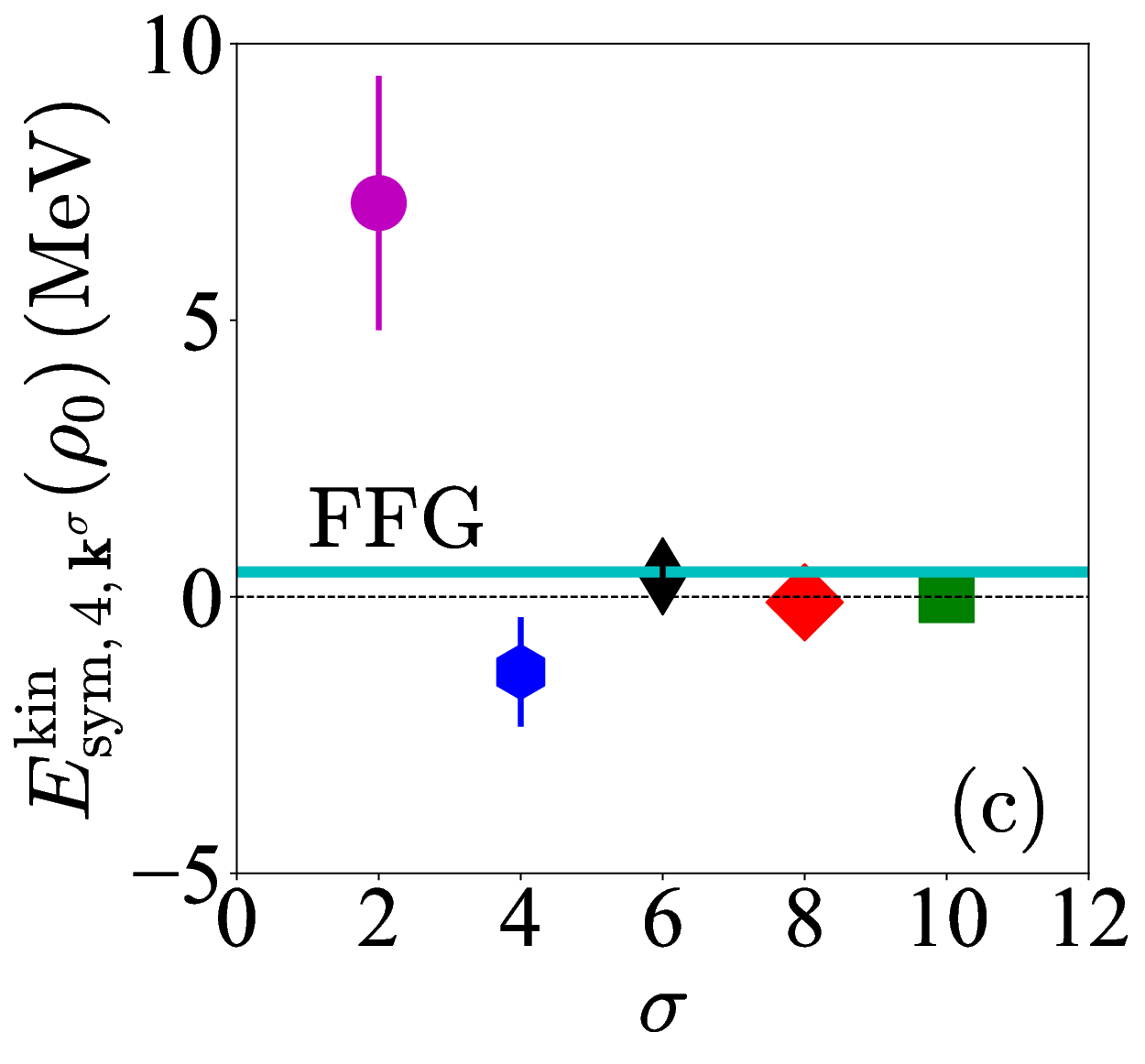}\quad
\includegraphics[width=4cm]{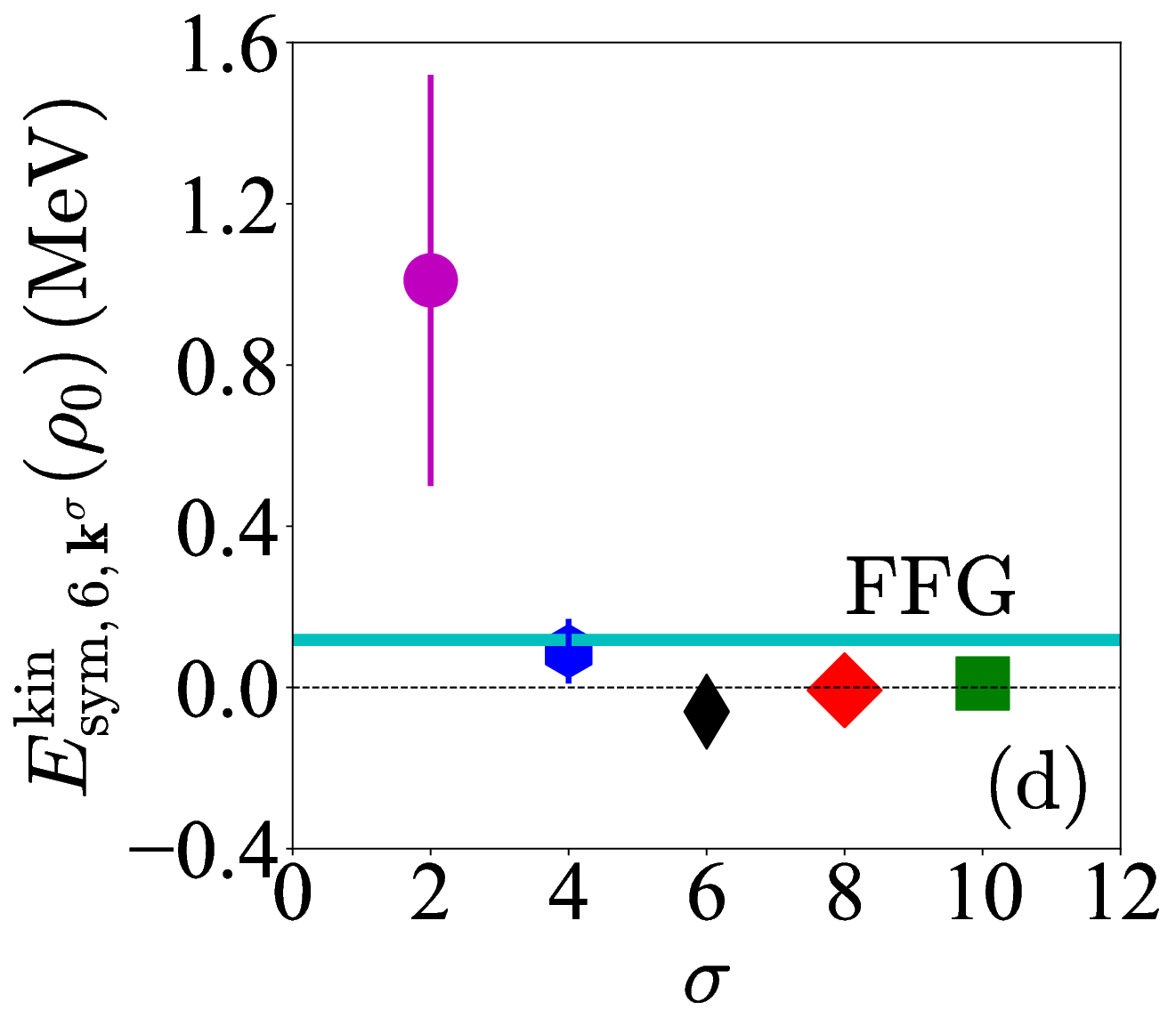}
\caption{(Color Online). Kinetic parts of the EOS of ANM with relativistic corrections for $E_0^{\rm{kin}}(\rho_0)$ (panel a), $E_{\rm{sym}}^{\rm{kin}}(\rho_0)$ (panel b), $E_{\rm{sym,4}}^{\rm{kin}}(\rho_0)$ (panel c), and $E_{\rm{sym,6}}^{\rm{kin}}(\rho_0)$ (panel d). Here $\rho_0\approx0.16\pm0.01\,\rm{fm}^{-3}$, $\sigma=2$ is the non-relativistic term while $\sigma=4,6,8,10$ correspond to the first four relativistic corrections from Eq.\,(\ref{def_nonexp}). Figures taken from Ref.\cite{CaiLi22PRCFFG}.}
\label{fig_RC}
\end{figure}

Using the parameters $\phi_0,\phi_1,C_0$ and $C_1$ introduced in Subsection \ref{sub_nk}, one can evaluate the magnitudes of all contributions to the kinetic EOS as well as their relativistic corrections in the presence of SRC. At saturation density $\rho_0\approx0.16\pm0.01\,\rm{fm}^{-3}$, the leading (non-relativistic) component is found to be
$E_{0,\v{k}^2}^{\rm{kin}}(\rho_0)\approx40.47\pm9.23\,\rm{MeV}$. The label ``$0,\v{k}^2$'' highlights that this term arises directly from the non-relativistic kinetic energy $\v{k}^2/2M_{\rm N}$, hence it does not represent a relativistic correction.
Proceeding to the higher orders in $\v{k}^{\sigma}$, the first relativistic correction, originating from the term ``$-\v{k}^4/8M_{\rm N}^3$'' (or equivalently ``$-\v{k}^2/4M_{\rm N}^2$'' after dividing by $\v{k}^2/2M_{\rm N}$ in Eq.\,(\ref{def_nonexp})), contributes
$E_{0,\v{k}^4}^{\rm{kin}}(\rho_0)\approx-1.68\pm1.11\,\rm{MeV}$. The subsequent three corrections coming from ``$\v{k}^6/16M_{\rm N}^5$'', ``$-5\v{k}^8/128M_{\rm N}^7$'', and ``$7\v{k}^{10}/256M_{\rm N}^9$'' provide
$0.21\pm0.25\,\rm{MeV}$, $-0.04\pm0.07\,\rm{MeV}$, and $0.01\pm0.02\,\rm{MeV}$, respectively. The smallness of the last value already indicates that including the first four relativistic corrections is sufficient to estimate the overall relativistic effects.
These results are summarized in panel (a) of FIG.\,\ref{fig_RC}, which also shows for comparison the non-relativistic FFG value of about $22.10\pm0.92\,\rm{MeV}$ (labeled as ``FFG''). The enhancement of $E_0^{\rm{kin}}$ induced by the SRC-generated HMT has long been recognized\cite{Cai15a}.
Summing the non-relativistic term and all four relativistic corrections gives $
E_0^{\rm{kin}}(\rho_0)\approx38.98\pm8.32\,\rm{MeV}$.
Relative to the purely non-relativistic value $40.47\,\rm{MeV}$, the reduction of about $1.49\,\rm{MeV}$ corresponds to a $\sim4\%$ correction.

Turning to the quadratic kinetic symmetry energy, the leading SRC-modified non-relativistic contribution is
$-13.80\pm9.57\,\rm{MeV}$. The first four relativistic corrections amount to
$2.11\pm1.87\,\rm{MeV}$, $-0.46\pm0.58\,\rm{MeV}$, $0.12\pm0.20\,\rm{MeV}$, and $-0.03\pm0.07\,\rm{MeV}$, respectively. Combining them yields $
E_{\rm{sym}}^{\rm{kin}}(\rho_0)\approx-12.01\pm8.23\,\rm{MeV}$.
The overall enhancement of about $1.79\,\rm{MeV}$ corresponds to a relative correction of roughly $13\%$, shown in panel (b) of FIG.\,\ref{fig_RC}. Both the uncorrected value ($-13.80\,\rm{MeV}$) and the corrected one ($-12.01\,\rm{MeV}$) are compatible with the result $-16.94\,\rm{MeV}$ obtained from a nonlinear RMF model including the HMT, see discussions given in Subsection \ref{sub_WaleckaSRC}.
We now consider the fourth- and sixth-order kinetic symmetry energies. Including the first four relativistic corrections in the presence of SRC gives $
E_{\rm{sym},4}^{\rm{kin}}(\rho_0)\approx6.06\pm1.77\,\rm{MeV}$ and $
E_{\rm{sym,6}}^{\rm{kin}}(\rho_0)\approx1.07\pm0.54\,\rm{MeV}$.
For $E_{\rm{sym},4}^{\rm{kin}}(\rho_0)$, the individual contributions are
$7.12\pm2.30\,\rm{MeV}$ (non-relativistic), $-1.36\pm0.99\,\rm{MeV}$, $0.37\pm0.44\,\rm{MeV}$, $-0.10\pm0.17\,\rm{MeV}$, and $0.03\pm0.06\,\rm{MeV}$.  
For $E_{\rm{sym,6}}^{\rm{kin}}(\rho_0)$, they are  
$1.01\pm0.51\,\rm{MeV}$ (non-relativistic), $0.09\pm0.08\,\rm{MeV}$, $-0.06\pm0.07\,\rm{MeV}$, $-0.007\pm0.05\,\rm{MeV}$, and $0.01\pm0.05\,\rm{MeV}$.  
These are displayed in panels (c) and (d) of FIG.\,\ref{fig_RC}.
The net relativistic modifications to the fourth- and sixth-order symmetry energies are therefore about $-1.06\,\rm{MeV}$ and $0.06\,\rm{MeV}$, respectively, corresponding to relative effects of roughly $15\%$ and $6\%$. The change in the sixth-order term is thus negligible when compared with its non-relativistic magnitude.

It is also interesting to compare the ratio $\Psi$ between the kinetic quartic and quadratic symmetry energies (as defined in Eq.\,(\ref{def-Psi-FFG})) with and without the relativistic corrections in the presence of SRC-induced high momentum nucleons. Based on the results presented above, the $\Psi$ changes from $\Psi\approx -7.12/13.80\approx-51.6\%$ (without) to $\Psi\approx-6.06/12.01\approx-50.4\%$ (with) the relativistic corrections. 
Moreover, considering the extreme case of PNM, the overall relativistic corrections can be evaluated by adding $E_0^{\rm{kin}}(\rho_0)$, $E_{\rm{sym}}^{\rm{kin}}(\rho_0)$, $E_{\rm{sym,4}}^{\rm{kin}}(\rho_0)$ and $E_{\rm{sym,6}}^{\rm{kin}}(\rho_0)$ all together. This sum is 34.8\,MeV for the non-relativistic FFG model and 34.1\,MeV by including the (first four) relativistic corrections, both are close to the non-relativistic FFG model prediction about $3k_{\rm{n}}^{2}/10M\approx35.1\,\rm{MeV}$ where $k_{\rm{n}}=(3\pi^2\rho)^{1/3}=2^{1/3}k_{\rm{F}}$ is the neutron Fermi momentum in the PNM. Thus, although the relativistic corrections to each kinetic energy term may be large or small, the overall effects are rather small (the relative effect is about 2\%), since some changes are positive while others are negative.
Particularly, for the four kinetic energies, only the (quadratic) symmetry energy is reduced considering the SRC-induced HMT, while the other three higher order symmetry energies are all enhanced, compared with their FFG predictions (i.e., without HMT), see FIG.\,\ref{fig_RC}. These results indicate clearly that the SRC-induced HMT is more fundamental and has much larger impacts on the ANM EOS than the relativistic corrections. Although the SRC effects are introduced through the HMT in the single nucleon momentum distribution function $n_{\v{k}}^J$ and affect apparently only the kinetic parts of ANM EOS in the models considered here, the SRC is fundamentally due to the tensor force in the neutron-proton isosinglet interaction channel\cite{Hen14}. 
Thus, in this sense, the finding is consistent with that in Subsection \ref{sub_WaleckaSRC}, namely that nucleon-nucleon interactions can modify the value of $\Psi$ significantly. It should also be noted that, once interaction effects beyond the FFG picture are incorporated, the ratio $\Psi$ is in principle allowed to become much larger than the values obtained here.

Finally, we discuss the effects of the $\beta_J$ on the kinetic EOS. Contributions from a finite $\beta_J$ to the first three terms of the kinetic EOS are
\begin{align}
\delta
E_{0}^{\rm{kin}}(\rho)=&\frac{3}{5}E_{\rm{F}}(\rho_0)\cdot\frac{4\beta_0}{35},\\
\delta
E_{\rm{sym}}^{\rm{kin}}(\rho)=&\frac{1}{3}E_{\rm{F}}(\rho_0)\cdot\frac{4\beta_0(1+3\beta_1)}{35},\\
\delta
E_{\rm{sym,4}}^{\rm{kin}}(\rho)=&\frac{1}{81}E_{\rm{F}}(\rho_0)\cdot\frac{4\beta_0(1-3\beta_1)}{35}.
\end{align}
With $\beta_0\approx-0.35$ (the maximum magnitude of $\beta_0$ as discussed in Subsection \ref{sub_Emass}), one can find that the maximal effect of $\beta_J$ produces negligible corrections on $E_{\rm{sym,4}}^{\rm{kin}}(\rho_0)$ and less than 2\,MeV on $E_{\rm{sym}}^{\rm{kin}}(\rho_0)$. Including these corrections, the final values are
$E_0^{\rm{kin}}(\rho_0)\approx39.77\pm8.13\,\rm{MeV}$,
$E_{\rm{sym}}^{\rm{kin}}(\rho_0)\approx-14.28\pm11.59\,\rm{MeV}$, and
$E_{\rm{sym,4}}^{\rm{kin}}(\rho_0)\approx7.18\pm2.52\,\rm{MeV}$\cite{Cai15a}. 
We want to point out that the approach has limitations. Since the momentum distribution parameters (Eq.\,(\ref{MDGen})) were fixed at saturation density using data and/or model calculations, their possible density dependence remains unexplored; the density dependence of the kinetic EOS arises solely from the Fermi energy.

\subsection{Incorporating SRC-HMT into the Nonlinear Walecka RMF and Modified-Gogny HF Models}\label{sub_WaleckaSRC}

\indent

In the previous subsections, the discussion focused on the SRC-HMT effects on the kinetic EOS while the effective nucleon interactions were not included. The next step is to examine how SRC can simultaneously modify both the kinetic and potential parts of the EOS of ANM, particularly the density dependence of the nuclear symmetry energy within a nonlinear relativistic-mean-field (RMF) model (the Walecka model) and a modified-Gogny energy functional. 
The applications of the SRC-HMT modified RMF model on NS physics would be discussed in Section \ref{SEC_NS}.
To incorporate the SRC-induced HMT into RMF models, the kinetic energy density and the corresponding pressure were modified according to (without $\beta_J$ parameter in Eq.\,(\ref{MDGen}))\cite{Cai16c}
\begin{align}
\varepsilon^{\rm{kin}}_{J} =&\frac{2}{(2\pi )^{3}}\int_{0}^{k_{\textrm{F}}^{J}}\textrm{d}
\textbf{k}\Delta_J\sqrt{|\textbf{k}|^{2}+{M_J^{\ast
2}}}\notag\\
&+\frac{2}{(2\pi )^{3}}\int_{k_{\rm{F}}^J}^{\phi_Jk_{\textrm{F}}^{J}}\textrm{d}
\textbf{k}C_J\left(\frac{k_{\rm{F}}^J}{|\v{k}|}\right)^4\sqrt{|\textbf{k}|^{2}+{M_J^{\ast 2}}},\label{EnDenKin}\\
P_{\rm{kin}}^{J}=&\frac{1}{3\pi ^{2}}\int_{0}^{k_{\textrm{F}}^{J}}\d k\Delta_J\frac{k^{4}}{%
\sqrt{k^{2}+{M_J^{\ast}}^{2}}}\notag\\
&+\frac{1}{3\pi ^{2}}\int_{k_{\rm{F}}^J}^{\phi_Jk_{\textrm{F}}^{J}}\d k C_J\left(\frac{k_{\rm{F}}^J}{k}\right)^4\frac{k^{4}}{%
\sqrt{k^{2}+{M_J^{\ast}}^{2}}}, \label{pressureKin}
\end{align}
following the general transformation for any function $f$ at zero temperature when the FFG step function is replaced by $n_{\v{k}}^J$ containing both depletion and the HMT,
\begin{equation}
\int_0^{k_{\rm{F}}^J}(\rm{FFG\ step\ function})f\d\v{k}\longrightarrow\int_0^{\phi_Jk_{\rm{F}}^J}n_{\v{k}}^J\,(\rm{HMT})f\d\v{k}.
\end{equation}

A conceptual inconsistency of this hybrid approach should be noted: the RMF framework is unable to self-consistently reproduce the phenomenological $n_{\v{k}}^J$ in Eq.\,(\ref{MDGen}) that is constrained by experiment. This issue is shared by most phenomenological mean-field models. Ideally, one would adjust the Lagrangian parameters to quantitatively reproduce the empirical $n_{\v{k}}^J$ within the same theory. Unfortunately, strong model dependence in predicting $n_{\v{k}}^J$, together with the limited experimental precision on the isospin dependence of SRC, continues to prevent a fully consistent description. As a result, direct implementation of the phenomenological $n_{\v{k}}^J$ while readjusting only the interaction sector provides a pragmatic although imperfect means to assess SRC effects on the EOS of ANM.

\begin{strip}
\hspace{0.25cm}
The kinetic symmetry energy in the RMF model with HMT is given by\cite{Cai16c}
\begin{align}
E_{\rm{sym}}^{\rm{kin}}(\rho)=&\frac{k_{\rm{F}}^2}{6E_{\rm{F}}^{\ast}}\left[1-3C_0\left(1-\frac{1}{\phi_0}\right)\right]
-3E_{\rm{F}}^{\ast}C_0\left[C_1\left(1-\frac{1}{\phi_0}\right)+\frac{\phi_1}{\phi_0}\right]\notag\\
&-\frac{9M_0^{\ast,4}}{8k_{\rm{F}}^3}\frac{C_0\phi_1(C_1-\phi_1)}{\phi_0}
\left[\frac{2k_{\rm{F}}}{M_0^{\ast}}\left(\left(\frac{k_{\rm{F}}}{M_0^{\ast}}\right)^2+1\right)^{3/2}
-\frac{k_{\rm{F}}}{M_0^{\ast}}\left(\left(\frac{k_{\rm{F}}}{M_0^{\ast}}\right)^2+1\right)^{1/2}-\rm{arcsinh}\left(\frac{k_{\rm{F}}}{M_0^{\ast}}\right)\right]\notag\\
&+\frac{2k_{\rm{F}}C_0(6C_1+1)}{3}\left[\rm{arcsinh}\left(\frac{\phi_0k_{\rm{F}}}{M_0^{\ast}}\right)-
\sqrt{1+\left(\frac{M_0^{\ast}}{\phi_0k_{\rm{F}}}\right)^2}
-\rm{arcsinh}\left(\frac{k_{\rm{F}}}{M_0^{\ast}}\right)+
\sqrt{1+\left(\frac{M_0^{\ast}}{k_{\rm{F}}}\right)^2}\right]\notag\\
&+\frac{3k_{\rm{F}}C_0}{2}\Bigg[\frac{(1+3\phi_1)^2}{9}\left(\frac{\phi_0k_{\rm{F}}}{F_{\rm{F}}^{\ast}}
-\frac{2F_{\rm{F}}^{\ast}}{\phi_0k_{\rm{F}}}\right)
+\frac{2F_{\rm{F}}^{\ast}(3\phi_1-1)}{9\phi_0k_{\rm{F}}}
-\frac{1}{9}\frac{k_{\rm{F}}}{E_{\rm{F}}^{\ast}}+\frac{4E_{\rm{F}}^{\ast}}{9k_{\rm{F}}}\Bigg]
+\frac{C_0(4+3C_1)}{3}\left[\frac{F_{\rm{F}}^{\ast}(1+3\phi_1)}{\phi_0}-E_{\rm{F}}^{\ast}\right],\label{EsymkinHMT}
\end{align}
under the Lagrangian\cite{Cai16c,Serot1986,Muller1996NPA,CuiY25}:
\begin{align}
\mathcal{L}_{\rm{Walecka}}=&\overline{\psi}\left[\gamma_{\mu}(i\partial^{\mu}-g_{\omega}\omega^{\mu}-g_{\rho}\vec{\rho}^{\mu}\cdot\vec{\tau})-(M_{\rm N}-g_{\sigma}\sigma)\right]\psi
+\frac{1}{2}\left(\partial_{\mu}\sigma\partial^{\mu}\sigma-m_{\sigma}^2\sigma^2\right)-\frac{1}{4}\omega_{\mu\nu}\omega^{\mu\nu}+\frac{1}{2}m_{\omega}^2\omega_{\mu}\omega^{\mu}\notag\\
&-\frac{1}{3}b_{\sigma}M_{\rm N}(g_{\sigma}\sigma)^3-\frac{1}{4}c_{\sigma}(g_{\sigma}\sigma)^4+\frac{1}{4}c_{\omega}(g_{\omega}^2\omega_{\mu}\omega^{\mu})^2
+\frac{1}{2}m_{\rho}^2\vec{{\rho}}_{\mu}\cdot\vec{{\rho}}^{\mu}
-\frac{1}{4}\vec{{\rho}}_{\mu\nu}\cdot\vec{{\rho}}^{\mu\nu}
+\frac{1}{2}\Lambda_{\rm V}g_{\rho}^2\vec{\rho}_{\mu}\cdot\vec{\rho}^{\mu}
\Lambda_{\rm{V}}g_{\omega}^2\omega_{\mu}\omega^{\mu}.\label{rmf_lag}
\end{align}
In this review, we use the terms ``Walecka model'' and ``RMF model'' interchangeably.
Moreover, in (\ref{EsymkinHMT}), we define $F_{\rm F}^{\ast}=\sqrt{M_0^{\ast,2}+p_{\rm F}^2}$ with $p_{\rm F}=\phi_0k_{\rm F}$\cite{Cai16c}.
This expression reduces to the familiar RMF result $E_{\rm{sym}}^{\rm{kin}}(\rho)=k_{\rm{F}}^2/6E_{\rm{F}}^\ast$ when $\phi_0=1$ and $\phi_1=0$. The presence of HMT makes $E_{\rm{sym}}^{\rm{kin}}(\rho)$ depend solely on $M_0^{\ast}\equiv M_{\rm D}^{\ast}$, and the reduction relative to the FFG case is substantial across the full range of physically reasonable effective masses. For instance, at $M_0^{\ast}/M_{\rm N}=0.6$, the kinetic symmetry energy becomes $E_{\rm{sym}}^{\rm{kin}}(\rho_0)\approx-16.94\pm13.66\,\rm{MeV}$ (see FIG.\,\ref{fig_xCaiFig2} for the $M_0^{\ast}$-dependence of $E_{\rm{sym}}^{\rm{kin}}(\rho_0)$), close to the corresponding non-relativistic values, as shown in FIG.\,\ref{fig_esym0}. The reduction of the kinetic symmetry energy with HMT is therefore a robust feature shared by both relativistic and non-relativistic treatments.
Similarly, the slope of the kinetic symmetry energy is\cite{Cai16c}:
\begin{align}
L^{\rm{kin}}(\rho)=&\left[\frac{k_{\rm{F}}^2({E}_{\rm{F}}^{\ast,2}+M_0^{\ast,2})}{6{E}_{\rm{F}}^{\ast,3}}
+\frac{g_{\sigma}k_{\rm{F}}^2M_0^{\ast}\rho}{2E_{\rm{F}}^{\ast,3}}\frac{\partial
{\overline{\sigma}}}{\partial\rho}\right]{}\left[1-3C_0\left(1-\frac{1}{\phi_0}\right)\right]
-\frac{9\rho}{E_{\rm{F}}^{\ast}}\left(\frac{\pi^2}{2k_{\rm{F}}}-g_{\sigma}M_0^{\ast}\frac{\partial
{\overline{\sigma}}}{\partial\rho}\right){}
C_0\left[C_1\left(1-\frac{1}{\phi_0}\right)+\frac{\phi_1}{\phi_0}\right]\notag\\
&-\frac{9C_0\phi_1(C_1-\phi_1)}{4\pi^2k_{\rm{F}}\phi_0E_{\rm{F}}^{\ast}}{}\Bigg[
\sqrt{1+\theta^2}{}\rm{arcsinh}\,\theta{}\left(\frac{3M_0^{\ast,5}\pi^2}{2k_{\rm{F}}^2}+4g_{\sigma}k_{\rm{F}}M_0^{\ast,4}\frac{\partial
{\overline{\sigma}}}{\partial\rho}\right)\notag\\
&\hspace*{2.cm}+\frac{\pi^2}{2k_{\rm{F}}^2}{}\left(2k_{\rm{F}}^5
-k_{\rm{F}}^3M_0^{\ast,2}-3k_{\rm{F}}M_0^{\ast,4}\right)
-4g_{\sigma}k_{\rm{F}}^2M_0^{\ast} E_{\rm{F}}^{\ast,2}\frac{\partial
{\overline{\sigma}}}{\partial\rho} \Bigg]\notag\\
&+\frac{2k_{\rm{F}}C_0(6C_1+1)}{3}{}\left[\rm{arcsinh}\,(\phi_0\theta)-\sqrt{1+\frac{1}{\phi_0^2\theta^2}}
-\rm{arcsinh}\,\theta+\sqrt{1+\frac{1}{\theta^2}}\right]\notag\\
&+{2k_{\rm{F}}\rho
C_0(6C_1+1)}{}\left(\frac{M_0^{\ast}\pi^2}{2k_{\rm{F}}^2}+g_{\sigma}k_{\rm{F}}\frac{\partial
{\overline{\sigma}}}{\partial\rho}\right)
\times\left(\frac{\phi_0}{F_{\rm{F}}^{\ast}M_0^{\ast}}
+\frac{M_0^{\ast}}{\phi_0^2k_{\rm{F}}^3\sqrt{1+\frac{\displaystyle1}{\displaystyle\phi_0^2\theta^2}}}
-\frac{1}{M_0^{\ast,2}\sqrt{1+\theta^2}}-\frac{1}{M_0^{\ast}k_{\rm{F}}\theta^2\sqrt{1+\frac{\displaystyle1}{\displaystyle\theta^2}}}\right)
\notag\\
&+\frac{3k_{\rm{F}}C_0}{2}{}\left[\frac{(1+3\phi_1)^2}{9}\left(\frac{\phi_0k_{\rm{F}}}{F_{\rm{F}}^{\ast}}
-\frac{2F_{\rm{F}}^{\ast}}{\phi_0k_{\rm{F}}}\right)+\frac{2F_{\rm{F}}^{\ast}(3\phi_1-1)}{9\phi_0k_{\rm{F}}}
-\frac{1}{9}\frac{k_{\rm{F}}}{E_{\rm{F}}^{\ast}}+\frac{4E_{\rm{F}}^{\ast}}{9k_{\rm{F}}}\right]\notag\\
&+\frac{9k_{\rm{F}}\rho
C_0}{2}{}\left(\frac{M_0^{\ast}\pi^2}{2k_{\rm{F}}^2}+g_{\sigma}k_{\rm{F}}\frac{\partial
{\overline{\sigma}}}{\partial\rho}\right){}\left[\frac{(1+3\phi_1)^2}{9}\left(\frac{\phi_0}{F_{\rm{F}}^{\ast}M_0^{\ast}}
-\frac{\phi_0^3\theta^2}{M_0^{\ast,2}(1+\phi_0^2\theta^2)^{3/2}}+\frac{2}{\phi_0k_{\rm{F}}^2\sqrt{1+\phi_0^2\theta^2}}\right)\right.\notag\\
&\hspace*{2cm}-\left.\frac{2(3\phi_1-1)}{9\phi_0k_{\rm{F}}^2\sqrt{1+\phi_0^2\theta^2}}-\frac{M_0^{\ast}}{9E_{\rm{F}}^{\ast,3}}-\frac{4}{9k_{\rm{F}}^2\sqrt{1+\theta^2}}
\right]\notag\\
&+{\rho C_0(4+3C_1)}{}\left[
\frac{\pi^2}{2k_{\rm{F}}^2}{}\left[\frac{(1+3\phi_1)\phi_0k_{\rm{F}}}{M_0^{\ast}\sqrt{1+\phi_0^2\theta^2}}
-\frac{k_{\rm{F}}}{M_0^{\ast}\sqrt{1+\theta^2}}\right]\right.\notag\\
&\hspace*{2.cm}-\left.g_{\sigma}\frac{\partial{\overline{\sigma}}
}{\partial\rho}{}\left[\frac{(1+3\phi_1)\sqrt{1+\phi_0^2\theta^2}}{\phi_0}
-\frac{(1+3\phi_1)\phi_0k_{\rm{F}}^2}{M_0^{\ast,2}\sqrt{1+\phi_0^2\theta^2}}
-\sqrt{1+\theta^2}+\frac{k_{\rm{F}}^2}{M_0^{\ast,2}
\sqrt{1+\theta^2}}\right]\right].
\label{Lkin}
\end{align}
here $\theta=k_{\rm{F}}/M_0^{\ast}$ (which plays the role of $\nu=k_{\rm F}/M_{\rm N}$ in the above), $\overline{\sigma}$ is the $\sigma$-field in SNM and $\partial \overline{\sigma}/\partial\rho$ is given by
\begin{equation}\label{f0withrho}
\frac{\partial
\overline{\sigma}}{\partial\rho}=\frac{g_{\sigma}}{R_{\sigma}}\left(\frac{\Delta_0M_0^{\ast}}{{E}_{\rm{F}}^{\ast}}+\frac{4\rho_{\rm{s},0}^{\rm{(2)}}}{3\rho}
+\overline{\Phi}\right),\quad \text{ with }\overline{\Phi}=C_0k_{\rm{F}}^2M_0^{\ast}\left(\frac{\phi_0}{p_{\rm{F}}^2F_{\rm{F}}^{\ast}}
-\frac{1}{k_{\rm{F}}^2E_{\rm{F}}}\right),
\end{equation}
with 
\begin{equation}
R_{\sigma}=m_{\sigma}^2+3g_{\sigma}^2\left(\frac{\rho_{\rm{s},0}^{\rm{(1)}}}{M_0^{\ast}}
-\frac{\Delta_0\rho}{{E}_{\rm{F}}^{\ast}}\right)+2b_{\sigma}Mg_{\sigma}^3\overline{\sigma}+3c_{\sigma}g_{\sigma}^4\overline{\sigma}^2
-g_{\sigma}^2\left(\frac{\rho_{\rm{s},0}^{\rm{(2)}}}{M_0^{\ast}}+\overline{\Psi}\right), \quad \text{ with }\overline{\Psi}=\frac{2C_0k_{\rm{F}}^4}{\pi^2}\left(\frac{1}{p_{\rm{F}}F_{\rm{F}}^{\ast}}-\frac{1}{k_{\rm{F}}E_{\rm{F}}^{\ast}}\right),
\end{equation}
and
\begin{equation}
\rho_{\rm{s},0}^{\rm{(1)}} =\frac{\Delta_0M_0^{\ast,3}}{\pi^2}
\left(\theta\sqrt{1+\theta^2}-\rm{arcsinh}\,\theta\right),~~\rho_{\rm{s},0}^{\rm{(2)}}
=\frac{2C_0k_{\rm{F}}^4}{\pi^2M_0^{\ast}}\left(\sqrt{1+\frac{1}{\theta^2}}-\sqrt{1+\frac{1}{\phi_0^2\theta^2}}\right)
,
\end{equation}
The sum of $\rho_{\rm{s},0}^{\rm{(1)}}$ and $\rho_{\rm{s},0}^{\rm{(2)}}$ gives the total scalar density:
\begin{align}\label{def-rhos0}
\rho_{\rm{s}}\equiv\rho_{\rm s,0}=\rho_{\rm{s},0}^{\rm{(1)}}+\rho_{\rm{s},0}^{\rm{(2)}}=\frac{\Delta_0 M_0^{\ast,3}}{\pi^2}
\left(\theta\sqrt{1+\theta^2}-\rm{arcsinh}\,\theta\right)
+\frac{2C_0 k_{\rm{F}}^4}{\pi^2 M_0^{\ast}}\left(\sqrt{1+\frac{1}{\theta^2}}-\sqrt{1+\frac{1}{\phi_0^2\theta^2}}\right).
\end{align}
In addition, the expression for the potential part of the slope parameter of the symmetry energy is given by\cite{Cai12PLB}:
\begin{equation}
L^{\rm{pot}}(\rho)=\frac{3g_{\rho}^2\rho}{2(m_{\rho}^2+\Lambda_{\rm{V}}g_{\omega}^2g_{\rho}^2\overline{\omega}_0^2)}
-\frac{3g_{\omega}^3g_{\rho}^4\Lambda_{\rm{V}}\overline{\omega}_0\rho^2}{(m_{\omega}^2+3c_{\omega}g_{\omega}^4\overline{\omega}_0)(m_{\rho}^2+\Lambda_{\rm{V}}g_{\omega}^2g_{\rho}^2\overline{\omega}_0^2)^2}.
\end{equation}
\end{strip}

\begin{figure}[h!]
\centering
\includegraphics[width=7.5cm]{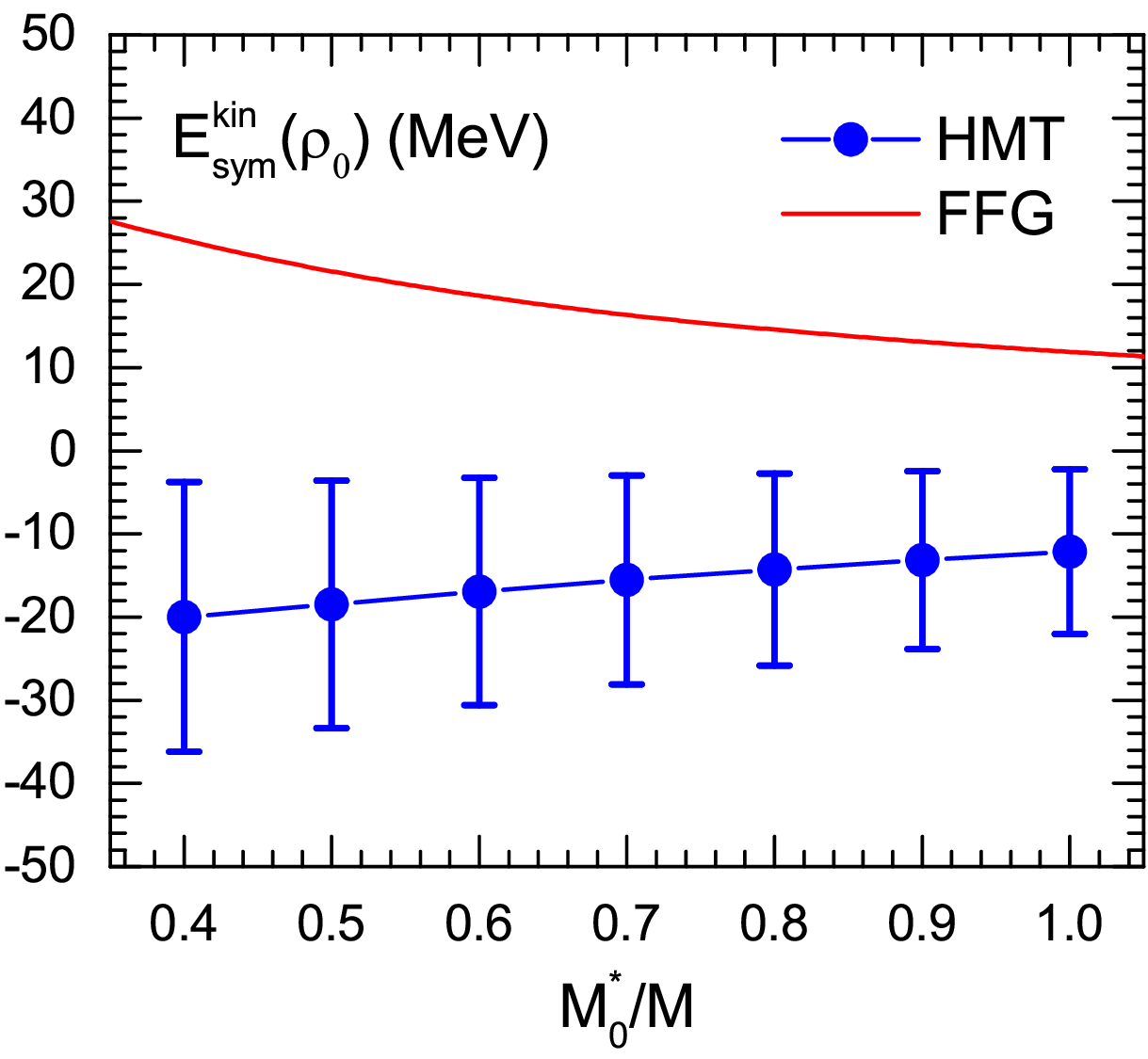}
\caption{(Color Online). Kinetic symmetry energy as a function of Dirac effective mass of nucleons in SNM in the RMF-FFG and RMF-HMT models at saturation density. Taken from Ref.\cite{Cai16c}.}\label{fig_xCaiFig2}
\end{figure}

\begin{figure}[h!]
\centering
\includegraphics[width=7.5cm]{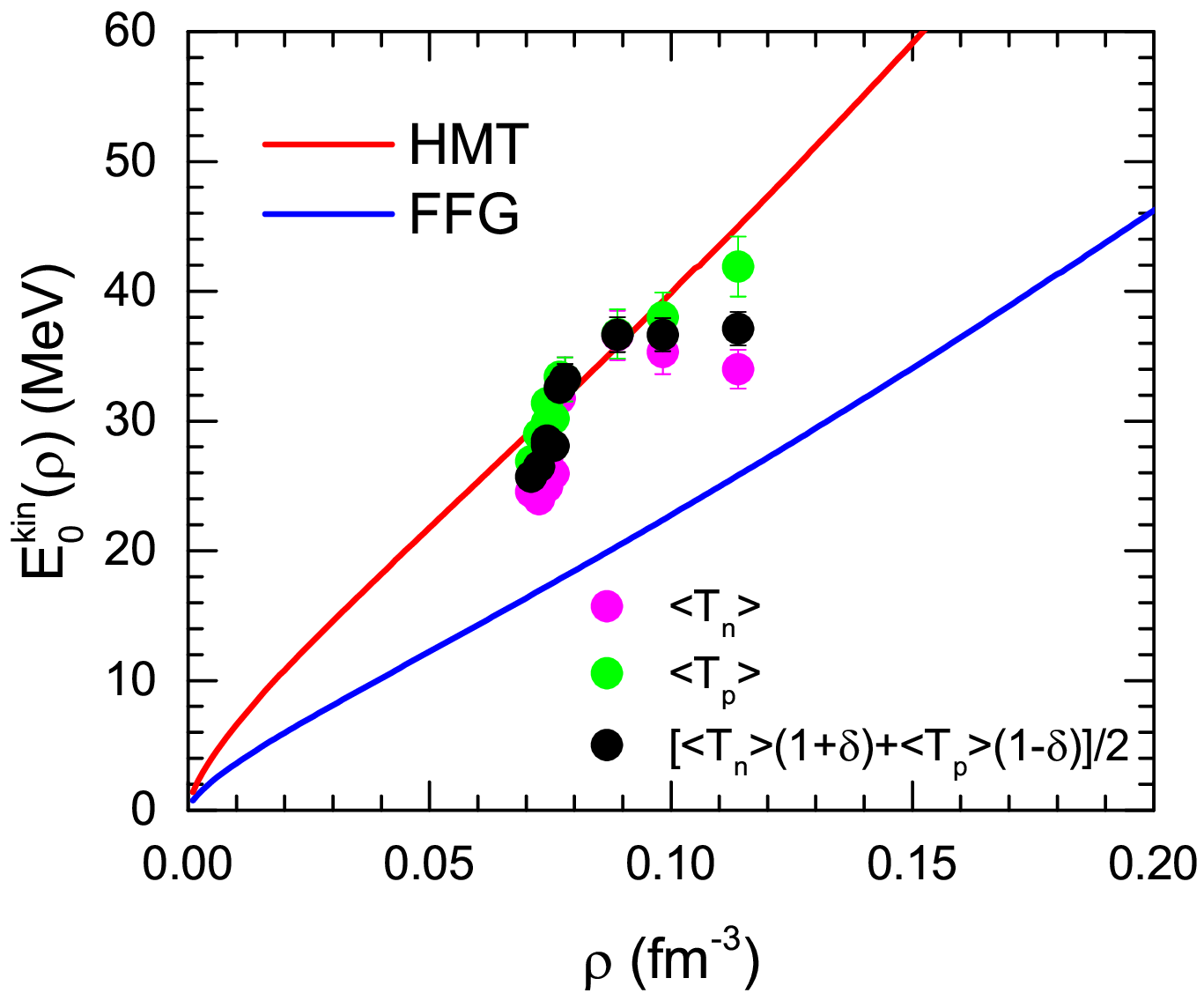}\\
\hspace{-0.3cm}\includegraphics[width=8.cm]{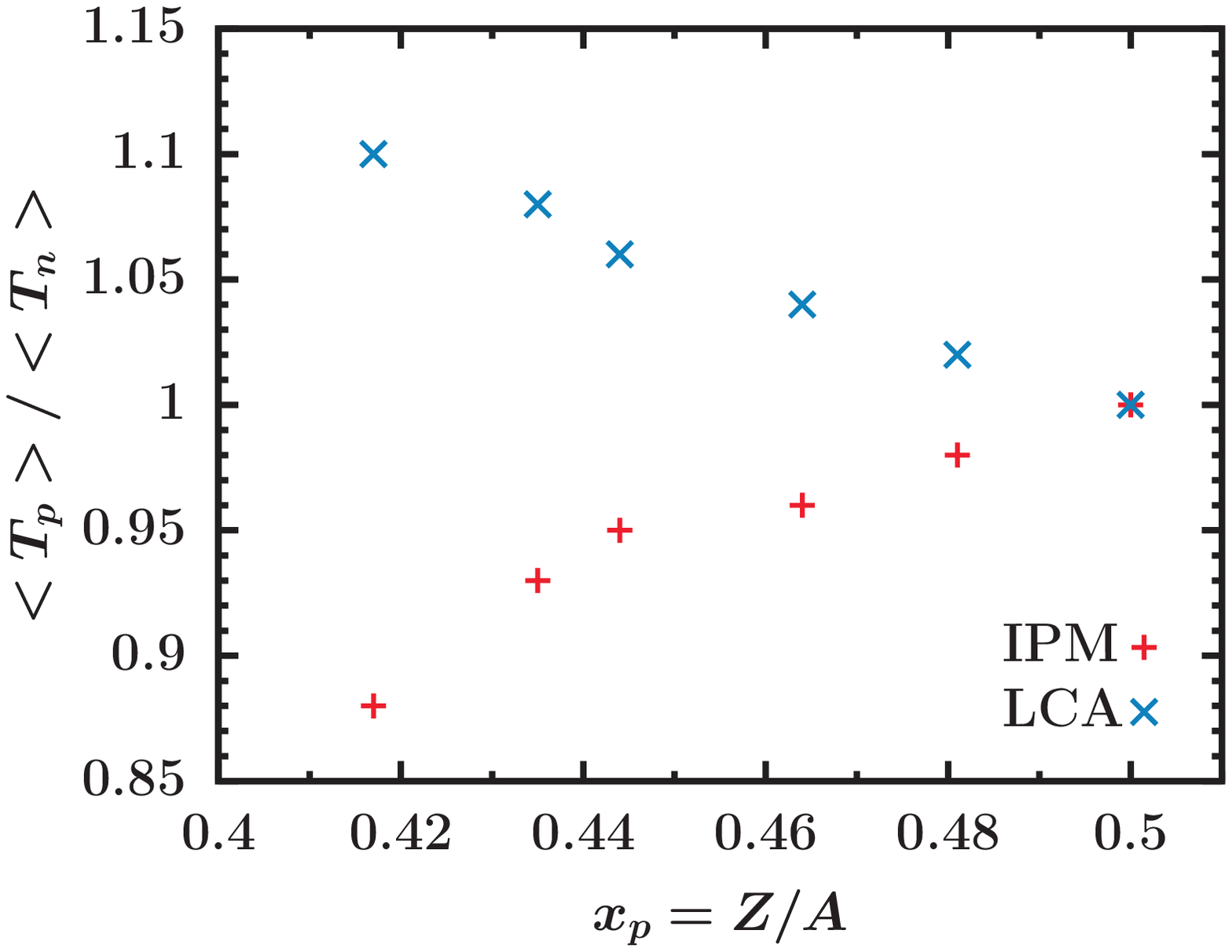}
\caption{(Color Online). Upper: kinetic EOS of SNM defined by Eq.\,(\ref{E0kin-RMF}), compared with data from the np dominance analysis. Lower: ratio of average proton to neutron kinetic energies as a function of proton fraction in the IPM and LCA. Figures taken from Refs.\cite{Cai16c,Ryc15}.}
\label{fig-Ekin0RMFHMT}
\end{figure}

One may further assess the reliability of the hybrid framework used here by comparing its predictions for the kinetic EOS with those obtained in other approaches\cite{Sar14,Ryc15}. In the nonlinear RMF model, the kinetic contribution to the EOS of SNM is defined as
\begin{equation}
\boxed{
E_0^{\rm{kin}}(\rho)\equiv\frac{1}{\rho}\frac{2}{(2\pi)^3}\int_0^{\phi_0k_{\rm{F}}}n_{\v{k}}^0\sqrt{\v{k}^2+M_0^{\ast,2}}\d\v{k}-M_0^{\ast},}
\label{E0kin-RMF}
\end{equation}
where $n_{\v{k}}^0$ denotes the nucleon momentum distribution in SNM. The HMT and FFG predictions for $E_0^{\rm{kin}}(\rho)$ are displayed in the upper panel of FIG.\,\ref{fig-Ekin0RMFHMT}.
Recent analyses of electron-nucleus scattering data using the np dominance picture\cite{Hen14,Sar14} have extracted average kinetic energies of neutrons and protons for C, Al, Fe and Pb (with uncertainties) and for $^{7,8,9}$Li, $^{9,10}$Be and $^{11}$B (without uncertainties). The empirical dependence of the average kinetic energy on the mass number $A$ can be mapped onto a density dependence through the well-established relation\cite{Cen09,Che11a,Roca13,Dan03,Dan09,Dan14,Zhang15}
\begin{equation}
\rho_{A}\approx\frac{\rho_0}{1+\alpha/\beta A^{1/3}},
\end{equation}
where $\alpha/\beta$ encodes the balance between surface and volume symmetry energies (see discussion in Subsection \ref{sub_kinEsym}). Using $\alpha/\beta\approx2.8$\cite{Dan03} for the relevant mass range, the average kinetic energy per nucleon,$ 
\langle T\rangle=[\langle T_{\rm{n}}\rangle(1+\delta)+\langle T_{\rm{p}}\rangle(1-\delta)]/2$,
calculated with both the HMT and FFG parameterizations is compared with the np dominance results in the upper panel of FIG.\,\ref{fig-Ekin0RMFHMT}. Within the parabolic approximation for ANM, $
E^{\rm{kin}}_{\rm{ANM}}(\rho)\approx
E^{\rm{kin}}_{0}(\rho)+\delta^2E^{\rm{kin}}_{\rm{sym}}(\rho)$,
even the most neutron-rich nucleus considered ($^{208}$Pb with $\delta^2\approx 0.045$) satisfies $
E^{\rm{kin}}_{\rm{ANM}}(\rho)\approx E_{0}^{\rm{kin}}(\rho)$,
indicating that the results shown in the upper panel of FIG.\,\ref{fig-Ekin0RMFHMT} effectively represent the kinetic EOS of SNM.
It is noteworthy that the RMF-HMT prediction agrees well with the np dominance extraction, whereas the FFG estimate lies about 40\% lower. It is well known that pure mean-field models underestimate spectroscopic factors from $^{7}$Li to $^{208}$Pb by approximately 30--40\% because they miss the occupation of high-momentum states generated by SRC\cite{Lap93}, shown in FIG.\,\ref{fig_SF}. The lower nucleon kinetic energies predicted by the RMF-FFG model relative to the RMF-HMT follow from the same underlying physics.
Another robust feature\cite{Hen14,Ryc15} is that protons carry more kinetic energy than neutrons in neutron-rich nuclei, particularly in the neutron-skin regions of heavy systems\cite{Cai16b} (to be discussed in Subsection \ref{sub_protonskin}). The np dominance picture predicts a growing difference between $\langle T_{\rm p}\rangle$ and $\langle T_{\rm n}\rangle$ as the isospin asymmetry increases. Within the low-order correlation operator approximation (LCA)\cite{Ryc15}, nucleon kinetic energies were evaluated by isolating contributions from central, spin-isospin and tensor correlations, enabling a detailed assessment of the relative strengths of nn, pp and pn pairs in the HMT. For momenta in the range $1.5\lesssim|\v{k}|\lesssim3\,\rm{fm}^{-1}$, the distribution is dominated by tensor-induced pn correlations, and the predicted pp-pn ratio in the HMT is consistent with exclusive two-nucleon knockout measurements showing np-SRC dominance\cite{Ryc15}.
As illustrated in the lower panel of FIG.\,\ref{fig-Ekin0RMFHMT}, the independent particle model (IPM) yields $\langle T_{\rm n}\rangle>\langle T_{\rm p}\rangle$, but this ordering is reversed once correlations are included. Furthermore, the difference $\langle T_{\rm p}\rangle-\langle T_{\rm n}\rangle$ increases approximately linearly as the proton fraction $x_{\rm p}$ decreases. For the most asymmetric nucleus studied, $^{48}\rm{Ca}$, the proton kinetic energy exceeds the neutron value by about 10\%, consistent with the findings in Ref.\cite{Hen14}; see also the magenta and green points.

\begin{figure}[h!]
\centering
\includegraphics[width=8.cm]{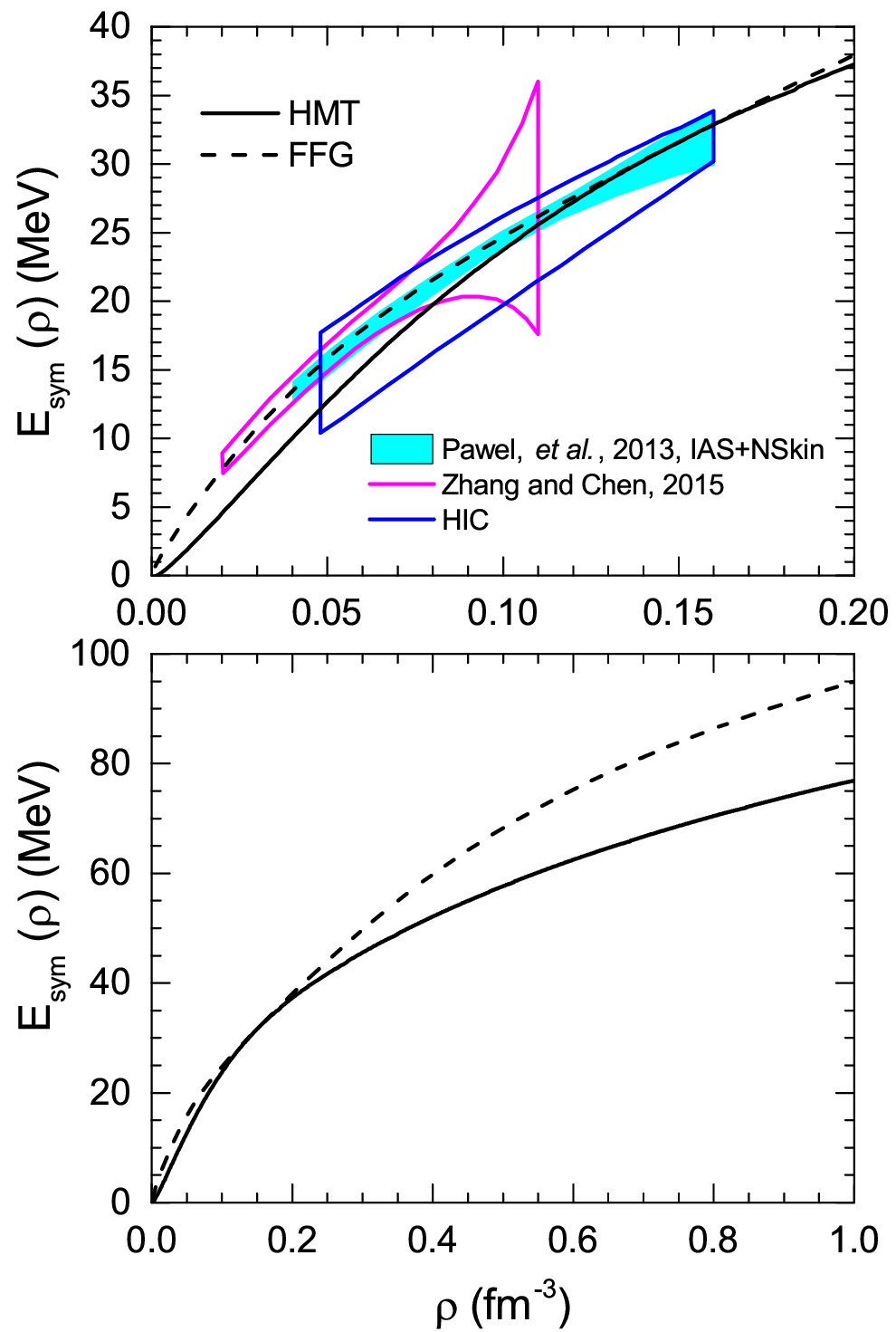}
\caption{(Color Online). Total symmetry energy versus density in the HMT and FFG scenarios, compared with several recent constraints. Figure taken from Ref.\cite{Cai16c}.}
\label{fig-EsymRMFHMT}
\end{figure}

Turning back to the RMF results, the SRC-reduced kinetic symmetry energy has important consequences for the density dependence of the total symmetry energy $E_{\rm{sym}}(\rho)$. Since $E_{\rm{sym}}(\rho_0)$ and $L(\rho_0)$ are both well constrained, a smaller kinetic contribution at $\rho_0$ necessarily implies a larger potential contribution. FIG.\,\ref{fig-EsymRMFHMT} compares the results obtained by fixing $M_0^{\ast}$, $E_0(\rho_0)$, $\rho_0$, $K_0$, $E_{\rm{sym}}(\rho_0)$ and $L(\rho_0)$ at their empirical values. The inclusion of HMT lowers $E_{\rm{sym}}(\rho)$ but makes its slope steeper at subsaturation densities; around $0.04$\,fm$^{-3}$ the decrease reaches $\sim30\%$, exceeding the current uncertainty band\cite{Zhang15}. At suprasaturation densities the reduction persists, amounting to $\sim25\%$ near $0.5$\,fm$^{-3}$. Overall, the curvature of $E_{\rm{sym}}(\rho)$ is substantially reduced, leading to a more concave density dependence.
An interesting feature is that the negative kinetic symmetry energy allows the construction of very soft or even decreasing symmetry energies at high densities by varying $L(\rho_0)$ while keeping $E_{\rm{sym}}(\rho_0)$ fixed\cite{LCCX18}. Such super-soft behaviors have been suggested to be favorable for explaining certain $\pi^-/\pi^+$ data in heavy-ion collisions\cite{Xiao09}.

\begin{figure}[h!]
\centering
  \includegraphics[width=9.cm]{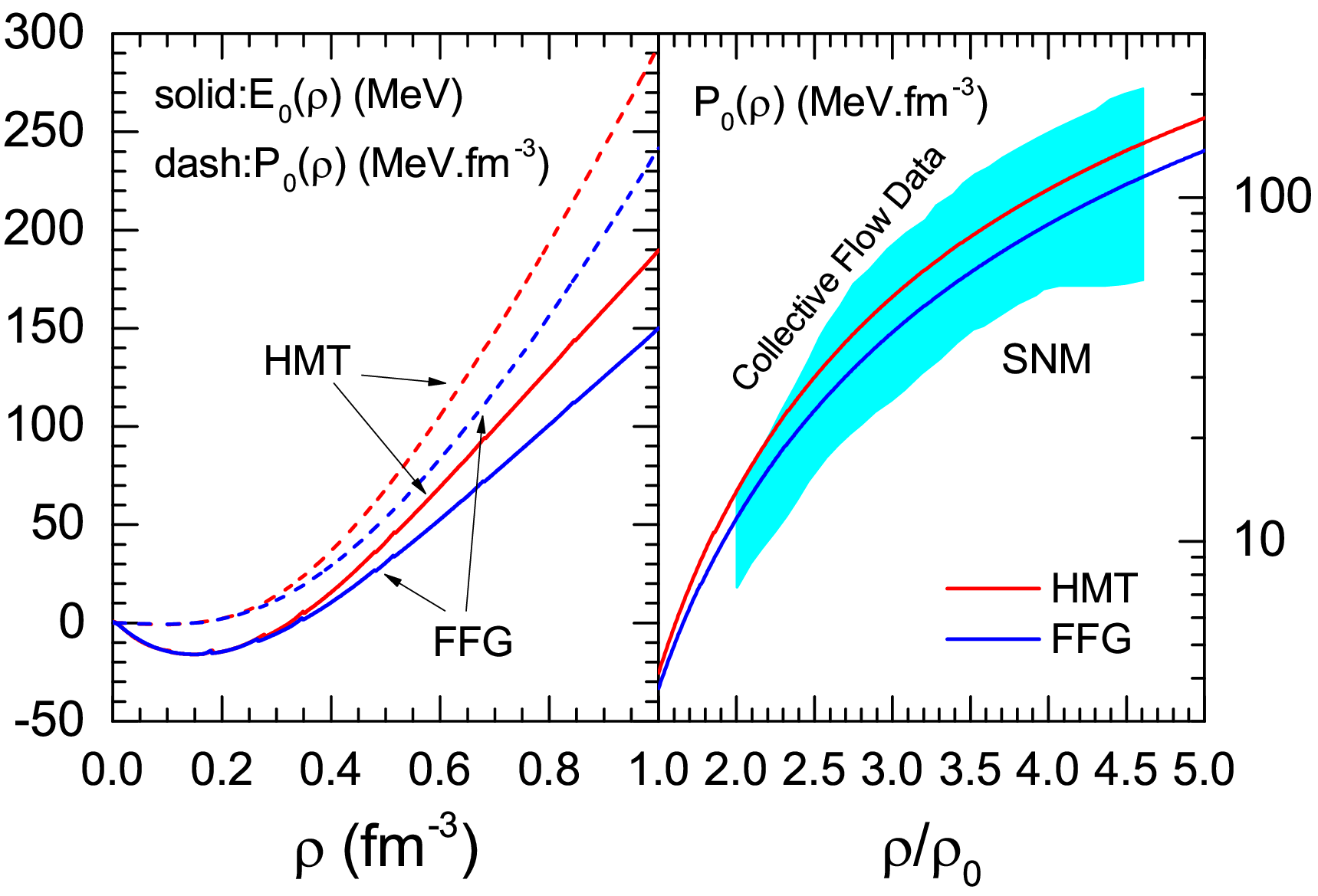}
  \caption{(Color Online). Left: the EOS of SNM and the pressure $P_0$ of SNM as functions of density for both the FFG and HMT models.
  Right: a comparison between the model pressure $P_0$ of SNM with the experimental constraints from analyzing nuclear collective flows in heavy ion collisions. Figure taken from Ref.\cite{Cai16c}.}
  \label{fig-E0RMFHMT}
\end{figure}

Besides the reduction of the symmetry energy due to SRC-induced HMT, the EOS of SNM in the modified Walecka model was also obtained in Ref.\cite{Cai16c}. To assess the HMT and FFG models, the left panel of FIG.\,\ref{fig-E0RMFHMT} shows the binding energy and pressure of SNM as functions of density; here the expressions for the EOS of SNM $E_0(\rho)=\varepsilon_0/\rho-M_{\rm N}$, the pressure $P_0(\rho)$ as well as the coefficient of incompressibility $K_0(\rho)$ are given by\cite{Cai16c}:
\begin{align}
\varepsilon_0(\rho)=&\frac{\Delta_0M_0^{\ast,4}}{\pi^2}\left[\frac{1}{4}\theta(1+\theta^2)^{3/2}-\frac{1}{8}\theta\sqrt{1+\theta^2}-\frac{1}{8}\rm{arcsinh}\,\theta\right]\notag\\
&+\frac{C_0k_{\rm{F}}^4}{\pi^2}\left[
V(\theta)-\sqrt{1+\frac{1}{\phi_0^2\theta^2}}+\sqrt{1+\frac{1}{\theta^2}}\right]\notag\\
&+\frac{1}{2}m_{\sigma}^2\overline{\sigma}^2+
+\frac{1}{3}b_{\sigma}M_{\rm N}g_{\sigma}^3\overline{\sigma}^3+\frac{1}{4}c_{\sigma}g_{\sigma}^4\overline{\sigma}^4\notag\\
&+\frac{1}{2}m_{\omega}^2\overline{\omega}_0^2
+\frac{3}{4}c_{\omega}(g_{\omega}\overline{\omega}_0)^3
,\label{cc2_eps0}\\
P_0(\rho)=&-\frac{1}{2}m_{\sigma}^2\overline{\sigma}^2-\frac{1}{3}b_{\sigma}M_{\rm N}g_{\sigma}^3\overline{\sigma}^3-\frac{1}{4}c_{\sigma}g_{\sigma}^4\overline{\sigma}^4
+\frac{1}{2}m_{\omega}^2\overline{\omega}_0^2\notag\\
&+\frac{1}{4}c_{\omega}g_{\omega}^4\overline{\omega}_0^4
+\frac{2\Delta_0M_0^{\ast,4}W(\theta)}{3\pi^2}
+\frac{2C_0k_{\rm{F}}^4V(\theta)}{3\pi^2},\label{cc2_P0}\\
K_0(\rho)=&-9\frac{\partial \overline{\sigma}}{\partial\rho}
\Bigg[\overline{\sigma}\left(m_{\sigma}^2+b_{\sigma}M_{\rm N}g_{\sigma}^3\overline{\sigma}+c_{\sigma}g_{\sigma}^4\overline{\sigma}^2\right)\notag\\
&\hspace{1.5cm}+\frac{8g_{\sigma}\Delta_0M_0^{\ast,3}W(\theta)}{3\pi^2}\Bigg]
+12C_0k_{\rm{F}}V(\theta)\notag\\
&+9\left[\frac{2\Delta_0k_{\rm{F}}^4}{3\pi^2M_0^{\ast}E_{\rm{F}}^{\ast}}
+\frac{2C_0k_{\rm{F}}^4}{3\pi^2M_0^{\ast}}\left(\frac{\phi_0}{F_{\rm{F}}^{\ast}}-\frac{1}{E_{\rm{F}}^{\ast}}\right)\right]\notag\\
&\hspace{0.5cm}\times\left(\frac{M_0^{\ast}\pi^2
}{2k_{\rm{F}}^2}+g_{\sigma}k_{\rm{F}}\frac{\partial
\overline{\sigma}}{\partial\rho}\right)+\frac{9\rho
g_{\omega}^2}{Q_{\omega}}-\frac{18P_0(\rho)}{\rho},\label{cc2_K0}
\end{align}
here $Q_\omega=
\overline{\omega}_0^2 + 3c_\omega g_\omega^4\overline{\omega}_0$ and $\partial\overline{\sigma}/\partial\rho$ is given in Eq.\,(\ref{f0withrho}), and the two functions $W(\theta)$ and $V(\theta)$ are\cite{Cai16c}:
\begin{align}
    W(\theta)=&\frac{1}{4}\theta^3\sqrt{1+\theta^2}
-\frac{3}{8}\theta\sqrt{1+\theta^2}+\frac{3}{8}\rm{arcsinh}\,\theta,\\
V(\theta)=&\rm{arcsinh}(\phi_0\theta)-\rm{arcsinh}\,\theta.
\end{align}

Interestingly, the HMT model predicts a stiffer EOS for SNM at supra-saturation densities compared to the FFG model, even though by construction both models share the same values of $M_0^{\ast}$, $\rho_0$, $E_0(\rho_0)$, and $K_0$. This is primarily due to the significant contribution of high-momentum nucleons to the kinetic EOS in the HMT model. Consequently, HMT is expected to affect the higher-order characteristic coefficients of SNM at $\rho_0$ relative to the FFG predictions.
Quantitatively, the third-order Taylor expansion coefficient of the EOS around $\rho_0$, i.e., the skewness $J_0$ of SNM
changes from $J_0^{\rm{FFG}}\approx -454\,\rm{MeV}$ in the FFG model to $J_0^{\rm{HMT}}\approx -266\,\rm{MeV}$ in the HMT model. Unfortunately, the current knowledge of $J_0$ is still too limited to provide a meaningful constraint. On the other hand, the pressure of SNM in the density range $2\rho_0\sim5\rho_0$ has been constrained experimentally via nuclear collective flows in heavy-ion collisions, shown as a cyan band in the right panel of FIG.\,\ref{fig-E0RMFHMT}. Despite the larger skewness in the HMT model, its EOS remains consistent with these empirical constraints, indicating that the current experimental uncertainty is too broad to distinguish HMT from FFG predictions.
Since the kinetic symmetry energy depends on the nucleon Dirac effective mass, which in turn is determined by the scalar baryon density $\rho_{\rm{s}}$, it is instructive to examine how the SRC-modified nucleon momentum distribution affects $\rho_{\rm{s}}$ and $M_0^{\ast}$. At low densities ($\theta \ll 1$), we have from Eq.\,(\ref{def-rhos0}) that
\begin{align}
\rho_{\rm{s}} \rightarrow& \frac{\Delta_0 M_0^{\ast,3}}{\pi^2}\frac{2}{3}\theta^3
+ \frac{2C_0 M_0^{\ast,3}}{\pi^2}\left(1-\frac{1}{\phi_0}\right)\theta^3\notag\\
= &\frac{2 M_0^{\ast,3}\theta^3}{3\pi^2}\left[\Delta_0 + 3C_0\left(1-\frac{1}{\phi_0}\right)\right] = \rho.
\end{align}
The next-order correction is $
M_0^{\ast,3}\theta^5(1+C_0\Phi_0)/5\pi^2$, 
which is negative, implying $\rho_{\rm{s}}<\rho$.
At the high-density limit $\rho\to\infty$, the $\sigma$-field saturates at $\overline{\sigma}^{\infty}\equiv \overline{\sigma}(\rho=\infty)=M_{\rm N}/g_{\sigma}$, so that the Dirac effective mass approaches zero. Correspondingly, the scalar density becomes
$
\rho_{\rm{s}}^{\infty}\equiv \rho_{\rm{s}}(\rho=\infty) = M_{\rm N}^3\left[(m_{\sigma}/g_{\sigma}M_{\rm N})^2 + b_{\sigma} + c_{\sigma}\right]$,
giving $\rho_{\rm{s}}^{\infty}(\rm{FFG}) \approx 3.29\,\rm{fm}^{-3}$ and $\rho_{\rm{s}}^{\infty}(\rm{HMT}) \approx 2.99\,\rm{fm}^{-3}$\cite{Cai16c}. Thus, $\rho_{\rm{s}}$ in the HMT model is always smaller than in the FFG model.
Actually, the nucleon Dirac effective masses in SNM for both FFG and HMT model are very similar, and this can be understood as follows. The effective mass is constrained at three points: $M_0^{\ast}(0)/M_{\rm N}=1$, $M_0^{\ast}(\rho_0)/M_{\rm N}\approx 0.6$, and $M_0^{\ast}(\infty)/M_{\rm N}=0$. Furthermore, Eq.\,(\ref{f0withrho}) ensures $\partial \overline{\sigma}/\partial \rho > 0$, so that $M_0^{\ast}/M_{\rm N}$ decreases monotonically over the full density range and is concave at high densities. Given these common constraints, it is not surprising that $M_0^{\ast}(\rho)/M_{\rm N}$ behaves very similarly in both models.

In the nonlinear RMF calculations, the HMT and FFG models are constructed to share the same symmetry energy $E_{\rm{sym}}(\rho_0)$ and slope parameter $L$, it is instructive to examine the curvature $K_{\rm{sym}}$ of the symmetry energy as a measure of how the HMT influences the total symmetry energy. The quantity $K_{\rm{sym}}$ also determines the isospin dependence of the incompressibility of ANM through $
K(\delta)\approx K_0+K_{\rm{sat,2}}\delta^2+\mathcal{O}(\delta^4)$,
where $K_{\rm{sat,2}}=K_{\rm{sym}}-6L-J_0L/K_0$\cite{Che09} represents the isospin-dependent contribution. Quantitatively, one finds $K_{\rm{sym}}^{\rm{FFG}}\approx -37\,\rm{MeV}$ and $K_{\rm{sym}}^{\rm{HMT}}\approx -274\,\rm{MeV}$. The corresponding values of the isospin coefficient of the incompressibility are $K_{\rm{sat,2}}^{\rm{FFG}}\approx -174\,\rm{MeV}$ and $K_{\rm{sat,2}}^{\rm{HMT}}\approx -470\,\rm{MeV}$, with the latter in very good agreement with the empirical estimate $K_{\rm{sat,2}}=-550\pm 100\,\rm{MeV}$\cite{Colo14}. Overall, the inclusion of the HMT makes the symmetry energy substantially more concave around saturation density, leading to a significantly stronger isospin dependence of the incompressibility of ANM compared with the FFG model.

\begin{figure}[h!]
\centering
\includegraphics[height=4.cm]{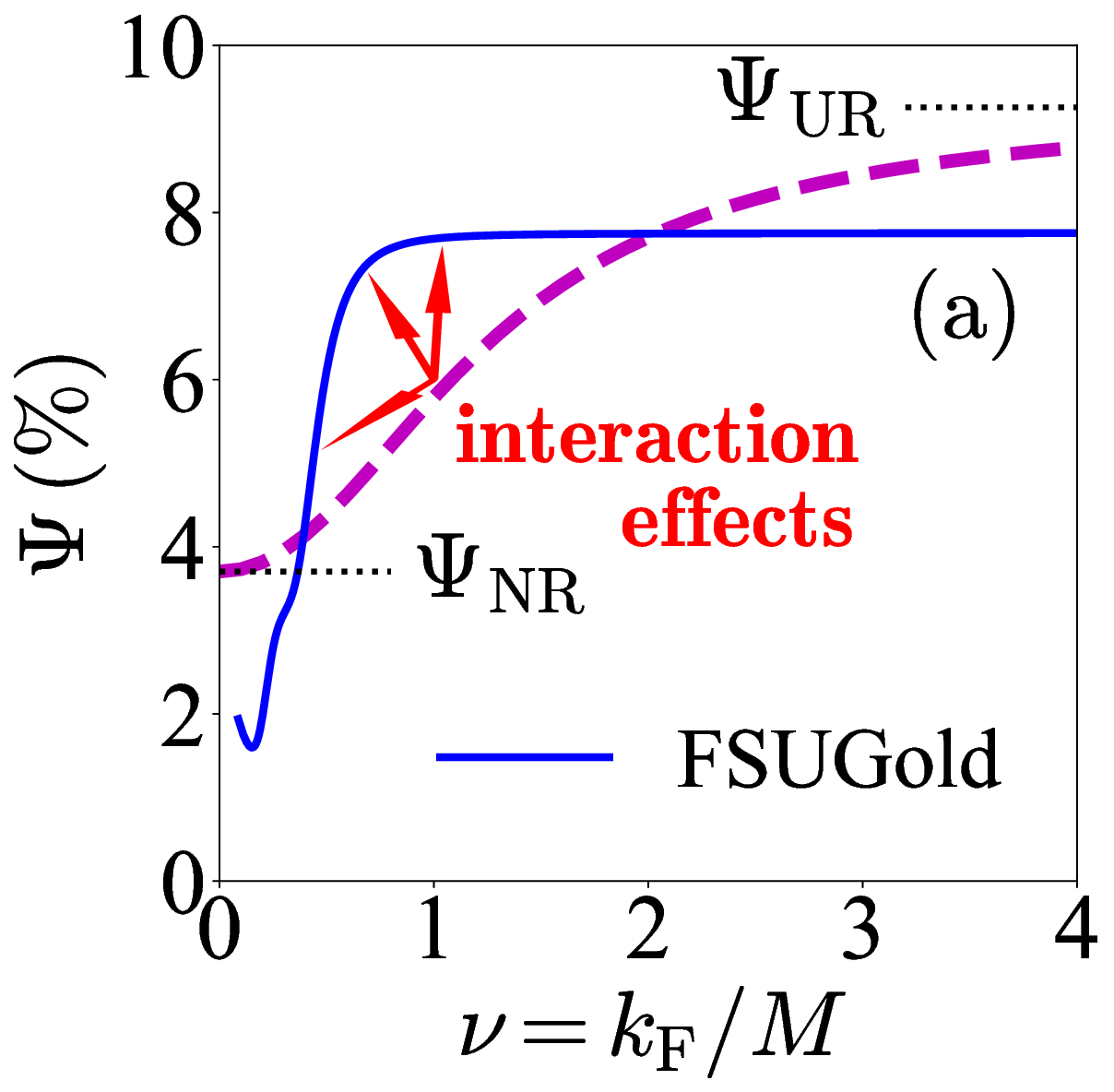}\quad
\includegraphics[height=4.cm]{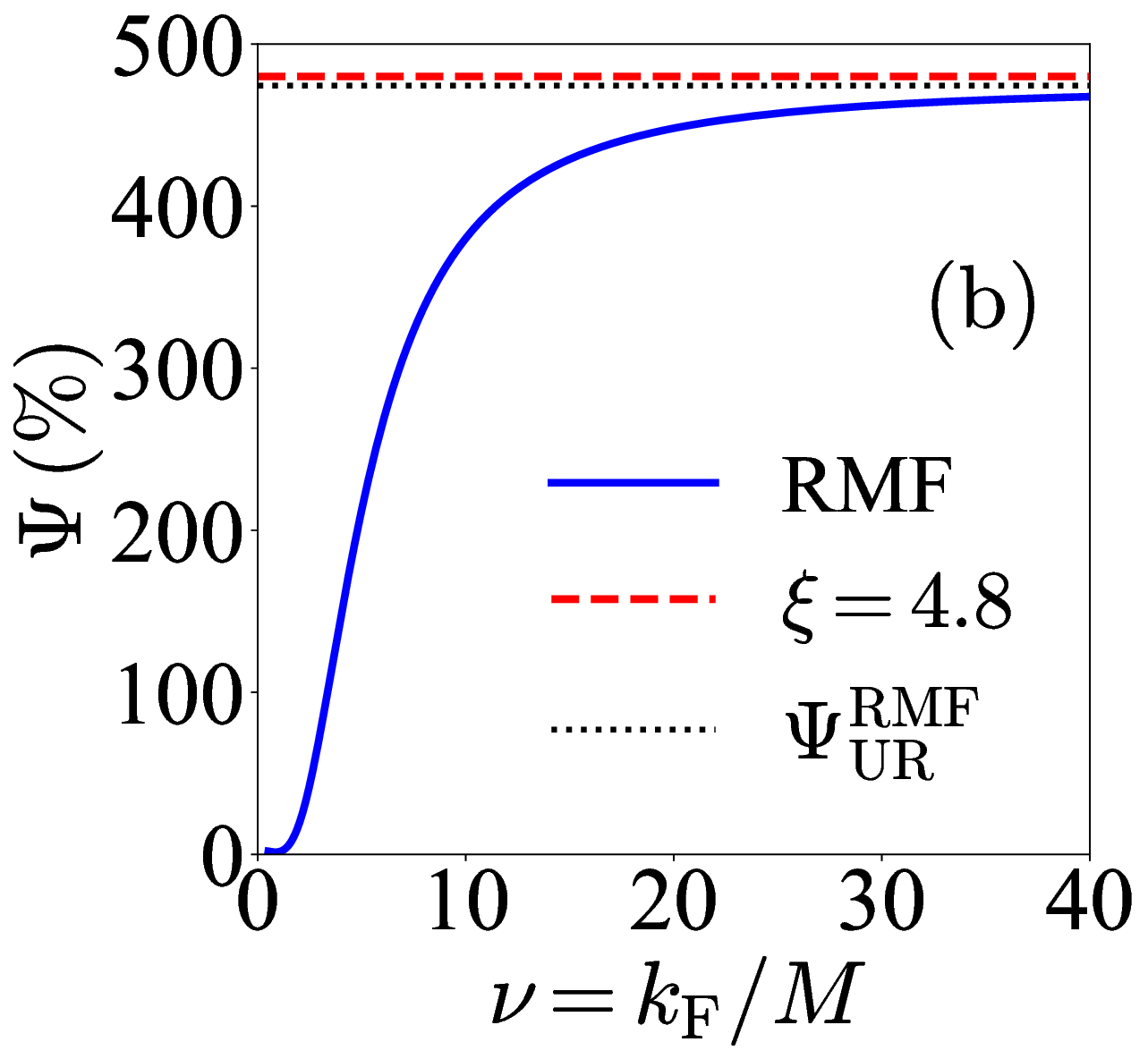}
\caption{(Color Online). Panel (a): The $\nu$ (equivalently $\rho$) dependence of the ratio $\Psi$ in the relativistic FFG model (red) compared with the nonlinear RMF model prediction (blue lines). 
Panel (b): results with a general RMF construction different from the FSUGold set. Here, $M\leftrightarrow M_{\rm N}$.
Figures taken from Ref.\cite{CaiLi22PRCFFG}.
}\label{fig_Psi_nu}
\end{figure}

The nonlinear RMF model allows us to study the ratio $\Psi$ of the quartic over quadratic symmetry energy even without including SRC-HMT effects. Specifically, the $\Psi$ obtained with the FSUGold parameter set\cite{Tod05} is displayed in the left panel of FIG.\,\ref{fig_Psi_nu} as the blue solid line, while the relativistic FFG prediction from Eq.\,(\ref{Psi_nu}) is shown as the dashed magenta line. It is clear that nuclear interactions significantly modify the FFG prediction (indicated by red arrows in panel (a)) at both low and high densities.
In the ultra-relativistic limit, the quadratic and quartic symmetry energies can be approximated as\cite{Cai12PRC-S4}:
\begin{align}
\boxed{
E_{\rm{sym}}(\rho) \approx \frac{k_{\rm{F}}}{6} + \frac{c_{\omega}^{2/3} \rho^{1/3}}{2 \Lambda_{\rm{V}}},~~
E_{\rm{sym,4}}(\rho) \approx \frac{5 k_{\rm{F}}}{324} + \frac{c_{\omega}^{5/3} \rho^{1/3}}{6 \Lambda_{\rm{V}}^2},}
\end{align}
both scaling as $\rho^{1/3}$\cite{Cai12PRC-S4}. Here, $c_\omega$ and $\Lambda_{\rm{V}}$ characterize the self-interaction of four $\omega$ mesons and the $\omega$-$\rho$ meson coupling in the nonlinear RMF Lagrangian $\mathcal{L}_{\rm{Walecka}}$ of Eq.\,(\ref{rmf_lag}) via $4^{-1} c_\omega g_\omega^4 (\omega_\mu \omega^\mu)^2$ and $2^{-1} \Lambda_{\rm{V}} g_\rho^2 g_\omega^2 \omega_\mu \omega^\mu \vec{\rho}_\nu \cdot \vec{\rho}^\nu$\cite{Cai12PRC-S4}, with $g_\omega$ and $g_\rho$ as nucleon-meson coupling constants. Consequently, the ratio in the ultra-relativistic limit is
\begin{equation}\label{rmf-case}
\Psi = \Psi^{\rm{RMF}}_{\rm{UR}} \equiv \frac{5 a/324 + c_\omega^{5/3}/6\Lambda_{\rm{V}}^2}{a/6 + c_\omega^{2/3}/2\Lambda_{\rm{V}}},
\end{equation}
where $a = (3 \pi^2 / 2)^{1/3}$\cite{Cai12PRC-S4}. Using FSUGold parameters $c_\omega = 0.01$ and $\Lambda_{\rm{V}} = 0.24$, one obtains $\Psi_{\rm{UR}}^{\rm{RMF}} \approx 7.8\%$, quickly reached for $\nu = k_{\rm F}/M_{\rm N} \gtrsim 1$. In both numerator and denominator, the first term is kinetic, the second potential. The potential-to-kinetic ratios are $(c_\omega^{2/3}/2\Lambda_{\rm{V}})/(a/6) \approx 0.24$ and $(c_\omega^{5/3}/6\Lambda_{\rm{V}}^2)/(5a/324) \approx 0.04$, showing kinetic dominance in FSUGold. If $c_\omega$ or $1/\Lambda_{\rm{V}}$ is small, the limit naturally approaches the FFG prediction $5/54$.

To explore interaction effects further, consider the theoretical scenario where the potential terms dominate, $c_\omega^{5/3}/6\Lambda_{\rm{V}}^2 \gg 5a/324$ and $c_\omega^{2/3}/2\Lambda_{\rm{V}} \gg a/6$. Then Eq.\,(\ref{rmf-case}) gives $\Psi \to \xi \equiv c_\omega / 3\Lambda_{\rm{V}}$, independent of density. In this limit, $E_{\rm{sym,4}}(\rho)$ can become comparable to or even exceed $E_{\rm{sym}}(\rho)$, demonstrating that nucleon-nucleon interactions can strongly modify the relative strengths of quadratic and quartic symmetry energies. Numerically, $\Psi$ could diverge if $\Lambda_{\rm{V}}$ is extremely small, although this occurs only at extremely high, theoretically inaccessible densities.
As an illustration, the right panel of FIG.\,\ref{fig_Psi_nu} shows a constructed RMF calculation for $\Psi$ (blue solid line) approaching both $\Psi_{\rm{UR}}^{\rm{RMF}}$ (black dotted line) and the constant $\xi = c_\omega / 3\Lambda_{\rm{V}} = 4.8$. Here, $c_\omega = 0.01$, while $\Lambda_{\rm{V}} = c_\omega / 3\xi \approx 7 \times 10^{-4}$, much smaller than the FSUGold set\cite{Tod05}. This gives $c_\omega^{2/3}/2\Lambda_{\rm{V}} \approx 33.42 \gg a/6 \approx 0.41$ and $c_\omega^{5/3}/6\Lambda_{\rm{V}}^2 \approx 160.41 \gg 5a/324 \approx 0.04$, confirming potential dominance. This corresponds to $\nu=k_{\rm F}/M_{\rm N} \sim 40$ or $\rho \sim 2.9 \times 10^6 \rho_0$, highlighting the theoretical nature. Thus, nucleon-nucleon interactions, combined with relativistic effects, strongly modify $\Psi$, whereas relativistic corrections alone have minimal impact.
In contrast, at ultra-low densities, the kinetic part dominates. The kinetic symmetry energy and the potential part expand as\cite{Cai12PRC-S4}
\begin{align}
E_{\rm{sym}}^{\rm{kin}}(\rho) \approx& \frac{k_{\rm F}^2}{6M_{\rm N}} \left[1 - \frac{1}{2} \left(\frac{k_{\rm F}}{M_{\rm N}}\right)^2 + \left(\frac{g_\sigma}{m_\sigma}\right)^2 \frac{\rho}{M_{\rm N}} \right] \sim b \nu^2,\\
E_{\rm{sym}}^{\rm{pot}}(\rho) \approx& \frac{\rho}{2} \frac{g_\rho^2}{m_\rho^2} \left[1 - \Lambda_{\rm{V}} \left(\frac{g_\rho}{m_\rho}\right)^2\left(\frac{g_\omega}{m_\omega}\right)^4 \rho^2 \right] \sim b' \nu^3,
\end{align}
respectively,
here $b$ and $b'$ are two constants;
so as $\rho \to 0$ we have $E_{\rm{sym}}^{\rm{pot}}(\rho)/E_{\rm{sym}}^{\rm{kin}}(\rho) \sim \nu \to 0$. Similarly, the fourth-order symmetry energy is dominated by its kinetic part, $E_{\rm{sym},4}^{\rm{kin}}(\rho)\approx k_{\rm{F}}^2/162M_{\rm N}+\cdots\sim c\nu^2+\cdots$, and similarly $E_{\rm{sym},4}^{\rm{pot}}(\rho)\approx-72^{-1}(g_{\sigma}/m_{\sigma})^2\rho\nu^4+\cdots\sim c'\nu^7+\cdots$, giving $E_{\rm{sym,4}}^{\rm{pot}}(\rho)/E_{\rm{sym,4}}^{\rm{kin}}(\rho)\approx (c'/c)\nu^5 \ll 1$. Therefore, at low densities, $\Psi$ is primarily determined by kinetic contributions.

\begin{figure}[h!]
\centering
  \includegraphics[width=7.cm]{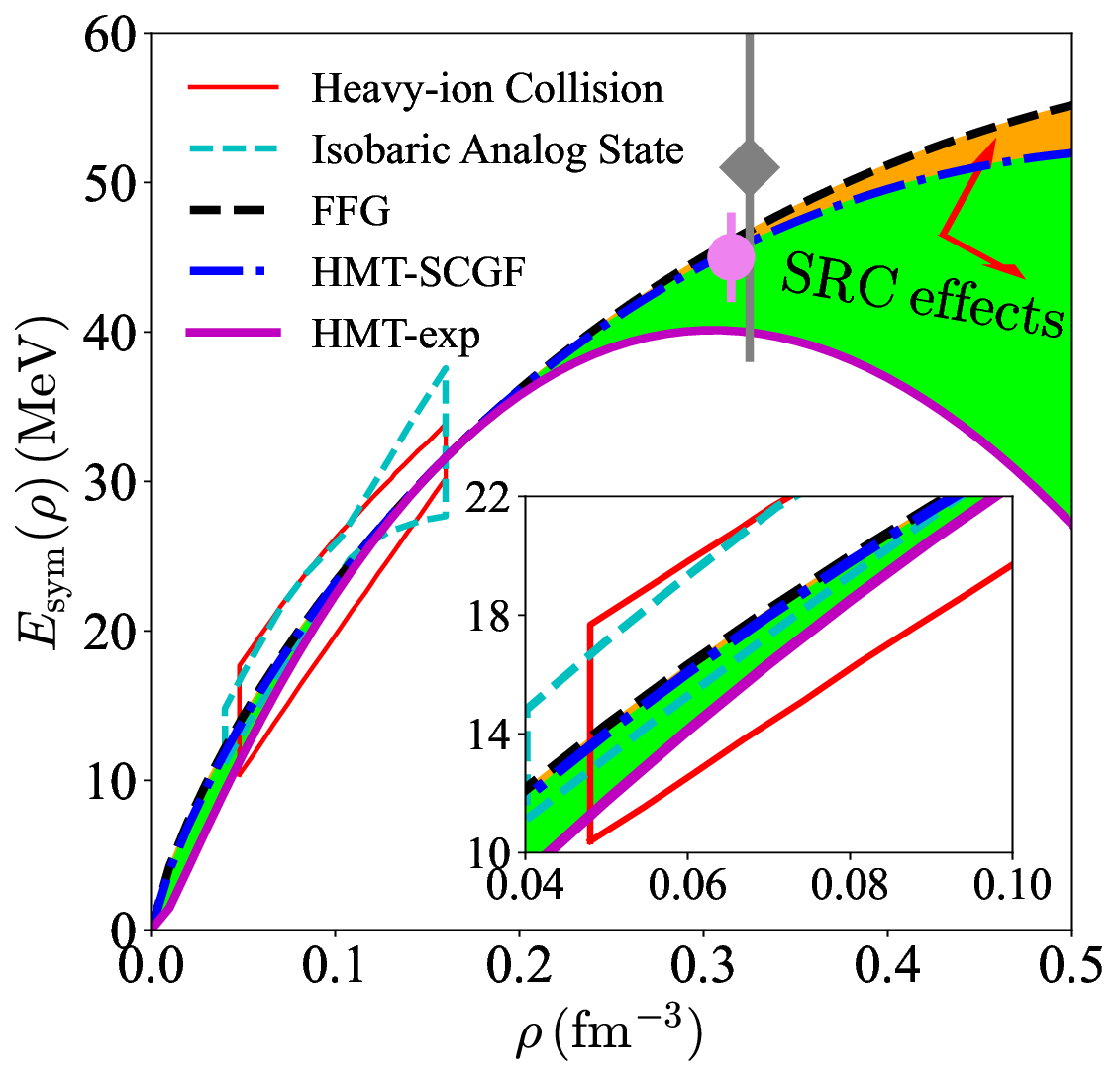}\\
  \hspace{-0.3cm}
  \includegraphics[width=7.2cm]{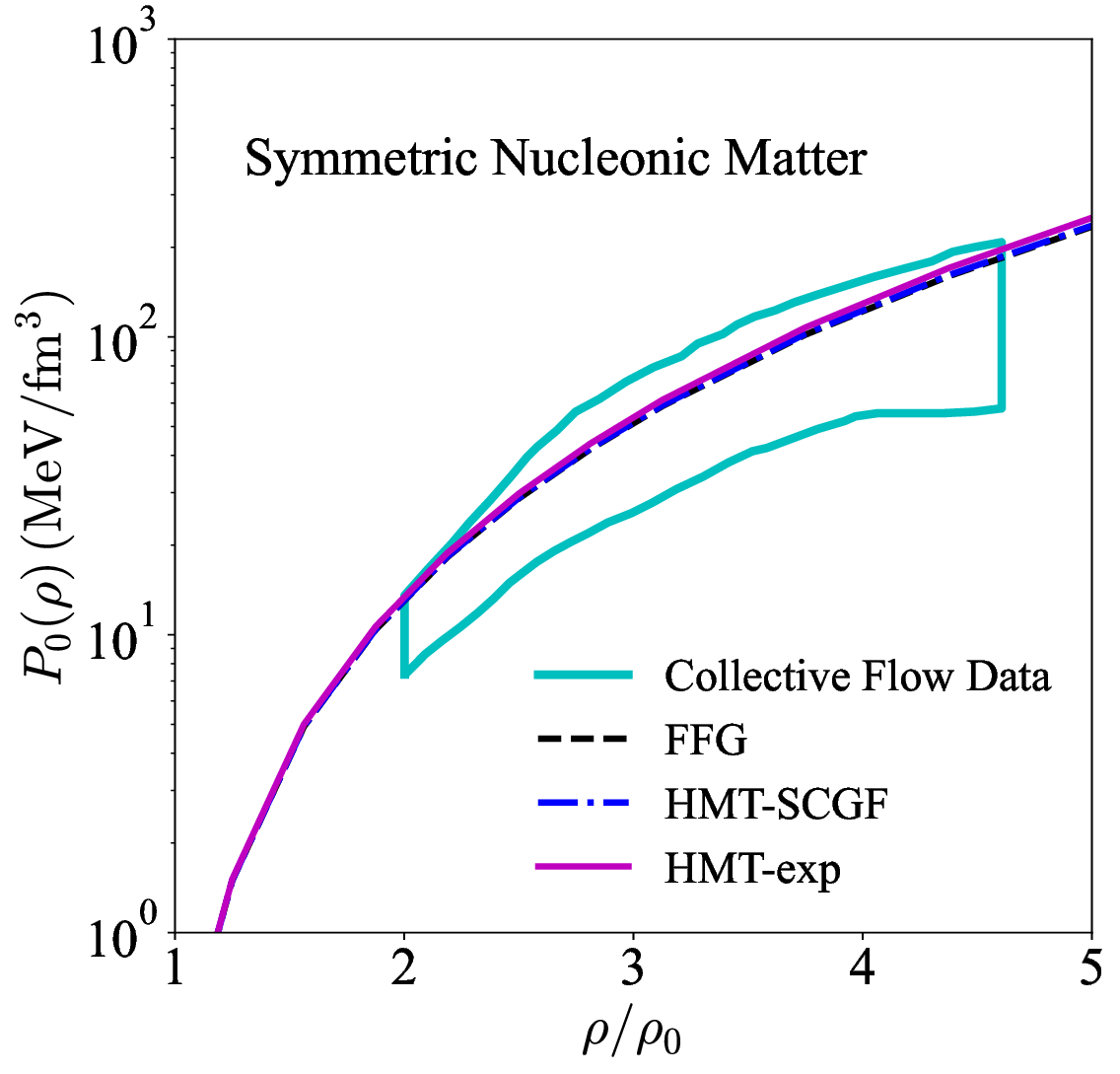}
  \caption{(Color Online). Upper: density dependence of the symmetry energy in the FFG model and HMT-SCGF/HMT-exp model. Constraints from heavy-ion collisions\cite{Tsa12} and the isobaric analog state studies\cite{Dan14} are shown for comparison. Lower: comparison between the pressure $P_0$ of SNM with the experimental constraints from analyzing nuclear collective flows in heavy ion collisions\cite{Dan02}. Figures taken from Ref.\cite{CaiLi22Gog}. }
  \label{fig_ab_Esym}
\end{figure}

\renewcommand*\tablename{\small TAB.}
\begin{table*}[h!]
\centering
\begin{tabular}{lr||cccc}
\hline\hline Quantity& Value & Coupling & FFG&HMT-SCGF&HMT-exp\\
\hline $\rho_0$ (fm$^{-3}$) & $0.16$ & $A_\ell^0$ $(\rm{MeV})$ &
$-266.2934$&$520.3611$&$146.6085$\\
\hline $E_0(\rho_0)$
$(\rm{MeV})$&$-16.0$&$A_{\rm{u}}^0$ $(\rm{MeV})$&$-86.8331$&$805.3082$&$1216.2500$\\
\hline $M_0^{\ast}/M_{\rm N}$ &$0.58$&$B$ $(\rm{MeV})$&$517.5297$&$-256.9850$&$-64.5669$\\
\hline $K_0$ ($\rm{MeV}$) &$230.0$&$C_\ell$ $(\rm{MeV})$&$-155.6406$&$-154.7508$&$-37.3249$\\
\hline $U_0(\rho_0,0)$ ($\rm{MeV}$)
&$-100.0$&$C_{\rm{u}}$ $(\rm{MeV})$&$-285.3256$&$-351.0989$&$-679.5379$\\
\hline $E_{\rm{sym}}(\rho_0)$ ($\rm{MeV}$)&$31.6$&$\sigma$&$1.0353$&$0.9273$&$0.6694$\\
\hline $L$ ($\rm{MeV}$)&$58.9$&$a$&$-5.4511$&$-5.0144$&$-4.1835$\\
\hline $U_{\rm{sym}}(\rho_0,1\,\rm{GeV})$ ($\rm{MeV}$)&$-20.0$&$x$&$0.6144$&$0.3774$&$-0.2355$\\
\hline\hline
\end{tabular}
\caption{\centering Coupling constants used in the two models and the corresponding empirical properties of ANM\cite{CaiLi22Gog}. Table taken from Ref.\cite{CaiLi22Gog}.}\label{tab_para} 
\end{table*}

As an example of non-relativistic calculations with momentum-dependent interaction, shown in FIG.\,\ref{fig_ab_Esym} are the results of SNM pressure and the $E_{\rm{sym}}(\rho)$ within a modified Gogny (with the MDI interaction \cite{Das2003PRC}) energy density functional (EDF) using the FFG, HMT-SCGF and HMT-exp parameter sets\cite{CaiLi22Gog}. In this study, several quantities were fixed at their empirical values, i.e., $M_0^{\ast}$, $E_0(\rho_0)$, $\rho_0$, $K_0$, $E_{\rm{sym}}(\rho_0)$, $L$, $U_0(\rho_0,0)$
and $U_{\rm{sym}}(\rho_0,1\,\rm{GeV})$, where $U_0(\rho,|\v{k}|)$ and $U_{\rm{sym}}(\rho,|\v{k}|)$ is the single-nucleon isoscalar and isovector potential, respectively, see TAB.\,\ref{tab_para} for the details on the parameters\cite{CaiLi22Gog}.
The energy per nucleon is given by\cite{Chen05,XuJ10,XuJ15},
\begin{align}\label{EDF}
E(\rho,\delta)=&\sum_{J=\rm{n,p}}\frac{1}{\rho_J}\int_0^{\infty}\frac{\v{k}^2}{2M_{\rm N}}n_{\v{k}}^J(\rho,\delta)\d\v{k}+\frac{A_\ell(\rho_{\rm{p}}^2+\rho_{\rm{n}}^2)}{2\rho\rho_0}\notag\\
&+\frac{A_{\rm{u}}\rho_{\rm{p}}\rho_{\rm{n}}}{\rho\rho_0}
+\frac{B}{\sigma+1}\left(\frac{\rho}{\rho_0}\right)^{\sigma}(1-x\delta^2),
\notag\\
&+\sum_{J,J'}\frac{C_{J,J'}}{\rho\rho_0}\int\d\v{k}\d\v{k}'f_J(\v{r},\v{k})f_{J'}(\v{r},\v{k}')\Omega(\v{k},\v{k}').
\end{align}
The relation between the phase space distribution and the single-nucleon momentum distribution function is 
\begin{equation}
f_J(\v{r},\v{k})=\frac{2}{h^3}n_{\v{k}}^J(\rho,\delta)=\frac{1}{4\pi^3}n_{\v{k}}^J(\rho,\delta),~~\hbar=1.
\end{equation}
In the FFG model, $f_J(\v{r},\v{k})=(1/4\pi^3)\Theta(k_{\rm{F}}^J-|\v{k}|)$. 
In the calculation, we use the regulator $\Omega$ of $
\Omega(\v{k},\v{k}')=1+{a}[({\v{k}\cdot\v{k}'}/{\Lambda^2})^2]^{1/4}
+{b}[({\v{k}\cdot\v{k}'}/{\Lambda^2})^2]^{1/6}$,
where $a$ and $b$ are two phenomenological parameters used\cite{CaiLi22Gog}, and $\Lambda$ the high-momentum cutoff. If $\Lambda$
is large compared to the momentum scale in studying, we then have $
\Omega(\v{k},\v{k}')\approx
1-{\v{k}^2}/{\Lambda^2}-{\v{k}'^2}/{\Lambda^2}+{2\v{k}\cdot\v{k}'}/{\Lambda^2}$.
This regulator function can be treated as a surrogate model of the original MDI regularization function.
If we change $b$
to $\xi b$, then after redefining $a'=a/\xi^{3/2}$ and
$\Lambda'=\Lambda/\xi^{3/2}$, we have $
\Omega(\v{k},\v{k}')=1+a'[(\v{k}\cdot\v{k}'/\Lambda'^2)^2]^{1/4}+b[(\v{k}\cdot\v{k}'/\Lambda'^2)^2]^{1/6}$, indicating the effect of $b$ can be absorbed into $\Lambda$ and $a$,
i.e., we have one degree of freedom to fix one of the three
parameters $a$, $b$ and $\Lambda$ without affecting the physical
quantities.
The SRC-induced reduction of
$E_{\rm{sym}}(\rho)$ within the non-relativistic EDF approach is qualitatively
consistent with the RMF-HMT result shown in FIG.\,\ref{fig-EsymRMFHMT}. Nevertheless,
since there is no explicit momentum dependence in the RMF model, the
corresponding reduction of $E_{\rm{sym}}(\rho)$ is smaller. Obviously, the
momentum-dependent interaction in the modified Gogny EDF makes the SRC-induced softening of the symmetry
energy at supra-saturation densities more evident, see the upper panel of FIG.\,\ref{fig_ab_Esym}.
The reduction on the symmetry energy at supra-saturation densities also generates corresponding reduction of the curvature coefficient $K_{\rm{sym}}$ of the symmetry energy, whose
value changes from $-109$\,MeV in the FFG model to about
$-121\,\rm{MeV}$/$-223$\,MeV in the HMT-SCGF/HMT-exp model.
While the $K_{\rm{sym}}\approx-223\,\rm{MeV}$ in the HMT-exp model is slightly smaller than today's fiducial constraint on it, these in the FFG and the HMT-SCGF model are very consistent.
Physically, the pressure of the neutron-rich nucleonic matter is
expected to decrease as some of the the nucleons tend to pair with
each other due to the SRC. If we write out the pressure, i.e.,
\begin{align}
P/3\rho\approx&L(\rho)\delta^2+3\rho\frac{\d E_0(\rho)}{\d\rho}\notag\\
\approx& L\delta^2\left(1+3\chi+\frac{9}{2}\chi^2\right)
+K_0\chi(1+3\chi)\notag\\
&+\frac{1}{2}\chi^2J_0+K_{\rm{sym}}\delta^2\chi(1+3\chi),
\end{align}
where $K_0$ and
$L$ are fixed in our study. The decreasing of the pressure $P$ (as well as a
slight increasing in $J_0$) needs that the $K_{\rm{sym}}$ should to
be correspondingly reduced. 
This explains the reduction of the $K_{\rm{sym}}$.
The HMT-induced reduction on the isospin incompressibility
coefficient $K_{\rm{sat},2}$ can be similarly analyzed, it changes from
$-365\,\rm{MeV}$ in the FFG model to about
$-378\,\rm{MeV}$ or $-492\,\rm{MeV}$ in the HMT-SCGF or the HMT-exp model. The maximum reduction is about 127\,MeV.
Similarly, the HMT models yield a slightly higher pressure for SNM at supra-saturation densities compared to the FFG model due to the enhanced kinetic contribution from high-momentum nucleons as shown in the lower panel of FIG.\,\ref{fig_ab_Esym}, yet all three models still produce pressures consistent with heavy-ion flow constraints\cite{Dan02}. Moreover, because the HMT-SCGF model contains fewer high-momentum nucleons than the HMT-exp model, its predicted pressure remains much closer to the FFG result.

\subsection{EOS of ANM in $d$ Dimensions with/without SRC-induced HMT Effects}\label{sub_dEOS}

\indent 

In this subsection, we briefly discuss the role of dimensionality in the EOS of SNM, with or without the effects of SRC-HMT effects. Dimensionality characterizes the number of independent degrees of freedom required to specify the state of a system. In physical contexts, it is typically defined as the minimum number of spatial coordinates needed to identify any point in the system. More broadly, dimensionality is a universal concept extending beyond conventional physics; for instance, the dimensionality of animal behavior can be defined as the minimum number of features of the past required to optimally predict the future\cite{Bialek2022}. Although everyday phenomena occur in three spatial dimensions, treating the spatial dimension $d$ as a continuous or discrete variable has historically been indispensable for developing modern theoretical methodologies and deepening our understanding of complex physical phenomena.

A prominent example is dimensional regularization in quantum field theory\cite{Collins1984}, in which divergent Feynman integrals are analytically continued to non-integer space–time dimensions $d+1$, enabling the first consistent renormalization of non-Abelian gauge fields\cite{Hooft1972}. Likewise, Onsager's exact solution of the two-dimensional square-lattice Ising model\cite{Onsager1944} revolutionized the study of phase transitions, laying foundations for low- and high-temperature expansions, universality, critical phenomena, and the renormalization group\cite{Stanley1971,Kardar2007}. In the past several decades, reduced-dimensional systems have become a central focus of condensed-matter and materials physics, with remarkable discoveries such as the two-dimensional electron gas\cite{Ando1982}, graphene\cite{Sarma2011}, and topological insulators\cite{Hasan2010,Qi2011}. The constant density of states in two dimensions\cite{Ashcroft1976} is one of the key features behind their unconventional behavior.

Recent experimental advances, including Feshbach resonances\cite{Chin2010}, laser cooling\cite{Schreck2021}, optical lattices and box potentials\cite{Vale2021,Navon2021}, and synthetic gauge fields\cite{Dalibard2011}, have made it possible to realize and probe strongly correlated quantum systems across one, two, and three dimensions\cite{Gio08RMP,Blo08RMP,BECBOOK}. These efforts have led to breakthroughs such as the realization of spin-orbit coupling in degenerate Fermi gases\cite{Zhai2015}, quantum-limited spin and sound transport in 2D systems\cite{Luciuk2017,Sommer2011,Bohlen2020,Ville2018,Patel2020,LiX2022,Christodoulou2021}, observation of non-Hermitian phase transitions\cite{Ozturk2021}, superfluidity and scale anomalies in 2D Fermi gases\cite{Sobirey2021,Murthy2019}, turbulence in low dimensions\cite{Navon2016,Navon2019,Gauthier2019,Johnstone2019}, generalized hydrodynamics in 1D Bose gases\cite{Malvania2021}, spin-charge separation in 1D Fermi gases\cite{Senaratne2022}, and the direct measurement of the EOS of a 2D photon gas\cite{Busley2022}. On the theoretical side, mixed-dimensional Fermi gases\cite{Nishida2011}, resonant Bose-Fermi mixtures\cite{Bertaina2013}, p-wave Fermi-Bose dimers in 2D\cite{Bazak2018}, and mappings between few- and many-body systems via higher-dimensional embeddings\cite{Guo2018} have revealed unexpected structures in low-dimensional quantum matter. The $\epsilon$-expansion approach to the unitary Fermi gas\cite{Nishida2006}, inspired by methods developed for critical phenomena\cite{Wilson1974,MaSK1976,Wallace1976}, further highlights the power of dimensional continuation.

The concept of dimensionality also plays fundamental roles beyond physics, notably in computational and data sciences, where high-dimensional data often possess effective low-dimensional structure. Techniques such as principal component analysis identify these low-dimensional manifolds\cite{Jolliffe2002}, while modern statistical learning theory relies critically on dimensional scaling\cite{Blum2020,FanJQ2020,Vershynin2018,Wainwright2018}.
Nuclear systems, finite nuclei and infinite nucleonic matter, are among the most complex many-body systems\cite{Fetter2003}. Their properties arise from intricate momentum-, density-, spin- and isospin-dependent nuclear interactions\cite{LiBA2008}, relativistic effects\cite{Walecka1974,Chin1977,Serot1986}, extreme astrophysical environments\cite{Shapiro1983}, finite-size effects, and, potentially, effective dimensionality. While nearly all nuclear many-body calculations assume three spatial dimensions, several physical scenarios naturally introduce reduced or quasi-reduced dimensional structures. Toroidal or bubble nuclei, as suggested by J. Wheeler and later explored theoretically and experimentally\cite{Siemens1967,Wong1972,Wong1973,Cao2019}, possess high symmetry that effectively lowers the dimensionality of nucleonic motion. Heavy-ion collisions can generate approximately 1D jet-like subsystems or 2D collective flow patterns\cite{LiBA2008}, while NS mergers exhibit strongly anisotropic emission\cite{Baiotti2017,Radice2020,Burns2020}. Dimensionality is also crucial for simulating neutrino-driven supernova explosions\cite{Burrows1990,Burrows2013,Burrows2021,Woosley1986,Janka2012}. Even within a NS, the crust's small thickness ${t\lesssim1\,\mathrm{km}}$ relative to stellar radius ${R\approx12\,\mathrm{km}}$ raises the question of whether it may be modeled as an effectively quasi-2D layer. Similarly, the stellar core, governed by the Tolman-Oppenheimer-Volkoff equation in a single radial coordinate, invites inquiry into whether it may be treated as an effectively 1D system.
These observations motivate a systematic investigation of nuclear quantities, including the EOS of ANM, in a general spatial dimension $d$. 

Exploring how the properties of nuclear matter evolve with $d$, with or without SRC-HMT effects, may reveal new structures, simplify existing treatments, and provide complementary routes to understanding dense matter in astrophysical and laboratory environments. 
In the following, we first study the EOS in 2D and that in general $d$ using the $\epsilon$-expansion technique, and then discuss how SRC-HMT effects may generate new features in the EOS of ANM\cite{Cai22-dD}.

{\bfseries Example 1.} The EOS of ANM in $d$D expressed in terms of the single nucleon potential as well as its momentum and isospin decomposition could be obtained by using the HVH theorem\cite{Hug58,XuC2011,ChenR2012,CaiBJ2012PLB,LiXH2013PLB,LiXH2015PLB}. In 2D, the Fermi momentum $k_{\rm{F}}$ is determined via the density $\rho$ as $
k_{\rm{F}}=(\pi \rho)^{1/2}$, and the dimensions of density and Fermi momentum scale as $[\rho]=[\rm{fm}]^{-2}=[\rm{MeV}]^2$. The corresponding formulas for the EOS of ANM are\cite{Cai22-dD}:
\begin{align}
E_0(\rho)=&\frac{\pi\rho}{4M_{\rm N}}+\frac{1}{\rho}\int_0^{\rho}U_0\left(f,k_{\rm{F}}^f\right)\d
f=\frac{\pi \rho}{4M_{\rm N}}\notag\\&+\frac{1}{\rho}\int_0^\rho
U_0\left(f,k_{\rm{F}}^f\right)\d
f,~~k_{\rm{F}}^f=(\pi f)^{1/2},\label{dd_E0_2}\\
E_{\rm{sym}}(\rho)=&\frac{\pi \rho}{4M_{\rm N}}+\left.\frac{\sqrt{\pi
\rho}}{4}\frac{\partial U_0}{\partial
|\v{k}|}\right|_{|\v{k}|=k_{\rm{F}}}+\frac{1}{2}U_{\rm{sym}}(\rho,k_{\rm{F}})\label{dd_Esym_2},\\
E_{\rm{sym},4}(\rho)=&\frac{k_{\rm{F}}}{64}\left(
\frac{\partial
U_{0}}{\partial|\v{k}|}-k_{\rm{F}}\frac{\partial^2U_{0}}{\partial|\v{k}|^2}
+\frac{k_{\rm{F}}^2}{3}\frac{\partial^3U_{0}}{\partial|\v{k}|^3}
\right)_{|\v{k}|=k_{\rm{F}}}\notag\\&+\frac{k_{\rm{F}}}{32}\left(\frac{\partial
U_{\rm{sym}}}{\partial|\v{k}|}+k_{\rm{F}}\frac{\partial^2U_{\rm{sym}}}{\partial|\v{k}|^2}\right)_{|\v{k}|=k_{\rm{F}}}\notag\\
&+\left.\frac{k_{\rm{F}}}{8}\frac{\partial
U_{\rm{sym},2}}{\partial|\v{k}|}\right|_{|\v{k}|=k_{\rm{F}}}
+\frac{1}{4}U_{\rm{sym},3}(\rho,k_{\rm{F}})\label{dd_Esym_4},\\
P_0(\rho)=&\frac{3\pi \rho^2}{4M_{\rm N}}-\int_0^\rho
U_0\left(f,k_{\rm{F}}^f\right)\d f+\rho
U_0(\rho,k_{\rm{F}}),\label{dd_p0_2}\\
K_0(\rho)=&9\rho\frac{\d U_0}{\d \rho}+\frac{18}{\rho}\int_0^\rho
U_0\left(f,k_{\rm{F}}^f\right)\d
f-18U_0(\rho,k_{\rm{F}}),\label{dd_K0}\\
J_0(\rho)=&27\rho^2\frac{\d^2U_0}{\d \rho^2}-81\rho\frac{\d U_0}{\d
\rho} \notag\\&-\frac{162}{\rho}\int_0^\rho U_0\left(f,k_{\rm{F}}^f\right)\d
f
+162U_0(\rho,k_{\rm{F}}),\label{dd_J0}\\
L(\rho)=&\frac{3\pi \rho}{4M_{\rm N}}+\frac{3\pi
\rho}{8}\left[\frac{\partial^2
U_0}{\partial|\v{k}|^2}+\frac{1}{\sqrt{\pi \rho}}\frac{\partial
U_0}{\partial|\v{k}|}\right]_{|\v{k}|=k_{\rm{F}}}\notag\\&+\left.\sqrt{\pi
\rho}\frac{\partial
U_{\rm{sym}}}{\partial|\v{k}|}\right|_{|\v{k}|=k_{\rm{F}}}
+\frac{3}{2}U_{\rm{sym}}(\rho,k_{\rm{F}})\notag\\&+3U_{\rm{sym},2}(\rho,k_{\rm{F}})\label{dd_L},\\
K^{\rm{SNM}}(\rho)=&9\frac{\partial P_0(\rho)}{\partial\rho}=\frac{9\pi \rho}{2M_{\rm N}}+9\rho\frac{\d U_0}{\d
\rho},\label{dd_KSNM}
\end{align}
see also Eqs.\,(\ref{4qEsymFF_1}) and (\ref{4qEsym4FF_1}) for the symmetry energies.
All the kinetic parts of the characteristic coefficients beyond the linear terms $L_0(\rho)$ and $L(\rho)$ are zero (including the coefficients $K_0(\rho),J_0(\rho)$ and $K_{\rm{sym}}(\rho)$), i.e., the EOS are totally represented by the effective potentials. Furthermore, the higher-order kinetic symmetry energies including the $E_{\rm{sym,4}}(\rho)$ are also zero (the sixth-order term is given by $E_{\rm{sym},6}^{\rm{kin}}(\rho)=(3d-2)(d-1)(d-2)(2d-1)k_{\rm{F}}^2/180d^5M_{\rm N}$) which vanishes exactly at $d=2$, indicating that the conventional parabolic approximation of the kinetic EOS of ANM is exact in 2D. Since the potential contribution to the EOS is essentially smaller than its kinetic part (especially at low densities), the absence of the kinetic part in the $E_{\rm{sym,4}}(\rho)$ already indicates the parabolic approximation should be good in 2D.

In 2D, the Fermi momentum $k_{\rm{F}}$ scales with the density as $\rho^{1/2}$, which implies that the kinetic symmetry energy is linear in density. Moreover, if the single-nucleon potential $U_0$ depends on momentum $k$ according to $U_0(\rho,k)\approx a+bk+ck^2+\cdots$ and $U_{\rm{sym}}(\rho)=U_{\rm{sym}}(\rho,k_{\rm{F}})\approx a'+b'k_{\rm{F}}+c'k_{\rm{F}}^2$, then $\partial U_0/\partial k\approx b+2ck+\cdots$ (similarly for the symmetry potential). Consequently, the density dependence of $E_0(\rho)$ or the symmetry energy can be expressed as
\begin{equation}
\sim f_1\sqrt{\rho}+f_2\rho+f_3\rho^{3/2}+\cdots,
\end{equation}
indicating that the effective expansion is based on $\rho^{1/2}$, in contrast to the 3D case where $k_{\rm{F}}\sim\rho^{1/3}$ serves as the expansion element.
To illustrate, consider $U_0$ as a function of density $\rho$ alone (i.e., neglecting momentum dependence), in a polynomial form,
\begin{equation}
U_0(\rho)=\sum_{i=1}c_ik_{\rm{F}}^i,~~k_{\rm{F}}=\sqrt{\pi\rho}.
\end{equation}
Then one finds
\begin{align}
&\frac{\d U_0}{\d\rho}=\frac{1}{2\rho}\sum_{i=1}ic_ik_{\rm{F}}^i,~
\frac{1}{\rho}\int_0^{\rho}U_0(f)\d f=2\sum_{i=1}\frac{c_i}{i+2}k_{\rm{F}}^i.
\end{align}
Accordingly, the EOS of SNM can be written as
\begin{equation}
E_0(\rho)=\frac{\pi\rho}{4M_{\rm N}}+2\sum_{i=1}\frac{c_i}{i+2}k_{\rm{F}}^i,
\end{equation}
while the pressure and incompressibility are
\begin{align}
P_0(\rho)=&\rho\left(\frac{3\pi\rho}{4M_{\rm N}}+\sum_{i=1}\frac{i}{i+2}c_ik_{\rm{F}}^i\right),\\
K_0(\rho)
=&9\sum_{i=1}\frac{i}{2}\frac{i-2}{i+2} c_ik_{\rm{F}}^i.
\end{align}

\begin{figure}[h!]
\centering
\includegraphics[width=7.5cm]{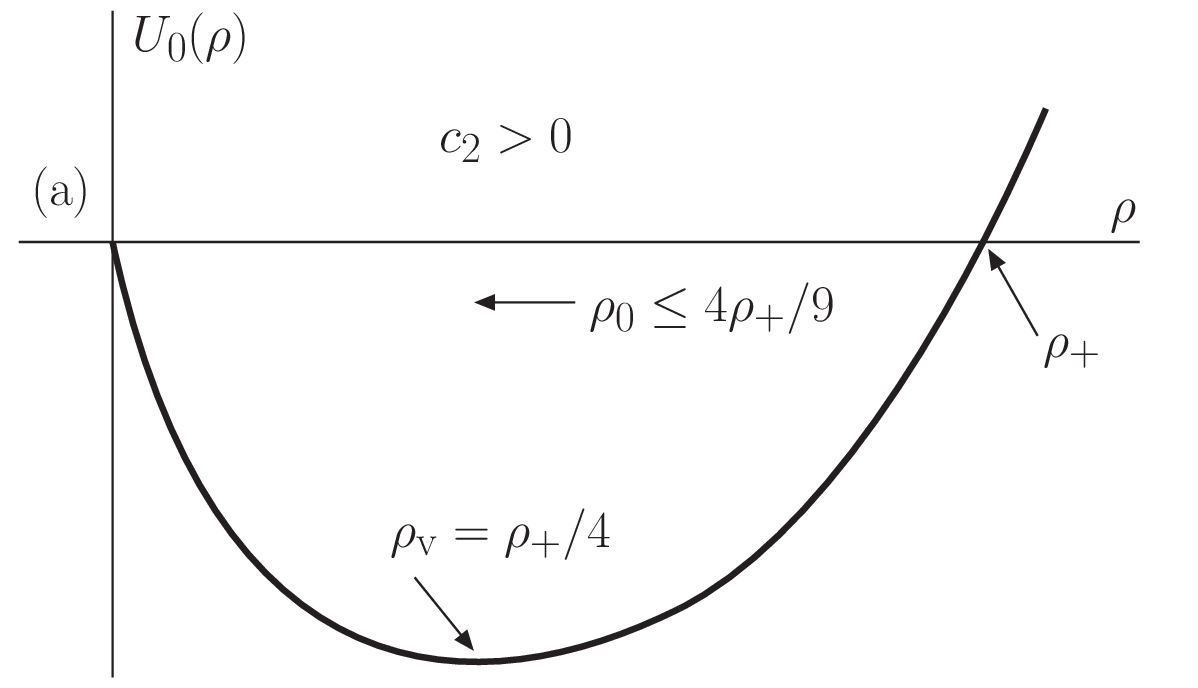}\\
\hspace{-0.1cm}
\includegraphics[width=7.5cm]{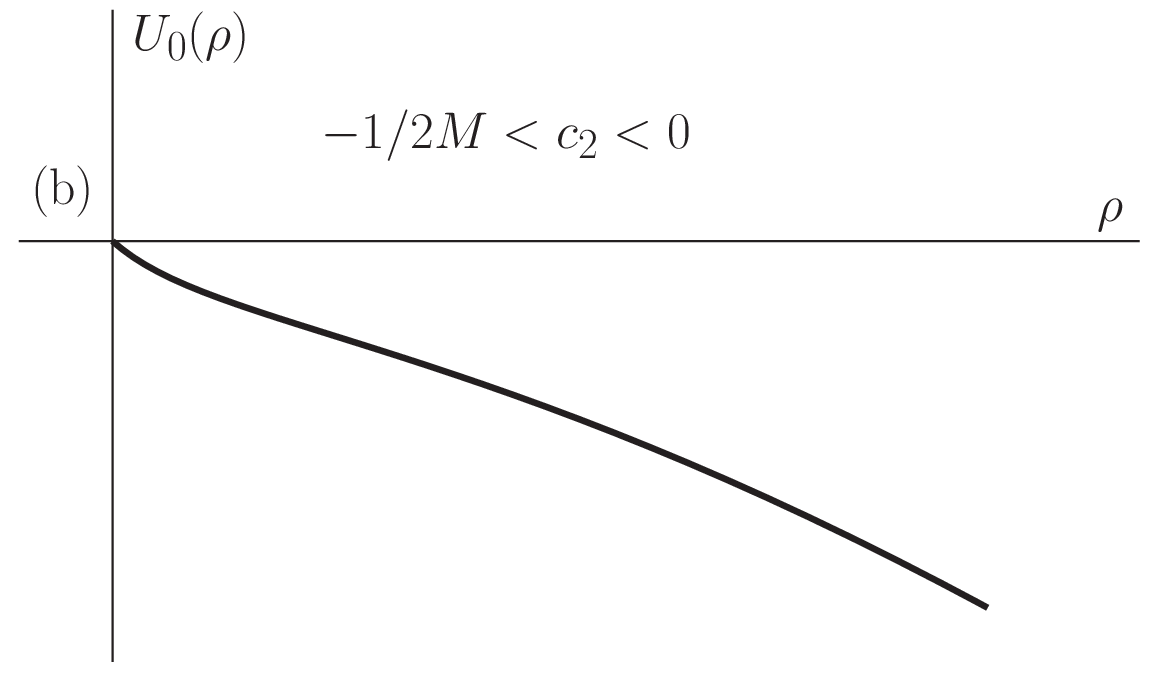}
\caption{Sketch of the single-nucleon potential $U_0(\rho)$ as a function of $\rho$ in 2D, here $M\leftrightarrow M_{\rm N}$.
Figures taken from Ref.\cite{Cai22-dD}.
}\label{fig_U0rho_p}
\end{figure}

If only the first two terms in the density expansion ($c_1$ and $c_2$) are retained, we then have
\begin{align}
E_0(\rho)=&\frac{2}{3}c_1\sqrt{\pi\rho}+\left(\frac{1}{4M_{\rm N}}+\frac{c_2}{2}\right)\pi\rho,\\
P_0(\rho)/\rho
=&\frac{1}{3}c_1\sqrt{\pi\rho}+\left(\frac{3}{4M_{\rm N}}+\frac{c_2}{2}\right)\pi\rho,\\
K_0(\rho)=&-\frac{3}{2}c_1\sqrt{\pi\rho}.
\end{align}
Assuming that the 2D nuclear system behaves qualitatively like the 3D one, i.e., $E_0(\rho)$ is negative at low densities and eventually becomes positive as density increases, we obtain the condition $c_1<0$, and also $1/4M_{\rm N}+c_2/2>0$ and $3/4M_{\rm N}+c_2/2>0$, implying $c_2>-1/2M_{\rm N}$. Notably, the first condition can alternatively be inferred from $K_0(\rho)$, which also predicts a $\sqrt{\rho}$-dependence of $K_0(\rho)$.
Furthermore, if one assumes that the potential $U_0$ eventually turns positive as $\rho$ increases, the coefficient $c_2$ must satisfy $c_2>0$. Under this assumption, three characteristic densities can be defined: the saturation density $\rho_0$ where $P_0(\rho_0)=0$, the density $\rho_{\rm{v}}$ corresponding to the minimum of $U_0$ with $\d U_0/\d\rho|_{\rho=\rho_{\rm{v}}}=0$, and the crossing density $\rho_+$ where $U_0(\rho_+)=0$,
\begin{equation}
\sqrt{\pi\rho_0}=-\frac{c_1}{3}\cdot\frac{1}{3/4M_{\rm N}+c_2/2},~
\sqrt{\pi\rho_{\rm{v}}}=-\frac{c_1}{2c_2},~
\sqrt{\pi\rho_+}=-\frac{c_1}{c_2}.
\end{equation}
Consequently,
\begin{equation}
\frac{\rho_+}{\rho_{\rm{v}}}=4,\quad \frac{\rho_+}{\rho_0}=\frac{9}{4}\left(1+\frac{3}{2M_{\rm N}c_2}\right)^2\geq\frac{9}{4}.
\end{equation}
These relations illustrate that $U_0$ is attractive at low densities, reaches a minimum at $\rho_{\rm{v}}$, and becomes repulsive above $\rho_+$. See FIG.\,\ref{fig_U0rho_p} for an illustration.

Next, we can examine the symmetry energy and fourth-order symmetry energy using Eqs.\,(\ref{dd_Esym_2}) and (\ref{dd_Esym_4}). Suppose the single-nucleon potential has a significant first-order symmetry potential $U_{\rm{sym}}$ while higher-order terms are negligible ($U_{\rm{sym,2}}\approx U_{\rm{sym,3}}\approx\cdots\approx0$), and the momentum-dependence is weak. Then one can approximate
\begin{align}
E_{\rm{sym}}(\rho)\approx&\frac{k_{\rm{F}}^2}{4M_{\rm N}}+\left.\frac{k_{\rm{F}}}{4}\frac{\partial U_0}{\partial|\v{k}|}\right|_{|\v{k}|=k_{\rm{F}}}
+\frac{1}{2}U_{\rm{sym}},\\
E_{\rm{sym,4}}(\rho)\approx&\left.\frac{k_{\rm{F}}}{64}\frac{\partial U_0}{\partial|\v{k}|}\right|_{|\v{k}|=k_{\rm{F}}}.
\end{align}
In 3D, using $M_{\rm{s}}^{\ast}(\rho_0)\gtrsim0.8M_{\rm N}$\cite{LCCX18} and $U_{\rm{sym}}\lesssim37\,\rm{MeV}$\cite{LCCX18,LiXH2013PLB,LiXH2015PLB}, the ratio
\begin{align}
\frac{E_{\rm{sym,4}}(\rho_0)}{E_{\rm{sym}}(\rho_0)}
\approx&\left.\frac{k_{\rm{F}}^2}{64}\left(\frac{1}{M_{\rm{s}}^{\ast}}-\frac{1}{M_{\rm N}}\right)\right/\left(\frac{k_{\rm{F}}^2}{4M_{\rm N}}+\frac{1}{2}U_{\rm{sym}}\right)\notag\\
\lesssim&\frac{1}{32}\left(\frac{M_{\rm N}}{M_{\rm{s}}^{\ast}}-1\right)\lesssim\frac{1}{128},
\end{align}
indicating that the parabolic approximation remains reasonable in 2D.
By comparison, in 1D, the symmetry energy and fourth-order symmetry energy are given as\cite{Cai22-dD}
\begin{align}
E_{\rm{sym}}(\rho)
=&\frac{k_{\rm{F}}^2}{2M_{\rm N}}+\left.\frac{k_{\rm{F}}^2}{2}\frac{\partial U_0}{\partial |\v{k}|}\right|_{|\v{k}|=k_{\rm{F}}}+\frac{1}{2}U_{\rm{sym}},\\
E_{\rm{sym,4}}(\rho)
=&\left.\frac{k_{\rm{F}}^3}{24}\frac{\partial^3U_0}{\partial |\v{k}|^3}\right|_{|\v{k}|=k_{\rm{F}}}+\left.\frac{k_{\rm{F}}^2}{8}\frac{\partial^2 U_{\rm{sym}}}{\partial |\v{k}|^2}\right|_{|\v{k}|=k_{\rm{F}}}
\notag\\&+\left.\frac{k_{\rm{F}}}{4}\frac{\partial U_{\rm{sym,2}}}{\partial |\v{k}|}\right|_{|\v{k}|=k_{\rm{F}}}+\frac{1}{4}U_{\rm{sym,3}},
\end{align}
with $k_{\rm{F}}=\pi\rho/4$, using the general formulas:
\begin{align}
E_{\rm{sym}}(\rho)=&\frac{1}{2d}\frac{k_{\rm{F}}^2}{M_{\rm N}}+\left.\frac{k_{\rm{F}}}{2d}\frac{\partial
U_0}{\partial
|\v{k}|}\right|_{|\v{k}|=k_{\rm{F}}}+\frac{1}{2}U_{\rm{sym}}(\rho,k_{\rm{F}})\label{4qEsymFF_1},\\
E_{\rm{sym},4}(\rho)=&\frac{1}{24d}\left(\frac{2}{d}-1\right)
\left(\frac{2}{d}-2\right)\frac{k_{\rm{F}}^2}{M_{\rm N}}
\notag\\&+\Bigg[\frac{k_{\rm{F}}^3}{24d^3}\frac{\partial^3U_{0}}{\partial|\v{k}|^3}
+\frac{k_{\rm{F}}^2}{8d^2}\left(\frac{1}{d}-1\right)\frac{\partial^2U_{0}}{\partial|\v{k}|^2}
\notag\\&\hspace{1cm}+\frac{k_{\rm{F}}}{24d}\left(\frac{1}{d}-1\right)
\left(\frac{1}{d}-2\right)\frac{\partial
U_{0}}{\partial|\v{k}|}\Bigg]_{|\v{k}|=k_{\rm{F}}}\notag\\
&+\left[\frac{k_{\rm{F}}^2}{8d^2}\frac{\partial^2U_{\rm{sym}}}{\partial|\v{k}|^2}
-\frac{k_{\rm{F}}}{8d}\left(\frac{1}{d}-1\right)\frac{\partial
U_{\rm{sym}}}{\partial|\v{k}|}\right]_{|\v{k}|=k_{\rm{F}}}
\notag\\&+\left.\frac{k_{\rm{F}}}{4d}\frac{\partial
U_{\rm{sym},2}}{\partial|\v{k}|}\right|_{|\v{k}|=k_{\rm{F}}}
+\frac{1}{4}U_{\rm{sym},3}(\rho,k_{\rm{F}})\label{4qEsym4FF_1}.
\end{align}
Under the same assumption of weak higher-order terms, the fourth-order symmetry energy can be approximated as zero, so the parabolic approximation works even better in 1D.

Finally, in the large-$d$ limit, due to small pre-factors like $1/24d,1/8d^2,1/24d^3$\cite{Cai22-dD}, etc., the fourth-order symmetry energy is expected to remain small. If the first-order symmetry potential dominates over the third-order term ($2^{-1}U_{\rm{sym}}>4^{-1}U_{\rm{sym,3}}$), then ${E_{\rm{sym,4}}(\rho_0)}/{E_{\rm{sym}}(\rho_0)}$ remains small, confirming the validity of the parabolic approximation. In the infinite-$d$ limit, one finds ${E_{\rm{sym}}(\rho_0)}/{E_{\rm{sym,4}}(\rho_0)}=U_{\rm{sym,3}}/2U_{\rm{sym}}\ll1$. Therefore, writing the EOS of ANM in a general dimension $d$ provides a clear understanding of the relative importance of different expansion terms and explains why the conventional parabolic approximation in 3D is not coincidental\cite{Cai22-dD}.
See similar discussions of Subsection \ref{sub_RFFG}.

{\bfseries Example 2.} We perturb the EOS of ANM in $d=3+\epsilon$ with $\epsilon$ being the perturbative dimension, we need to expand the relevant quantities assuming $\epsilon$ is small.
The expansion of the function $a(d)=[2^{d-2}\pi^{d/2}\Gamma({d}/{2}+1)]^{1/d}$ around $\epsilon=0$ is 
\begin{align}
a(3+\epsilon)\approx({3\pi^2}/{2})^{1/3}\left(1+\sigma
\epsilon\right),\end{align} where,
\begin{equation}\label{defff_sig}
\sigma=\frac{4}{9}-\frac{\gamma_{\rm{E}}}{6}+\frac{1}{18}\ln\left(\frac{4}{9\pi}\right)
\approx0.2396,\end{equation}
where $\gamma_{\rm{E}}\approx0.5773$ is Euler's constant.
The Fermi momentum $k_{\rm{F}}$ in $3+\epsilon$ dimensions to linear order of $\epsilon$ is $
k_{\rm{F}}\approx \overline{k}_{\rm{F}}(1+\sigma
\epsilon)$, where $
\overline{k}_{\rm{F}}=({3\pi^2\rho}/{2})^{1/3}$ is the conventional 3D Fermi momentum (indicated by the ``$\overline{~~}$'' over the quantity).
The final expressions for relevant quantities are\cite{Cai22-dD}

\begin{strip}
\begin{align}
{E}_0(\rho)=&\frac{3\overline{k}_{\rm{F}}^2}{10M_{\rm N}}+
\frac{1}{\rho}\int_0^\rho
U_0\left(f,\overline{k}_{\rm{F}}^f\right)\d
f+\epsilon
\left[\frac{3\overline{k}_{\rm{F}}^2}{5M_{\rm N}}\left(\sigma+\frac{1}{15}\right)+\frac{\sigma}{\rho}\int_0^{\rho}\left(
\left.\frac{\partial
U_0}{\partial|\v{k}|}\right|_{|\v{k}|=\overline{k}^f_{\rm{F}}}\cdot\overline{k}^f_{\rm{F}}\right)\d
f\right],\label{ddef_E0}\\
P_0(\rho)=&\frac{\overline{k}_{\rm{F}}^2\rho}{5M_{\rm N}}+
\int_0^\rho U_0\left(f,\overline{k}_{\rm{F}}^f\right)\d
f+\rho U_0(\rho,\overline{k}_{\rm{F}})\notag\\
&+\epsilon\left[\frac{\rho\overline{k}_{\rm{F}}^2}{10M_{\rm N}}\left(2\sigma-\frac{1}{5}\right)
-\sigma\int_0^{\rho}\left( \left.\frac{\partial
U_0}{\partial|\v{k}|}\right|_{|\v{k}|=\overline{k}^f_{\rm{F}}}\cdot\overline{k}^f_{\rm{F}}\right)\d
f+\rho\sigma\left.\frac{\partial U_0}{\partial
|\v{k}|}\right|_{|\v{k}|=\overline{k}_{\rm{F}}}\cdot\overline{k}_{\rm{F}}\right],\label{ddef_P0}\\
K_0(\rho)=&-\frac{3\overline{k}_{\rm{F}}^2}{5M_{\rm N}}
+9\rho\left(\frac{\partial U_0}{\partial
\rho}+\left.\frac{\overline{k}_{\rm{F}}}{3\rho}\frac{\partial
U_0}{\partial|\v{k}|}\right|_{|\v{k}|=\overline{k}_{\rm{F}}}\right)
+\frac{18}{\rho}\int_0^\rho
U_0\left(f,\overline{k}_{\rm{F}}^f\right)\d
f-18U_0(\rho,\overline{k}_{\rm{F}})\notag\\
&+\epsilon\left[3\sigma\frac{\partial^2
U_0}{\partial|\v{k}|^2}
\cdot\overline{k}_{\rm{F}}^2-\frac{3\overline{k}_{\rm{F}}^2}{5M_{\rm N}}\left(\frac{7}{15}+2\sigma\right)+\frac{18\sigma}{\rho}\int_0^{\rho}\left(
\left.\frac{\partial
U_0}{\partial|\v{k}|}\right|_{|\v{k}|=\overline{k}^f_{\rm{F}}}\cdot\overline{k}^f_{\rm{F}}\right)\d
f-15\sigma\frac{\partial
U_0}{\partial|\v{k}|}\cdot\overline{k}_{\rm{F}}
\right]_{|\v{k}|=\overline{k}_{\rm{F}}},\label{ddef_K0}\\
E_{\rm{sym}}(\rho)=&\frac{\overline{k}_{\rm{F}}^2}{6M_{\rm N}}+\left.\frac{\overline{k}_{\rm{F}}}{6}
\frac{\partial U_0}{\partial
|\v{k}|}\right|_{|\v{k}|=\overline{k}_{\rm{F}}}+\frac{1}{2}U_{\rm{sym}}(\rho,\overline{k}_{\rm{F}})\notag\\
&+\epsilon\left[\frac{\overline{k}_{\rm{F}}^2}{6M_{\rm N}}\left(2\sigma-\frac{1}{3}\right)
+\frac{\overline{k}_{\rm{F}}}{6}\left({\displaystyle\sigma\overline{k}_{\rm{F}}\frac{\partial^2
U_0}{\partial
|\v{k}|^2} }+\left(\sigma-\frac{1}{3}\right)\frac{\partial
U_0}{\partial
|\v{k}|}\right)
+\frac{\sigma}{2}\frac{\partial U_{\rm{sym}}}{\partial
|\v{k}|}\cdot\overline{k}_{\rm{F}}\right]_{|\v{k}|=\overline{k}_{\rm{F}}},\label{ddef_Esym}\\
L(\rho)=&\frac{\overline{k}_{\rm{F}}^2}{3M_{\rm N}}+\left(\frac{\overline{k}_{\rm{F}}^2}{6}\frac{\partial^2
U_0}{\partial|\v{k}|^2}+\frac{\overline{k}_{\rm{F}}}{6}\frac{\partial
U_0}{\partial|\v{k}|}\right)_{|\v{k}|=\overline{k}_{\rm{F}}}+\left.\overline{k}_{\rm{F}}\frac{\partial
U_{\rm{sym}}}{\partial|\v{k}|}\right|_{|\v{k}|=\overline{k}_{\rm{F}}}
+\frac{3}{2}U_{\rm{sym}}(\rho,\overline{k}_{\rm{F}})
+3U_{\rm{sym},2}(\rho,\overline{k}_{\rm{F}})\notag\\
&+\epsilon\left[\left(2\sigma-\frac{2}{3}\right)\left(\frac{\overline{k}_{\rm{F}}^2}{3M_{\rm N}}+
\frac{\overline{k}_{\rm{F}}^2}{6}\frac{\partial^2
U_0}{\partial|\v{k}|^2}\right)+\frac{\sigma}{6}\frac{\partial^3
U_0}{\partial|\v{k}|^3}\cdot\overline{k}_{\rm{F}}^3+
\frac{\sigma}{6}\frac{\partial^2
U_0}{\partial|\v{k}|^2}\cdot\overline{k}_{\rm{F}}^2+
\frac{1}{6}\left(\sigma-\frac{2}{3}\right)
\frac{\partial
U_0}{\partial|\v{k}|}\cdot \overline{k}_{\rm{F}}\right.
\notag\\
&\hspace*{1.5cm}\left.+\left(\frac{5}{2}\sigma-\frac{1}{3}\right)\frac{\partial
U_{\rm{sym}}}{\partial|\v{k}|}\cdot\overline{k}_{\rm{F}}+
\sigma\frac{\partial^2
U_{\rm{sym}}}{\partial|\v{k}|^2}\cdot\overline{k}_{\rm{F}}^2
+
3\sigma\frac{\partial
U_{\rm{sym,2}}}{\partial|\v{k}|}\cdot\overline{k}_{\rm{F}}\right]_{|\v{k}|=\overline{k}_{\rm{F}}}.\label{ddef_L}
\end{align}
\end{strip}

These expressions are physically intuitive and useful since all the quantities involved are those in 3D, where they are well known and most of them are tightly constrained either theoretically or experimentally. Let us now turn to the situation in which a nonzero $\epsilon$ is applied and examine several main features. In addition to the kinetic contribution, a non-trivial part of the EOS of SNM arises from integrating over the momentum dependence of the potential $U_0$, i.e., $(\sigma/\rho)\int_0^{\rho}[\overline{k}_{\rm{F}}^f\cdot\partial U_0/\partial|\v{k}|]_{|\v{k}|=\overline{k}^f_{\rm{F}}}\d f$. From nuclear optical model analyses of nucleon-nucleus scattering data, one knows that the nucleon isoscalar potential $U_0$ increases with momentum\cite{LiXH2013PLB,LiXH2015PLB,Hama1990}; thus the above integral is positive since $\sigma>0$. Consequently, the linear $\epsilon$-term contributes a positive correction to the unperturbed EOS of SNM, as shown in Eq.\,(\ref{ddef_E0}). Extrapolating the 3D EOS of SNM to lower (higher) dimensions therefore reduces (enhances) its magnitude, with 2D SNM being more strongly bound than its 3D counterpart.
For the symmetry energy, we first neglect the momentum dependence of the symmetry potential and the second derivative $\partial^2U_0/\partial|\v{k}|^2$. Under these simplifications, the $\epsilon$-term of Eq.\,(\ref{ddef_Esym}) becomes
\begin{align}\label{ckj-1}
&\epsilon\left[\frac{\overline{k}_{\rm{F}}^2}{6M_{\rm N}}\left(2\sigma-\frac{1}{3}\right)
+\frac{\overline{k}_{\rm{F}}}{6}\frac{\partial
U_0}{\partial
|\v{k}|}\left(\sigma-\frac{1}{3}\right)_{|\v{k}|=\overline{k}_{\rm{F}}}
\right]\notag\\
=&\frac{\overline{k}_{\rm{F}}^2}{6M_{\rm N}}\left[\frac{M_{\rm N}}{M_{\rm{s}}^{\ast}}\left(\sigma-\frac{1}{3}\right)+\sigma\right]\epsilon\approx
\frac{\overline{k}_{\rm{F}}^2}{6M_{\rm N}}\left(0.24-0.09\frac{M_{\rm N}}{M_{\rm{s}}^{\ast}}\right)\epsilon,
\end{align}
where $M_{\rm{s}}^{\ast}\gtrsim0.8M_{\rm N}$\cite{LCCX18} is the scalar Landau effective mass. Since the bracket is very likely positive, the linear $\epsilon$-correction becomes negative for $\epsilon<0$, implying that the symmetry energy in 2D is reduced (similar to the EOS of SNM). However, because the symmetry potential $U_{\rm{sym}}$ still carries large uncertainties, especially regarding its momentum dependence\cite{LiXH2013PLB,LiXH2015PLB,LCCX18}, the final implications of Eq.\,(\ref{ckj-1}) and of the term $(\sigma/2)[\overline{k}_{\rm{F}}\cdot\partial U_{\rm{sym}}/\partial|\v{k}|]_{|\v{k}|=\overline{k}_{\rm{F}}}$ remain inconclusive and require further analysis.
To illustrate the sensitivity, we rewrite the linear $\epsilon$-term in Eq.\,(\ref{ddef_Esym}) as
\begin{equation}\label{edf}
\Pi=\sigma\left(\frac{\overline{k}_{\rm{F}}^2}{3M_{\rm N}}+\frac{\overline{k}_{\rm{F}}}{6}\frac{\partial U_0}{\partial|\v{k}|}
+\frac{\overline{k}_{\rm{F}}^2}{6}\frac{\partial^2U_0}{\partial|\v{k}|^2}
+\frac{\overline{k}_{\rm{F}}}{2}\frac{\partial U_{\rm{sym}}}{\partial|\v{k}|}
\right)_{|\v{k}|=\overline{k}_{\rm{F}}}-\frac{\overline{k}_{\rm{F}}^2}{18M^{\ast}_{\rm{s}}},
\end{equation}
and adopt the optical-model fitted values in Ref.\cite{LiXH2013PLB}. These give ${\overline{k}_{\rm{F}}^2}/{3M_{\rm N}}+6^{-1}[{\overline{k}_{\rm{F}}}\cdot{\partial U_0}/{\partial|\v{k}|}]_{|\v{k}|=\overline{k}_{\rm{F}}}\approx37.7\,\rm{MeV}$, $6^{-1}[{\overline{k}_{\rm{F}}^2}\cdot{\partial^2U_0}/{\partial|\v{k}|^2}]_{|\v{k}|=\overline{k}_{\rm{F}}}\approx-2.3\,\rm{MeV}$, and $[\overline{k}_{\rm{F}}\cdot\partial U_{\rm{sym}}/\partial|\v{k}|]_{|\v{k}|=\overline{k}_{\rm{F}}}\approx-46.0\,\rm{MeV}$. As a result, $\Pi\approx(12.4\sigma-5.1)\epsilon\,\rm{MeV}\approx-2.1\epsilon\,\rm{MeV}$, suggesting that the symmetry energy becomes enhanced when the dimension is perturbed downward. Nevertheless, even a small change in $[\overline{k}_{\rm{F}}\cdot\partial U_{\rm{sym}}/\partial|\v{k}|]_{|\v{k}|=\overline{k}_{\rm{F}}}$ could flip the sign of $\Pi$. More accurate constraints on $U_{\rm{sym}}$ and/or higher-order $\epsilon$-calculations are thus crucial.

The perturbative expression Eq.\,(\ref{ddef_Esym}) can be rewritten as
\begin{align}
E_{\rm{sym}}(\rho)=&\frac{\overline{k}_{\rm{F}}^2}{6M_{\rm N}}\left[1+\epsilon\left(2\sigma-\frac{1}{3}\right)\right]\notag\\
&+\left.\frac{\overline{k}_{\rm{F}}}{6}
\left[1+\epsilon\left(2\sigma-\frac{1}{3}\right)\right]\frac{\partial U_0}{\partial
|\v{k}|}\right|_{|\v{k}|=\overline{k}_{\rm{F}}}
+\frac{1}{2}U_{\rm{sym}}(\rho,\overline{k}_{\rm{F}})\notag\\&+\epsilon\sigma\left(
\frac{\overline{k}_{\rm{F}}^2}{6}\frac{\partial^2U_0}{\partial|\v{k}|^2}
+\frac{\overline{k}_{\rm{F}}}{2}\frac{\partial U_{\rm{sym}}}{\partial|\v{k}|}-\frac{\overline{k}_{\rm{F}}}{6}\frac{\partial U_0}{\partial|\v{k}|}\right)_{|\v{k}|=\overline{k}_{\rm{F}}}.
\end{align}
To streamline the structure, we introduce the effective mass $M_{\rm{eff}}$ (relative to the bare mass), the effective isoscalar potential $U_0^{\rm{eff}}$, and the effective symmetry potential $U_{\rm{sym}}^{\rm{eff}}$:
\begin{align}
M_{\rm{eff}}=&M_{\rm N}\left[1+\epsilon\left(2\sigma-\frac{1}{3}\right)\right]^{-1},\\
U_0^{\rm{eff}}(\rho,|\v{k}|)=&\left[1+\epsilon\left(2\sigma-\frac{1}{3}\right)\right] U_0,\\
U_{\rm{sym}}^{\rm{eff}}(\rho,|\v{k}|)=&U_{\rm{sym}}(\rho,|\v{k}|)\notag\\
&\hspace{-1.5cm}+2\epsilon\sigma\left(
\frac{\overline{k}_{\rm{F}}^2}{6}\frac{\partial^2U_0}{\partial|\v{k}|^2}+\frac{\overline{k}_{\rm{F}}}{2}\frac{\partial U_{\rm{sym}}}{\partial|\v{k}|}
-\frac{\overline{k}_{\rm{F}}}{6}\frac{\partial U_0}{\partial|\v{k}|}\right)_{|\v{k}|=\overline{k}_{\rm{F}}}.
\end{align}
These quantities all depend on the perturbative parameter $\epsilon$. With these definitions, the symmetry energy becomes
\begin{empheq}[box=\fbox]{align}\label{ddef_Esym_eff}
E_{\rm{sym}}(\rho)=&\frac{\overline{k}_{\rm{F}}^2}{6M_{\rm{eff}}}
\left.+\frac{\overline{k}_{\rm{F}}}{6}\frac{\partial U_0^{\rm{eff}}}{\partial |\v{k}|}\right|_{|\v{k}|=\overline{k}_{\rm{F}}}+\frac{1}{2}U^{\rm{eff}}_{\rm{sym}}(\rho,\overline{k}_{\rm{F}})\notag\\
=&\frac{\overline{k}_{\rm{F}}^2}{6M_{\rm{s},\rm{eff}}^{\ast}}+\frac{1}{2}U^{\rm{eff}}_{\rm{sym}}(\rho,\overline{k}_{\rm{F}}),
\end{empheq}
where $M_{\rm{s},\rm{eff}}^{\ast}/M_{\rm{eff}}=[1+(M_{\rm{eff}}/|\v{k}|)\partial U_0^{\rm{eff}}/\partial|\v{k}|]^{-1}_{|\v{k}|=k_{\rm{F}}}$ is the scalar Landau effective mass based on $M_{\rm{eff}}$. Remarkably, (\ref{ddef_Esym_eff}) preserves the same functional form as the unperturbed symmetry energy in 3D. For the special case of 2D ($\epsilon=-1$), we have $M_{\rm{eff}}>M_{\rm N}$ and similarly $0<\partial U_0^{\rm{eff}}/\partial|\v{k}|<\partial U_0/\partial|\v{k}|$, since $-1<\epsilon(2\sigma-1/3)<0$. Thus Eq.\,(\ref{ddef_Esym_eff}) indicates that, in 2D, the nucleon behaves as if it moves with a heavier mass in a less momentum-dependent potential compared with 3D. The effective form Eq.\,(\ref{ddef_Esym_eff}) therefore offers a convenient starting point for studying $E_{\rm{sym}}(\rho)$ in reduced dimensions.

To explore the performance of the $\epsilon$-expansion, we next compare it with exact $d$-dimensional expressions for the kinetic contributions. A natural issue is how to define the Fermi momentum in $d$D. We adopt two schemes: (a) The Fermi momentum is taken to be the same in all dimensions (e.g., $k_{\rm{F}}\approx200$-300\,MeV), eliminating the need for perturbations in $k_{\rm{F}}$; the factor $\sigma$ is then unnecessary; (b) Alternatively, we fix the 3D density (as in the previous perturbative treatment) and relate $\rho_d$ in $d$D to it by $\rho_d=\rho^{d/3}$, so that $k_{\rm{F}}\sim\rho^{1/3}$ holds in any dimension. In this scheme, $\sigma$ must be retained.

For the total EOS, the kinetic trend under dimensional perturbation is relatively transparent. Let us extrapolate the kinetic EOS of SNM and the kinetic symmetry energy to other dimensions and compare with the exact results. For $\epsilon=-1$ (2D), we obtain
\begin{align}
E_0^{\rm{kin}}(\rho)\approx&\frac{3\overline{k}_{\rm{F}}^2}{10M_{\rm N}}\left(1-2\sigma-\frac{2}{15}\right)=\frac{3\overline{k}_{\rm{F}}^2}{10M_{\rm N}}\left(\frac{13}{15}-2\sigma\right).
\end{align}
If $\overline{k}_{\rm{F}}$ is assumed equal in all dimensions and hence $\sigma$ is set to zero, the ratio of this approximation to the exact 2D kinetic EOS $k_{\rm{F}}^2/4M_{\rm N}$ becomes $26/25\approx1.04$, indicating that the perturbation from $d=3$ to $d=2$ is reasonably accurate. Similar extrapolation to $d=4$ ($\epsilon=1$) and $d=1$ ($\epsilon=-2$) yields ratios $51/50\approx1.02$ and $33/25\approx1.32$, respectively, confirming that the approximation worsens as $|\epsilon|$ increases.
For the kinetic symmetry energy, we have as $\sigma\to0$,
\begin{equation}
E_{\rm{sym}}^{\rm{kin}}(\rho)=\frac{\overline{k}_{\rm{F}}^2}{6M_{\rm N}}
\left[1+\left(2\sigma-\frac{1}{3}\right)\epsilon\right]
\to\frac{\overline{k}_{\rm{F}}^2}{6M_{\rm N}}\left(1-\frac{\epsilon}{3}\right).
\end{equation}
Thus for 2D this gives $2\overline{k}_{\rm{F}}^2/9M_{\rm N}$, close to the exact $k_{\rm{F}}^2/4M_{\rm N}$, with ratio $8/9\approx0.89$. For 4D ($\epsilon=1$) and 1D ($\epsilon=-2$), the ratios using (\ref{4qEsymFF_1}) are again $8/9\approx0.89$ and $5/9\approx0.56$, confirming the method's effectiveness for the kinetic part.

If instead the Fermi momenta differ across dimensions according to $\rho_d=\rho^{d/3}$, then $\sigma$ must remain. For 2D, where $k_{\rm{F}}=(\pi \rho_2)^{1/2}$, the ratio becomes
\begin{equation}\label{ia-1}
\left.\frac{3\overline{k}_{\rm{F}}^2}{10M_{\rm N}}\left(\frac{13}{15}-2\sigma\right)\right/\frac{k_{\rm{F}}^2}{4M_{\rm N}}
=\frac{6}{5}\pi^{1/3}\left(\frac{3}{2}\right)^{2/3}\left(\frac{13}{15}-2\sigma\right)\approx0.89,
\end{equation}
where the perturbative 3D kinetic EOS gives about 8.56\,MeV and the exact 2D value is about 9.60\,MeV. For 4D ($\epsilon=1$) and 1D ($\epsilon=-2$), the corresponding kinetic EOS values lead to ratios $\approx0.98$ and $-3.96$, respectively. While the extrapolation to 1D is clearly unreasonable, the absolute magnitudes are still comparable.
For the kinetic symmetry energy, the ratio becomes
\begin{equation}\label{ia-2}
\pi^{1/3}\left(\frac{3}{2}\right)^{5/2}\left(\frac{4}{3}-2\sigma\right)
\approx1.09,
\end{equation}
corresponding to 10.49\,MeV from the perturbative expression versus 9.60\,MeV from the exact 2D value. Again, the $\epsilon$-expansion performs well. For reference, the 4D and 1D cases give ratios 1.04 and 2.31, respectively.

We emphasize that all perturbative calculations of the EOS of ANM presented above are performed in conventional 3D space. However, there is no fundamental reason why the reference dimension must be fixed at $3$. Indeed, several examples in many-body and quantum field theories show that the reference dimension can naturally differ from $3$\cite{Nishida2006,Wilson1974}. Motivated by this, we generalize the correction factor $\sigma$ to an arbitrary reference dimension $d_{\rm f}$ and examine how the corresponding $\epsilon$-expansion modifies the symmetry energy in $d_{\rm f}$ dimensions.
The generalized form of $\sigma$ reads
\begin{align}\label{def_sigf}
\sigma(d_{\rm{f}})=&\frac{1}{2d_{\rm{f}}^2}\left(d_{\rm{f}}\Psi\left(\frac{d_{\rm{f}}}{2}\right)
+2(d_{\rm{f}}+2)\ln 2 + d_{\rm{f}}\ln\pi + 2\right)\notag\\
&- \frac{1}{d_{\rm{f}}^2}\ln\left( 2^{d_{\rm{f}}} \pi^{d_{\rm{f}}/2} \Gamma\left(\frac{d_{\rm{f}}}{2}+1\right) \right),
\end{align}
where $\Psi(x)\equiv \d\ln\Gamma(x)/\d x$ is the digamma function. The conventional 3D value is recovered as $\sigma(3)\equiv\sigma\approx0.2396$, see Eq.\,(\ref{defff_sig}).
As an illustration, the symmetry energy to linear $\epsilon$-order, neglecting the second derivative of $U_0$ with respect to $|\mathbf{k}|$, becomes then:
\begin{align}
&E_{\rm{sym}}(\rho)
\approx\frac{k_{\rm{F}}^2}{2 d_{\rm f} M_{\rm N}}
+ \left.\frac{k_{\rm{F}}}{2 d_{\rm f}} \frac{\partial U_0}{\partial |\mathbf{k}|}\right|_{|\mathbf{k}|=k_{\rm F}}
+ \frac{1}{2} U_{\rm{sym}}(\rho,k_{\rm{F}}) \notag\\
&+ \epsilon \Bigg[
\left(\sigma_{\rm f}-\frac{1}{2 d_{\rm f}}\right) \frac{k_{\rm F}^2}{d_{\rm f} M_{\rm s}^{\ast}}
+ \frac{k_{\rm F}}{2} \sigma_{\rm f} 
\left( \frac{\partial U_{\rm{sym}}}{\partial |\mathbf{k}|} - \frac{1}{d_{\rm f}} \frac{\partial U_0}{\partial |\mathbf{k}|} \right)_{|\mathbf{k}|=k_{\rm F}}
\Bigg],
\end{align}
where $\sigma_{\rm f} \equiv \sigma(d_{\rm f})$ and the second line is a direct generalization of Eq.\,(\ref{edf}).  
Assuming the momentum dependence of the single-particle potentials remains qualitatively the same as in 3D, i.e.,
$\partial U_{\rm{sym}}/\partial |\mathbf{k}|<0$ and $\partial U_0/\partial |\mathbf{k}|>0$, the first $\epsilon$-order term $
(\sigma_{\rm f}-1/{2 d_{\rm f}}){k_{\rm F}^2}/{d_{\rm f} M_{\rm s}^{\ast}}$
decreases (but remains positive) as $d_{\rm f}$ increases. The second term eventually saturates to $
({k_{\rm F}}/{2})\sigma_{\rm f} {\partial U_{\rm{sym}}}/{\partial |\mathbf{k}|}$,
so that in the large-$d_{\rm f}$ limit the density dependence of the perturbed symmetry energy is dominated by the momentum dependence of the symmetry potential. Given $\partial U_{\rm{sym}}/\partial |\mathbf{k}|<0$, an upward perturbation ($\epsilon>0$) tends to reduce $E_{\rm{sym}}$ relative to its $d_{\rm f}$-dimensional value.
For the special case $d_{\rm f}=1$, the symmetry energy becomes
\begin{align}
&E_{\rm{sym}}(\rho)
\approx \frac{k_{\rm F}^2}{2 M_{\rm N}} 
+ \left.\frac{k_{\rm F}}{2} \frac{\partial U_0}{\partial |\mathbf{k}|}\right|_{|\mathbf{k}|=k_{\rm F}}
+ \frac{1}{2} U_{\rm{sym}}(\rho,k_{\rm F}) \notag\\
&+ \epsilon \left[
\frac{k_{\rm F}^2}{2 M_{\rm s}^{\ast}} (\sigma_1 - 1)
+ \sigma_1 \frac{k_{\rm F}}{2} \left( \frac{k_{\rm F}}{M_{\rm N}} + \frac{\partial U_{\rm{sym}}}{\partial |\mathbf{k}|} \right)_{|\mathbf{k}|=k_{\rm F}}
\right].
\end{align}
Since $\sigma_1>1$ and the combination $
[{k_{\rm F}}/{M_{\rm N}} + {\partial U_{\rm{sym}}}/{\partial |\mathbf{k}|}]_{|\mathbf{k}|=k_{\rm F}} > 0$
as in 3D\cite{LiXH2013PLB}, both $\epsilon$-order contributions are positive. Consequently, an upward perturbation ($\epsilon>0$) enhances $E_{\rm{sym}}(\rho)$.

These results demonstrate that dimensionality can qualitatively change the conclusions regarding the EOS of ANM. Furthermore, the relatively smaller values of $\sigma_{\rm f}$ for $d_{\rm f}\approx 2$ or $3$ compared to $\sigma_1 \approx 1.5253$ indicate that $\epsilon$-expansion around $d_{\rm f}=1$ may be challenging in certain applications.
Although the linear-order $\epsilon$-expansion based on 3D provides reasonably accurate estimates for both the kinetic EOS of SNM and the kinetic symmetry energy compared with exact results, further studies are needed to consolidate its validity. It will be important to extend the expansion to higher orders (since the breakdown at $\epsilon=-2$ from 3D base already signals the limits of the linear approximation) and to incorporate effective NN interactions. 

Specifically, by perturbing the symmetry energy in 3D to second order in $\epsilon = d - 3$, a new term at order $\epsilon^2$ appears:
\begin{align}
&\epsilon^2
\Bigg[
\frac{\sigma^2\overline{k}_{\rm{F}}^2}{12M_{\rm N}}\left(
\overline{k}_{\rm{F}}M_{\rm N}\frac{\partial^3U_0}{\partial|\v{k}|^3}+2M_{\rm N}\frac{\partial^2U_0}{\partial|\v{k}|^2}+3M_{\rm N}\frac{\partial^2U_{\rm{sym}}}{\partial|\v{k}|^2}+2\right)
\notag\\
&-\frac{\sigma\overline{k}_{\rm{F}}^2}{18M_{\rm N}}\left(M_{\rm N}
\frac{\partial^2U_0}{\partial|\v{k}|^2}+\frac{M_{\rm N}}{\overline{k}_{\rm{F}}}\frac{\partial U_0}{\partial|\v{k}|}+2\right)+\frac{\overline{k}_{\rm{F}}^2}{54M_{\rm N}}\left(\frac{M_{\rm N}}{\overline{k}_{\rm{F}}}\frac{\partial U_0}{\partial|\v{k}|}+1\right)
\notag\\
&
+\frac{\sigma'\overline{k}_{\rm{F}}^2}{6M_{\rm N}}\left(
M_{\rm N}\frac{\partial^2U_0}{\partial|\v{k}|^2}+\frac{M_{\rm N}}{\overline{k}_{\rm{F}}}\frac{\partial U_0}{\partial |\v{k}|}+\frac{3M_{\rm N}}{\overline{k}_{\rm{F}}}
\frac{\partial U_{\rm{sym}}}{\partial |\v{k}|}+2\right)
\Bigg]_{|\v{k}|=\overline{k}_{\rm{F}}},
\label{ddef_Esym-o2}
\end{align}
this generalizes Eq.\,(\ref{ddef_Esym}), here $\sigma' \approx -0.0307$ characterizes the second-order correction of the Fermi momentum via 
$k_{\rm F} \approx \overline{k}_{\rm F} ( 1 + \sigma \epsilon + \sigma' \epsilon^2 )$.  
TAB.\,\ref{Tab_2DEsym-o2} lists the contributions from different models to the symmetry energy at order $\epsilon^2$\cite{Cai22-dD}.
In the ImMDI model, for instance, if the $y$ parameter is chosen as $y \approx -115$\,MeV\cite{XuJ15}, all $\epsilon^2$ contributions to $E_{\rm{sym}}(\rho_0)$ are negative. A general feature, however, is that the second-order contribution does not change the qualitative conclusion from the linear-order term. For example, in the Skyrme model\cite{WangR2018}, although the second-order contribution is negative, the overall result remains positive due to the dominant linear term.  
This demonstrates that the second-order term indeed acts as a small ``perturbation'' even when $\epsilon = -1$ is applied for the 2D symmetry energy. Similarly, one can define an effective form of the 2D symmetry energy as in Eq.\,(\ref{ddef_Esym_eff}).

\begin{table*}[h!]
\centering
\begin{tabular}{c|c|c|c|c}\hline
&$\epsilon$-contribution &$\epsilon^2$-contribution&sign of $\epsilon^2$-term&reference\\
\hline\hline
Skyrme&$+2.2\epsilon\,\rm{MeV}$&$-0.82\epsilon^2\,\rm{MeV}$&negative&Ref.\cite{WangR2018}\\\hline
MDI&$-1.1\epsilon\,\rm{MeV}$&$-0.92\epsilon^2\,\rm{MeV}$&negative&Ref.\cite{Das2003PRC}\\\hline
ImMDI&$4.4\epsilon\,\rm{MeV}+0.14y\epsilon$&$5.3\epsilon^2\,\rm{MeV}+0.05y\epsilon^2$&undetermined&Ref.\cite{XuJ15}\\\hline
\end{tabular}
\caption{\centering Second order of $\epsilon$ correction to the symmetry energy from different models. Table taken from Ref.\cite{Cai22-dD}.}\label{Tab_2DEsym-o2}
\end{table*}

If one neglects the second- and third-order momentum dependence of $U_0$, i.e., $\partial^2 U_0 / \partial |\mathbf{k}|^2 \approx 0$ and $\partial^3 U_0 / \partial |\mathbf{k}|^3 \approx 0$, as well as the second-order momentum dependence of $U_{\rm{sym}}$, i.e., $\partial^2 U_{\rm{sym}} / \partial |\mathbf{k}|^2 \approx 0$, then the symmetry energy perturbed to order $\epsilon^2$ can be approximated as\cite{Cai22-dD}:
\begin{align}
&E_{\rm{sym}}(\rho) 
= \frac{\overline{k}_{\rm F}^2}{6M_{\rm N}} 
+ \left. \frac{\overline{k}_{\rm F}}{6} \frac{\partial U_0}{\partial |\mathbf{k}|} \right|_{|\mathbf{k}|=\overline{k}_{\rm F}} 
+ \frac{1}{2} U_{\rm{sym}}(\rho, \overline{k}_{\rm F}) \notag\\
& + \epsilon \left[
\frac{\overline{k}_{\rm F}^2}{6M_{\rm N}} \left( 2\sigma - \frac{1}{3} \right) 
+ \frac{\overline{k}_{\rm F}}{6} \left( \sigma - \frac{1}{3} \right) \frac{\partial U_0}{\partial |\mathbf{k}|} 
+ \frac{\sigma}{2} \frac{\partial U_{\rm{sym}}}{\partial |\mathbf{k}|} \cdot \overline{k}_{\rm F} 
\right]_{|\mathbf{k}|=\overline{k}_{\rm F}} \notag\\
& + \epsilon^2 \Bigg[
\frac{\sigma^2 \overline{k}_{\rm F}^2}{6M_{\rm N}} 
- \frac{\sigma \overline{k}_{\rm F}^2}{18 M_{\rm N}} \left( \frac{M_{\rm N}}{\overline{k}_{\rm F}} \frac{\partial U_0}{\partial |\mathbf{k}|} + 2 \right) \notag\\
&\hspace{1cm} 
+ \frac{\sigma' \overline{k}_{\rm F}^2}{6 M_{\rm N}} \left( \frac{M_{\rm N}}{\overline{k}_{\rm F}} \frac{\partial U_0}{\partial |\mathbf{k}|} + \frac{3M_{\rm N}}{\overline{k}_{\rm F}} \frac{\partial U_{\rm{sym}}}{\partial |\mathbf{k}|} + 2 \right) 
\notag\\
&\hspace{1cm}+ \frac{\overline{k}_{\rm F}^2}{54 M_{\rm N}} \left( \frac{M_{\rm N}}{\overline{k}_{\rm F}} \frac{\partial U_0}{\partial |\mathbf{k}|} + 1 \right)
\Bigg]_{|\mathbf{k}|=\overline{k}_{\rm F}}.
\label{ddef_Esym-o2-1}
\end{align}
For the four models listed in TAB.\,\ref{Tab_2DEsym-o2}, the $\epsilon^2$-order contributions are $-0.46 \epsilon^2$, $-0.36 \epsilon^2$ and  $8.2 \epsilon^2 - 0.003 y \epsilon^2$, respectively.  
In addition, the correction to the kinetic symmetry energy in 2D at order $\epsilon^2$ is given by 
$[\overline{k}_{\rm F}^2 / (54 M_{\rm N})] \cdot [(3\sigma-1)^2 + 18\sigma'] \approx -0.65 \, \rm{MeV}$, 
leading to $E_{\rm{sym}}^{\rm{kin}}(\rho_0) \approx 9.84 \, \rm{MeV}$.  
Compared with the linear-order estimate of approximately $10.49 \, \rm{MeV}$ (see estimate of Eq.\,(\ref{ia-2})), the second-order correction brings the result much closer to the exact value of about $9.60 \, \rm{MeV}$. For $|\epsilon| \lesssim 1$, higher-order corrections (i.e., beyond $\epsilon^2$) are expected to be small.

{\bfseries SRC-HMT Effects on Kinetic EOS in $d$ Dimensions.}
Now we combine the ideas of the above two examples, namely the EOS of ANM in general $d$ dimensions and the method of $\epsilon$-expansion, to study the effects of SRC-HMT on the kinetic EOS. When generalizing $n_{\mathbf{k}}$ to arbitrary dimensions, certain assumptions on the involved parameters are necessary. The contact coefficient $C_J$, explored so far in nuclear systems using different theoretical approaches, or in ultra-cold gases with various trapping potentials at zero or finite temperatures, all exhibit a remarkable universality\cite{Bulgac2005,Kuhnle2010PRL,Stewart2010PRL,Drut2011PRL,Kuhnle2011PRL,Wild2012PRL,Sagi2012PRL,
Frohlich2012PRL,Hoinka2013PRL,Vignolo2013PRL,
Rossi2018PRL,Yao2018PRL,Mukherjee2019PRL, Carcy2019PRL, Jensen2020PRL,Pit2016}. We therefore assume that $C_J$ (including its $C_0$ and $C_1$ components) is also (approximately) universal with respect to the spatial dimension $d$.  
On the other hand, to our knowledge, there are no fundamental constraints on the high-momentum cutoffs in dimensions other than $3$. We thus assume that they are also universal and investigate the kinetic EOS of ANM based on these assumptions. Although the SRC-induced HMT affects only the kinetic EOS (through the single-nucleon momentum distribution function) in the present study, it can be regarded as an effective nucleon-nucleon interaction.

For nucleons in ANM in $d$ dimensions, the normalization relation of the HMT parameters should be generalized as:
\begin{equation}\label{def_DCPanm}
\Delta_J+\frac{d C_J}{d-4}\left(\phi_J^{d-4}-1\right)=1,
\end{equation}
where the second term on the left-hand side corresponds to $x_J^{\rm{HMT}}$ in $d$ dimensions. 
A notable feature is that $\phi_J$ in $d=1,2,3$ can be mathematically taken to be infinite, whereas in $d=4$ it has a natural upper bound, $\phi_J \leq \exp(1/4C_J)$.  
Similarly, the kinetic EOS of ANM in $d$ dimensions is given by
\begin{align}\label{kef_EOSANM}
&E^{\rm{kin}}(\rho,\delta)=\sum_{J=\rm{n,p}}\frac{d}{d+2}\frac{k_{\rm{F}}^2}{2M_{\rm N}}\left(1+\tau_3^J\delta\right)^{1+2/d}\notag\\
&\times\left[1+C_J\left(\frac{8}{(d-2)(d-4)}+\frac{d+2}{d-2}\phi^{d-2}_J
-\frac{d}{d-4}\phi^{d-4}_J\right)\right].
\end{align}
The dependence of $E^{\rm{kin}}(\rho,\delta)$ on the spatial dimension $d$ is non-trivial due to the intricate interplay among $C_J$, $\phi_J$, and $d$.

According to the general formula (\ref{kef_EOSANM}), the kinetic part of the SNM EOS in $d$D can be directly obtained as
\begin{align}
&E_0^{\rm{kin}}(\rho)=\frac{d}{d+2}\frac{k_{\rm{F}}^2}{2M_{\rm N}}\notag\\&\times
\left[1+C_0\left(\frac{8}{(d-2)(d-4)}+\frac{d+2}{d-2}\phi^{d-2}_0
-\frac{d}{d-4}\phi^{d-4}_0\right)\right]\notag\\
&\equiv\frac{k_{\rm{F}}^2}{2M_{\rm N}}\Lambda(d,C_0,\phi_0),\label{def_LAM}
\end{align}
where the last relation defines the function $\Lambda(d,C_0,\phi_0)$, which encodes the dependence of the EOS on the dimension $d$. We note that $\Lambda(d,C_0,\phi_0)$ apparently diverges at $d=d_{\rm{f}}=2$ and/or $d=d_{\rm{f}}=4$. Nevertheless, away from the poles with $d=d_{\rm{f}}+\epsilon$, the diverging terms cancel completely and naturally in the $\epsilon$-expansion. For example, in the 2D case, one obtains the $\epsilon$-expansion
\begin{align}
&\frac{8}{(d-2)(d-4)}+\frac{d+2}{d-2}\phi^{d-2}_0
-\frac{d}{d-4}\phi^{d-4}_0\notag\\
\approx&-1+4\ln\phi_0+\frac{1}{\phi_0^2}+\epsilon\left(1+\ln\phi_0+2\ln^2\phi_0+\frac{\ln\phi_0+1}{\phi_0^2}\right),
\end{align}
to order $\epsilon=d-2$, using $a^x\approx 1+x\ln a+2^{-1}x^2\ln^2 a$ for small $x$.
Expanding the EOS of SNM around the dimension $d_{\rm{f}}$ (with $d_{\rm{f}}=1,2,3,4$) to order $\epsilon$ yields the corresponding expression $E_{0,d_{\rm{f}}=1\sim4}^{\rm{kin}}(\rho)$\cite{Cai22-dD}. The $\epsilon^0$-order terms of these expansions represent the EOS of SNM in each dimension:
\begin{align}
E_{0,{(\rm{1D})}}^{\rm{kin}}(\rho)=&\frac{k_{\rm{F}}^2}{6M_{\rm N}}
\left[1+C_0\left(\frac{8}{3}-\frac{3}{\phi_0}+\frac{1}{3\phi_0^3}\right)\right],\label{E-0-1-kin}\\
E_{0,{(\rm{2D})}}^{\rm{kin}}(\rho)=&\frac{k_{\rm{F}}^2}{4M_{\rm N}}\left[1+C_0\left(4\ln \phi_0+\frac{1}{\phi_0^2}-1\right)\right],\label{E-0-2-kin}\\
E_{0,{(\rm{3D})}}^{\rm{kin}}(\rho)=&\frac{3}{5}\frac{k_{\rm{F}}^2}{2M_{\rm N}}
\left[1+C_0\left(5\phi_0+\frac{3}{\phi_0}-8\right)\right],\label{E-0-3-kin}\\
E_{0,{(\rm{4D})}}^{\rm{kin}}(\rho)=&\frac{k_{\rm{F}}^2}{3M_{\rm N}}\left[1+C_0\left(3\phi_0^2-4\ln\phi_0-3\right)\right],\label{E-0-4-kin}
\end{align}
where the nucleon Fermi momentum $k_{\rm{F}}$ in $d$ dimensions is related to the nucleon density $\rho$\cite{Cai22-dD}.

\begin{figure}[h!]
\centering
 \includegraphics[height=7.cm]{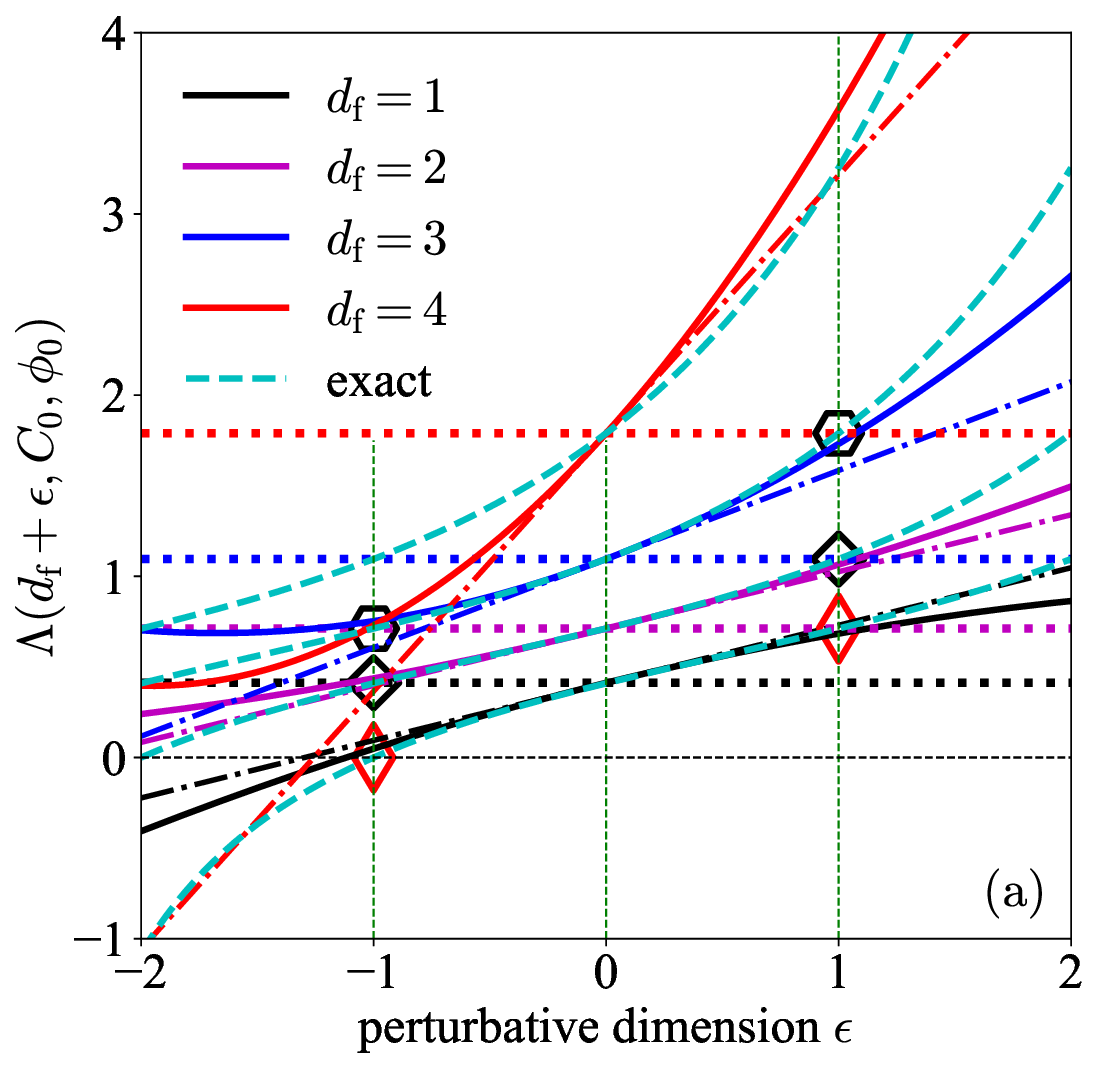}\\[0.25cm]
 \hspace{-0.3cm}
 \includegraphics[height=7.cm]{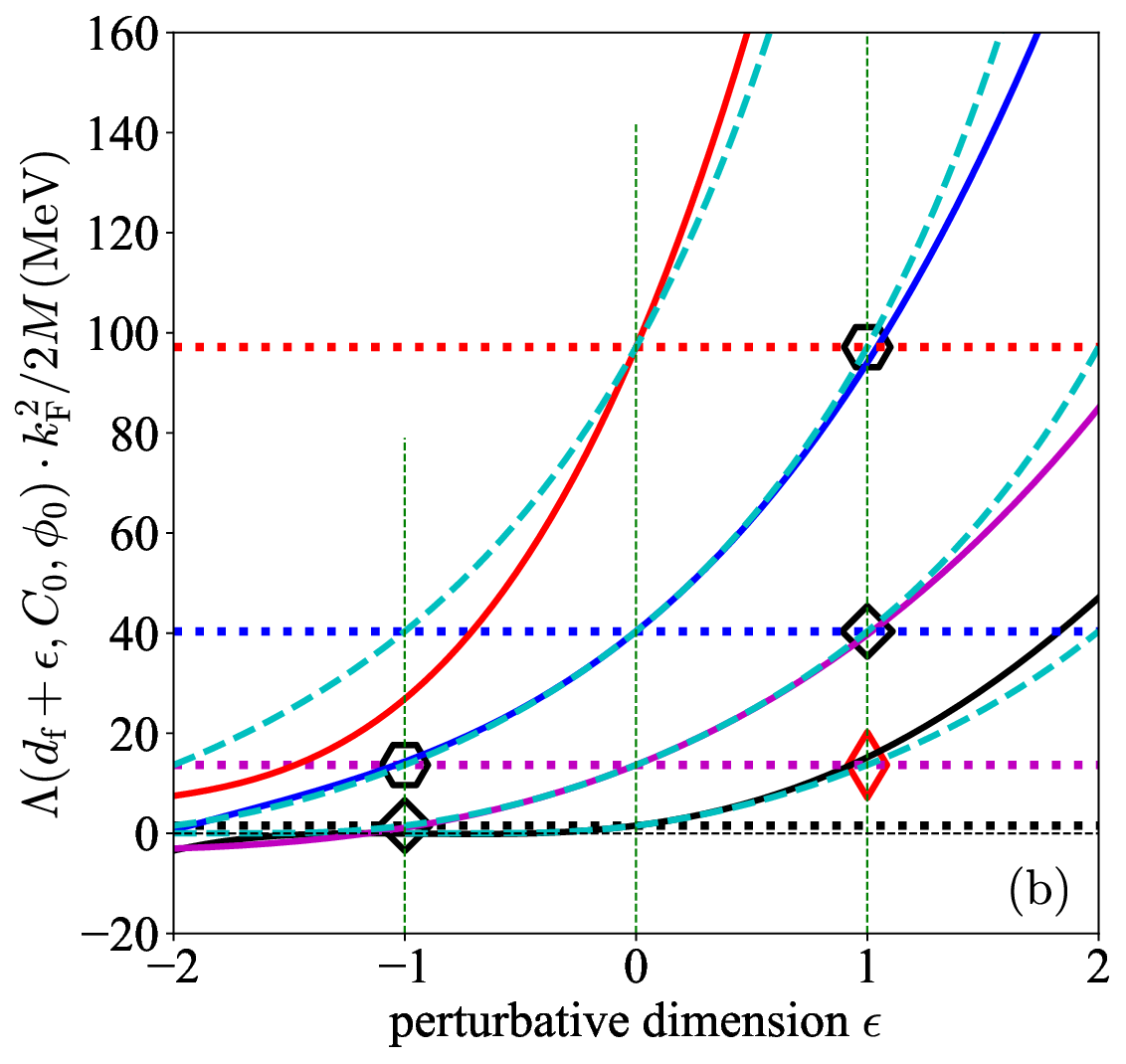}
\caption{(Color Online). Perturbative calculations of $\Lambda(d_{\rm{f}}+\epsilon,C_0,\phi_0)$ (upper) and the kinetic EOS $E_0^{\rm{kin}}(\rho_0)$ (lower) for different $d_{\rm{f}}$. Only second-order results are shown in the lower panel. Figures from Ref.\cite{Cai22-dD}.}
\label{fig_pert-ep}
\end{figure}

In the upper panel of FIG.\,\ref{fig_pert-ep}, the values of $\Lambda(d,C_0,\phi_0)$ from perturbative calculations are shown. The dotted, dash-dotted, and solid lines correspond to zeroth-order, first-order, and second-order approximations, respectively, while the dashed cyan lines represent exact results for each $d_{\rm{f}}$. It is clear that the difference between first- and second-order results is smaller than that between zeroth- and first-order approximations (e.g., for $-1\lesssim\epsilon\lesssim1$), and this difference increases with $d$. This indicates convergence of the $\epsilon$-expansion, even when $\epsilon$ takes sizable values like 1 or $-1$. Results starting from $d_{\rm{f}}=1,2,3$ are all reasonable. For example, starting from $d_{\rm{f}}=1$ with $\epsilon=1$ or $\epsilon=2$ gives the $\Lambda$ function in 2D or 3D, see the dash-dotted black line in FIG.\,\ref{fig_pert-ep}. In particular, $\Lambda(1+1,C_0,\phi_0)\approx0.728$ to linear order and $\Lambda(1+1,C_0,\phi_0)\approx0.683$ to quadratic order, while the exact 2D value is 0.713, using $C_0\approx0.161$ and $\phi_0\approx2.38$\cite{Cai15a}. Starting from $d_{\rm{f}}=2$ and taking $\epsilon=-1$ or $\epsilon=1$ (to order $\epsilon^2$) gives reasonable approximations for 1D and 3D, see the dash-dotted magenta line. Similarly, starting from $d_{\rm{f}}=3$ yields reliable results (blue lines). In particular, $d_{\rm{f}}=3$ and $\epsilon=-1$ gives $\Lambda(3-1,C_0,\phi_0)\approx0.607$ and $0.754$ for linear and quadratic orders, respectively, close to the exact value 0.713. However, starting from $d_{\rm{f}}=4$ with $\epsilon=-1$ does not yield effective results. Overall, perturbations behave well for $1\lesssim d_{\rm{f}}\lesssim 3$ and $-1\lesssim\epsilon\lesssim1$.

Next, we consider the Fermi energy $E_{\rm{F}}=k_{\rm{F}}^2/2M_{\rm N}$ in $d$ dimensions. For 2D, $k_{\rm{F}}^{(\rm{2D})}=\sqrt{\pi\rho_2}$ with $\rho_2=\rho^{2/3}$. Taking $\rho\approx0.16\,\rm{fm}^{-3}$, we have $k_{\rm{F}}^{(\rm{2D})}\approx189.88\,\rm{MeV}$ and $E_{\rm{F}}^{(\rm{2D})}\approx19.20\,\rm{MeV}$, giving $E_{0,(\rm{2D})}^{\rm{kin}}(\rho_0)\approx E_{\rm{F}}^{(2\rm{D})}\Lambda(2,C_0,\phi_0)\approx13.69\,\rm{MeV}$. Starting from 1D, $k_{\rm{F}}^{(\rm{1D})}=\pi\rho_1/4\approx84.13\,\rm{MeV}$ with $\rho_1=\rho^{1/3}$, and $k_{\rm{F}}^{(\rm{2D})}\approx k_{\rm{F}}^{(\rm{1D})}(1+\sigma_1\epsilon)\approx212.46\,\rm{MeV}$ with $\epsilon=1$ and $\sigma_1\equiv\sigma(1)\approx1.5253$ (see Eq.\,(\ref{def_sigf})), giving $E_{0,(\rm{2D})}^{\rm{kin}}(\rho_0)\approx17.50\,\rm{MeV}$ using $\Lambda(1+\epsilon,C_0,\phi_0)\approx0.728$. Starting from 3D, $k_{\rm{F}}^{(\rm{2D})}\approx k_{\rm{F}}^{(\rm{3D})}(1-\sigma)\approx200.02\,\rm{MeV}$ with $\sigma\approx0.2396$ (see Eq.\,(\ref{defff_sig})) and $\Lambda(3-1,C_0,\phi_0)\approx0.607$, giving $E_{0,(\rm{2D})}^{\rm{kin}}(\rho_0)\approx12.93\,\rm{MeV}$. Hence, both 1D and 3D perturbative expansions are effective, with better accuracy starting from 3D due to smaller $\sigma(d_{\rm{f}})$.

As an application of the $\epsilon$-expansion, we estimate the 2D kinetic EOS as $\phi_0\to\infty$ using the 1D kinetic EOS. The dependence of $\Lambda(d,C_0,\phi_0)$ on $\phi_0$ is given in Ref.\cite{Cai22-dD}. Analytically, we have
\begin{align}\label{def-Lf}
\frac{\partial\Lambda(d,C_0,\phi_0)}{\partial\phi_0}
=&-\frac{C_0d\phi_0^{d-3}}{d+2}\frac{1}{d+2}\left(\frac{d}{d+2}\frac{1}{\phi_0^{2}}-1\right).
\end{align}
Since $\phi_0\geq1$, this derivative is always positive, indicating $\Lambda(d,C_0,\phi_0)$ increases with $\phi_0$ regardless of $d$. The kinetic EOS diverges if $\phi_0\to\infty$, except in 1D. The divergence order increases with $d$: logarithmic in 2D, linear in 3D, quadratic in 4D. More generally, for $d\ge3$, the divergence order is $d-2$, as in Eq.\,(\ref{def_LAM}). In 1D, the high-momentum part contributes $8C_0/3$ (see Eq.\,(\ref{E-0-1-kin}).
Taking $\phi_0\to\infty$ in $E_{0,d_{\rm f}=2}^{\rm{kin}}(\rho)$\cite{Cai22-dD} gives to linear order in $\epsilon$:
\begin{align}\label{dE_eps}
E_{0,d_{\rm{f}}=1,\epsilon}^{\rm{kin}}(\rho)
\approx& \frac{k_{\rm{F}}^{(\rm{1D}),2}(1+2\sigma_1\epsilon)}{6M_{\rm N}}
\left[1+\frac{8}{3}C_0+\left(\frac{2}{3}
+\frac{16}{3}C_0
\right)\epsilon\right]\notag\\
&+\sum\rm{term\;approaches\;zero\;as\;}\phi_0\to\infty.
\end{align}
Consequently,
\begin{align}
E_{0,\rm{(2D)}}^{\rm{kin},\phi_0\to\infty}(\rho)\approx&
E_{0,d_{\rm{f}}=1,\epsilon=1}^{\rm{kin}}(\rho)\notag\\
\approx&\frac{k_{\rm{F}}^{(\rm{1D}),2}(1+2\sigma_1)}{6M_{\rm N}}
\left(\frac{5}{3}+8C_0
\right)\approx 15.04\,\rm{MeV}.
\end{align}
This is slightly larger than the exact $E_{\rm{F}}^{(2\rm{D})}\Lambda(2,C_0,\phi_0)\approx13.69\,\rm{MeV}$ with $\phi_0\approx2.38$. The FFG model predicts $E_{0,\rm{(2D)}}^{\rm{kin}}(\rho)\approx k_{\rm{F}}^{(\rm{2D}),2}/4M_{\rm N}=\pi\rho_2/4M_{\rm N}\approx9.60\,\rm{MeV}$.
Interestingly, applying $\epsilon=2$ in Eq.\,(\ref{dE_eps}) gives $E_{0,\rm{(3D)}}^{\rm{kin},\phi_0\to\infty}(\rho)\approx39.97\,\rm{MeV}$, close to the 3D value of $\sim 40.45$\,MeV\cite{Cai15a} with $\phi_0\approx2.38$ and $C_0\approx0.161$. Physically, $\phi_0$ cannot be extended to infinity in 3D because the $k^{-4}$ HMT breaks down at high momenta due to higher-order correlations such as three-nucleon interactions\cite{Hen17RMP,Egi06,Fom12,Fa17,Ye2018PRC,Sargsian2019PRC}. Other forms of HMT in $n_{\v{k}}^0$ should then be considered.
These results indicate that the $\epsilon$-expansion from a reference dimension $d_{\rm{f}}$ (e.g., 1D) is reasonable and effective. However, as the expansion order $n$ increases, extrapolation becomes less reliable, since $(1+1)^n=2^n$ can become very large/divergent.

Once the nucleon momentum distribution $n_{\mathbf{k}}^0$ is specified, one can study its moments, which characterize various aspects of nucleon motion. A particularly useful quantity is the variance of the nucleon momentum, $\langle k^2\rangle - \langle k\rangle^2$, which describes the strength of momentum fluctuations. To compare across different dimensions, we define the relative nucleon momentum fluctuation in SNM in $d$ dimensions as\cite{Cai22-dD}:
\begin{align}
\Upsilon_k(d) \equiv&\sqrt{\frac{\langle k^2\rangle - \langle k\rangle^2}{\langle k\rangle^2}} = \sqrt{\frac{\langle k^2\rangle}{\langle k\rangle^2} - 1},\\
\langle \v{k}^{\gamma}\rangle=&\left.\int n_{\v{k}}^0\v{k}^{\gamma}\d^d\v{k}\right/\int n_{\v{k}}^0\d^d\v{k}.\end{align}
For our analysis, only $\gamma = 1$ and $\gamma = 2$ are relevant:
\begin{align}\label{def-flc}
&\frac{\langle k\rangle}{k_{\rm F}} = \frac{d}{d+1}\notag\\
&\times\left[ 1 + C_0 \left( \frac{4}{(d-3)(d-4)} + \frac{d+1}{d-3}\phi_0^{d-3} - \frac{d}{d-4}\phi_0^{d-4} \right) \right],\\
&\frac{\langle k^2 \rangle}{k_{\rm F}^2} = \frac{d}{d+2} \notag\\
&\times\left[ 1 + C_0 \left( \frac{8}{(d-2)(d-4)} + \frac{d+2}{d-2}\phi_0^{d-2} - \frac{d}{d-4}\phi_0^{d-4} \right) \right].
\end{align}
For example, we have in 3D:
\begin{align}
\Upsilon_k^{\rm{HMT}}(3) =& \sqrt{ \frac{16 \phi_0}{15} \frac{5 C_0 \phi_0^2 - 8 C_0 \phi_0 + 3 C_0 + \phi_0}{4 C_0 \phi_0 \ln\phi_0 - 3 C_0 \phi_0 + 3 C_0 + \phi_0} - 1 } \notag\\
\to \Upsilon_k^{\rm{FFG}}(3) = &\sqrt{\frac{1}{15}} \text{ as } \phi_0\to1.
\end{align}
For the FFG model, one has
\begin{equation}
\Upsilon_k^{\rm{FFG}}(d) = \frac{1}{\sqrt{d^2 + 2d}}, 
\end{equation}
with, for example, $\Upsilon_k^{\rm{FFG}}(3) \approx 0.258$ and $\Upsilon_k^{\rm{FFG}}(2) \approx 0.354$.

\begin{figure}[h!]
\centering
 \includegraphics[height=7.cm]{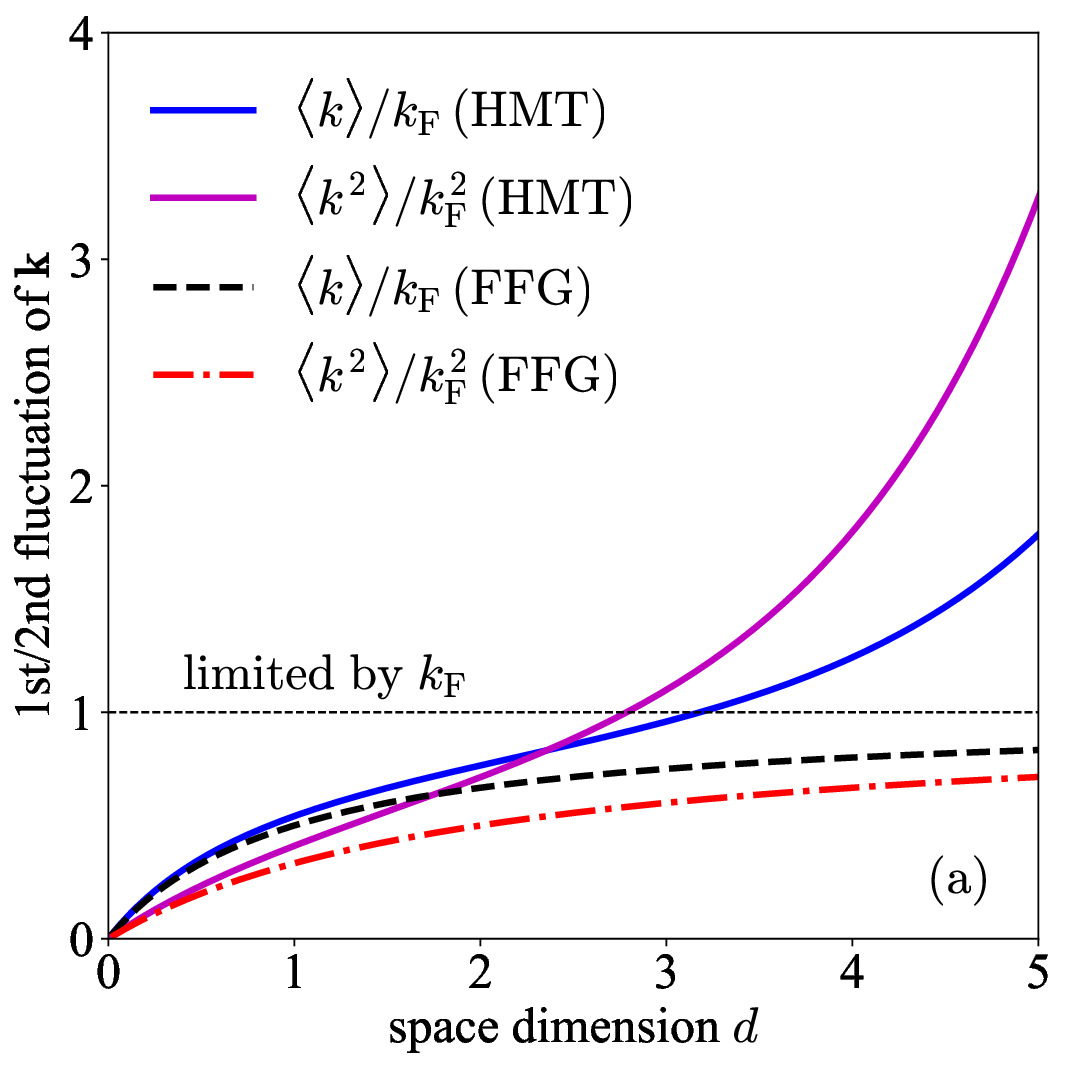}\\
 \hspace{-0.3cm}
 \includegraphics[height=7.cm]{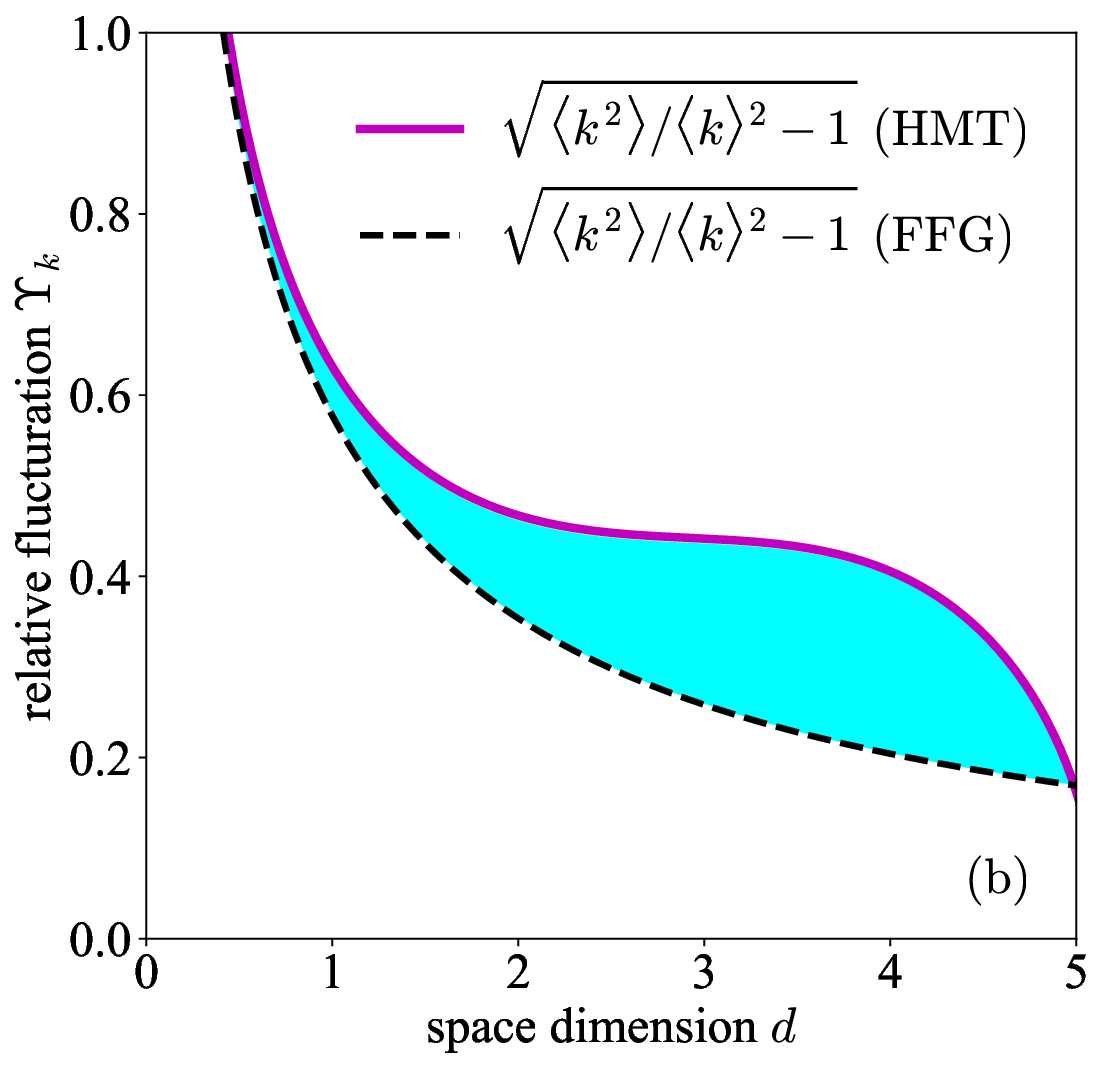}
\caption{(Color Online). Fluctuation of the momentum $\Upsilon_k$ at saturation density $\rho_0$ as a function of $d$. Figure taken from Ref.\cite{Cai22-dD}.}
\label{fig_fluc-d}
\end{figure}

FIG.\,\ref{fig_fluc-d} presents the first and second moments of the momentum $\mathbf{k}$ as functions of the spatial dimension $d$ (upper panel), along with the relative momentum fluctuation $\Upsilon_k$ (lower panel), for both the FFG and HMT models. From the figure, one observes: (a) as $d$ decreases, both the first and second moments of $k$ diminish, while the relative fluctuation $\Upsilon_k$ increases. This indicates that nucleonic matter in lower dimensions exhibits larger relative momentum fluctuations compared to higher-dimensional systems, for both the FFG and HMT cases; (b) when the SRC-induced HMT is included, the fluctuations at a given $d$ are enhanced, showing that interactions increase momentum fluctuations by allowing nucleons to occupy high-momentum states above the Fermi surface. In the FFG model, the moments $\langle k^{\gamma} \rangle^{1/\gamma}$ are constrained by the Fermi momentum $k_{\rm F}$, as nucleons cannot exceed the Fermi surface. The stronger fluctuation enhancement in low dimensions due to interactions is consistent with the Mermin--Wagner theorem\cite{Mermin1966}, which states that continuous symmetries cannot be spontaneously broken at finite temperatures in systems with short-range interactions for $d \leq 2$. Physically, this implies that long-range fluctuations can arise with relatively low energy cost. Furthermore, the fluctuation $\Upsilon_k$ in low dimensions approaches its FFG value, although both remain quantitatively large.

By expanding the kinetic EOS of ANM and extracting the coefficient of the $\delta^2$ term, we obtain the kinetic symmetry energy; similarly, the expressions for the fourth-order kinetic symmetry energy can also be derived. Analytical expressions for both the kinetic symmetry energy and the fourth-order kinetic symmetry energy in general $d$ dimensions are available\cite{Cai22-dD}. 
In the limit of high-momentum cutoff parameters $\phi_0=1$ and $\phi_1=0$, all kinetic symmetry energies reduce to the FFG predictions, e.g., $E_{\rm{sym}}^{\rm{kin}}(\rho)={k_{\rm{F}}^2}/{2dM_{\rm N}}$, see Eq.\,(\ref{4qEsymFF_1}). Moreover, in dimensions $d=1$ and $d=2$, $E_{\rm{sym},4}^{\rm{kin}}(\rho)$ vanishes, consistent with the FFG expression Eq.\,(\ref{4qEsym4FF_1}).

\begin{figure}[h!]
\centering
 \includegraphics[height=3.3cm]{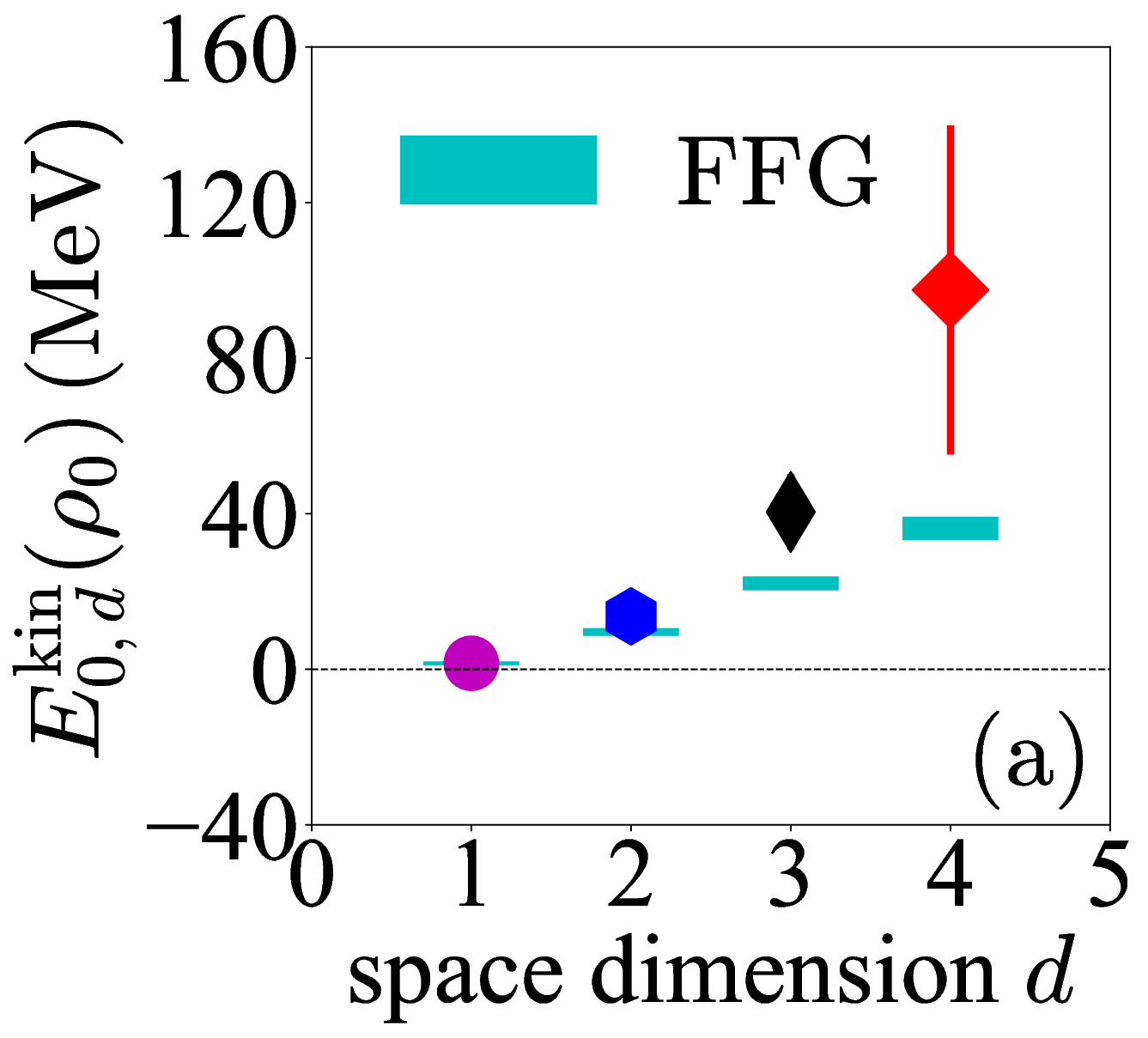}\quad
 \includegraphics[height=3.3cm]{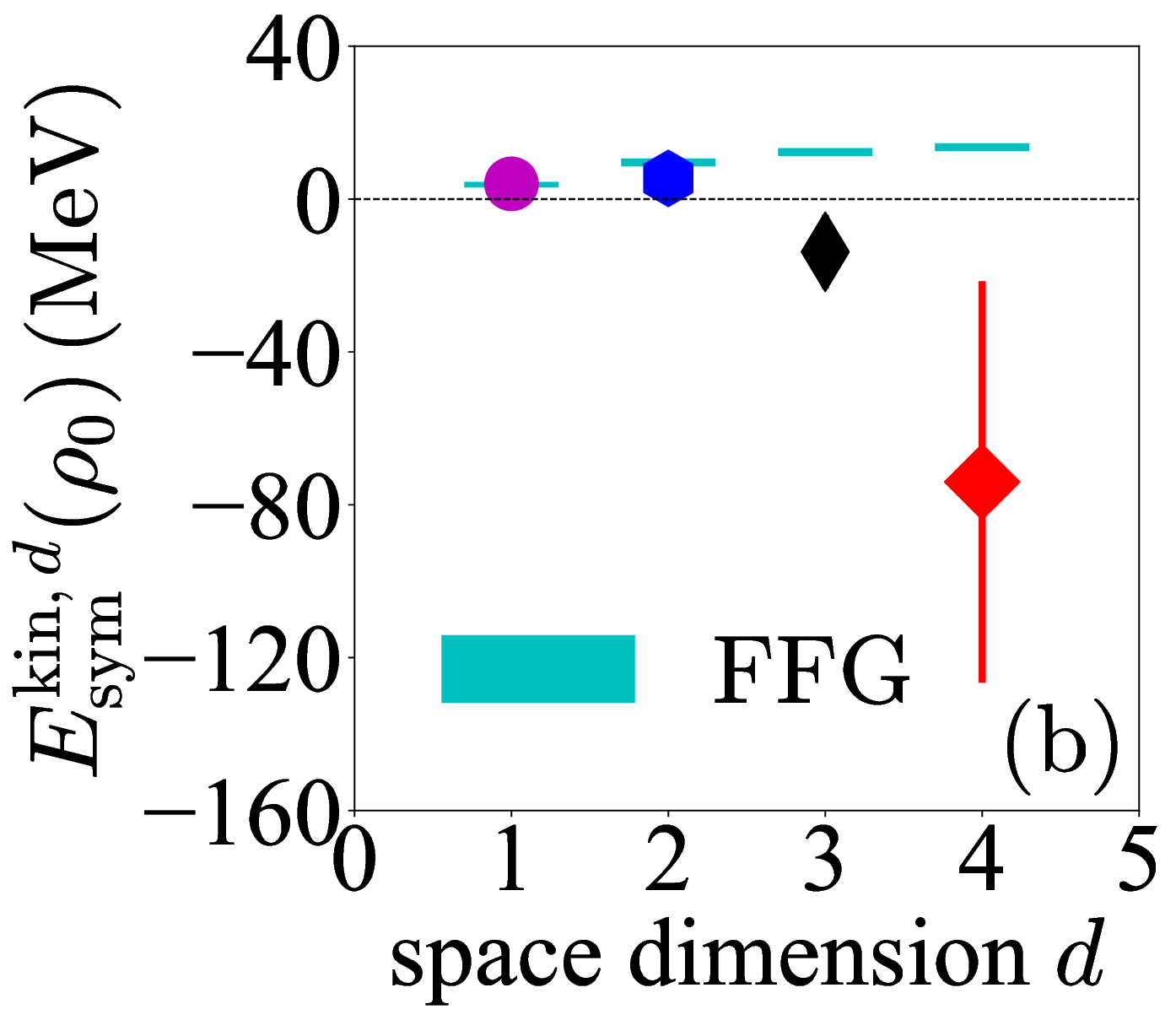}\\[0.25cm]
 \hspace{0.1cm}
 \includegraphics[height=3.3cm]{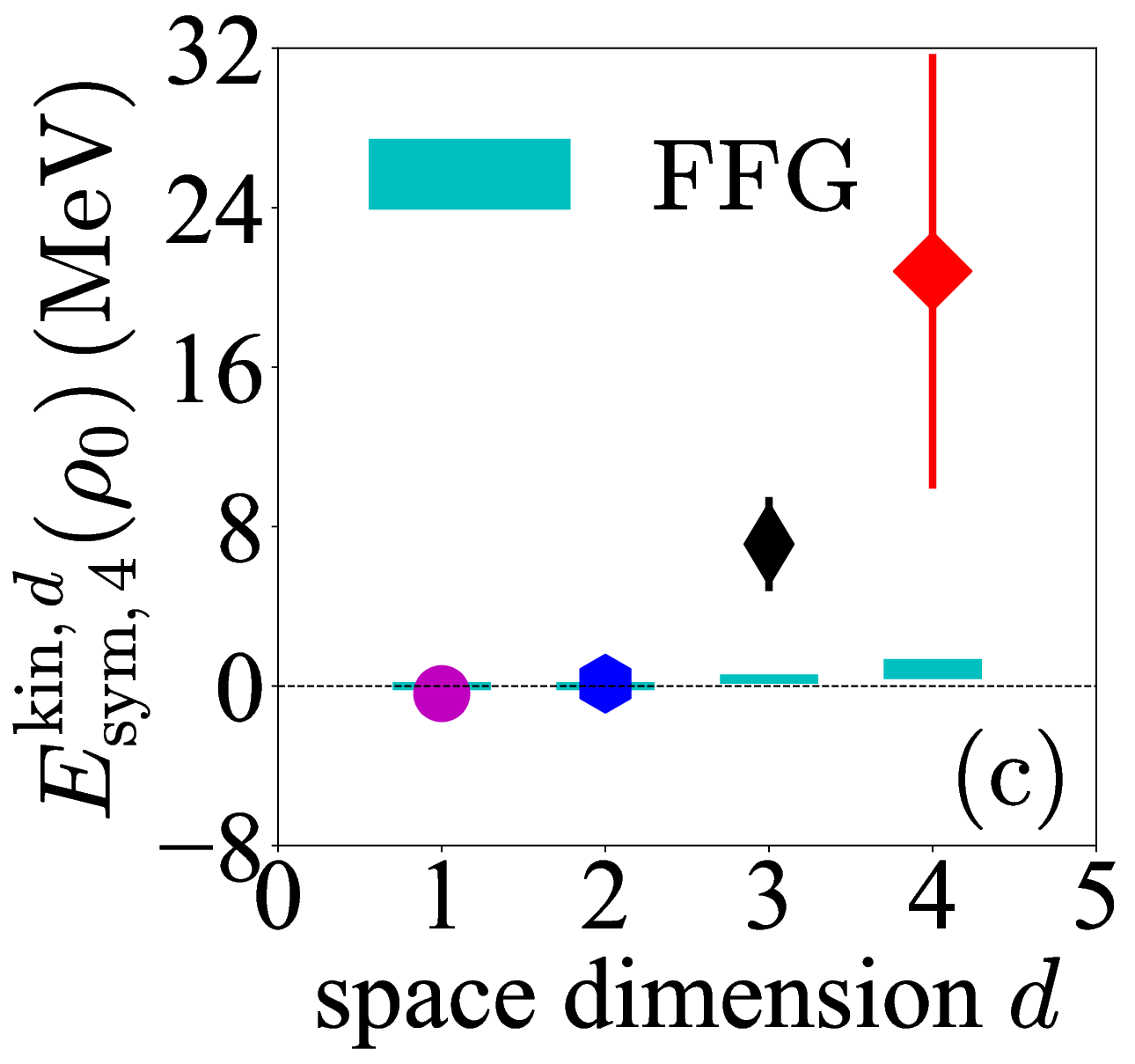}\qquad
 \hspace{-0.1cm}
 \includegraphics[height=3.3cm]{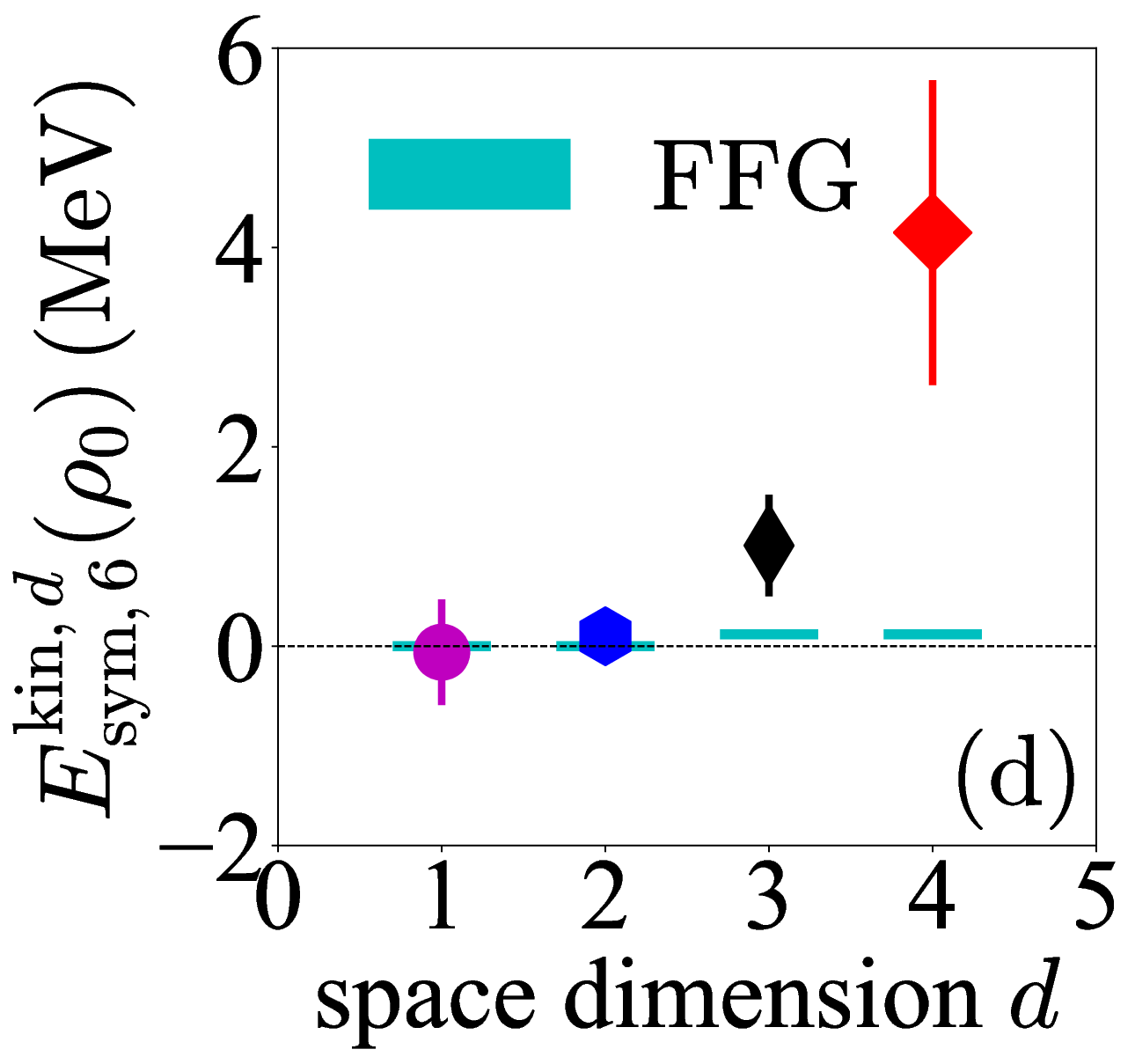}
\caption{(Color Online). Kinetic EOS of ANM in different dimensions. Figures taken from Ref.\cite{Cai22-dD}.}
\label{fig_kinetic-E-d}
\end{figure}

\begin{figure}[h!]
\centering
 \includegraphics[height=6.5cm]{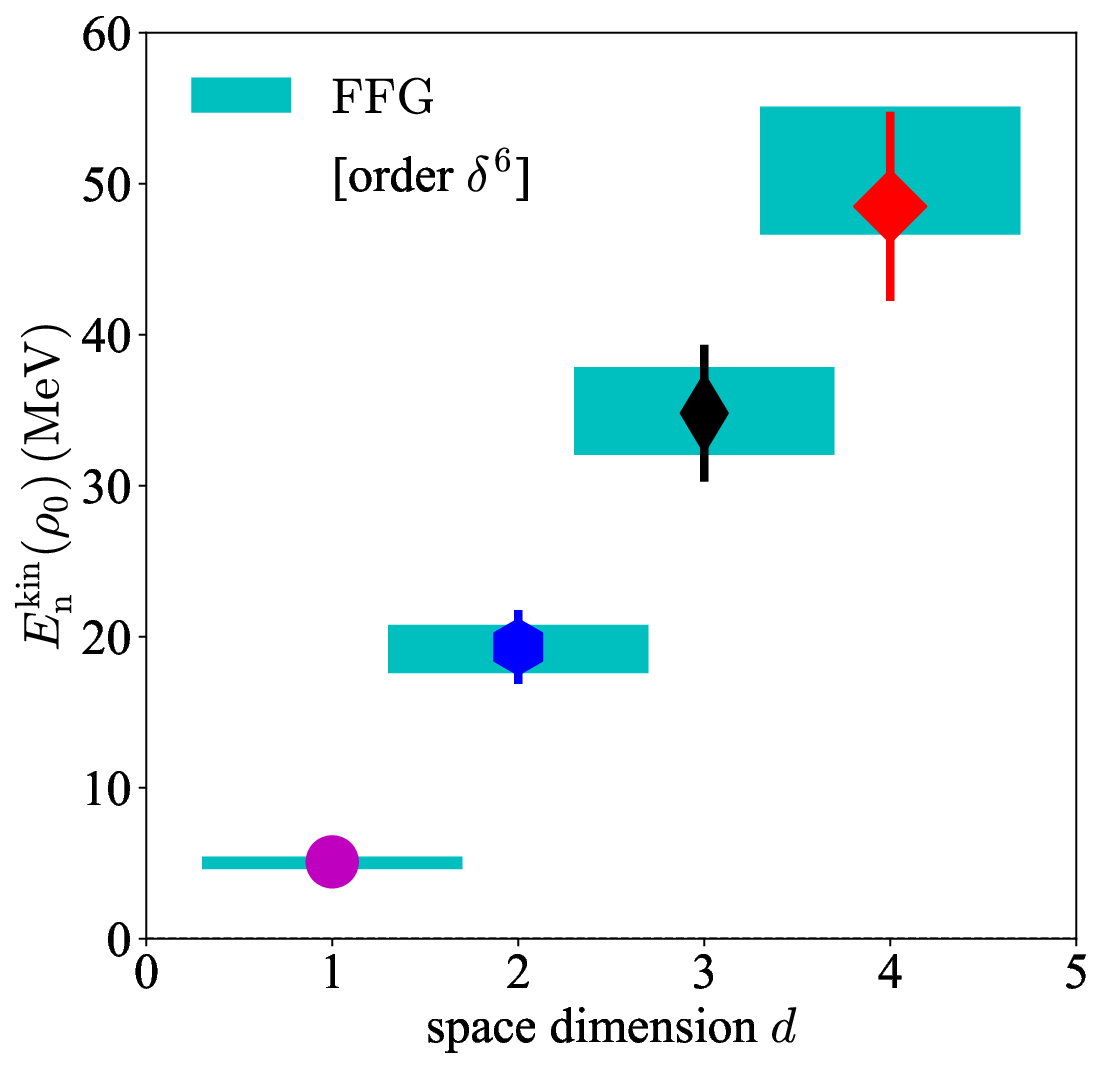}
\caption{(Color Online). Kinetic EOS of PNM to order $\delta^6$ in the FFG/$\rm{HMT}$ model. Figure taken from Ref.\cite{Cai22-dD}.}
\label{fig_Enkin-d}
\end{figure}

\begin{figure}[h!]
\centering
\hspace{-0.3cm}
 \includegraphics[height=7.5cm]{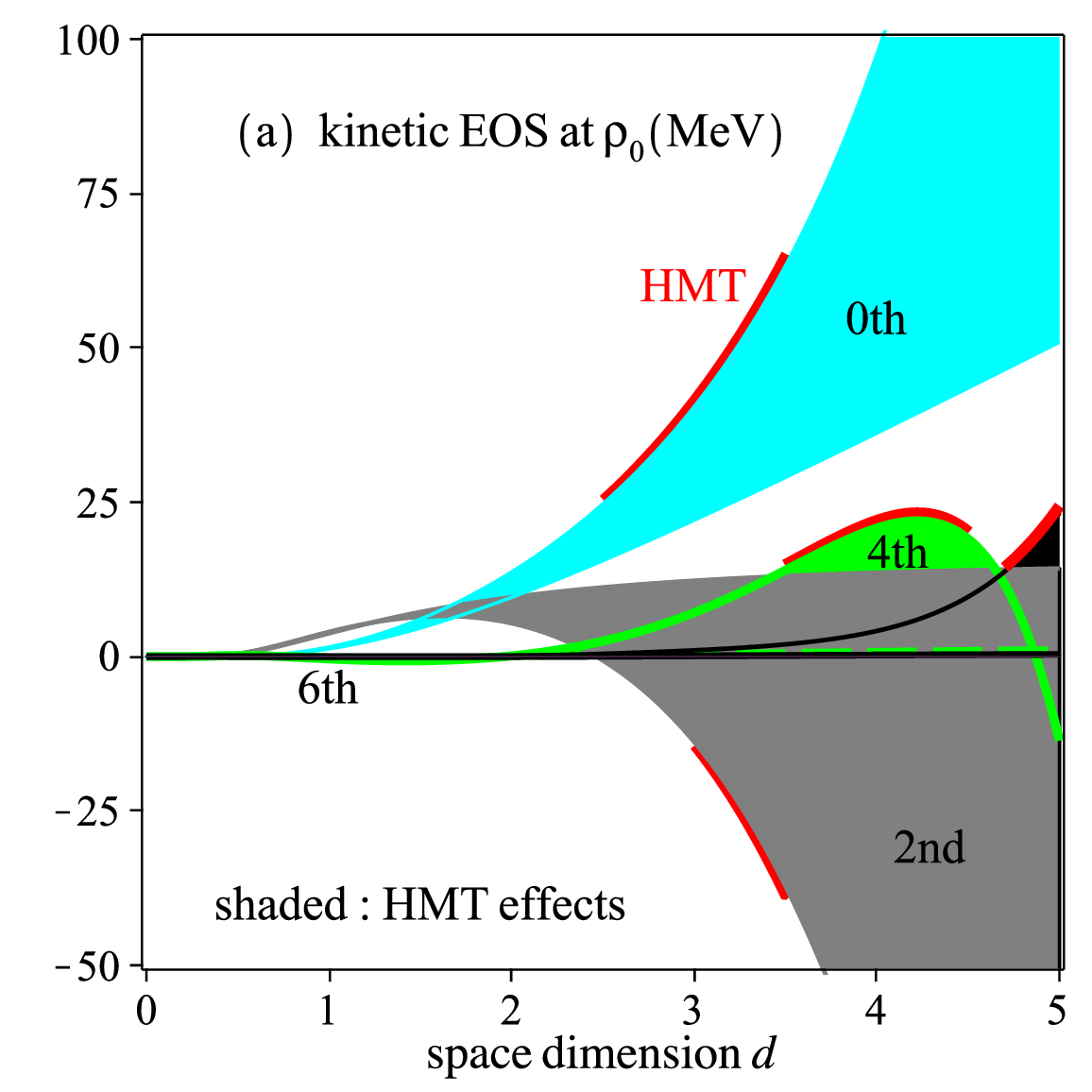}\qquad
 \includegraphics[height=7.5cm]{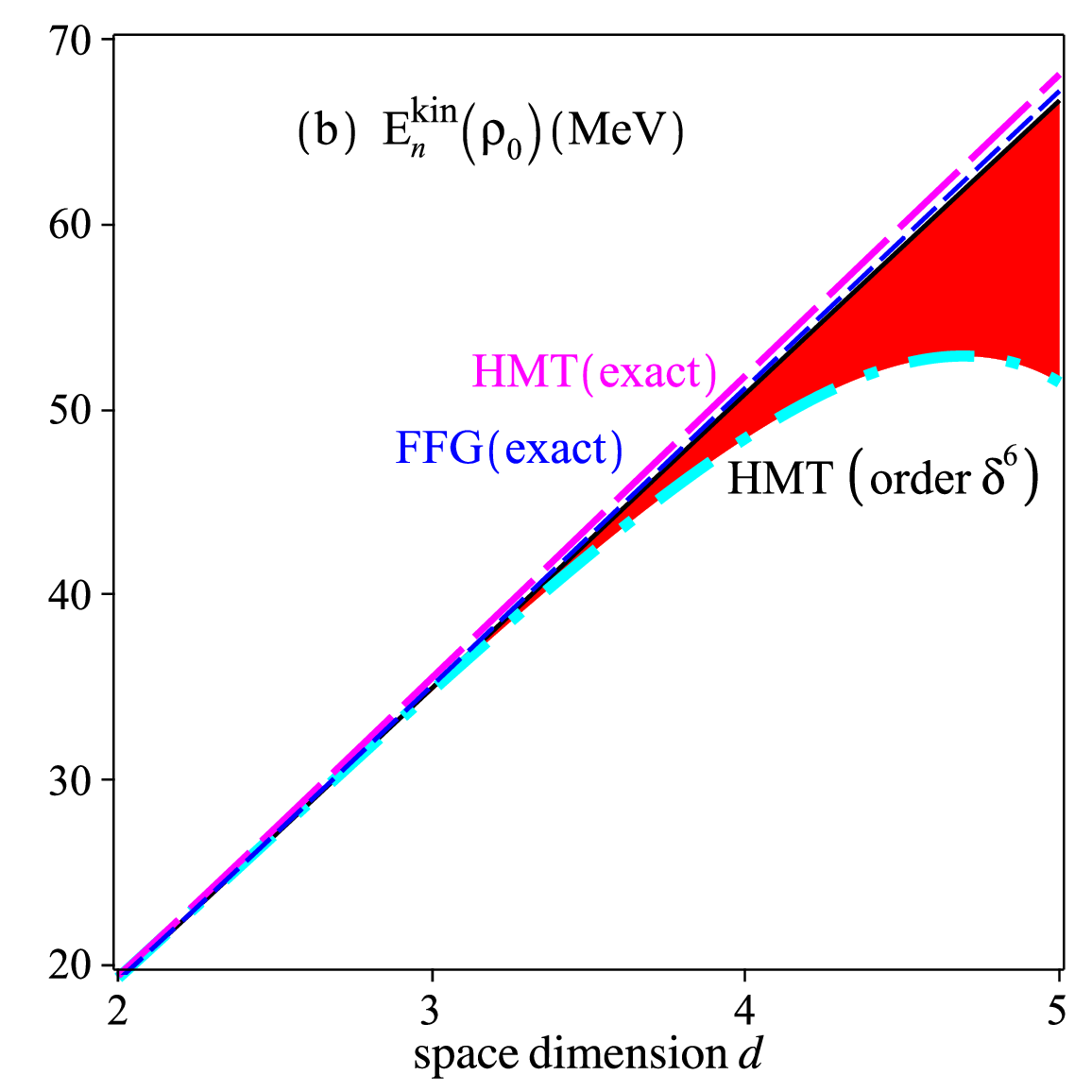}
\caption{(Color Online). Kinetic EOS of ANM as a function of $d$ at saturation density $\rho_0$.
The upper boundary of the red area in the lower panel corresponds to the FFG prediction (to order $\delta^6$), while the blue dash line shows the exact kinetic EOS of PNM from the FFG model. Figures taken from Ref.\cite{Cai22-dD}.}
\label{fig_Ekin024-d}
\end{figure}

For comparison, the FFG model predicts the kinetic EOSs as $E_0^{\rm{kin}}(\rho)=dk_{\rm{F}}^2/(d+2)$, $E_{\rm{sym}}^{\rm{kin}}(\rho)=k_{\rm{F}}^2/2dM_{\rm N}$ as well as $E_{\rm{sym,4}}^{\rm{kin}}(\rho)=(2/d-1)(2/d-2)k_{\rm{F}}^2/24dM_{\rm N}$ and $E_{\rm{sym},6}^{\rm{kin},d}(\rho)=(3d-2)(d-1)(d-2)(2d-1)k_{\rm{F}}^2/180d^5M_{\rm N}$, yielding numerically $E_{0}^{\rm{kin},d}(\rho_0)\lesssim36.20\,\rm{MeV}$, $E_{\rm{sym}}^{\rm{kin},d}(\rho_0)\lesssim13.57\,\rm{MeV}$, $E_{\rm{sym},4}^{\rm{kin},d}(\rho_0)\lesssim0.85\,\rm{MeV}$, and $E_{\rm{sym},6}^{\rm{kin},d}(\rho_0)\lesssim0.12\,\rm{MeV}$ for $1\le d\le4$. FIG.\,\ref{fig_kinetic-E-d} compares the SRC-induced $\rm{HMT}$ kinetic EOS components with the FFG predictions. For $d=1$ and $d=2$, all $\rm{HMT}$ components are very close to FFG results, with the fourth- and sixth-order symmetry energies exactly zero in the FFG model and remaining nearly identical in the $\rm{HMT}$ model. At higher dimensions, the $\rm{HMT}$ effects become more significant. FIG.\,\ref{fig_Enkin-d} shows the PNM kinetic EOS to order $\delta^6$ in FFG and $\rm{HMT}$ models, indicating minimal differences for $d\lesssim3$ due to the low HMT fraction in PNM\cite{Cai15a}.
Treating $d$ as a continuous variable allows us to plot the different-order terms of the kinetic EOS versus $d$, as shown in the upper panel of FIG.\,\ref{fig_Ekin024-d}. The red segment in each shaded area includes the SRC-induced $\rm{HMT}$. The fourth- and sixth-order kinetic symmetry energies are smaller than the SNM kinetic EOS and the quadratic kinetic symmetry energy, with $E_{\rm{sym},4}^{\rm{kin},d}(\rho_0)$ peaking around $d\approx4$ (see also the third panel of FIG.\,\ref{fig_kinetic-E-d}). The SNM kinetic EOS, quadratic, fourth-, and sixth-order symmetry energies are all analytical in $d$ as $d\to0$, and are continuous at $d=2$ and/or $d=4$, showing that the apparent singularities in the general expressions are artificial. For $d\lesssim2$, all kinetic EOSs deviate only slightly from the FFG predictions, consistent with the relative momentum fluctuations $\Upsilon_k$ (lower panel of FIG.\,\ref{fig_fluc-d}).
The lower panel of FIG.\,\ref{fig_Ekin024-d} shows the corresponding PNM kinetic EOS results. The blue dash line and magenta long dash line represent the exact FFG and $\rm{HMT}$ kinetic EOS, respectively. The red region indicates the approximation to order $\delta^6$, with upper and lower boundaries corresponding to FFG and $\rm{HMT}$. The kinetic EOS of PNM is weakly affected by the $\rm{HMT}$ for $d\lesssim3$, reflecting the small HMT fraction in PNM.

Examining several limiting cases, we first note that in 3D the exact FFG model gives 
$E_{\rm{n}}^{\rm{kin}}(\rho_0)=2^{1/3}\cdot 3k_{\rm{F}}^2/10M_{\rm N}\approx35.1\,\rm{MeV}$. As the dimension $d$ approaches zero, the kinetic energy behaves as
\begin{align}
E_{\rm{n}}^{\rm{kin}}(\rho)
\to&\frac{k_{\rm{F}}^2}{2M_{\rm N}}\frac{d}{d+2}\cdot2^{1+2/d}\left[1+C_{\rm{n}}\left(1-\frac{1}{\phi_{\rm{n}}^2}\right)\right]\notag\\
\approx&\frac{k_{\rm{F}}^2}{2M_{\rm N}}\frac{d}{d+2}\left(2+\frac{4\ln2}{d}\right)
\left[1+C_{\rm{n}}\left(1-\frac{1}{\phi_{\rm{n}}^2}\right)\right]\notag\\
\approx&\frac{k_{\rm{F}}^2\ln 2}{M_{\rm N}}\left[1+C_{\rm{n}}\left(1-\frac{1}{\phi_{\rm{n}}^2}\right)\right]\to0,
\end{align}
since $k_{\rm{F}}\approx\rho_d2^{1-2/d}\sqrt{\pi}\to0$ as $d\to0$, with $C_{\rm{n}}=C_0(1+C_1)$ and $\phi_{\rm{n}}=\phi_0(1+\phi_1)$. In four dimensions, the exact $\rm{HMT}$ prediction gives $E_{\rm{n}}^{\rm{kin}}(\rho_0)\approx51.8\,\rm{MeV}$, while the $\delta^6$ approximation yields $48.5\,\rm{MeV}$, indicating that higher-order kinetic symmetry energies remain significant at higher dimensions. In one dimension, taking the high-momentum cutoff $\phi_0\to\infty$ introduces an additional $\rm{HMT}$ contribution to the kinetic symmetry energy, given by $k_{\rm{F}}^2/2M_{\rm N}\cdot[8C_0(1+C_1)/3]$\cite{Cai22-dD}, which is positive since $1+C_1>0$. The corresponding contribution to the fourth-order symmetry energy is $4C_0C_1k_{\rm{F}}^2/9M_{\rm N}$\cite{Cai22-dD}, which is generally negative if $C_1<0$. When combining the contributions from the SNM, quadratic, and fourth-order symmetry energies, one finds
\begin{equation}
\Delta E^{\rm{kin}}_{\rm{n},(\rm{1D})}(\rho)=({16k_{\rm{F}}^2}/{9M_{\rm N}})\cdot C_{\rm{n}},~~\rm{as}~~\phi_0\to\infty.
\end{equation}
This then gives the 1D PNM EOS as $E_{\rm{n}}(\rho)=2k_{\rm{F}}^2/3M_{\rm N}+({16k_{\rm{F}}^2}/{9M_{\rm N}})\cdot C_{\rm{n}}$, since the sixth- and higher-order symmetry energies vanish in this limit. Numerically, using the conventional 3D saturation density $\rho_0\approx0.16\,\rm{fm}^{-3}$, one obtains $E_{\rm{n}}(\rho_0)\approx5.03\,\rm{MeV}$ and $\Delta E^{\rm{kin}}_{\rm{n},(\rm{1D})}(\rho_0)\approx1.62\,\rm{MeV}$, which corresponds to an effective 1D density of $\rho_0^{1/3}\approx0.54\,\rm{fm}^{-1}$.

In this subsection, we study the kinetic EOS of ANM in $d$ dimensions, taking into account SRC-induced HMT effects. Complementing conventional approaches, we employ the $\epsilon$-expansion, treating the spatial dimension $d$ as a continuous variable and introducing $\epsilon = d - d_{\rm f}$ as a small perturbative parameter. Historically developed for critical phenomena in the 1970s\cite{Wilson1974,MaSK1976,Wallace1976}, the $\epsilon$-expansion has also been successfully applied in modern contexts, such as the EOS of a unitary Fermi gas, accurately predicting Bertsch's parameter about 0.367 in agreement with experiment\cite{Nishida2006,Arnold2007,Ku2012,Gio08RMP,Blo08RMP}. Applying this method to nuclear matter, we demonstrate that starting from the conventional 3D EOS, one can reasonably approximate the kinetic EOS of both symmetric and pure neutron matter in lower dimensions, including 2D and 1D, with higher-order $\epsilon$-corrections to the symmetry energy generally small. The analysis shows that in reduced dimensions, the nucleon specific energy and pressure decrease, the symmetry energy is softened, and the system behaves increasingly like a FFG, especially when SRC-induced HMTs are included, with predictions from the FFG and HMT models converging for $d \lesssim 3$. Moreover, the parabolic approximation remains valid across dimensions, and the conventional 3D EOS already encapsulates substantial information relevant for near-3D systems. While this subsection is exploratory in nature, it highlights the utility of the $\epsilon$-expansion in bridging knowledge from 3D to low-dimensional nuclear matter and provides analytical formulas for nucleon-specific energy, pressure, incompressibility, skewness, symmetry energy, and higher-order symmetry coefficients in general $d$D. Many open questions remain, including the systematic incorporation of nucleon-nucleon interactions in $d$D, the precise role of momentum-dependent potentials on the symmetry energy, and potential applications of low-dimensional EOSs in astrophysical contexts such as NS crusts or sub-systems in heavy-ion collisions\cite{LiXH2013PLB,LiXH2015PLB,Dan02,Hama1990,LCCX18,Pawel2014,Tews2013,Zarro2009,Rahaman2014,Misner1973}.

There is currently no systematic study on the EOS of ANM in a general dimension $d$, partially due to experimental limitations, theoretical challenges, or lack of interest. While our work presented here is exploratory, it demonstrates the potential usefulness of low-dimensional analyses. Important future directions include the self-consistent incorporation of nucleon-nucleon interactions in $d$D, development of appropriate principles or symmetry constraints, and exploration of the HVH theorem in the presence of HMTs, where the chemical potential is no longer trivially related to the single-nucleon energy\cite{AGD}. Given the close similarity between low-dimensional systems and FFG behavior, higher-order $\epsilon$-expansions may help extract information relevant for the 3D EOS, including correlations between the symmetry energy and its slope parameter. It would also be interesting to explore relativistic effects in combination with nucleon interactions in $d$D and to clarify the momentum dependence of symmetry potentials. Finally, potential applications of low-dimensional EOSs are numerous and exciting, such as modeling the quasi-2D crust or quasi-1D core of NSs and certain sub-systems in heavy-ion collisions. These applications may require modifications to conventional equations, such as the Tolman--Oppenheimer--Volkoff equation, to account for dimensional effects. Overall, the $\epsilon$-expansion provides a practical framework for connecting 3D nuclear matter properties to low-dimensional systems, offering insights into both theoretical structure and potential astrophysical applications, and we anticipate it will stimulate further theoretical and phenomenological investigations.

\subsection{Estimate on the Kinetic EOS due to an Extra HMT of the form $k^{-\sigma}$ with $\sigma>4$}\label{sub_k6_EOSkin}

\indent

In Subsection \ref{sub_nk}, we discussed that the HMT in nuclear systems may deviate from the $k^{-4}$ form\cite{Ant88}, as also illustrated in FIG.\,\ref{fig_k4n} and the related discussion around Eq.\,(\ref{for-kn6}). In this subsection, we investigate this issue by assuming that the single-nucleon momentum distribution in SNM takes the form\footnote{For simplicity, we focus only on SNM in this subsection and denote the distribution as $n_{\v{k}}$, which corresponds to $n_{\v{k}}^0$ in general. Similarly, $\Delta\leftrightarrow\Delta_0$, $C\leftrightarrow C_0$, $D\leftrightarrow D_0$ and $\phi\leftrightarrow\phi_0$.}  
\begin{equation}\label{def-nk-D}
n_{\v{k}}=
\begin{cases}
\Delta, & k \leq k_{\rm F},\\
C \left(k_{\rm F}/k\right)^4, & k_{\rm F} \leq k \leq \phi k_{\rm F},\\
D \left(k_{\rm F}/k\right)^{\sigma}, & k \geq \phi k_{\rm F}.
\end{cases}
\end{equation}
As previously noted, the standard $k^{-4}$ behavior of the HMT does not necessarily persist at very high momenta (e.g., $k \gtrsim 2.5 k_{\rm F}$), where the distribution may instead follow a modified trend. Choosing $\sigma=6$ provides an effective parametrization that incorporates more intricate nucleon-nucleon interactions\cite{Miller18PLB,Ant88}, potentially including contributions from three-body correlations.
Here $\sigma \geq 6$ is taken as an integer, and according to Eq.\,(\ref{LS-nk}), only even values of $\sigma$ appear\cite{Miller18PLB}. In the analysis below, we employ the conventional contact parameter $C$ with its established value $0.16 \pm 0.015$, while the parameter $D$ represents the contribution from the higher-order contact term.

In the presence of the $D$-term contribution, the high-momentum fraction in SNM takes the form
\begin{equation}
x_{\rm{HMT}}=3C\left(1-\frac{1}{\phi}\right)+\frac{3D}{\sigma-3}\phi^{3-\sigma},
\end{equation}
and it satisfies $\Delta+x_{\rm{HMT}}=1$. Inverting this relation gives the expression for the coefficient $D$ in terms of $C$, $\phi$ and $\sigma$:
\begin{equation}\label{def_D}
D=\frac{1}{3}\frac{\sigma-3}{\phi^{3-\sigma}}\left[x_{\rm{HMT}}-3C\left(1-\frac{1}{\phi}\right)\right].
\end{equation}
While the high-momentum cutoff $\phi$ is in principle a free parameter, one can constrain its physically reasonable range. Since the coefficient $D$, which characterizes the part of the HMT beyond Tan's $k^{-4}$ form\cite{Tan08-a,Tan08-b,Tan08-c}, should naturally be positive, a first constraint arises,
\begin{equation}\label{def_phi1}
\phi\leq\phi_{\max}\equiv\left(1-\frac{x_{\rm{HMT}}}{3C}\right)^{-1},
\end{equation}
assuming $x_{\rm{HMT}}/3C<1$. For instance, taking $x_{\rm{HMT}}\approx28\%$ and $C\approx0.16$ gives $x_{\rm{HMT}}/3C\approx0.583$ and hence $\phi_{\max}\approx2.4$, the value used in the preceding subsections. At the transition momentum $k=\phi k_{\rm{F}}$, the momentum distribution takes the two values $C/\phi^4$ and $D/\phi^{\sigma}$ from the two branches below and above $\phi k_{\rm{F}}$. Requiring that $n_{\v{k}}$ decrease with $k$ leads to $C/\phi^4\geq D/\phi^{\sigma}$, which provides a second constraint,
\begin{equation}\label{def_phi2}
\phi\geq \phi_{\min}\equiv \frac{\sigma-4}{\sigma-3}\cdot\left(1-\frac{x_{\rm{HMT}}}{3C}\right)^{-1}
=\frac{\sigma-4}{\sigma-3}\cdot\phi_{\max}.
\end{equation}
Together, (\ref{def_phi1}) and (\ref{def_phi2}) restrict the cutoff to an effective range $
\phi_{\min}\leq\phi\leq\phi_{\max}$, with $\delta\phi\equiv\phi_{\max}-\phi_{\min}=\phi_{\max}/(\sigma-3)$.

\begin{figure}[h!]
\centering
\includegraphics[width=7.cm]{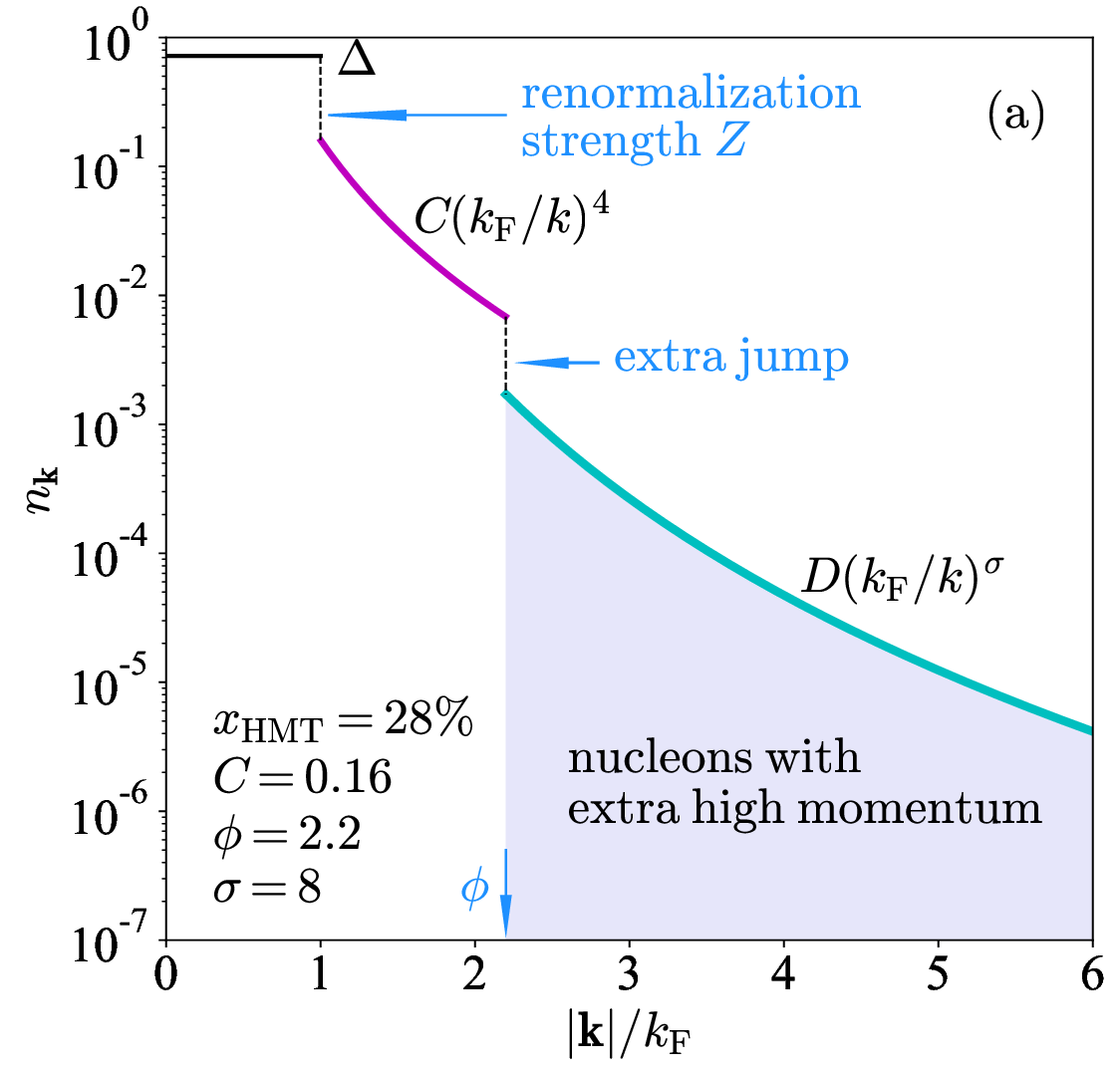}\\
\includegraphics[width=7.cm]{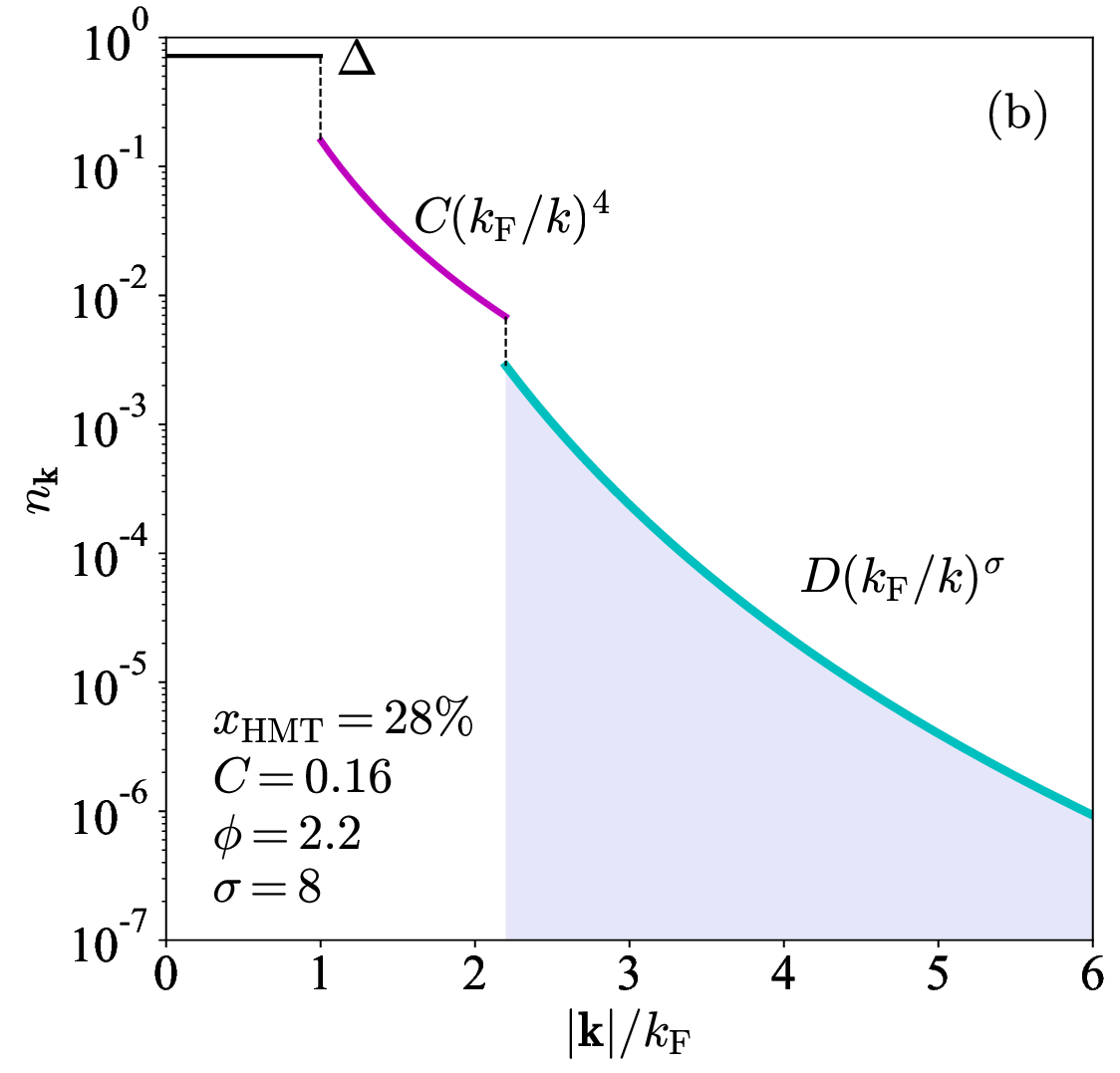}
\caption{(Color Online). Momentum distribution $n_{\v{k}}^0$ with two different values of $\sigma$, $C\approx0.16$ and $x_{\rm{HMT}}\approx28\%$ are fixed.}
\label{fig_nk0}
\end{figure}

The upper bound $\phi_{\max}$ is independent of $\sigma$, whereas the lower bound $\phi_{\min}$ depends on it. For example, $\sigma=6$ gives $2/3\leq\phi/\phi_{\max}\leq1$, and $\sigma=8$ gives $4/5\leq\phi/\phi_{\max}\leq1$. As $\sigma$ increases, the allowed interval for $\phi$ narrows as $\phi_{\max}/(\sigma-3)$. FIG.\,\ref{fig_nk0} illustrates the single-nucleon momentum distribution for two values of $\sigma$, using $x_{\rm{HMT}}\approx28\%$, $C\approx0.16$\cite{Cai15a}, and a representative $\phi=2.2$ for demonstration. Here the normalization strength at the Fermi surface is $Z=\Delta-C$, inversely proportional to the effective mass\cite{Migdal57}, and the discontinuity at $k=\phi k_{\rm{F}}$ is $Y=C/\phi^4-D/\phi^{\sigma}$, which we required to be non-negative when deriving $\phi_{\min}$. With fixed $\sigma$, the coefficient $D$ increases monotonically as $\phi$ decreases from $\phi_{\max}$ to $\phi_{\min}$ as shown by Eq.\,(\ref{def_D}); for fixed $\phi$, $D$ is larger for larger $\sigma$. For example, $1.6\lesssim\phi\lesssim2.4$ for $\sigma=6$, and $1.92\lesssim\phi\lesssim2.4$ for $\sigma=8$ (using $C\approx0.16$ and $x_{\rm{HMT}}\approx28\%$). The high-momentum fraction contributed specifically by the $D$-term is
\begin{equation}
x_{\rm{HMT}}^{(D)}=1-x_{\rm{HMT}}^{(C)}=1-3C\left(1-\phi^{-1}\right),
\end{equation}
which does not depend on $\sigma$.
The kinetic EOS of SNM is
\begin{align}\label{E0kin-2}
E_0^{\rm{kin}}(\rho)=&\frac{1}{\rho}\frac{4}{(2\pi)^3}\int_0^{\infty}\frac{\v{k}^2}{2M_{\rm N}} n_{\v{k}}\d\v{k}\notag\\
=&\frac{3}{5}\frac{k_{\rm{F}}^2}{2M_{\rm N}}\Bigg[
1-x_{\rm{HMT}}+5C(\phi-1)\notag\\
&\hspace{0.cm}+\frac{5\phi^2}{3}\frac{\sigma-3}{\sigma-5}\left[x_{\rm{HMT}}-3C\left(1-\frac{1}{\phi}\right)\right]\Bigg],
\end{align}
which becomes divergent as $\phi\to\infty$ if $\sigma\leq5$. This expression is used at densities around $\rho\approx\rho_0\approx0.16\,\rm{fm}^{-3}$, and its dependence on $\phi$ is non-trivial, with the allowable range of $\phi$ itself depending on $\sigma$.
The lower cutoff $\phi_{\min}$ given in Eq.\,(\ref{def_phi2}) can also be obtained directly from the kinetic EOS. Since (\ref{def_D}) shows that $D$ increases as $\phi$ decreases, the kinetic EOS also increases accordingly. The kinetic EOS therefore reaches its maximum when the $D$-branch of $n_{\v k}$ becomes linked to the $C$-branch, corresponding to the condition
\begin{equation}
\left.\frac{\partial}{\partial\phi}E_0^{\rm{kin}}(\rho_0)\right|_{\phi=\phi_{\min}}=0.
\end{equation}
This leads to $\phi_{\min}=[(\sigma-4)/(\sigma-3)]\cdot\phi_{\max}$, see Eq.\,(\ref{def_phi2}), and for $\sigma\geq6$ one has
\begin{equation}
\left.\frac{\partial^2}{\partial\phi^2}E_0^{\rm{kin}}(\rho_0)\right|_{\phi=\phi_{\min}}
=-\frac{3\overline{k}_{\rm{F}}^2}{M_{\rm N}}\frac{C}{\phi_{\max}}\frac{\sigma-3}{\sigma-5}<0,
\end{equation}
with $\overline{k}_{\rm{F}}=k_{\rm{F}}(\rho_0)$, confirming that $\phi_{\min}$ indeed corresponds to the maximum of $E_0^{\rm{kin}}(\rho_0)$. In addition to the representative $\sigma=6$ case, we also consider $\sigma=8$, which has appeared in an earlier work\cite{Ant07PRC}.

\begin{figure}[h!]
\centering
\includegraphics[width=7.cm]{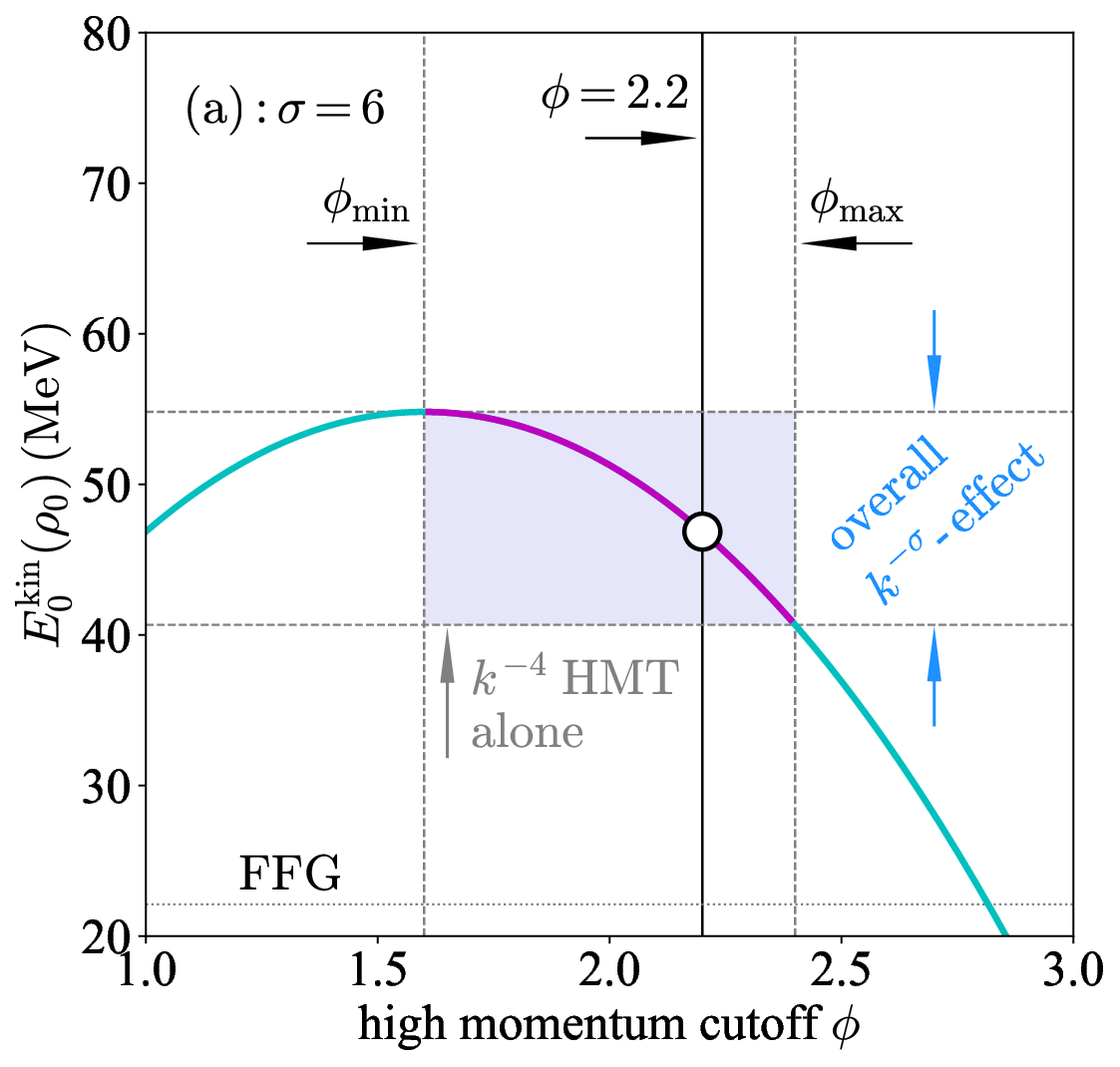}\\
\includegraphics[width=7.cm]{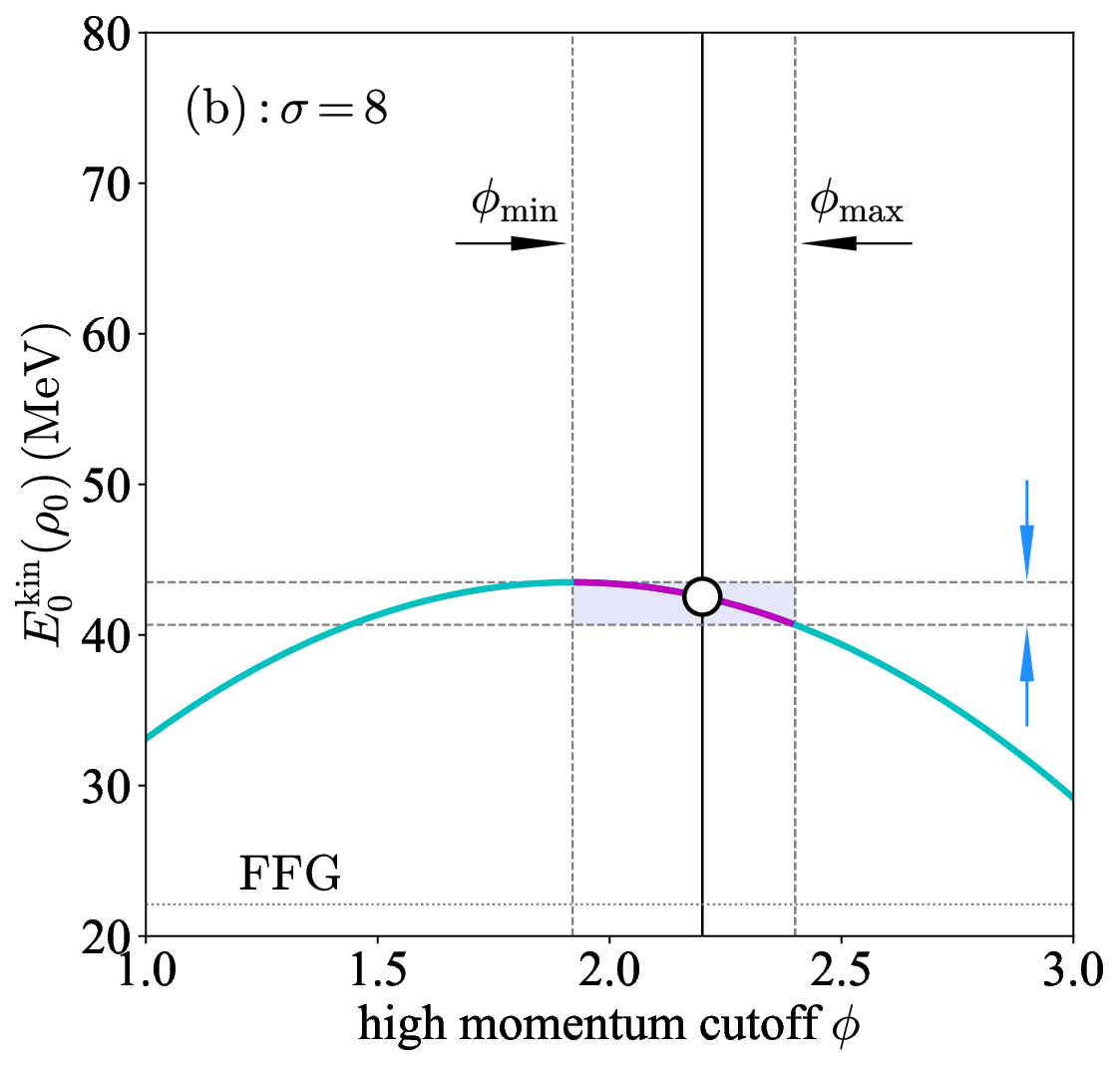}
\caption{(Color Online). Kinetic EOS of SNM at $\rho_0$ as a function of $\phi$ corresponding to FIG.\,\ref{fig_nk0}, $C\approx0.16,x_{\rm{HMT}}\approx28\%$ are fixed.}
\label{fig_E0}
\end{figure}

The dependence of $E_0^{\rm{kin}}(\rho_0)$ on the high-momentum fraction $x_{\rm{HMT}}$ is linear,
\begin{equation}
\frac{\partial}{\partial x_{\rm{HMT}}}E_0^{\rm{kin}}(\rho_0)
=\frac{3}{5}\frac{\overline{k}_{\rm{F}}^2}{2M_{\rm N}}\left(-1+\frac{5\phi^2}{3}\frac{\sigma-3}{\sigma-5}\right).
\end{equation}
The negative ``$-1$'' term shows that increasing $x_{\rm{HMT}}$ reduces the contribution from the zero-momentum depletion ($\Delta=1-x_{\rm{HMT}}$), while the positive term $5\phi^2/3\cdot(\sigma-3)/(\sigma-5)$ shows that the $k^{-\sigma}$ tail increases the kinetic energy. The part from the conventional $k^{-4}$ tail characterized by $5C(\phi-1)$ is independent of $x_{\rm{HMT}}$. The net effect is positive, i.e., $\partial E_0^{\rm{kin}}(\rho_0)/\partial x_{\rm{HMT}}>0$, since typically $\phi^2\gtrsim2$ and $(\sigma-3)/(\sigma-5)>1$ for $\sigma\geq6$. A smaller $x_{\rm{HMT}}$ gives a smaller $\phi_{\max}$, e.g., $\phi_{\max}=2$ for $x_{\rm{HMT}}\approx24\%$ with $C\approx0.16$, which implies $\phi_{\min}\approx1.33$ for $\sigma=6$ and $\phi_{\min}=1.6$ for $\sigma=8$. The corresponding kinetic EOS values span $26.3\,\rm{MeV}\lesssim E_0^{\rm{kin}}(\rho_0)\lesssim67.4\,\rm{MeV}$ for $24\%\lesssim x_{\rm{HMT}}\lesssim32\%$ with $\sigma=6$, and $31.5\,\rm{MeV}\lesssim E_0^{\rm{kin}}(\rho_0)\lesssim53.5\,\rm{MeV}$ with $\sigma=8$. Thus extending $x_{\rm{HMT}}$ from 24\% to 32\% changes the kinetic EOS by approximately $41.1\,\rm{MeV}$ for $\sigma=6$ or $22.0\,\rm{MeV}$ for $\sigma=8$. Although $E_0^{\rm{kin}}(\rho_0)$ at $x_{\rm{HMT}}\approx24\%$ is larger for $\sigma=8$, the slope $\partial E_0^{\rm{kin}}(\rho_0)/\partial x_{\rm{HMT}}$ is larger for $\sigma=6$.
For fixed $x_{\rm{HMT}}$ and $\phi$, the kinetic EOS decreases as $\sigma$ increases, which follows from the factor $(\sigma-3)/(\sigma-5)$ in Eq.\,(\ref{E0kin-2}). Its limit as $\sigma\to\infty$ is 1, and its maximum for $\sigma\geq6$ is 3 at $\sigma=6$. In the limit $\sigma\to\infty$, one has $\phi_{\min}\to\phi_{\max}=1/(1-x_{\rm{HMT}}/3C)$ and $\delta\phi\to0$, so $\phi$ is fixed to $\phi_{\max}$ and the contribution from the extra HMT vanishes. These features appear in FIG.\,\ref{fig_E0}, where for example $E_0^{\rm{kin}}(\rho_0)\approx46.9\,\rm{MeV}$ for $\sigma=6$ with $\phi=2.2$. The corresponding ranges are $40.7\,\rm{MeV}\lesssim E_0^{\rm{kin}}(\rho)\lesssim54.8\,\rm{MeV}$ for $\sigma=6$ and $40.7\,\rm{MeV}\lesssim E_0^{\rm{kin}}(\rho)\lesssim43.5\,\rm{MeV}$ for $\sigma=8$, where the lower value $40.7\,\rm{MeV}$ corresponds to the pure $k^{-4}$ tail. The extra HMT always enhances the kinetic EOS because it allows integration to arbitrarily large momenta without divergence.

\begin{figure}[h!]
\centering
 \includegraphics[width=7.cm]{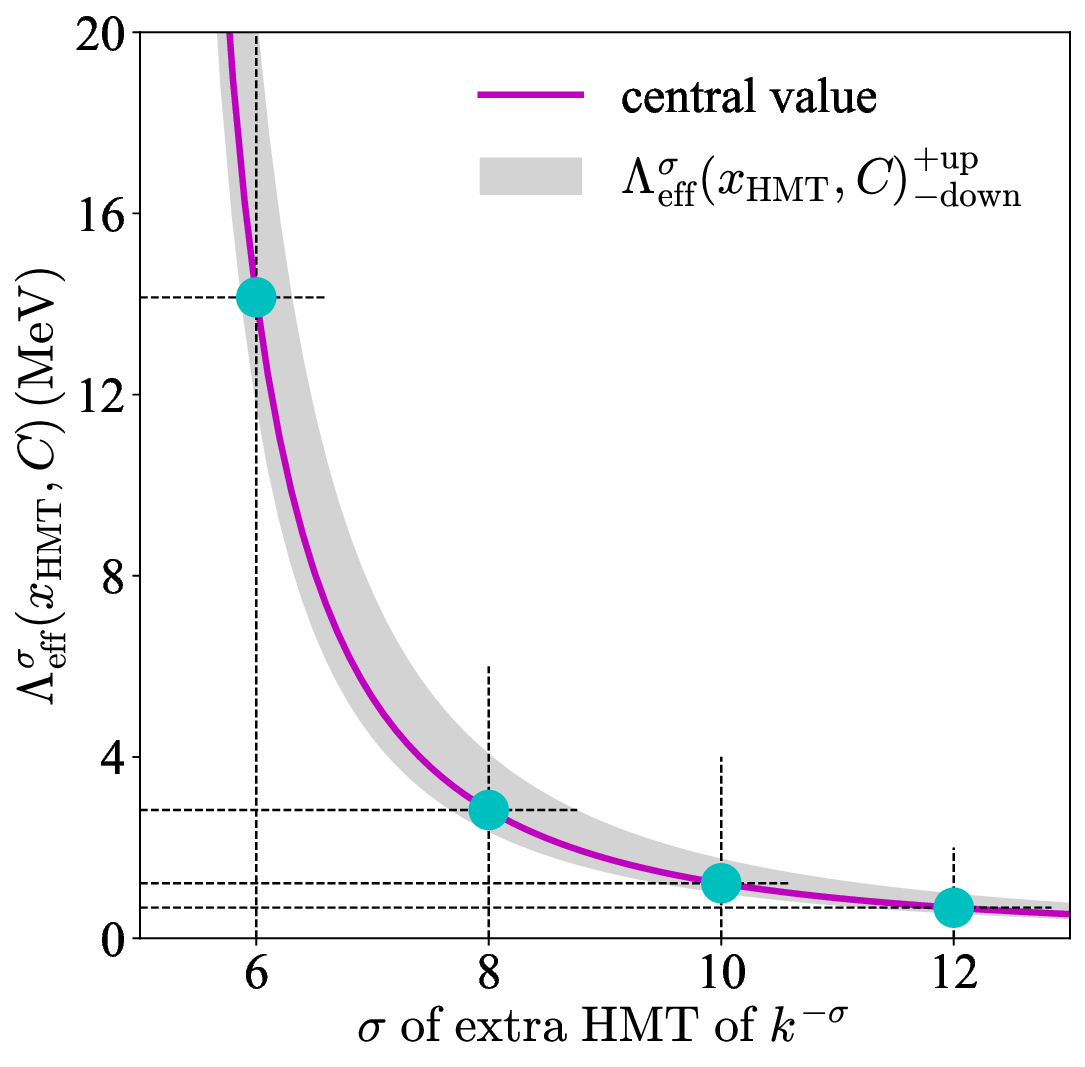}
\caption{(Color Online). Effective contribution $\Lambda_{\rm{eff}}(\sigma,C)$ to the kinetic EOS from $k^{-\sigma}$ tail as a function of $\sigma$, $C\approx0.16\pm0.015,x_{\rm{HMT}}\approx28\%\pm4\%$.}
\label{fig_DE-sigma}
\end{figure}

An important quantity is the correction on the $E_0^{\rm{kin}}(\rho_0)$ due to the extra HMT of the form $k^{-\sigma}$, compared with the $E_0^{\rm{kin}}(\rho_0)$ adopting only the conventional $k^{-4}$ HMT.
The effective contribution $\Lambda_{\rm{eff}}^{\sigma}(x_{\rm{HMT}},C)
\equiv E_0^{\rm{kin}}(\rho_0,\phi_{\min})- E_0^{\rm{kin}}(\rho_0,\phi_{\max})$ could be written explicitly as,
\begin{equation}
\Lambda_{\rm{eff}}^{\sigma}(x_{\rm{HMT}},C)
=\frac{1}{2}\left(\frac{1}{\sigma-5}-\frac{1}{\sigma-3}\right)\frac{9C^2}{3C-x_{\rm{HMT}}}\frac{\overline{k}_{\rm{F}}^2}{2M_{\rm N}}.
\end{equation}
Consequently, the effective contribution from the $k^{-\sigma}$ is about $\Lambda_{\rm{eff}}^{\sigma}(x_{\rm{HMT}},C)=54.8\,\rm{MeV}-40.7\,\rm{MeV}\approx14.1\,\rm{MeV}$, see the upper panel of FIG.\,\ref{fig_E0} (indicated by the light-blue arrows), since as the high momentum cutoff $\phi$ varies the corresponding decomposition of the kinetic EOS also varies.
Similarly, the effective contribution from the $k^{-\sigma}$ tail with $\sigma=8$ is about $\Lambda_{\rm{eff}}^{\sigma}(x_{\rm{HMT}},C)=43.5\,\rm{MeV}-40.7\,\rm{MeV}\approx2.8\,\rm{MeV}$, see the lower panel of FIG.\,\ref{fig_E0}. Naturally one has $\lim_{\sigma\to\infty}\Lambda_{\rm{eff}}^{\sigma}(x_{\rm{HMT}},C)=0\,\rm{MeV}$, see FIG.\,\ref{fig_DE-sigma}.
Considering the range spanned by $C$ and that by $x_{\rm{HMT}}$, one can find that the contribution $14.1\,\rm{MeV}$ for $\sigma=6$ is not the maximum.
In order to find the maximum effects from the extra HMT on the $E_0^{\rm{kin}}(\rho_0)$, we notice the derivatives,
\begin{align}
\frac{\partial}{\partial x_{\rm{HMT}}}\left(\frac{C^2}{3C-x_{\rm{HMT}}}\right)=&\frac{C^2}{(3C-x_{\rm{HMT}})^2},\\
\frac{\partial}{\partial C}\left(\frac{C^2}{3C-x_{\rm{HMT}}}\right)=&\frac{2(3C-2x_{\rm{HMT}})}{(3C-x_{\rm{HMT}})^2},
\end{align}
where $C^2/(3C-x_{\rm{HMT}})$ is the relevant factor appearing in $\Lambda_{\rm{eff}}^{\sigma}(x_{\rm{HMT}},C)$.
The first relation tells that $x_{\rm{HMT}}\approx32\%$ should be taken for maximizing $\Lambda_{\rm{eff}}^\sigma(x_{\rm{HMT}},C)$, while the second relation indicates that the derivative is negative as $3C-0.64<0$ for $0.145\leq C\leq 0.175$\cite{Cai15a}.  Consequently $C\approx0.145$ should be adopted for maximizing $\Lambda_{\rm{eff}}^{\sigma}(x_{\rm{HMT}},C)$.
Thus, by taking $C\approx0.145$ and $x_{\rm{HMT}}\approx32\%$, one obtains $\Lambda_{\rm{eff}}^{6}(x_{\rm{HMT}},C)\lesssim\Lambda_{\rm{eff}}^{6}(0.32,0.145)\approx20.2\,\rm{MeV}$ with $59.6\,\rm{MeV}\lesssim E_0^{\rm{kin}}(\rho_0)\lesssim 79.8\,\rm{MeV}$ and $2.52\lesssim\phi\lesssim3.78$.
By comparing with the $E_0^{\rm{kin}}(\rho_0)\approx59.6\,\rm{MeV}$ only adopting the $k^{-4}$ HMT, the enhancement is about $20.2/59.6\approx33.9\%$ (which is not the maximum relative enhancement).
Similarly, we have for $\sigma=8$ that $59.6\,\rm{MeV}\lesssim E_0^{\rm{kin}}(\rho_0)\lesssim 63.7\,\rm{MeV}$, $3.03\lesssim\phi\lesssim3.78$ and $\Lambda_{\rm{eff}}^{8}(x_{\rm{HMT}},C)\lesssim\Lambda_{\rm{eff}}^{\sigma}(0.32,0.145)\approx4.1\,\rm{MeV}$.
See FIG.\,\ref{fig_DE-sigma} for the uncertainties on the $\Lambda_{\rm{eff}}^{\sigma}(x_{\rm{HMT}},C)$, i.e., $\Lambda_{\rm{eff}}^{\sigma}(x_{\rm{HMT}},C)_{-\rm{down}}^{+\rm{up}}$, e.g., we have $\Lambda_{\rm{eff}}^{6}(x_{\rm{HMT}},C)\approx14.1_{-2.2}^{+6.1}\,\rm{MeV}$ and $\Lambda_{\rm{eff}}^{8}(x_{\rm{HMT}},C)\approx2.8_{-0.4}^{+1.3}\,\rm{MeV}$.

\begin{figure}[h!]
\centering
 \includegraphics[width=8.5cm]{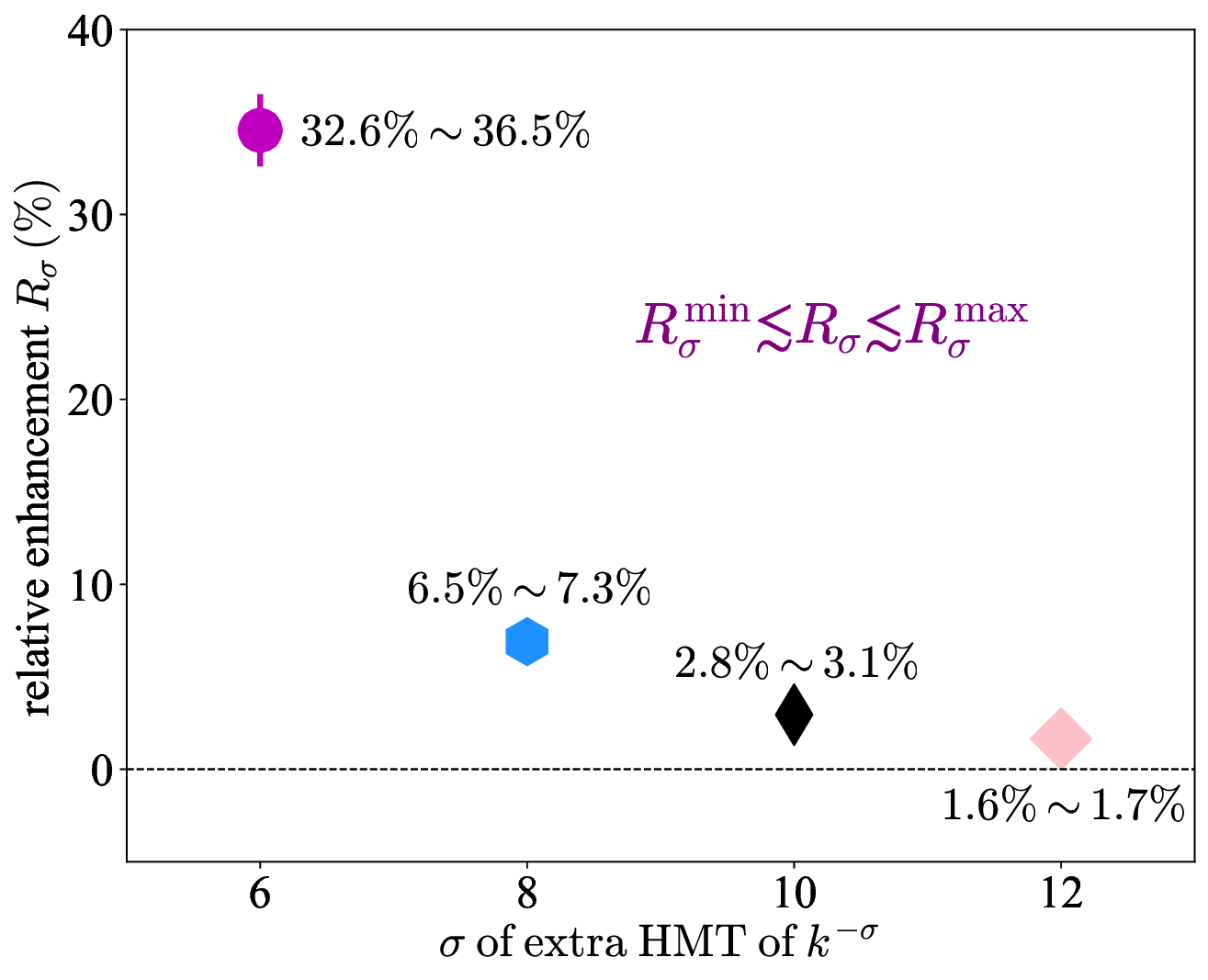}
\caption{(Color Online). Relative enhancement $R_{\sigma}$ as a function of $\sigma$, $0.145\lesssim C\lesssim 0.175,24\%\lesssim x_{\rm{HMT}}\lesssim 32\%$.}
\label{fig_Lameff-R}
\end{figure}

Similarly, in order to analyze the relative contribution of the extra HMT (compared with the kinetic EOS only with the conventional HMT of the form $k^{-4}$), one needs to consider the ratio between the $E_0^{\rm{kin},(D)}(\rho_0)$ and the kinetic EOS with only the HMT of the form $k^{-4}$, i.e.,
\begin{align}
R_{\sigma}\equiv&{E_0^{\rm{kin},(D)}(\rho_0)}\left/\left[\frac{3k_{\rm{F}}^2}{10M_{\rm N}}\cdot\left(1-x_{\rm{HMT}}+5C(\phi_{\max}-1)\right)\right]\right.\notag\\
\sim&\frac{C^2}{3C-x_{\rm{HMT}}}\frac{1}{1-x_{\rm{HMT}}+5C(\phi_{\max}-1)},
\end{align}
where $\phi_{\max}=3C/(3C-x_{\rm{HMT}})$.
Straightforward calculations give $\partial R_{\sigma}/\partial C>0$ and $\partial R_{\sigma}/\partial x_{\rm{HMT}}>0$ for $0.145\lesssim C\lesssim 0.175$ and $24\%\lesssim x_{\rm{HMT}}\lesssim32\%$, indicating in order to obtain the maximum value of $R_{\sigma}$, one needs to take $x_{\rm{HMT}}\approx32\%$ as well as $C\approx0.175$.
The corresponding results are $R_{\sigma}^{\max}\approx36.5\%$ with $45.2\,\rm{MeV}\lesssim E_0^{\rm{kin}}(\rho_0)\lesssim 61.7\,\rm{MeV}$ and $\Lambda_{\rm{eff}}^\sigma(x_{\rm{HMT}},0.175)\approx16.5\,\rm{MeV}$ for $\sigma=6$, and $R_{\sigma}^{\max}\approx7.3\%$ with $45.2\,\rm{MeV}\lesssim E_0^{\rm{kin}}(\rho_0)\lesssim 48.5\,\rm{MeV}$ and $\Lambda_{\rm{eff}}^\sigma(x_{\rm{HMT}},0.175)\approx3.3\,\rm{MeV}$ for $\sigma=8$.
It is obvious that the absolute value of $\Lambda_{\rm{eff}}^\sigma(x_{\rm{HMT}},0.175)$ is smaller than that of $\Lambda_{\rm{eff}}^\sigma(x_{\rm{HMT}},0.145)$, but the relative contribution is maximized.
The reason is simple: when $C$ and $x_{\rm{HMT}}$ vary, the $E_0^{\rm{kin}}(\rho_0)$ adopting only the $k^{-4}$ HMT may also change correspondingly.
FIG.\,\ref{fig_Lameff-R} shows the relative enhancement $R_{\sigma}$ as a function of $\sigma$ of the extra HMT in the form $k^{-\sigma}$, here $R_{\sigma}^{\max}$ is the $R_{\sigma}$ under $C\approx0.175$ and $x_{\rm{HMT}}\approx32\%$ while $R_{\sigma}^{\min}$ is the $R_{\sigma}$ under $C\approx0.145$ and $x_{\rm{HMT}}\approx24\%$.
Numerically, one has $32.6\%\lesssim R_{\sigma}\lesssim36.5\%$ for $\sigma=6$ and $6.5\%\lesssim R_{\sigma}\lesssim 7.3\%$ for $\sigma=8$.
The results for $\sigma=10$ and $\sigma=12$ on $R_{\sigma}$ are also shown in FIG.\,\ref{fig_Lameff-R}.
Similarly the maximum absolute contribution for $\sigma=10$ and $\sigma=12$ could also be obtained, and they are 1.7\,MeV and 1.0\,MeV, respectively.
As discussed in Subsection \ref{sub_nk}, the maximum high momentum cutoff $\phi_{\max}$ for the $k^{-4}$ form of the contact may physically be less then 2.
If $\phi_{\max}\approx1.5$, then $x_{\rm{HMT}}/C\approx1$ since $\phi_{\max}=1/(1-x_{\rm{HMT}}/3C)$. If one artificially adopts $x_{\rm{HMT}}\approx28\%\pm4\%$ and the approximated relation $x_{\rm{HMT}}/C\approx1$, then $44.1\%\lesssim R_{\sigma}\lesssim 54.1\%$ for $\sigma=6$ while $8.8\%\lesssim R_{\sigma}\lesssim 10.8\%$ for $\sigma=8$, both are sizable. The contribution from the $k^{-6}$ HMT is even comparable to that from the conventional $k^{-4}$ HMT in this case. On the other hand, if a smaller $x_{\rm{HMT}}$ is adopted, e.g., the one from the self-consistent Green's function theories, about $x_{\rm{HMT}}\approx10\%\pm5\%$ (together with $\phi_{\max}\approx1.5$), the $R_{\sigma}$ for $\sigma=6$ is found to be about $11.6\%\lesssim R_{\sigma}\lesssim 30.6\%$, and $2.3\%\lesssim R_{\sigma}\lesssim 6.1\%$ for $\sigma=8$.

\begin{figure}[h!]
\centering
 \includegraphics[width=7.0cm]{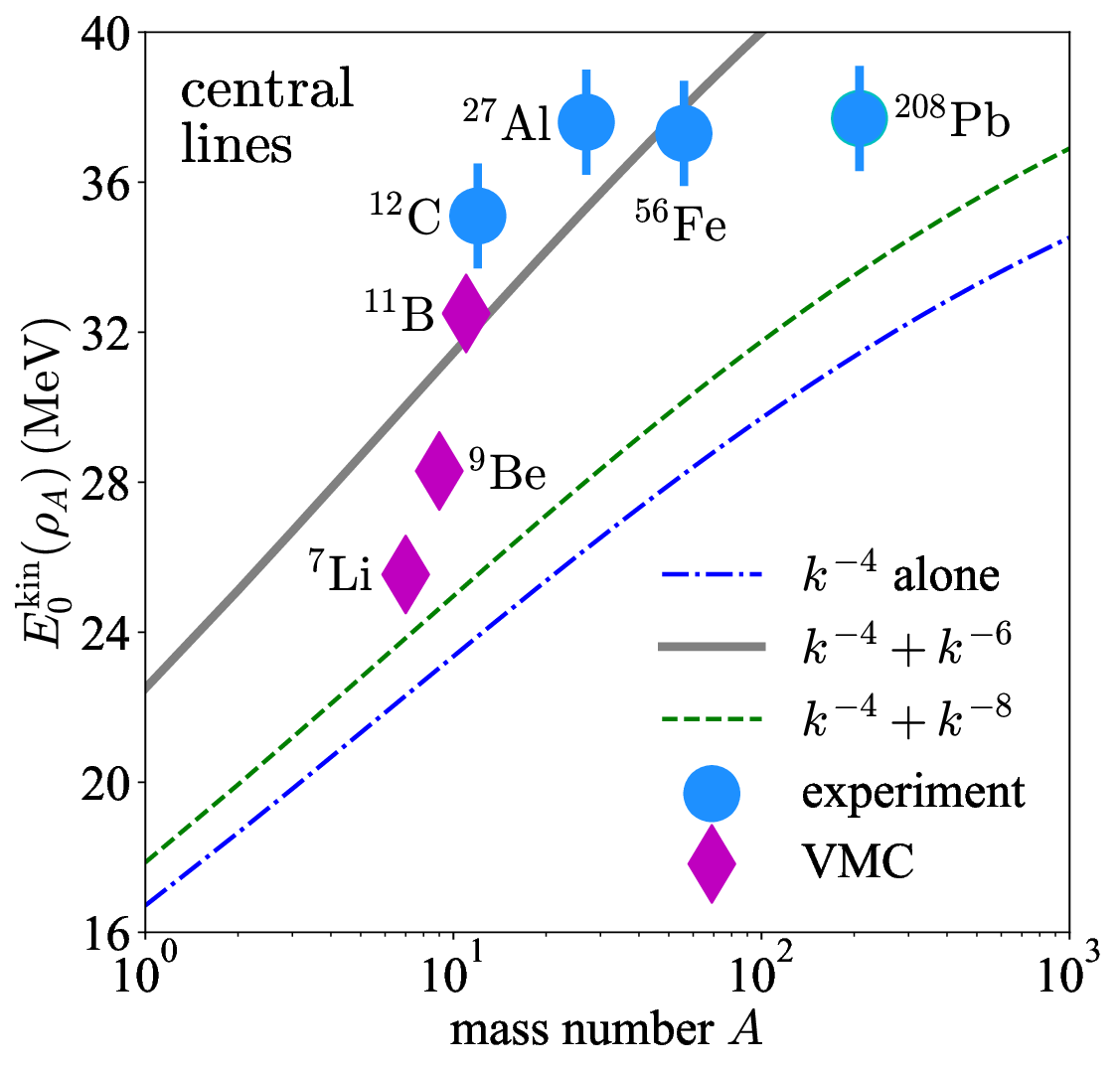}
\caption{(Color Online). Comparison on the average kinetic energy for finite nuclei with experiments (cyan solid circles) and those from VMC calculations (red diamonds), where $C\approx0.16$ and $x_{\rm{HMT}}\approx28\%$ are used at $\rho_0\approx0.16\,\rm{fm}^{-3}$, $\rho_A/\rho_0=(1+\alpha/\beta A^{1/3})^{-1}$, $\phi_{\min}$ is adopted for $k^{-6}$ or $k^{-8}$ HMTs.}
\label{fig_HMT-sigma-EXP}
\end{figure}

\begin{figure}[h!]
\centering
 \includegraphics[width=7.cm]{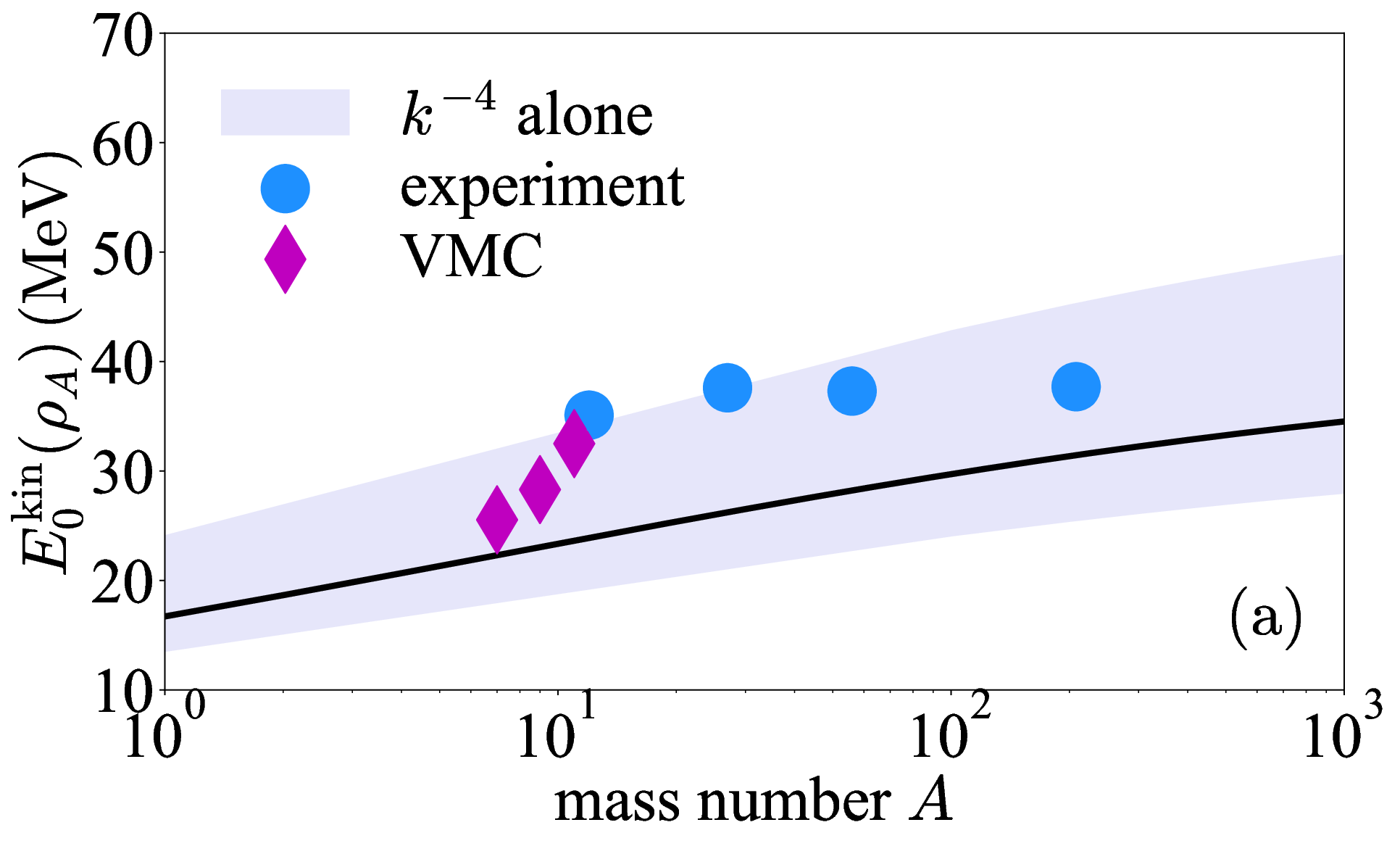}\\
 \includegraphics[width=7.cm]{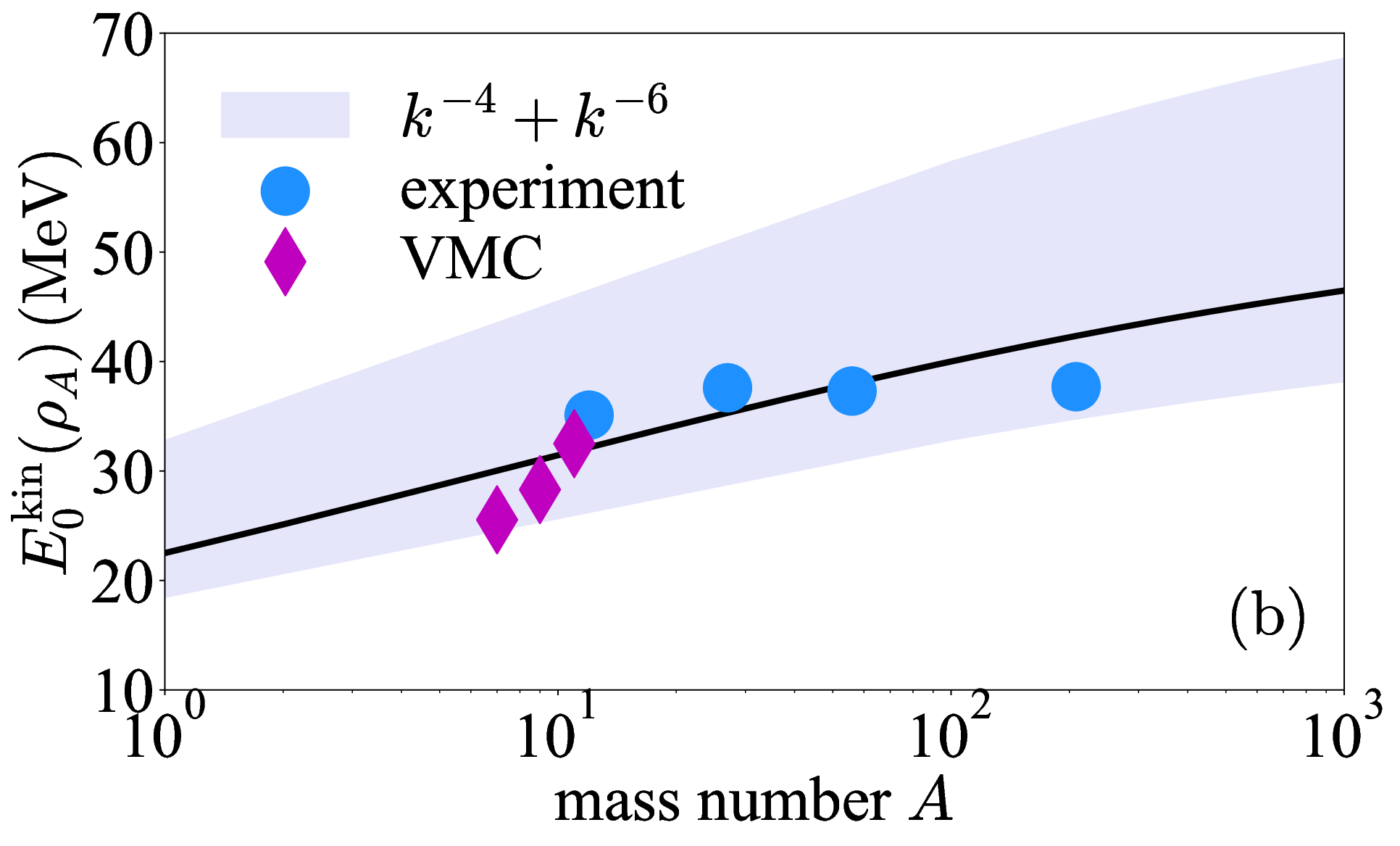}\\
 \includegraphics[width=7.cm]{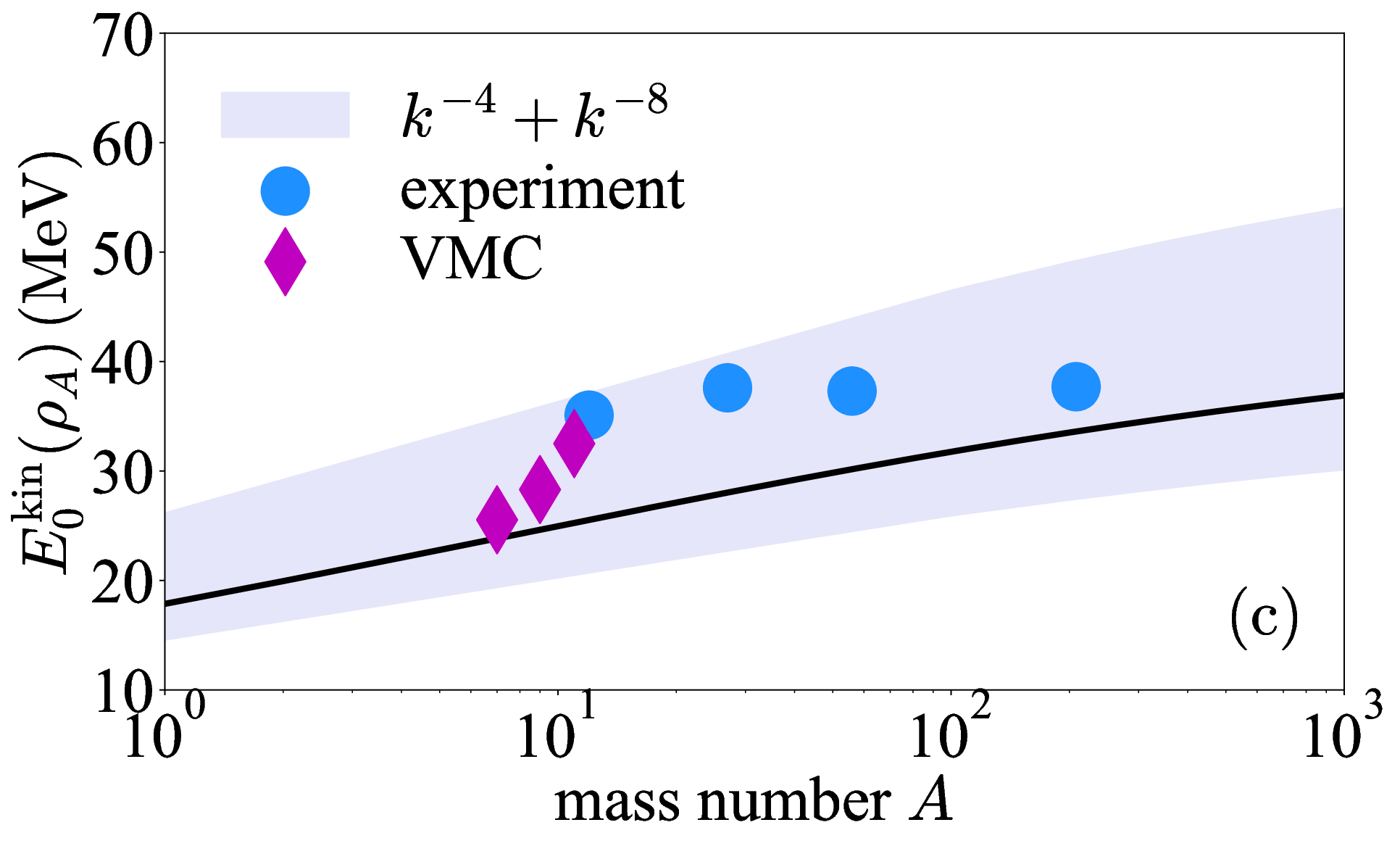}
\caption{(Color Online). Same as FIG.\,\ref{fig_HMT-sigma-EXP} but consider the uncertainties on $C$ and $x_{\rm{HMT}}$, i.e., $C\approx0.16\pm0.015$ and $x_{\rm{HMT}}\approx28\%\pm4\%$.}
\label{fig_HMT-sigma-EXP-band}
\end{figure}

In order to further study the extra HMT of the form $k^{-\sigma}$, we compare the predictions on the kinetic energy of finite nuclei with those extracted from experiments, including $^{12}\rm{C}$, $^{27}\rm{Al}$, $^{56}\rm{Fe}$ and $^{208}\rm{Pb}$\cite{Hen14} as well as from VMC calculations\cite{Wir14}, including $^7\rm{Li}$, $^9\rm{Be}$ and $^{11}\rm{B}$.
The average nucleon kinetic energy in finite nucleus is obtained as $\langle T\rangle=[\langle T_{\rm{n}}\rangle(1+\delta)+\langle T_{\rm{p}}\rangle(1-\delta)]/2$, where $\langle T_{\rm{p}}\rangle$ and $\langle T_{\rm{n}}\rangle$ are the average kinetic energies for protons and neutrons\cite{Hen14}.
For finite nucleus, the isospin asymmetry $\delta$ is generally small, this means one can approximate the kinetic EOS $E_{\rm{ANM}}^{\rm{kin}}(\rho_A)\approx E_0^{\rm{kin}}(\rho_A)+\delta^2E_{\rm{sym}}^{\rm{kin}}(\rho_A)$ simply as $E_0^{\rm{kin}}(\rho_A)$, here the (average-finite-nucleus) density $\rho_A$ could be obtained from $\rho_0$ via $\rho_A/\rho_0\approx (1+\alpha/A^{1/3})^{-1}$ with $\alpha/\beta\approx2.8$\cite{Cai16c}, see Subsection \ref{sub_WaleckaSRC} for details on these quantities.
The comparison is shown in FIG.\,\ref{fig_HMT-sigma-EXP} where the central values for $C$ and $x_{\rm{HMT}}$, i.e., $C\approx0.16$ and $x_{\rm{HMT}}\approx28\%$ are adopted at $\rho_0\approx0.16\,\rm{fm}^{-3}$, and the discrete points are shown for $\langle T\rangle$.
From the figure one can find that the kinetic EOS with HMT in the form of $k^{-4}$ alone slightly under-estimates the experimental data\cite{Cai16c}, and in addition, the extra HMT taking the form of $k^{-6}$ performs better than the HMT with the form $k^{-8}$.
If one considers the uncertainties on the parameters $C$ and $x_{\rm{HMT}}$, i.e., $C\approx0.16\pm0.015$ and $x_{\rm{HMT}}\approx28\%\pm4\%$, the results are shown in FIG.\,\ref{fig_HMT-sigma-EXP-band}.
Finally, even if a small value for $\alpha/\beta\approx2.0$ is adopted\cite{Dan03}, the prediction from the model including the extra HMT with form of $k^{-6}$ or $k^{-8}$ performs better than the model with the conventional $k^{-4}$ HMT alone.

Finally, we emphasize that the treatment of $n_{\v k}$ in Eq.\,(\ref{def-nk-D}) is merely a pragmatic approximation, since the momentum dependence of the distribution at large $k$ is very complicated. Nevertheless, it provides a useful estimate of how the HMT $k^{-\sigma}$ (with $\sigma>4$) affects the kinetic energy in finite nuclei\cite{Hen14}; in particular, such a tail avoids the divergence of $E^{\rm{kin}}_0(\rho)$ when performing the integration. Very similarly, one may introduce multiple extra high-momentum components into the distribution function, namely
\begin{equation}\label{def_nk0DD}
n_{\v{k}}=\left\{
\begin{array}{ll}
\Delta,&k\leq k_{\rm F},\\
C(k_{\rm F}/k)^4,&k_{\rm F}\leq k\leq \phi k_{\rm F},\\
D_1(k_{\rm F}/k)^{\sigma_1},&\phi k_{\rm F}\leq k\leq \phi_1 k_{\rm F},\\
D_2(k_{\rm F}/k)^{\sigma_2},&\phi_1 k_{\rm F}\leq k\leq \phi_2 k_{\rm F},\\
\cdots,&
\end{array}
\right.
\end{equation}
with the contact coefficients $D_1,D_2,\cdots$ and cutoffs $\phi,\phi_1,\phi_2,\cdots$, where typically $\sigma_1=6$, $\sigma_2=8$, $\cdots$. The resulting contribution to the kinetic EOS from Eq.\,(\ref{def_nk0DD}) is smaller than that obtained by adopting only a single $\sigma=6$ tail, but larger than that from a single $\sigma=8$ tail. In particular, we find that the contribution from the $k^{-6}$ HMT can be as large as about $20.2\,\rm{MeV}$, while that from the $k^{-8}$ HMT can reach about $4.1\,\rm{MeV}$, based on the empirical $x_{\rm{HMT}}\approx28\%\pm4\%$ and $C\approx0.16\pm0.015$. Thus, maximally the correction due to the $k^{-\sigma}$ with $6\leq\sigma\leq8$ on the $E_0^{\rm{kin}}(\rho_0)$ is about $4.1\sim20.2\,\rm{MeV}$.

\section{SRC-HMT Impacts on Heavy-ion Reactions or Scattering Experiments}\label{SEC_HIC}

\indent 

In this section, we discuss the effects of SRC and HMT on various phenomena in low- and intermediate-energy heavy-ion collisions. In Subsection \ref{sub_protonskin}, we examine the coexistence of a proton-skin in momentum space and a neutron-skin in coordinate space in neutron-rich nuclei, a feature that naturally arises from Liouville's theorem in statistical physics and the uncertainty principle of quantum mechanics. Subsection \ref{sub_HIC} focuses on several isospin-sensitive observables in heavy-ion reactions, such as yield ratios and particle flows, highlighting the influence of both the proton-skin and neutron-skin. In Subsection \ref{sub_threshold}, we discuss sub-threshold particle production in $\rm{p}A$ collisions with explicit consideration of SRC-HMT effects. Finally, Subsection \ref{sub_gamma} discusses the determination of the HMT fraction in finite nuclei via neutron-proton bremsstrahlung photon emission in low-energy reactions, offering a novel approach to probe SRC physics in finite nuclear systems.

\subsection{Coexistence of Proton-skin in $k$-space and Neutron-skin in $r$-space}\label{sub_protonskin}

\indent

As discussed in previous sections, the SRC between a neutron-proton pair is about 18-20 times stronger than that between two protons\cite{Hen14,Sub08}, and the HMT in PNM was therefore estimated to be about 1-2\%\cite{Hen15a}. Although the precise size of the HMT remains uncertain, a qualitative consensus has been reached; in the previous sections, we use the HMT-exp and HMT-SCGF parameter sets\cite{Cai16b} to illustrate several relevant issues.
The resulting neutron fractions in the HMT as a function of $\delta$ at $\rho_0$ for the two parameter sets are compared in the right panel of FIG.\,\ref{fig_xCaiFig1}. The corresponding reduced nucleon momentum distributions (normalized to unity at zero momentum) for $\delta\approx0.21$ and $\delta\approx0.5$ are shown in the left and middle panels, respectively. Clearly, relative to the center of momentum space, nucleonic matter exhibits a distinct proton skin in momentum space, whose thickness grows with the isospin asymmetry $\delta$ at a rate that depends on the adopted HMT parameters\cite{Cai16b}.
The neutron skins in coordinate space and proton skins in momentum space coexist in heavy nuclei, and their correlation is governed by Liouville's theorem and the Heisenberg uncertainty principle\cite{Cai16b}. A careful study of both skins simultaneously may help to better constrain the underlying isovector interaction. In the following discussions in this subsection, we adopt the HMT-exp parameter set unless specified otherwise.

\begin{figure}[h!]
\centering
\includegraphics[width=9.cm]{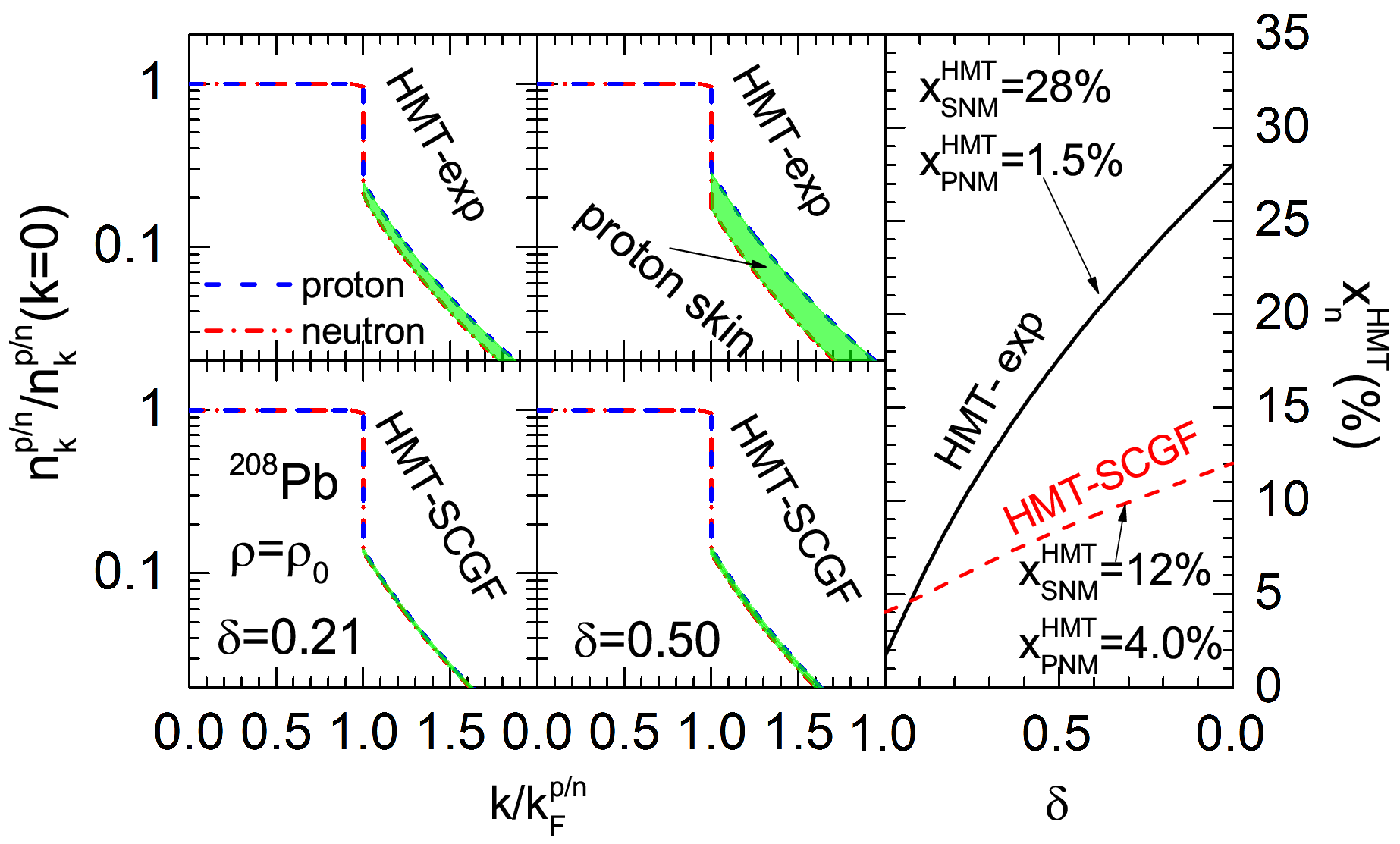}
\caption{(Color Online) Reduced nucleon momentum distribution of neutron-rich nucleonic matter with an isospin asymmetry of $\delta\approx0.21$ (left) and $\delta\approx0.50$ (middle), and the fraction of neutrons in the HMT as a function of $\delta$ using the HMT-exp and HMT-SCGF parameter sets (right). Taken from Ref.\cite{Cai16b}.}
\label{fig_xCaiFig1}
\end{figure}

An extended Thomas--Fermi method (denoted as $\rm{ETF}^+$) was adopted in Ref.\cite{Cai16b}. In the original ETF framework, which is a semi-classical approximation to the Hartree-Fock theory, the nucleon kinetic energy density profile in nuclei is given by\cite{Cai16b}
\begin{equation}\label{epskin1}
\boxed{
\varepsilon_J^{\rm{kin}}(r)=\frac{1}{2M_{\rm N}}\left[\alpha_J^{\infty}\rho_J^{5/3}+\eta_J\frac{1}{36}\frac{(\nabla\rho_J)^2}{\rho_J}+\frac{1}{3}\Delta\rho_J\right],}
\end{equation}
obtained by truncating the Wigner--Kirkwood expansion of the Block density matrix at order $\hbar^2$\cite{Bra85,Kri79,Cam80,Bar83}. Importantly, this relationship is independent of the interactions involved. The first term, originally with $\alpha_J^{\infty}=(3/5)(3\pi^2)^{2/3}$, represents the bulk contribution as if nucleons are in infinite nuclear matter with a step-function momentum distribution. The second term, with $\eta_J=1$, was originally proposed by Weizs\"acker\cite{Bra85} and is highly sensitive to surface properties. In practice, $\eta_J$ is often adjusted to account for higher-order terms and other missing effects, reproducing nuclear binding energies or predictions from more microscopic theories. In the following, we fix $\eta_J$ by fitting the average kinetic energies of protons and neutrons extracted from SRC experiments at JLab\cite{Hen14}. The last term, involving a Laplacian, is generally small as nuclear surfaces are typically smooth (so the second-order derivative is small).

With the SRC-modified single-nucleon momentum distribution function of Eq.\,(\ref{MDGen}), the kinetic energy density in infinite matter becomes\cite{Cai16b}
\begin{equation}\label{epskin}
\frac{2}{(2\pi)^3}\int_0^{\phi_J k_{\rm{F}}^J} \frac{\v{k}^2}{2M_{\rm N}} n_{\v{k}}^J \d\v{k} = \frac{1}{2M_{\rm N}}\frac{3}{5}(3\pi^2)^{2/3} \rho_J^{5/3} \Phi_J,
\end{equation}
where $\Phi_J = 1 + C_J(5\phi_J + 3/\phi_J - 8) > 1$ depends on the properties of the HMT, see Eq.\,(\ref{def-Phi0}) for the factor in SNM. Consequently, the bulk factor $\alpha_J^{\infty}$ in Eq.\,(\ref{epskin1}) is enhanced by the SRC factor $\Phi_J$, i.e., $\alpha_J^{\infty} = (3/5)(3\pi^2)^{2/3} \Phi_J$. In neutron-rich systems, because more protons are depleted from the Fermi sea to populate the HMT, the bulk kinetic energy density is enhanced more for protons than neutrons. For $^{208}\rm{Pb}$, we find $\Phi_{\rm{p}}\approx2.09\pm0.50$ and $\Phi_{\rm{n}}\approx1.60\pm0.33$. 
To quantify the proton-skin in momentum space, one defines\cite{Cai16b}
\begin{equation}
\boxed{
\Delta E_{\rm{pn}}^{\rm{kin}} \equiv \langle E_{\rm{p}}^{\rm{kin}}\rangle - \langle E_{\rm{n}}^{\rm{kin}}\rangle,\quad
\langle E_J^{\rm{kin}}\rangle = \frac{\int_0^\infty \varepsilon_J^{\rm{kin}}(r)\d\v{r}}{\int_0^\infty \rho_J(r)\d\v{r}} \equiv \frac{\langle k_J^2 \rangle}{2M_{\rm N}},}
\end{equation}
where the average kinetic energies are given here.
For the nucleon density profile $\rho_J(r)$, we adopt the widely used 2-parameter Fermi (2pF) distribution\cite{Jon14,Cai16b}, 
$
\rho_J(r) = \rho_0^J [1 + \exp(({r - c_J)}/{a_J})]^{-1}$,
where $c_J$ and $a_J$ are the half-density radius and diffuseness parameter, respectively, and $\rho_0^J$ is determined by normalization. Using the experimental average kinetic energy $\langle E_{\rm{p}}^{\rm{kin}}\rangle \approx 41.9\,\rm{MeV}$\cite{Hen14}, one finds $\eta_{\rm{p}}\approx26.7$ for $^{208}\rm{Pb}$. For neutrons, with the constraints $\langle E_{\rm{n}}^{\rm{kin}}\rangle \approx 34.0\,\rm{MeV}$\cite{Hen14} and $\Delta r_{\rm{np}} \approx 0.159\,\rm{fm}$\cite{Dan14}, the unknowns $a_{\rm{n}}$, $c_{\rm{n}}$, and $\eta_{\rm{n}}$ yield $\eta_{\rm{n}}\approx 7.2 \sim 9.2$, with an average of $\eta_{\rm{n}}\approx8.2$\cite{Cai16b}.  

\begin{figure}[h!]
\centering
\includegraphics[width=8.5cm]{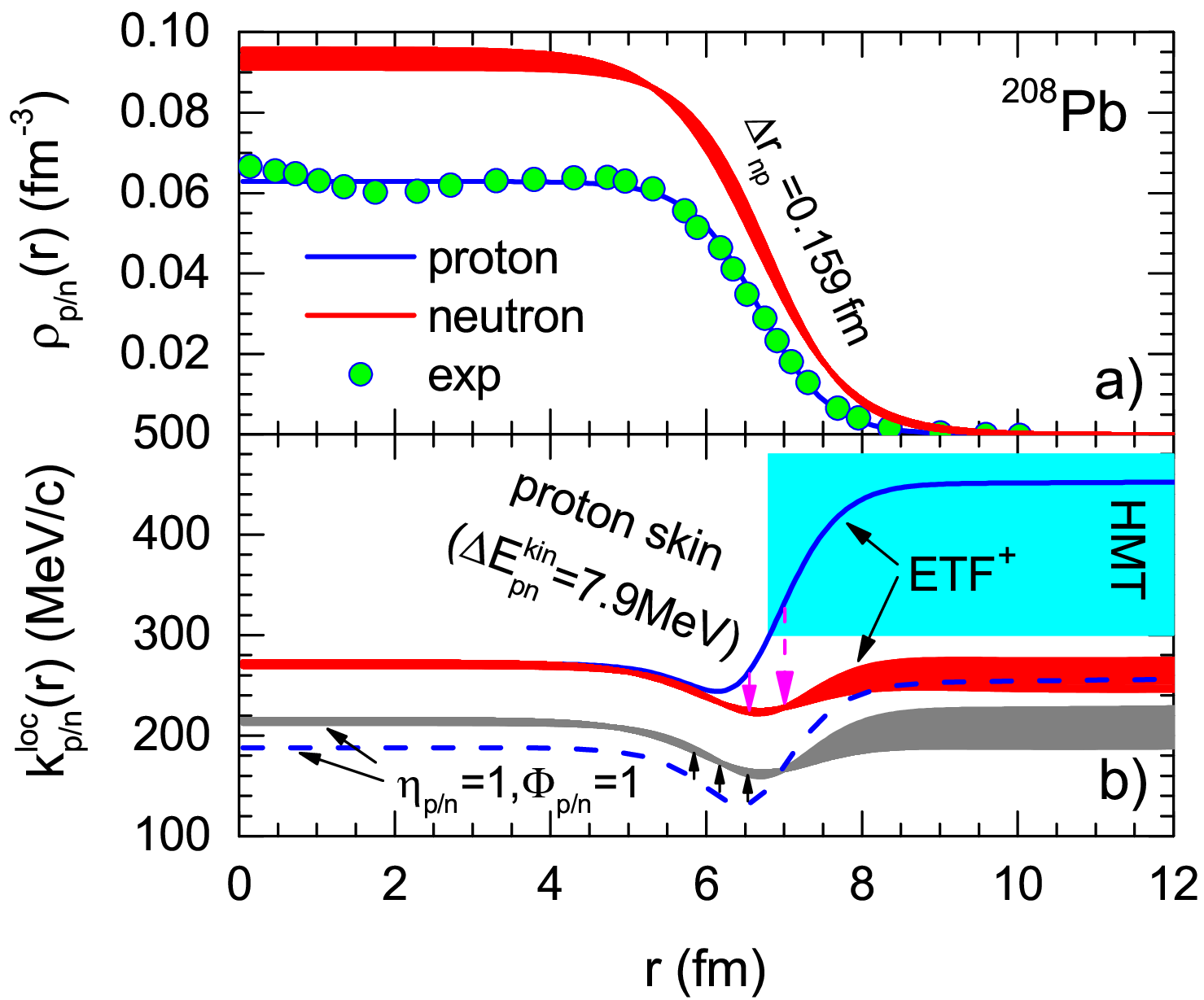}\\[0.25cm]
\hspace{-0.1cm}
\includegraphics[width=8.5cm]{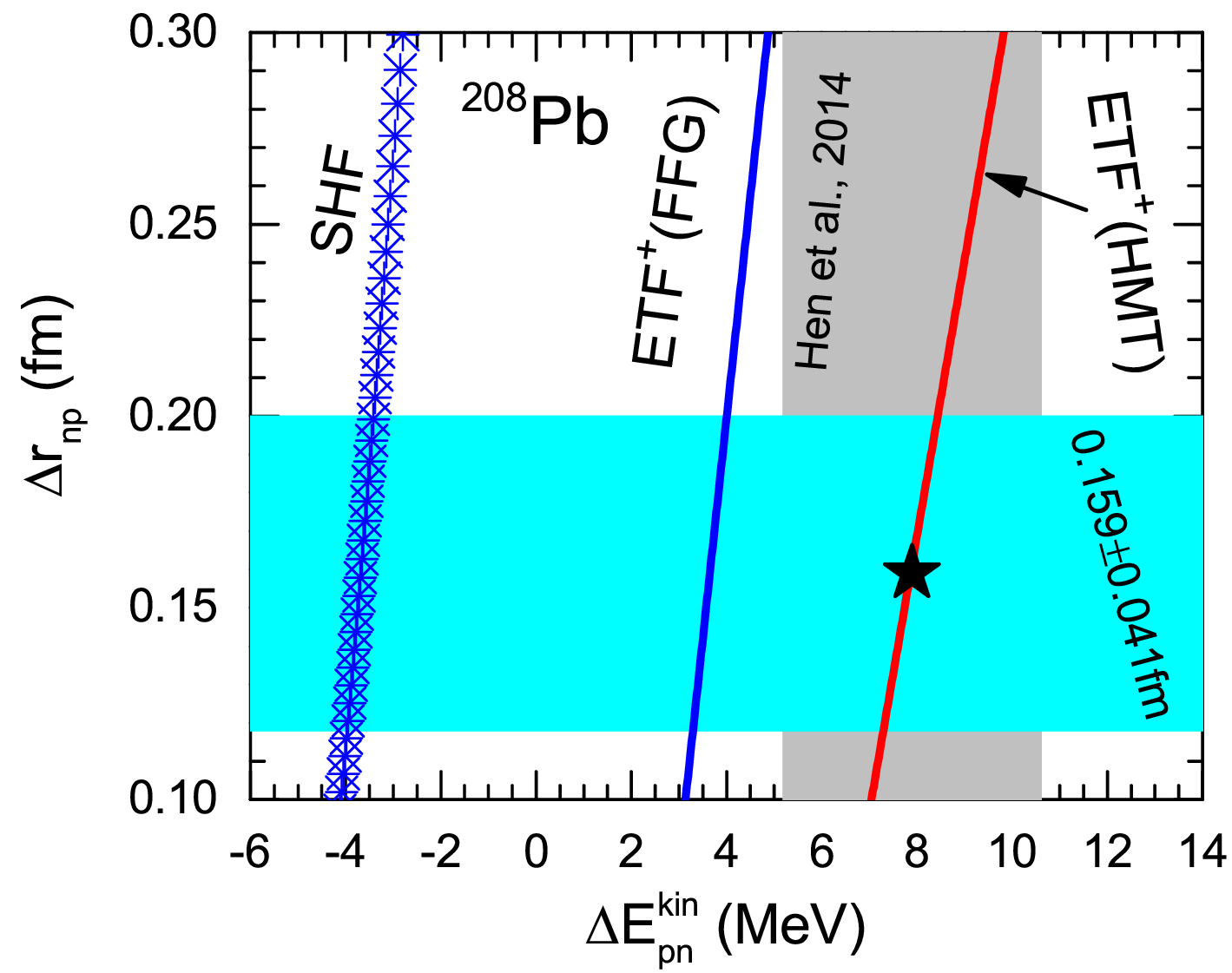}
\caption{(Color Online). Upper: density profile and local momentum in $^{208}\rm{Pb}$ under $\Delta r_{\rm{np}}\approx 0.159\,\rm{fm}$\cite{Dan14} and $\Delta E_{\rm{pn}}^{\rm{kin}} \approx 7.9\,\rm{MeV}$\cite{Hen14}. Lower: correlation between neutron-skin in r-space $\Delta r_{\rm{np}}$ and proton-skin in k-space $\Delta E_{\rm{pn}}^{\rm{kin}}$ for $^{208}\rm{Pb}$ within the ETF$^+$ approach with (HMT) and without (FFG) SRC effects. The black star represents the central cross point of the constraints on both $\Delta r_{\rm{np}}$ and $\Delta E_{\rm{pn}}^{\rm{kin}}$. Taken from Ref.\cite{Cai16b} where the HMT-exp set is adopted in the calculations.}
\label{fig_Fig24-ps}
\end{figure}

In the upper panel of FIG.\,\ref{fig_Fig24-ps}, the density profile and the local momentum defined via $k_J^{\rm{loc},2}(r)/2M_{\rm N} = \varepsilon_J^{\rm{kin}}(r)/\rho_J(r)$ are shown. Two main features emerge: (a) high-momentum nucleons with $300\,\rm{MeV} \lesssim |\v{k}| \lesssim 600\,\rm{MeV}$\cite{Hen14} are mainly distributed near the neutron skin; and (b) in the neutron skin, the local momentum of protons exceeds that of neutrons, implying faster proton motion in this region.  
Without the HMT factor $\Phi_{\rm{p/n}}$ and $\eta_{\rm{p/n}}$, $k_{\rm{p}}^{\rm{loc}}(r)$ is smaller than $k_{\rm{n}}^{\rm{loc}}(r)$ in the center and bulk skin region ($R \approx 1.12 A^{1/3}\,\rm{fm} \approx 6.6\,\rm{fm}$), reversing only in the surface skin region. In this region, $k_{\rm{p/n}}(r)\approx(1/72M_{\rm N})(\nabla\rho_{\rm{p/n}}/\rho_{\rm{p/n}})^2 \approx 1/(72 M_{\rm N} a_{\rm{p/n}}^2)$, so $k_{\rm{p}}^{\rm{loc}}(r) > k_{\rm{n}}^{\rm{loc}}(r)$ for $a_{\rm{p}} < a_{\rm{n}}$, but this is sensitive to small changes in $a_{\rm{p/n}}$. 
Using the Skyrme-Hartree-Fock method, FIG.\,\ref{fig_Guo23-nk} shows the similar average local density and momentum of nucleons in $^{208}$Pb\cite{Guo23PRC}. 
The lower panel of FIG.\,\ref{fig_Fig24-ps} shows correlations between neutron-skin $\Delta r_{\rm{np}}$ and proton-skin $\Delta E_{\rm{pn}}^{\rm{kin}}$ in $^{208}\rm{Pb}$ using ETF$^+$ with HMT or FFG, along with SHF predictions. Several observations are noteworthy: (a) SHF and ETF$^+$ (FFG) predictions fail to satisfy the combined constraints on both skins. Measuring both neutron and proton skins provides a more stringent test of nuclear models; (b) $\Delta r_{\rm{np}}$ and $\Delta E_{\rm{pn}}^{\rm{kin}}$ are approximately linearly correlated. Along the red line satisfying both constraints, a precise measurement of one skin determines the other, assuming ETF$^+$ (HMT) is correct.

\begin{figure}[h!]
\centering
\includegraphics[width=9.cm]{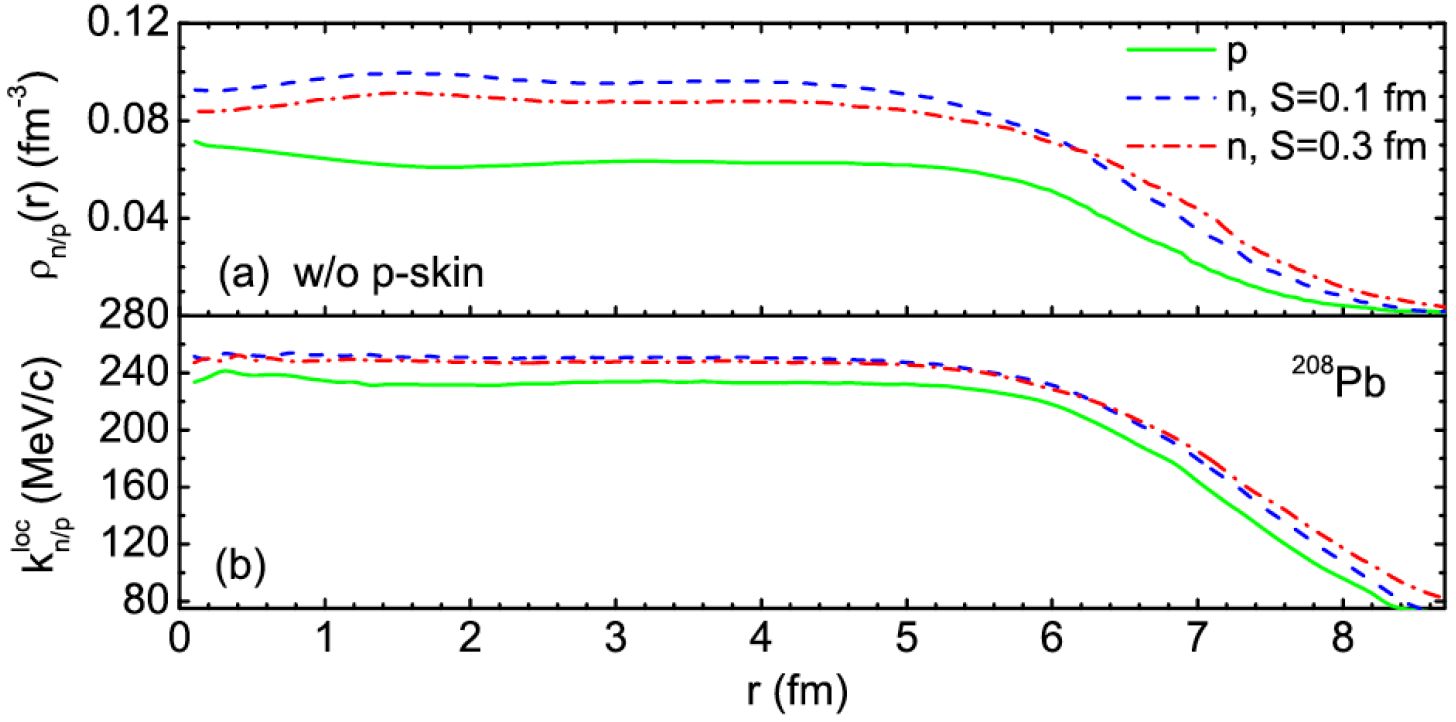}\\[0.25cm]
\includegraphics[width=9.cm]{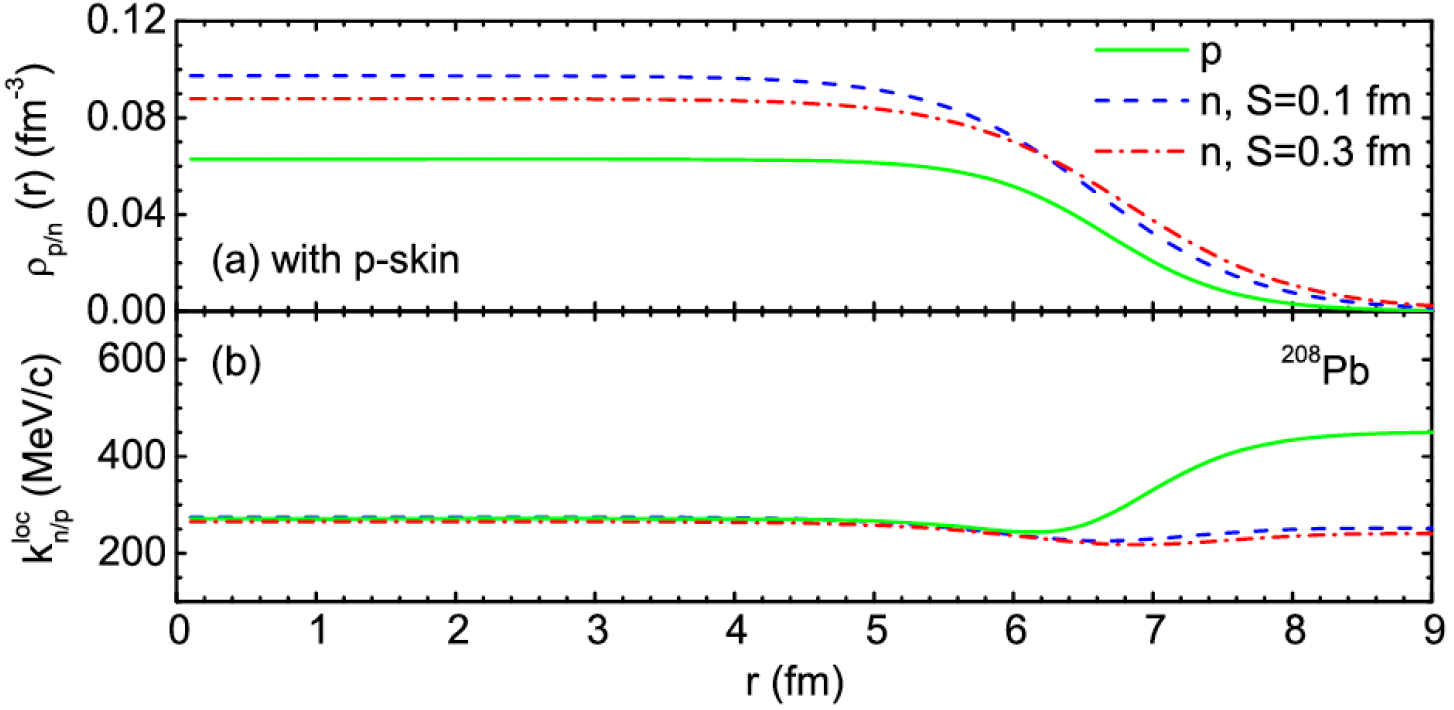}
\caption{(Color Online). 
Upper: input local density (a) and calculated momentum distributions (b) for nucleons in $^{208}$Pb using the Skyrme functional, and for neutrons with neutron skin thicknesses $S =\langle\Delta r_{\rm{np}}\rangle= 0.1$\,fm and $S =\langle\Delta r_{\rm{np}}\rangle= 0.3$\,fm. In this case, the nucleon kinetic-energy density of the colliding nuclei includes only the bulk contribution. 
Lower: same as the upper panel, but now including the surface term in the nucleon kinetic-energy density of the colliding nuclei.
Figure taken from Ref.\cite{Guo23PRC}.}
\label{fig_Guo23-nk}
\end{figure}

Physically, neutron skins in r-space and proton skins in k-space coexist and are intrinsically correlated by Liouville's theorem and Heisenberg's uncertainty principle, so precise measurements of either can constrain the other and improve understanding of nuclear surfaces in full phase space.
The normalization condition is $[{2}/{(2\pi)^3}]\int f(\v{r},\v{k})\d\v{r}\d\v{k}=N$, where $f(\v{r},\v{k})$ is the phase-space distribution function and $N$ the total number of particles.\footnote{In nuclear systems, $N\leftrightarrow A$, with $A$ the total nucleon number.} For a uniform distribution $f(\v{r},\v{k})=s_{\v{r}}n_{\v{k}}=\Theta(R-|\v{r}|)\Theta(k_{\rm{F}}-|\v{k}|)$, one finds $Rk_{\rm{F}}=({9\pi N}/{4})^{1/3}$, and similarly
\begin{align}
\langle r^2\rangle \langle k^2\rangle=&\left[\left.{\int_0^Rr^4\d r}\right/{\int_0^Rr^2\d r}\right]\left[\left.{\int_0^{k_{\rm{F}}}k^4\d k}\right/{\int_0^{k_{\rm{F}}}k^2\d k}\right]\notag\\
=&\frac{9}{25}R^2k_{\rm{F}}^2=\frac{4}{25\pi}\left(\frac{9\pi}{4}\right)^{5/3}N^{2/3},
\end{align}
and more generally,
\begin{equation}
\langle r^\ell\rangle\langle k^\ell\rangle=\left(\frac{3}{\ell+3}\right)^2R^\ell k_{\rm{F}}^\ell.
\end{equation}
For general distributions, the normalization conditions for $s_{\v{r}}$ and $n_{\v{k}}$ are $\int s_{\v{r}}\d\v{r}=4\pi R^3/3$ and $[{2}/{(2\pi)^3}]\int n_{\v{k}}\d\v{k}={k_{\rm{F}}^3}/{3\pi^2}$, which define the sharp radius $R$ and Fermi momentum $k_{\rm{F}}$. In the HMT case, assuming $s_{\v{r}}$ is uniform, we have (for $\ell\neq1$)
\begin{align}
\langle r^\ell\rangle\langle k^\ell\rangle
=&\left(\frac{3}{\ell+3}\right)^2R^\ell k_{\rm{F}}^\ell\left[1+C_0\left(\frac{\ell+3}{\ell-1}\phi_0^{\ell-1}+\frac{3}{\phi_0}-\frac{4\ell}{\ell-1}\right)\right].
\end{align}
The limit $\ell\to1$ could be obtained by this expression appropriately.
Liouville's theorem ensures that $Rk_{\rm{F}}$ is fixed regardless of the forms of $s_{\v{r}}$ and $n_{\v{k}}$, but the average $\langle r^\ell\rangle\langle k^\ell\rangle$ depends on $\ell$ and the shapes of $s_{\v{r}}$ and $n_{\v{k}}$.
For a nucleon in symmetric nuclear matter, Liouville's theorem reads
\begin{empheq}[box=\fbox]{align} &\mbox{(spin factor)}\times\frac{\mbox{(r-vol)}\times\mbox{(k-vol)}}{h^3}=\#\mbox{ of particle}\notag\\
\leftrightarrow 
&2\times\left.{\displaystyle\frac{4}{3}\pi\langle r\rangle^3\times\frac{4}{3}\pi\langle k\rangle^3}\right/{(2\pi\hbar)^3}=N,
\end{empheq}
where $\langle r\rangle$ and $\langle k\rangle$ are effective radii in r- and k-space, respectively. Therefore, $\langle r\rangle\langle k\rangle\sim \rm{const.}$; together with the uncertainty principle $\delta r\delta k\sim \rm{const.}$, we obtain
\begin{equation}
\langle r^2\rangle \langle k^2\rangle=\langle r\rangle^2\langle
k\rangle^2\left[ 1+\left(\frac{\delta r}{\langle r\rangle}\right)^2
+\left(\frac{\delta k}{\langle k\rangle}\right)^2 \right]+(\delta
r)^2(\delta k)^2.
\end{equation}
If the fluctuations are small, $\delta r\ll\langle r\rangle$ and $\delta k\ll\langle k\rangle$, then $\langle r^2\rangle \langle k^2\rangle \approx \langle r\rangle^2\langle k\rangle^2+(\delta r)^2(\delta k)^2$ remains nearly constant.
For the 2pF distribution in r-space, we have $
\langle r\rangle\approx({3c}/{4})(1+\pi^2\chi^2)$ and $\langle r^2\rangle\approx({3c^2}/{5})(1+{7}\pi^2\chi^2/3)$ where $\chi=a/c$,
so that $
{\delta r}/{\langle r\rangle}\approx0.26(1+8\pi^2\chi^2/3)$ as well as $
({\delta r}/{\langle r\rangle})^2\approx0.07(1+16\pi^2\chi^2/3)$.
For k-space including HMT, we have:
\begin{align}
\langle k^2\rangle=&\frac{3}{5}k_{\rm{F}}^2\left[1+C_0\left(5\phi_0+\frac{3}{\phi_0}-8\right)\right],\\
\langle k\rangle=&\frac{3}{4}k_{\rm{F}}\left[1+C_0\left(4\ln\phi_0+\frac{3}{\phi_0}-3\right)\right],
\end{align}
see equations of (\ref{def-flc}), numerically giving $\delta k/\langle k\rangle\approx0.18$.
For ANM, the phase-space distribution $f_J(\v{r},\v{k})$ of nucleon type $J$ is normalized as $[2/(2\pi)^3]\int f_J(\v{r},\v{k})\d\v{r}\d\v{k}=N_J$, with $N_{\rm n}=N$ and $N_{\rm p}=Z$. Assuming a uniform distribution, $f_J(\v{r},\v{k})=\Theta(R-|\v{r}|)\Theta(K-|\v{k}|)$, we have $H_J\equiv\langle r_J^2\rangle\langle k_J^2\rangle=(4/25\pi)(9\pi/4)^{5/3}N_J^{2/3}$, which depends only on $N_J$.
In realistic cases, several parameters in $\rho_J(r)$, such as $\Lambda_{\rm{av}}^J$ (average/bulk properties) and $\Lambda_{\rm{df}}^J$ (diffuseness properties), characterize the profile. By dimensional analysis, the dimensionless product $H_J=\langle r_J^2\rangle\langle k_J^2\rangle$ depends only on the ratio $\Sigma_J = \Lambda_{\rm{df}}^J / \Lambda_{\rm{av}}^J$:
\begin{align} H_J\approx&\vartheta_0(N_J)+\vartheta_1(N_J)\left(\frac{\Lambda_{\rm{df}}^J}{\Lambda_{\rm{av}}^J}\right) +\vartheta_2(N_J)\left(\frac{\Lambda_{\rm{df}}^J}{\Lambda_{\rm{av}}^J}\right)^2+\cdots, \end{align}
where $\vartheta_0$, $\vartheta_1$ and $\vartheta_2$ are dimensionless functions of $N_J$. Consequently, $H_J=\langle r_J^2\rangle\langle k_J^2\rangle$ keeps almost constant due to the smallness of the factor $\Lambda_{\rm{df}}^J/\Lambda_{\rm{av}}^J$. 
Specifically, by expanding $H_{\rm{n}}$ around
$\sigma_{\rm{n}}=a_{\rm{n}}/c_{\rm{n}}=0$, we have
$H_{\rm{n}}\approx(4/25\pi)(9\pi/4)^{5/3}N^{2/3}[1-2.4\sigma_{\rm{n}}+18.7\sigma_{\rm{n}}^2+\cdots]\approx1.33N^{2/3}[1-2.4\sigma_{\rm{n}}+18.7\sigma_{\rm{n}}^2+\cdots]$; and moreover
$\Delta r_{\rm{np}}\approx(H_{\rm{n}}/8M_{\rm N})^{1/2}\langle
E_{\rm{p}}^{\rm{kin}}\rangle^{-3/2}\Delta E_{\rm{pn}}^{\rm{kin}}
+[(H_{\rm{n}}/2M_{\rm N})^{1/2}-(H_{\rm{p}}/2M_{\rm N})^{1/2}]\langle
E_{\rm{p}}^{\rm{kin}}\rangle^{-1/2}$.
The constancy of $H_J$ naturally follows from these expressions (see Ref.\cite{Cai16b} for more examples).

This situation is very similar to the case in fluid mechanics: the velocity distribution $\v{v}$ of a body in a fluid is given by $\v{v}=u\v{f}(\v{r}/\ell,\rm{R})$, where $\v{r}$ is the position of the body, and $\rm{R}=\rho u\ell/\eta$ is the Reynolds number constructed from the density $\rho$, the linear dimension of the problem $\ell$, the velocity of the main stream $u$ and the shear viscosity $\eta$. Then in two different flows of the same type, the velocities $\v{v}/u$ are the same functions of the ratio $\v{r}/\ell$ if the Reynolds number is the same for each flow. Furthermore, if the Reynolds number is small, Navier-Stokes equation gives the drag force on a sphere moving slowly in a fluid as $F\approx6\pi \eta ua(1+3\rm{R}/8)$ where $a$ is the radius of the sphere. The factor $6\pi\eta ua$ depending on the properties of the fluid and the sphere macroscopically is independent of the detailed structure of the sphere (corresponding to the parametrization of the density profile of the nucleon), which is like the number $N_J$ in the above analysis for $\langle r_J^2\rangle\langle k_J^2\rangle=H_J$.

\begin{figure*}[h!]
\centering
\includegraphics[width=16.cm]{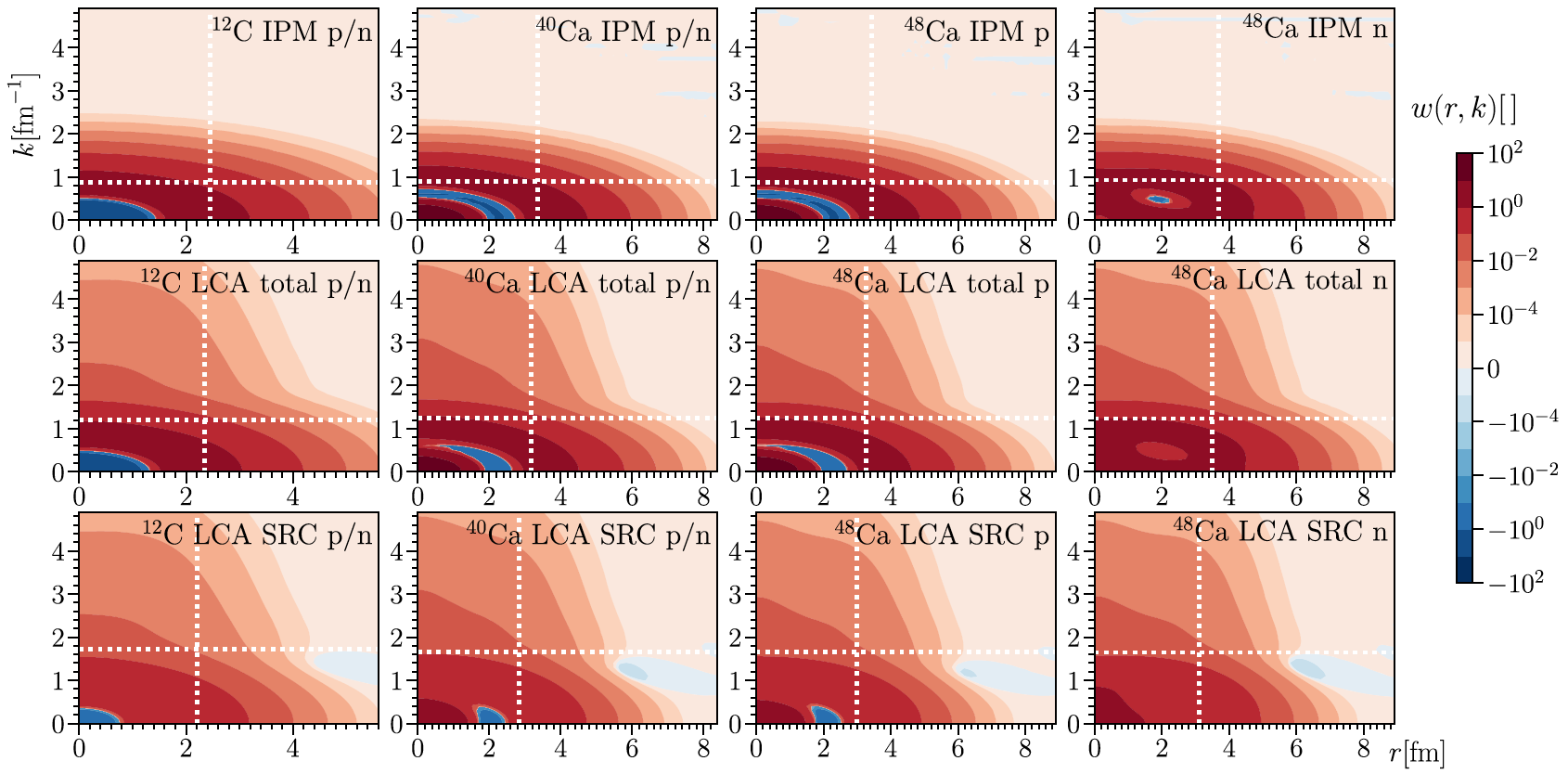}
\caption{(Color Online). Two-dimensional proton and neutron Wigner distributions \(w(r,k)\) for 
\(^{12}\mathrm{C}\), \(^{40}\mathrm{Ca}\), and \(^{48}\mathrm{Ca}\) computed in the IPM and LCA. 
Top: IPM; middle: full LCA; bottom: SRC contribution. White lines indicate \(r_{\mathrm{rms}}\) and \(k_{\mathrm{rms}}\).
Figure taken from Ref.\cite{Cosyn21PLB}.}
\label{fig_wigner-1}
\end{figure*}

\begin{figure}[h!]
\centering
\includegraphics[width=8.9cm]{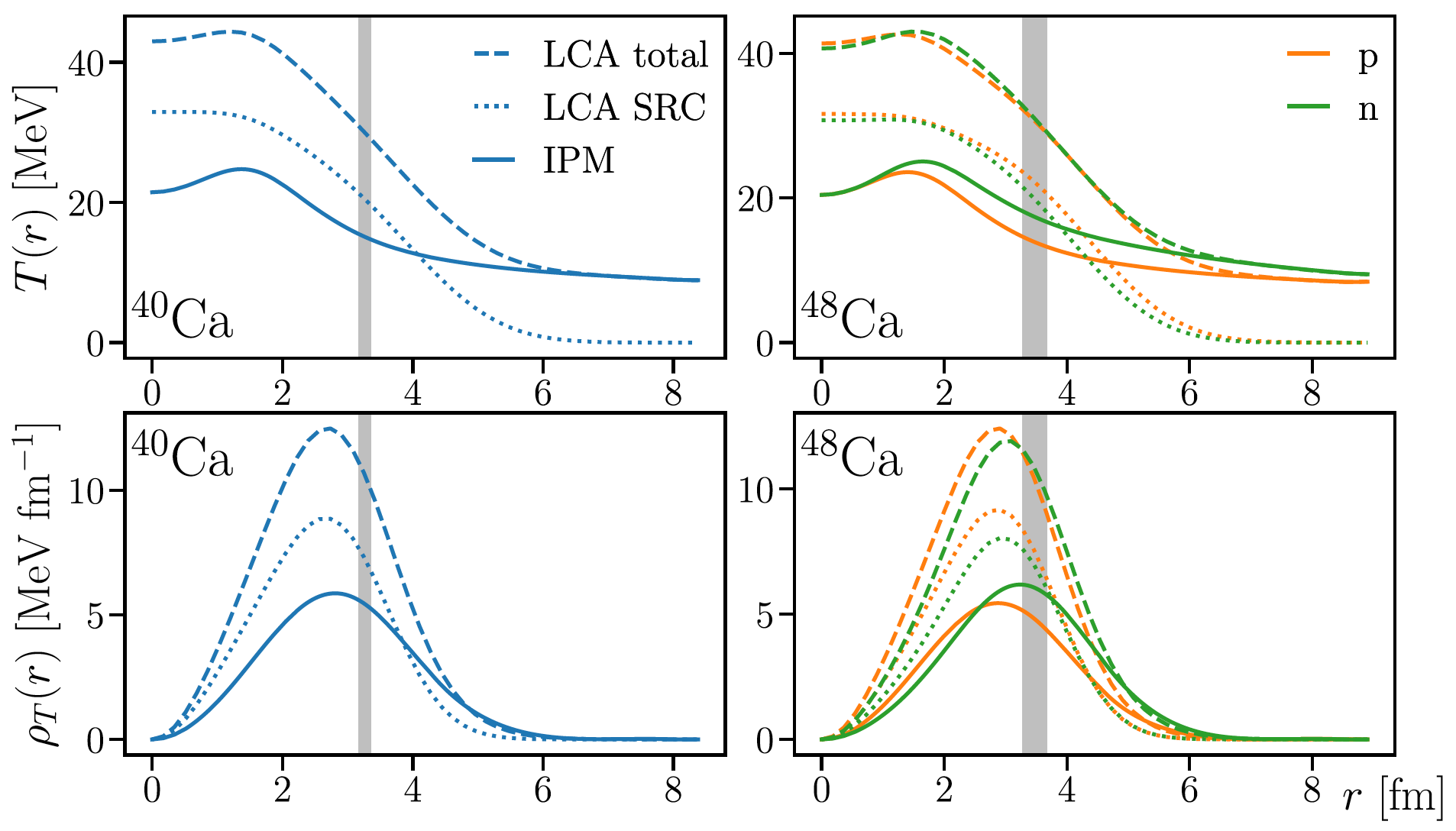}\\[0.5cm]
\hspace{0.2cm}
\includegraphics[width=9cm]{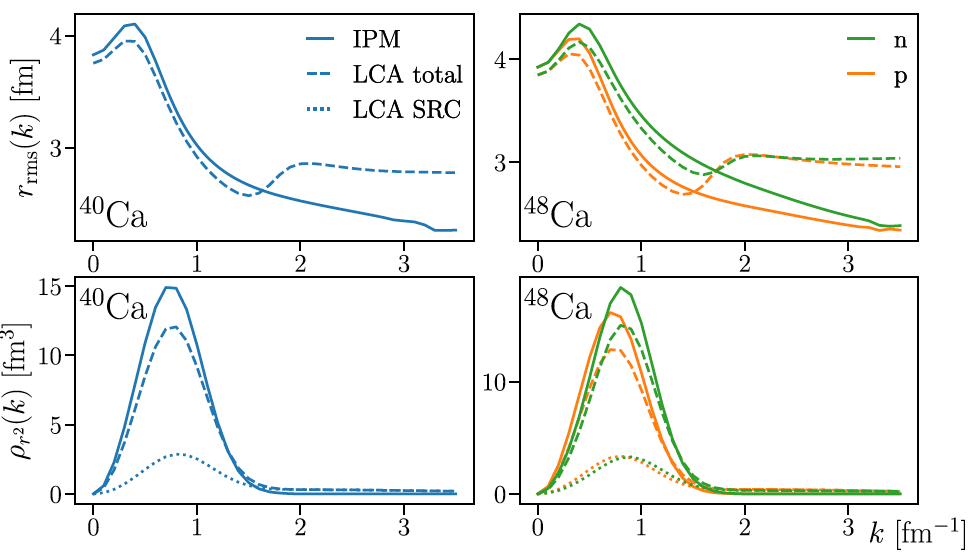}
\caption{(Color Online). 
Upper: $^{40}$Ca and $^{48}$Ca radial dependence of the kinetic energy expectation value $T(r)$ (top row) and the kinetic energy density $\rho_T(r)$ (bottom row). Proton and neutron results are identical in $^{40}$Ca. Lower: as in the upper panels, but now showing the momentum dependence of $r_{\rm{rms}}(k)$ (top row) and the density $\rho_{r^2}(k)$ (bottom row). Figures taken from Ref.\cite{Cosyn21PLB}.}
\label{fig_wigner-2}
\end{figure}

\begin{figure*}[h!]
\centering
\includegraphics[width=15.cm]{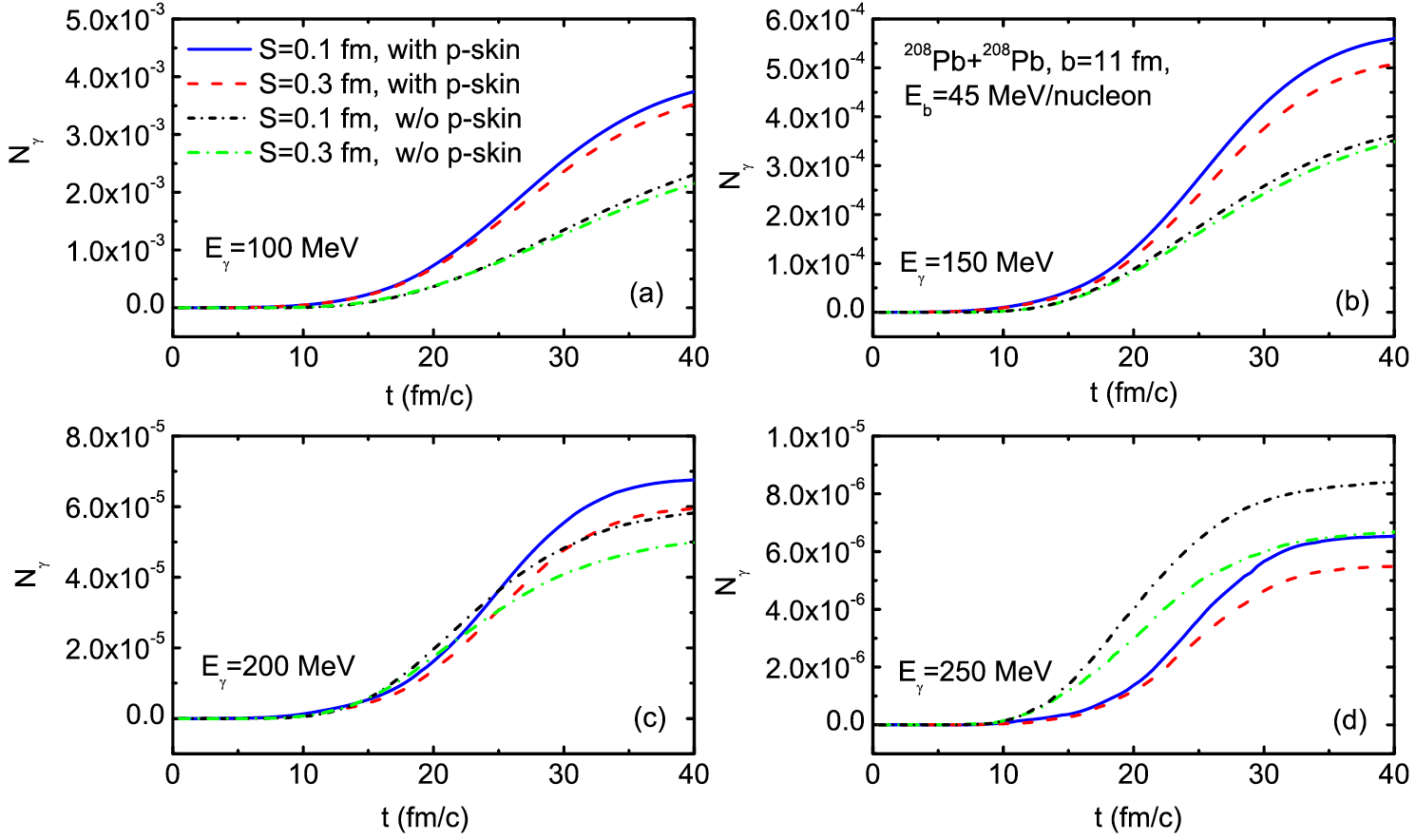}
\caption{(Color Online). Effects of the neutron skin in coordinate ($r$) space and proton skin in momentum ($k$) space on the multiplicity of hard photons with energies $E_\gamma = 100$, 150, 200, and 250\,MeV in peripheral $^{208}$Pb+$^{208}$Pb collisions at a beam energy of 45\,MeV/nucleon. Figure taken from Ref.\cite{Guo23PRC}.}
\label{fig_Guo23-Ngamma}
\end{figure*}

The density and momentum distribution functions are very fundamental concepts. In Ref.\cite{Cosyn21PLB}, the Wigner distribution $w(\v{r},\v{k})$ as a quasi-probability representation of nucleon dynamics in simultaneous coordinate-momentum space is studied. It begins with the spectral decomposition of the quantum Wigner operator\cite{Cosyn21PLB},
\begin{align}
\widehat w(\v{r},\v{k})=&\frac{1}{(2\pi)^3}\int \d \v{x}e^{i \v{k}\cdot \v{x}} 
\left| \v{r}-\frac{\v{x}}{2}\right\rangle \left\langle \v{r}+\frac{\v{x}}{2}\right|\notag\\
=&\frac{1}{(2\pi)^3}\int \d \v{q} e^{-i \v{q}\cdot \v{r}}
\left| \v{k}+\frac{\v{q}}{2}\right\rangle \left\langle \v{k}-\frac{\v{q}}{2}\right|,
\end{align}
and defines the Wigner quasi-probability distribution for a pure state $|\psi\rangle$ as
$
w(\v{r},\v{k})=\langle \psi|\widehat w(\v{r},\v{k})|\psi\rangle$.
Integrating $\widehat w(\v{r},\v{k})$ over $\v{r}$ or $\v{k}$ yields the standard one-body momentum and spatial density operators,
$
\widehat n(\v{k})=\int \d \v{r}\widehat w(\v{r},\v{k})=|\v{k}\rangle\langle \v{k}|$ and $
\widehat\rho(\v{r})=\int \d \v{k} \widehat w(\v{r},\v{k})=|\v{r}\rangle\langle \v{r}|$, respectively.
Expectation values of general operators $\widehat{\mathcal{O}}$ can be represented as phase-space averages, $
\langle \widehat{\mathcal{O}}\rangle=\int \d \v{r}\d \v{k}w(\v{r},\v{k})O(\v{r},\v{k})$,
where $O(\v{r},\v{k})$ is the Wigner transform of $\widehat{\mathcal{O}}$,
\begin{align}
O(\v{r},\v{k})&=\int \d \v{x}e^{-i \v{k}\cdot \v{x}} 
\left\langle \v{r}-\frac{\v{x}}{2}\right| \widehat{\mathcal{O}} \left| \v{r}+\frac{\v{x}}{2}\right\rangle\notag\\
&=\int \d \v{q}e^{i \v{q}\cdot \v{r}} 
\left\langle \v{k}+\frac{\v{q}}{2}\right| \widehat{\mathcal{O}} \left|\v{k}-\frac{\v{q}}{2}\right\rangle .
\end{align}
Assuming spherical symmetry, the study adopts the angle-averaged Wigner distribution, $
w(r,k)\equiv \int \d\Omega_{\v r}\d\Omega_{\v k}w(\v{r},\v{k})$,
which is normalized as $
\int r^2 \d r \int k^2 \d kw(r,k)=A$.

Within the lowest-order correlation operator approximation (LCA), SRCs generate many-body operators when acting on Slater determinants, which are truncated at the two-body level. This leads to a decomposition of the Wigner distribution into four isospin pair channels as $
w(r,k)=[w_{\rm{pp}}(r,k)+w_{\rm{pn}}(r,k)]
+[w_{\rm{nn}}(r,k)+w_{\rm{np}}(r,k)]$,
where each $w_{ij}(r,k)$ includes an uncorrelated mean-field component and a correlated SRC component.
The Wigner distribution is then used to compute quasi-expectation values of the kinetic-energy operator $T_{\v k}=\v{k}^2/2M_{\rm N}$ and the rms-radius operator $\v{r}^2$ as functions of position and momentum\cite{Cosyn21PLB}:
\begin{align}
T(r)=&\langle T_{\v k}\rangle=\left.{\int k^2 \d k\frac{\v{k}^2}{2M_{\rm N}}w(r,k)}\right/
{\int k^2 \d kw(r,k)}, \\
r^2_{\rm{rms}}(k)=&\langle \v{r}^2\rangle=\left.{\int r^4 \d r w(r,k)}\right/
{\int r^2 \d r w(r,k)}.
\end{align}
Because the Wigner distribution is not positive definite, these local quasi-expectation values can become negative.
The authors further construct properly normalized distributions,
\begin{align}
\rho_T(r)\equiv&
\left(r^2\left/\int r^2\d r\right.\right)\cdot
\left.{\int k^2 \d k\frac{\v{k}^2}{2M_{\rm N}}w(r,k)}\right/
{\int k^2 \d kw(r,k)},\\
\rho_{r^2}(k)\equiv&\left(k^2\left/\int k^2\d r\right.\right)
\cdot\left.{\int r^4 \d r w(r,k)}\right/
{\int r^2 \d rw(r,k)},
\end{align}
which encode the spatial structure of the kinetic energy and the momentum structure of the rms radius, while integrating to the correct scalar observables. Consequently, $\int \d r\rho_T(r)=\langle {T}_{\v k}\rangle$ and $\int \d k\rho_{r^2}(k)=\langle \v{r}^2\rangle$.

FIG.\,\ref{fig_wigner-1} illustrates the two-dimensional Wigner distribution \(w(r,k)\) obtained in the LCA framework\cite{Cosyn21PLB}. The results have been validated through several checks, including consistency with independently computed momentum distributions and rms radii, as well as the correct normalization of all isospin combinations. The remaining sub-percent deviations originate from the truncation of the quantum-number summation and the use of finite numerical grids. Physically, the figure shows that the IPM \(w(r,k)\) spans the full radial range and a well-defined momentum region common to all nuclei studied, whereas SRCs generate a pronounced high-momentum tail that is largely confined to the nuclear interior. This behavior reflects the fact that, in the LCA, SRCs arise predominantly from correlation operators acting on nodeless relative \(S\)-pairs, which peak at \(r < r_{\mathrm{rms}}\) and become negligible for \(r \gtrsim 2 r_{\mathrm{rms}}\). Consequently, the SRC contribution to \(w(r,k)\) is shifted to smaller radii, even though non-negligible components also appear at \(k < k_{\mathrm{F}}\). The normalization of the SRC sector (e.g., about \(1.8/6\) nucleons in \(^{12}\mathrm{C}\)) therefore cannot be directly connected to experimental extractions of the number of SRC pairs, which refer exclusively to high-momentum components. For all nuclei, the SRC part dominates the LCA result for \(k > 2~\mathrm{fm}^{-1}\) at any \(r\), despite its rapid falloff at large radii. Although this raises questions regarding experimental access to the nuclear interior, earlier studies have shown that reactions probing correlated pairs remain sensitive to interior dynamics even after accounting for strong final-state interactions. The Wigner distributions also exhibit small negative regions at low \(r\) and \(k\), characteristic of quantum effects, and in \(^{48}\mathrm{Ca}\) a neutron skin is clearly visible. Finally, compared to the deuteron, the SRC-induced momentum tails in finite nuclei extend over a larger momentum range and appear smoother, an effect attributable to center-of-mass motion and the coherent contribution of many nucleon pairs.

FIG.\,\ref{fig_wigner-2} shows the IPM and LCA results for the combinations $[T(r), \rho_{T}(r)]$ and $[r_{\rm{rms}}(k), \rho_{r^2}(k)]$ for $^{40}$Ca and $^{48}$Ca, with a focus on the effects of SRCs and the role of the neutron excess in $^{48}$Ca\cite{Cosyn21PLB}. The IPM results, using the harmonic oscillator frequency, are compared with LCA calculations. For $^{48}$Ca, IPM neutrons have higher kinetic energy than protons at all radii, while SRC contributions in LCA reverse this trend, making protons more energetic in the interior (roughly at $r \lesssim 1.8\,\rm{fm}$) and neutrons slightly more energetic at intermediate radii. The radial kinetic energy density $\rho_T(r)$ shows that IPM protons contribute more to the bulk kinetic energy in the interior, with LCA enhancing this effect, resulting in overall more energetic protons. The momentum-space radii $r_{\rm{rms}}(k)$ plateau around $k \approx 2\,\rm{fm}^{-1}$, with neutrons larger than protons below the Fermi momentum, while tensor correlations make proton and neutron radii nearly equal for $1.5\,\rm{fm}^{-1} \lesssim k \lesssim 2.5\,\rm{fm}^{-1}$. High-momentum nucleons contribute minimally to the rms radius, even in LCA. The total kinetic energies and rms radii obtained from phase-space integration of $w(r,k)$ agree within 10\% with ab initio AV18 calculations for $^{40}$Ca. For $^{48}$Ca, all LCA variants predict a proton kinetic energy $\sim 1$\,MeV higher than neutrons, a ``kinetic energy inversion''\cite{Sar14}, with SRCs reducing the neutron skin by 5-10\%. Overall, SRCs strongly affect kinetic energies (uncertainties 2-3\,MeV) and moderately reduce radii (3-6\%), while the largest uncertainty for radii arises from the choice of harmonic oscillator parameters, and the kinetic energies are slightly less sensitive to this choice (1-2\,MeV).

As a short summary, in their work\cite{Cosyn21PLB}, the authors extend the study of nuclear SRCs beyond the traditional momentum-space perspective by incorporating a spatial perspective through nuclear Wigner distributions $w(r,k)$. For $^{12}$C, $^{40}$Ca, and $^{48}$Ca, they find that SRCs affect $w(r,k)$ over the full momentum range while being confined to the deep nuclear interior. From these distributions, they compute kinetic energies and point-nucleon radii, showing that SRCs nearly double proton and neutron kinetic energies and induce a proton-neutron asymmetry in neutron-rich nuclei such as $^{48}$Ca, where SRCs reduce the neutron skin by 5-10\%. They further emphasize that the Wigner distributions provide valuable input for semi-classical transport calculations and suggest that models including both long- and short-range correlations could lead to measurable changes in nuclear radii.

\subsection{Particle Production, Flows and Yield Ratios in Heavy-ion Reactions}\label{sub_HIC}

\indent

Building on the previous discussion of proton skins, Ref.\cite{Guo23PRC} investigated hard-photon emission from neutron-proton bremsstrahlung in $^{208}$Pb+$^{208}$Pb collisions near the Fermi energy, focusing on the roles of neutron skins in coordinate ($r$) space and proton skins in momentum ($k$) space on the time evolution at different beam energies. They found that direct hard-photon emission is sensitive to neutron skins, with stronger effects for higher-energy photons, while proton skins have an even larger impact, indicating that proton-skin effects must be considered when extracting neutron skin thickness from transport-model predictions.
Specifically, FIG.\,\ref{fig_Guo23-Ngamma} shows the effects of proton skins in $k$ space and neutron skins in $r$-space on the time evolution of hard photons with $E_\gamma = 100$, 150, 200, and 250\,MeV in peripheral $^{208}$Pb+$^{208}$Pb collisions at $E_{\rm b} = 45$\,MeV/nucleon and an impact parameter of 11\,fm. Photon production decreases for thicker neutron skins, especially for higher-energy photons, because the larger neutron densities reduce emission via incoherent $\rm{p}+\rm{n} \to \rm{p}+\rm{n}+\gamma$ bremsstrahlung. In contrast, proton skins enhance hard-photon emission more strongly than neutron skins due to the surface term in the nucleon kinetic-energy density\cite{Cai16b}, which increases local proton momentum in the nuclear surface and thus the emission probability. The hardest photons ($E_\gamma = 250$\,MeV) likely originate from multiple scatterings of nucleons below the Fermi momentum rather than from high-momentum tail collisions, and are therefore insensitive to the surface term and SRCs.
Other quantities are also studied in Ref.\cite{Guo23PRC}, including the angular distributions, the cross section as well as the transverse momentum dependence of photons. For example, the transverse momentum dependence of the photons is shown in FIG.\,\ref{fig_Guo23-TM}; the spectra exhibit typical exponential shapes for different neutron and proton skins. Notably, when the proton skin in $k$-space is included, the influence of varying neutron skins on the transverse momentum spectrum becomes more pronounced, particularly at high transverse momentum.
The hard photon emission problem with SRC-HMT was also studied previously in Ref.\cite{Yong3}.

\begin{figure}[h!]
\centering
\includegraphics[width=7.cm]{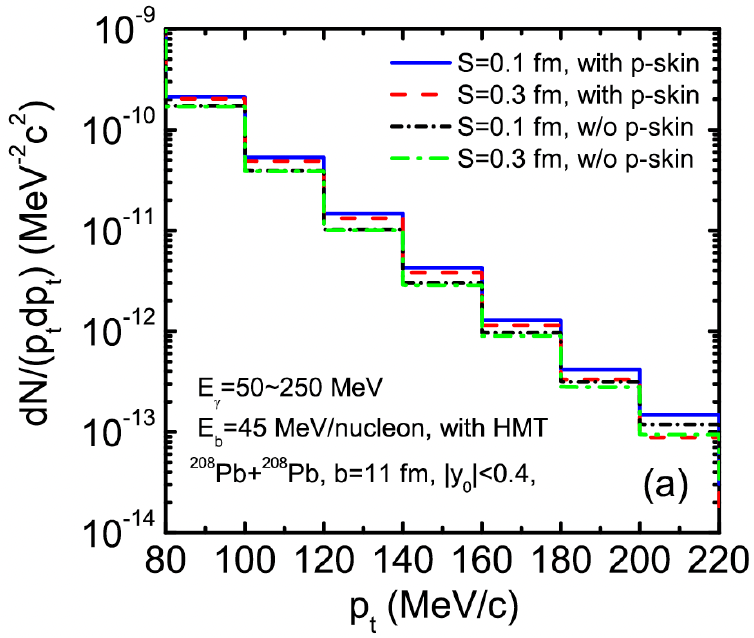}\\[0.25cm]
\includegraphics[width=6.5cm]{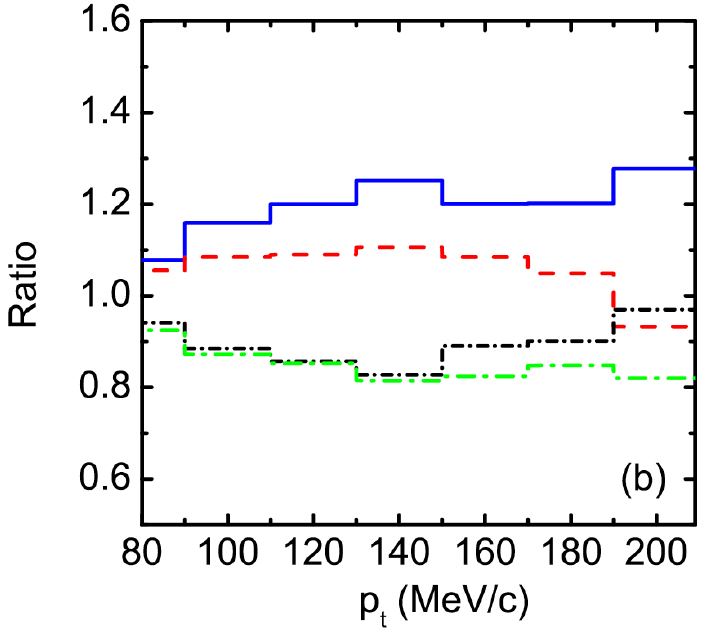}
\caption{(Color Online). Effects of the neutron skin in coordinate ($r$) space and proton skin in momentum ($k$) space on (a) the transverse momentum dependence of hard photons and (b) the ratios of each transverse momentum spectrum to the average spectrum for four cases, in peripheral $^{208}$Pb+$^{208}$Pb collisions at $E_{\rm b} = 45$\,MeV/nucleon. Figure taken from Ref.\cite{Guo23PRC}.}
\label{fig_Guo23-TM}
\end{figure}

\begin{figure}[h!]
\centering
\includegraphics[width=9.cm]{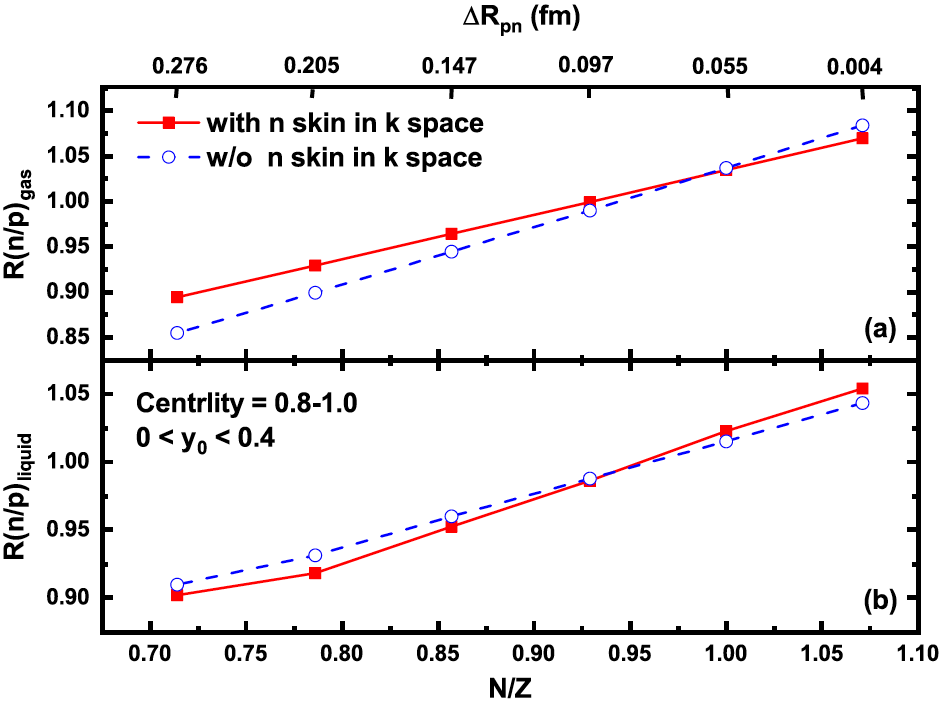}
\caption{(Color Online). Projectile $N/Z$ dependence of the neutron-to-proton yield ratios $R(\rm{n/p})$ in the gas (a) and liquid (b) phases from peripheral $^{48,50,52,54,56,58}$Ni+$^{58}$Ni collisions at a beam energy of 400\,MeV/nucleon, shown with and without the neutron skin in momentum ($k$) space. The proton-skin thickness $\Delta R_{\rm{pn}}$ (in $r$ space) dependence of $R(\rm{n/p})$ in both phases is indicated on the top $x$-axis. Figure taken from Ref.\cite{Guo24PRC}.}
\label{fig_Guo24-Rnp}
\end{figure}

It is important to note that the studies discussed here have limitations and that further work is needed, as pointed out by Ref.\cite{Guo23PRC}. Beyond the effects of neutron skins in coordinate-space and proton skins in momentum-space, many other factors in the initial state of the projectile and target can influence hard-photon observables in transport-model simulations. These include the size, shape, and isospin dependence of the SRC-HMT, the momentum and density dependence of the symmetry potential and the associated neutron-proton effective mass splitting, and in-medium NN collision cross sections. While some of these uncertainties have been studied individually in previous works, systematic uncertainties remain that are not yet fully quantified. A comprehensive analysis to assess the uncertainties and correlations among these key variables, using approaches such as covariance analysis or Bayesian inference once sufficient experimental data are available, would be highly valuable.

In a recent study, the authors of Ref.\cite{Guo23NPA} employed an isospin- and momentum-dependent Boltzmann--Uehling--Uhlenbeck (IBUU) transport model \cite{Li:1997rc,Li:1997px,Li:2003zg,Li:2003ts} with the Gogny-type energy density functional \cite{Das2003PRC} to examine the effects of nucleon-nucleon SRCs and nuclear symmetry energy on n/p yield ratios in four reaction systems, namely $^{94}$Kr+$^{94}$Kr, $^{94}$Zr+$^{94}$Zr, $^{94}$Mo+$^{94}$Mo, and $^{94}$Ru+$^{94}$Ru. They found that the impacts of the symmetry energy and the SRC-induced HMT on the kinetic-energy dependence of the n/p yield ratio are essentially the same across all four systems. Notably, the HMT significantly enhances the double n/p ratio between neutron-rich and neutron-deficient systems ($^{94}$Kr+$^{94}$Kr vs $^{94}$Ru+$^{94}$Ru), producing an effect roughly twice as large as that of the symmetry energy at high kinetic energies. In contrast, the isospin-fractionation, defined as the double n/p ratio between gas and liquid phases, is insensitive to HMT but remains strongly dependent on the symmetry energy. This suggests that the dependence of isospin-fractionation on the system's isospin asymmetry could serve as a robust experimental observable for probing the nuclear symmetry energy.

\begin{figure}[h!]
\centering
\resizebox{0.45\textwidth}{!}{
  \includegraphics[width=7cm]{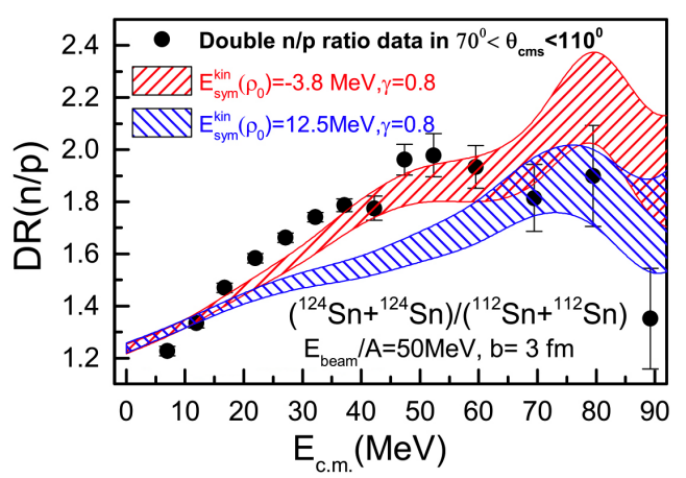}
  }
  \resizebox{0.43\textwidth}{!}{
  \includegraphics[width=10cm]{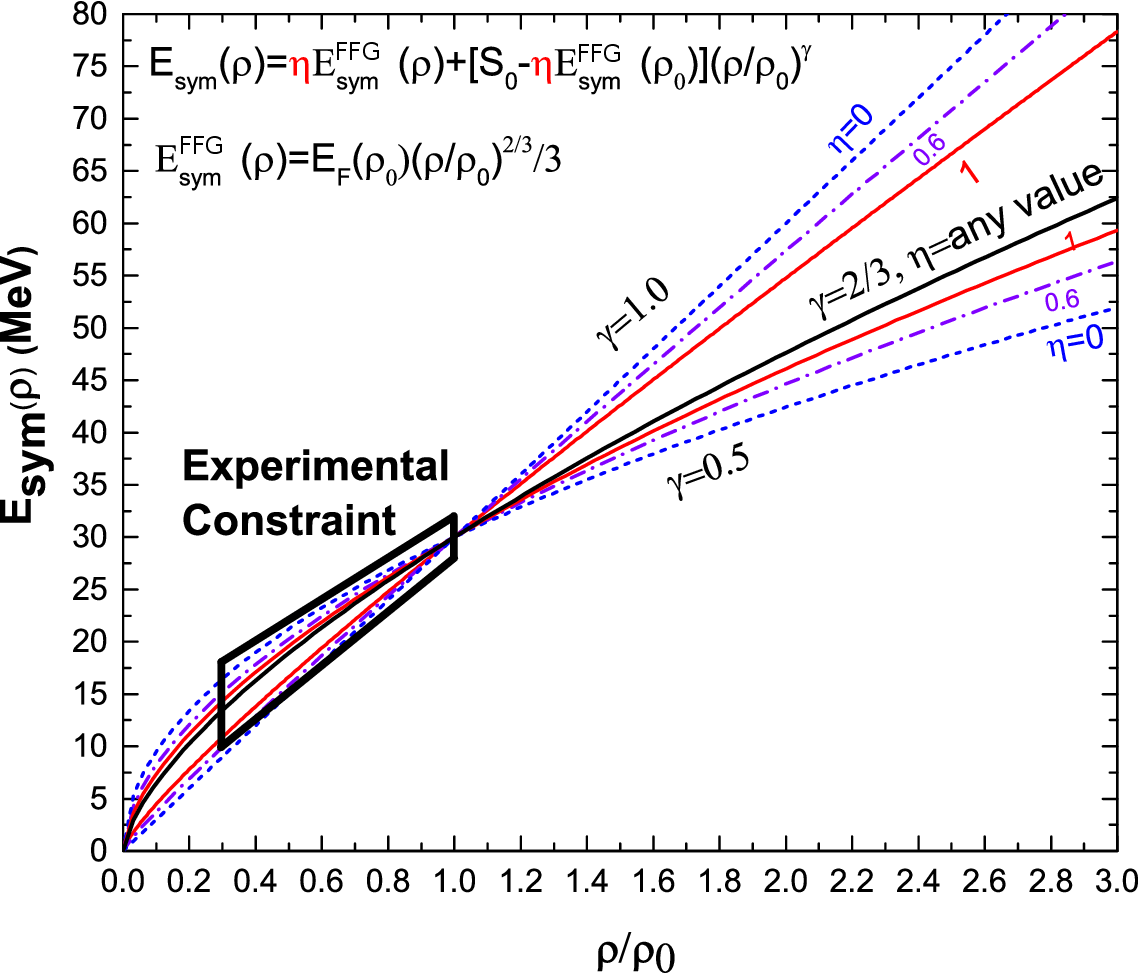}
  }
  \caption{(Color Online). Upper: calculated double ratio of free
neutron/protons in the two reactions in comparison with the
MSU data with different kinetic symmetry energy; figure taken from Ref.\cite{Hen15b}. Lower: freedom of diving total symmetry energy into its kinetic and potential parts within the existing constraints; figure taken from Ref.\cite{Li15PRC}.}
  \label{Esym-eta}
\end{figure}

\begin{figure}[h!]
\centering
  \includegraphics[width=8.5cm]{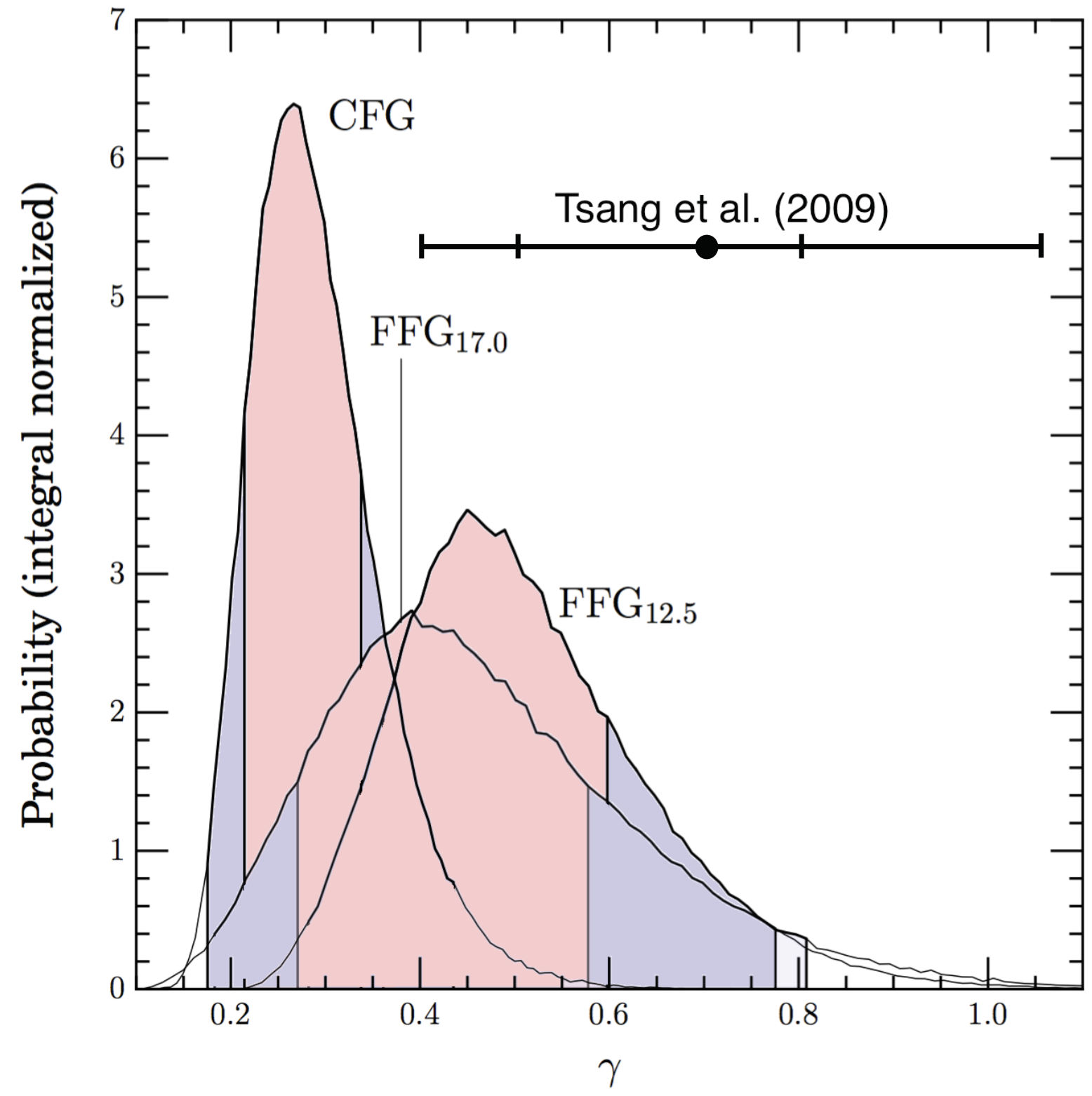}
  \caption{(Color Online). Bayesian analysis on the $\gamma$ parameter using NS observables in the correlated Fermi gas (CFG) and FFG models.  Figure taken from Ref.\cite{Hen16-gamma}.}
  \label{fig_Hen-gamma}
\end{figure}

\begin{figure*}[h!]
\centering
  \includegraphics[height=5.1cm]{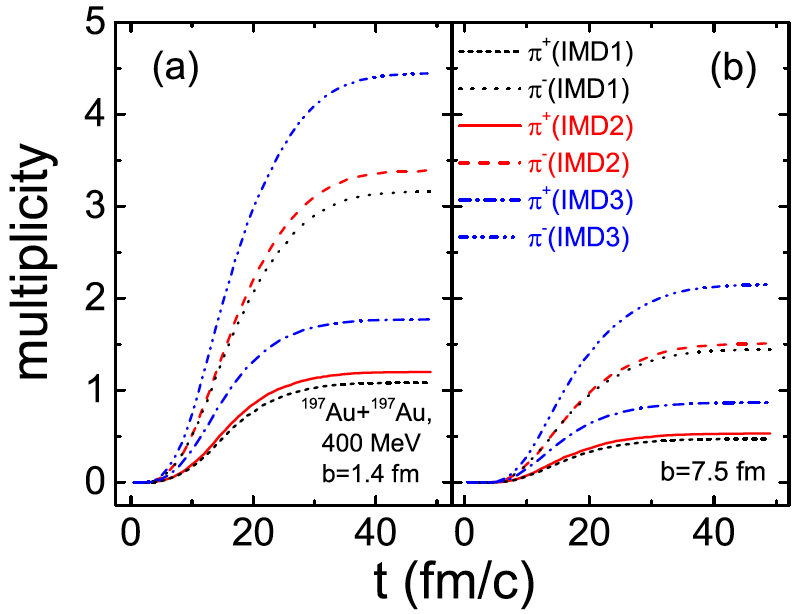}\qquad
  \includegraphics[height=5.cm]{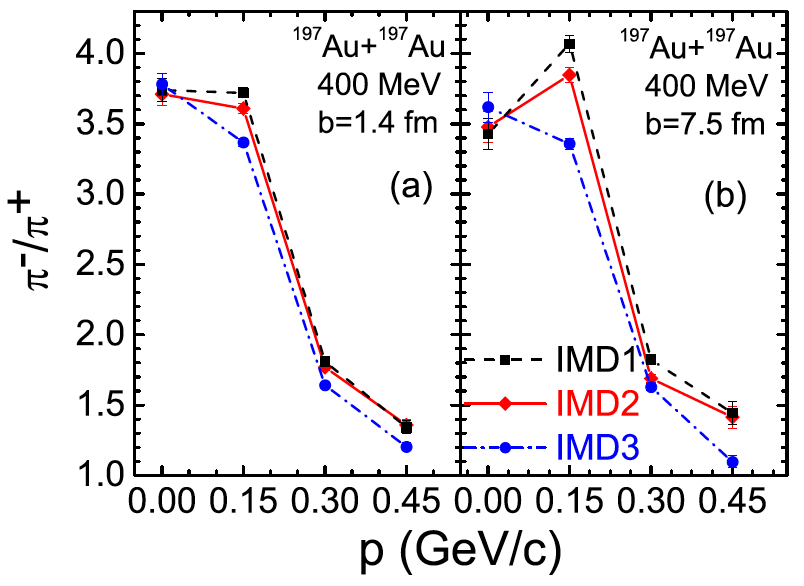}\\[0.5cm]
\hspace{-0.3cm}  
  \includegraphics[height=5.cm]{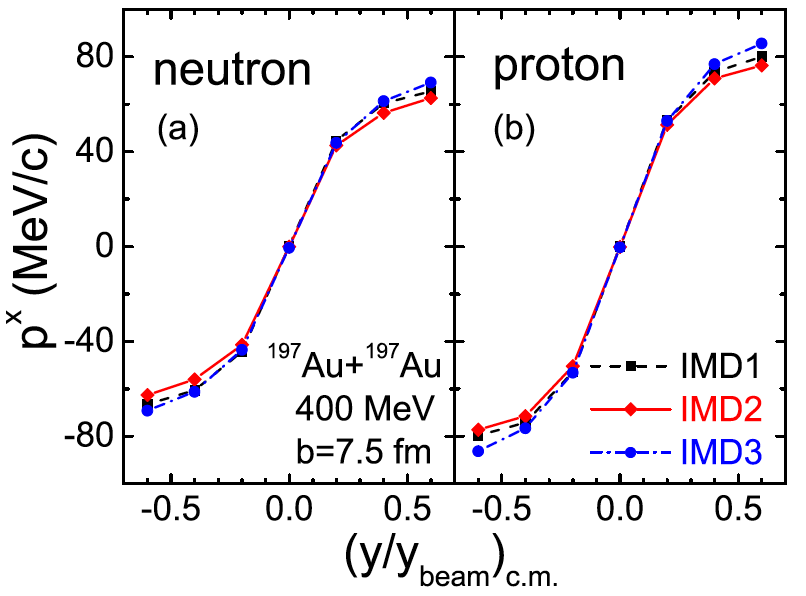}\qquad
  \includegraphics[height=5.2cm]{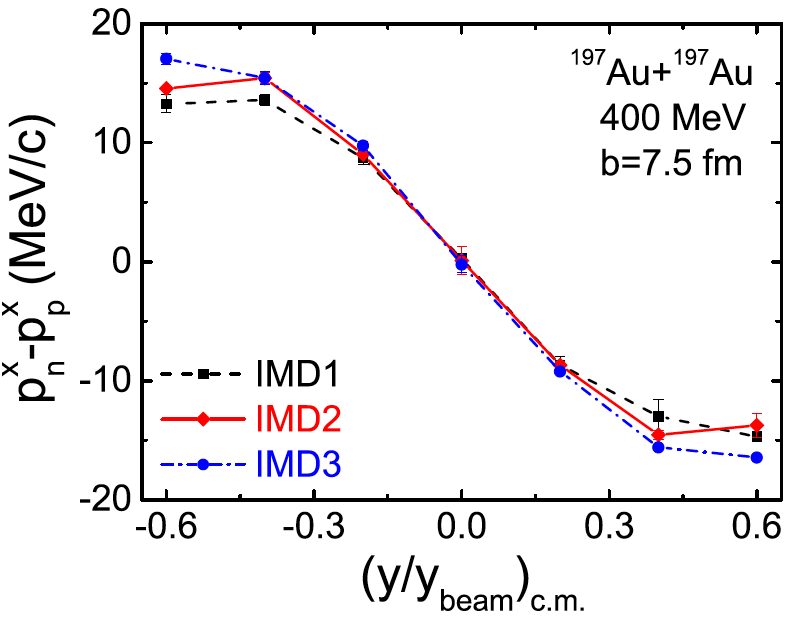}\\[0.5cm]
\hspace{-0.4cm}  
  \includegraphics[height=5.cm]{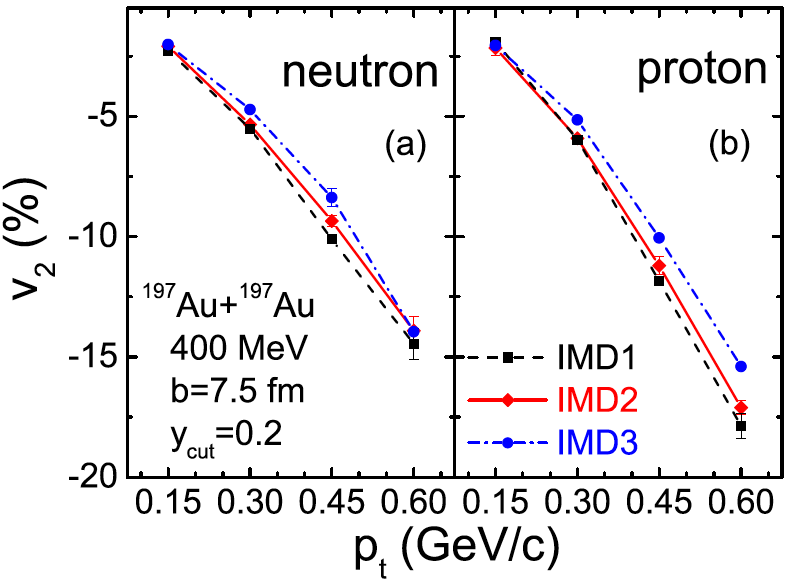}\qquad
  \includegraphics[height=5.2cm]{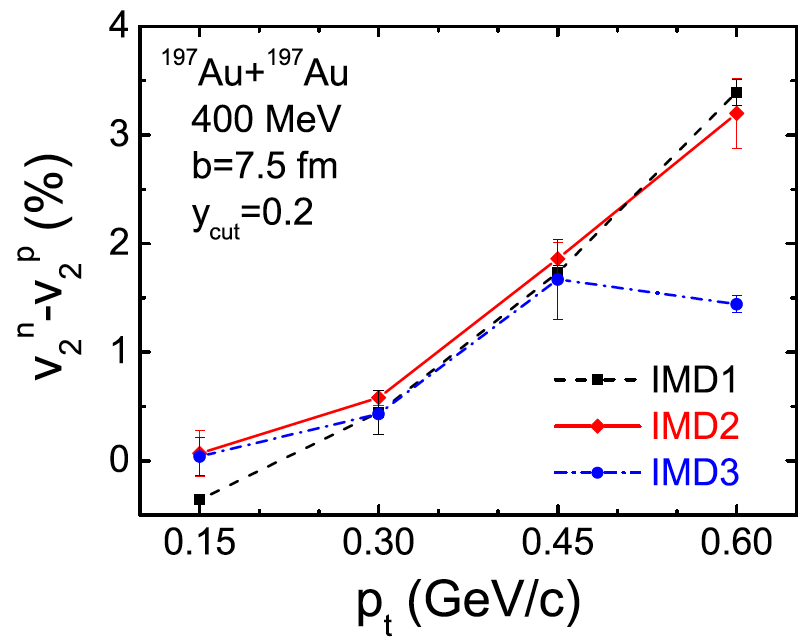}
  \caption{(Color Online). First row: yields of $\pi^{-}$ and $\pi^{+}$ as functions of time for different initializations of the nucleon momentum distribution in (a) central and (b) semi-central ${}^{197}\mathrm{Au}+{}^{197}\mathrm{Au}$ reactions at an incident beam energy of 400\,MeV/nucleon (left) and the $\pi^{-}/\pi^{+}$ ratio as a function of momentum for the same three initializations in (a) central and (b) semi-central collisions at 400\,MeV/nucleon (right).
  Second row: nucleon transverse flow $p^{x}$ and (left) the relative flow $p_{x}^{\rm{n}}-p_{y}^{\rm{p}}$ (right).
Third row: nucleon elliptic flow $v_2$ as a function of transverse
momentum $p_{\rm t}$ (left) and the relative nucleon elliptic flow $v^{\rm n}_2-v_2^{\rm p}$ (right). Figures taken from Ref.\cite{Yang18PRC}.}
  \label{fig_Yang18PRC}
\end{figure*}

\begin{figure*}[h!]
\centering
  \includegraphics[width=15.cm]{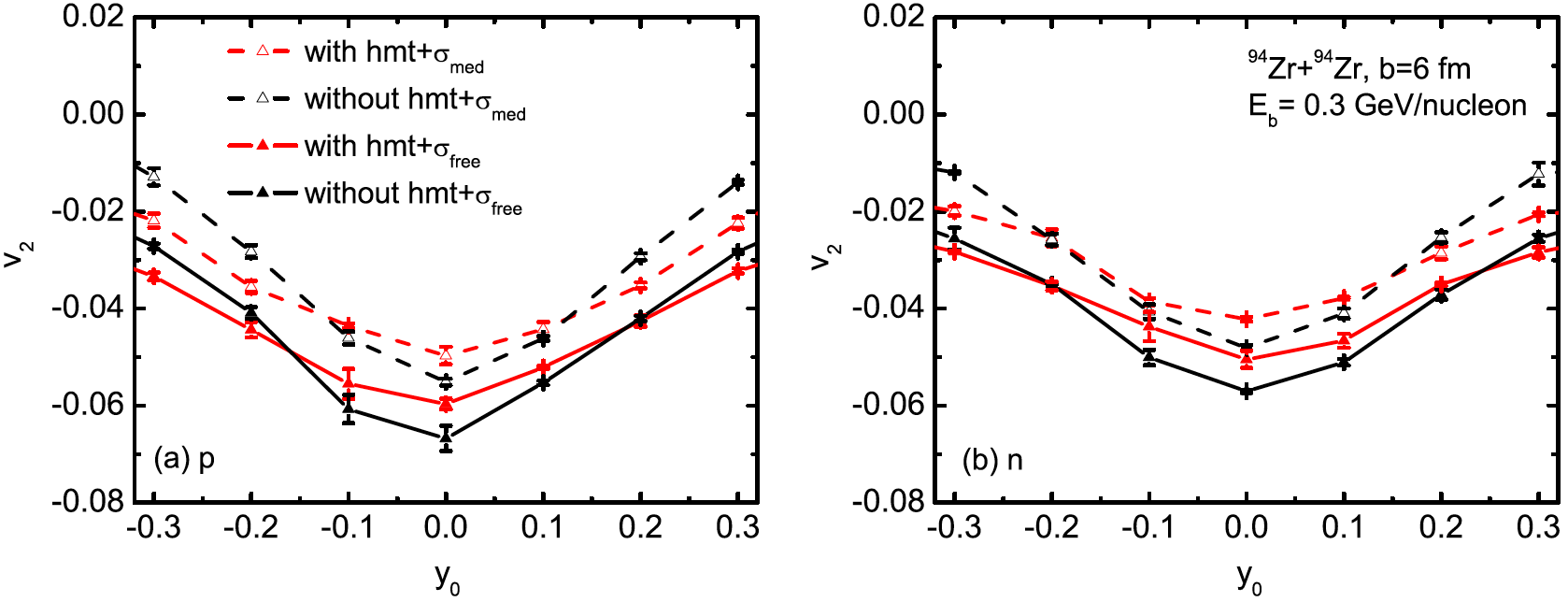}\\[0.5cm]
  \hspace{0.4cm}
   \includegraphics[width=15.2cm]{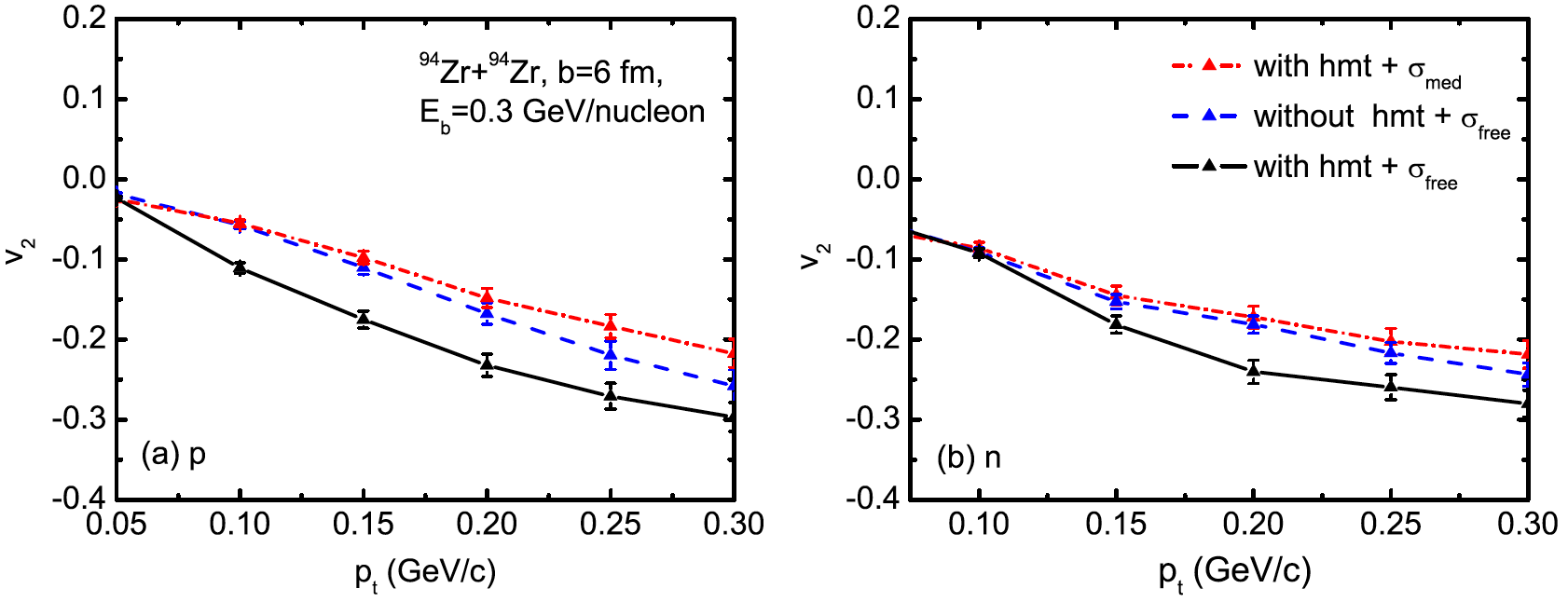}
  \caption{(Color Online). Upper: elliptic flow $v_{2}$ of free (a) protons and (b) neutrons in peripheral $^{94}\mathrm{Zr}{+^{94}}\mathrm{Zr}$ collisions at $E_{\rm b}=0.3\,\mathrm{GeV}/\mathrm{nucleon}$ as a function of the reduced rapidity $y_{0}$. 
 Lower: similar as the first row but showing $v_2$ as a function of the nucleon transverse momentum. Figures taken from Ref.\cite{Guo25PRC}.}
  \label{fig_Guo25-v2}
\end{figure*}

Extending these ideas to neutron-deficient Ni isotopes ($^{48,50,52,54,56,58}$Ni+$^{58}$Ni at 400\,MeV/nucleon)\cite{Guo24PRC}, SRCs induced by the tensor force in the isosinglet neutron-proton channel were incorporated via an extended Thomas--Fermi approximation (ETF$^+$)\cite{Cai16b} for the nucleon kinetic-energy density, producing a neutron skin in momentum-space. The study\cite{Guo23PRC} using the Gogny-type approach revealed that the yield ratios of $R(\rm{n/p})$ from different Ni projectiles follows a negative exponential trend with proton skin thickness, while inclusion of the neutron skin in $k$-space reduces the attenuation of the yield ratio. This underscores the need to account for both SRCs and neutron skins in momentum-space when deducing proton skin thickness from isotopic comparisons. Moreover, the $N/Z$ dependence of $R(\rm{n/p})$ in the gas phase at projectile-like mid-rapidity effectively reflects the surface properties of neutron-deficient nuclei, making it a useful observable for studying nuclear surface features near the $\beta$-stability line.
Specifically, in FIG.\,\ref{fig_Guo24-Rnp}, the top $x$-axis shows the dependence of $R(\rm{n/p})$ on proton-skin thickness $\Delta R_{\rm{pn}}$ (in $r$-space) for cases with and without the neutron skin in momentum ($k$) space. Both gas- and liquid-phase yield ratios decrease with increasing $\Delta R_{\rm{pn}}$, indicating that Ni projectiles with thicker proton skins produce smaller $R(\rm{n/p})$. This reflects the fact that projectile-like fragments originate mainly from the projectile portion of the overlap zone, so the yield ratios are close to the projectile's $N/Z$. For proton-rich isotopes, a thicker proton skin implies fewer neutrons, reducing $R(\rm{n/p})$. The yield ratios decrease approximately exponentially with $\Delta R_{\rm{pn}}$, and the attenuation is reduced when including the neutron skin in $k$-space, highlighting its importance when deducing proton-skin thicknesses from mid-rapidity $R(\rm{n/p})$ in peripheral collisions.

Actually, the SRC-induced reduction of the kinetic symmetry energy relative to the FFG prediction significantly impacts the understanding of the origin of the symmetry energy and affects various isovector observables in heavy-ion collisions\cite{Hen15b,Hen15a,Li15PRC,Yong1,Yong2,Yong3,Yong4,Zhang19,ShenL22,XBS25}, such as the free n/p and $\pi^-/\pi^+$ ratios. In transport simulations and phenomenological EOS constructions, it is customary to parameterize the symmetry energy as a sum of the FFG kinetic term and an interaction contribution:
\begin{equation}\label{esym}
E_{\rm{sym}}(\rho)=E_{\rm{sym}}^{\rm{kin,FFG}}(\rho)+\left[S_0-E_{\rm{sym}}^{\rm{kin,FFG}}(\rho_0)\right]\left(\frac{\rho}{\rho_0}\right)^{\gamma},
\end{equation}
where $E_{\rm{sym}}^{\rm{kin,FFG}}(\rho)\approx 12.5(\rho/\rho_0)^{2/3}\equiv E^{\rm{kin}}_{\rm{sym}}(\rho)$, $S_0$ is the total symmetry energy at $\rho_0$, and $\gamma$ is a parameter to be extracted from isospin-sensitive observables. The conventional use of $E_{\rm{sym}}^{\rm{kin,FFG}}(\rho_0)\approx 12.5$\,MeV is questioned, particularly in heavy-ion simulations, since only the nucleon symmetry potential (corresponding to the potential part of the symmetry energy) enters the transport equations directly, see, e.g., Ref. \cite{Li:1997rc}. Under fixed $S_0$, the choice of kinetic symmetry energy affects the extracted $\gamma$ from data fitting. Current experimental constraints near and below $\rho_0$ do not restrict the division of the total symmetry energy, see FIG.\,\ref{Esym-eta}.
To account for possible reductions of kinetic symmetry energy, a phenomenological factor $\eta$ can be introduced:
\begin{equation}\label{esym-reduced}
E_{\rm{sym}}(\rho)=\eta E_{\rm{sym}}^{\rm{kin,FFG}}(\rho)+\left[S_0-\eta  E_{\rm{sym}}^{\rm{kin,FFG}}(\rho_0)\right]\left(\frac{\rho}{\rho_0}\right)^{\gamma}.
\end{equation}
Since $E_{\rm{sym}}^{\rm{kin,FFG}}(\rho)\sim (\rho/\rho_0)^{2/3}$, $\eta$ can take any value if $\gamma=2/3$, yielding $E_{\rm{sym}}(\rho)=S_0 (\rho/\rho_0)^{2/3}$. For $0.5\lesssim \gamma \lesssim 1$, $\eta$ between 0 and 1 is still consistent with existing constraints\cite{Li15PRC}. 
Therefore, a reduction of the quasi-nucleon kinetic symmetry energy relative to the FFG value is not forbidden. Moreover, the slope of the symmetry energy at $\rho_0$ is given by:
\begin{align}
    L\equiv \left.3\rho_0\frac{\d E_{\rm{sym}}}{\d\rho}\right|_{\rho=\rho_0}=&2\eta L_{\rm{kin}}^{\rm{FFG}}+3\gamma\left[S_0-\eta  E_{\rm{sym}}^{\rm{kin,FFG}}(\rho_0)\right]\notag\\
    =&(4-3\gamma)\eta E_{\rm{sym}}^{\rm{kin,FFG}}(\rho_0)+3\gamma S_0,
\end{align}
the $\eta$ parameter affects the $L$.
Recent analyses using nonlinear RMF models indicate $\gamma\approx 0.25\pm0.05$ based on maximum NS mass and the $E_{\rm{sym}}(\rho_0)$-$L$ correlation\cite{Dut18}.
A small $\gamma\approx0.3$ was also obtained in Ref.\cite{Hen16-gamma} using a Bayesian analysis on the NS observable, as shown in FIG.\,\ref{fig_Hen-gamma}.
In addition, the $\gamma$ parameter in QCD sum rules incorporating the twist-4 condensates was shown to be about 0.17\cite{Cai19-QSR2}, see FIG.\,\ref{fig_QCDSR} for the density dependence of the $E_{\rm{sym}}(\rho)$ of the set QCDSR-3 (where twist-4 effects are included).

\begin{figure*}[h!]
\centering
  \includegraphics[width=17.5cm]{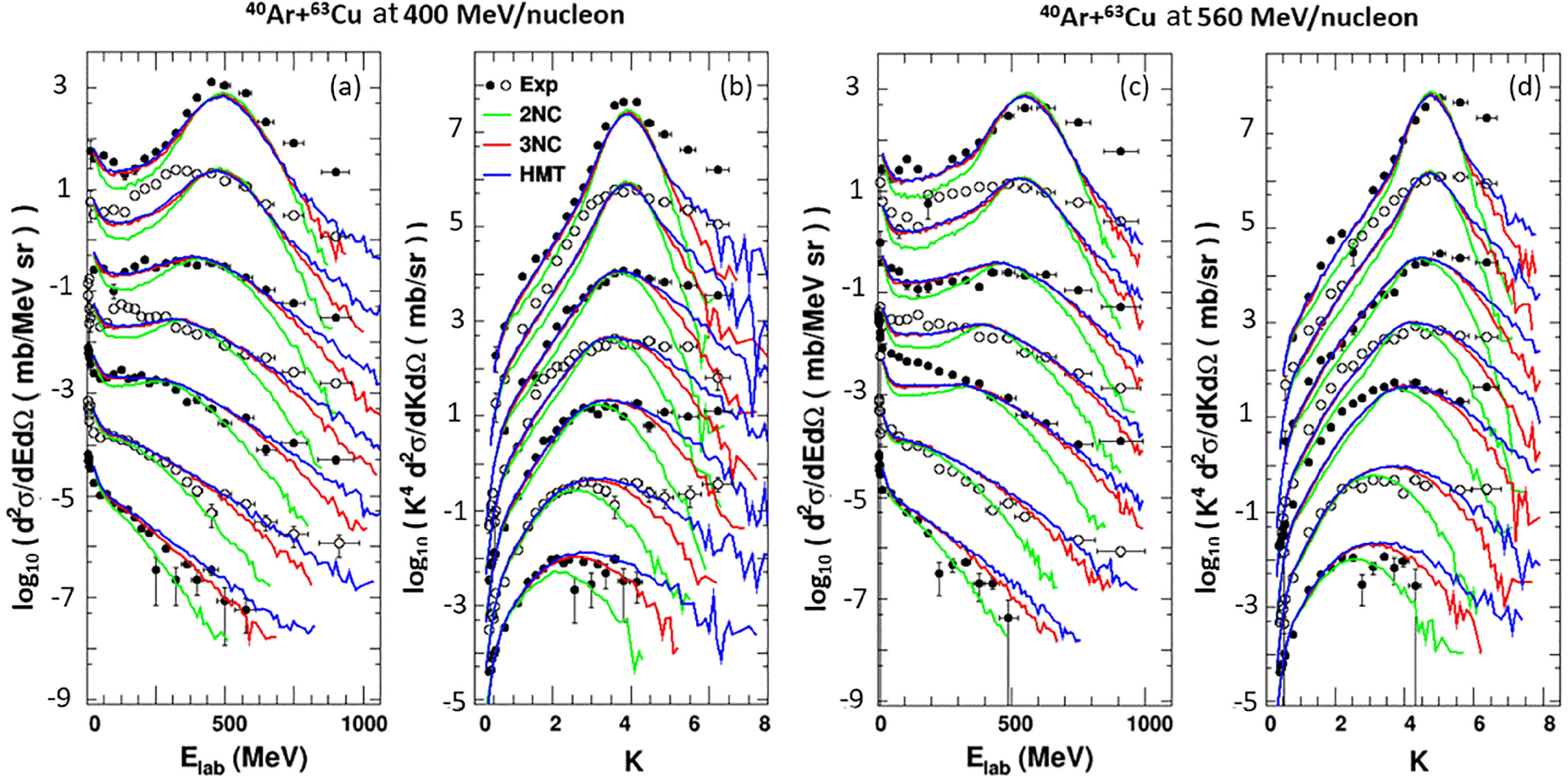}\\[0.5cm]
  \hspace{.1cm}
  \includegraphics[width=17.5cm]{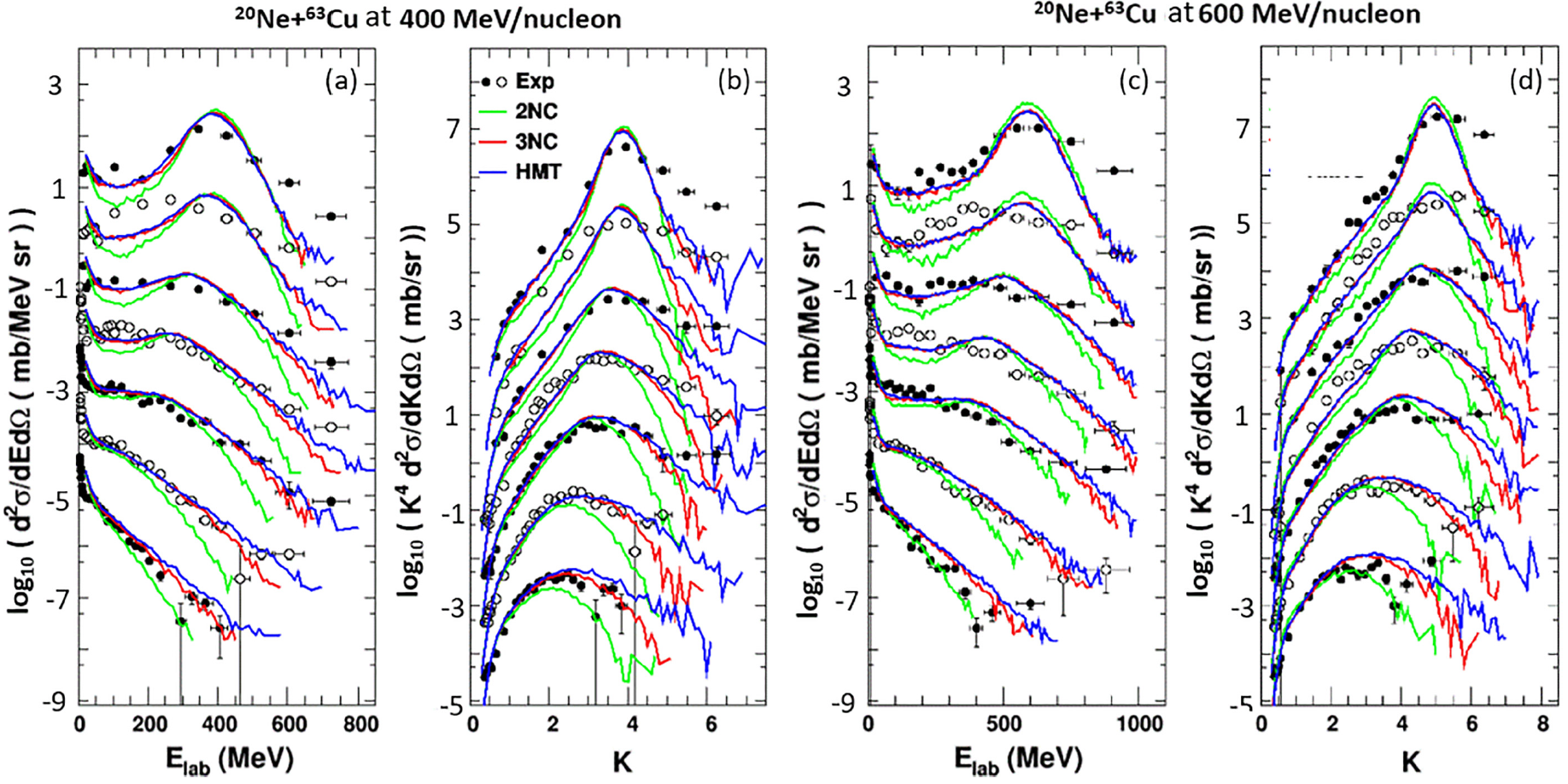}
  \caption{(Color Online). Upper: neutron double-differential cross sections for $^{40}$Ar + Cu collisions are shown at $E_{\mathrm{inc}}/A = 400$\,MeV (left) and 560\,MeV (right).  
For each energy, panels (a) and (c) display the neutron energy spectra, while panels (b) and (d) show the $k^{4}$-weighted momentum spectra.  
In all panels, spectra for $\theta_{\mathrm{lab}} = 5^\circ, 10^\circ, 20^\circ, 30^\circ, 40^\circ, 60^\circ$, and $80^\circ$ are plotted from top to bottom.  
Bottom: Plots similar to those in the upper panel, but for $^{20}\rm{Ne}$+Cu collisions at $E_{\rm{inc}}/A = 400$\,MeV on (a) and (b) and at 600\,MeV on (c) and (d). Figures taken from Ref.\cite{Wada25PRC}.}
  \label{fig_Wada25}
\end{figure*}

Similarly, the effects of differences in the HMTs of the nucleon momentum distributions in the colliding nuclei are investigated for the ${}^{197}\mathrm{Au}+{}^{197}\mathrm{Au}$ reactions at a beam energy of 400\,MeV/nucleon using the IBUU model. It is found that the nucleon transverse and elliptic flows, as well as the free n/p ratio at low momenta, are only weakly sensitive to the specific form of the HMT. In contrast, the free n/p ratio at high momenta, the yields of $\pi^{-}$ and $\pi^{+}$, and particularly the $\pi^{-}/\pi^{+}$ ratio around the Coulomb peak, exhibit a pronounced sensitivity to the details of the HMT. When combined with experimental measurements from rare-isotope beam facilities worldwide, these studies may provide deeper insights into nuclear SRCs in heavy nuclei and in nuclear matter. See FIG.\,\ref{fig_Yang18PRC} for the results where three different parametrizations for the HMT are adopted in the study.
Similar investigations were also executed in Ref.\cite{Wang17PRC,Wei20NST}, see FIG.\,\ref{fig_Guo25-v2}.
The elliptic flow $v_2$ was studied using the same systems of Ref.\cite{Guo24PRC} by considering the in-medium NN cross sections besides the SRC-HMT. The elliptic flows of protons and neutrons exhibit strong sensitivity to both the in-medium NN cross sections and the presence of HMT, in peripheral $^{94}\mathrm{Zr}+{^{94}}\mathrm{Zr}$ collisions at $E_{\rm b}=0.3\,\mathrm{GeV}/\text{nucleon}$. Increasing the in-medium cross sections reduces $v_{2}$ at mid-rapidity ($y_{0}=0$), as more frequent collisions enhance squeeze-out (negative $v_{2}$) dynamics. When HMTs are included, $v_{2}$ displays a characteristic pattern: it increases slightly for $|y_{0}| \lesssim 0.15$ and decreases for $|y_{0}| \gtrsim 0.15$. This behavior reflects the rapidity dependence of anisotropic nucleon emission. The SRC-induced HMT depletes low-rapidity nucleons, weakening emission anisotropy and thus reducing $v_{2}$ around mid-rapidity. At the same time, high-momentum nucleons preferentially populate larger rapidities, experience more NN collisions, and consequently exhibit stronger out-of-plane emission.
In addition, the transverse-momentum dependence of the elliptic flow $v_{2}$ for free protons and neutrons reveals a clear interplay between in-medium cross sections and HMT. The impact of the in-medium cross sections is pronounced for both species, while the inclusion of HMT systematically enhances the magnitude of $v_{2}$. Moreover, the HMT-induced increase is stronger for protons than for neutrons, reflecting that in a neutron-rich system protons are more strongly influenced by SRCs.
Other quantities like the  nuclear stopping powers as well as their ratio were also studied in Ref.\cite{Guo25PRC}.

\begin{figure}[h!]
\centering
  \includegraphics[width=9.cm]{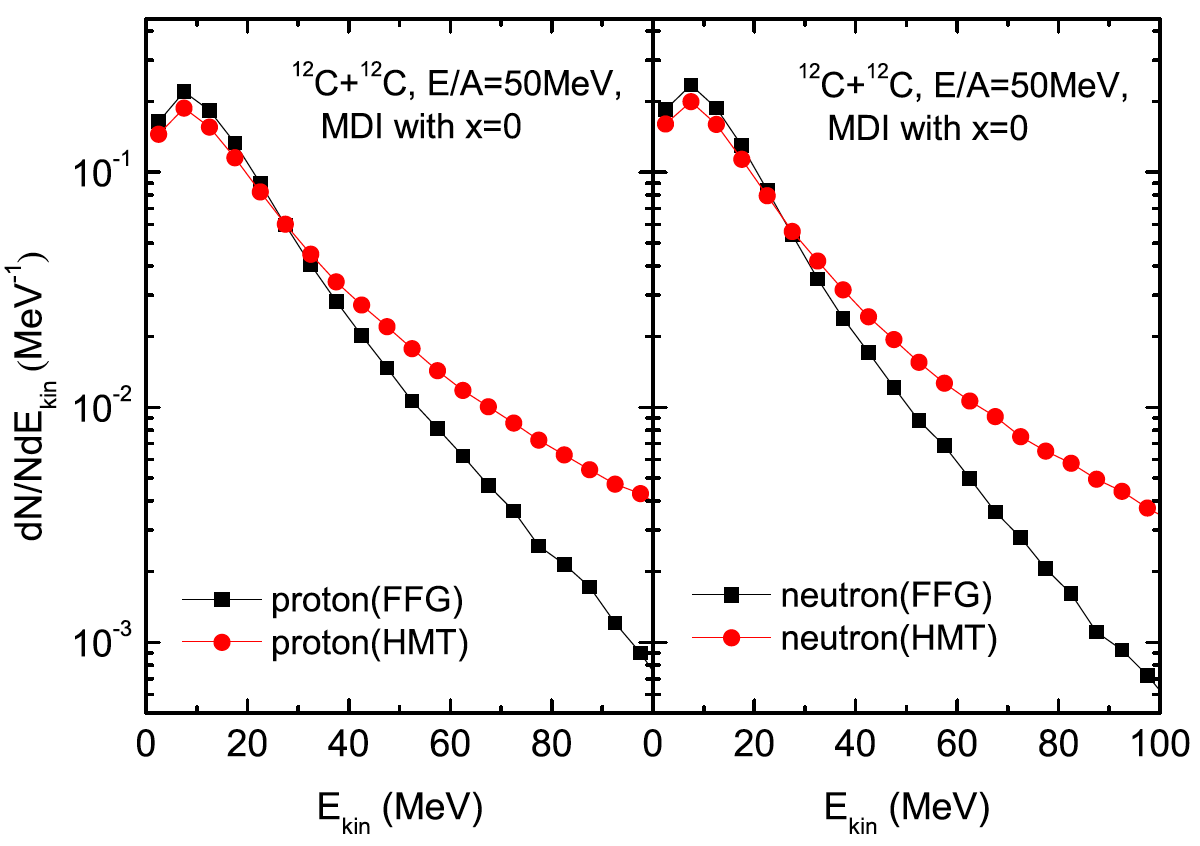}
  \caption{(Color Online). Effect of the HMT on the probability distributions of free proton and neutron kinetic energies in the $^{12}$C + $^{12}$C central reaction with the MDI interaction at a beam energy of 50\,MeV/nucleon.  Figures taken from Ref.\cite{Wang17PRC}.}
  \label{fig_HMT-em}
\end{figure}

\begin{figure}[h!]
\centering
  \includegraphics[width=7.cm]{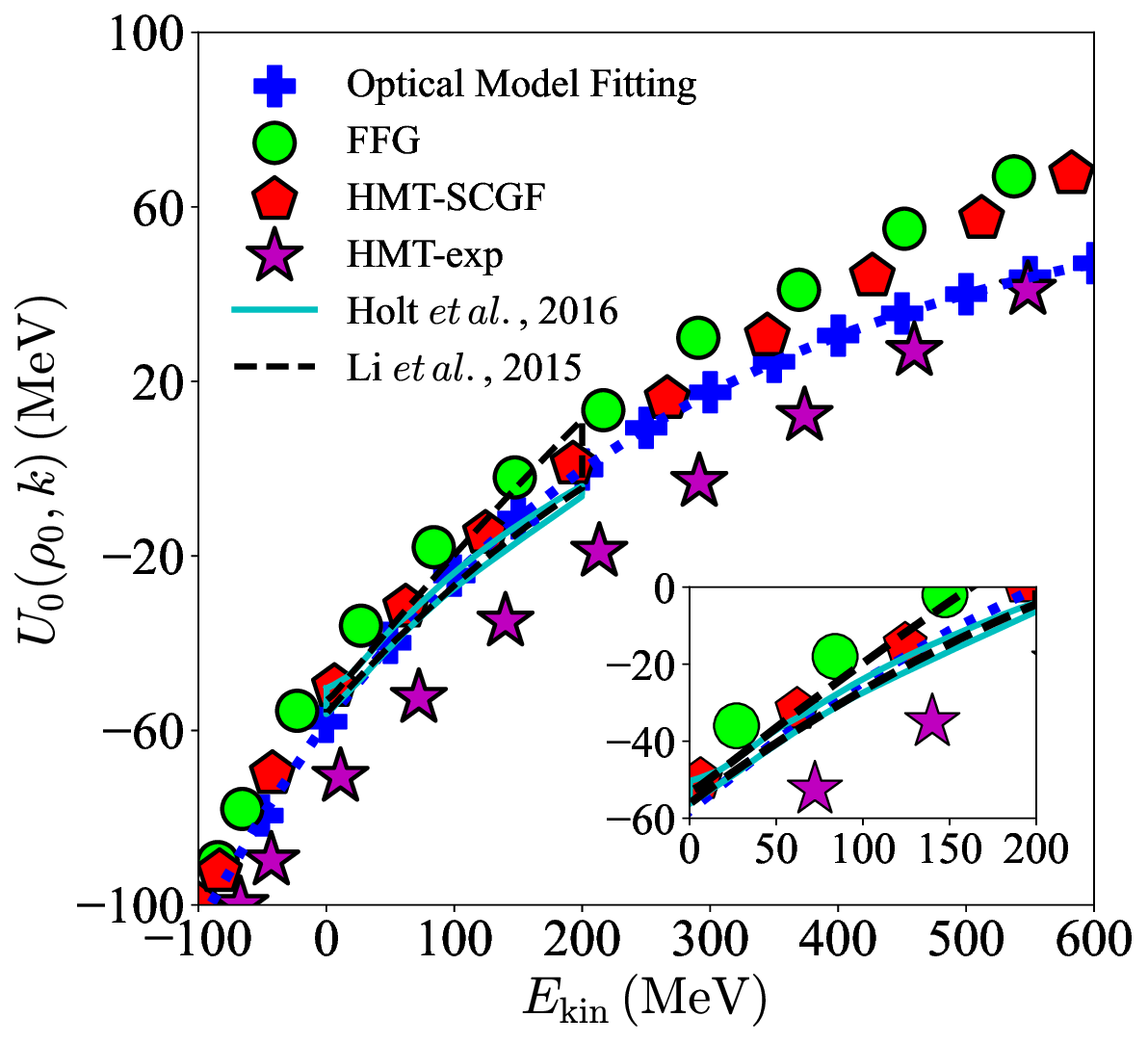}\\[0.25cm]
  \includegraphics[width=4.3cm]{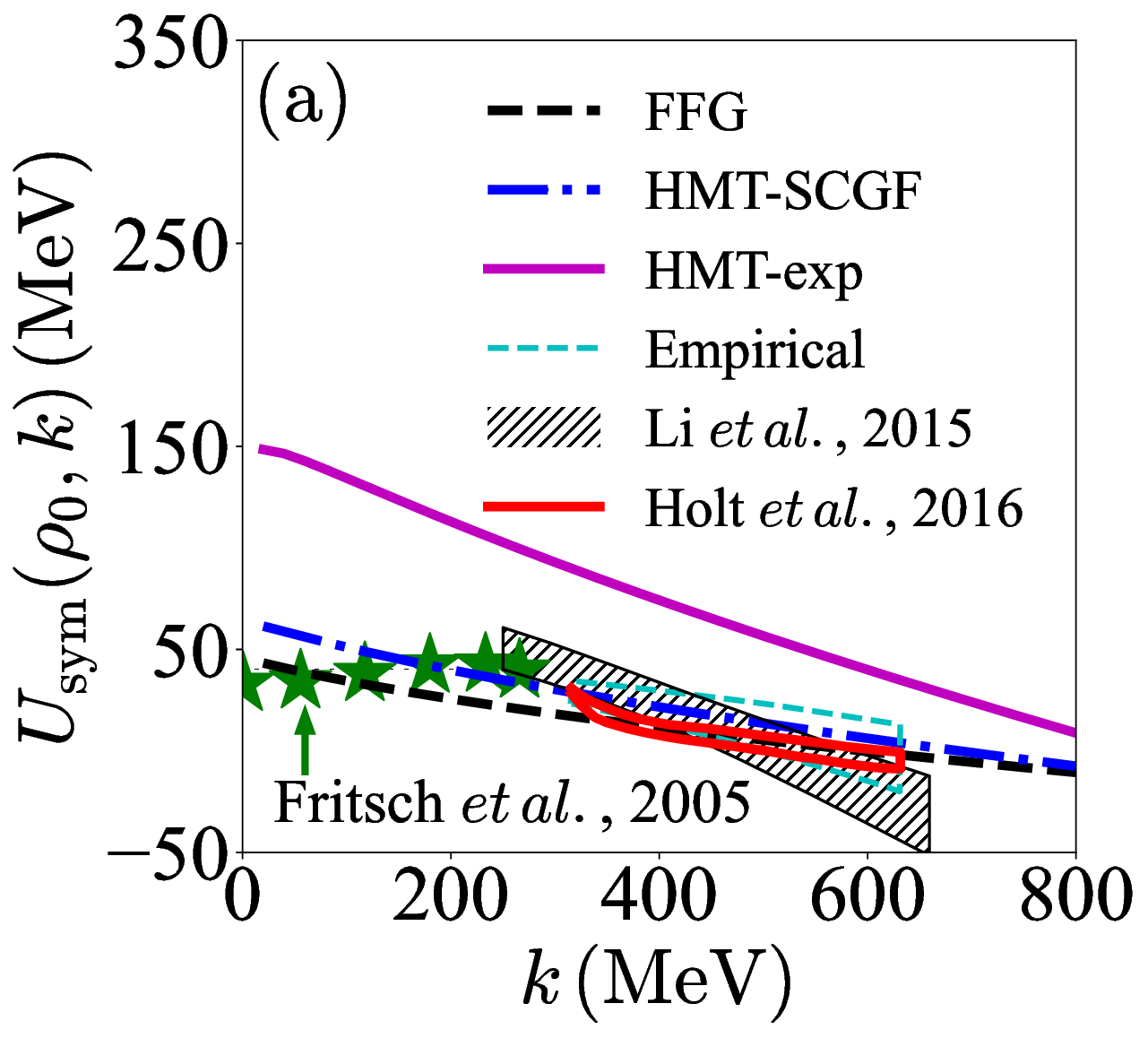}\quad
  \includegraphics[width=4.3cm]{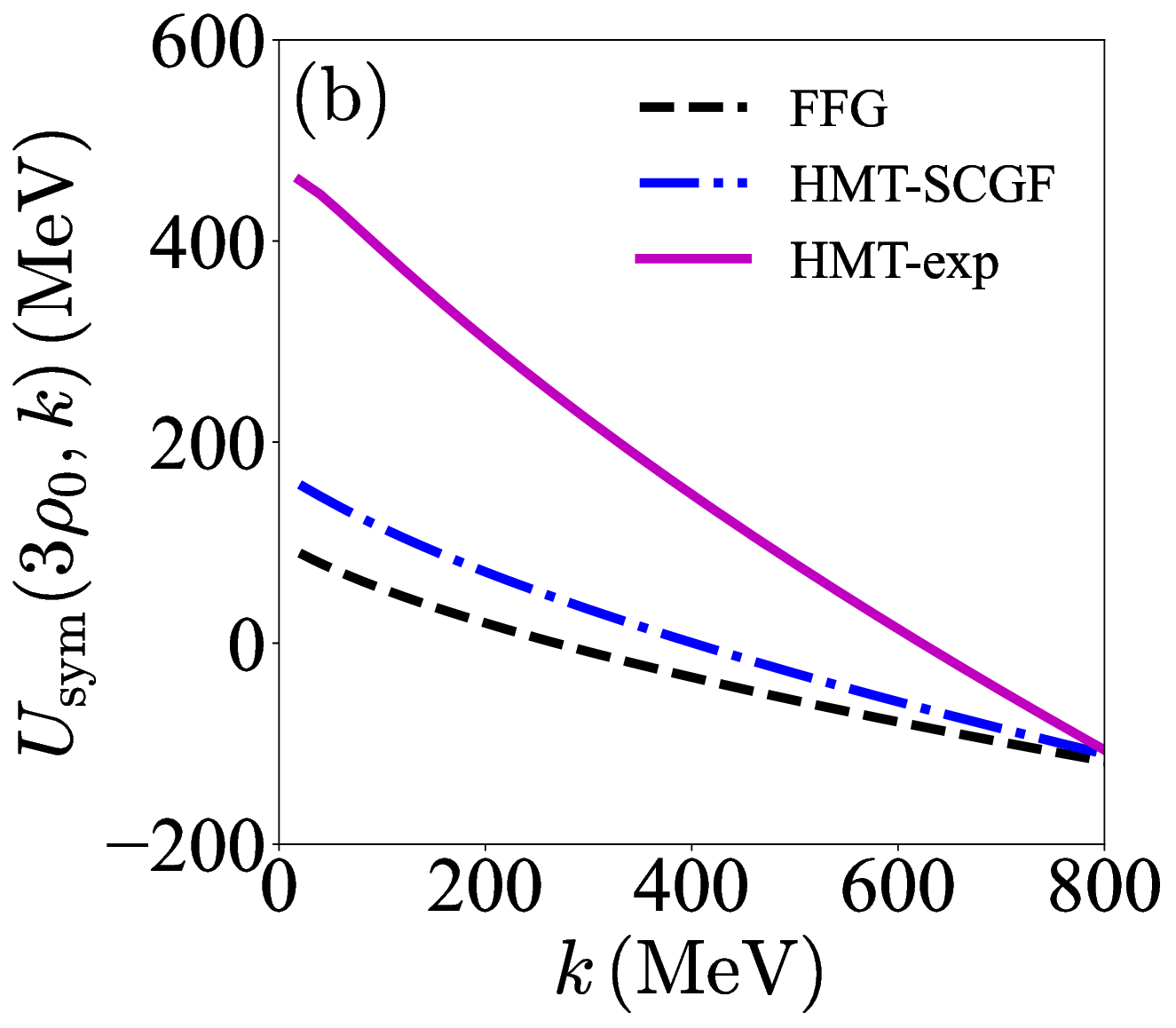}
\caption{(Color Online). Single nucleon potential in SNM as a function of kinetic energy in the FFG model and
  HMT models (upper) and the symmetry potential at two reference densities (lower). Figures taken from Ref.\cite{CaiLi22Gog}.}
  \label{fig_ab_U0}
\end{figure}

In Ref.\cite{Wada25PRC}, the authors investigates whether high-energy neutron emission in $^{40}\mathrm{Ar}$ and $^{20}\mathrm{Ne}$ on Cu reactions at $E_{\mathrm{inc}}/A=400$-$600$\,MeV carries signatures of nucleon-nucleon SRCs. Using available neutron spectra\cite{Iwata01PRC}, the authors employ semi-relativistic antisymmetrized molecular dynamics (SrAMD) with a three-nucleon collision mechanism in which nucleons receive boosts including both Fermi motion and an SRC-related HMT. Forward-angle neutron spectra exhibit high-energy components above projectile velocity, but these are attributed to nonphysical origins. In contrast, at angles $\theta_{\mathrm{lab}}\gtrsim 60^{\circ}$, the high-energy tails originate at sub-projectile velocities and are well reproduced by SrAMD only when HMTs are included, indicating sensitivity to SRC-induced high-momentum components. This suggests that SRCs can influence high-energy neutron emission in heavy-ion collisions, although limitations in data accuracy and statistics prevent firm conclusions. While HMTs enhance the high-energy spectral tails, the SrAMD calculations do not show the expected flattening in the $k^{4}$-weighted neutron spectra.
Specifically, the HMT of the nucleon momentum distribution with a $k^{-4}$ form, were investigated using SrAMD simulations for the $^{40}$Ar + Cu reaction at $E_{\rm{inc}}/A = 400$ and 560\,MeV, and compared with Iwata's experimental neutron double-differential cross sections. While the experimental data at $\theta_{\rm{lab}} \lesssim 30^\circ$ show significant enhancements above the projectile velocity, these irregularities are largely attributed to non-physical origins. The SrAMD simulations, namely 2NC (SrAMD/D), 3NC (SrAMD/D-3NC), and HMT (SrAMD/D-3NC-HMT)\cite{Wada25PRC}, show that the inclusion of 3NC enhances the high-energy tails at forward angles, while the additional HMT contribution slightly increases the tails but does not produce a flattening in the $k^4$-weighted momentum spectra, as shown in the upper panel of FIG.\,\ref{fig_Wada25}. At low energies and larger angles ($\theta_{\rm{lab}} \gtrsim 40^\circ$), 3NC and HMT reproduce the data reasonably well, whereas 2NC underpredicts the yield. Overall, although the $k^{-4}$ HMT enhances the high-momentum components, it does not generate the expected flattening in the $k^4 \d^2\sigma/\d k \d\Omega$ spectra, suggesting that the experimental high-momentum features are largely dominated by non-physics-related contributions.
In the lower panel of FIG.\,\ref{fig_Wada25}, the neutron double-differential cross sections for the $^{20}$Ne + Cu reaction at $E_{\rm{inc}}/A = 400$\,MeV (left) and 600\,MeV (right) are shown. The overall features of both the experimental data and SrAMD simulations closely resemble those of the $^{40}$Ar + Cu reactions shown in the first row. Minor enhancements over the simulations appear at $E_{\rm{lab}} \sim 100$-200\,MeV for $\theta = 10^\circ$, while the bumps at $\theta = 30^\circ$ are less pronounced and those at $\theta = 40^\circ$ are absent at 600\,MeV. High-energy tail enhancements between 2NC and 3NC are visible at $\theta \lesssim 40^\circ$ for both incident energies, whereas the additional effect of the HMT is slightly reduced compared to the $^{40}$Ar case, indicating that heavier projectiles are more effective in probing HMT contributions. Similar trends are observed in the $k^4$-weighted momentum spectra, consistent with the $^{40}$Ar results.
Overall, the correlation between the flattening of the high-energy tails and SRCs in the $k^4 \d^2\sigma/\d k \d\Omega$ spectra is not supported by the SrAMD/D-3NC-HMT simulations, although noticeable enhancements of the high-energy tails are observed when incorporating the HMT with a $k^{-4}$ power distribution in the Fermi boost.

On the other hand, Ref.\cite{Wang17PRC} found that the HMT strongly affects nucleon emission. In particular, the probability of emitting free nucleons, especially high-energy ones, is sensitive to the high-momentum components of the nuclear momentum distribution. Compared to the FGG case, the HMT enhances the emission of both high-energy protons and neutrons in symmetric and asymmetric nuclear systems.
FIG.\,\ref{fig_HMT-em} shows the probability distributions of free proton and neutron kinetic energies in the $^{12}$C + $^{12}$C central reaction with the MDI interaction \cite{Das2003PRC} at a beam energy of 50\,MeV/nucleon.

To study isospin-sensitive processes and observables, it is essential to construct physically reasonable optical potentials. In Ref.\cite{CaiLi22Gog}, this is explored using a Gogny-type model with momentum-dependent interactions, both with and without the HMT-SRC. The upper panel of FIG.\,\ref{fig_ab_U0} shows the single-nucleon potential in SNM, \(U_0\), as a function of the kinetic energy $
E_{\rm{kin}} = \sqrt{|\v{k}|^2 + M_{\rm N}^2} + U_0(\rho_0,|\v{k}|) - M_{\rm N}$,
with \(U_0(\rho,|\v{k}|)\) given by\cite{CaiLi22Gog}
\begin{align}
U_0(\rho,|\v{k}|) = &\frac{1}{2}A_{\rm{tot}}\left(\frac{\rho}{\rho_0}\right) + B\left(\frac{\rho}{\rho_0}\right)^{\sigma} + C_{\rm{tot}}\left(\frac{\rho}{\rho_0}\right)\notag\\
&\times\left[1 + \frac{4a}{7}\Pi_1\left(\frac{|\v{k}| k_{\rm{F}}}{\Lambda^2}\right)^{1/2}+ \frac{27b}{40}\Pi_2\left(\frac{|\v{k}| k_{\rm{F}}}{\Lambda^2}\right)^{1/3}\right],
\end{align}
here $\Pi_1$ and $\Pi_2$ are two functions characterizing the HMT\cite{CaiLi22Gog}.
Because the SRC-HMT slightly enhances the kinetic part of the EOS of SNM (lower panel of FIG.\,\ref{fig_ab_Esym}), the potential part must be correspondingly reduced to maintain the total EOS at saturation density within empirical constraints. As the difference in the fraction of high-momentum nucleons between SNM and PNM increases, the kinetic EOS of SNM becomes more enhanced, resulting in a softer potential; that is, \(U_0\) in the HMT-exp model is smaller than in the HMT-SCGF model.  
Also shown in the upper panel of FIG.\,\ref{fig_ab_U0} are predictions for \(U_0(\rho_0,|\v{k}|)\) from other approaches, including global relativistic fits to electron-scattering data up to \(\sim 1\,\rm{GeV}\)\cite{Hama1990} (blue ``+''), neutron optical model predictions up to \(\sim 200\,\rm{MeV}\)\cite{LiXH2015PLB} (dashed black band), and chiral effective field theory calculations\cite{Holt2016} (cyan band). It is clear that the HMT-SCGF and HMT-exp models yield results consistent with these approaches for kinetic energies \(\lesssim 600\,\rm{MeV}\).  
Similarly, the momentum dependence of the symmetry potential \(U_{\rm{sym}}(\rho,|\v{k}|)\) at saturation density (lower-left panel) and three times saturation density (lower-right panel) is shown in FIG.\,\ref{fig_ab_U0}. The uncertainties in the symmetry potential are larger than those in \(U_0\), reflecting the poorly known isospin dependence of the EOS of ANM. While the HMT-exp prediction for \(U_{\rm{sym}}(\rho_0,|\v{k}|)\) shows some deviation from microscopic predictions, the FFG and HMT-SCGF models yield results in good agreement with optical model fits\cite{LiXH2015PLB}. Interestingly, because no direct constraints exist on the symmetry potential at higher densities, e.g., \(\rho = 3\rho_0\), intermediate-energy heavy-ion collisions may provide valuable insights into these questions and, in turn, offer a means to test the physics of the SRC-induced HMT.

\subsection{Sub-threshold Particle Production in $\rm{p}A$ Collisions with SRC-HMT}\label{sub_threshold}

\indent
It has long been expected that SRC-HMT effects can affect significantly subthreshold particle productions in nuclear reactions. Very interestingly, in a very recent work \cite{Reich25}, the authors investigated the production of multi-strange baryons and mesons in proton-nucleus collisions at sub-threshold beam energies. They attribute this enhanced production to SRC between nucleons, which effectively increase the center-of-mass energy of individual nucleon-nucleon collisions. By incorporating SRC, the authors show that the probability for sub-threshold particle production can increase by up to three orders of magnitude compared to a simple Fermi gas model, and they benchmark their approach by calculating the $\Xi^-$ multiplicity in p+Nb collisions, finding good agreement with HADES data. 

\begin{figure}[h!]
\centering
  \includegraphics[height=8.cm]{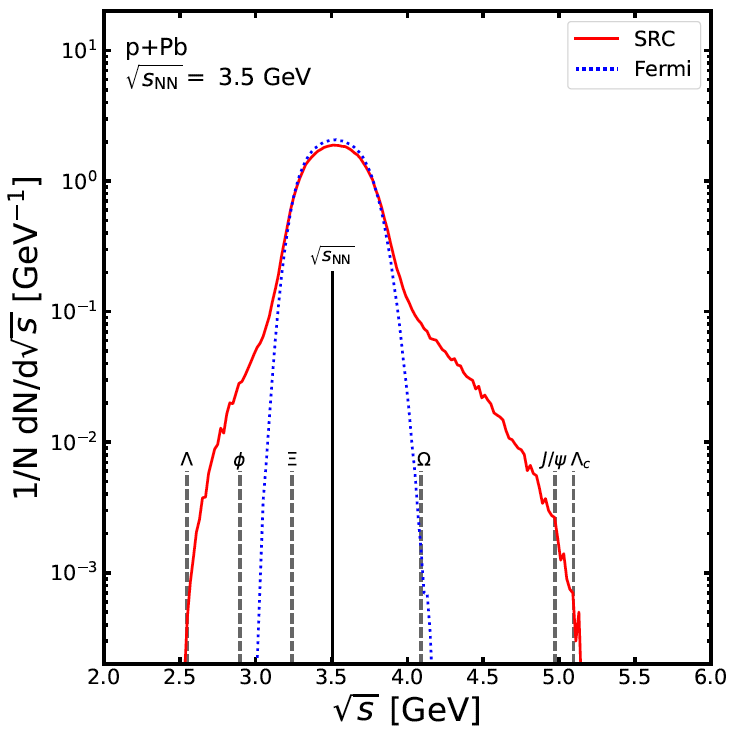}\\[0.25cm]
  \includegraphics[height=8.cm]{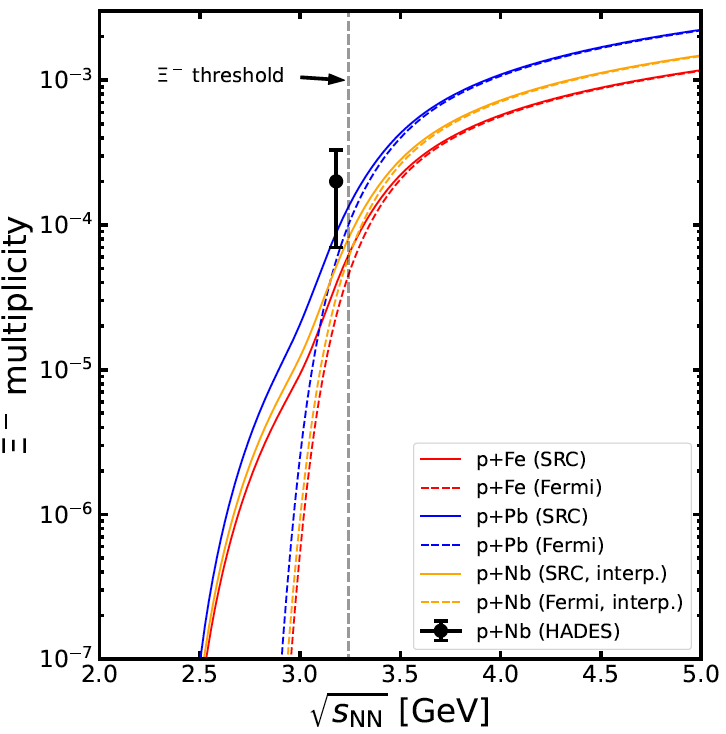}
  \caption{(Color Online). Upper: the normalized collision energy spectrum $\d N/\d\sqrt{s}$ is shown for a nominal nucleon-nucleon collision energy of $\sqrt{s_{\rm{NN}}} = 3.5$\,GeV in p+Pb collisions. The red solid curve corresponds to the nucleus including SRC, while the blue dotted line represents the nucleus without SRC. The solid vertical line indicates the nominal collision energy, and the dashed vertical lines mark selected hadron production thresholds.
Lower: the multiplicity of $\Xi^-$ baryons as a function of the nucleon-nucleon center-of-mass energy $\sqrt{s_{\rm{NN}}}$ in proton+nucleus reactions. The solid curves show the results including SRC, while the dashed lines correspond to nuclei described by a Fermi distribution only. The HADES measurement\cite{HADES} (black circle with error bar) was obtained in p+Nb collisions. The dashed vertical line indicates the production threshold energy. Figures taken from Ref.\cite{Reich25}.}
  \label{fig_Reich25}
\end{figure}

In Ref.\cite{Reich25}, the single nucleon momentum distribution in a nucleus is separated into a Fermi-like part and a HMT due to SRC, $
n(k) = n_0(k) + n_1(k)$, where $n_0(k)$ represents the Fermi component (HMT-free) and $n_1(k)$ the SRC-induced HMT.
The momentum distribution function of Ref.\cite{Cio96} is adopted in their calculations\cite{Reich25}.
Using this distribution, the normalized spectrum of nucleon-nucleon center-of-mass energies $\sqrt{s}$ in a proton-nucleus collision with projectile momentum $p_{\rm{lab}}$ is\cite{Reich25} 
\begin{equation}
\left.\frac{\d N(\sqrt{s})}{\d\sqrt{s}}\right|_{p_{\rm{lab}}} = \int \d^3k \frac{\d^3 n(k)}{\d^3k} \delta\left((k^\mu + p^\mu)^2 - s \right),
\end{equation}
where $k^\mu$ and $p^\mu$ are the four-momenta of the target and projectile nucleons. The probability that a collision exceeds a given threshold $\sqrt{s_{\rm{thr}}}$ is\cite{Reich25}
\begin{equation}
P(\sqrt{s} \ge \sqrt{s_{\rm{thr}}}) = \int_{\sqrt{s_{\rm{thr}}}}^{\infty} \d\sqrt{s} \frac{\d N}{\d\sqrt{s}} \equiv \frac{N_{\rm{coll}}^{\sqrt{s} \ge \sqrt{s_{\rm{thr}}}}}{N_{\rm{coll}}^{\rm{tot}}}.
\end{equation}
This formalism allows one to quantify the enhancement of sub-threshold particle production in p$A$ collisions due to SRC and HMT effects.

The upper panel of FIG.\,\ref{fig_Reich25} shows the normalized collision energy spectrum $\d N/\d\sqrt{s}$, where the red solid curve corresponds to scattering on a Pb nucleus including SRC, and the blue dotted line corresponds to a Pb nucleus described by a Fermi distribution only. The solid vertical line indicates the nominal collision energy, while the dashed vertical lines mark selected hadron production thresholds.  
As pointed out by the authors of Ref.\cite{Reich25}, one can clearly observe the impact of SRC on the available center-of-mass energy for particle production. The coupling of the incoming proton to a nucleon from the deuteron-like HMT leads to a broader distribution of $\sqrt{s}$. In nuclei with SRC, the distribution exhibits an enhanced tail at large $\sqrt{s}$ compared to the Fermi-only case, while also extending to smaller $\sqrt{s}$ due to the random orientation of nucleons. As a result, the production of hadrons such as $\Omega$ states is strongly enhanced when SRC are included, relative to the Fermi-only scenario.
The lower panel of FIG.\,\ref{fig_Reich25} shows the $\Xi^-$ multiplicity as a function of $\sqrt{s_{\rm{NN}}}$ in proton-nucleus reactions. The HADES measurement for p+Nb is indicated by a black circle with error bar\cite{HADES}. Results from p+Fe (red) and p+Pb (blue) reactions have been interpolated to the p+Nb case (orange). The dashed vertical line indicates the production threshold.  
One can clearly see that deep below threshold, the estimated hadron multiplicity is strongly enhanced by SRC in the nuclear momentum distribution, with the effect becoming more pronounced further below threshold. The HADES measurement, however, lies only slightly below threshold, where the difference between Fermi and SRC calculations is comparable to the experimental uncertainty. Upcoming experiments at other facilities, e.g., FAIR, will allow measurements of $\Xi$ production much deeper below threshold, providing a direct probe of SRC in strong interactions and a natural explanation for deep sub-threshold particle production.  
This calculation is particularly interesting because the only measured sub-threshold hadron multiplicity in a proton-nucleus reaction comes from HADES\cite{HADES}, which reported the $\Xi^-$ yield in p+Nb collisions at a beam energy of 3.5\,GeV (corresponding to $\sqrt{s_{\rm{NN}}} = 3.18$\,GeV), about 61\,MeV below the elementary $\Xi$ production threshold. In their model, the $\Xi$ multiplicity is obtained by multiplying the $\sqrt{s}$ spectrum with the production cross section $\sigma_{\rm{pp} \to \Xi^- \rm{X}}(\sqrt{s})$ and integrating over $\sqrt{s}$, with the normalization of the $\sqrt{s}$ distribution fixed to match a Glauber calculation of the number of binary collisions.

Studies like this play an important role in advancing our understanding of nucleon-nucleon SRC and HMT effects in dense matter. They provide both motivation and concrete guidance for investigating these phenomena in intermediate-energy heavy-ion collisions where dense matter is created. Such work helps bridge experimental observations with theoretical modeling, highlighting key SRC-HMT features in dense matter that should be explored in future studies.

In the future, the treatment of SRC and HMT effects in heavy-ion collision studies could be further improved by incorporating them in a more self-consistent manner, for example through the Wigner function approach. This would allow for a more realistic description of nucleon momentum distributions and their influence on the reaction dynamics. Additionally, it is important to identify and employ experimental probes that are directly sensitive to these correlations, enabling more accurate and reliable constraints on theoretical models. Together, these efforts can help establish a clearer connection between microscopic nuclear structure and observables in intermediate-energy heavy-ion reactions.

\subsection{Experimental Extraction of HMT Fraction using Hard Photons from Heavy-Ion Collisions}\label{sub_gamma}

\indent 
An experimental investigation into SRC-HMT physics was carried out recently by the Xiao Group at Tsinghua University by measuring neutron-proton bremsstrahlung $\gamma$-rays in $^{124}\mathrm{Sn}+{^{124}}\mathrm{Sn}$ collisions at 25\,MeV/u at the CSHINE/HIRFL facility in Lanzhou \cite{JHXu25PRR,JHXu25xx}. They demonstrated clearly that electromagnetic radiations during $\rm{np}\to\rm{np}\gamma$ scattering (FIG.\,\ref{fig_npgamma}) carries a clear imprint of the underlying SRC-HMT physics. In particular, they experimentally verified the physics picture that when the colliding nucleons originate from the SRC-induced HMTs, their more energetic relative motion naturally generates a harder $\gamma$ spectrum than one would expect from a purely mean-field momentum distribution. This mechanism provides a physically clean and intuitive path for accessing SRC through hard photons, which are not distorted by final-state strong interactions compared to hadronic observables.

\begin{figure}[h!]
\centering
  \includegraphics[width=9.cm]{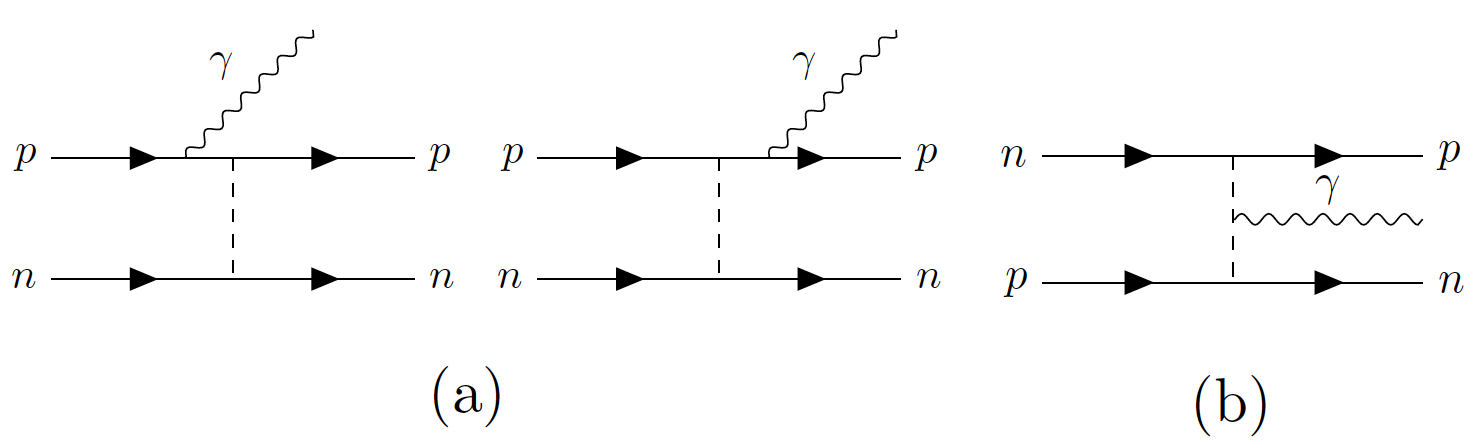}
  \caption{Feynman diagrams of bremsstrahlung photons in $\rm{np}\to\rm{np}\gamma$ scattering from (a) external lines and (b) internal lines. Figure taken from Ref.\cite{JHXu25xx}.
}
  \label{fig_npgamma}
\end{figure}

\begin{figure}[h!]
\centering
  \includegraphics[height=9.cm]{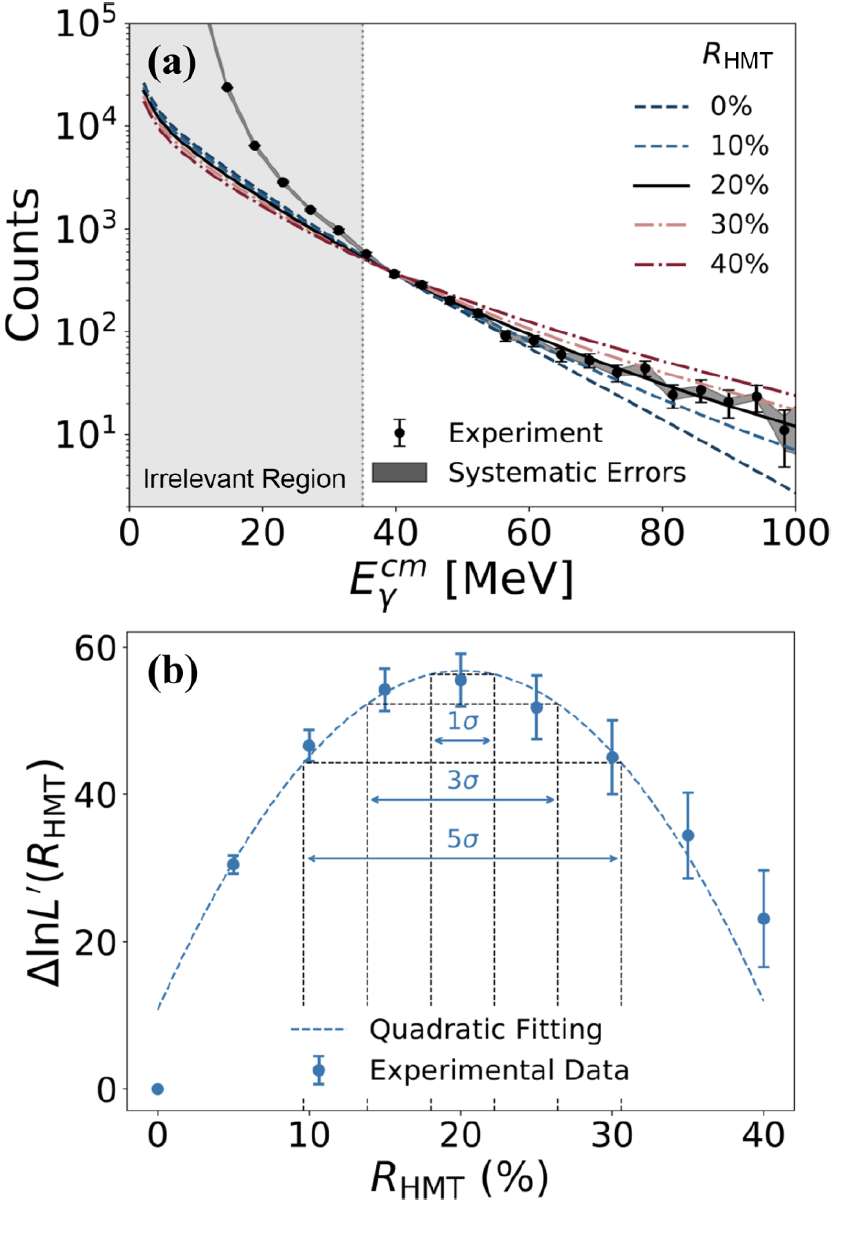}\\[0.25cm]
  \hspace{-0.7cm}
  \includegraphics[height=9.cm]{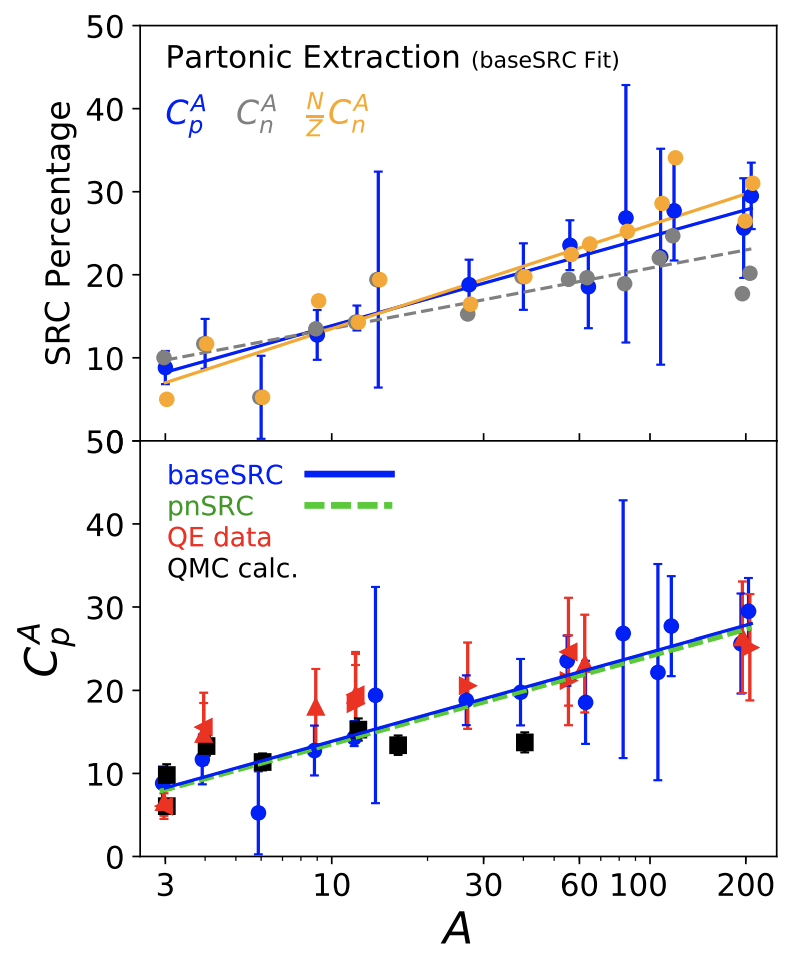}
  \caption{(Color Online). Upper (panels (a) and (b)): comparison of the experimental $\gamma$ spectrum (black dots) in the CM frame with various detector-filtered theoretical curves, where statistical (systematic) uncertainties are shown as vertical bars (gray bands). The likelihood distribution as a function of $R_{\rm{HMT}}$ is shown in the lower part, with a quadratic fit indicated by the dashed curve. Error bars represent the standard deviation of likelihood values obtained by varying the calibration and detector response data sets. Figure taken from Ref.\cite{JHXu25PRR}. Lower: nuclear structure parameters $C^A_{\rm p}$, $C^A_{\rm n}$ and $(N/Z) C^A_{\rm n}$ defined in Ref.\cite{Denn24PRL} for the base SRC fit. Uncertainties are shown for $C^A_{\rm p}$, similar in size for the other quantities. Figure taken from Ref.\cite{Denn24PRL}.
}
  \label{fig_Xiao}
\end{figure}

To translate the qualitative picture into quantitative information, they performed IBUU transport model calculations using the MDI potential \cite{Das2003PRC} and compared with their measured $\gamma$ spectra. The SRC-HMT physics enters the IBUU model through the initialization of the nucleon momentum distributions in the two colliding nuclei. The resulting hard photon spectra, after being filtered through the CSHINE detector response, allow the extraction of an HMT fraction of about $R_{\rm{HMT}}\approx20\pm3\%$ in $^{124}$Sn, as shown in the upper panel of FIG.\,\ref{fig_Xiao}. 
In their analysis, the likelihood function
\begin{eqnarray}
    \sum_{i=1}^{\rm{range}}n_i\ln p_i(R_{\rm{HMT}})
\end{eqnarray}
is used\cite{JHXu25PRR}; here $n_i$ denotes the count in the $i$th experimental bin, summed over all bins within the statistical analysis interval. The quantity $p_i$ represents the probability that the theoretical model predictions fall into the corresponding histogram bin for a given $R_{\rm{HMT}}$ within the same statistical analysis interval.
The result on $R_{\rm{HMT}}$ is remarkably consistent with the SRC systematics established in high-energy lepton-scattering experiments, suggesting that the characteristic high-momentum structure generated by SRC remains robust across widely differing reaction mechanisms and energy domains. The agreement also highlights the importance of implementing realistic nucleon momentum distributions and nuclear structure information in initializing transport model simulations of heavy-ion collisions, especially when describing hard-photon or deeply sub-threshold particle productions and other reaction channels sensitive to large relative momenta. 

Moreover, these results suggest a promising alternative approach to probe the nucleon parton distribution functions in nuclei (nPDFs) via low-energy heavy-ion collisions. Within the collinear factorization framework\cite{Denn24PRL}, the nPDF can be expressed as a linear combination of free-nucleon PDFs and contributions from SRC pairs, with the fraction determined by the neutron and proton content in SRCs. For example, adopting the HMT fraction of 20\% in $^{124}$Sn\cite{JHXu25PRR} yields $C^{124}_{\rm n} \approx 16\%$ and $C^{124}_{\rm p} \approx 24\%$, which is quite consistent with high-energy data\cite{Denn24PRL} as shown in the lower panel of FIG.\,\ref{fig_Xiao}. 

Beyond the quantitative result, the broader significance of this work lies in demonstrating that low-energy heavy-ion collisions can meaningfully constrain SRC-HMT fractions. The hard $\gamma$-ray spectrum is shown to be dominated by contributions from SRC-induced high-momentum nucleons in the two colliding nuclei, turning electromagnetic emission into a useful and complementary probe alongside traditional knockout measurements. This reinforces the idea that SRC phenomena permeate a wide range of nuclear observables and that their signatures are accessible even in environments far from the high-energy electron-scattering domain where they were first established.
Similar works using neutron-proton bremsstrahlung photons are previously studied by the same group, e.g., in $^{86}\rm{Kr}$+$^{124}\rm{Sn}$ at 25\,MeV/nucleon\cite{YHQin24PLB-a,JHXu24PLB,SDW24}.
Taken as a whole, the study of hard photons from heavy-ion reactions represents a gentle yet conceptually compelling step toward expanding the experimental toolkit available for SRC investigations. By leveraging the capabilities of the CSHINE facility, the authors provide an alternative and experimentally clean approach to studying the high-momentum structure of nuclei in heavy-ion reactions. The method is promising for future work, especially in view of its potential applications to other correlation-sensitive processes in intermediate-energy nuclear reactions, and will likely motivate further exploration of SRC-driven high-momentum dynamics in a broader range of nuclear systems.

\section{Neutron Star Properties with SRC-HMT in $n(k)$}\label{SEC_NS}

\indent 

This section is devoted to exploring the effects of SRC and HMT on selected NS properties. In Subsection \ref{sub_NScooling}, we examine how SRC-HMT influence the NS M-R relations and the tidal deformabilities, using the nonlinear RMF model as an example. Subsection \ref{sub_NScooling} further investigates how SRCs, particularly through the Migdal--Luttinger jump $Z$, affect NS cooling patterns, highlighting recent advances in understanding these modifications. In Subsection \ref{sub_CC}, we address the core-crust transition in NSs, including the possible formation/disappear of nuclear pasta phases. Finally, Subsection \ref{sub_DM} briefly discusses the combined effects of SRC-HMT and dark-matter interactions on NS structures.

\subsection{NS M-R Relations and Tidal Deformabilities}\label{sub_NSMR}

\indent

The SRC-induced HMT, particularly its isospin dependence, is expected to affect the properties of cold NSs, since the neutron-rich environment enhances the high-momentum component of protons, which can play a significant role.
The upper panel of FIG.\,\ref{fig-NSMR-CAI} shows the EOSs of NS matter obtained within the FFG and HMT models\cite{Cai16c}. The similarity of the two EOSs suggests that their corresponding M-R relations will not differ dramatically. In the lower panel, the M-R relations from the two models are compared. The NS mass is obtained by integrating the TOV equations (adopting $c=1$)\cite{TOV39-1,TOV39-2}:
\begin{align}
\frac{\d P}{\d r}=&-\frac{GM\varepsilon}{r^2}\left(1+\frac{P}{\varepsilon}\right)\left(1+\frac{4\pi r^3P}{M}\right)
\left(1-\frac{2GM}{r}\right)^{-1},\\
\frac{\d M}{\d r}=&4\pi r^2\varepsilon,
\end{align}
here the mass $M=M(r)$, pressure $P=P(r)$ and energy density $\varepsilon=\varepsilon(r)$ are functions of the distance $r$ from NS center. The TOV equations are obviously nonlinear. They are traditionally solved numerically by selecting a central pressure to start the integration towards the surface of a NS of radius $R$ where the pressure $P(R)=0$ for a given input EOS $P(\varepsilon)$.
The NS mass is then $
    M_{\rm{NS}}\equiv M(R)=\int_0^R4\pi r^2\varepsilon(r)\d r$.
As discussed earlier in Subsection \ref{sub_WaleckaSRC}, the skewness of SNM mainly characterizes the high-density behavior of the EOS of SNM. The SRC-induced HMT increases the skewness of SNM, thereby hardening the EOS of NS matter. In contrast, the effect of the symmetry energy on the M-R relation is relatively small in RMF models. Consequently, the enhanced skewness $J_0$ due to the HMT increases the maximum mass of NSs, as shown in the lower panel. Quantitatively, the maximum masses in the HMT and FFG models are $M_{\rm{NS}}^{\max}\approx 1.87\,M_\odot$ and $M_{\rm{NS}}^{\max}\approx 1.74\,M_\odot$, with corresponding radii of approximately $11.21$\,km and $10.89$\,km, respectively. The relative increase in the maximum mass is about 8\%. Although the HMT EOS still predicts a maximum mass below the observational data, it moves closer to the observed values\cite{Riley19,Miller19,Riley21,Miller21,Fon21,Choud24,Reard24,Ditt24,Salmi22,Salmi24,Vin24}. It was pointed out in Ref.\cite{Hen16-gamma} that, while NS observables are sensitive to the high-density behavior of the symmetry energy, they are not sufficient to distinguish between the CFG and FFG models, see FIG.\,\ref{fig_Hen-NS}. This suggests that the NS EOS obtained from Bayesian analyses of NS observations is robust and largely insensitive to the specific nuclear model used for the kinetic part of the symmetry energy. Such insensitivity, or ``blindness'' of the TOV equations is consistent with a recent analysis employing the dimensionless formulation as the starting point\cite{CaiLi25-IPADTOV}.

\begin{figure}[h!]
\centering
 \includegraphics[height=7.cm]{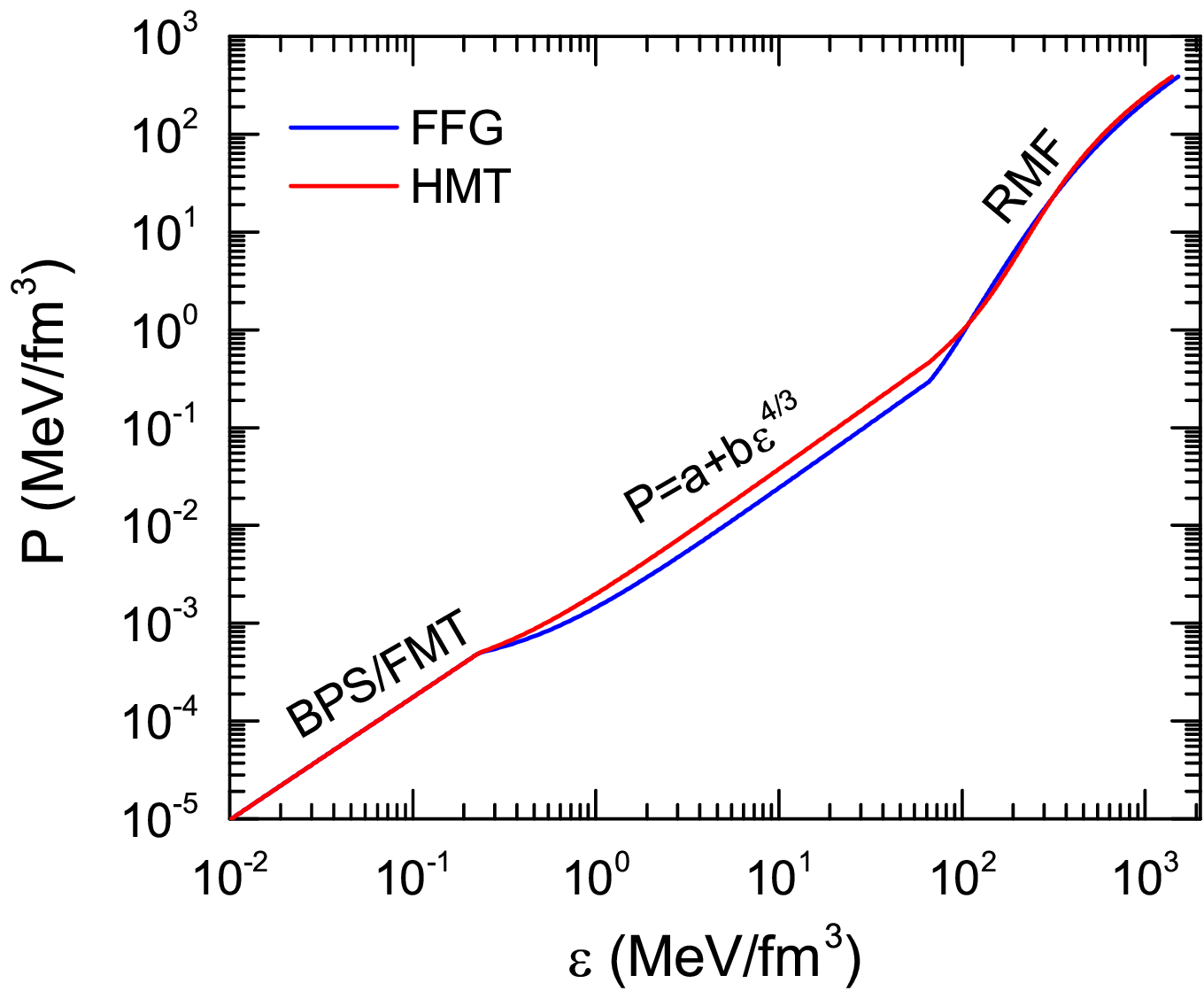}\\[0.25cm]
 \hspace{0.4cm}
 \includegraphics[height=7cm]{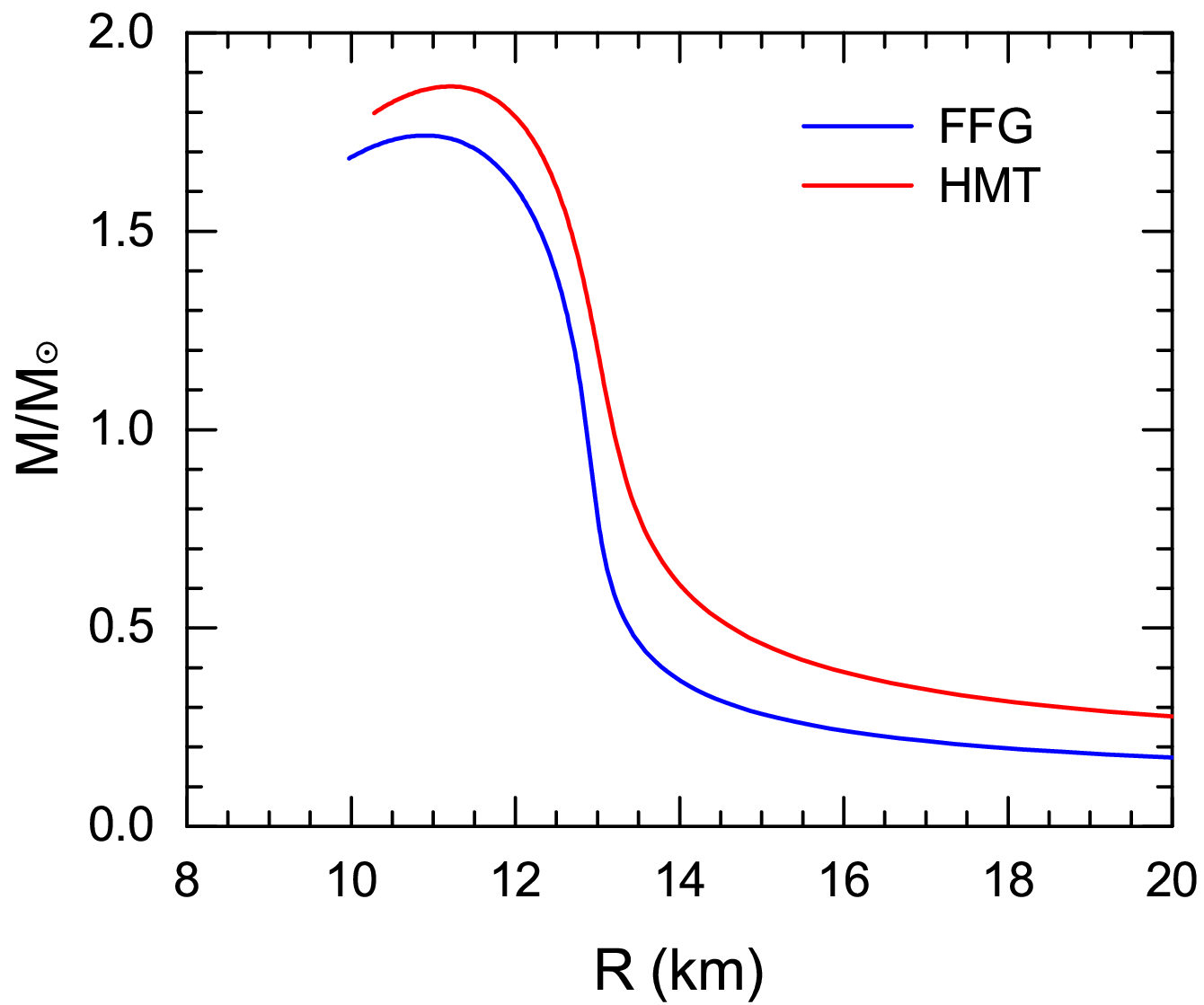}
\caption{(Color Online). Upper: the EOS of NS matter. Lower: the NS M-R relation obtained by integrating the TOV equation under the EOS of NS matter in the FFG and HMT models, respectively. Figures taken from Ref.\cite{Cai16c}.}
  \label{fig-NSMR-CAI}
\end{figure}

\begin{figure}[h!]
\centering
  \includegraphics[width=8.cm]{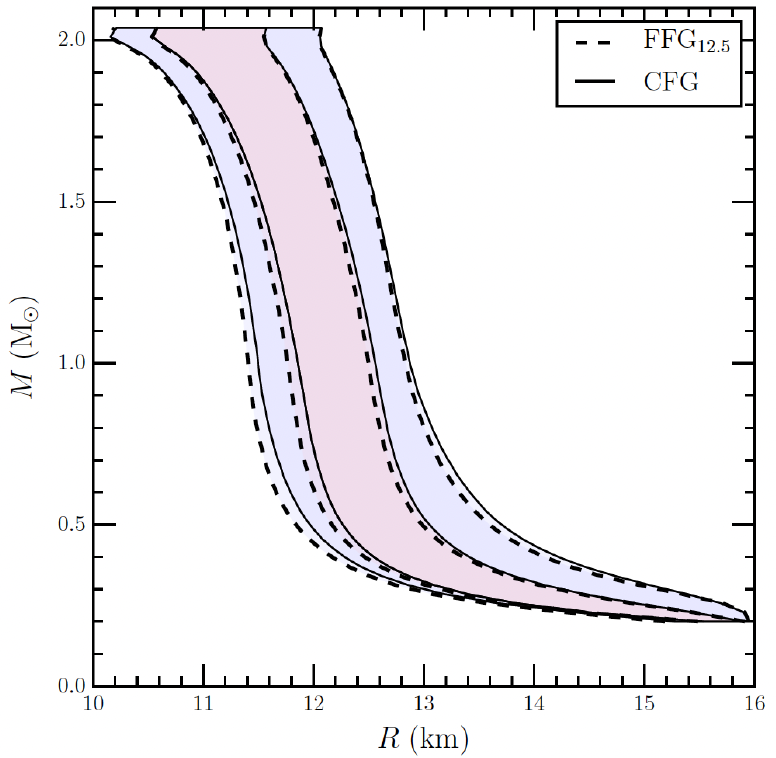}\\[0.25cm]
  \includegraphics[width=8.cm]{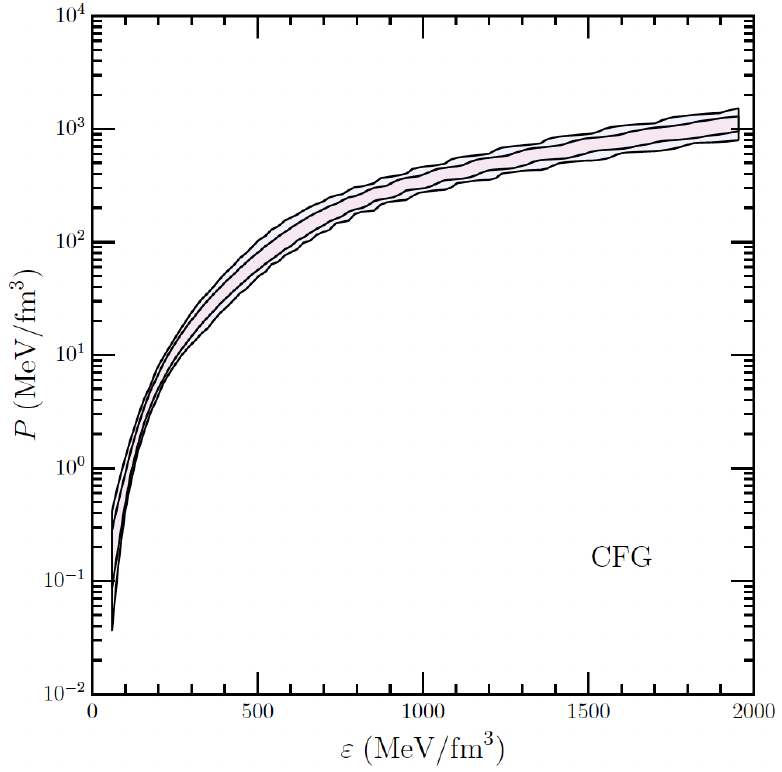}
\caption{(Color Online). Extracted M-R (upper) and NS EOS (lower) for the CFG/HMT and FFG  models, using a Bayesian analysis. Figures taken from Ref.\cite{Hen16-gamma}.}
  \label{fig_Hen-NS}
\end{figure}

\begin{figure}[h!]
\centering
  \includegraphics[width=9.cm]{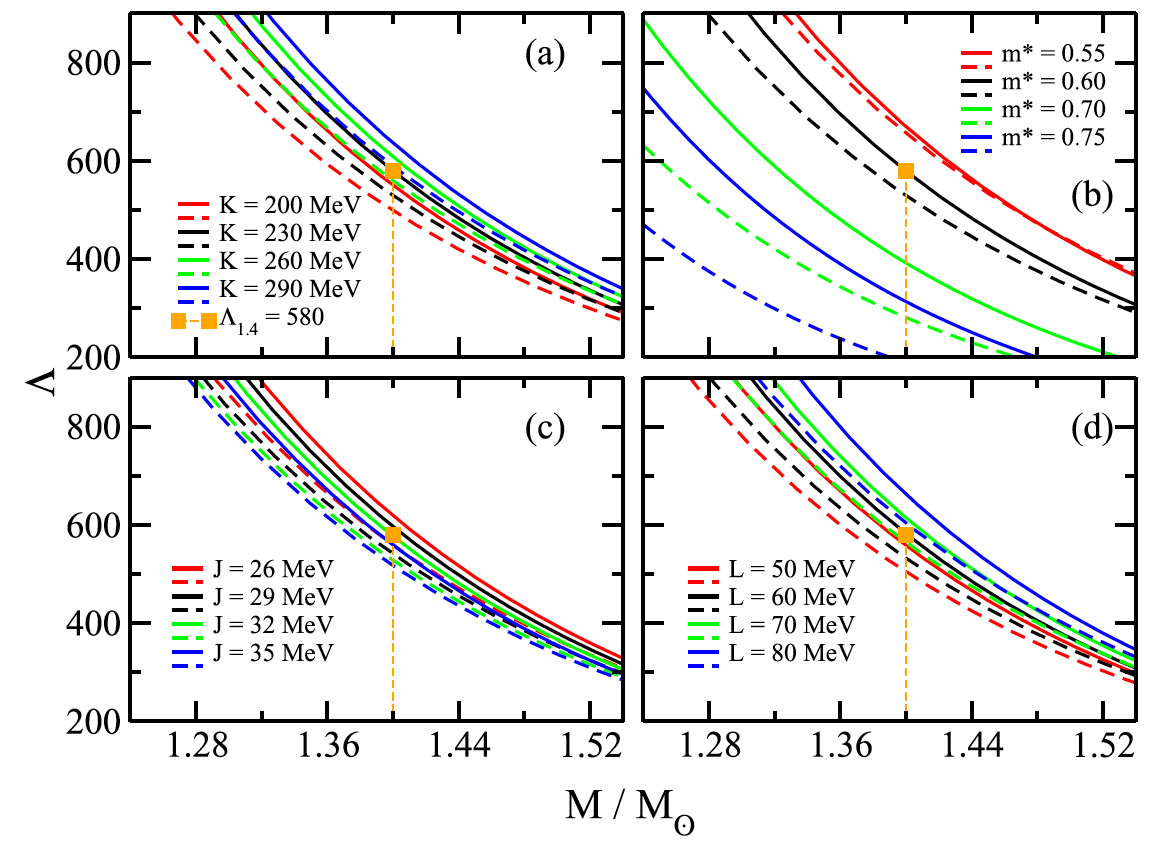}\\[0.25cm]
  \includegraphics[width=9.cm]{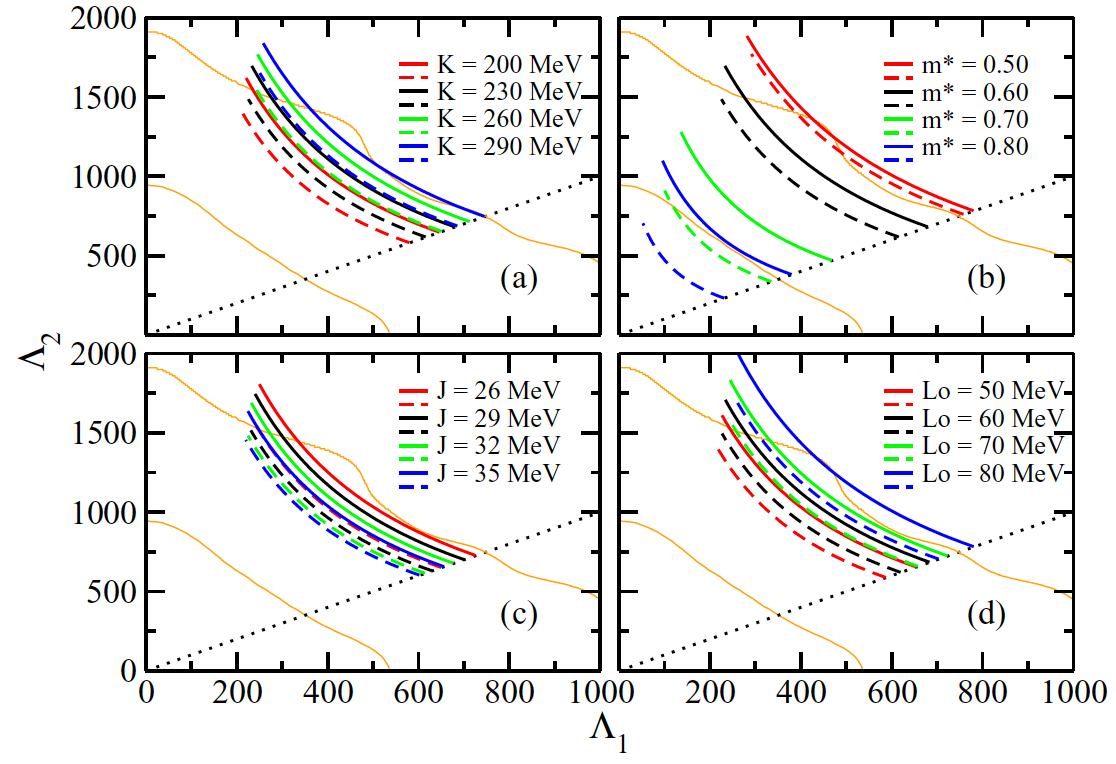}
\caption{(Color Online). Upper: the $\Lambda$ as a function of the NS mass for different parametrizations of the RMF model with (dashed lines) and without (full lines) SRC included. Lower: same as the upprt panel but for $\Lambda_1\times\Lambda_2$.  Figures taken from Ref.\cite{Sou20PRC}.}  \label{fig-Sou20PRC-Lambda}
\end{figure}

\begin{figure}[h!]
\centering
  \includegraphics[width=8.5cm]{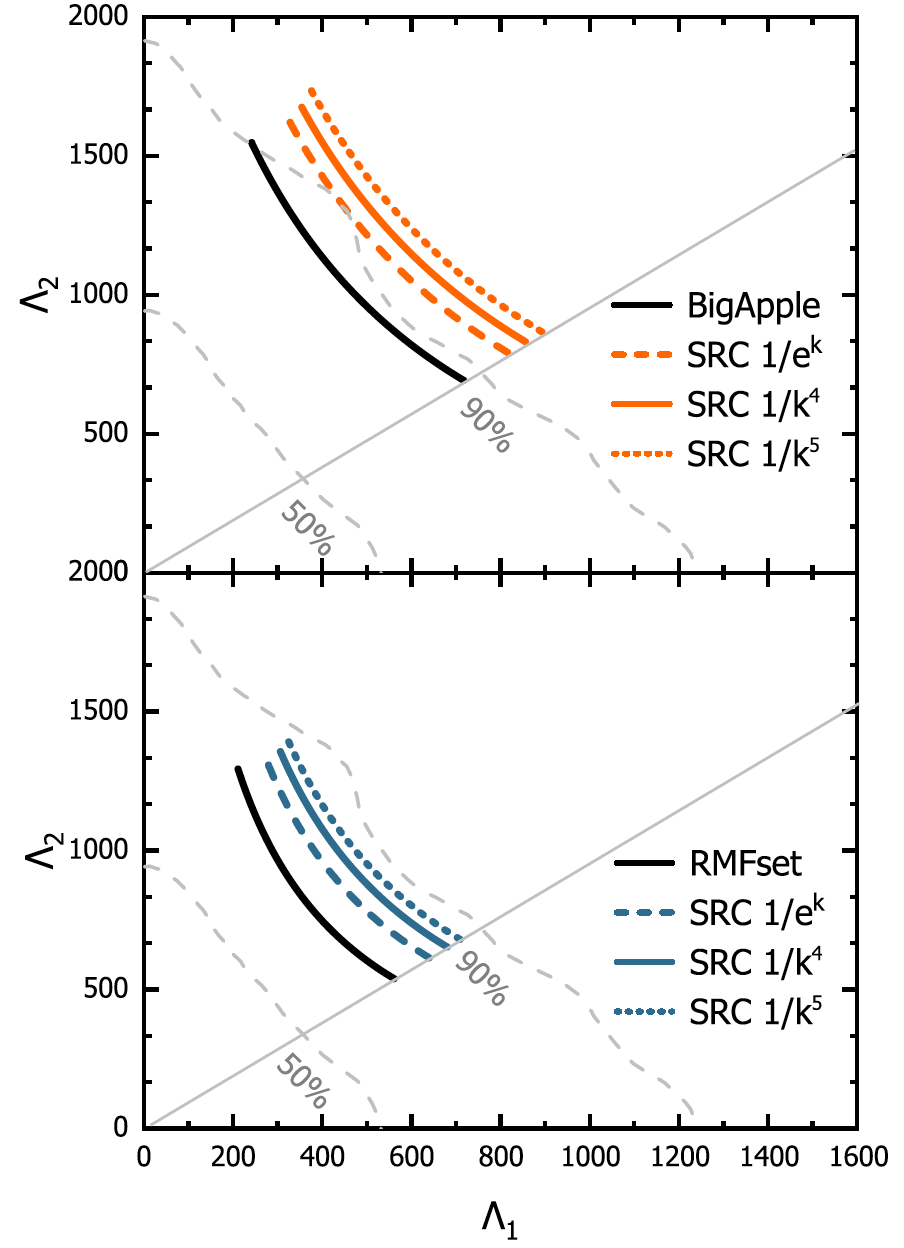}
\caption{(Color Online). The dimensionless tidal deformability $\Lambda$ obtained in a Walecka model adopting three different types of HMT. Figure taken from Ref.\cite{Hong25PLB}.}  \label{fig_Hong-Lambda}
\end{figure}

The enhancement of the NS maximum mass due to the SRC-HMT effects was also obtained in Ref.\cite{Sou20PRC}; besides that, they also found that the SRC-HMT has a sizable impact on the NS deformability $\Lambda$, shown in FIG.\,\ref{fig-Sou20PRC-Lambda}.
The $n(k)$ with the isospin structure of Ref.\cite{Cai16b} was adopted in these studies for deformability\cite{Sou20PRC}.
The (dimensionless) tidal deformability\cite{Hinderer2008} \begin{equation}
\Lambda=(2/3)k_2\xi^{-5}
\end{equation} 
provides very useful constraint on NS EOS\cite{Abbott2017,Abbott2018,Abbott2020-a,Abbott2020}; here the quadrupolar tidal Love number $k_2$ is determined by ($\xi=M_{\rm{NS}}/R$ is the NS compactness)
\begin{align}
k_2=&\frac{8}{5}\xi^5(1-2\xi)^2\left[2-y_R+2\xi\left(y_R-1\right)\right]
\big\{6\xi\left[2-y_R+\xi\left(5y_R-8\right)\right]\notag\\
&\hspace*{1cm}+4\xi^3\left[13-11y_R+\xi\left(3y_R-2\right)+2\xi^2\left(1+y_R\right)\right]\notag\\
&\hspace*{1cm}+3\left(1-2\xi\right)^2\ln\left(1-2\xi\right)
\left[2-y_R+2\xi\left(y_R-1\right)\right]\big\},
\end{align}
and the function $y(r)$ satisfies the differential equation (prime is taken with respect to $r$):
\begin{equation}\label{ode_y}
r y'+y^2+ye^{\lambda}\left[1+4\pi r^2(P-\varepsilon)\right]+r^2Q=0,
\end{equation}
with
\begin{equation}
Q=4\pi e^{\lambda}\left(5\varepsilon+9P+\frac{\varepsilon+P}{\d P/\d\varepsilon}\right)
-\frac{6e^{\lambda}}{r^2}-\nu'^2.
\end{equation}
In this differential equation, $\nu(r)$ and $\lambda(r)$ are the time and space components of the spacetime metric; the function
$y(r)$ satisfies the boundary condition $y(0)=2$ and $y_R$ is its value at the radius $R$.

In their analysis\cite{Sou20PRC}, it was found that SRCs improve the agreement of the model with the GW170817 constraints. In particular, they found that the tidal deformability $\Lambda_{1.4}$ decreases when SRC effects are included and the $\Lambda_2$-$\Lambda_1$ curve shifts toward the inner region of the LIGO-Virgo Collaboration (LVC) data. In  their calculations, they adopted the mass range $1.365 \le m_1/M_\odot \le 1.60$\cite{Abbott2017} for the primary NS and computed the companion mass using the chirp-mass relation $(m_1 m_2)^{3/5}/(m_1 + m_2)^{1/5} = 1.188\,M_\odot$\cite{Abbott2018}. They also tested whether the SRC effects depend on the reference parametrization by generating additional models through independent variation of each bulk parameter. This procedure isolates the effect of the parameter being modified. The corresponding results are shown in FIG.\,\ref{fig-Sou20PRC-Lambda} where four parametrizations are generated with fixed values $\rho_0 = 0.15$\,fm$^{-3}$, $E_0(\rho_0) = -16.0$\,MeV, $M^\ast_{\rm D}/M_{\rm N} = 0.60$, $E_{\rm{sym}}(\rho_0) = 31.6$\,MeV, and $L= 58.9$\,MeV\cite{Cai16b}, while $K_0$ is varied as indicated. For each parametrization, the impact of SRCs is evaluated, and similar behavior is observed: SRCs systematically reduce $\Lambda_{1.4}$ and shift the $\Lambda_1\times\Lambda_2$ pair toward the LVC data. The smallest SRC effect occurs for the parametrization with $M^\ast_{\rm D} = 0.55$, whereas the impact becomes more pronounced for larger effective masses. Since $M^\ast_{\rm D} = M_{\rm N} - g_\sigma \overline{\sigma}$\cite{Cai16b}, this indicates that the attractive interaction mediated by the scalar $\sigma$ field plays the dominant role in the SRC-induced modifications. This trend is consistently reflected in the tidal deformability results.
The dimensionless and dimensional tidal deformabilities were also investigated in Ref.\cite{Hong25PLB} within a nonlinear RMF framework that incorporates the HMT. In addition to the conventional $k^{-4}$-type HMT, their study considered two alternative parameterizations, the $k^{-5}$-form and an exponential $e^{-k}$-form, as illustrated in FIG.\,\ref{fig_Hong-Lambda}. Across both models, all three HMT prescriptions lead to very similar trends: the resulting tidal deformabilities are systematically larger than those obtained for traditional NSs without HMT effects. Notably, although the $e^{-k}$-form is phenomenological, it avoids the divergence encountered in the kinetic-energy integral,
\begin{eqnarray}
\int_{k_{\rm F}}^{\infty} e^{-k/k_{\rm F}}\frac{\v{k}^2}{2M_{\rm N}}\mathrm{d}\v{k}
\sim\int_{k_{\rm F}}^{\infty}\d k k^4e^{-k/k_{\rm F}}
\sim \frac{k_{\rm F}^5}{\rm{e}},
\end{eqnarray}
which remains finite.
Since the exponential function decreases much faster than polynomials, it is closer to the original framework without a HMT.

The nonlinear RMF model including hyperonic degrees of freedom with SRC-induced HMT effects was investigated in Ref.\cite{LiZ19NPA}, where the HMT corrections were incorporated to ensure consistency between the model and the established physics of nucleon-nucleon SRCs.
In addition, the HMT form, especially its isospin structure, is different from the one adopted by Refs.\cite{Cai16b,Sou20PRC}.
The resulting properties of nucleonic and hyperonic stars were then systematically examined. 
With the inclusion of HMT effects, the EOS of nucleonic matter becomes appreciably softer, leading to a smaller NS maximum mass, whereas the maximum mass of hyperonic stars is generally enhanced when SRC-HMT effects are incorporated. While these findings are intriguing, no possible physical explanations for the contrasting behaviors were provided in their work.
In hyperonic stars, the HMT corrections strongly influence the particle fractions and significantly shift the threshold densities for hyperon formation. Notably, they found that once HMT effects are accounted for, the M-R relations of both nucleonic and hyperonic stars become compatible with the stringent constraints extracted from recent gravitational-wave observations.
In Ref.\cite{Gaut25}, the authors revisited the properties of NSs in the presence of SRC-HMT, following the framework of Refs.\cite{Cai16b,Sou20PRC}. In their analysis, empirical bulk quantities such as $K_0$, $L$, and others are not fixed; instead, the model parameters are constrained directly by selected experimental data. As a result, the impact of SRC-HMT on the NS maximum mass is not universal: in some parametrizations the maximum mass is enhanced, whereas in others it is reduced.
The SRC-HMT effects on NS M-R relations were also studied in Ref.\cite{LuH22NPA}, where the fractions of np and nn pairs in NS matter were emphasize and investigated based on specific model assumptions. Since the available HMT information is mainly extracted from experimental data on finite nuclei, where the relevant densities are at or below saturation, extending these constraints to the high-density regime of NSs inevitably requires additional model input. This limitation represents a major challenge in exploring NS properties using SRC-induced HMTs.

\begin{figure}[h!]
\centering
  \includegraphics[height=6.5cm]{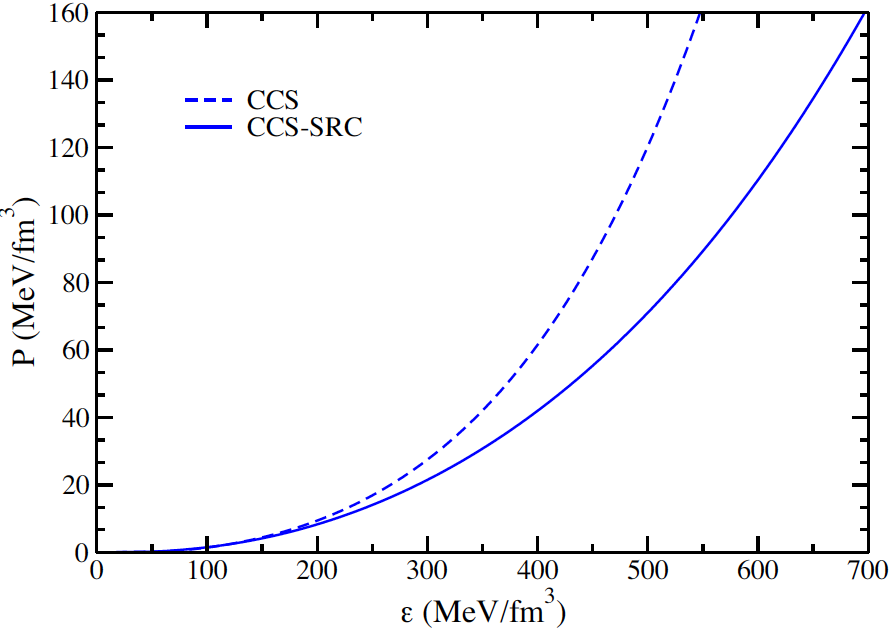}\\[0.25cm]
  \hspace{-0.2cm}
  \includegraphics[height=6.5cm]{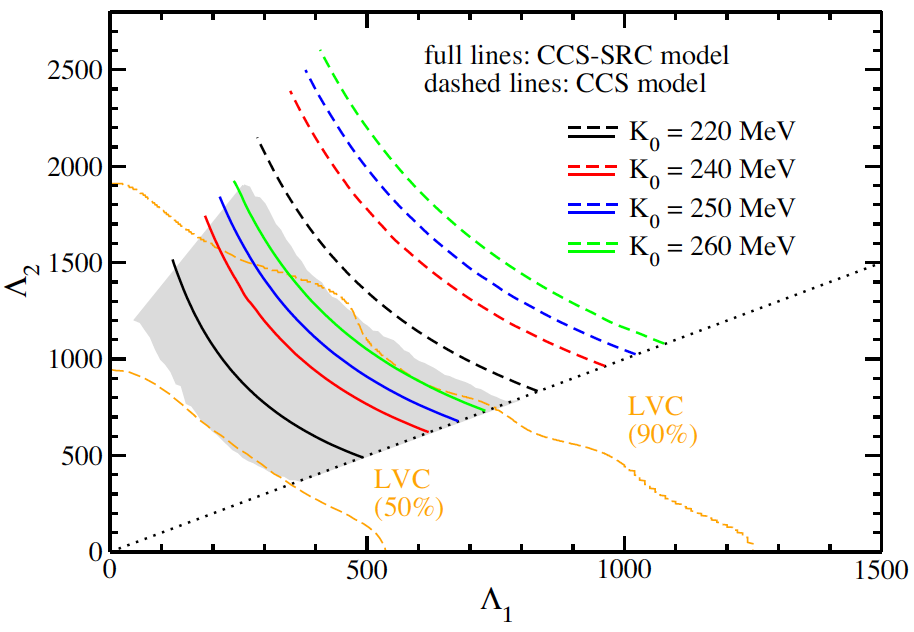}
\caption{(Color Online). The NS M-R relations (upper) and the dimensionless tidal deformabilities (lower) from the CCS-SRC model.
Figures taken from Ref.\cite{Rod23}.}  \label{fig_Rod23}
\end{figure}

\begin{figure}[h!]
\centering
 \includegraphics[width=6.5cm]{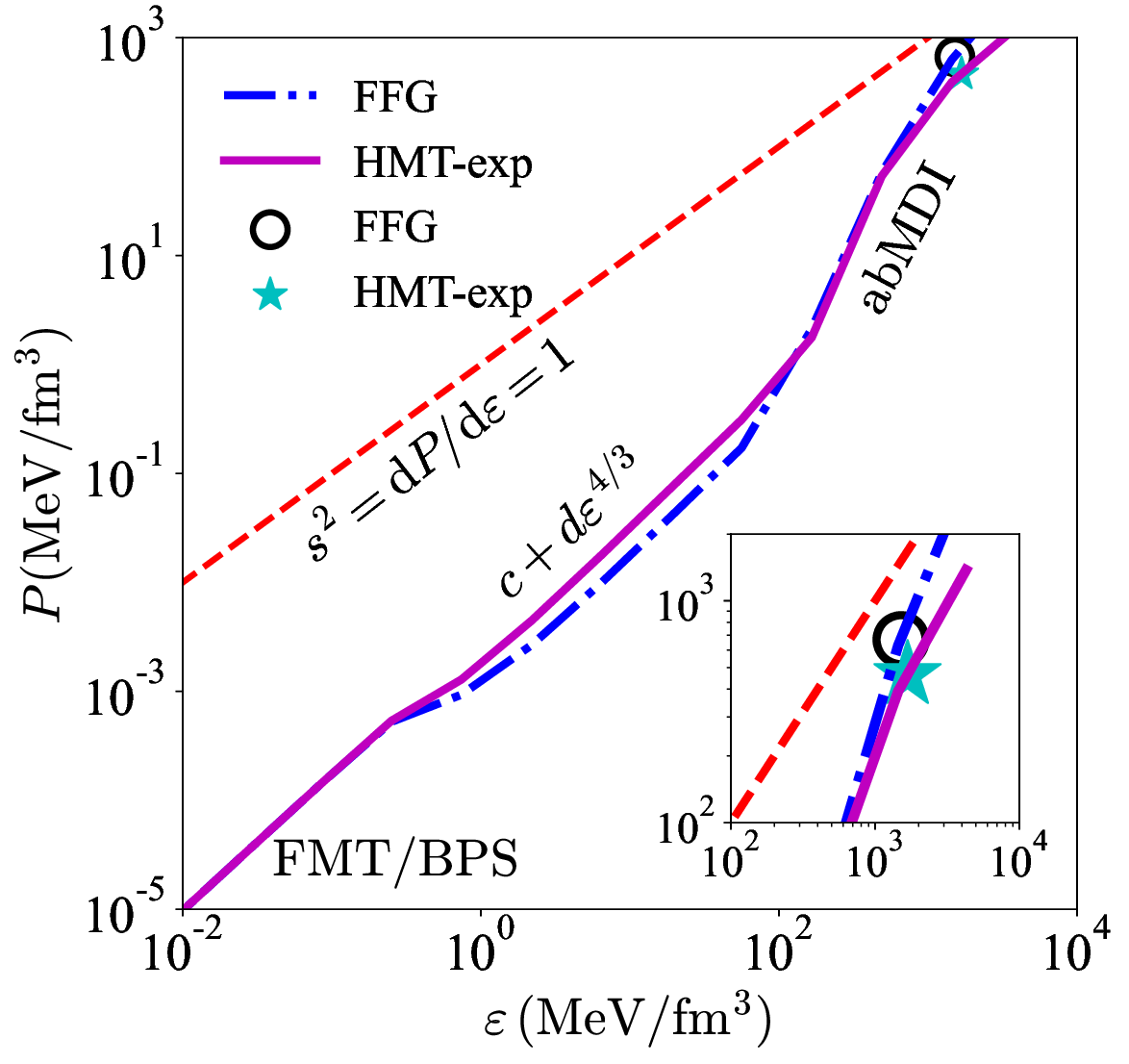}\\[0.25cm]
 \hspace{0.2cm}
 \includegraphics[width=6.5cm]{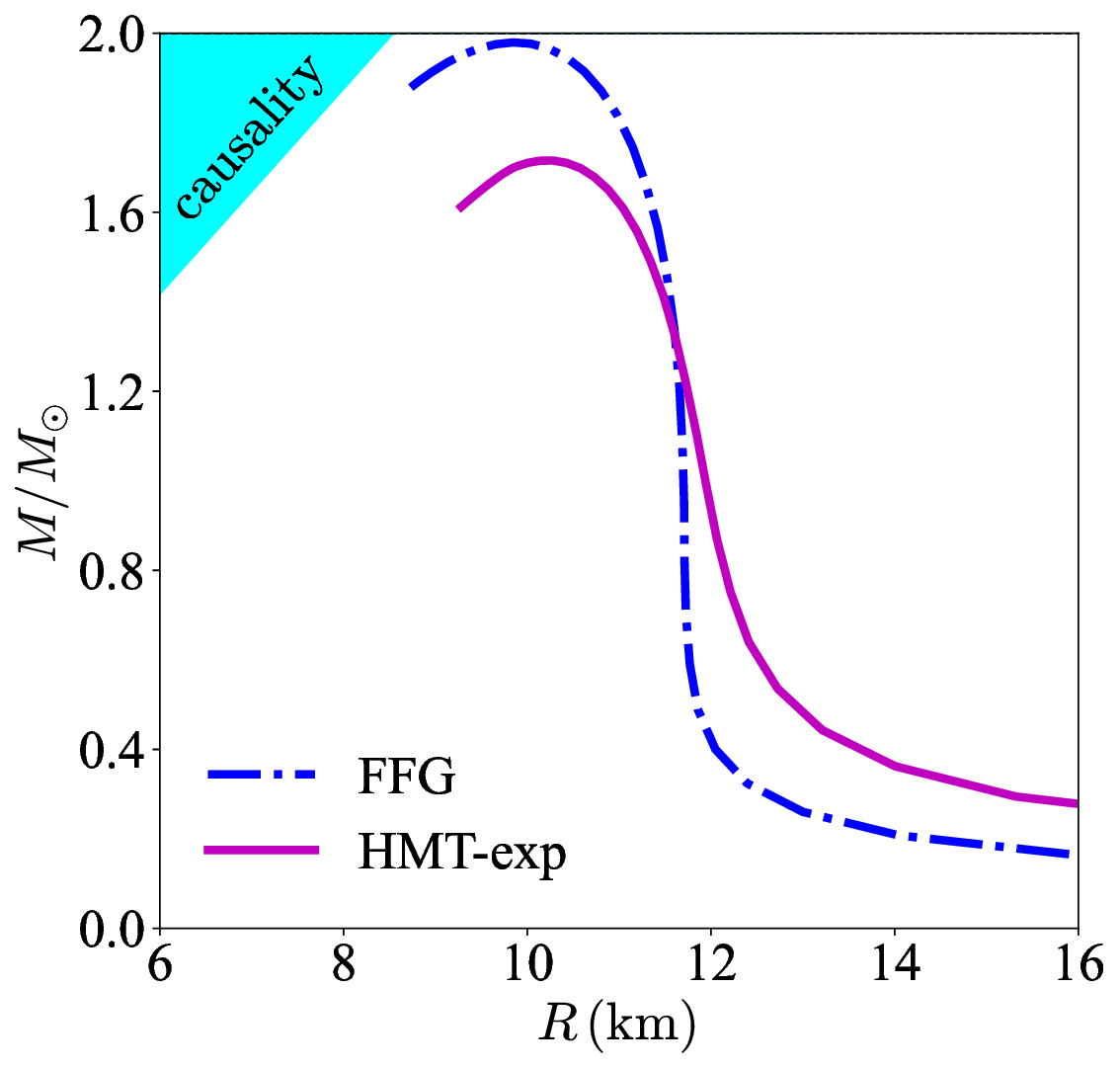}
  \caption{(Color Online). The $\beta$-stable EOS of NS matter (upper) and the M-R relation (lower) obtained using a Gogny-type model. Figures taken from Ref.\cite{CaiLi22Gog}.}
  \label{fig_MR-Gogny}
\end{figure}

\begin{figure*}[h!]
\centering
  \includegraphics[width=14.cm]{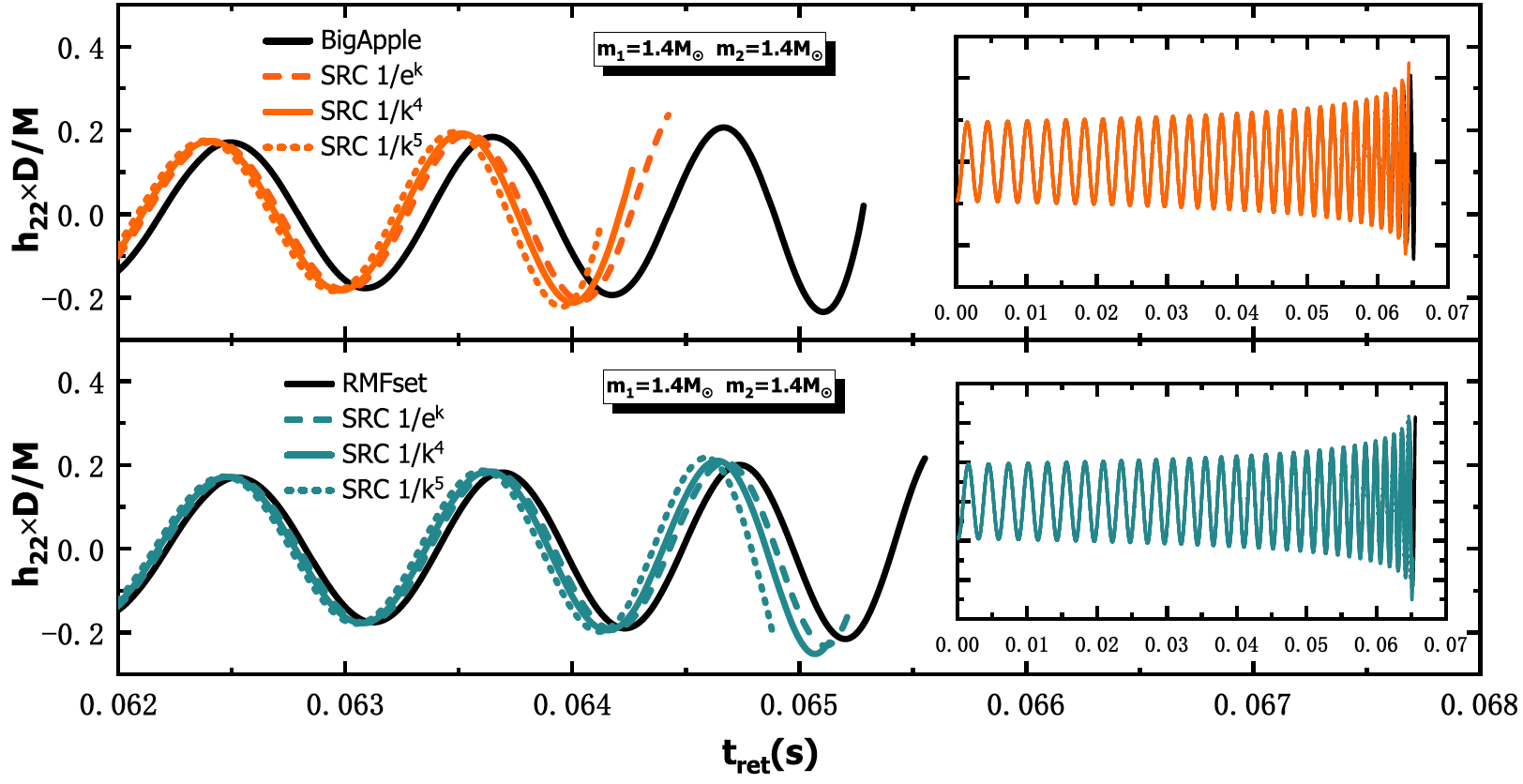}
\caption{(Color Online). Gravitational-wave amplitude $h_{22}$ as a function of the retarded time $t_{\rm{ret}}$ for $1.4\,M_\odot$ equal-mass binary NSs under different SRC scenarios.
Figure taken from Ref.\cite{Hong25PLB}.}  \label{fig_Hong-GW}
\end{figure*}

In a recent study, the properties of NSs were investigated using a van der Waals-type EOS that incorporates the effects of SRC-HMT\cite{Rod23}. In this framework, the attractive and repulsive components of the nucleon-nucleon interaction are treated as density-dependent functions. The repulsive term is modeled using the Clausius--Carnahan--Starling (CCS) prescription, while the attractive term is constructed in a form that reproduces the characteristic structure of the Clausius real-gas model.
In particular, the energy density takes the form:
\begin{align}
    \varepsilon(\rho)=&\left[1-\rho B(\rho)\right]\left[\varepsilon_{\rm{p},\ast}^{\rm{kin}}(\rho)+\varepsilon_{\rm{n},\ast}^{\rm{kin}}(\rho)\right]\notag\\
    &-\rho^2A(\rho)+d(1-2x_{\rm p})^2\rho^2,
\end{align}
where $A(\rho)$ and $B(\rho)$ are two density-dependent functions characterizing the van der Waals interaction, $d$ is a model parameter, $x_{\rm p}=\rho_{\rm p}/\rho$ is the proton faction.
The kinetic part of the energy density is given similarly as that of Ref.\cite{Cai16b}, namely
\begin{align}
    \varepsilon_{J,\ast}^{\rm{kin}}(\rho)=&
    \frac{g\Delta_J}{2\pi^2}\int_0^{k_{\rm F}^{J,\ast}}\d kk^2\sqrt{k^2+M_{\rm N}^2}\notag\\
    &+\frac{gC_Jk_{\rm F}^{J,\ast,4}}{2\pi^2}
    \int_{k_{\rm F}^{J,\ast}}^{\phi_Jk_{\rm F}^{J,\ast}}\d k\frac{\sqrt{k^2+M_{\rm N}^2}}{k^2},
\end{align}
here $g=2$ is the degeneracy factor for p or n.
The effective nucleon momentum $k_{\rm F}^{J,\ast}$ is related to the baryon density $\rho_J$ by
\begin{eqnarray}
    k_{\rm F}^{J,\ast}=\sqrt[3]{{6\pi^2\rho_J^\ast}/{g}},\quad\text{
    with }~\rho_J^\ast=\frac{\rho_J}{1-\rho B(\rho)}.
\end{eqnarray}
It is worth noting that, within this model, no effective-mass correction appears in the single nucleon energy. In contrast to the Walecka-type RMF framework\cite{Cai16b}, in which the nucleon mass is shifted as $M_{\rm N} \rightarrow M_{\rm N} - g_{\sigma}\overline{\sigma}$, the present van der Waals-type construction retains the bare nucleon mass in the kinetic term.
The pressure could be calculated via $P=\rho^2\partial(\varepsilon/\rho)/\partial\rho$, i.e.,
\begin{align}
    P(\rho)=&P_{\rm{p},\ast}^{\rm{kin}}(\rho)+P_{\rm{n},\ast}^{\rm{kin}}(\rho)-\rho^2A(\rho)\notag\\
    &+\rho\Sigma(\rho,x_{\rm{p}})+d(1-2x_{\rm p})^2\rho^2,
\end{align}
here $\Sigma(\rho,x_{\rm{p}})$ is the rearrangement term (due to the presence of $A$ and $B$ functions)\cite{Rod23}, and the kinetic part is
\begin{align}
    P_{J,\ast}^{\rm{kin}}(\rho)=&\frac{g\Delta_J}{6\pi^2}\int_0^{k_{\rm F}^{J,\ast}}\d k\frac{k^4}{\sqrt{k^2+M_{\rm N}^2}}\notag\\
&    +\frac{gC_Jk_{\rm F}^{J,\ast,4}}{6\pi^2}\int_{k_{\rm F}^{\ast,J}}^{\phi_Jk_{\rm F}^{J,\ast}}\frac{\d k}{\sqrt{k^2+M_{\rm N}^2}}.
\end{align}
If $\phi_J\to1$ and $\Delta_J\to1$ are taken, the above expressions naturally reduce to the FFG model ones; while for the CCS-SRC model, the HMT takes the form of Ref.\cite{Cai16b}. The kinetic symmetry energy is also derived in Ref.\cite{Rod23} by adopting the similar expression given in Ref.\cite{Cai16b}.
Unlike the nonlinear RMF model, the van der Waals-type CCS-SRC model predicts a softer EOS for stellar matter (upper panel of FIG.\,\ref{fig_Rod23}) and shifts the $\Lambda_1 \times \Lambda_2$ curves toward the region compatible with LIGO-Virgo data from the GW170817 event (lower panel of FIG.\,\ref{fig_Rod23}). This behavior arises primarily because SRCs reduce the values of both dimensionless tidal deformabilities. As discussed in Ref.\cite{Rod23}, the quartic self-interaction term $c_\omega (\omega_\mu \omega^\mu)^2$ in the nonlinear RMF Lagrangian is responsible for the opposite effect of SRC-HMT on NS M-R relations, whereby SRC-HMT enhances the maximum mass compared with the CCS model.
The similar behavior is observed in a Gogny-type model with momentum-dependent interactions\cite{CaiLi22Gog}, as illustrated in FIG.\,\ref{fig_MR-Gogny}.

Besides the NS M-R relations and tidal deformabilities, other quantities such as radial and non-radial oscillations, as well as the gravitational-wave frequency and amplitude, have also been investigated in the presence of SRC-HMT effects\cite{Hong22CPC,Hong23CQG,Hong25PLB}, further enriching our understanding of SRC impacts on NS physics.
For example, FIG.\,\ref{fig_Hong-GW} presents the strain amplitude of the dominant quadrupole mode $h_{22}$ of gravitational waveform in the inspiral phase as a function of the retarded time $t_{\rm{ret}}$. The inset traces the full inspiral evolution, showing a steady increase in the gravitational-wave amplitude that peaks near the merger. One can see that SRCs shorten the overall inspiral duration. More importantly, with SRC effects included, phase deviations accumulate over time and become most pronounced in the final cycles relative to a traditional binary NS system.
These features can be qualitatively understood from the M-R relations and tidal deformabilities\cite{Hong25PLB}. The inclusion of SRCs generally leads to larger NS radii (see FIG.\,\ref{fig-NSMR-CAI}), which enhance the tidal response to the companion, thereby accelerating the inspiral and reducing the retarded time. When translated into the gravitational waveform, this modification of the inspiral dynamics produces increasingly large phase deviations compared with waveforms from traditional systems\cite{Hong25PLB}.

\subsection{Proton Fraction $x_{\rm p}$, Migdal--Luttinger $Z$-factor, and their Role in NS Cooling}\label{sub_NScooling}

\indent 

The SRC-induced HMT can significantly modify the proton fraction in NSs, and correspondingly influence the cooling behavior of protoneutron stars. The chemical composition of the NS is determined by the requirement of charge neutrality and equilibrium with respect to the weak interaction. From the binding energy per nucleon, one can calculate the proton fraction for $\beta$-stable matter as found in the interior of NSs. For neutrino-free $\beta$-stable matter, the chemical equilibrium for the reactions $\rm{n}\rightarrow\rm{p}+\rm{e}^{-}+\overline{\nu}_{\rm{e}}$ and $\rm{p}+\rm{e}^{-}\rightarrow \rm{n}+\nu_{\rm{e}}$ requires $
\mu_{\rm{e}}=\mu_{\rm{n}}-\mu_{\rm{p}}\approx 4E_{\rm{sym}}(\rho)\delta+8E_{\rm{sym,4}}(\rho)\delta^3+12E_{\rm{sym},6}(\rho)\delta^5+\cdots$, where $E_{\rm{sym},6}(\rho)\equiv 120^{-1}\partial^6E(\rho,\delta)/\partial\delta^6|_{\delta=0}$ is the sixth-order symmetry energy. For relativistic degenerate electrons, 
$
\mu_{\rm{e}}=[m_{\rm{e}}^{2}+(3\pi^{2}\rho x_{\rm{e}})^{2/3}]^{1/2}\approx (3\pi^{2}\rho x_{\rm{e}})^{1/3}$,
where $m_{\rm{e}}\approx 0.511\,\rm{MeV}$ is the electron mass, and $x_{\rm{p}}=x_{\rm{e}}$ because of charge neutrality. Just above the nuclear matter density at which $\mu_{\rm{e}}$ exceeds the muon mass $m_{\mu}\approx 105.7\,\rm{MeV}$, the reactions $\rm{e}^{-}\rightarrow \mu^{-}+\nu_{\rm{e}}+\overline{\nu}_{\mu}$, $\rm{p}+\mu^{-}\rightarrow \rm{n}+\nu_{\mu}$ and $\rm{n}\rightarrow\rm{p}+\mu^{-}+\overline{\nu}_{\mu}$ are energetically allowed so that both electrons and muons are present in $\beta$-stable matter. This alters the $\beta$-stability condition to $
\mu_{\rm{e}}=\mu_{\mu}=[m_{\mu}^{2}+(3\pi^{2}\rho x_{\mu})^{2/3}]^{1/2}$ together with $ x_{\rm{p}}=x_{\rm{e}}+x_{\mu}$.
Generally, when the distribution function $n_{\mathbf{k}}^J(\rho,\delta)$ has a low- (high-) momentum depletion (tail), the nucleon chemical potential $\mu_{\rm{n,p}}\equiv\partial\varepsilon/\partial\rho_{\rm{n,p}}$ is not given by $k_{\rm{F}}^{\rm{n/p},2}/2M_{\rm N}+U_{\rm{n/p}}(\rho,\delta,k_{\rm{F}}^{\rm{n/p}})$, i.e., it cannot be obtained by simply evaluating the single-nucleon potential at $k_{\rm F}^J$. Actually, determining the relation between the nucleon chemical potential and the single-nucleon potential for a general $n_{\mathbf{k}}^J(\rho,\delta)$ is a fundamental problem in nuclear many-body theory\cite{AGD}.

\begin{figure}[h!]
\centering
  \hspace{0.5cm}
 \includegraphics[width=8.5cm]{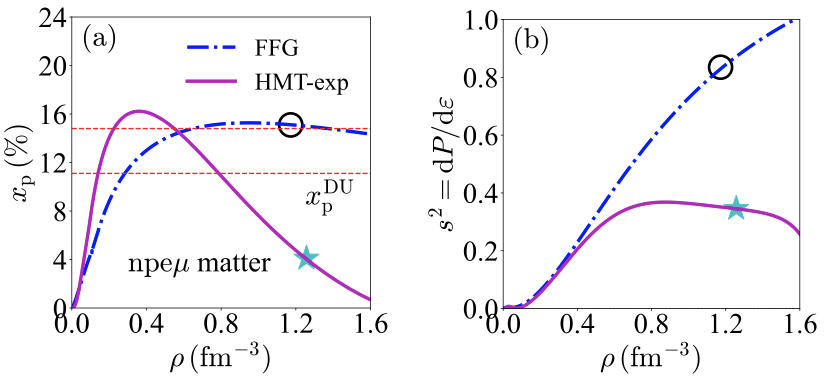}\\[0.25cm]
 \includegraphics[width=9.cm]{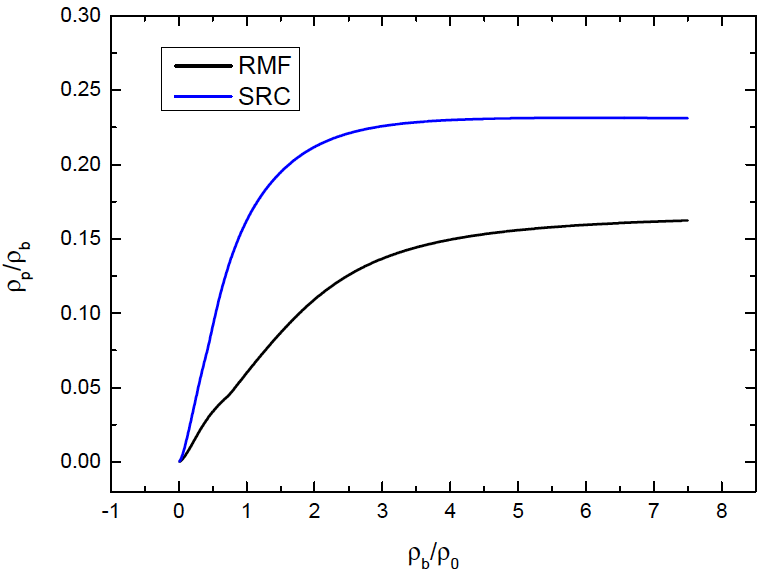}
 \vspace{-0.2cm}
  \caption{(Color Online). Upper: the proton fraction $x_{\rm{p}}$ (panel (a)) and the square of the speed of sound $s^2$ (panel (b)) in NS matter under the FFG and the HMT-exp sets using the Gogny-type model.
  The black circle and the cyan star correspond to the central densities of the NS in the two models. Figures taken from Ref.\cite{CaiLi22Gog}.
  Lower: the $x_{\rm p}$ from the nonlinear RMF model; figure taken from Ref.\cite{Sou20xx}. }
  \label{fig_abMDI-xp}
\end{figure}

For example, in the Gogny-type model, the single-nucleon potential with  a momentum-dependent part is given by,
\begin{align}\label{Gen-U}
U_J(\rho,\delta,|\v{k}|)=&\frac{A_\ell\rho_J}{\rho_0}+\frac{A_{\rm{u}}\rho_{\overline{J}}}{\rho_0}
+B\left(\frac{\rho}{\rho_0}\right)^{\sigma}(1-x\delta^2)\notag\\
&-4x\tau_3^J\frac{B}{\sigma+1}\frac{\rho^{\sigma-1}}{\rho_0^{\sigma}}\delta\rho_{\overline{J}}\notag\\
&+\sum_{J'}\frac{2C_{J,J'}}{\rho_0}\int\d\v{k}'f_{J'}(\v{r},\v{k}')\Omega(\v{k},\v{k}'),
\end{align}
and the energy density is much involved\cite{CaiLi22Gog}.
Nonetheless, one can calculate the proton fraction directly from $\varepsilon(\rho,x_{\rm p})$ numerically. Shown in the panel (a) of FIG.\,\ref{fig_abMDI-xp} is the proton fraction $x_{\rm p}$ in NS matter as a function of density for the FFG and HMT-exp models. Since the HMT-SCGF result is very close to that of the FFG model, we focus on the comparison between the FFG and HMT-exp predictions. The most notable feature is that, owing to the reduced high-density symmetry energy in the HMT model, the proton fraction decreases rapidly above a certain density. Consequently, the isospin asymmetry $\delta$ in the HMT-exp model approaches unity much faster than in the FFG case, underscoring the key role of the symmetry energy in the NS core. This behavior has important implications for cooling of NSs and the associated neutrino emission\cite{Lat91PRL}.
Quantitatively, the critical proton fraction for the direct Urca (DU) process is $
x_{\rm p}^{\rm{DU}}\approx{1}/[{1+(1+x_{\rm e}^{1/3})^3}]$,
where $x_{\rm e}\equiv\rho_{\rm e}/\rho_{\rm p}$ ranges from $1$ to $0.5$, giving $x_{\rm p}^{\rm{DU}}\approx 11.1\%\sim14.8\%$\cite{Lat91PRL}. These thresholds correspond to the red dashed lines in panel (a) of FIG.\,\ref{fig_abMDI-xp}. The black circle and cyan star mark the central densities of NSs in the two models, where the DU process is allowed in the FFG model but forbidden in the HMT-exp model, according to this conventional picture. Actually, the SRC-HMT may significantly modify the proton fraction threshold, allowing the DU process to occur easier.
The softening of the NS matter EOS in the Gogny-type model can also have manifestation on the speed of sound squared (SSS) defined by $s^2=\d P/\d\varepsilon$, which reflects the stiffness of the EOS, the $s^2$ is shown in panel (b) of FIG.\,\ref{fig_abMDI-xp}. The HMT-exp model yields a much softer EOS than the FFG model above approximately $2.5\rho_0$. Since the ANM EOS can be written as $E(\rho,\delta)\approx E_0(\rho)+E_{\rm{sym}}(\rho)\delta^2+\cdots$, the reduced symmetry energy and enhanced $\delta^2$ together soften the EOS at high densities. The net effect, however, depends on how much the symmetric-matter EOS $E_0(\rho)$ is stiffened by SRC/HMT. The results clearly indicate that the SRC/HMT-induced reduction of the symmetry energy outweighs their stiffening effect on $E_0(\rho)$, leading to an overall softer EOS in the HMT-exp model. 
While this feature exists in the Gogny-type interaction, it does not generally hold for other models with SRC-HMT effects; for example, in the nonlinear Walecka model with a HMT, the $\beta$-stable EOS is hardened\cite{Cai16b} although its symmetry energy is softened at supra-saturation densities.
In order to calculate the proton fraction in the nonlinear Walecka model, one needs to use the expression for the chemical potential\cite{Sou20xx,Sou20PRC}:
\begin{align}
    \mu_J=&3C_J\left(\mu_J^{\rm{kin}}
    -\frac{1}{\phi_J}\sqrt{\phi_J^2k_{\rm{F}}^{J,2}+M_{\rm{D}}^{\ast,2}}
    \right)\notag\\
    &+4C_Jk_{\rm F}^J\ln\left(
    \frac{\phi_Jk_{\rm F}^J+\sqrt{\phi_J^2k_{\rm{F}}^{J,2}+M_{\rm{D}}^{\ast,2}}}{k_{\rm F}^J+\sqrt{k_{\rm{F}}^{J,2}+M_{\rm{D}}^{\ast,2}}}
\right)\notag\\
&+\Delta_J\mu_J^{\rm{kin}}+g_\omega\overline{\omega}_0+\tau_3^Jg_\rho\overline{\rho}_0^{(3)},\label{mu-app}
\end{align}
adopting the HMT form of Ref.\cite{Cai16b}, here $\mu_J^{\rm{kin}}=\sqrt{k_{\rm F}^{J,2}+M_{\rm D}^{\ast,2}}$ and $M_{\rm D}^{\ast}=M_{\rm N}-g_\sigma\overline{\sigma}$ is the nucleon Dirac mass\cite{Qin24}.
With $\Delta_J\to1$ and $\phi_J\to1$, this $\mu_J$ reduces to $\sqrt{k_{\rm F}^{J,2}+M_{\rm D}^{\ast,2}}+g_\omega\overline{\omega}_0+\tau_3^Jg_\rho\overline{\rho}_0^{(3)}$, the familiar expression in the FFG model.
The lower panel of FIG.\,\ref{fig_abMDI-xp} shows the proton fraction $x_{\rm p}$ as a function of baryon density.
It is necessary to point out that Eq.\,(\ref{mu-app}) for the chemical potential corresponds to an approximate form and the full expression for it is given recently in Ref.\cite{Lour25xx}.

\begin{figure}[h!]
\centering
   \includegraphics[width=7.5cm]{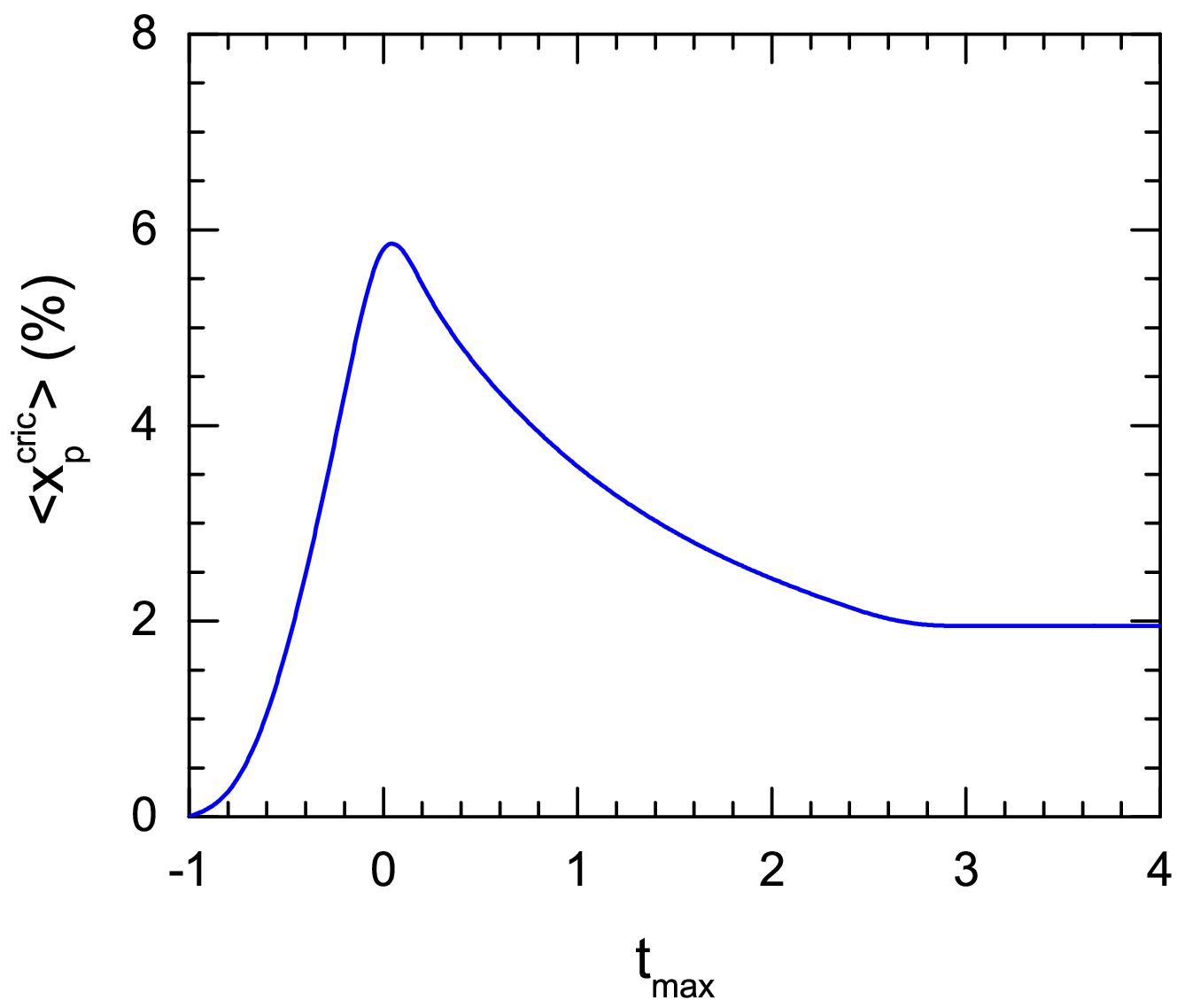}
 \caption{(Color Online). The critical proton fraction as a function of $t_{\max}$ for the occurrence of the DU process. Figure taken from Ref.\cite{LCCX18}.}\label{xpcProb1}
\end{figure}

\begin{figure*}[h!]
\centering
  \includegraphics[width=17.cm]{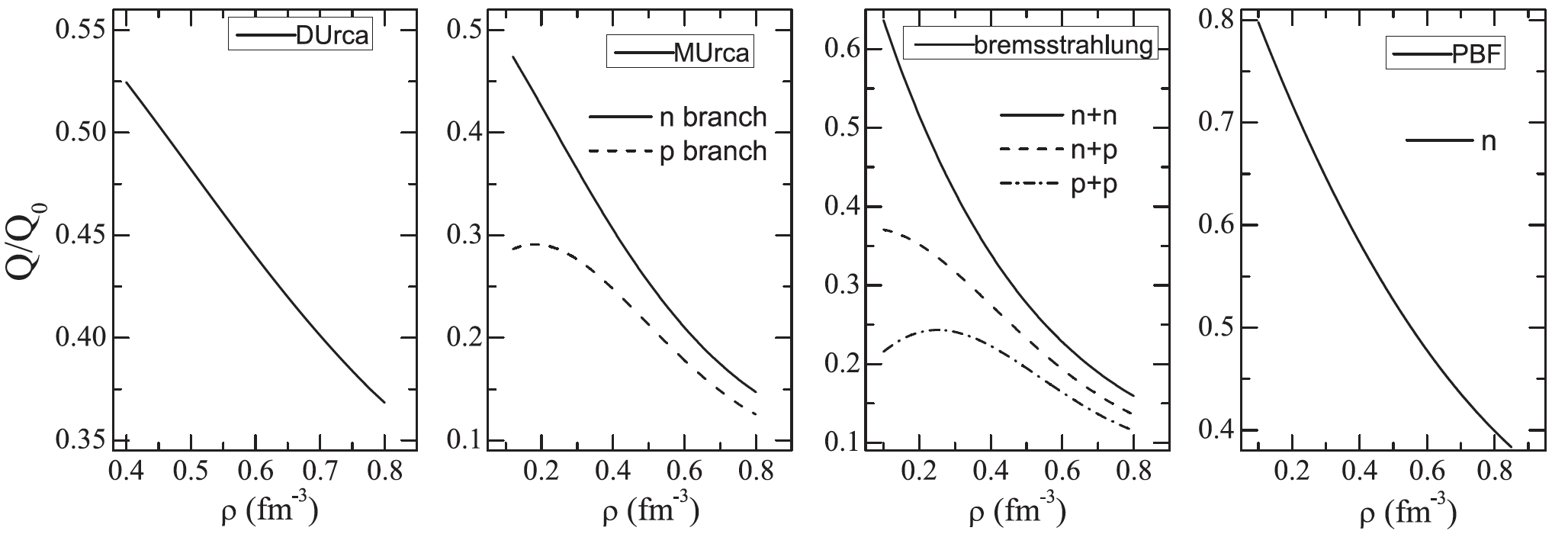}
\caption{(Color Online). $Q/Q_0$ for the DU process, modified Urca process (MU), NN bremsstrahlung and PBF processes as a function of density $\rho$ in $\beta$-stable NS matter. Figure taken from Ref.\cite{Dong16ApJ}. }  \label{fig_Dong-Q}
\end{figure*}

\begin{figure}[h!]
\centering
  \hspace{-0.5cm}
  \includegraphics[width=7.5cm]{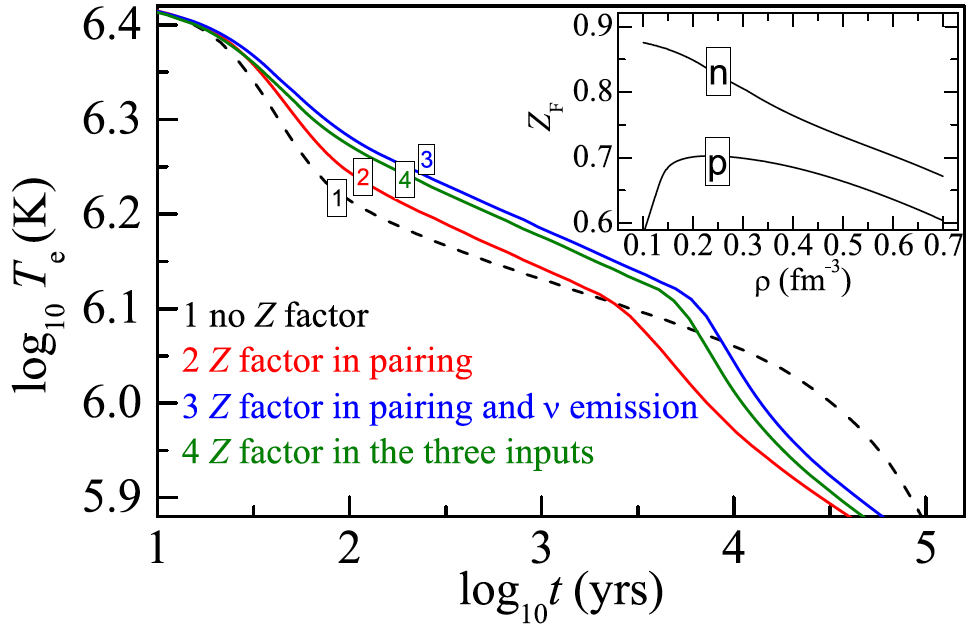}\\[0.25cm]
  \includegraphics[width=8.5cm]{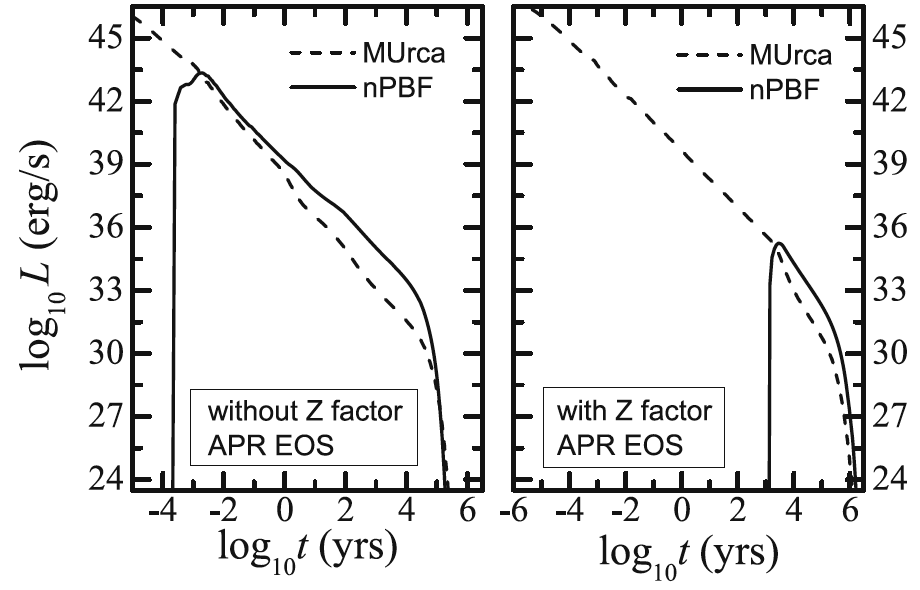}
\caption{(Color Online). Upper: cooling curves of a canonical NS, with the stellar structure constructed using the APR EOS. Results are shown for calculations with or without any $Z$ factors. The inset shows the neutron and proton $Z$ factors as a function of density in $\beta$-stable matter. Figure taken from Ref.\cite{Dong16ApJ}. Lower: the neutrino luminosity as a function of NS age from the PBF for neutron $^{3}\rm{PF}_2$ pairing, using the APR EOS. Figure taken from Ref.\cite{Dong24EPJA}.
 }  \label{fig_Dong-Temp}
\end{figure}

The SRC-HMT may influence the critical proton fraction for the fast DU process\cite{Fra08}. To our knowledge, there is still no community consensus on the dominating cooling mechanism of protoneutron stars, particularly because the limited observational data available tend to support the slower modified URCA process, whereas some earlier studies that included SRC effects suggested that the critical density for direct URCA could be very low\cite{LCCX18}. Here, we aim to provide some insight by discussing, in a more quantitative yet purely kinematic manner, the effects of the SRC-induced HMT on the critical proton fraction $x_{\rm p}^{\rm{cric}}$, while a fully consistent treatment of both $x_{\rm p}^{\rm{cric}}$ and the corresponding emissivity would preferably require using spectral function technique\cite{Sed24PRL}.
In the presence of HMT in the npe matter of protoneutron stars, there is not only additional phase space below the Fermi surface for new particles to occupy, but particles in the HMT also possess higher momenta and energies that can be shared with other particles. The critical momentum-conservation condition for the DU process $\rm{n}\to\rm{p}+\rm{e}^{-}+\overline{\nu}_{\rm{e}^-}$ to occur then generally requires 
$k_{\rm{F}}^{\rm{n}}+\epsilon_{\rm{F}}^{\rm{n}}\leq k_{\rm{F}}^{\rm{p}}+\epsilon_{\rm{F}}^{\rm{p}}+k_{\rm{F}}^{\rm{e}}$, 
where $\epsilon_{\rm{F}}^J$ quantifies the nucleon momentum in excess of its Fermi momentum when SRCs are included. Charge neutrality implies 
$k_{\rm{F}}^{\rm{p}}=k_{\rm{F}}^{\rm{e}}$, we have
$k_{\rm{F}}^{\rm{n}}(1+t_{\rm{n}})\leq k_{\rm{F}}^{\rm{p}}(2+t_{\rm{p}})$, 
where $t_J=\epsilon_{\rm{F}}^J/k_{\rm{F}}^J$ with $-1\leq t_J\leq \phi_J$. Using the relation 
$k_{\rm{F}}^J=\text{const.}\times\rho^{1/3}(1+\tau_3^J\delta)^{1/3}$, 
one then obtains the general condition for the DU process in the presence of SRC-generated HMT:
\begin{equation}\label{E1}
\boxed{
x_{\rm{p}}\geq
x_{\rm{p}}^{\rm{cric}}(t_{\rm{p}},t_{\rm{n}})
\equiv\frac{(t_{\rm{n}}+1)^3}{(t_{\rm{n}}+1)^3+(t_{\rm{p}}+2)^3}.}
\end{equation}
In the FFG limit with $t_{\rm{p}}=t_{\rm{n}}=0$, this condition reduces to the standard result 
$x_{\rm{p}}^{\rm{cric}}(0,0)=1/9\approx 11\%$ 
commonly used in the literature. In the presence of SRC-induced HMT and low-momentum depletion, however, DU processes may occur with 
$x_{\rm{p}}^{\rm{cric}}$ either smaller or larger than $1/9$, depending on the specific momentum configuration of the participating nucleons, since particles may come from the HMT or from deep within the Fermi sea.
To estimate the probability for a given DU channel, notice that it is the product of the momentum distributions in the initial and final states\cite{LCCX18},
\begin{align}
\mathcal{P}_{\rm{pn}}\equiv \mathcal{P}(t_{\rm{p}},t_{\rm{n}})=
\begin{cases}
\Delta_{\rm{n}}(1-\Delta_{\rm{p}}),\\
\Delta_{\rm{n}}[1-{C_{\rm{p}}}/{(t_{\rm{p}}+1)^4}],\\
[{C_{\rm{n}}}/{(t_{\rm{n}}+1)^4}](1-\Delta_{\rm{p}}),\\
[{C_{\rm{n}}}/{(t_{\rm{n}}+1)^4}][1-{C_{\rm{p}}}/{(t_{\rm{p}}+1)^4}],
\end{cases}
\end{align}
here the branches are defined for $-1\leq t_{\rm{n}}\leq 0,\ -1\leq t_{\rm{p}}\leq 0$; $-1\leq t_{\rm{n}}\leq 0,\ 0<t_{\rm{p}}\leq \phi_{\rm{p}}-1$; $0<t_{\rm{n}}\leq \phi_{\rm{n}}-1,\ -1\leq t_{\rm{p}}\leq 0$ and $0<t_{\rm{n}}\leq \phi_{\rm{n}}-1,\ 0<t_{\rm{p}}\leq \phi_{\rm{p}}-1$, respectively.
For instance, the first case corresponds to a neutron below its Fermi surface decaying into a proton also below its Fermi surface. The channel 
$-1\leq t_{\rm{n}}\leq 0,\ 0<t_{\rm{p}}\leq\phi_{\rm{p}}-1$, 
where the neutron lies below its Fermi surface but the proton occupies a state above its Fermi surface, clearly has the largest probability. According to Eq.~(\ref{E1}), this channel yields a critical $x_{\rm p}$ much smaller than $1/9$.
The statistical average of the critical proton fraction is
\begin{equation}
\langle x_{\rm{p}}^{\rm{cric}}\rangle\equiv
\frac{\sum_{i_1,i_2}\mathcal{P}(i_iH,i_2H)\,x_{\rm{p}}^{\rm{cric}}(i_1H,i_2H)}
{\sum_{i_1,i_2}\mathcal{P}(i_1H,i_2H)},
\end{equation}
where $i_1H=t_{\rm{p}}^{i_1}$ and $i_2H=t_{\rm{n}}^{i_2}$ vary from $-1$ to a maximum of 
$t_{\max}\lesssim t_{\max}^{\rm{th}}\equiv\phi_{\max}-1=\phi_0(1-\phi_1)-1\approx 2.71$. 
Here $H$ denotes the step size of the two-dimensional momentum-space lattice. FIG.\,\ref{xpcProb1} illustrates $x_{\rm{p}}^{\rm{cric}}$ as a function of $t_{\max}$, indicating that DU process with threshold proton fractions approaching zero is possible. Overall, the HMT and low-momentum depletion significantly reduce the averaged critical proton fraction below $11\%$. Since $\langle x_{\rm{p}}^{\rm{cric}}\rangle$ saturates near $t_{\max}\approx 2.6<t_{\max}^{\rm{th}}$, the final averaged value is approximately $2\%$, implying DU process is statistically much easier to occur. This conclusion is consistent with earlier findings based on different arguments\cite{Fra08}.

However, recent work\cite{Dong16ApJ} proposed that the neutrino emissivity of the direct URCA process is reduced by a factor $Z_{\rm{F}}^{\rm{p}}Z_{\rm{F}}^{\rm{n}}$ compared to the FFG model, where $Z_{\rm{F}}^J$ is the Migdal--Luttinger jump. In models with SRC-induced HMT, the depletion of the Fermi surface and large values of $C_J$ significantly reduce this factor. In particular, they showed that the neutrino emissivity $Q$ modifies as\cite{Dong16ApJ}
\begin{align}
&\rm{n}+\to\rm{p}+l+\overline{\nu}_l,\;\rm{p}+l\to\rm{n}+\nu_l\;\text{(DU)}:\notag\\
&\hspace{0.5cm}Q/Q_0=Z_{\rm F}^{\rm n}Z_{\rm F}^{\rm p};\\
&\rm{n}+\rm{N}\to\rm{p}+\rm{N}+l+\overline{\nu}_l,\;\rm{p}+\rm{N}+l\to\rm{n}+\nu_l\;\text{(MU)}:\notag\\
&\hspace{0.5cm}Q^{\rm{n}}/Q_0=Z_{\rm F}^{\rm n,3}Z_{\rm F}^{\rm p},~Q^{\rm{p}}/Q_0=Z_{\rm F}^{\rm n}Z_{\rm F}^{\rm p,3};\\
&\rm{N}+\rm{N}\to\rm{N}+\rm{N}+\nu+\overline{\nu}\;\text{(NN bremsstrahlung)}:\notag\\
&\hspace{0.5cm}
Q^{\rm{nn}}/Q_0=Z_{\rm F}^{\rm n,4},~Q^{\rm{pp}}/Q_0=Z_{\rm F}^{\rm p,4},~Q^{\rm{np}}/Q_0=Z_{\rm F}^{\rm n,2}Z_{\rm F}^{\rm p,2};
\\
&\rm{N}+\rm{N}\to[\rm{NN}]+\nu+\overline{\nu}\;\text{(PBF)}:\notag\\
&\hspace{0.5cm}Q^{\rm{n}}/Q_0=Z_{\rm F}^{\rm n,2},
\end{align}
where the subscript ``0'' marks the corresponding case without the $Z$-factor. Besides the DU process, here the modified Urca process (MU), the nucleon-nucleon (NN) bremsstrahlung process and the Cooper pair breaking and formation (PBF) are also listed\cite{Dong16ApJ}.
FIG.\,\ref{fig_Dong-Q} shows the calculated $Q/Q_0$ for the DU, MU,  NN bremsstrahlung and the PBF processes as a function of density $\rho$. The DU process occurs only when the proton fraction exceeds a critical threshold, which depends on the high-density behavior of the symmetry energy. A stiff symmetry energy, such as that predicted by Brueckner theory, results in a lower threshold.
Their calculations indicate that the neutrino emissivity is reduced by more than $50\%$, in contrast with the conclusion of Ref.\cite{Fra08}. For the MU processes, the neutrino emissivities are reduced by more than $50\%$ for the neutron branch and more than $70\%$ for the proton branch. Since the proton $Z$-factor is smaller than the neutron one (see inset of the upper panel of FIG.\,\ref{fig_Dong-Temp}), the proton branch of $Q_{\rm{MU}}$ is more strongly suppressed by Fermi surface depletion. At high densities, $Q_{\rm{MU}}$ can be reduced by an order of magnitude. Similarly, the NN bremsstrahlung processes are also more strongly suppressed at high densities. The computed $Q/Q_0$ for these processes is insensitive to the presence of superfluidity.
Using a softer symmetry energy, such as that from APR EOS\cite{Akmal1998}, yields similar results for the MU, NN bremsstrahlung, and PBF processes compared with Brueckner theory, but the DU threshold is significantly higher due to the reduced symmetry energy\cite{Dong16ApJ}.
To proceed, they calculated the cooling of canonical NSs within the minimal cooling paradigm, adopting the APR EOS\cite{Akmal1998}, the basic equations are
\begin{align}
&\frac{1}{4\pi r^2 e^{2\Phi}} \sqrt{1-\frac{2GM}{r}}\frac{\partial}{\partial r} \left( e^{2\Phi} L \right) = - Q_\nu - \frac{C_\nu}{ e^{\Phi}} \frac{\partial T}{\partial t},\\
&L = - 4\pi r^2 \kappa e^{-\Phi} \sqrt{1-\frac{2GM}{r}} \frac{\partial (T e^{\Phi})}{\partial r}, \label{eq:energy_transport}
\end{align}
where $\Phi$ is the metric function,  $Q_\nu$ is the neutrino emissivity, $C_\nu$ and $\kappa$ are the specific heat capacity and the thermal conductivity, respectively; here\cite{Wang24PRC}
\begin{equation}
\boxed{
    C_\nu=\sum_{J=\rm{n,p}}C_\nu^J=\sum_{J=\rm{n,p}}Z_{\rm F}^Jk_{\rm{B}}^2T{M_{\rm N}^{J,\ast} k_{\rm F}^J}/{3\hbar^3},}
\end{equation}
with $M_{\rm N}^{J,\ast}$ denoting the nucleon Landau effective mass.\footnote{We use $M_{\rm D}^{\ast}$ to denote the (relativistic) nucleon Dirac effective mass in this review, see Ref.\cite{LCCX18} for the difference and relation between these quantities.}
The local luminosity $L$ and temperature $T$ depend on the radial coordinate $r$ and time $t$.
Their results showed that including the $Z$-factor suppresses the neutron $^3\rm{P}_2$ superfluidity and lowers its critical temperature $T_{\rm c}$. As a result, NS cooling is initially slowed during the first $3\times 10^3$ years, but once the stellar temperature falls below $T_{\rm c}$, the PBF process becomes active and accelerates the cooling, being more efficient than the MU processes, see the upper panel of FIG.\,\ref{fig_Dong-Temp}.
The $Z$-factor also reduces the neutrino emissivity from MU, NN bremsstrahlung, and PBF processes, as well as the NS heat capacity, lowering the thermal energy. These combined effects initially retard cooling, but later slightly enhance it due to the reduced thermal energy. Overall, the weak neutron $^3\rm{P}_2$ superfluidity, strongly quenched by Fermi surface depletion, cannot significantly suppress neutrino emissivity or heat capacity. Their results demonstrate that the $Z$-factor-induced Fermi surface depletion is an essential factor for accurately modeling NS cooling.
The neutrino luminosity as a function of NS age due to the PBF process is shown in the lower panel of FIG.\,\ref{fig_Dong-Temp}.

\begin{figure}[h!]
\centering
  \includegraphics[width=9cm]{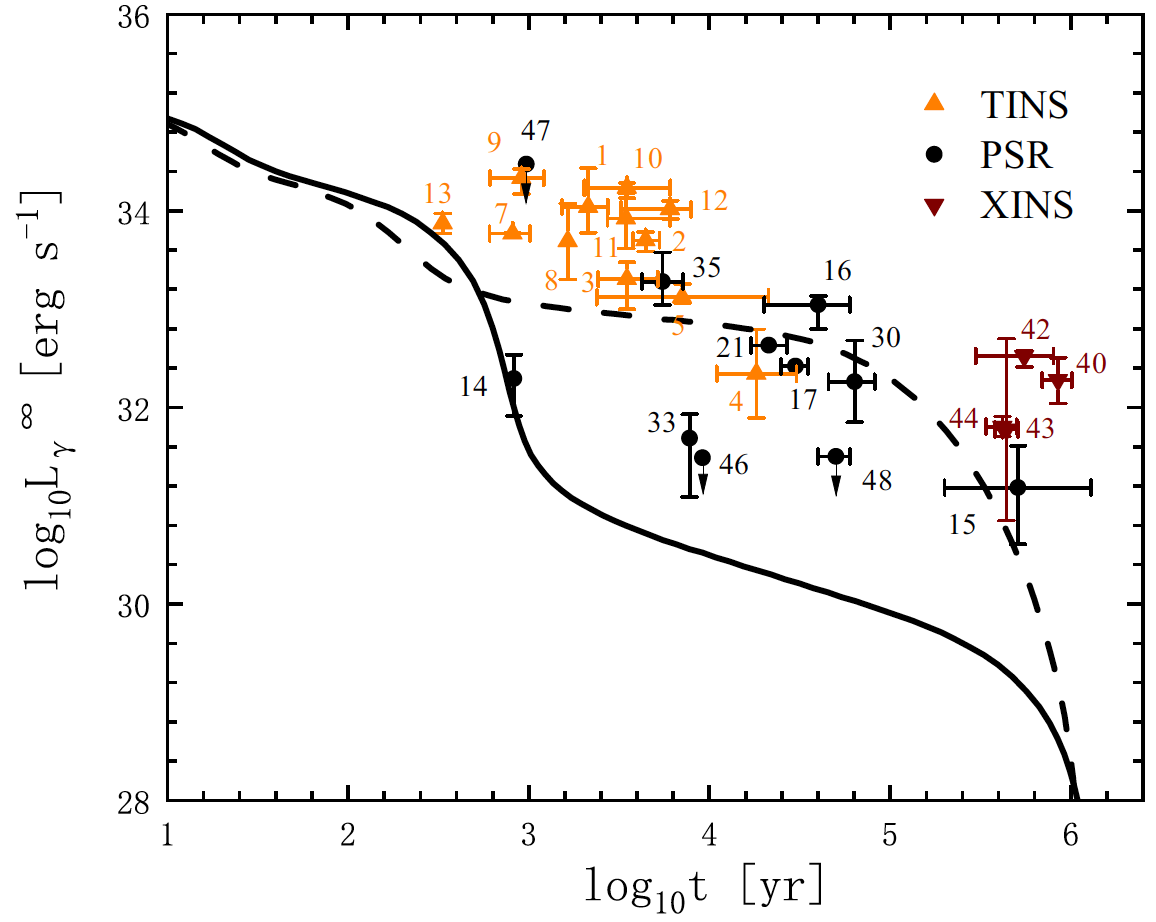}\\[0.25cm]
  \includegraphics[width=9cm]{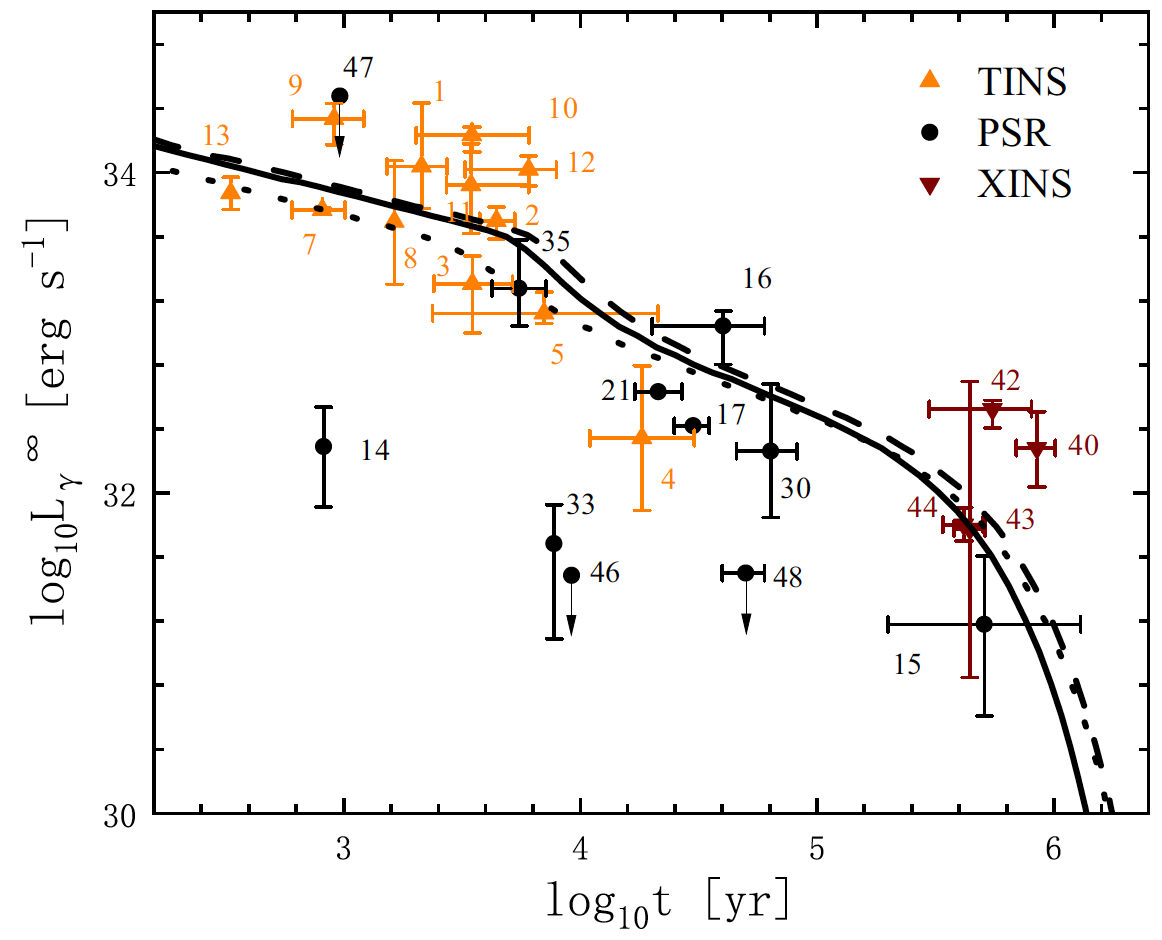}
\caption{(Color Online). Upper: comparison of predicted cooling curves with the observed data for $2.0\,M_\odot$ NSs constructed using the APR EOS, without (dashed line) and with (solid line) the effect of the $Z$-factor. Lower: same as the upper panel, but for canonical NSs. Figures taken from Ref.\cite{Wang24PRC}.
}  \label{fig_Wang-12}
\end{figure}

\begin{figure}[h!]
\centering
  \includegraphics[width=9.cm]{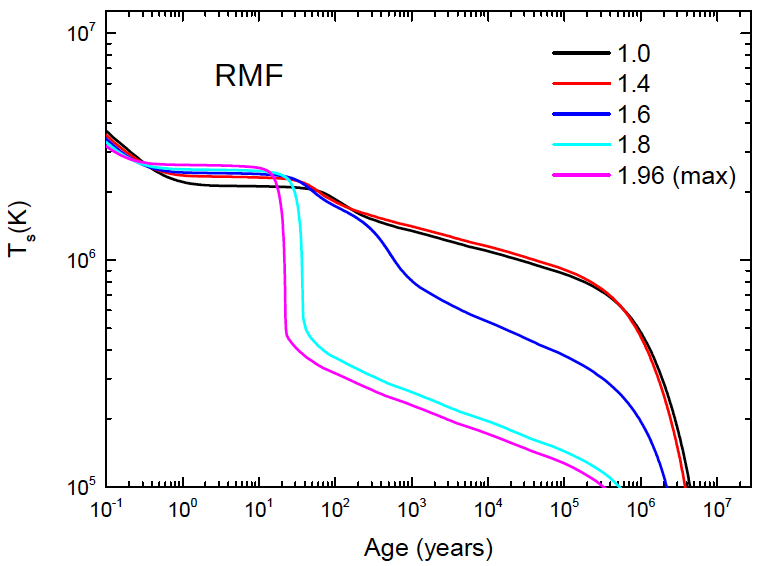}\\[0.25cm]
  \includegraphics[width=9.cm]{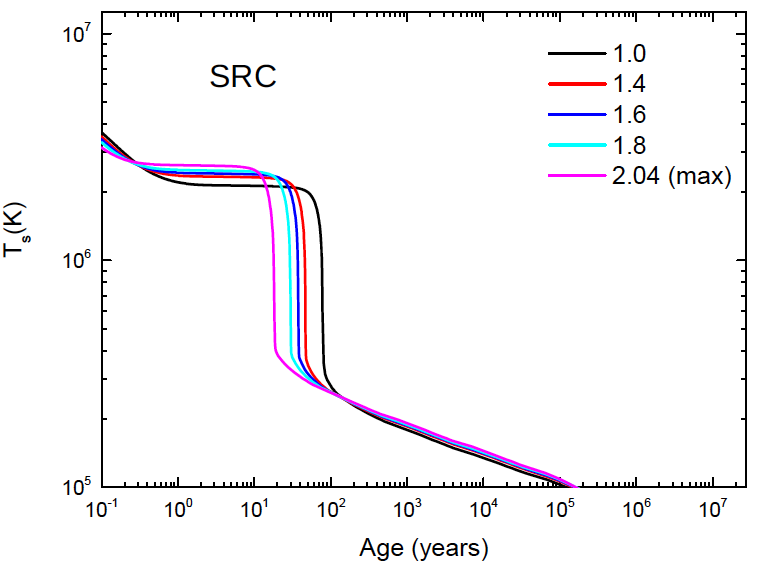}
\caption{(Color Online). Red-shifted surface temperature evolution for the RMF model without (upper) or with (lower) SRC-HMT. The labels indicate the mass of each NS in units of solar mass. Figures taken from Ref.\cite{Sou20xx}.}  \label{fig_Sou-12}
\end{figure}

In Ref.\cite{Wang24PRC}, the authors calculated the cooling curves of NSs using the APR EOS\cite{Akmal1998} while including the effects of the $Z$-factor (SRC) on superfluid gaps, neutrino emissivity, and heat capacity. A key finding is the strong dependence of cooling behavior on the NS mass. For low- and medium-mass NSs (roughly $M_{\rm{NS}}\lesssim1.8M_{\odot}$), the DU process is not activated, so cooling proceeds relatively slowly, as shown in the lower panel of FIG.\,\ref{fig_Wang-12}. In contrast, massive NSs can undergo rapid cooling via the DU process, with even a small increase in mass enough to trigger a sudden transition to the fast cooling regime (upper panel). The $Z$-factor further modifies the cooling by suppressing proton $^1\rm{S}_0$ and neutron $^3\rm{PF}_2$ superfluidity, reducing neutrino emission in MU, NN bremsstrahlung  and PBF processes during the early neutrino cooling era, while slightly enhancing cooling in the later photon cooling era due to reduced heat capacity\cite{Wang24PRC}.
The mass dependence of the cooling curves is essential for matching observations: the thermal luminosities of NSs increase with mass up to about 1.8$M_\odot$\cite{Wang24PRC}, and most observed data can be reproduced when this effect (mass-dependence) is considered. These results underscore that SRCs and the associated $Z$-factor are critical for accurate modeling of NS cooling, particularly in understanding why massive stars cool much faster than their lighter counterparts and how small mass differences can lead to abrupt transitions in cooling behavior.
A similar, though not entirely consistent, trend is found using a nonlinear RMF model. While the bulk properties of the models are comparable, the cooling behavior differs somewhat. In the RMF model without SRC\cite{Sou20xx}, stars up to about 1.6$M_\odot$ cool slowly, and only more massive NSs can activate the DU process, leading to rapid cooling. When SRC are included, the DU process becomes allowed for a wider range of stars, generally accelerating cooling, though the precise mass thresholds and cooling rates show some variation compared with the SRC model, as shown in FIG.\,\ref{fig_Sou-12}. This highlights that SRCs have a strong, but model-dependent, effect on the thermal evolution of NSs.
As a final comment, the effects of HMT nucleons, not accounted for in Refs.\cite{Dong16ApJ,Wang24PRC}, may enhance the neutrino emissivity\cite{Fra08}.  
For example, the $Z$-factor should in principle be calculated in a self-consistent manner once the nucleon-nucleon interaction scheme is adopted, which is clearly not the case in either the BHF or the APR EOS approach.  Therefore, the net impact of the SRC-modified momentum distribution on both the critical density for the DU process and the cooling rate of protoneutron stars remains uncertain.
Besides the DU process in NS cooling simulations, other related quantities, such as transport coefficients, have also been investigated in the presence of SRC-HMT; see, e.g., Refs.\cite{LiuCX24,ShangXL20} for discussions.

\begin{figure}[h!]
\centering
  \includegraphics[height=6.5cm]{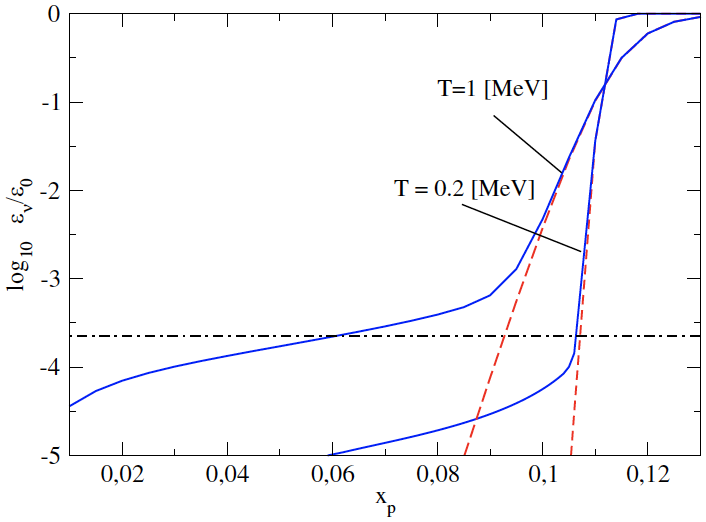}\\[0.25cm]
  \includegraphics[height=6.5cm]{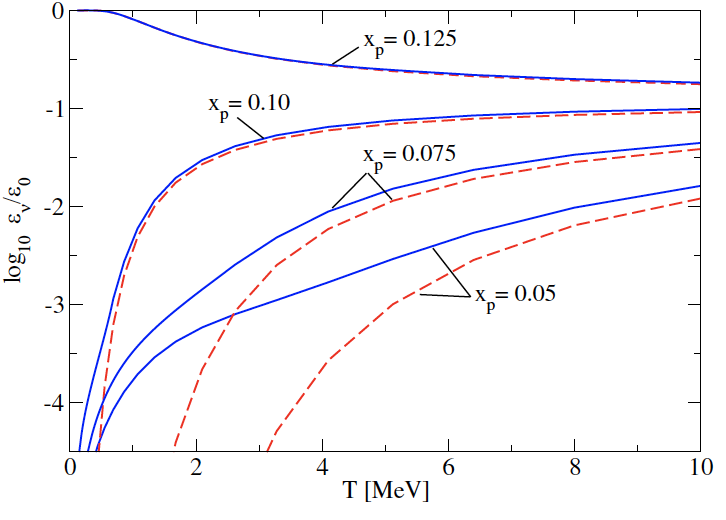}
\caption{(Color Online).
Upper: the non-Fermi-liquid Urca emissivity in units of $\epsilon_0$ for $T = 1$\,MeV and $T = 0.2$\,MeV at a density of $\rho \approx1.5\rho_0$, as a function of the proton fraction $x_{\rm p}$. The dashed lines show the DU process, i.e., the quasi-particle limit corresponding to a vanishing with $\gamma_{\rm p} = 0$. The Fermi-liquid  one-pion-$\rho$ exchange based MU emissivity, $\epsilon_{\rm{MU}}/\epsilon_0$, for $T = 1$\,MeV is shown by the dot-dashed horizontal line. Lower: the non-Fermi-liquid Urca emissivity in units of $\epsilon_0$ for $x_{\rm p} = 0.05$, $0.10$, and $0.125$ at $\rho \approx 1.5\rho_0$ as a function of temperature $T$. The solid lines show the full result, whereas the dashed lines correspond to the quasi-particle limit $\gamma_{\rm p} = 0$. Figures taken from Ref.\cite{Sed24PRL}.
}  \label{fig_Sed12}
\end{figure}

A recent development in solving the NS cooling problem is the work of Sedrakian (2024)\cite{Sed24PRL}, who revisited the long-standing issue of neutrino emission via the Urca process by incorporating SRCs into the microscopic description of dense matter, in the spectral function approach. Due to the SRC-induced HMT in the nucleon momentum distribution, the proton component in dense matter is no longer adequately described as a Landau Fermi liquid with a sharp Fermi surface, as shown in FIG.\,\ref{fig_nkZ}. Instead, Ref.\cite{Sed24PRL} employs a full proton spectral function with finite width $\gamma_{\rm p}\sim T^2[1+(E_{\v k}-\mu_{\rm p})^2/\pi^2T^2]$, derived from nucleon self-energies, thereby going beyond the usual zero-width quasi-particle approximation that underlies traditional calculations of the DU emissivity.
This modification has important thermodynamic and kinematic consequences. In the conventional Fermi-liquid picture, the momentum conservation ``triangle condition'' enforces a strong proton-fraction threshold, typically $x_{\rm p} \gtrsim 0.11$, below which the DU process is prohibited. Once the SRC-induced broadening of the proton spectral function is taken into account, this sharp threshold disappears: the Urca emissivity becomes a smooth function of $x_{\rm p}$ and remains significant even at low proton fractions; see the upper panel of FIG.\,\ref{fig_Sed12}. Moreover, Ref.\cite{Sed24PRL} also investigates the temperature dependence of the non-Fermi-liquid Urca rate. As shown in the lower panel of FIG.\,\ref{fig_Sed12}, the quasi-particle-limit calculation exhibits the familiar behavior that DU cooling is kinematically allowed and efficient only when the proton fraction is sufficiently large (e.g., \ $x_{\rm p}\approx0.125$). For smaller proton fractions ($x_{\rm p}\approx0.10$, $0.075$, and $0.05$), the quasi-particle emissivity remains strongly suppressed at low temperatures and becomes visible only through modest thermal activation around $T\sim 1\,\mathrm{MeV}$. In contrast, once the finite proton width induced by SRCs is included, the non-Fermi-liquid Urca emissivity decreases much more slowly with temperature and no longer vanishes in the low-temperature limit. The SRC-induced off-shellness of the proton thus smooths out both the proton-fraction and temperature thresholds that characterize the DU process in the Fermi-liquid framework.
These results provide a unified many-body interpretation in which the traditional distinction between DU and MU processes arises mainly from the quasi-particle approximation rather than from a fundamental separation of physical mechanisms. The astrophysical implications are also important: the standard paradigm of NS cooling, in which only sufficiently massive NSs with large proton fractions cool rapidly via DU process while low-mass NSs cool slowly via MU process, may need revision. If SRC-induced proton widths are a generic feature of dense matter, then efficient neutrino cooling can occur over a much wider range of stellar masses and interior compositions, smoothing the classical dichotomy between ``slow'' and ``fast'' cooling families. Furthermore, this approach directly connects NS thermal evolution with modern nuclear many-body theory and experimental evidence for SRCs in finite nuclei, establishing a framework in which neutrino emission properties can be traced to the microscopic dynamics of strong nuclear forces.

\subsection{Core-crust Transition and Low-density Structure(s)}\label{sub_CC}

\indent

Nuclear ``pasta'' phases are expected to appear in the inner crust of NSs where the baryon density reaches roughly about $0.3\sim0.7\rho_0$, with $\rho_0\approx0.16\,\mathrm{fm^{-3}}$ denoting the saturation density of SNM. In this transition region between the spherical nuclei of the outer crust and the uniform nucleonic fluid of the core, competition between the nuclear surface energy and the long-range Coulomb interaction leads to a sequence of exotic, nonspherical geometries. Depending on the local density and proton fraction, the matter may organize into rod-like ``spaghetti'', slab-like ``lasagna'' as well as inverted structures such as cylindrical or spherical bubbles. These phases typically extend over several hundred meters in depth and may possess characteristic length scales of tens of femtometers in their microscopic structure. Nuclear pasta is believed to influence various NS observables, including thermal conductivity, impurity scattering, neutrino transport, and crustal elasticity. See Refs.\cite{Ravenhall83,Hashimoto84,Oyamatsu93,Horowitz15,NewtonStone09} for more related discussions.

\begin{figure}[h!]
\centering
  \includegraphics[height=13.cm]{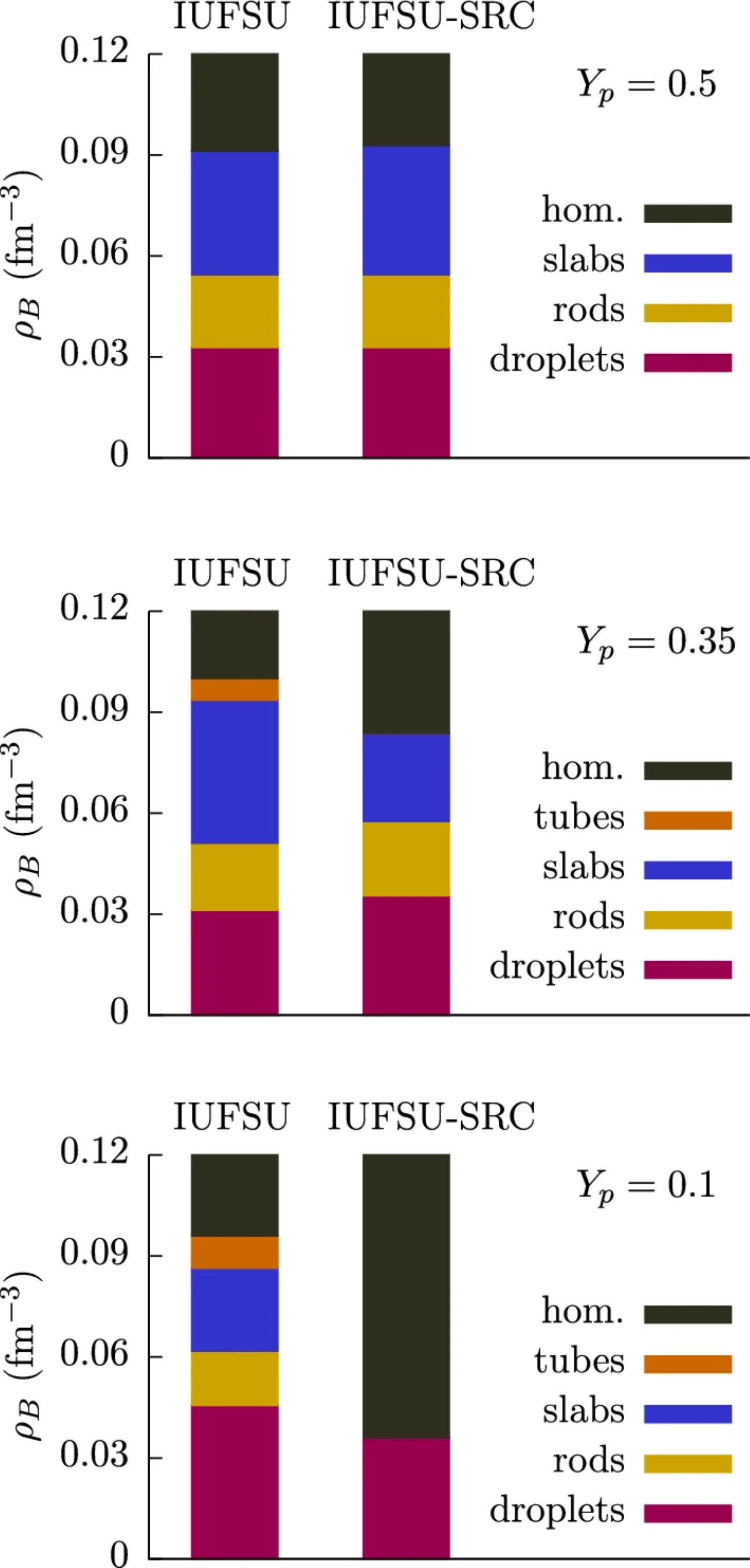}
\caption{(Color Online). Baryon density ranges in which each pasta geometry is the energetically favored configuration for the IUFSU  model\cite{IUFSU}, shown for proton fractions $x_{\rm p}\equiv Y_{\rm p} = 0.5$, $0.35$, and $0.1$ (upper, center, and lower, respectively). Figures taken from Ref.\cite{Peli23EPJA}.
}  \label{fig_Peli}
\end{figure}

In a recent work \cite{Peli23EPJA}, the authors studied the effects of SRC-HMT on the formation of pasta phases based on the nonlinear Walecka model\cite{Cai16b,Sou20xx,Sou20PRC}.
Specifically, they first explored how SRCs affect homogeneous nuclear matter using three RMF parameterizations: IUFSU\cite{IUFSU}, FSU2R\cite{FSU2R} and NL3\cite{NL3}, the latter chosen because of its much stiffer symmetry energy. Including SRCs changes the kinetic contributions to the energy density, pressure, and scalar density. As a result, each model must be reparametrized to continue reproducing basic nuclear properties. They then calculated the nuclear pasta phase using the coexistence-phase approximation, imposing the usual Gibbs conditions for pressure and chemical equilibrium. A crucial ingredient in this method is the surface tension. For IUFSU and NL3, they adopted Thomas--Fermi based fits, while for FSU2R they used a fixed value $\sigma_0 \approx 1.2\,\mathrm{MeV\,fm^{-2}}$. Although this is a simplified choice, they showed that reasonable variations of the surface tension mainly shift the pasta-uniform transition density and do not change the qualitative trend: SRCs tend to reduce the density range where pasta exists. In their treatment, SRCs do not directly modify the Coulomb or surface energies, but they do affect the symmetry energy, which in turn influences when pasta phases appear. The geometries considered include droplets and bubbles (3D), rods and tubes (2D), and slabs (1D). By comparing the free energies of all structures with that of uniform matter, they determined the most favorable configuration at each density. Their overall conclusion is that SRCs significantly shrink the pasta region in very asymmetric matter and may even eliminate all but the simple droplet structures at zero temperature.

FIG.\,\ref{fig_Peli} illustrates several examples from Ref.\cite{Peli23EPJA}, showing results for the IUFSU model without and with SRCs. The impact of SRCs becomes more pronounced as the isospin asymmetry increases. For symmetric matter, the extent of the pasta region and the sequence of geometries remain nearly unchanged when SRCs are included. For $Y_{\rm p} = 0.35$, however, noticeable differences arise: the pasta region becomes smaller and the tube geometry disappears entirely once SRCs are incorporated. The effect is even stronger at $Y_{\rm p} = 0.1$, the case most relevant for the inner crust of NSs. In this strongly neutron-rich regime, the pasta phase shrinks significantly, leaving only the droplet configuration.
Based on these results and earlier studies, the authors pointed out two likely implications for NS crusts. First, as the temperature increases to only a few keV, the pasta structures may disappear altogether when SRCs are included, since thermal effects tend to destroy the long-range ordering of the pasta\cite{Horowitz15}. Second, if the pasta phase indeed vanishes, calculations of transport properties\cite{Peli23xx} could be substantially simplified, because the loss of nonspherical clusters would remove anisotropies in electron-ion collision frequencies. At the same time, the authors noted that SRCs themselves will also modify transport coefficients, adding further complexity to an already challenging problem.

\begin{figure}[h!]
\centering
 \includegraphics[height=3.cm]{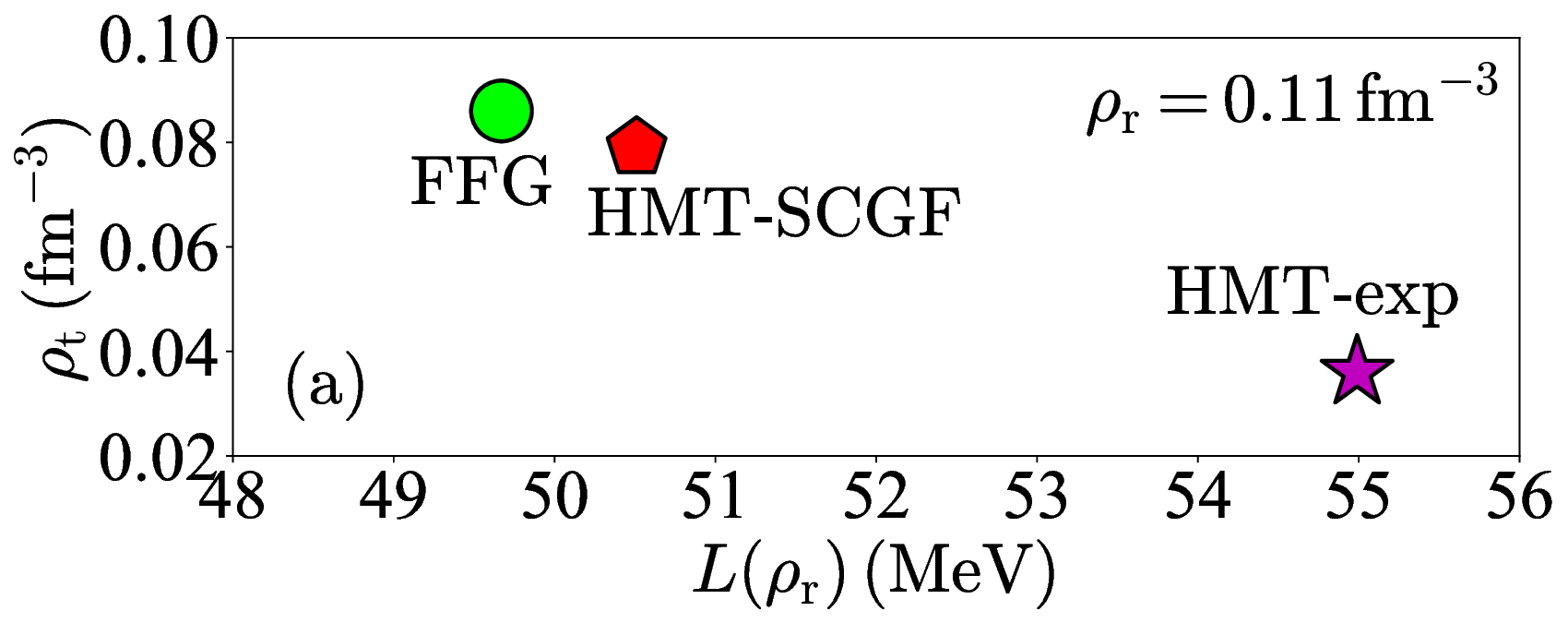}\\
 \hspace{0.cm}
 \includegraphics[height=3.cm]{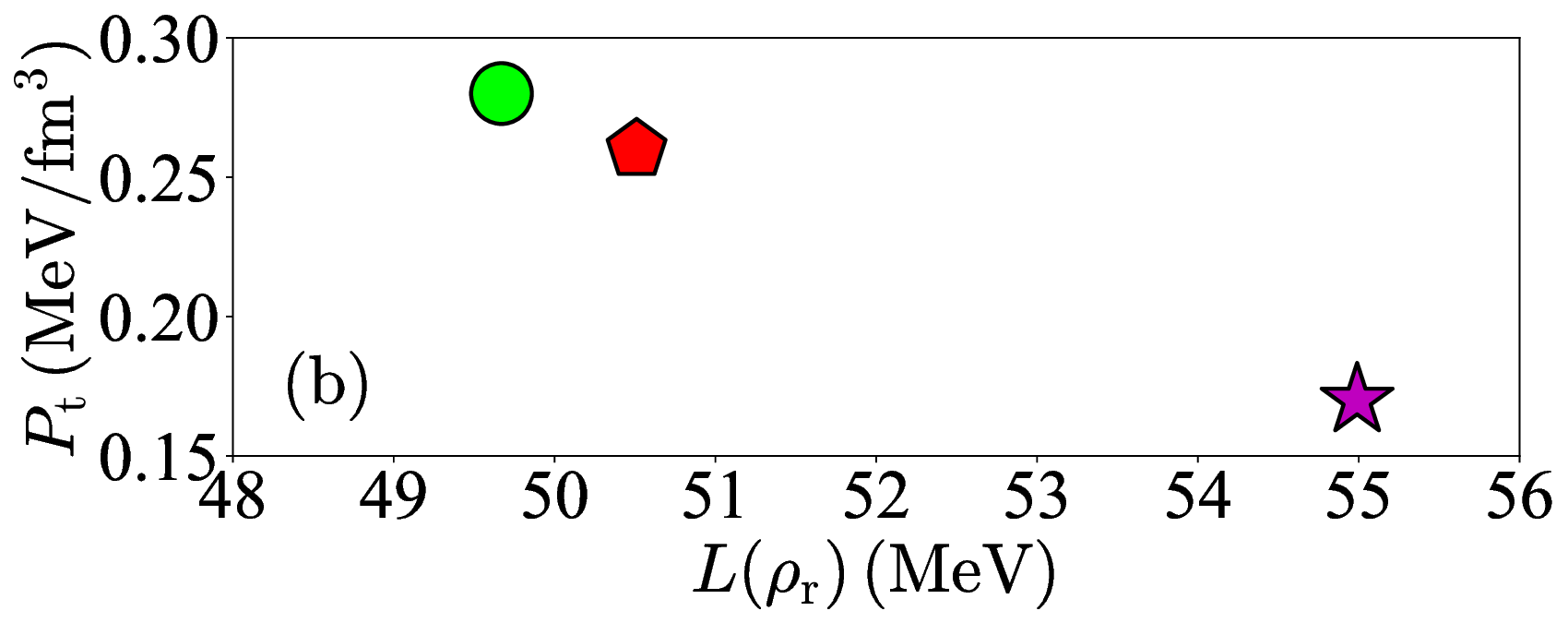}
  \caption{(Color Online). Correlation between the core-crust transition density $\rho_{\rm t}$ (upper) and the transition pressure $P_{\rm t}$ (lower) in $\beta$-stable NS matter
  and the slope parameter $L(\rho_{\rm{r}})$ of the symmetry energy at $\rho_{\rm{r}}\approx0.11\,\rm{fm}^{-3}$. Figures taken from Ref.\cite{CaiLi22Gog}.}
  \label{fig_ab_rhot}
\end{figure}

In fact, the reduction of the nuclear symmetry energy due to the SRC-HMT can significantly influence the core-crust transition density $\rho_{\rm t}$, which in turn affects the onset and extent of the pasta phase and, consequently, the transport properties in the NS crust. To illustrate this effect, we examine how SRC-HMT modifies the transition density. As an illustration of using the Gogny-type EDF\cite{CaiLi22Gog} to investigate SRC-HMT effects on NSs, we study the corresponding core-crust transition density $\rho_{\rm t}$, together with the transition pressure $P_{\rm t}$, in cold NSs.
A commonly used method to determine $\rho_{\rm t}$ is the thermodynamical approach\cite{CaiLi22Gog}, which requires the system to satisfy the intrinsic stability condition ${U}_{\rm{ther}}(\rho) > 0$\cite{XuJ09PRC}, or (since $x_{\rm p}=x_{\rm p}(\rho)$):
\begin{align}
{U}_{\rm{ther}}(\rho) =& \rho^2 \left(\frac{\partial^2 E(\rho, x_{\rm p})}{\partial x_{\rm p}^2}\right)^{-1} \cdot
\left[ \frac{\partial \mu_{\rm n}}{\partial \rho_{\rm n}} \frac{\partial \mu_{\rm p}}{\partial \rho_{\rm p}} - \left(\frac{\partial \mu_{\rm n}}{\partial \rho_{\rm p}}\right)^2 \right]>0,
\label{Vther}
\end{align}
which defines the region of thermodynamical instability in $\beta$-equilibrium NS matter. Here, $E(\rho, x_{\rm p})$ is the energy per nucleon, and the total pressure is $P = P_{\rm N} + P_{\rm e}$, with contributions from nucleons ($P_{\rm N}$) and electrons ($P_{\rm e}$). At these low densities, muons are typically absent. The electron energy density and pressure can be obtained within a FFG model\cite{XuJ09PRC}:
\begin{align}
\varepsilon_{\rm e}(\rho, \delta) &= \eta_{\rm e} \Phi_{\rm e}(t_{\rm e}), 
\end{align}
with
\begin{align}
\eta_{\rm e} =& \frac{m_{\rm e}}{8\pi^2 \lambda_{\rm e}^3}, ~\lambda_{\rm e} = m_{\rm e}^{-1},~  t_{\rm e} = \lambda_{\rm e} (3\pi^2 \rho_{\rm e})^{1/3}, \\
\Phi_{\rm e}(t_{\rm e}) =& t_{\rm e}\left(1 + 2 t_{\rm e}^2\right) \sqrt{1 + t_{\rm e}^2} - \ln\left(t_{\rm e} + \sqrt{1 + t_{\rm e}^2}\right),
\end{align}
here $\mu_{\rm e} = \sqrt{k_{\rm e}^2 + m_{\rm e}^2} \approx k_{\rm e}$ and $k_{\rm e} = (3\pi^2 \rho_{\rm e})^{1/3}$, by neglecting the electron mass $m_{\rm e}$. The core–crust transition occurs at the density where Eq.\,(\ref{Vther}) is first violated.  
Using this method, the transition density in the FFG model is found to be about $\rho_{\rm t} \approx 0.086\,\rm{fm}^{-3}$, while in the HMT models it reduces to $\rho_{\rm t} \approx 0.079\,\rm{fm}^{-3}$ (HMT-SCGF) and $\rho_{\rm t} \approx 0.036\,\rm{fm}^{-3}$ (HMT-exp). Corresponding transition pressures are $P_{\rm t} \approx 0.28\,\rm{MeV}/\rm{fm}^{-3}$ (FFG), $0.26\,\rm{MeV}/\rm{fm}^{-3}$ (HMT-SCGF) and $0.17\,\rm{MeV}/\rm{fm}^{-3}$ (HMT-exp)\cite{CaiLi22Gog}, respectively. Thus, the SRC-induced HMT can substantially reduce the core–crust transition density, by as much as $\sim 58\%$ in the HMT-exp model.  
This reduction is linked to the behavior of the symmetry energy $E_{\rm{sym}}(\rho)$ at sub-saturation densities. As seen in the upper panel of FIG.\,\ref{fig_ab_Esym}, the symmetry energy is effectively hardened in this density range, leading to an enhanced slope parameter $L(\rho)$ for $\rho \lesssim \rho_0$. A larger $L(\rho)$ at sub-saturation densities tends to decrease the core-crust transition density\cite{XuJ09PRC}, as also illustrated in FIG.\,\ref{fig_ab_rhot}. The anti-correlation between $\delta x = x_{\rm{HMT}}^{\rm{SNM}} - x_{\rm{HMT}}^{\rm{PNM}}$, which characterizes the strength of the SRC-induced HMT, and the transition density $\rho_{\rm t}$ provides an important connection for understanding the inner-crust thickness and may have significant implications for heat transport in NSs\cite{Cac08}.

\subsection{Dark Matter Mixed SRC-HMT Effects on NS Properties}\label{sub_DM}

\indent 

The study of dark matter (DM) particles or components in NSs has emerged as a rapidly developing research area at the intersection of nuclear physics, astrophysics, and particle physics, see, e.g., Refs.\cite{Pan17,John18,Nelson19,Das21,Liu23DM,Zhang25DM,Sun24DM,Raj18,Gara19,Bell20,Bell21,Liu25DM,Hipp23DM,Kumar24,Maha25,Raf1,Raf2}. Although the microscopic nature of DM remains elusive\cite{Cire24,Rosz18,Ferr25DM,Bert10}, its potential interactions with baryons and leptons can leave observable imprints on the structure, dynamics, and thermal evolution of NSs. Once captured gravitationally, DM particles may accumulate, thermalize, and form a self-gravitating core, or interact with dense nuclear matter through weak, scalar, or vector couplings, thereby modifying the underlying EOS. In most scenarios, the presence of DM introduces additional energy density with relatively small pressure, or opens new degrees of freedom, effectively softening the EOS and altering key macroscopic properties of NSs, including the M-R relation, tidal deformability, and maximum stable mass. DM accumulation can also influence the star's thermal evolution by contributing additional heating sources, modifying neutrino emission, or affecting transport and conductivity in both the crust and core\cite{Bram25}. Depending on the specific DM model, ranging from weakly interacting massive particles (WIMPs) and asymmetric DM to bosonic condensates, these effects may be substantial enough to trigger gravitational collapse or leave signatures detectable through pulsar timing, cooling observations, or gravitational-wave measurements. With the advent of high-precision multi-messenger observations, including $\sim 2\,M_\odot$ pulsars and constraints from NS mergers\cite{Abbott2017,Abbott2018,Riley19,Miller19,Riley21,Miller21,Fon21,Choud24,Reard24,Ditt24,Salmi22,Salmi24,Vin24}, NSs have become unique astrophysical laboratories for probing DM properties far beyond the reach of terrestrial experiments, see Refs.\cite{Bram25,Grip25} for recent reviews.
Importantly, the impact of DM on NS structure and evolution cannot be fully understood without considering the underlying nuclear microphysics. In particular, SRC among nucleons, which modify momentum distributions and the symmetry energy, play a crucial role in determining the stiffness of the nuclear EOS and transport properties. The interplay between SRC and DM accumulation can therefore influence both the onset of DM-induced softening and the thermal and transport behavior of the star.

\begin{figure}[h!]
\centering
 \includegraphics[height=6.3cm]{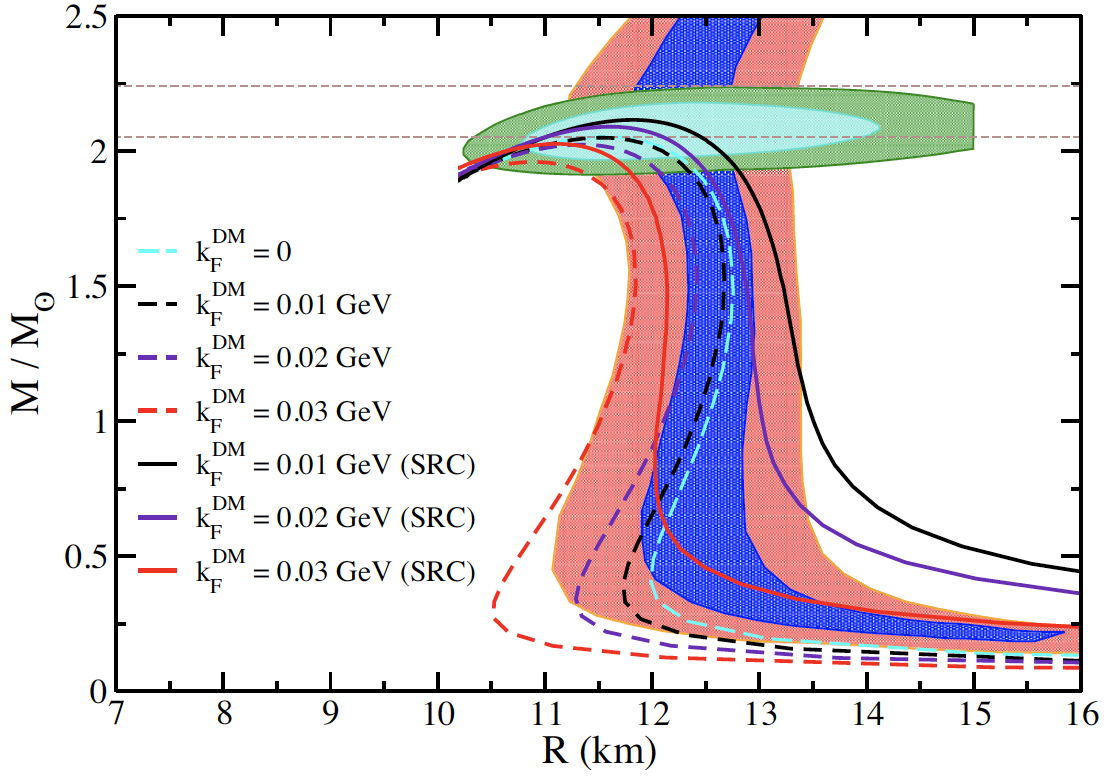}\\[0.25cm]
 \hspace{0.1cm}
 \includegraphics[height=6.3cm]{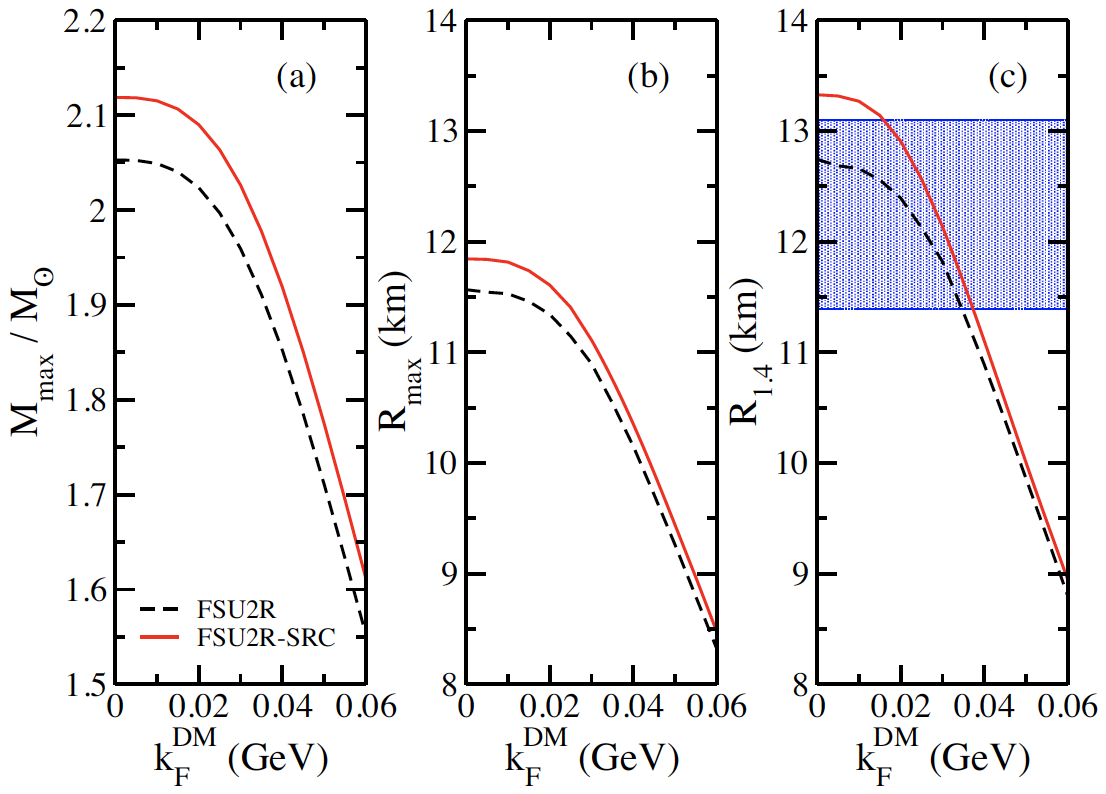}
  \caption{(Color Online). Upper: NS M-R relation in the DM-RMF model with and without SRC, for various $k_{\rm F}^{\rm{DM}}$.
Lower (three panels): the maximum mass, radius, and $R_{1.4}$ of canonical NSs as functions of $k_{\rm F}^{\rm{DM}}$. Observational constraints are plotted by shaded region in these panels. Figures taken from Ref.\cite{Lour22PRD}.}
  \label{fig_DM-NS-12}
\end{figure}

The DM components can be incorporated into the Walecka model by the following Lagrangian\cite{Lour22PRD}:
\begin{empheq}[box=\fbox]{align}
\mathcal{L}_{\rm{DM}}=&\overline{\chi}\left[i\gamma_\mu\partial^\mu-M_\chi-y\rm{H}\right]\chi\notag\\
&+\frac{1}{2}\partial_\mu\rm{H}\partial^\mu\rm{H}-\frac{1}{2}m_{\rm H}^2\rm{H}^2+\frac{fM_{\rm N}}{v}\rm{H}\overline{\psi}\psi,
\end{empheq}
where $\chi$ is the (Fermonic) dark-matter field with its mass $M_\chi$, $\rm{H}$ is the Higgs field, $v\approx246\,\rm{GeV}$ is the Higgs vacuum expectation value and $m_{\rm{H}}\approx125\,\rm{GeV}$ is the mass of Higgs particle; the Higgs field couples to $\chi$ and $\psi$ in the form $\rm{H}\overline{\chi}\chi$ or $\rm{H}\overline{\psi}\psi$, $f$ and $y$ are two coupling constants of the model.
The full model is described by the Lagrangian $\mathcal{L}=\mathcal{L}_{\rm{DM}}+\mathcal{L}_{\rm{Walecka}}$ with the latter given by Eq.\,(\ref{rmf_lag}).
Besides the conventional equations for the Walecka model, a few new equations related to the DM components could be obtained straightforwardly, e.g., the equation of motion for Higgs field $m_{\rm H}^2\rm{H}=y\rho_{\rm s}^{\rm{DM}}+fM_{\rm N}\rho_{\rm s}/v$ where $\rho_{\rm s}^{\rm{DM}}$ is the scalar density for DM:
\begin{equation}
    \rho_{\rm s}^{\rm{DM}}=\langle\overline{\chi}\chi\rangle=\frac{g_\chi M_\chi^\ast}{2\pi^2}
    \int_0^{k_{\rm F}^{\rm{DM}}}\d k\frac{k^2}{\sqrt{k^2+M_\chi^{\ast,2}}},
\end{equation}
where $M_\chi^{\ast}=M_{\chi}-y\rm{H}$ is the effective mass for DM particles; $g_\chi=2$ is the degeneracy factor.
The DM particle Fermi momentum is determined as $k_{\rm F}^{\rm{DM},3}/3\pi^2=\rho_{\rm{DM}}$.
The kinetic energy density and the kinetic pressure of the DM components are similarly obtained
\begin{align}
    \varepsilon_{\rm{DM}}^{\rm{kin}}(\rho)
    =&\frac{g_\chi}{2\pi^2}\int_0^{k_{\rm F}^{\rm{DM}}}\d k{k^2}{\sqrt{k^2+M_\chi^{\ast,2}}},\\
    P_{\rm{DM}}^{\rm{kin}}(\rho)
    =&\frac{g_\chi}{6\pi^2}\int_0^{k_{\rm F}^{\rm{DM}}}\d k\frac{k^4}{\sqrt{k^2+M_\chi^{\ast,2}}}.
\end{align}
For massive DM particles with $M_\chi \gg M_{\rm N}\gg k_{\rm{F}}^{\rm{DM}}$, the pressure contributed by the DM components is vanishingly small $P_{\rm{DM}}^{\rm{kin}}\sim k_{\rm{F}}^{\rm{DM},5}/M_\chi^\ast\approx0$, while the energy density can be significant. As a result, the EOS of NS matter, including both nuclear matter and DM, is effectively softened, leading to reduced maximum masses and smaller radii for a given stellar configuration.
See FIG.\,\ref{fig_DM-NS-12} for the corresponding predictions in the DM-RMF model considering different values of $k_{\rm F}^{\rm{DM}}$\cite{Lour22PRD}, similar issues were also studied in Ref.\cite{Hong23CQG}.
When including the SRC-induced HMT in the nucleon momentum distribution, the maximum NS mass is increased, as shown in FIG.\,\ref{fig_DM-NS-12}. Thus, the DM component and the SRC-HMT together exert a combined influence on the NS M-R relation.

\begin{figure}[h!]
\centering
 \includegraphics[height=6.cm]{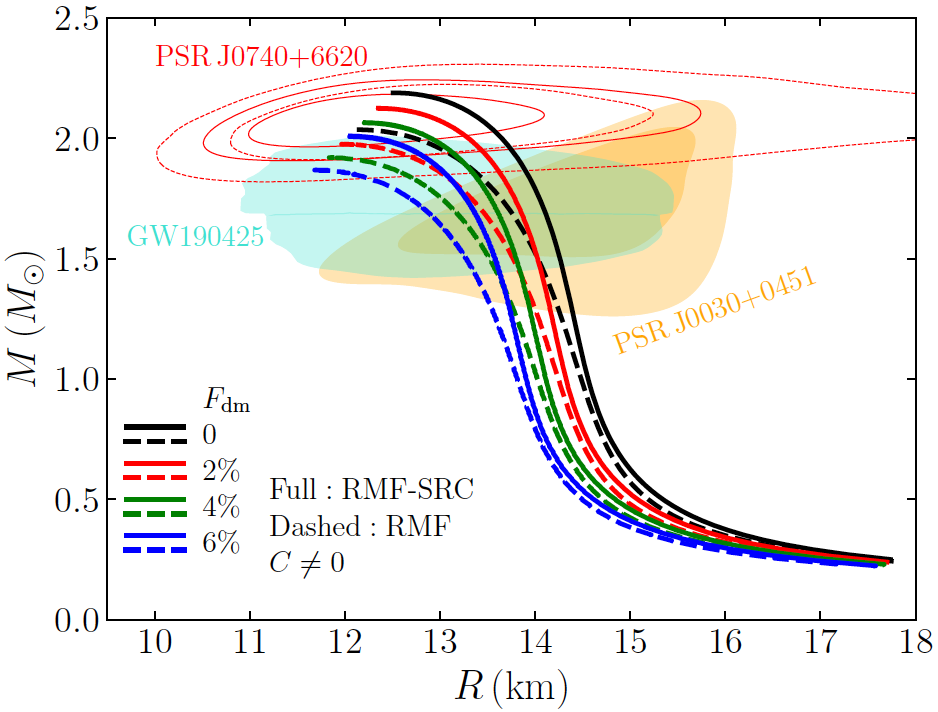}\\[0.25cm]
 \includegraphics[height=6.cm]{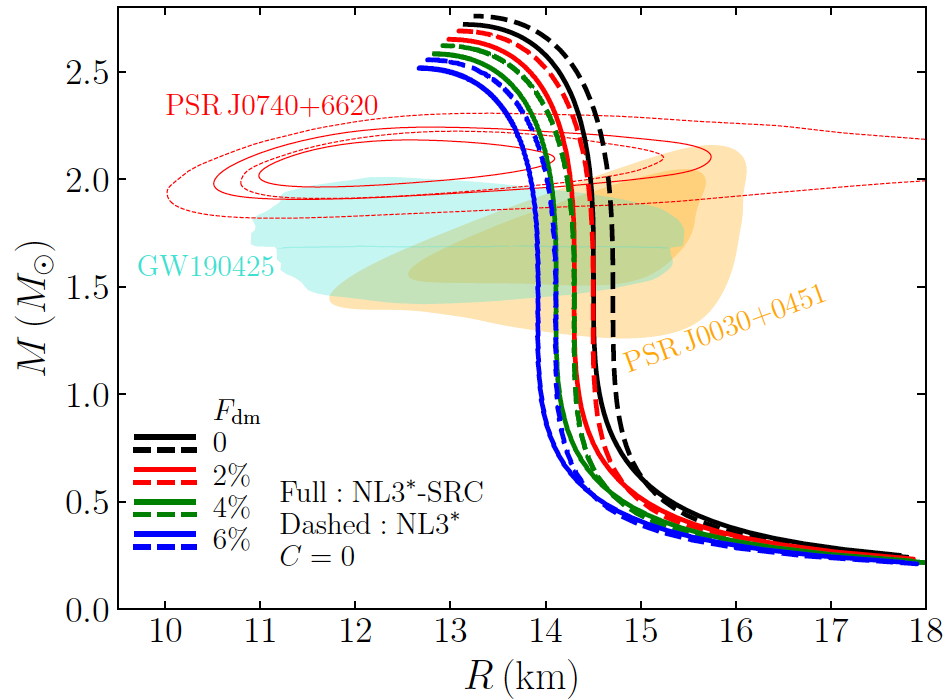}
  \caption{(Color Online). The mixed influence of DM and SRC-HMT on the NS M-R relations in models with (upper) or without (lower) the $\omega$-field self-interaction $\sim C(g_\omega^2\omega_{\mu}\omega^{\mu})^2$.
   Figures taken from Ref.\cite{Lour25xx}.}
  \label{fig_DMRMF}
\end{figure}

In Ref.\cite{Lour22PRD}, a self-interaction term 
\(\sim C (g_\omega^2 \, \omega_\mu \omega^\mu)^2=c_\omega (g_\omega^2 \, \omega_\mu \omega^\mu)^2\) in the Lagrangian\cite{Muller1996NPA} is introduced, see Eq.\,(\ref{rmf_lag}), which dominates the enhancement effects of the SRC-HMT. Later, in Ref.\cite{Lour25xx}, the role of this term is further investigated in the DM-RMF model. In particular, if the self-interaction is absent (as shown in the lower panel of FIG.\,\ref{fig_DMRMF}), the SRC-HMT tends to reduce the maximum NS mass, contrary to the model including the self-interaction term (upper panel).
In these calculations, the NS M-R relation is determined by a two-fluid formalism of the TOV equations: each NS is assumed to contain a fixed dark matter fraction $
F_{\rm{dm}} = {M_{\rm{DM}}}/{M_{\rm{NS}}}$.
The two-fluid TOV equations are solved by specifying the central pressures and energy densities and assuming a proportionality $\ell$ between the dark and visible energy densities\cite{Xiang14PRC}. For each \(\ell\), the resulting fraction \(F_{\rm{dm}}\) is computed, and only the \((M_{\rm{NS}},R)\) pairs satisfying the prescribed fraction are retained. This procedure yields the M-R relation with a fixed dark matter content.
Besides the Higgs field\cite{Lour22PRD}, one can similarly consider a dark vector field \(V^\mu\). In this case, the dark matter component acquires an additional contribution $
2^{-1}({g_V}/{m_V})^2 \rho_{\rm{DM}}^2$, to both the energy density and the pressure, where \(g_V\) and \(m_V\) are the \(V\)-nucleon coupling constant and the mass of the \(V\) field, respectively. 
In this scenario, the DM pressure may no longer be vanishingly small\cite{Lour25xx}.
Besides the NS M-R relations, other observables such as the tidal deformabilities, moment of inertia as well as NS crust properties were also studied in the nonlinear RMF model with both SRC-HMT and DM components, see Ref.\cite{Dutra22MNRAS} for more details.

While models including both DM and SRC-HMT effects can successfully reproduce many of the observed NS properties, some aspects remain less certain. In particular, the physical basis for the assumed DM distribution and interactions in two-fluid RMF models is largely phenomenological, and the detailed mechanism by which DM is captured and accumulates inside NSs is still an open question. Similarly, although SRC-HMT effects modify the nuclear EOS in meaningful ways, the interplay between DM and nucleonic correlations has not yet been explored from a fully microscopic perspective. Therefore, while the combined ``DM''+``SRC-HMT'' framework provides valuable insights and is broadly consistent with astrophysical data, further studies are needed to strengthen the underlying physical justification and clarify the role of DM in dense neutron-rich matter.

\section{Summary, Open Questions and Future Perspectives}\label{SEC_OUTLOOK}

\indent

This review has summarized recent progress in understanding how nucleon short-range correlations and the associated high-momentum tail in the single-nucleon momentum distribution influence the EOS of dense nuclear matter, with broad implications for neutron stars and heavy-ion collisions (HICs). In the latter, SRC-HMT effects manifest through particle production, yield ratios, and collective flow observables. In dense matter existing in NSs and/or HICs, they modify kinetic contributions to the energy density, pressure, and symmetry energy, thereby affecting the stiffness of its EOS. In particular, they affect the core-crust transition density and the internal structure of NSs. We have also examined some of their observational consequences in astrophysics, including mass-radius relations of NSs, tidal deformabilities, and cooling behavior of protoneutron stars, with/without additional particle components such as dark matter. Collectively, these studies emphasize the deep interplay between microscopic nuclear correlations and macroscopic stellar properties. Given our limited knowledge in this highly complicated yet compelling field, the review presented here is inevitably incomplete and may reflect unintended biases on some specific issues.

Although major progress has been achieved in the past decades, many facets of SRCs and HMTs remain uncertain and deserve further investigation. Below we outline several necessary refinements and open problems whose resolution would advance our understanding of SRC-HMT physics in NSs and HICs, as well as in nuclear many-body theory more broadly.

\begin{enumerate}[label=(\alph*),leftmargin=*]

\item \textit{Form of the HMT.}
While the $k^{-4}$ scaling of the HMT is experimentally well established, a fully self-consistent theoretical derivation that incorporates realistic nucleon-nucleon interactions remains lacking. Such a derivation is essential for reliable predictions of SRC effects in finite nuclei, asymmetric nuclear matter, and NSs. Finite-size effects may further modify the HMT in nuclei\cite{Ant88} compared with ultra-cold atomic systems, where Feshbach resonances technique\cite{Chin10RMP} allow tuning of the scattering length. A deeper microscopic understanding of the HMT will strengthen the connection between nuclear correlations and macroscopic observables in NSs and HICs, while enriching our knowledge of strongly correlated quantum systems\cite{AGD}.

\item \textit{Isospin structure of the HMT.}
The isospin dependence of the HMT\cite{Ant07PRC} is critical for describing isospin-sensitive observables in HICs\cite{LCCX18} and for understanding the properties of highly asymmetric matter in NSs. Progress requires both high-precision experiments and theoretical developments. Rare-isotope beam facilities such as FRIB, FAIR, and HIAF will enable systematic studies of SRCs across a broad isospin range, providing important constraints for nuclear theory and astrophysics.

\item \textit{Spectral function and the HVH theorem.}
Deriving the momentum distribution $n(k)$ from the nucleon spectral function $S(\mathbf{k},E)$ and understanding $S(\mathbf{k},E)$ itself\cite{Dick25B} are essential for extending the Hugenholtz--Van Hove (HVH) theorem\cite{Hug58} to cases where SRCs and HMTs are present. At the mean-field level, the nuclear symmetry energy is related to the nucleon effective mass and symmetry potential through $E_{\rm{sym}}(\rho)=k_{\rm F}^2/6M^{\ast}_0 + 2^{-1}U_{\rm{sym}}(\rho)$. Once SRC-induced HMTs are included, this relation becomes nontrivial\cite{Gal58}. Establishing the generalized HVH relation would help constrain the EOS directly from experimental nucleon potentials and elucidate the microscopic origin of EOS uncertainties\cite{AGD}.

\item \textit{Phase-space consistency and momentum-coordinate duality.}
Achieving a self-consistent determination of nucleon momentum and coordinate distributions from the underlying phase-space distribution\cite{Cosyn21PLB} remains a key challenge. An intriguing example is the coexistence of a proton-skin in $k$-space and a neutron-skin in $r$-space\cite{Cai16b}. Progress here will shed new light on neutron-skin thickness\cite{Qiu25,Yue24,Yue22}, probed in PREX\cite{PREX} and CREX\cite{CREX} experiments at JLab, and link it to SRCs measured at various facilities around the world.

\item {\it HMT at supra-saturation densities.} Experimentally, the SRC-induced HMT has been confirmed primarily near nuclear saturation density using finite-nucleus probes\cite{Hen14}. Its generalization to higher densities, relevant for NS interiors\cite{Shapiro1983}, relies on assumptions that remain to be critically examined. In other words, an important open question is whether one can reliably constrain HMT properties, such as the neutron-proton or proton-proton SRC ratio at high densities, using self-consistent theoretical approaches or experimentally validated methods.

\item \textit{Maintaining HMT effects throughout reaction dynamics.}
In HICs, SRC-HMT physics affects the initial state\cite{Reich25,LiP25}, but how to consistently propagate its influence during the full dynamical evolution remains unresolved. A self-consistent treatment (with the off-shell effects) must incorporate HMT influences together with in-medium cross sections, nucleon mean fields, and temperature\cite{Buss12}. This demands new theoretical and computational strategies.

\item {\it Mean free path and the dimer-like picture.} The nucleon momentum distribution $n(k)$ can significantly affect scattering mechanisms in heavy-ion collisions through two key aspects. First, the depletion of low-momentum states and the presence of the HMT modify the statistical restrictions imposed by Pauli blocking, thereby altering phase-space availability during nucleon-nucleon scatterings. Second, the strong tensor-correlations between neutrons and protons may induce an effective ``dimer-like'' behavior, influencing the microscopic collision dynamics beyond the conventional independent-particle picture. As a result, the nucleon mean free path in ANM\cite{Neg81,Fang14PRC}, along with other transport properties such as conductivities and viscosities\cite{SXLi11PRC,XuJ13PLB,CLZhou13PRC,LiuC22,Deng24PPNP,Ma23,Deng22PRC,Deng16PRC}, is expected to exhibit qualitatively new features. 

\item \textit{Three-nucleon SRCs.}
Most previous studies focus on two-body SRCs, but growing evidence for three-nucleon (3N) SRCs\cite{2023EPJA...59..205F,Fa17,Ye18,Sargsian2019PRC,Hen17RMP,Li22Nature,Li24PLB} suggests they may significantly modify the HMT, isospin structure, and transport properties of dense matter. A key challenge is to consistently incorporate 3N-SRCs into existing theoretical models. In particular, understanding how 3N correlations coherently influence transport processes in heavy-ion collisions and the structure, composition, and thermal properties of NSs is an open and promising direction.

\item {\it HMT on neutrino transport.} The SRC-induced HMT may play a significant role in quantum neutrino transport in dense matter environments\cite{Volpe24RMP}, such as NS interiors or core-collapse supernovae\cite{Bur2021,Burrows2013,Burns2020, Burrows1990,Burrows2021,Woosley1986,Janka2012}. By modifying nucleon momentum distributions, SRCs can influence neutrino scattering rates, mean free paths, and energy emission, thereby affecting NS cooling, thermal evolution, and overall neutrino transport phenomena\cite{Yak01,Haen2007}. Developing a self-consistent treatment of these effects could provide deeper insights into dense-matter physics and observable neutrino signals from astrophysical sources.

\item \textit{SRCs and quantum entanglement.}
Although conceptually distinct, SRCs and quantum entanglement\cite{Bulgac2023-a,Bulgac2023-b} both encode forms of non-classical
correlations in nuclear systems. A systematic exploration of their
interplay may uncover new perspectives on the emergence of many-body
nuclear structure and deepen our understanding of correlated quantum
matter\cite{WangSM23,Shang25NST}.

\item {\it Low-dimensional systems.} The role of SRCs and the HMT in low-dimensional or constrained nuclear systems remains largely unexplored. Theoretical studies, such as the $\epsilon$-expansion and renormalization group treatments\cite{Wilson1974}, could provide guidance on how correlations manifest in systems with reduced spatial dimensions. Insights may also be gained by drawing analogies with ultra-cold atomic systems, where advanced experimental techniques allow realizations of quasi-one- and quasi-two-dimensional configurations. In nuclear physics, confined geometries, such as the toroidal structure in $^{12}$C\cite{Siemens1967,Wong1972,Wong1973,Cao2019} or clustered configurations in finite nuclei\cite{WangR25,ZhouB19,HuangBS25}, offer concrete examples of low-dimensional systems where SRCs and high-momentum components may behave differently than in bulk three-dimensional matter.

\item {\it NS asteroseismology.} SRC-HMT effects modify the dense-matter EOS and therefore essentially influence the frequencies and damping times of various NS oscillation modes.
Since asteroseismology\cite{Cox1980,Unno1989,Aerts2010,Aerts21RMP} interprets observed oscillation frequencies to probe NS interiors\cite{Lai1994,Andersson98,Kokkotas1999,Haskell21,Kumar25,Sag20,ZhouY14,Wen13}, incorporating SRC-HMT effects can refine constraints on mass, radius and crust thickness content. Moreover, SRC-induced modifications of the core-crust transition and composition may alter crust-core coupling, mode splitting, and gravitational-wave emission during stellar perturbations or magnetar flares\cite{Yak24}. Understanding these effects is crucial for connecting microscopic correlations to observable NS oscillation properties.

\end{enumerate}

Beyond their role in the EOS and NS structure, SRCs and HMTs shape a wide array of nuclear and astrophysical phenomena. In neutrino-nucleus interactions\cite{Ruso18PPNP,Pandey25,Acker25Nature,Athar23PPNP,Athar2020B}, the spectral function $S(\mathbf{k},E)$, with its SRC-generated high-momentum and high-removal-energy components, significantly affects both charged- and neutral-current responses\cite{Reddy98PRD,Reddy99PRC}. This has implications for oscillation experiments, supernova dynamics\cite{Wilson74nt}, and neutrino transport in NS interiors\cite{Haen2007,Yak01}.

SRCs also play a key role in neutrinoless double-beta ($0\nu\beta\beta$) decay. Their modification of short-range two-body operators affects nuclear matrix elements (NMEs)\cite{Men11PRL,Men14PRC}, the unitarity of the CKM matrix\cite{Cond22PRC}, and neutrino--nucleus scattering\cite{Cuyck16PRC}. SRC-driven nuclear effects also impact neutral-current deep inelastic scattering and may contribute to resolving the NuTeV anomaly\cite{Yang23PRD}. Connections among isospin-dependent SRCs, $0\nu\beta\beta$ NMEs, CE$\nu$NS\cite{Akimov17,Adam25PRL}, and neutron skins\cite{Kos21,Hof20,Huang19PRD,Huang22PRD,Ding24NST,An24,Yan24,MaCW24,ZhouJ24} further illustrate the unifying role of SRC physics across nuclear structure and neutrino physics.

The characteristic $k^{-4}$ momentum tail in nuclear systems parallels that of unitary Fermi gases\cite{Gio08RMP,Blo08RMP}, where its strength is governed by the two-body ``contact''\cite{Tan08-a,Tan08-b,Tan08-c,SLZhang17}. Similar behavior appears in Bose systems\cite{Smith14,Mak14}. The Efimov effect\cite{Efimov1973,Endo25} induces an additional three-body contact in unitary Bose gases, modifying the large-$k$ distribution\cite{Smith14}. Recent work shows that p-wave resonant interactions also exhibit universal behavior\cite{Yos15,MHe16}, suggesting that new HMT scaling laws, such as $k^{-2}$, may emerge at high density where higher partial waves contribute substantially.
Nuclear systems add further richness due to large but finite scattering lengths, tensor forces, and isospin asymmetry. Pure neutron matter at low density approaches the unitary regime, and its energy satisfies universal relations involving the Bertsch parameter $\xi$\cite{Bishop2002}, as well as finite-range parameters $\zeta$ and $\nu$\cite{Bulg05}. These parameters enter Tan's adiabatic sweep theorem and determine the neutron contact. Analytical approaches based on the Nishida--Son $\epsilon$-expansion\cite{Nishida2006} may illuminate their physical origin. Moreover, bosonic degrees of freedom such as pions and kaons may also exhibit short-range contact behavior in dense matter, raising the possibility of HMT signatures in their momentum distributions.

Taken as a whole, these developments reveal a coherent picture: SRCs and HMTs form a microscopic bridge linking electron scattering, dense-matter EOS, neutrino physics, double-beta decay, HIC dynamics, NS phenomenology, and universal features of strongly interacting quantum systems. Progress in any single area, from ab-initio neutron-matter calculations to cold-atom experiments, will have far-reaching implications for our understanding of matter under extreme conditions.

\section*{Acknowledgment}

\indent 

We would like to thank Rui Wang for helpful discussions.
This work was supported in part by the National Natural Science Foundation
of China under contract No. 12547102, the U.S. Department of Energy, Office of Science, under Award Number DE-SC0013702, the CUSTIPEN (China-U.S. Theory Institute for Physics with
Exotic Nuclei) under the US Department of Energy Grant No. DE-SC0009971.

\bibliographystyle{unsrt}


{\small
\begin{spacing}{0.85}
\balance
\bibliography{Ref}
\end{spacing}
}

}
\end{document}